\documentclass[10pt,english,pdftex,a4paper]{book}

\newcommand{\licTitle}{Weigh them all!}
\newcommand{\licAuthor}{Sunny Vagnozzi}
\newcommand{\licYear}{2019}

\newif\ifShowHalfPage
\newif\ifShowLayout
\newif\ifShowGrid
\newif\ifSpaper
\newif\ifSclip
\newif\ifIncludePDFs

  \Scliptrue            
  \ShowHalfPagetrue     

\usepackage{lmodern}

\usepackage[T1]{fontenc}
\usepackage[utf8]{inputenc}
\usepackage{dcolumn}
\usepackage{multirow}

\usepackage{geometry}
\usepackage{color}
\usepackage{babel}
\usepackage{longtable}
\usepackage{float}
\usepackage{mathtools}
\usepackage{enumitem}
\usepackage{amsmath}
\usepackage{amsthm}
\usepackage{amssymb}
\usepackage{mathdots}
\usepackage{stmaryrd}
\usepackage{xcolor}
\usepackage{epigraph}

\makeatletter
\renewcommand{\@chapapp}{}
\newenvironment{chapquote}[2][2em]
  {\setlength{\@tempdima}{#1}%
   \def\chapquote@author{#2}%
   \parshape 1 \@tempdima \dimexpr\textwidth-2\@tempdima\relax%
   \itshape}
  {\par\normalfont\hfill--\ \chapquote@author\hspace*{\@tempdima}\par\bigskip}
\makeatother

\usepackage[unicode=true,colorlinks=true]{hyperref}

\hypersetup{
  pdftitle           = {\licTitle},
  pdfauthor          = {\licAuthor},
  pdfsubject         = {Doctoral Thesis in Theoretical Physics},
  bookmarks          = true,
  bookmarksnumbered  = true,
  bookmarksopen      = true,
  bookmarksopenlevel = 1,
  breaklinks         = true,
  pdfborder          = {0 0 0},
  pdfborderstyle     = {},
  backref            = false,
  colorlinks         = true,
  linkcolor          = blue,
  urlcolor           = blue,
  citecolor          = blue,
  anchorcolor        = green,
  pdfstartview       = {Fit},
  pdfpagelayout      = {TwoPageRight},
  linktocpage, pagebackref, hyperfigures
}

\ifSpaper
  \hypersetup{colorlinks=true}
\fi

\ifSpaper
\fi

\ifShowLayout
  \usepackage[noframe]{showframe} 
  \usepackage{layouts}
  \makeatletter
  \let\FontSize=\f@size
  \makeatother
\fi

\usepackage[{sf,bf}]{titlesec}

\usepackage[margin=10pt,labelfont=bf]{caption} 

\usepackage{cite} 

\pdfmapfile{+rsfso.map}  
\DeclareSymbolFont{rsfso}{U}{rsfso}{m}{n}
\DeclareSymbolFontAlphabet{\mathscr}{rsfso}

\usepackage{xfrac}  

\usepackage{tensind} 
\tensordelimiter{?}

\usepackage{xcolor}
\usepackage{colortbl}

\newcommand{\bSe}{\begin{subequations}} 
\newcommand{\eSe}{\end{subequations}}

\newcommand{\bWe}{\begin{widetext}} 
\newcommand{\eWe}{\end{widetext}}

\allowdisplaybreaks

\usepackage{tikz,amsmath,amssymb,bm,color,cancel}
\usetikzlibrary{shapes,arrows}
\usetikzlibrary{calc}
\usetikzlibrary{positioning}
\usetikzlibrary{decorations.pathreplacing}

\usepackage{lastpage}

\usepackage{fancyhdr}
\usepackage[iso]{datetime}

\fancypagestyle{empty}
{
    \fancyhf{} 
}

\fancypagestyle{plain}
{
    \fancyhf{} 
    \fancyfoot[LE,RO]{\thepage}
    \fancyfoot[RE,LO]{}
    \renewcommand{\headrulewidth}{0pt}
    \renewcommand{\footrulewidth}{0pt}
}

\renewcommand{\headrulewidth}{0pt}
\renewcommand{\footrulewidth}{0pt}

\frontmatter

\usepackage[titletoc]{appendix}

\usepackage[nottoc,notlof,notlot]{tocbibind}

\usepackage{emptypage}

\numberwithin{equation}{chapter}

\numberwithin{figure}{chapter}

\numberwithin{table}{chapter}

\usepackage{pdfpages}
\usepackage{ifoddpage}

\usepackage{tabularx}

\newlength{\InnerEdge}
\newlength{\OuterEdge}
\newcommand{\FrameEdgeColor}{blue!30}
\newcommand{\ShowSubGridColor}{gray!15}
\newcommand{\ShowGridColor}{gray!25}
\newcommand{\ShowGridTextColor}{gray!50}

\ifSpaper
\else
\fi

\ifSpaper
\else
\fi

\newcommand{\MarkFrameEdges}{%
  \begin{tikzpicture}[x=1mm, y=1mm, remember picture, overlay]
      \checkoddpage\ifoddpage
        \draw [dashed,\FrameEdgeColor] 
          ($(current page.south west) + (\InnerEdge,0)$) -- 
          ($(current page.south west) + (\InnerEdge,\paperheight)$);
        \draw [dashed,\FrameEdgeColor] 
          ($(current page.south east) + (-\OuterEdge,0)$) -- 
          ($(current page.south east) + (-\OuterEdge,\paperheight)$);
      \else
        \draw [dashed,\FrameEdgeColor] 
          ($(current page.south east) + (-\InnerEdge,0)$) -- 
          ($(current page.south east) + (-\InnerEdge,\paperheight)$);
        \draw [dashed,\FrameEdgeColor] 
          ($(current page.south west) + (\OuterEdge,0)$) -- 
          ($(current page.south west) + (\OuterEdge,\paperheight)$);
      \fi
  \end{tikzpicture}
}

\newcommand{\ShowGrid}{%
  \begin{tikzpicture}[x=1mm, y=1mm, remember picture, overlay]
      \checkoddpage\ifoddpage
        \draw[step=1,\ShowSubGridColor,very thin] 
            ($(current page.south west) + (0,0)$) grid 
            ($(current page.south west) + (\paperwidth,\paperheight)$);
        \draw[step=10,\ShowGridColor,very thin] 
            ($(current page.south west) + (0,0)$) grid 
            ($(current page.south west) + (\paperwidth,\paperheight)$);
        \foreach \i in {1,...,30} {
            \node [anchor=west,align=right,\ShowGridTextColor] 
            at ($(current page.south west) + (1,\i * 10)$) {$\i$};
        }
        \foreach \i in {1,...,21} {
            \node [anchor=south,align=center,\ShowGridTextColor] 
            at ($(current page.south west) + (\i * 10,1)$) {$\i$};
        }
        \draw [thin,red] (current page.south west)
           -- ($(current page.south west) + (10,10)$);
      \else
        \draw[step=1,\ShowSubGridColor,very thin] 
            ($(current page.south east) + (0,0)$) grid 
            ($(current page.south east) + (-\paperwidth,\paperheight)$);
        \draw[step=10,\ShowGridColor,very thin] 
            ($(current page.south east) + (0,0)$) grid 
            ($(current page.south east) + (-\paperwidth,\paperheight)$);
        \foreach \i in {1,...,30} {
            \node [anchor=east,align=right,\ShowGridTextColor] 
            at ($(current page.south east) + (-1,\i * 10)$) {$\i$};
        }
        \foreach \i in {1,...,21} {
            \node [anchor=south,align=center,\ShowGridTextColor] 
            at ($(current page.south east) + (-\i * 10,1)$) {$\i$};
        }
        \draw [thin,red] (current page.south east)
           -- ($(current page.south east) + (-10,10)$);
      \fi
  \end{tikzpicture}
}

\ifx\ShowFrameColor\unknown\else
  \renewcommand*\ShowFrameColor{\color{red!10}}
  \AddToShipoutPictureBG{\ShowFramePicture} 
  \ifShowGrid
    \AddToShipoutPictureBG{\ShowGrid}
  \fi
  \AddToShipoutPictureFG{\MarkFrameEdges}
\fi

\makeatletter

\newlength{\TopOffset}
\newlength{\ThumbBoxH}
\newlength{\ThumbBoxW}
\newlength{\ThumbBoxX}
\newlength{\ThumbBoxLargeX}
\newlength{\ThumbBoxY}
\newlength{\OuterMargin}

\newcommand{\ThumbBoxColor}{black!30}

\ifSpaper
\else
\fi

\newcommand{\labelPaper}[1]{%
   \let\orglabel\label
   \let\label\@gobble
   \edef\@currentlabel{#1\unskip}%
   \let\label\orglabel
}

\newcounter{PaperSubFolio}

\ifShowLayout
  \tikzset{folioFill/.style={red!10}}
\else
  \tikzset{folioFill/.style={white,fill=white}}
\fi

\makeatother

\makeatletter
\newcommand{\lofwithouttitle}{\@starttoc{lof}}
\newcommand{\lotwithouttitle}{\@starttoc{lot}}
\newcommand{\tocwithouttitle}{\@starttoc{toc}}
\makeatother

\usepackage{tocloft}

\cftpagenumbersoff{part}

\newcommand{\mnu}{M_{\nu}}

\begin{document}

\frontmatter


\pagestyle{empty}
\thispagestyle{empty}

\ifSpaper
  \includepdf[pages = {-}, scale = 0.93, offset = -5mm 1.5mm]
    {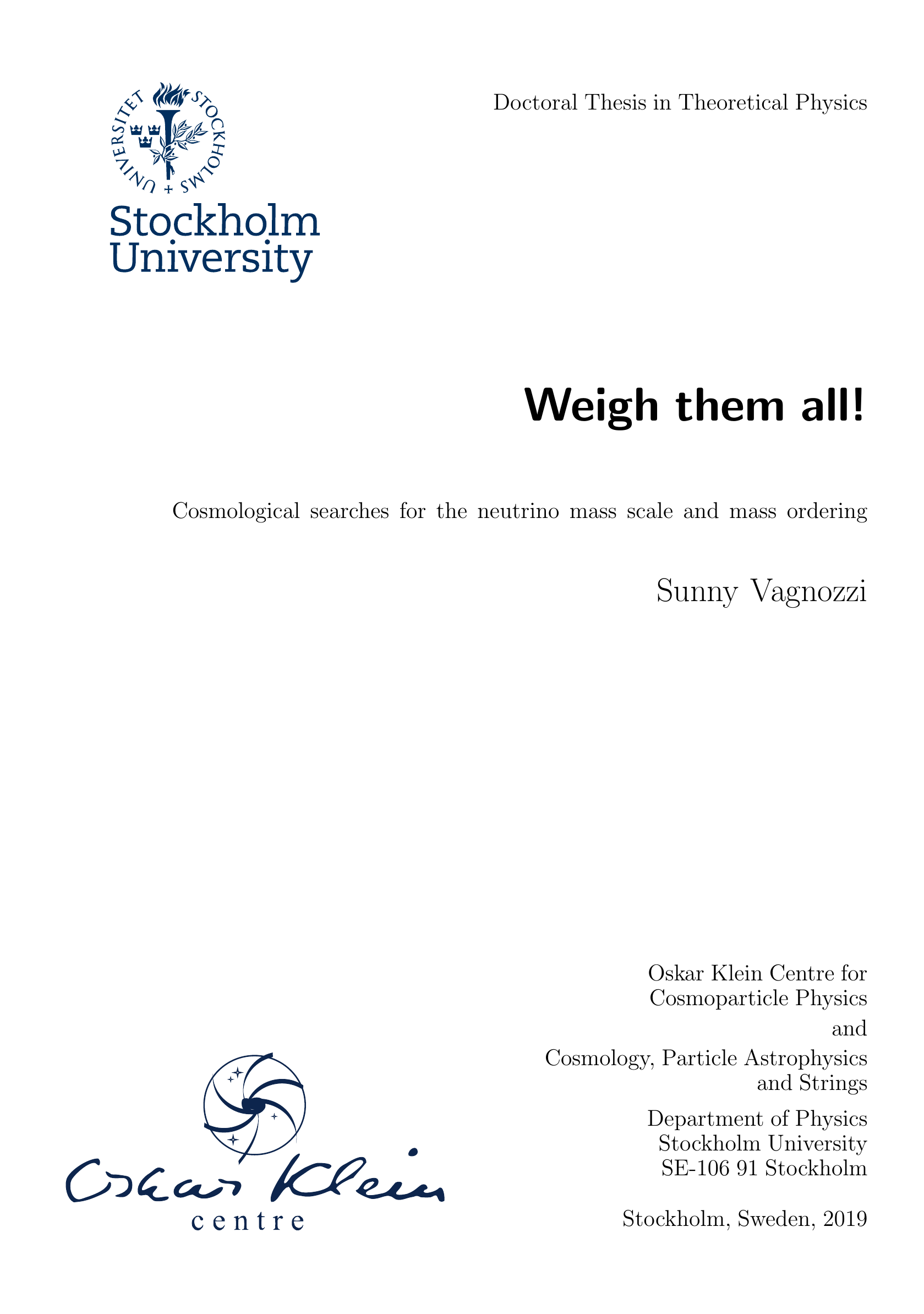}
\else
  \includepdf[pages = {-}, scale = 0.828, offset = -2mm 4mm]
    {title-page.pdf}
\fi


\clearpage
~\vspace{183mm}


\ifSpaper
  \includepdf[pages = {-}, scale = 0.93, offset = -5mm 1.5mm]
    {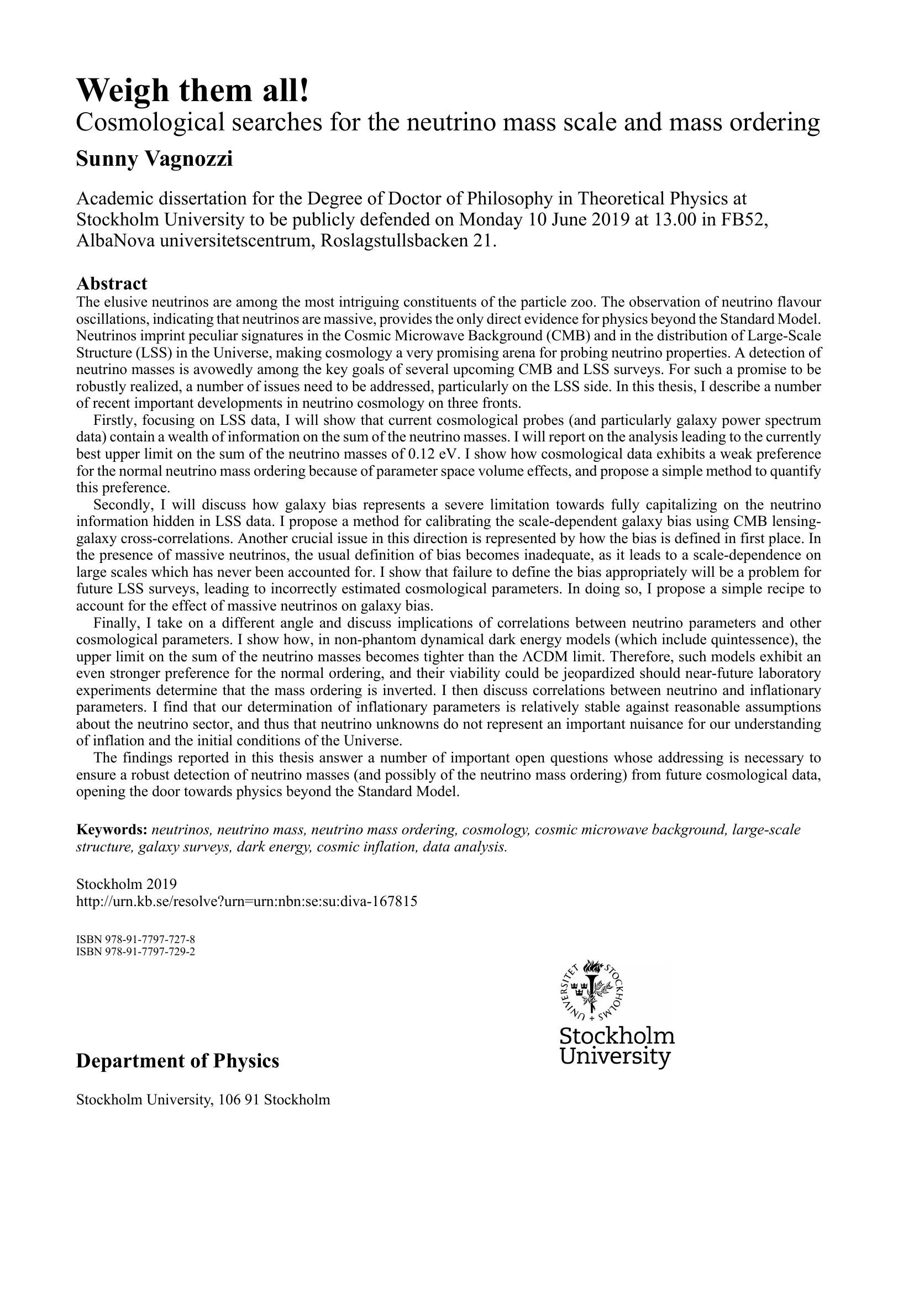}
\else
  \includepdf[pages = {-}, scale = 0.828, offset = -2mm 4mm]
    {page.pdf}
\fi

\noindent
  Printed: \today\\[5mm]
  pp. i--xli,
     1--154,
     \copyright~2019 by \licAuthor\\[1mm]
  Typeset in pdf\LaTeX
  
\ifSpaper
  \includepdf[pages = {-}, scale = 0.93, offset = -5mm 1.5mm]
    {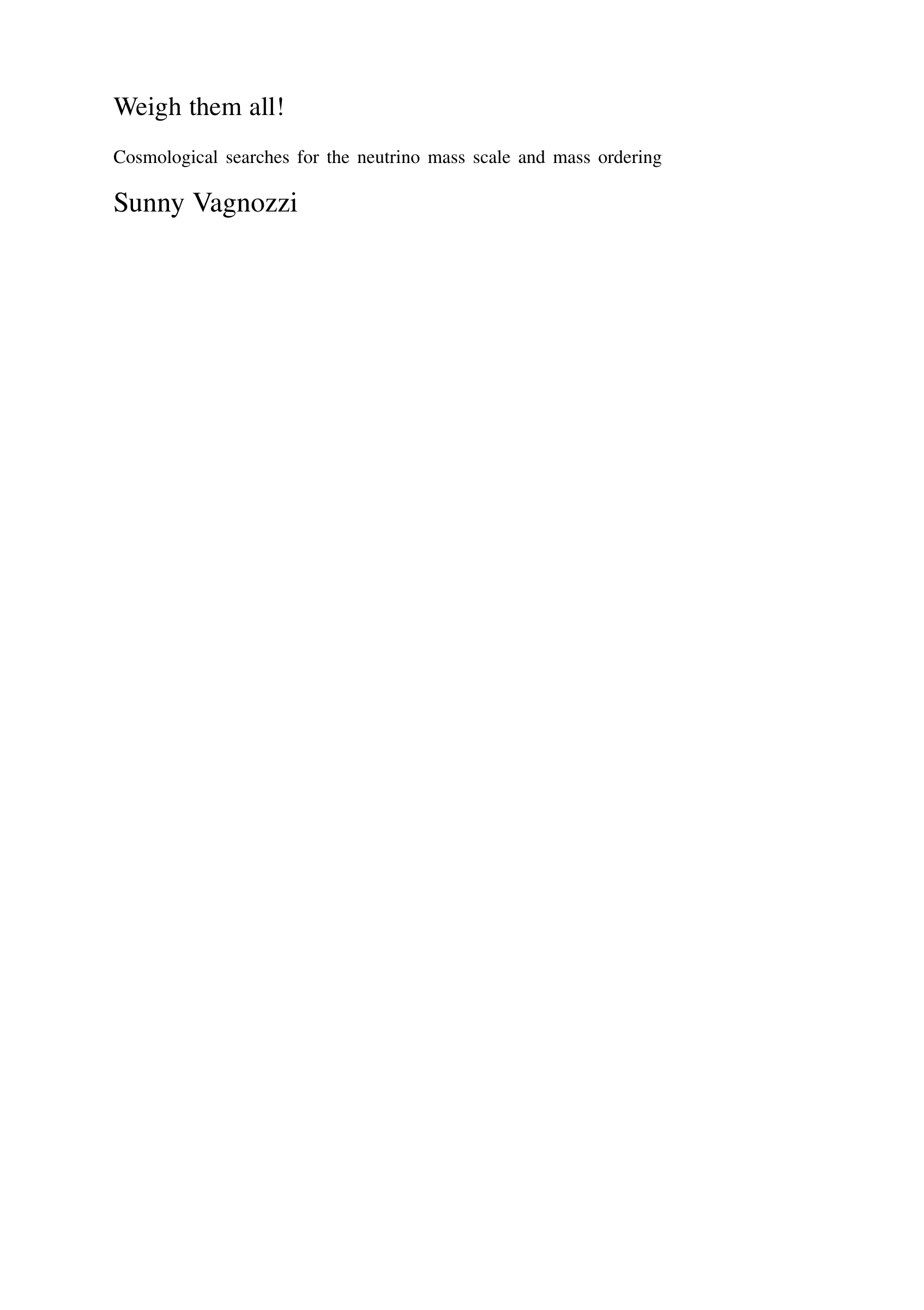}
\else
  \includepdf[pages = {-}, scale = 0.828, offset = -2mm 4mm]
    {half-page.pdf}
\fi

\cleardoublepage


\fancypagestyle{plain}
{
    \fancyhf{}
    \fancyfoot[LE,RO]{\thepage}
    \fancyfoot[RE,LO]{}
    \renewcommand{\headrulewidth}{0pt}
    \renewcommand{\footrulewidth}{0pt}
}

\fancyhf{}
\fancyhead[LE,RO]{\thepage}
\fancyhead[RE]{\sc\leftmark}
\fancyhead[LO]{\sc\rightmark}

\chapter*{}
 
\topskip0pt
\vspace*{\fill}
\large \mbox{}\hfill\textsl{\textit{A Cristina, il mio Universo}} \normalsize
\vspace*{\fill}

\chapter*{}
 
\topskip0pt
\vspace*{\fill}
\noindent \textsl{Neutrinos, they are very small \newline
They have no charge and have no mass \newline
And do not interact at all. \newline
The earth is just a silly ball \newline
To them, through which they simply pass, \newline
Like dustmaids down a drafty hall \newline
Or photons through a sheet of glass. \newline
They snub the most exquisite gas, \newline
Ignore the most substantial wall, \newline
Cold-shoulder steel and sounding brass, \newline
Insult the stallion in his stall, \newline
And, scorning barriers of class, \newline
Infiltrate you and me! Like tall \newline
And painless guillotines, they fall \newline
Down through our heads into the grass. \newline
At night, they enter at Nepal \newline
And pierce the lover and his lass \newline
From underneath the bed—you call \newline
It wonderful; I call it crass.} \vskip 0.5 cm
--\textit{Cosmic Gall}, John Updike (1960) \vskip 1.5 cm

\noindent \textsl{Neutrinos...win the minimalist contest: zero charge, zero radius, and very possibly zero mass} \vskip 0.5 cm
--In \textit{The God Particle: If the Universe is the Answer, What is the Question?}, Leon M. Lederman and Dick Teresi (1993), p. xiii \vskip 1.5 cm

\noindent \textsl{Neutrinos have mass? I didn't even know they were Catholic!} \vskip 0.5 cm
--Robert Langdon to Vittoria Vetra in \textit{Angels and Demons}, Dan Brown (2000), p. 476
\vspace*{\fill}

\chapter*{Abstract}
\addcontentsline{toc}{chapter}{Abstract}

The elusive neutrinos are among the most intriguing constituents of the particle zoo. The observation of neutrino flavour oscillations, indicating that neutrinos are massive, provides the only direct evidence for physics beyond the Standard Model. Neutrinos imprint peculiar signatures in the Cosmic Microwave Background (CMB) and in the distribution of Large-Scale Structure (LSS) in the Universe, making cosmology a very promising arena for probing neutrino properties. A detection of neutrino masses is avowedly among the key goals of several upcoming CMB and LSS surveys. For such a promise to be robustly realized, a number of issues need to be addressed, particularly on the LSS side. In this thesis, I describe a number of recent important developments in neutrino cosmology on three fronts. \newline
Firstly, focusing on LSS data, I will show that current cosmological probes (and particularly galaxy power spectrum data) contain a wealth of information on the sum of the neutrino masses. I will report on the analysis leading to the currently best upper limit on the sum of the neutrino masses of $0.12\,{\rm eV}$. I show how cosmological data exhibits a weak preference for the normal neutrino mass ordering because of parameter space volume effects, and propose a simple method to quantify this preference. \newline
Secondly, I will discuss how galaxy bias represents a severe limitation towards fully capitalizing on the neutrino information hidden in LSS data. I propose a method for calibrating the scale-dependent galaxy bias using CMB lensing-galaxy cross-correlations. Another crucial issue in this direction is represented by how the bias is defined in first place. In the presence of massive neutrinos, the usual definition of bias becomes inadequate, as it leads to a scale-dependence on large scales which has never been accounted for. I show that failure to define the bias appropriately will be a problem for future LSS surveys, leading to incorrectly estimated cosmological parameters. In doing so, I propose a simple recipe to account for the effect of massive neutrinos on galaxy bias. \newline
Finally, I take on a different angle and discuss implications of correlations between neutrino parameters and other cosmological parameters. I show how, in non-phantom dynamical dark energy models (which include quintessence), the upper limit on the sum of the neutrino masses becomes tighter than the $\Lambda$CDM limit. Therefore, such models exhibit an even stronger preference for the normal ordering, and their viability could be jeopardized should near-future laboratory experiments determine that the mass ordering is inverted. I then discuss correlations between neutrino and inflationary parameters. I find that our determination of inflationary parameters is relatively stable against reasonable assumptions about the neutrino sector, and thus that neutrino unknowns do not represent an important nuisance for our understanding of inflation and the initial conditions of the Universe. \newline
The findings reported in this thesis answer a number of important open questions whose addressing is necessary to ensure a robust detection of neutrino masses (and possibly of the neutrino mass ordering) from future cosmological data, opening the door towards physics beyond the Standard Model.

\chapter*{Svensk sammanfattning}
\addcontentsline{toc}{chapter}{Svensk sammanfattning}

De sv\r{a}rf\r{a}ngade neutrinerna \"{a}r bland de mest f\"{a}ngslande best\r{a}ndsdelarna i partiklarnas zoo. Observationen av neutrinooscillationer, som tyder p\r{a} att neutriner har massa, utg\"{o}r det enda direkta beviset f\"{o}r fysik ut\"{o}ver Standardmodellen. Neutriner l\"{a}mnar annorlunda signaturer i den kosmiska bakgrundsstr\r{a}lningen (CMB) och i f\"{o}rdelningen av Universums storskaliga struktur (LSS), vilka g\"{o}r kosmologi till en mycket lovande arena f\"{o}r att unders\"{o}ka neutrinernas egenskaper. Att uppt\"{a}cka neutrinomassorna \"{a}r ocks\r{a} bland de viktigaste m\r{a}len f\"{o}r flera kommande CMB- och LSS-experiment. F\"{o}r att det h\"{a}r l\"{o}ftet ska realiseras m\r{a}ste ett antal fr\r{a}gor behandlas, s\"{a}rskilt p\r{a} LSS-sidan. I denna avhandling beskriver jag ett antal nya viktiga utvecklingar i neutrinokosmologi p\r{a} tre fronter. \newline
F\"{o}r det f\"{o}rsta, med fokus p\r{a} LSS-data, kommer jag att visa att nuvarande kosmologiska unders\"{o}kningar inneh\r{a}ller en stor m\"{a}ngd information om summan av neutrinomassorna. Jag kommer att beskriva analysen som leder till den f\"{o}r n\"{a}rvarande b\"{a}sta \"{o}vre gr\"{a}nsen f\"{o}r summan av neutrinomassorna av $0.12\,{\rm eV}$. Jag visar hur kosmologiska data indikerar en svag preferens f\"{o}r den normala neutrino massordningen (d\"{a}r man har tv\r{a} l\"{a}tta neutriner och en tyngre neutrino, i motsats till den omv\"{a}nda massordningen med en l\"{a}tt neutrino och tv\r{a} tunga neutriner) och l\"{a}gger fram en enkel metod f\"{o}r att kvantifiera denna preferens. \newline
D\"{a}refter kommer jag att diskutera hur galax-``bias'' starkt begr\"{a}nsar m\"{o}jligheten f\"{o}r att fullt ut utnyttja all information om neutriner som \"{a}r dold i LSS-data. Jag l\"{a}gger fram en metod f\"{o}r att kalibrera det skalaberoende galaxbiaset genom att anv\"{a}nda korskorrelationer mellan CMB-linsning och galaxer. En annan viktig fr\r{a}ga i det h\"{a}r sammanhanget \"{a}r hur biaset fr\r{a}n b\"{o}rjan definieras. N\"{a}rvaron av massiva neutriner g\"{o}r den vanliga definitionen av biaset bristf\"{a}llig, eftersom det leder till att galaxbiaset blir skalaberoende p\r{a} stora skalor, n\r{a}got som aldrig tidigare har beaktats. Jag visar att om galaxbiaset inte definieras p\r{a} ett korrekt s\"{a}tt kommer det att ge problem f\"{o}r framtida LSS-experiment, eftersom det leder till felaktiga uppskattningar av de kosmologiska parametrarna. Jag presenterar ocks\r{a} ett enkelt recept f\"{o}r att beakta massiva neutrinernas effekt p\r{a} galaxbiaset. \newline
Slutligen tar jag en annan infallsvinkel och diskuterar konsekvenserna av korrelationer mellan neutrinoparametrar och andra kosmologiska parametrar. Jag visar hur den \"{o}vre gr\"{a}nsen f\"{o}r summan av neutrinomassorna blir str\"{a}ngare \"{a}n $\Lambda$CDMs \"{o}vre gr\"{a}ns i icke-fantom dynamiska m\"{o}rk energi modeller (som inkluderar kvintessens). D\"{a}rf\"{o}r uppvisar s\r{a}dana modeller en \"{a}nnu starkare preferens f\"{o}r den normala massordningen och deras giltighet kan \"{a}ventyras om labexperiment i n\"{a}ra framtid skulle uppt\"{a}cka att massordningen \"{a}r omv\"{a}nd. Till sist diskuterar jag korrelationer mellan neutrino- och inflationsparametrar. Jag finner att v\r{a}ra uppskattningar av inflationsparametrarna \"{a}r relativt stabilt mot rimliga antaganden om neutrinosektorn, och s\r{a}lunda att neutrinerok\"{a}nda inte utg\"{o}r en stor k\"{a}lla till os\"{a}kerhet f\"{o}r v\r{a}r f\"{o}rst\r{a}else av inflationen och av Universums initiala f\"{o}rh\r{a}llandena. \newline
Denna avhandlings resultat svarar p\r{a} viktiga \"{o}ppna fr\r{a}gor vars svar kr\"{a}vs f\"{o}r att s\"{a}kerst\"{a}lla en robust detektion av neutrinomassorna (och m\"{o}jligen av massordningen) fr\r{a}n framtida kosmologiska data, vilket skulle kunna \"{o}ppna d\"{o}rren mot fysik ut\"{o}ver Standardmodellen.

\chapter*{List of Papers}
\addcontentsline{toc}{chapter}{List of papers}
\markboth{List of papers}{List of paper}

\noindent The following papers are included in the thesis. They are referred to by their Roman numerals in the text.

\begin{description}[labelindent=2mm,labelwidth=10mm,leftmargin=12mm,labelsep=0mm]

\item [{I}]
\textbf{Sunny Vagnozzi}, Elena Giusarma, Olga Mena, Katherine Freese, Martina Gerbino, Shirley Ho \& Massimiliano Lattanzi,
\emph{Unveiling $\nu$ secrets with cosmological data: neutrino masses and mass hierarchy},\newline
\href{http://doi.org/10.1103/PhysRevD.96.123503}{Phys.~Rev.~D \textbf{96} (2017) 123503} [\href{https://arxiv.org/abs/1701.08172}{\texttt{arXiv:1701.08172}}]

\item [{II}]
Elena Giusarma, \textbf{Sunny Vagnozzi}, Shirley Ho, Simone Ferraro, Katherine Freese, Rocky Kamen-Rubio \& Kam-Biu Luk,
\emph{Scale-dependent galaxy bias, CMB lensing-galaxy cross-correlation, and neutrino masses},\newline
\href{http://doi.org/10.1103/PhysRevD.98.123526}{Phys.~Rev.~D \textbf{98} (2018) 123526} [\href{https://arxiv.org/abs/1802.08694}{\texttt{arXiv:1802.08694}}]

\item [{III}]
\textbf{Sunny Vagnozzi}, Thejs Brinckmann, Maria Archidiacono, Katherine Freese, Martina Gerbino, Julien Lesgourgues \& Tim Sprenger,
\emph{Bias due to neutrinos must not uncorrect'd go},\newline
\href{http://doi.org/10.1088/1475-7516/2018/09/001}{JCAP \textbf{1809} (2018) 001} [\href{https://arxiv.org/abs/1807.04672}{\texttt{arXiv:1807.04672}}]

\item [{IV}] 
\textbf{Sunny Vagnozzi}, Suhail Dhawan, Martina Gerbino, Katherine Freese, Ariel Goobar \& Olga Mena,
\emph{Constraints on the sum of the neutrino masses in dynamical dark energy models with $w(z)\geq-1$ are tighter than those obtained in $\Lambda$CDM},\newline
\href{http://doi.org/10.1103/PhysRevD.98.083501}{Phys.~Rev.~D \textbf{98} (2018) 083501} [\href{https://arxiv.org/abs/1801.08553}{\texttt{arXiv:1801.08553}}]

\item [{V}] 
Martina Gerbino, Katherine Freese, \textbf{Sunny Vagnozzi}, Massimiliano Lattanzi, Olga Mena, Elena Giusarma \& Shirley Ho,
\emph{Impact of neutrino properties on the estimation of inflationary parameters from current and future observations},\newline
\href{http://doi.org/10.1103/PhysRevD.95.043512}{Phys.~Rev.~D \textbf{95} (2017) 043512} [\href{https://arxiv.org/abs/1610.08830}{\texttt{arXiv:1610.08830}}]

\end{description}

\vspace{2ex}

\noindent The following are additional papers I worked on during my PhD but which did not fit within the storyline of this thesis and are not included. They are quoted as ordinary references in the main text where appropriate.

\begin{description}

\item [VI]
Ratbay Myrzakulov, Lorenzo Sebastiani, \textbf{Sunny Vagnozzi} \& Sergio Zerbini,
\emph{Static spherically symmetric solutions in mimetic gravity: rotation curves and wormholes},
\href{https://doi.org/10.1088/0264-9381/33/12/125005}{Class.~Quant.~Grav. \textbf{33} (2016) 125005} [\href{https://arxiv.org/abs/1510.02284}{\texttt{arXiv:1510.02284}}]

\item [VII]
Guido Cognola, Ratbay Myrzakulov, Lorenzo Sebastiani, \textbf{Sunny Vagnozzi} \& Sergio Zerbini,
\emph{Covariant Ho\v{r}ava-like and mimetic Horndeski gravity: cosmological solutions and perturbations},
\href{https://doi.org/10.1088/0264-9381/33/22/225014}{Class.~Quant.~Grav. \textbf{33} (2016) 225014} [\href{https://arxiv.org/abs/1601.00102}{\texttt{arXiv:1601.00102}}]

\item [VIII]
Robert Foot \& \textbf{Sunny Vagnozzi},
\emph{Solving the small-scale structure puzzles with dissipative dark matter},
\href{https://doi.org/10.1088/1475-7516/2016/07/013}{JCAP \textbf{1607} (2016) 013} [\href{https://arxiv.org/abs/1602.02467}{\texttt{arXiv:1602.02467}}]

\item [IX]
\textbf{Sunny Vagnozzi}, Katherine Freese \& Thomas H. Zurbuchen,
\emph{Solar models in light of new high metallicity measurements from solar wind data},
\href{https://doi.org/10.3847/1538-4357/aa6931}{Astrophys.~J. \textbf{839} (2017) no.~01, 55} [\href{https://arxiv.org/abs/1603.05960}{\texttt{arXiv:1603.05960}}]

\item [X]
Elena Giusarma, Martina Gerbino, Olga Mena, \textbf{Sunny Vagnozzi}, Shirley Ho \& Katherine Freese,
\emph{Improvement of cosmological neutrino mass bounds},
\href{https://doi.org/10.1103/PhysRevD.94.083522}{Phys.~Rev.~D \textbf{94} (2016) 083522} [\href{https://arxiv.org/abs/1605.04320}{\texttt{arXiv:1605.04320}}]

\item [XI]
Lorenzo Sebastiani, \textbf{Sunny Vagnozzi} \& Ratbay Myrzakulov,
\emph{Mimetic gravity: a review of recent developments and applications to cosmology and astrophysics},
\href{https://doi.org/10.1155/2017/3156915}{Adv.~High Energy Phys. \textbf{2017} (2017) 3156915} [\href{https://arxiv.org/abs/1612.08661}{\texttt{arXiv:1612.08661}}]

\item [XII]
Thomas Schwetz, Katherine Freese, Martina Gerbino, Elena Giusarma, Steen Hannestad, Massimiliano Lattanzi, Olga Mena \& \textbf{Sunny Vagnozzi},
\emph{Comment on ``Strong evidence for the normal neutrino hierarchy''},
[\href{https://arxiv.org/abs/1703.04585}{\texttt{arXiv:1703.04585}}]

\item [XIII]
\textbf{Sunny Vagnozzi},
\emph{New solar metallicity measurements},
\href{https://www.mdpi.com/2218-2004/7/2/41}{Atoms \textbf{7} (2019) 41} (\href{http://inspirehep.net/record/1520918/files/1589498_175-180.pdf}{Proceedings of the 51st Rencontres de Moriond, Cosmology Session, ARISF (2016), 175}) [\href{https://arxiv.org/abs/1703.10834}{\texttt{arXiv:1703.10834}}]

\item [XIV]
\textbf{Sunny Vagnozzi},
\emph{Recovering a MOND-like acceleration law in mimetic gravity},
\href{https://doi.org/10.1088/1361-6382/aa838b}{Class.~Quant.~Grav. \textbf{34} (2017) 185006} [\href{https://arxiv.org/abs/1708.00603}{\texttt{arXiv:1708.00603}}]

\item [XV]
Luca Visinelli, Nadia Bolis \& \textbf{Sunny Vagnozzi},
\emph{Brane-world extra dimensions in light of GW170817},
\href{https://doi.org/10.1103/PhysRevD.97.064039}{Phys.~Rev.~D \textbf{97} (2018) 064039} [\href{https://arxiv.org/abs/1711.06628}{\texttt{arXiv:1711.06628}}]

\item [XVI]
Jibitesh Dutta, Wompherdeiki Khyllep, Emmanuel N. Saridakis, Nicola Tamanini \& \textbf{Sunny Vagnozzi},
\emph{Cosmological dynamics of mimetic gravity},
\href{https://doi.org/10.1088/1475-7516/2018/02/041}{JCAP \textbf{1802} (2018) 041} [\href{https://arxiv.org/abs/1711.07290}{\texttt{arXiv:1711.07290}}]

\item [XVII]
Alessandro Casalino, Massimiliano Rinaldi, Lorenzo Sebastiani \& \textbf{Sunny Vagnozzi},
\emph{Mimicking dark matter and dark energy in a mimetic model compatible with GW170817},
\href{https://doi.org/10.1016/j.dark.2018.10.001}{Phys.~Dark Univ. \textbf{22} (2018) 018} [\href{https://arxiv.org/abs/1803.02620}{\texttt{arXiv:1803.02620}}]

\item [XVIII]
Weiqiang Yang, Supriya Pan, Eleonora Di Valentino, Rafael C. Nunes, \textbf{Sunny Vagnozzi} \& David F. Mota,
\emph{Tale of stable interacting dark energy, observational signatures, and the $H_0$ tension},
\href{https://doi.org/10.1088/1475-7516/2018/09/019}{JCAP \textbf{1809} (2018) 018} [\href{https://arxiv.org/abs/1805.08252}{\texttt{arXiv:1805.08252}}]

\item [XIX]
William H. Kinney, \textbf{Sunny Vagnozzi} \& Luca Visinelli,
\emph{The zoo plot meets the swampland: mutual (in)consistency of single-field inflation, string conjectures, and cosmological data}, to appear in \href{https://iopscience.iop.org/article/10.1088/1361-6382/ab1d87}{Class.~Quant.~Grav.} [\href{https://arxiv.org/abs/1808.06424}{\texttt{arXiv:1808.06424}}]

\item [XX]
Peter Ade \textit{et al.} (incl. \textbf{Sunny Vagnozzi}) for the \textit{Simons Observatory} collaboration,
\emph{The Simons Observatory: science goals and forecasts},
\href{https://doi.org/10.1088/1475-7516/2019/02/056}{JCAP \textbf{1902} (2019) 056} [\href{https://arxiv.org/abs/1808.07445}{\texttt{arXiv:1808.07445}}]

\item [XXI]
Luca Visinelli \& \textbf{Sunny Vagnozzi},
\emph{Cosmological window onto the string axiverse and the supersymmetry breaking scale},
\href{https://doi.org/10.1103/PhysRevD.99.063517}{Phys.~Rev.~D \textbf{99} (2019) 063517} [\href{https://arxiv.org/abs/1809.06382}{\texttt{arXiv:1809.06382}}]

\item [XXII]
Alessandro Casalino, Massimiliano Rinaldi, Lorenzo Sebastiani \& \textbf{Sunny Vagnozzi},
\emph{Alive and well: mimetic gravity and a higher-order extension in light of GW170817},
\href{https://doi.org/10.1088/1361-6382/aaf1fd}{Class.~Quant.~Grav. \textbf{36} (2019) 017001} [\href{https://arxiv.org/abs/1811.06830}{\texttt{arXiv:1811.06830}}]

\item [XXIII]
Cosimo Bambi, Katherine Freese, \textbf{Sunny Vagnozzi} \& Luca Visinelli,
\emph{Testing the rotational nature of the supermassive object M87* from the circularity and size of its first image},
submitted to \href{https://journals.aps.org/prd/}{Phys. Rev. D} [\href{https://arxiv.org/abs/1904.12983}{\texttt{arXiv:1904.12983}}]

\end{description}

\chapter*{Acknowledgements}

\addcontentsline{toc}{chapter}{Acknowledgements}
\markboth{Acknowledgements}{Acknowledgements}

My first and foremost thank you goes to my advisor Katie Freese. Working with you has been challenging but also great fun. Thank you for all the energy, experience, and passion you put into training me as a scientist, for always leaving me enormous independence in pursuing my research interests and ideas, and for always pushing me to do my best. Next, I cannot express how much I am grateful to my \textit{de facto} co-advisors Shirley Ho and Olga Mena. Thank you for all the time and passion you put in mentoring me, even though I was not officially your student. I am also extremely grateful to my official co-advisors Lars Bergstr\"{o}m and Joakim Edsj\"{o} for always having their doors open whenever I needed help or advice, and particularly for valuable help both on the scientific and practical sides when preparing for my PhD defense. Thanks are also due to Alessandra Silvestri for having agreed to be my opponent at my PhD defense (and apologies for forcing you to read this beast!).

People often ask me how I managed to write so many papers during my PhD. The honest answer is that I was extremely lucky and privileged to have awesome collaborators. I want to express my huge thanks to two collaborators who stick out particularly among the crowd: Martina Gerbino and Elena Giusarma have been my \textit{de facto} day-to-day mentors, and I could not have asked for better postdocs to mentor me. Thank you for teaching me how science is done in practice, for your infinite patience, and for bringing a bit of italianity (or should I say laziality?) in my everyday work routine.

I thank all my other collaborators and co-authors, whose input has been invaluable in my research and from whom I have learned a great deal. In rigorously alphabetical order (by last name): Maria Archidiacono, Cosimo Bambi, Thejs Brinckmann, Nadia Bolis, Alessandro Casalino, Guido Cognola, Pablo Fern\'{a}ndez de Salas, Suhail Dhawan, Eleonora Di Valentino, Jibitesh Dutta, Mads Frandsen, Simone Ferraro, Ariel Goobar, Steffen Hagstotz, Steen Hannestad, Robert Foot, Rocky Kamen-Rubio, Wompher Khyllep, Will Kinney, Massi Lattanzi, Julien Lesgourgues, Kam-Biu Luk, David Mota, Ratbay Myrzakulov, Rafael Nunes, Supriya Pan, Max Rinaldi, Manos Saridakis, Subir Sarkar, Thomas Schwetz, Lorenzo Sebastiani, Ian Shoemaker, Tim Sprenger, Nicola Tamanini, Luca Visinelli, Weiqiang Yang, Sergio Zerbini, Thomas Zurbuchen, and all my collaborators in the \textit{Simons Observatory} collaboration (especially, again, Martina Gerbino).

My stay at the OKC has been amazing thanks to a large number of people, who have made science an extremely enjoyable adventure. Special thanks to Sebastian Baum, Andrea Chiappo, Adri Duivenvoorden, Pablo Fern\'{a}ndez de Salas, Martina Gerbino, Ariel Goobar, J\'{o}n Gudmundsson, Steffen Hagstotz, Fawad Hassan, Edvard M\"{o}rtsell, Francesco Torsello, Janina Renk, Doug Spolyar, Luca Visinelli, and Axel Widmark for their friendship, company over lunch or a drink, for the good times spent sharing our office (especially Adri and Janina and, for a much shorter time, Sebastian and Francesco), och f\"{o}r att alltid ha haft en \"{o}ppen d\"{o}rr f\"{o}r att diskutera fysik och \"{o}va min svenska (Ariel och Edvard). Jag \"{a}r v\"{a}ldigt tacksam mot Vetenskapsr\r{a}det f\"{o}r att ha gjort det m\"{o}jligt f\"{o}r mig att arbeta i en s\r{a} prestigefylld institution som OKC. And of course I apologize if I inadvertently left someone out!

Ringrazio anche tutti i miei amici e colleghi ``trentini'': Lorenzo Andreoli, Dante Bonolis, Andrea Endrizzi, Lorenzo Festa, Davide Gualdi, Vittorio Ghirardini, Alan Hubert, Paolo Mori, Matteo Puel, Ilenia Salvadori, e Daniela Scardi. L'Universit\`{a} di Trento \`{e} sempre stata e sar\`{a} sempre per me la mia prima ``casa accademica''. Per questo ringrazio Max Rinaldi, Lorenzo Sebastiani, e Sergio Zerbini per avermi sempre fatto sentire bentornato l\`{i}, nonch\'{e} per le molte interessanti discussioni, collaborazioni, e inviti a visitare nel corso di questi anni. Tak ogs\r{a} til Amel Durakovi\'{c} for altid at v\ae re en konstant kilde til ekstremt interessante diskussioner og ideer til projekter (s\r{a}vel som m\ae rkelige dansk-svenske samtaler). And thanks to my Aussie friends Callum Jones, Brian Le, and Alex Millar, and to Vitali Halenka, for your friendship and our many interesting conversations over the years.

There is, of course, life outside of physics. My father, mother, and brother, have been a constant source of unconditional support and encouragement. Thank you so much for all the troubles you had to endure, for always having an open door, and for your being a continuous source of wisdom. This would not have been possible without you. Grazie anche a Claudio e Mariella, Leda e Massimo, ai miei amici d'infanzia Cecilia, Daniele, Davide, Riccardo, Francesco I., e Francesco T., e ai miei futuri suoceri Elisabetta e Mauro, per tutti i bei momenti passati insieme ogni volta che torno in Italia. La mia passione per il violino, e il mio amore incondizionato per la Juve (nonostante in occasione delle due finali di Champions perse mi abbia fatto dannare) e il Latina (una menzione speciale al gruppo MLM) mi hanno aiutato a rimanere sano in tutti questi anni, anche nelle occasioni in cui lo stress da lavoro diventava schiacciante.

E infine, last but absolutely not least, grazie con tutto il cuore alla mia futura moglie Cristina. Grazie per il tuo infinito amore, compagnia, e incrollabile supporto in ogni momento della giornata. Grazie di ogni momento passato insieme, dal primo istante la mattina all'ultimo la sera, e di tutti i momenti che verranno. Grazie di essere stata al mio fianco per tutte le interminabili sere mentre scrivevo questa tesi, e mentre lavoravo per scrivere gli articoli qui inclusi. \`{E} a te che dedico questo lavoro.

\vspace{5mm}

\vspace{5mm}
\noindent \begin{flushright}
\licAuthor\\
Stockholm, \today
\par\end{flushright}

\chapter*{Preface}
\addcontentsline{toc}{chapter}{Preface}
\markboth{Preface}{Preface}

This thesis deals with recent developments in the quest towards using cosmological observations to determine properties of the elusive particles known as neutrinos, with a particular focus on their mass and mass ordering. The fact that neutrinos are massive represents \textit{the only direct evidence for physics beyond the Standard Model}, while the three neutrinos remain to date the only particles of the Standard Model of unknown mass. Disclosing the neutrino mass scale would unlock the door for physics beyond the Standard Model, likely operating at energy scales we can only ever dream of reaching on Earth.

Cosmological observations, particularly observations of the large-scale structure of the Universe, have long been known to have the potential to measure the sum of the neutrino masses. In a very simplified picture, reaching this tremendous achievement would consist of at least two steps. The first step would be to make sure we address a number of difficulties associated with the use of large-scale structure data, or at least keep them under control. The second step would be to actually convince the cosmology and non-cosmology communities that we have genuinely detected neutrino masses, and not something else which can mimic their effect. The papers included in this thesis work towards achieving both the first (Paper~I, Paper~II, and Paper~III) and, at least in part, the second goal (Paper~IV, Paper~V).

The main aim of this thesis is to put the included papers into the broader context for non-experts. The physics required to fully understand the included papers span a very broad range of topics within the field of cosmology, ranging from the complex statistical mechanics (equilibrium and non-equilibrium) underlying the Cosmic Microwave Background and more generally the early Universe, to galaxy bias (a topic of research still very much under development and definitely not as well understood as we would like), dark energy, cosmic inflation, as well as non-cosmology topics such as neutrino oscillation experiments.

With the above in mind, it is certainly not feasible to provide a pedagogical introduction to all these topics, and in most cases the included papers contain introductory sections (written mostly by myself) which are quite self-contained. Therefore, the first part of my thesis will intentionally only provide an introductory review to the topics discussed in the papers, going deeper into the technical details only whenever strictly necessary. Rather, my aim is to focus on providing the context within which the work was done. On the other hand, I aim to make up for this deficiency in depth by providing (or at least attempting to provide) a very broad coverage in my bibliography, wherein the reader will find excellent references for a more in-depth and pedagogical/technical coverage of the topics discussed. The same holds for my results: Chapter~\ref{chap:6} of the thesis itself will only summarize my results, and the interested and expert reader is invited to read the included papers alongside the thesis to get a deeper understanding of the results and their implications.

\chapter*{Thesis plan}
\addcontentsline{toc}{chapter}{Thesis plan}
\markboth{Thesis plan}{Thesis plan}

This thesis is divided into two parts: the first part provides an introduction to the field of cosmology, with a focus on neutrino cosmology, in order to put my work in context. The first part also provides summaries of my work. The second part provides the included papers.

In the first part, Chapter~\ref{chap:1} provides a layman introduction to the current status of cosmology and the importance of neutrinos, setting the scene for the rest of the thesis: ideally, it should be understandable to the general public. Chapter~\ref{chap:2} provides a brief introduction to the Standard Model of particle physics, and a more detailed introduction to the Standard Model of cosmology (the $\Lambda$CDM model). Chapter~\ref{chap:3} provides an overview of a number of concepts in modern cosmology useful for understanding the subsequent Chapters, in particular the thermal history of the Universe. Chapter~\ref{chap:4} presents a review of modern cosmological observations, inevitably biased towards the observations this thesis will focus on: Cosmic Microwave Background (CMB) and Large-Scale Structure (LSS). The same Chapter is devoted to an account of how massive neutrinos impact CMB and LSS observations, and therefore of how one can use the latter to constrain neutrino properties. Chapter~\ref{chap:5} then introduces some basic data analysis and statistics tools widely used in cosmology and, in particular, in deriving the results presented in Chapter~\ref{chap:6}. Finally, Chapter~\ref{chap:7} provides a conclusive summary and outlook on future directions.

The second part provides five included papers. I recommend reading them alongside Chapter~\ref{chap:6}, as they effectively integrate the discussion therein. Paper~I (Chapter~\ref{sec:paper1}) discusses cosmological limits on neutrino masses and the neutrino mass ordering using state-of-the-art datasets, highlighting important issues which need to be addressed if progress is to be made. A better understanding of galaxy bias, and its scale-dependence, is highlighted as a particularly pressing concern. This problem is partially addressed in Paper~II (Chapter~\ref{sec:paper2}), where we propose a new method to calibrate the scale-dependent galaxy bias, based on cross-correlations between CMB lensing and galaxy maps. A related issue is addressed in Paper~III (Chapter~\ref{sec:paper3}), where we highlight the importance of defining the galaxy bias in the presence of massive neutrinos in a meaningful way, a subtlety which had not been appreciated so far. The final two papers deal with the issue of degeneracies, \textit{i.e.} the fact that different cosmological parameters (among which neutrino masses) can have comparable effects on cosmological observations and hence it is sometimes difficult to disentangle the individual effects. As a result, our upper limits on neutrino masses usually degrade when relaxing our assumptions on the underlying cosmological model, and hence our ignorance on other parameters affects what we learn about neutrinos and vice-versa. In Paper~IV (Chapter~\ref{sec:paper4}) we argue that this is not always the case, highlighting an important example where we relax the assumption that dark energy should consist of a simple cosmological constant. Finally, in Paper~V (Chapter~\ref{sec:paper5}) we tackle the reverse problem, namely whether our ignorance of neutrino properties can affect what we learn about the rest of the Universe. We focused on what we learn about cosmic inflation, which supposedly occurred in the very early instants of the Universe and set the initial conditions for the hot Big Bang theory.

\chapter*{Contribution to papers}
\addcontentsline{toc}{chapter}{Contribution to papers}
\markboth{Contribution to papers}{Contribution to papers}

\begin{description}

\item [Paper~I.] The idea for this work came from me, and I designed the entirety of the study. I developed and coded up the BOSS DR12 $P(k)$ likelihood, with help from Elena and Martina. I was also responsible for running all the MCMC chains, with very useful assistance when necessary from Elena and Olga, and wrote the paper myself. I produced Figs.~2, 3, 4, and 5, whereas Elena produced Fig.~1 and Massimiliano produced Figs.~6 and 7. The whole group took part in discussing the methods, interpreting the results, and revising the paper.

\item [{Paper~II.}] The idea for this work came from Shirley. After that, Elena and I developed it to its final version, contributing in equal amount and benefiting from many illuminating discussions with Simone. In particular, I developed and coded up the $C_{\ell}^{\kappa g}$ likelihood for the CMB lensing-galaxy cross-correlation measurements, while Elena ran all the MCMC chains and produced the plots. I wrote most of the paper myself, with Elena taking care of the rest of the writing. The whole group took part in discussing the methods, interpreting the results, and revising the paper.

\item [{Paper~III.}] The idea for this work came from me, and I designed most of the study in collaboration with Thejs and Martina. The mock Euclid likelihood had already been developed by Tim, whereas Maria developed the modified version of \texttt{CLASS} to calculate $P_{cb}$. The MCMC chains were run by myself and Thejs, and I produced all the plots. I wrote most of the paper myself, with Thejs and Julien taking care of small portions of the writing. The whole group took part in discussing the methods, interpreting the results, and revising the paper.

\item [{Paper~IV.}] The idea for this work came from me upon discussing with Suhail, and I designed most of the study in collaboration with Suhail and Martina. I ran all the MCMC chains, produced all the plots, and wrote the paper myself. The whole group took part in discussing the methods, interpreting the results (with important contributions from Katie, Ariel, and Olga), and revising the paper.

\item [{Paper~V.}] The idea for this work came from Katie after several discussions with Martina and myself. After that, it was Martina and I who developed it to its final version. The MCMC chains where run by myself and Martina, and Martina produced all the plots. I produced the theoretical predictions for the inflationary models in Figs.~7, 8, 9, 10, 11, and 12. Martina wrote most of the first two-thirds of the paper, while I wrote most of the latter third. Olga and Massimiliano proposed the idea of looking at low-reheating scenarios and making forecasts for future data. The whole group took part in discussing the methods, interpreting the results, and revising the paper.

\end{description}

\chapter*{Abbreviations}
\addcontentsline{toc}{chapter}{Abbreviations}
\markboth{Abbreviations}{Abbreviations}

\noindent \bgroup\def\arraystretch{1.2}%
\begin{longtable}[l]{ll}
  
BAO & Baryon Acoustic Oscillations \tabularnewline
BBN & Big Bang Nucleosynthesis \tabularnewline
BE & Bose-Einstein \tabularnewline
BOSS & Baryon Oscillation Spectroscopic Survey \tabularnewline
BSM & Beyond the Standard Model \tabularnewline
CDM & Cold Dark Matter \tabularnewline
CKM & Cabibbo-Kobayashi-Maskawa \tabularnewline
C.L. & Confidence level \tabularnewline
CMB & Cosmic Microwave Background \tabularnewline
CNB & Cosmic Neutrino Background \tabularnewline
CPL & Chevallier-Polarski-Linder \tabularnewline
COBE & Cosmic Background Explorer \tabularnewline
DDE & Dynamical dark energy \tabularnewline
DE & Dark Energy \tabularnewline
DES & Dark Energy Survey \tabularnewline
DESI & Dark Energy Spectroscopic Instrument \tabularnewline
DM & Dark Matter \tabularnewline
DR & Data Release \tabularnewline
DUNE & Deep Underground Neutrino Experiment \tabularnewline
eBOSS & Extended Baryon Oscillation Spectroscopic Survey \tabularnewline
EISW & Early integrated Sachs-Wolfe \tabularnewline
EoS & Equation of state \tabularnewline
EW & Electro-weak \tabularnewline
FD & Fermi-Dirac \tabularnewline
FIRAS & Far Infrared Absolute Spectrophotometer \tabularnewline
FKP & Feldman-Kaiser-Peacock \tabularnewline
FLRW & Friedmann-Lema\^{i}tre-Robertson-Walker \tabularnewline
GR & General Relativity \tabularnewline
GW & Gravitational wave \tabularnewline
HFI & High Frequency Instrument \tabularnewline
\texttt{IO} & Inverted neutrino mass ordering \tabularnewline
ISW & Integrated Sachs-Wolfe \tabularnewline
KamLAND & Kamioka Liquid Scintillator Antineutrino Detector \tabularnewline
LFI & Low Frequency Instrument \tabularnewline
LISW & Late integrated Sachs-Wolfe \tabularnewline
LSS & Large-scale structure \tabularnewline
LSST & Large Synoptic Space Telescope \tabularnewline
MCMC & Markov Chain Monte Carlo \tabularnewline
MGS & Main Galaxy Sample \tabularnewline
MSW & Mikheyev-Smirnov-Wolfenstein \tabularnewline
NISDB & Neutrino-induced scale-dependent bias \tabularnewline
\texttt{NO} & Normal neutrino mass ordering \tabularnewline
NO$\nu$A & Neutrinos at the main injector off-axis $\nu_e$ appearance \tabularnewline
NPDDE & Non-phantom dynamical dark energy \tabularnewline
PCA & Principal component analysis \tabularnewline
PMNS & Pontecorvo-Maki-Nakagawa-Sakata \tabularnewline
QCD & Quantum chromodynamics \tabularnewline
RSD & Redshift-space distortions \tabularnewline
SDSS & Sloan Digital Sky Survey \tabularnewline
SM & Standard Model of Particle Physics\tabularnewline
SNe1a & Type 1a Supernovae \tabularnewline
SNO & Sudbury Neutrino Observatory \tabularnewline
SPHEREx & Spectro-Photometer for the History of the Universe, Epoch of Reionization, \tabularnewline
 & and Ices Explorer \tabularnewline
T2K & Tokai to Kamioka \tabularnewline
UV & Ultraviolet \tabularnewline
WFIRST & Wide Field Infrared Survey Telescope \tabularnewline
WMAP & Wilkinson Microwave Anisotropy Probe \tabularnewline
$\Lambda$CDM & $\Lambda$-cold dark matter (standard model of cosmology) \tabularnewline
2dfGRS & 2-degree field galaxy redshift survey \tabularnewline
6dFGS & 6-degree field galaxy survey \tabularnewline

\end{longtable}\egroup

\chapter*{Notation}
\addcontentsline{toc}{chapter}{Notation}
\markboth{Notation}{Notation}

Certain symbols have more than one meaning, which depends on the context. These symbols are marked by ``(\textbf{context})''

\noindent \bgroup\def\arraystretch{1.2}%
\begin{longtable}[l]{ll}

$a$ & Scale factor/scale-independent bias factor (\textbf{context}) \tabularnewline
$a_{lm}$ & Coefficients of the decomposition of $\Theta$ in spherical harmonics \tabularnewline
$a_{\rm nr}$ & Scale factor at $z_{\rm nr}$ \tabularnewline
$a_0$ & Scale factor today (usually normalized to $1$) \tabularnewline
$A_L$ & Phenomenological parameter governing the amplitude of CMB lensing \tabularnewline
$A_s$ & Amplitude of primordial scalar power spectrum \tabularnewline
$b$ & Galaxy bias \tabularnewline
$b_{\rm auto}$ & Galaxy bias in auto-correlation \tabularnewline
$b_{cb}$ & Galaxy bias defined with respect to the cold dark matter+baryons field \tabularnewline
$b_{\rm cross}$ & Galaxy bias in cross-correlation \tabularnewline
$B_{ij}$ & Bayes factor of model $i$ with respect to model $j$ \tabularnewline
$b_m$ & Galaxy bias defined with respect to the total matter field \tabularnewline
$c$ & Scale-dependent bias factor in cross-correlation \tabularnewline
$c_s$ & Speed of sound \tabularnewline
$C_{\ell}^{BB}$ & CMB B-mode polarization anisotropy angular power spectrum \tabularnewline
$C_{\ell}^{EE}$ & CMB E-mode polarization anisotropy angular power spectrum \tabularnewline
$C_{\ell}^{TE}$ & CMB temperature-E-mode polarization anisotropy angular cross-power spectrum \tabularnewline
$C_{\ell}^{TT}$ & CMB temperature anisotropy angular power spectrum \tabularnewline
$C_{\ell}^{\kappa g}$ & CMB lensing convergence-galaxy angular cross-power spectrum \tabularnewline
$C_{\ell}^{\phi\phi}$ & CMB lensing potential power spectrum \tabularnewline
$c_{\nu}$ & Neutrino speed \tabularnewline
${\cal C}$ & Collision operator \tabularnewline
$d$ & Scale-dependent bias factor in auto-correlation \tabularnewline
$\boldsymbol{d}$ & Data \tabularnewline
$d_R^i$ & Right-handed down quark singlet \tabularnewline
$d_V$ & Volume distance \tabularnewline
$d\sigma_T/d\Omega$ & Thomson scattering differential cross section \tabularnewline
$D_{\ell}$ & $\ell(\ell+1)C_{\ell}$ \tabularnewline
$e_R^i$ & Right-handed electron singlet \tabularnewline
${\cal E}(\boldsymbol{d})$ & Bayesian evidence/marginal likelihood \tabularnewline
$E(z)$ & Normalized expansion rate $E(z) \equiv H(z)/H_0$ \tabularnewline
$f$ & Distribution function \tabularnewline
$f_{cb}$ & Growth rate of the cold dark matter+baryons power spectrum \tabularnewline
$f_m$ & Growth rate of the matter power spectrum \tabularnewline
$f_{\nu}$ & Fraction of the matter density parameter in neutrinos $f_{\nu} \equiv \Omega_{\nu}/\Omega_m$ \tabularnewline
$g_i$ & Internal degrees of freedom of species $i$ \tabularnewline
$g_{\star}$ & Effective number of relativistic degrees of freedom \tabularnewline
$g_{\star}^s$ & Effective number of entropy degrees of freedom \tabularnewline
$G_F$ & Fermi constant \tabularnewline
$G_{\mu \nu}$ & Einstein tensor \tabularnewline
$h$ & Reduced Hubble constant \tabularnewline
$H$ & Hubble parameter at a given redshift/neutral Hydrogen (\textbf{context}) \tabularnewline
$H_0$ & Hubble constant \tabularnewline
$k$ & FLRW metric curvature/wavenumber (\textbf{context}) \tabularnewline
$k_{\rm eq}$ & Wavenumber of perturbation entering the horizon at $z_{\rm eq}$ \tabularnewline
$k_{\rm fs}$ & Neutrino free-streaming wavenumber \tabularnewline
$k_n$ & Wavenumber of the $n$-th CMB acoustic peak \tabularnewline
$k_{\rm nr}$ & Wavenumber of perturbation entering the horizon at $z_{\rm nr}$ \tabularnewline
$k_{sd}$ & Wavenumber at which scale-dependent bias becomes important \tabularnewline
$L_L^i$ & Left-handed lepton doublet \tabularnewline
${\cal L}$ & Liouville operator \tabularnewline
${\cal L}(\boldsymbol{d} \vert \boldsymbol{\theta})$ & Likelihood \tabularnewline
${\cal L}_{\rm SM}$ & Standard Model Lagrangian \tabularnewline
$\ell$ & Multipole \tabularnewline
$\ell_n$ & Multipole of the $n$-th CMB acoustic peak \tabularnewline
$m_i$ & Mass of species $i$ \tabularnewline
$m_{\rm light}$ & Mass of lightest neutrino eigenstate \tabularnewline
$m_s^{\rm eff}$ & Effective sterile neutrino mass \tabularnewline
$M_{\nu}$ & Sum of the three active neutrino masses \tabularnewline
$n_e$ & Number density of free electrons \tabularnewline
$n_i$ & Number density of species $i$ \tabularnewline
$n_{\rm run}$ & Running of the scalar spectral index $dn_s/d\ln k$ \tabularnewline
$n_{\rm runrun}$ & Running of the running of the scalar spectral index $dn_{\rm run}/d\ln k$ \tabularnewline
$n_s$ & Tilt of primordial scalar power spectrum (scalar spectral index) \tabularnewline
$N_{\rm eff}$ & Effective number of relativistic degrees of freedom \tabularnewline
$N_{\star}$ & Number of \textit{e}-folds of cosmic inflation \tabularnewline
$p$ & Momentum/probability (\textbf{context}) \tabularnewline
$p(\boldsymbol{\theta} \vert \boldsymbol{d})$ & Posterior distribution \tabularnewline
$P_{cb}(k)$ & Cold dark matter+baryons power spectrum \tabularnewline
$P_i$ & Pressure of species $i$ \tabularnewline
$P(k)$ & Matter power spectrum \tabularnewline
$P_g(k)$ & Galaxy power spectrum \tabularnewline
$P_{{\rm HF}\nu}$(k) & Non-linear power spectrum from \texttt{Halofit} calibrated to massive neutrinos \tabularnewline
$P_{mg}(k)$ & Matter-galaxy cross-power spectrum \tabularnewline
$P_{\rm prim}(k)$ & Primordial power spectrum of matter fluctuations \tabularnewline
$P_{\cal R}(k)$ & Primordial power spectrum of ${\cal R}$ \tabularnewline
${\cal P}_{\cal R}$ & Dimensionless primordial power spectrum of ${\cal R}$ \tabularnewline
$P^{\rm shot}$ & Shot noise \tabularnewline
${\cal P}(\boldsymbol{\theta})$ & Prior distribution \tabularnewline
$Q_L^i$ & Left-handed quark doublet \tabularnewline
$q(\boldsymbol{\theta^{\star}} \vert \boldsymbol{\theta})$ & Proposal distribution for Metropolis-Hastings algorithm \tabularnewline
$R$ & Baryon-to-photon momentum density ratio \tabularnewline
$r$ & Tensor-to-scalar ratio evaluated at the pivot scale $k=0.05\,{\rm Mpc}^{-1}$ \tabularnewline
$r_d$ & Damping scale \tabularnewline
$r_{\rm fs}$ & Neutrino free-streaming horizon \tabularnewline
$r_s$ & Comoving sound horizon \tabularnewline
$s_i$ & Entropy density of species $i$ \tabularnewline
$t$ & Time \tabularnewline
$T$ & Temperature of the Universe (photon temperature) \tabularnewline
$T_{\rm CMB}$ & CMB temperature today \tabularnewline
$T(k)$ & Transfer function \tabularnewline
$T_{\mu \nu}$ & Stress-energy tensor \tabularnewline
$T_{\nu}$ & Effective neutrino temperature \tabularnewline
$T_{\nu,{\rm dec}}$ & Neutrino decoupling temperature \tabularnewline
$u_R^i$ & Right-handed up quark singlet \tabularnewline
$U_{ij}$ & PMNS matrix \tabularnewline
$w$ & Dark energy equation of state \tabularnewline
$w_0$ & Dark energy EoS today (CPL parametrization) \tabularnewline
$w_a$ & Minus derivative of dark energy EoS with respect to scale factor (CPL parametrization) \tabularnewline
$W^{\kappa}$ & Kernel for CMB lensing \tabularnewline
$Y_{lm}$ & Spherical harmonics \tabularnewline
$Y_p$ & Primordial Helium fraction \tabularnewline
$z$ & Redshift \tabularnewline
$z_{\rm dec}$ & Redshift of decoupling \tabularnewline
$z_{\rm drag}$ & Redshift of baryon drag \tabularnewline
$z_{\rm eff}$ & Effective redshift \tabularnewline
$z_{\rm eq}$ & Redshift of matter-radiation equality \tabularnewline
$z_{\rm nr}$ & Redshift of neutrino non-relativistic transition \tabularnewline
$z_{\rm re}$ & Redshift of reionization \tabularnewline
$z_{\Lambda}$ & Redshift of matter-$\Lambda$ equality \tabularnewline
$\alpha$ & $\alpha \equiv [1+7/8(4/11)^{4/3}N_{\rm eff}] \approx (1+0.2271N_{\rm eff})$ \tabularnewline
$\Gamma$ & Reaction rate \tabularnewline
$\delta$ & Dirac Delta \tabularnewline
$\delta_i$ & Overdensity of species $i$ \tabularnewline
$\Delta m_{21}^2$ & Solar mass-squared splitting \tabularnewline
$\vert \Delta m_{31}^2 \vert$ & Atmospheric mass-squared splitting \tabularnewline
$\eta$ & Baryon-to-photon ratio \tabularnewline
$\boldsymbol{\theta}$ & Parameter vector \tabularnewline
$\theta_d$ & Angular size of the damping scale \tabularnewline
$\theta_n$ & Angular size of the $n$-th CMB acoustic peak \tabularnewline
$\theta_s$ & Angular size of the first CMB acoustic peak \tabularnewline
$\Theta$ & CMB temperature anisotropies/Heaviside step function (\textbf{context}) \tabularnewline
$\kappa$ & CMB lensing convergence \tabularnewline
$\lambda$ & Wavelength \tabularnewline
$\lambda_{\rm fs}$ & Neutrino free-streaming scale \tabularnewline
$\Lambda$ & Cosmological constant \tabularnewline
$\nu_i$ & Neutrino mass eigenstates ($i=1,2,3$) \tabularnewline
$\nu_{\alpha}$ & Neutrino flavour eigenstates ($\alpha=e,\mu,\tau$) \tabularnewline
$\xi(r)$ & Galaxy 2-point correlation function \tabularnewline
$\rho_{\rm crit}$ & Critical energy density of the Universe today \tabularnewline
$\rho_i$ & Energy density of species $i$ \tabularnewline
$\sigma_T$ & Thomson scattering cross section \tabularnewline
$\sigma_8$ & Amplitude of matter fluctuations averaged on a sphere of radius $8\,h^{-1}{\rm Mpc}$ \tabularnewline
$\tau$ & Optical depth to reionization \tabularnewline
$\phi$ & Inflaton/gravitational potential/lensing potential/quintessence field (\textbf{context}) \tabularnewline
$\Phi$ & Higgs doublet \tabularnewline
$\chi$ & Comoving distance to a given redshift \tabularnewline
$\chi_h$ & Comoving particle horizon at a given redshift \tabularnewline
$\chi_{\star}$ & Comoving distance to $z_{\rm dec}$ \tabularnewline
$\Psi$ & Gravitational potential \tabularnewline
$\omega_b$ & Physical density parameter of baryons \tabularnewline
$\omega_c$ & Physical density parameter of cold dark matter \tabularnewline
$\omega_k$ & Physical density parameter associated to curvature \tabularnewline
$\omega_m$  & Physical density parameter of matter \tabularnewline
$\omega_r$ & Physical density parameter of radiation \tabularnewline
$\omega_{\gamma}$ & Physical density parameter of photons \tabularnewline
$\omega_{\nu}$ & Physical density parameter of neutrinos \tabularnewline
$\omega_{\Lambda}$ & Physical density parameter of $\Lambda$ \tabularnewline
$\Omega_b$ & Density parameter of baryons \tabularnewline
$\Omega_c$ & Density parameter of cold dark matter \tabularnewline
$\Omega_k$ & Density parameter associated to curvature \tabularnewline
$\Omega_m$ & Density parameter of matter \tabularnewline
$\Omega_r$ & Density parameter of radiation \tabularnewline
$\Omega_{\gamma}$ & Density parameter of photons \tabularnewline
$\Omega_{\nu}$ & Density parameter of neutrinos \tabularnewline
$\Omega_{\Lambda}$ & Density parameter of $\Lambda$ \tabularnewline

\end{longtable}\egroup

\cleardoublepage

\pdfbookmark[1]{Table of Contents}{indexname}

\chapter*{Contents}
\markboth{Contents}{Contents}
\addcontentsline{toc}{chapter}{Contents}
\tocwithouttitle

\chapter*{Illustrations}
\markboth{Illustrations}{Illustrations}
\addcontentsline{toc}{chapter}{Illustrations}
\vspace{0.2em}

\subsection*{List of Figures}\vspace{-0.5em}
\lofwithouttitle\vspace{0.2em}

\subsection*{List of Tables}\vspace{-0.5em}
\lotwithouttitle

\cleardoublepage

\label{lastPageOfFrontMatter}
\cleardoublepage

\renewcommand{\thepage}{\roman{page}}
\setcounter{page}{1}

\renewcommand{\thechapter}{\arabic{chapter}}
\setcounter{chapter}{0}

\mainmatter

\pagestyle{fancy}
\newcommand{\ChapterThumb}{}

\fancypagestyle{plain}
{
    \fancyhf{}
    \fancyfoot[LE,RO]{\thepage}
    \fancyfoot[RE,LO]{\small\textcolor{black!35}{
        \ifSpaper\else ~\\[-1ex] \fi
        \textcolor{black!20}{\rule{75mm}{0.4pt}}~\\
        \licAuthor,
        \emph{\licTitle},
        SU \licYear
        }
    }
    \renewcommand{\headrulewidth}{0pt}
    \renewcommand{\footrulewidth}{0pt}
    \fancyhead[CE]{\ChapterThumb}
    \fancyhead[CO]{\ChapterThumb}
}

\chapter{Introduction}
\label{chap:1}

\begin{chapquote}{Wolfgang Pauli (after having postulated the existence of the neutrino, 1930)}
``I have done a terrible thing, I have postulated a particle that cannot be detected.''
\end{chapquote}

\section{Cosmology, the dark Universe, and neutrinos}
\label{sec:cosmology}

\textit{What are we made of}? \textit{Where do we come from}? \textit{Where are we going}? These are probably among the most fundamental questions one can come up with, and have tormented mankind since the dawn of days. Remarkably, the field of cosmology is tasked with the responsibility of providing answers to the modern versions of these three questions: \textit{What is the Universe made of}? \textit{What are the initial conditions of the Universe}? \textit{How will the Universe evolve}?

Even more remarkably, we have a semi-decent idea of how to answer these questions, although several crucial gaps remain. We know that most of the Universe is \textit{not} made up of stuff we know and love (dubbed \textit{baryonic matter}), but rather of invisible dark matter and dark energy. The question of their composition and origin, however, remains well open. As for the initial conditions of the Universe, we have good reason to believe that when the Universe was just a fraction of a second old, it underwent a period of accelerated expansion which goes under the name of \textit{inflation} (what happened before, however, remains a mystery, at least until we have a complete theory of quantum gravity). Presumably, inflation set up the seeds which later grew under gravity to form the structure we observe today: galaxies, clusters, and the whole cosmic web in its beauty. And finally, we believe that a mysterious dark energy is driving the current accelerated expansion of the Universe, and the nature of the dark energy will determine the fate of the Universe.

Besides cosmology, particle physics is also tasked with the responsibility of answering the first question (and, to some extent, the other two). The Standard Model of particle physics provides a remarkable description of most experimental results to date...with one notable exception. Surely the reader will have heard about neutrinos, ghostly particles permeating the world and constantly hurtling past us, and yet extremely elusive and hard to detect. We know that neutrinos come in three ``flavours'' ($\nu_e$, $\nu_{\mu}$, and $\nu_{\tau}$), and that as they propagate they can switch among different flavours. This is a phenomenon known as neutrino oscillations, whose discovery was awarded the 2015 Nobel Prize in Physics. Neutrino oscillations can only occur if neutrinos have mass. However, the Standard Model of Particle Physics predicts that neutrinos are massless. Neutrino masses are therefore the only direct evidence for physics beyond the Standard Model, the quest for which is extremely hot in particle physics nowadays. Unraveling the neutrino mass scale would likely shed light on physics operating at energy scales we can only ever dream of reaching on Earth, and would be a feat of indescribable impact. But it's not easy...

Enter cosmology. Neutrinos are very peculiar particles, as we shall see in this thesis, and their distinctive behaviour imprints equally unique signatures in cosmological observations. Two types of observations, in particular, are crucial in this sense. One is the Cosmic Microwave Background (CMB), a left-over radiation from the Big Bang and the oldest light reaching us from a time when the Universe was ``only'' 380000 years old (for comparison, the Universe is now about 14 billion years old). Another important set of observations is constituted by the large-scale structure (LSS), in particular how galaxies in the Universe are distributed and cluster with each other. The physics of neutrinos creates subtle correlations among the positions of various points in the CMB and among the positions of the millions of galaxies in the sky.

Until a few decades ago, cosmology was not considered a ``real'' science, because observations were hard to come by and those few observations we had were of poor quality. The situation has now drastically changed. We have immense amounts of data, of extraordinary quality. Inside this data is a colossal treasure of information on the content of the Universe, its origin, its fate, and the ghostly neutrinos. However, analysing the data is becoming ever more challenging, and as the data grows in quantity, quality, and complexity, these challenges only keep growing.

At the time I started my PhD, three things soon became clear to me. The first was that understanding the properties of neutrinos, and in particular their masses, was an exciting problem which would only have kept getting hotter. The second was that cosmology and in particular data from the LSS provides an extraordinary route towards achieving this goal. The third was that there were still several crucial open questions in the field and in particular in the use of LSS data, questions which needed to be answered if we wanted to make real progress. Getting a bit technical, some of these questions included: understanding if and how cosmology can determine the neutrino mass ordering (normal or inverted); understanding how to properly define galaxy bias, and hence analyse galaxy clustering data, in the presence of massive neutrinos; devising wiser ways of calibrating galaxy bias; and so on. At this point, there was really no questioning the fact that I was going to focus my thesis work on understanding how to hunt neutrinos in cosmology, and how to make the most out of current and future CMB and LSS data.

With this in mind, in my thesis I will describe a number of recent important developments in the field of neutrino cosmology, focusing on advances I either led or gave decisive contributions to. Despite their elusive nature and their limited contribution to the energy budget, neutrinos are an extremely important component of the Universe. A very limited amount of neutrinos is sufficient to completely reshape the Universe, and hence revealing their properties will partly address the ``\textit{What are we made of}?'' question. However, in my thesis I have also tied the question of the neutrino unknowns to the ``\textit{Where do we come from?}'' and ``\textit{Where are we going?}'' questions.

In my thesis I have addressed some of the open questions I outlined above. As always in research, answering questions has led to more questions, which I have tried my best to answer. Some of the questions I have addressed in my thesis are the following:
\begin{itemize}
\item What do the positions and subtle correlations between the positions of millions of galaxies in the sky tell us about the neutrino masses? In Paper~I, we looked at millions of galaxies and found that neutrinos can weigh at most about $10^{-37}\,{\rm kg}$. I always find it impressive that by looking at such huge objects in the sky we can probe mass scales that small. This is currently the best limit on the sum of the neutrino masses, and resulted in our work being cited in the 2018 Review of Particle Physics~\cite{Tanabashi:2018oca}.
\item Can cosmological data tell apart the two neutrino mass orderings (normal and inverted ordering, \textit{i.e.} whether we have two light neutrinos and one heavier neutrino, or one light neutrino and two heavier neutrinos), and if so how? We answered this question in Paper~I.
\item Can we find a wiser way of calibrating galaxy bias, perhaps using the lensing of the CMB? We devised a simple way for doing so in Paper~II.
\item What is the proper way of defining galaxy bias itself, when massive neutrinos are present? Have people been defining it incorrectly, and does this mistake have an important effect? In Paper~III we found that the answer is yes, and devised a simple way for correcting this mistake.
\item Can neutrinos tell us something about dark energy, and hence the fate of the Universe? In Paper~IV, quite unexpectedly we found that the answer is yes. We showed that if future underground detectors find that the neutrino mass ordering is inverted, dark energy would likely have to be of \textit{phantom} nature, which could result in the final fate of the Universe being a Big Rip.
\item Can our ignorance about neutrino properties bias the conclusions we draw about inflation and hence the initial conditions of our Universe? Fortunately, in Paper~V we found that the answer is mostly no.
\end{itemize}

\section{Outline of the thesis}
\label{sec:outline}

My thesis is outlined as follows. I set the stage for the play in Chapter~\ref{chap:2} by providing an overview of the Standard Model of particle physics as well as the Standard Model of cosmology, the $\Lambda$CDM model. Next, in Chapter~\ref{chap:3}, I provide an overview of the main concepts and equations in physical cosmology, which will be useful in understanding the role played by neutrinos during the evolution of the Universe. In Chapter~\ref{chap:4}, I first discuss the physics of massive neutrinos, before explaining how their behaviour throughout the evolution of the Universe is expected to leave peculiar signatures. I then describe the main cosmological observations, focusing on CMB and LSS data, and discuss the signatures of massive neutrinos in these observations. In Chapter~\ref{chap:5}, I discuss statistical tools which will turn out to be useful when attempting to analyse cosmological data to study neutrino properties. The heart of this thesis is Chapter~\ref{chap:6}, where I describe the results of the five included papers, addressing the points I outlined previously at the end of Chapter~\ref{sec:cosmology}. Finally, in Chapter~\ref{chap:7} I summarize my results and provide an outlook for future work.

Before starting, I need to warn the reader about one particular point. It has not been feasible to provide a pedagogical introduction to all the involved topics. Therefore, Chapters~\ref{chap:2} through~\ref{chap:5} will be rather introductory in nature, with my aim being more that of providing the context within which my work was done. Often (especially in the context of CMB and LSS observations), I will discuss the physics at a heuristic level. Anticipating that most of my readers will not be experts on the subject, my aim has been that of endowing the reader with the intuition necessary to grasp why cosmology works the way it works. I often refer the reader to pedagogical/technical and seminal references wherein the topics in question are covered in greater depth. I suggest that the reader interested in going deeper into a particular topic consult these references.

\chapter{Standard Models and what lies beyond}
\label{chap:2}

\begin{chapquote}{Hamlet to Horatio in \textit{Hamlet}, William Shakespeare (1603), 1.5.167-8}
``There are more things in heaven and earth, Horatio, than are dreamt of in your philosophy.''
\end{chapquote}

The backbone of particle physics and cosmology consists of two \textit{Standard Models}, providing the mathematical description of these two fields. In the case of particle physics, the Standard Model is usually referred to as the Standard Model of Particle Physics (SM), whereas the standard model of cosmology is usually referred to as the concordance $\Lambda$CDM model. While the two have provided an astonishingly accurate description of almost all physical phenomena to date, both in the laboratory and in the Universe, indications persist that physics beyond the Standard Model(s) is needed for a more complete description of Nature. In this sense neutrinos, the protagonists of this thesis, represent a key example: the SM predicts neutrinos to be massless (or rather, was constructed in such a way that neutrinos are massless), whereas the observation of flavour oscillations has convincingly determined that neutrinos are massive, with the sum of the masses of the three neutrinos $M_{\nu}$ being at least $0.06\,{\rm eV}$~\cite{GonzalezGarcia:2012sz,Gonzalez-Garcia:2014bfa,Gonzalez-Garcia:2015qrr,Esteban:2016qun,deSalas:2017kay,deSalas:2018bym}. Similarly, in the concordance $\Lambda$CDM model the value of $M_{\nu}$ is fixed to $0.06\,{\rm eV}$ by hand: the truth is that we don't know what the value of $M_{\nu}$ is, and near-future cosmological observations hold the promise of a first convincing detection of neutrino masses~\cite{Hannestad:2002cn,Joudaki:2011nw,Carbone:2011by,Hamann:2012fe,
Allison:2015qca,Archidiacono:2016lnv,Boyle:2017lzt,Sprenger:2018tdb,
Mishra-Sharma:2018ykh,Brinckmann:2018owf,Kreisch:2018var,Ade:2018sbj,Yu:2018tem,
Boyle:2018rva,Hanany:2019lle}. Beyond neutrino masses, a host of other measurements/observations hint at the existence of physics beyond the Standard Model(s), albeit at a statistical significance which is in most cases mild at best (see e.g.~\cite{Bhattacharyya:2008ez,Lykken:2010mc,Allanach:2016yth,Rosenfeld:2017ksi,Graverini:2018riw} for the SM and e.g.~\cite{Joyce:2014kja,Bull:2015stt,Freedman:2017yms,DiValentino:2017gzb,Douspis:2018xlj} for the $\Lambda$CDM model). Still, it is hard to believe that the SM and the $\Lambda$CDM model are the end of the story, and many (including me) are of the opinion that the in the coming years we might finally get a convincing glimpse of physics beyond the Standard Model(s).

In this Chapter, I will provide a very brief review of the Standard Models of particle physics and cosmology. Notice that the literature is full of well-written, extensive, up-to-date introductions to particle physics and cosmology which do justice to the two subjects way more than this Chapter. I will avoid being technical, with the aim of simply setting the stage for the rest of the thesis, and providing an useful introduction to the main concepts and tools necessary to understand the rest of the thesis at a high level.

\section{The Standard Model of Particle Physics}
\label{sec:sm}

The mathematical description of the SM, whose current formulation was finalized in the 1970s, is based on a special type of quantum field theories known as gauge theories: such theories are described by a Lagrangian invariant under local transformations generated by the elements of a symmetry group (or product of symmetry groups). To ensure gauge invariance, it is necessary to include vector fields known as gauge fields into the Lagrangian (more precisely, derivatives are upgraded to covariant derivatives involving these gauge fields). Each symmetry group of the Lagrangian can then be interpreted as describing a force between particles, whose force carriers are the gauge fields. For pedagogical introductions to the SM, I refer the reader to classic textbooks such as~\cite{Mandl:1985bg,Cheng:1985bj,Kane:1987gb,Aitchison:1989bs,Collins:1989kn,
Donoghue:1992dd,Cottingham:2007zz,Griffiths:2008zz,Langacker:2010zza,Schwartz:2013pla}.

The SM is a chiral gauge theory, formulated in terms of separate left- and right-handed chiral components of the fermion matter fields. The mathematical description of the SM is based on the gauge group $SU(3)_c \times SU(2)_L \times U(1)_Y$, where the $SU(3)_c$ part describes the strong force (and correspondingly the theory of quantum chromodynamics - QCD), whereas the $SU(2)_L \times U(1)_Y$ part describes the electroweak (EW) interactions. In a rather symbolic form which hides a lot of dust under the carpet, the SM Lagrangian is given by:
\begin{eqnarray}
{\cal L}_{\rm SM} = -\frac{1}{4}F_{\mu \nu}F^{\mu \nu}+i\bar{\Psi}\gamma^{\mu}D_{\mu}\Psi + D_{\mu}\Phi D^{\mu}\Phi - V(\Phi) - Y^{ij}\bar{\Psi}_i\Phi\Psi_j\,.
\label{eq:sm}
\end{eqnarray}
The first term includes the kinetic terms for the gauge fields (through their field-strengths $F_{\mu \nu}$), the second term includes the kinetic terms for the matter fields (symbolically denoted by $\Psi$) and their couplings to the gauge fields. The third term is the kinetic term for the Higgs field $\Phi$ (and specifies its interactions with gauge bosons), whereas the fourth term is the Higgs potential which gives rise to the Higgs mechanism and hence to EW symmetry breaking, wherein the $SU(2)_L \times U(1)_Y$ symmetry is broken down to the $U(1)_{\rm em}$ subgroup (with $_{\rm em}$ standing for ``electromagnetism'').

The matter content of the SM is arranged into left-handed $SU(2)$ quark doublets [$Q_L^i = (u_L^i,d_L^i)$, with $_L$ for left-handed and $i=1,2,3$ running over the three generations] and lepton doublets [$L_L^i = (e_L^i,\nu_L^i)$], and right-handed singlets $u_R^i$, $d_R^i$, and $e_R^i$. The last term in Eq.~(\ref{eq:sm}) is the Yukawa interaction term, which couples the left-handed fermion doublets with the right-handed fermion singlets through the Higgs doublet. Upon EW symmetry breaking, the Yukawa interaction term gives mass to the charged leptons and quarks.

Importantly, the SM matter content does not include right-handed neutrino fields $\nu_R^i$. Therefore, the Yukawa interaction term cannot generate masses for the neutrinos. This is no coincidence, rather occurs by construction. At the time the SM was formulated, there only existed upper limits on $\nu_e$ of about $200\,{\rm eV}$, much smaller than the next lightest known fermion, the electron whose mass is about $0.5\,{\rm MeV}$. Therefore, the SM was constructed to accommodate massless neutrinos. However, when in 1998 the SuperKamiokande atmospheric neutrino experiment detected neutrino oscillations (possible only if two out of the three neutrino mass eigenstates are massive, as we will discuss later in Chapter~\ref{subsec:oscillations})~\cite{Fukuda:1998mi}, it became clear that the picture had to be enlarged to allow for neutrino masses. Several approaches to give mass to neutrinos in Beyond the Standard Model (BSM) scenarios exist: for a very incomplete list of seminal papers and reviews, which does not do justice to the wide literature of well-motivated models, see e.g.~\cite{Minkowski:1977sc,Yanagida:1979as,GellMann:1980vs,Mohapatra:1979ia,
Schechter:1980gr,Magg:1980ut,Lazarides:1980nt,Mohapatra:1980yp,Wetterich:1981bx,
Foot:1988aq,Ma:1998dn,Ma:2002pf,Zee:1985rj,Zee:1985id,Babu:1988ki,Asaka:2005an,
Asaka:2005pn,Boyarsky:2009ix,Boucenna:2014zba,King:2015aea,King:2003jb,Cai:2017jrq,
Mohapatra:2006gs,Altarelli:2010gt,Meloni:2017cig}. See also~\cite{Capozzi:2018ubv} for a recent review of unknowns in the neutrino sector.

At any rate, it is clear that the absence of a mechanism for providing mass to the neutrinos is among the most important shortcomings of the SM. Conversely, the observation that neutrinos have mass is \textit{the only direct evidence for physics beyond the Standard Model}, presumably operating at extremely high energy scales (which could explain the smallness of neutrino masses). As such, there is no doubt that shedding light on the neutrino mass scale would open the door towards new physics, and the impact such a feat would have cannot be understated. In fact, unveiling the neutrino mass scale (as well as the mass ordering, an aspect of the neutrino mass spectrum which we will return to in Chapter~\ref{subsec:oscillations}) is avowedly among the key goals of several experimental efforts, both in the lab and in cosmology. 

Cosmological observations appear to be a very promising avenue towards unraveling the neutrino mass scale and possibly the mass ordering. This possibility constitutes the main topic and thread of this thesis. I therefore now continue by providing a brief overview of the Standard Model of Cosmology, the $\Lambda$CDM model.

\section{The Standard Model of Cosmology}
\label{sec:lcdm}

\subsection{A brief history of cosmology}
\label{subsec:historycosmology}

Physical cosmology is a relatively new branch of science, born less than a hundred years prior to the time of writing. In 1929, while working at Mount Wilson Observatory, a young astronomer named Edwin Hubble was measuring the relation between the recession velocities of galaxies and their distances from Earth: surprisingly, he found a linear relation between these quantities, implying that farther galaxies move away from us faster~\cite{Hubble:1929ig}. This relation became known as Hubble's law, and was consistent with a solution to Einstein's equations found earlier in 1927 by the astronomer and priest Georges Lema\^{i}tre, describing an expanding Universe~\cite{Lemaitre:1927zz}. When winding back the tape of the expanding Universe, we see that in the past the Universe must have been in a much hotter and denser state. At the time, most astronomers were strong supporters of the steady state Universe, and the idea of an expanding Universe was greeted with much skepticism: during a 1949 BBC radio broadcast, astronomer Fred Hoyle referred to Lema\^{i}tre's theory as the ``Big Bang theory'', a name which was meant to be sarcastic. Meanwhile, already as early as in 1933, Fritz Zwicky realized that a substantial amount of \textit{dark matter} (DM) was needed to reconcile the observed motions of galaxies within the Coma Cluster with the inferred amount of luminous matter~\cite{Zwicky:1933gu}.

As time went by, the Big Bang theory started gradually gaining support, especially in light of two definite predictions it made. The first was the prediction for the abundance of light elements in an expanding Universe, carried out in the famous 1948 $\alpha\beta\gamma$ paper~\cite{Alpher:1948ve}, which correctly predicted the relative abundance of Hydrogen and Helium in the Universe. The second was the prediction of the existence of the Cosmic Microwave Background (CMB), a bath of left-over photons from the Big Bang~\cite{Alpher:1948we}. The CMB was eventually discovered, rather serendipitously, by Penzias and Wilson in 1965~\cite{Penzias:1965wn}, whereas in the same issue of \textit{ApJ} another paper correctly interpreted their observation as being the first detection of the CMB~\cite{Dicke:1965zz}. Observations of the CMB continued over the coming years, culminating with the first precise measurement of its black-body spectrum from the Far Infrared Absolute Spectrophotometer (FIRAS) instrument on board the Cosmic Background Explorer (COBE) satellite~\cite{Fixsen:1996nj}. In 1992, COBE was also the first experiment to detect anisotropies in the CMB~\cite{Smoot:1992td}.~\footnote{A number of other CMB experiments were launched during those and subsequent years, but it is fair to say that two stand out particularly among the others: the Wilkinson Microwave Anisotropy Probe (WMAP), operating between 2001 and 2010, played a crucial role in definitely establishing the current concordance $\Lambda$CDM model~\cite{Bennett:2003ba,Spergel:2006hy,Bennett:2012zja}. The \textit{Planck} satellite has instead mapped the CMB sky to exquisite accuracy and is currently providing the tightest constraints on cosmological parameters from a single experiment~\cite{Planck:2006aa,Ade:2013sjv,Ade:2013kta,Ade:2013zuv,
Adam:2015rua,Ade:2015xua,Aghanim:2015xee,Akrami:2018vks,Aghanim:2018eyx}.}

Meanwhile, evidence for the existence of dark components in our Universe kept growing. In the 1970s, seminal works by Vera Rubin~\cite{Rubin:1970zza,Rubin:1978kmz,Rubin:1980zd,Rubin:1982kyu,Rubin:1985ze}, along with upper limits on the amplitude of temperature anisotropies in the CMB, provided strong support for the existence of the DM already theorized by Zwicky in the 1930s. By the end of the 1990s, two independent groups led by Riess and Perlmutter used Type Ia Supernovae (SNeIa) to demonstrate that the Universe is accelerating~\cite{Riess:1998cb,Perlmutter:1998np}, thus requiring some form of \textit{dark energy} (DE), possibly in the form of a cosmological constant $\Lambda$~\cite{Peebles:2002gy,Huterer:2017buf}, or requiring modifications of gravity~\cite{Nojiri:2006ri,Silvestri:2009hh,Tsujikawa:2010zza,
Clifton:2011jh,Joyce:2016vqv}.

Besides CMB and SNeIa, a number of other observational probes began flourishing especially in the early 2000s. A special mention goes to probes of the large-scale structure (LSS), particularly galaxy redshift surveys. A crucial role in the development of galaxy redshift surveys was played by the Sloan Digital Sky Survey (SDSS)~\cite{York:2000gk}: in 2005, SDSS was the first survey to detect baryon acoustic oscillations (BAOs) in the LSS~\cite{Eisenstein:2005su}, a signature of primordial sound waves ringing in the early Universe, from an epoch prior to the formation of the CMB. 

Recently, the first detection of gravitational waves (GWs)~\cite{Abbott:2016blz,Abbott:2016nmj} by the LIGO collaboration~\cite{Abramovici:1992ah,Abbott:2007kv,Evans:2016mbw} has opened an unprecedented window onto the Universe, and has inaugurated the era of multi-messenger astronomy thanks to the first coincident detection of GW and electromagnetic signal with the GW170817 and GRB170817A events~\cite{TheLIGOScientific:2017qsa,GBM:2017lvd,Monitor:2017mdv}. The GW events detected so far have already been used to place extremely important constraints on cosmological theories (see e.g.~\cite{TheLIGOScientific:2016src,Blas:2016qmn,Creminelli:2017sry,Sakstein:2017xjx,
Ezquiaga:2017ekz,Baker:2017hug,Boran:2017rdn,Green:2017qcv,Nojiri:2017hai,Arai:2017hxj,
Jana:2017ost,Amendola:2017orw,Visinelli:2017bny,Crisostomi:2017lbg,Langlois:2017dyl,
Gumrukcuoglu:2017ijh,Kreisch:2017uet,Bartolo:2017ibw,Dima:2017pwp,Pardo:2018ipy,
Casalino:2018tcd,Jana:2018djs,Ezquiaga:2018btd,Abbott:2018lct,Casalino:2018wnc}).~\footnote{See also e.g.~\cite{Raveri:2014eea,Lombriser:2015sxa,Lombriser:2016yzn,Bettoni:2016mij,
Alonso:2016suf} for important early works in this direction.} The prospect of using future GW events to constrain cosmology appear extremely promising, see e.g.~\cite{Schutz:1986gp,Nissanke:2009kt,Tamanini:2016zlh,Caprini:2016qxs,Cai:2016sby,
Cai:2017yww,Chen:2017rfc,Feeney:2018mkj,Wang:2018lun,Belgacem:2018lbp,
DiValentino:2018jbh,Nunes:2018evm,Du:2018tia,Calcagni:2019kzo,Yang:2019bpr}. Another cosmological probe expected to be particularly important in the coming years is the 21-cm line~\cite{Furlanetto:2006jb,Morales:2009gs,Pritchard:2011xb,Munoz:2015eqa,
Villaescusa-Navarro:2016kbz,Obuljen:2017jiy,Barkana:2018lgd,Barkana:2018cct,
Fraser:2018acy,Munoz:2018jwq,Kovetz:2018zan,Bowman:2018yin,Nebrin:2018vqt,Munoz:2019fkt,
Munoz:2019rhi}.

\subsection{Basics of physical cosmology}
\label{subsec:basics}

The standard model of cosmology is the mathematical framework describing the Universe on the largest observable scales. Its lies on two cornerstones: the first is Einstein's theory of General Relativity (GR)~\cite{Einstein:1916vd}. The second is an assumption known as \textit{cosmological principle}, stating that the Universe is homogeneous and isotropic on large scales. For pedagogical references on cosmology, see e.g.~\cite{Bergstrom:1999kd,Dodelson:2003ft,Mukhanov:2005sc,Durrer:2008eom,
Weinberg:2008zzc,Lesgourgues:2018ncw}. The essence of GR is encapsulated in the Einstein field equations~\cite{Einstein:1916vd} (see~\cite{Carroll:2004st} for one of the best pedagogical resources on GR):
\begin{eqnarray}
G_{\mu \nu} = 8\pi GT_{\mu \nu}\,.
\label{eq:einstein}
\end{eqnarray}
The left-hand side of Eq.~(\ref{eq:einstein}) contains the Einstein tensor $G_{\mu \nu}$, and describes the geometrical properties of spacetime, whereas the right-hand side contains the stress-energy tensor $T_{\mu \nu}$ which includes contributions from the various sources of matter and energy residing in the spacetime. The general form of a metric respecting the cosmological principle is known as the Friedmann-Lema\^{i}tre-Robertson-Walker (FLRW) metric, and is described by the following line-element~\cite{Friedmann:1924bb,Lemaitre:1931zza,Robertson:1935zz,Walker}:
\begin{eqnarray}
ds^2 = dt^2-a^2(t) \left [ \frac{dr^2}{1-kr^2}+r^2(d\theta^2 + \sin^2\theta d\phi^2) \right ]\,,
\label{eq:flrw}
\end{eqnarray}
where $t$ is time, $r$, $\theta$, and $\phi$ are the usual spherical coordinates, and $k$ is the curvature parameter which determines the overall geometry of the Universe. The function $a(t)$ is known as the scale factor, and describes the expansion (or contraction) of the Universe. Taking $d_{\rm ini}$ to be the distance between two objects at some reference time $t_{\rm ini}$, then assuming the objects have no peculiar velocity, at a later time $t$ their distance will be given by $d(t) = a(t)d_{\rm ini}/a_{\rm ini}$. It is common practice to normalize the scale factor to take the value $1$ today: $a_0=1$ (the subscript $_0$ usually refers to quantities evaluated today).

The time evolution of the scale factor can be determined by solving the Einstein equations, and consequently will depend on the matter/energy content of the Universe. One can make progress assuming that $T_{\mu \nu}$ in Eq.~(\ref{eq:einstein}) takes the form $T_{\mu \nu} = {\rm diag}(\rho,p,p,p)$ describing a perfect fluid with energy density $\rho$ and pressure $p$. Inserting this into Eq.~(\ref{eq:einstein}), with $G_{\mu \nu}$ computed from the FLRW metric with line element given by Eq.~(\ref{eq:flrw}), one arrives (through what is a very lengthy but classic exercise done in basically any graduate-level cosmology course~\cite{Carroll:2004st}!) at the following equations for the scale factor known as Friedmann equations:
\begin{eqnarray}
\label{eq:friedmann1}
\left ( \frac{\dot{a}}{a} \right )^2 + \frac{k}{a^2} = \frac{8\pi G}{3}\rho\,, \\
\label{eq:friedmann2}
\frac{\ddot{a}}{a} = -\frac{4\pi G}{3}(\rho + 3p)\,,
\end{eqnarray}
with the dot (double dot) denoting a time derivative (second time derivative). Another useful but not independent equation, known as the continuity equation, follows from energy-momentum conservation $\nabla_{\mu}T^{\mu \nu}=0$ and reads:
\begin{eqnarray}
\dot{\rho}+3H(\rho+p)=0\,,
\label{eq:continuity}
\end{eqnarray}
where the quantity $H \equiv \dot{a}/a$ describes the expansion rate of the Universe, and is usually referred to as Hubble parameter $H(t)$. The value of the Hubble parameter today, $H_0$, is instead typically called Hubble constant. The reduced Hubble constant $h$ is given by the Hubble constant expressed in units of $100\,{\rm km}\,{\rm s}^{-1}\,{\rm Mpc}^{-1}$: $h \equiv H_0/100\,{\rm km}\,{\rm s}^{-1}\,{\rm Mpc}^{-1}$. The bulk of the game reduces to specifying the matter/energy content of the Universe, \textit{i.e.} the $\rho$ and $p$ on the right-hand sides of the two Friedmann equations (and we will return in more detail to this in Chapter~\ref{sec:elementary}).

\begin{figure}
\centering
\includegraphics[width=0.6\linewidth]{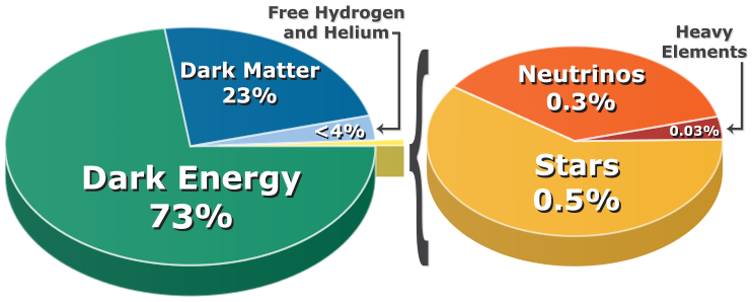}
\caption{Pie chart representing the energy budget of the Universe today, as we believe we understand it: less than $5\%$ is in the form of matter we are familiar with, dubbed baryonic matter. Credits: The Conversation~\cite{TheConversation:2017ghw}.}
\label{fig:pie_chart}
\end{figure}

Here's where things start to get interesting though. It turns out that, in order to match observations, much of what we need to introduce on the right-hand sides of Eqs.~(\ref{eq:friedmann1},\ref{eq:friedmann2}) is ``dark'': in terms of energy budget, about $23\%$ of the budget resides in a mysterious form of dark matter responsible for the formation of structure in the Universe and for explaining the motion of galaxies and clusters, whereas about $73\%$ of the budget is in an even more mysterious form of \textit{dark energy} (DE) responsible for the late-time accelerated expansion of the Universe, first discovered in 1998. Only $\approx 4\%$ of the energy budget of the Universe is in the form of matter we know and love, usually referred to as ``baryonic matter''. See Fig.~\ref{fig:pie_chart} for a pie chart representation of the Universe's energy budget.

The dark energy component appears to be well described by a cosmological constant $\Lambda$, which can be accounted for by adding a term $\Lambda g_{\mu \nu}$ to the left-hand side of Eq.~(\ref{eq:einstein}). Einstein originally introduced this term in his equations to obtain a static solution~\cite{Einstein:1917ce}, before later calling it his ``biggest blunder'' following the discovery that the Universe is expanding. The discovery of the Universe's acceleration in 1998 resuscitated the idea of the cosmological constant. While providing an excellent fit to observations, a cosmological constant appears to be very problematic from a fundamental physics point of view, an issue which is reflected in the \textit{cosmological constant problem} (see e.g.~\cite{Weinberg:1988cp,Carroll:1991mt,Carroll:2000fy,Weinberg:2000yb,
Sahni:2002kh,Peebles:2002gy,Padmanabhan:2002ji,Nobbenhuis:2004wn,Polchinski:2006gy} for reviews). Dark matter, on the other hand, appears to be well described by ``cold'' (\textit{i.e.} non-relativistic) particles. The combination of the cosmological constant $\Lambda$, and cold dark matter (CDM), is at the origin of the standard model of cosmology being dubbed the $\Lambda$CDM model.

At present, we do not know what the correct underlying models for DM and DE are, and a wide variety of models have been proposed in the literature. It is fair to say that the general consensus in the field is that DM should consist of a cold particle. Various models of particle DM have been proposed, see e.g.~\cite{Peccei:1977hh,Wilczek:1977pj,Hu:2000ke,Dodelson:1993je,McDonald:1993ex,
Jungman:1995df,Kusenko:1997si,Moroi:1999zb,Burgess:2000yq,Covi:2001nw,
TuckerSmith:2001hy,Servant:2002aq,Cheng:2002ej,Foot:2004wz,Cirelli:2005uq,
Pospelov:2007mp,ArkaniHamed:2008qn,Sikivie:2009qn,Kaplan:2009ag,Visinelli:2009zm,
Visinelli:2009bg,Beltran:2010ww,Feng:2011vu,Arias:2012az,Tulin:2013teo,Petraki:2013wwa,
Zurek:2013wia,Visinelli:2014twa,Baldes:2015lka,Visinelli:2015wha,Blennow:2015hzp,
Visinelli:2016hzy,Escudero:2016tzx,Escudero:2016ksa,Baum:2016oow,Blennow:2016gde,
Visinelli:2017imh,Brinckmann:2017uve,Foot:2017dgx,Boucenna:2017ghj,Hui:2016ltb,
Baum:2017enm,Foot:2018qpw,Escudero:2018thh,Escudero:2018fwn,Sokolenko:2018noz,
Akarsu:2018aro,Elor:2018twp,Kumar:2019gfl,Dvorkin:2019zdi} for a very incomplete list of references examining particle DM models and their phenomenology, and e.g.~\cite{Drukier:1986tm,Spergel:1999mh,Oikonomou:2006mh,Goodman:2010ku,Aartsen:2012kia,
Agnese:2013rvf,Viel:2013apy,Bernabei:2013xsa,Visinelli:2013fia,
Daylan:2014rsa,Visinelli:2015eka,Freese:2015mta,Blennow:2015oea,Ackermann:2015zua,
Giesen:2015ufa,Blennow:2015gta,Poulin:2015pna,Blennow:2015yca,Bird:2016dcv,
Munoz:2016tmg,Carr:2016drx,Akerib:2016vxi,Escudero:2016gzx,
TheMADMAXWorkingGroup:2016hpc,Escudero:2016kpw,Millar:2016cjp,Gariazzo:2017pzb,
Aprile:2017iyp,Poulin:2017bwe,Millar:2017eoc,Kumar:2017bpv,Baum:2017kfa,
Escudero:2017yia,Visinelli:2017ooc,Visinelli:2017qga,Zumalacarregui:2017qqd,
Munoz:2018pzp,Kumar:2018yhh,Baum:2018ekm,Knirck:2018knd,Baum:2018tfw,Visinelli:2018zif,
Visinelli:2018wza,Drukier:2018pdy,Edwards:2019puy,
Baum:2018sxd,Baum:2018lua,Wu:2019nhd,Ramberg:2019dgi,Lawson:2019brd} for ideas and developments concerning experimental and observational tests of these models. However, in principle DM could be the manifestation of a breakdown of GR, and a wide variety of modified gravity models accounting for DM have been proposed in the literature, see e.g.~\cite{Milgrom:1983ca,Mannheim:1988dj,Mannheim:1992vj,Bento:2003dj,Zhang:2004gc,
Moffat:2004bm,Arik:2005ir,Moffat:2005si,Capozziello:2006ph,Sereno:2006mw,Zlosnik:2006zu,
Nojiri:2006gh,Cognola:2006sp,Brownstein:2007sr,Saffari:2007xc,Kahya:2007zy,
Boehmer:2007kx,Nojiri:2008nt,Mukohyama:2009mz,Lim:2010yk,Sebastiani:2010ct,
Capozziello:2012qt,Chamseddine:2013kea,Chamseddine:2014vna,Boehmer:2014ipa,
Mirzagholi:2014ifa,Tamanini:2015iia,Myrzakulov:2015nqa,Ramazanov:2015pha,
Guendelman:2015rea,Myrzakulov:2015kda,Salzano:2016udu,Babichev:2016bxi,
Zaregonbadi:2016xna,Rinaldi:2016oqp,Verlinde:2016toy,Sebastiani:2016ras,
Marttens:2017njo,Calmet:2017voc,Hirano:2017zox,Koutsoumbas:2017fxp,Vagnozzi:2017ilo,
Marzola:2017lbt,Dutta:2017fjw,Frandsen:2018ftj,Zlosnik:2018qvg,Barrientos:2018giw,
Burrage:2018zuj,Lisanti:2018qam,Odintsov:2019mlf} for an incomplete list of such models and their observational tests. As for dark energy, the situation is even more uncertain, and a broad array of models have been proposed, involving either new fields or modifications to gravity. See e.g.~\cite{Horndeski:1974wa,Bilic:2001cg,Freese:2002sq,Bento:2002ps,Sahni:2002dx,
Li:2004rb,Elizalde:2004mq,Nojiri:2005vv,Kolb:2005da,Nojiri:2005jg,Alnes:2005rw,
Cognola:2006eg,Copeland:2006wr,Amendola:2006kh,Carroll:2006jn,
Amendola:2006we,Hu:2007nk,Pogosian:2007sw,Tsujikawa:2007xu,Saridakis:2007wx,
Jhingan:2008ym,Gavela:2009cy,Saridakis:2009bv,Zumalacarregui:2010wj,
Dent:2011zz,Elizalde:2011ds,Geng:2011aj,Bamba:2012cp,Bamba:2012qi,
Pan:2012ki,Myrzakulov:2013hca,Pan:2013rha,Gleyzes:2013ooa,Cognola:2013fva,
Maggiore:2014sia,Rinaldi:2014yta,Rinaldi:2015iza,Rabochaya:2015haa,Wang:2016lxa,Elizalde:2017mrn,
Saridakis:2017rdo,Bahamonde:2017ize,Bahamonde:2018miw,Saridakis:2018unr,
Odintsov:2018nch,Marsh:2018kub,Odintsov:2018zai,Langlois:2018dxi,Barros:2019rdv,
Paliathanasis:2019hbi} for a very incomplete list of proposed models of dark energy, and e.g.~\cite{Huterer:1998qv,Perlmutter:1999jt,Huterer:2000mj,Hannestad:2002ur,
Melchiorri:2002ux,Blake:2003rh,Wang:2004ru,Hannestad:2004cb,
Blomqvist:2008ud,Zhao:2009fn,Giannantonio:2009gi,Sherwin:2011gv,
Hojjati:2011ix,Zumalacarregui:2012pq,Bellini:2012qn,Zumalacarregui:2012us,
Bartolo:2013ws,Salvatelli:2013wra,Bellini:2014fua,Archidiacono:2014msa,Baum:2014rka,
Ade:2015rim,Bellini:2015oua,Frusciante:2015maa,Alam:2015rsa,Bellini:2015xja,
Bianchini:2015iaa,Hu:2016zrh,Nunes:2016dlj,Pogosian:2016pwr,Pan:2016ngu,Raveri:2017qvt,
Dhawan:2017leu,Yang:2017zjs,Yang:2017alx,Dhawan:2017kft,Guo:2017deu,
Peirone:2017vcq,Peirone:2017ywi,Pan:2017zoh,Garcia-Garcia:2018hlc,Poulin:2018dzj,Espejo:2018hxa,Visinelli:2018utg,Abbott:2018xao,
Brush:2018dhg,Contigiani:2018hbn,Yang:2018qec,Zucca:2019xhg,Bambi:2019tjh} for works examining observational constraints on dark energy models and/or modifications to gravity.

\subsection{A sneak peek at the concordance $\Lambda$CDM model}
\label{subsec:concordancelcdmmodel}

The set of theoretical equations governing the evolution of the Universe (including those we saw so far, and others to be discussed in more detail in Chapter~\ref{chap:3}), in combination with a set of six parameters allowing for a simple and physically motivated comparison between observations and theory, forms the backbone of the concordance $\Lambda$CDM model. We usually refer to this model as \textit{concordance} model because different observational probes of appear to point to consistent values for these six fundamental parameters (alongside other derived parameters).~\footnote{This overall concordance holds modulo a number of mild tensions which overall do not (yet) undermine the consistency of the model itself. See e.g.~\cite{Efstathiou:2013via,Raveri:2015maa,Joudaki:2016mvz,Kitching:2016hvn,
Riess:2016jrr,Grandis:2016fwl,DiValentino:2016hlg,Poulin:2016nat,Qing-Guo:2016ykt,Bernal:2016gxb,Ko:2016uft,Joudaki:2016kym,Zhao:2017cud,Kumar:2017dnp,
Zhao:2017urm,Camera:2019vbp,DiValentino:2017iww,Sola:2017znb,Feeney:2017sgx,
Efstathiou:2017rgv,Yang:2017ccc,DiValentino:2017rcr,DiValentino:2017oaw,Pan:2017ent,
Abbott:2017smn,An:2017crg,Benetti:2017juy,Feng:2017usu,Renzi:2017cbg,
Riess:2018uxu,Mortsell:2018mfj,Nunes:2018xbm,Poulin:2018zxs,Yang:2018ubt,Yang:2018euj,
Adhikari:2018wnk,Raveri:2018wln,DiValentino:2018wum,Yang:2018xlt,DEramo:2018vss,
Guo:2018ans,Yang:2018uae,Yang:2018qmz,DiValentino:2018gcu,Poulin:2018cxd,Kumar:2019wfs,
Vattis:2019efj,Pan:2019jqh,Agrawal:2019lmo,Yang:2019jwn} for an incomplete list of recent papers discussing these tensions and possible solutions.}

The six parameters of the $\Lambda$CDM model include two parameters quantifying the amount of baryons and the amount of cold dark matter, two parameters describing the power spectrum of primordial scalar fluctuations, one parameter describing the overall geometry of the Universe (more precisely, the angular scale under which BAOs appear in the CMB, which is related to the geometry of the Universe), and one parameter describing the amount of reionization the Universe experienced due to the formation of the first stars. These parameters will be described in more detail in Chapter~\ref{sec:concordance}. In the following Chapter, I will provide a more detailed (but still brief) overview of physical cosmology, including a brief history of the Universe.

\chapter{Overview of physical cosmology}
\label{chap:3}

\begin{chapquote}{Carl Sagan in \textit{Cosmos: A Personal Voyage}, Episode 10: ``The Edge of Forever'' (1980)}
``Cosmology brings us face to face with the deepest mysteries, questions that were once treated only in religion and myth.''
\end{chapquote}

In Chapter~\ref{sec:lcdm}, we have seen how the Universe on the largest observable scales is described by Einstein's equations of General Relativity, the FLRW metric, and correspondingly the Friedmann equations, Eqs.~(\ref{eq:friedmann1},\ref{eq:friedmann2}). In this Chapter, I will provide a more detailed (but still rather brief) picture of physical cosmology and the (thermal) history of the Universe, starting from the equations we have seen in Chapter~\ref{sec:lcdm} and elementary notions of thermodynamics and statistical mechanics. More pedagogical and in-depth treatments of the topics covered here can be found in classical cosmology texbooks, including e.g.~\cite{Bergstrom:1999kd,Dodelson:2003ft,Mukhanov:2005sc,Durrer:2008eom,
Weinberg:2008zzc,Lesgourgues:2018ncw}.

\section{Elementary notions of cosmology}
\label{sec:elementary}

To make progress, we have to specify the matter/energy content of the Universe, \textit{i.e.} the right-hand sides of Eqs.~(\ref{eq:friedmann1},\ref{eq:friedmann2}). We will assume that the Universe is filled with fluid(s) whose relation between pressure $p$ and energy density $\rho$ takes the form:
\begin{eqnarray}
p = w\rho\,,
\label{eq:eos}
\end{eqnarray}
where the constant $w$ is called equation of state (EoS). It is trivial to solve the continuity equation Eq.~(\ref{eq:continuity}) and show that, for a Universe filled with a single fluid with EoS $w$, the energy density evolves as a function of scale factor as:
\begin{eqnarray}
\rho(a) \propto a^{-3(1+w)}\,.
\label{eq:rhooft}
\end{eqnarray}
Similarly, the scale factor in the same Universe evolves as follows [which can be easily shown by solving either one of Eqs.~(\ref{eq:friedmann1},\ref{eq:friedmann2})]:
\begin{eqnarray}
  a(t) \propto \begin{cases}
      t^{\frac{2}{3(1+w)}} & w \neq -1 \\
      e^{H_0t} & w=-1
    \end{cases}\,.
\label{eq:aoft}
\end{eqnarray}
where $H_0$ denotes the Hubble parameter today (Hubble constant).

\begin{figure}[!t]
\centering
\includegraphics[width=1.0\textwidth]{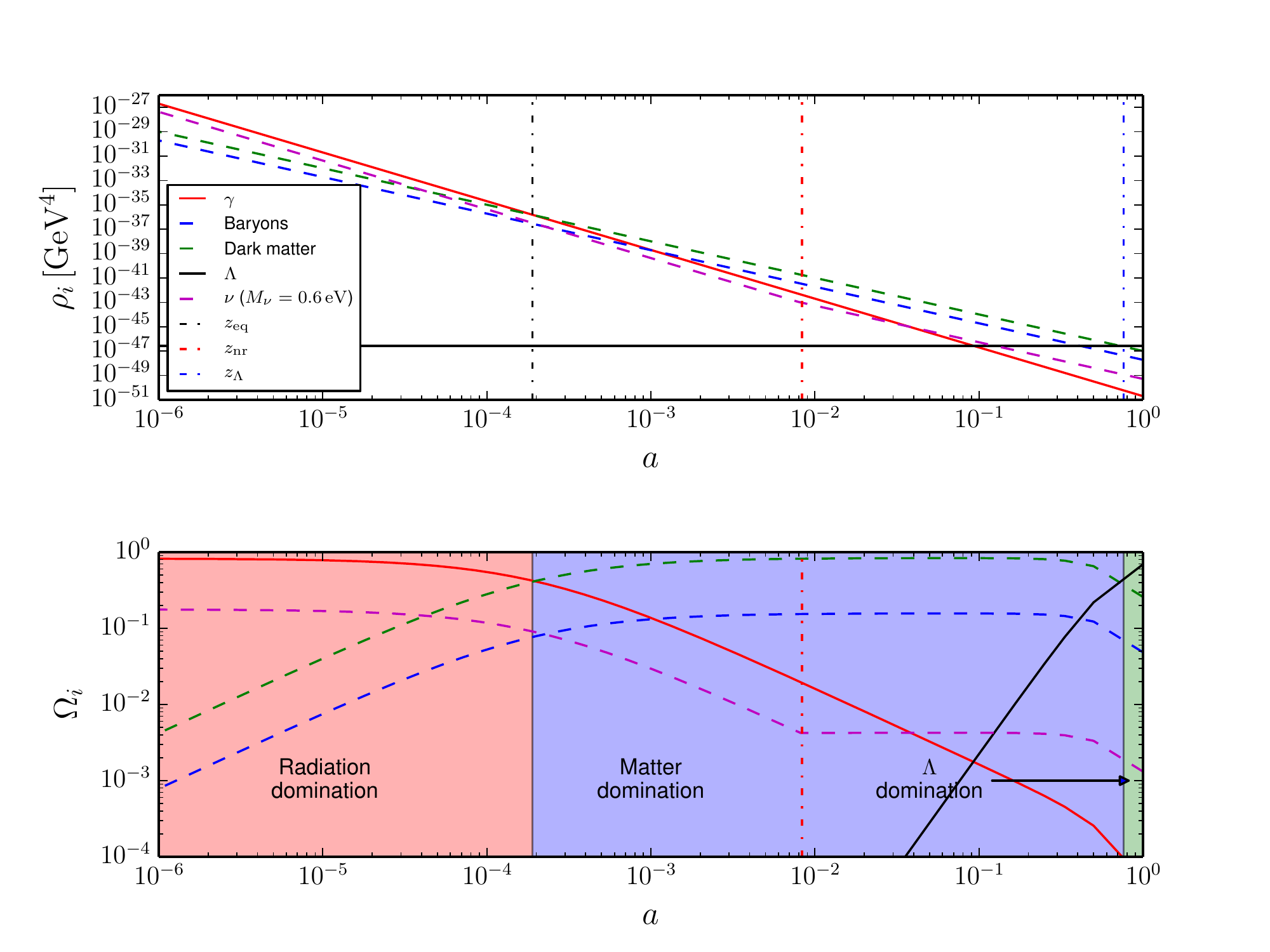}
\caption{Evolution of energy density and density parameters of the various components of the Universe. \textit{Upper panel}: evolution of the energy densities $\rho_i$, in ${\rm GeV}^4$, of photons (red solid curve), baryons (blue dashed curve), dark matter (green dashed curve), the cosmological constant (black solid curve), and massive neutrinos (with $M_{\nu}=0.06\,{\rm eV}$, purple dashed curve) as a function of scale factor $a$. The three vertical lines denote the redshift of matter-radiation equality (black dot-dashed line), the redshift of non-relativistic transition of massive neutrinos (red dot-dashed line), and the redshift of matter-$\Lambda$ equality (blue dot-dashed line). \textit{Lower panel}: evolution of the density parameters $\Omega_i$ for the various species, with the same color coding as the upper panel. In addition, the red, blue, and green shaded regions denote the eras of radiation, matter, and $\Lambda$ domination.}
\label{fig:energy}
\end{figure}

It is then useful to classify the components making up the cosmic inventory according to their EoS:
\begin{itemize}
\item \textbf{Radiation}: radiation has $w=1/3$, therefore from Eq.~(\ref{eq:rhooft}) and Eq.~(\ref{eq:aoft}) we find that $\rho(a) \propto a^{-4}$ and $a(t) \propto \sqrt{t}$. Photons contribute to the radiation energy density, and so do neutrinos at early times. The radiation energy density decreases with the scale factor as $a^{-4}$ since three powers of $a$ account for the expansion of the Universe, whereas one power of $a$ accounts for the fact that the radiation loses energy (it is \textit{redshifted}) due to its wavelength stretching as the Universe expands.
\item \textbf{Matter}: matter has $w=0$, therefore from Eq.~(\ref{eq:rhooft}) and Eq.~(\ref{eq:aoft}) we find that $\rho(a) \propto a^{-3}$ and $a(t) \propto t^{2/3}$. Baryons and cold dark matter contribute to the matter energy density, and so do neutrinos at late times.
\item \textbf{Dark energy}: the cosmological constant in the Friedmann equations is equivalent to a fluid with $w=-1$. Therefore, its energy density stays constant even as the Universe expands, and its presence leads to an exponential expansion. Beyond the cosmological constant, a simple phenomenological parametrization of the physics underlying cosmic acceleration is that of a more general dark energy component with constant EoS $w \neq -1$. As long as $w<-1/3$, such a fluid can drive cosmic acceleration. In this case, one finds that $\rho(a) \propto a^{-3(1+w)}$. Finally, for a more generic dark energy component with time-varying EoS $w(a)$, one finds that $\rho(a) \propto a^{-3}\exp \left [ -3\int_{1}^{a}da'\,w(a')/a' \right ]$.
\end{itemize}
Given the way the energy densities of these three different components scale as a function of scale factor or time, we can expect that radiation dominated the energy budget of the Universe early on. At some point (known as \textit{matter-radiation equality}), the energy density of matter was equal to that of radiation, and from that point on matter took on the scene. Finally, at very late times, the energy density in dark energy became larger than that of matter, leading to the accelerated expansion we see today. A visual representation of how the different components of the Universe take over at different times can be seen in the upper panel of Fig.~\ref{fig:energy}, where I plot the evolution of the energy densities $\rho_x$ for each species $x$ (photons, dark matter, baryons, cosmological constant, neutrinos). As we shall see later, massive neutrinos, the protagonists of this thesis, behave distinctly to the point that they escape the cosmic inventory classification given above. At early times, when the Universe was very hot and dense, neutrinos were relativistic and behaved as radiation. At late times, neutrinos instead become non-relativistic and contribute to the matter budget of the Universe. We will return to this important point later, as it underlies one of the most peculiar signatures of massive neutrinos in cosmological observations.

It it convenient to define the critical energy density $\rho_{\rm crit}$ as the current energy density required for the Universe to be flat [$k=0$ in of Eq.~(\ref{eq:friedmann1})]:
\begin{eqnarray}
\rho_{\rm crit} \equiv \frac{3H_0^2}{8\pi G}\,.
\label{eq:rhoc}
\end{eqnarray}
Following this definition, for any given species $x$ with energy density today $\rho_{x,0}$, we can define the density parameter $\Omega_x$ as $\Omega_x \equiv \rho_{x,0}/\rho_{\rm crit}$. For the cosmological constant $\Lambda$ we define $\Omega_{\Lambda} = \Lambda/(3H_0^2)$, while the curvature component can be seen as having an effective density parameter $\Omega_k = -k/(a_0H_0^2)$. A visual representation of the evolution with time of the density parameters $\Omega_x$ for each species $x$ (photons, dark matter, baryons, cosmological constant, neutrinos) is shown in the lower panel of Fig.~\ref{fig:energy}.~\footnote{Notice that the density parameters are defined at the present time, but can naturally be extended to be time-dependent, as long as one considers the time dependence of the density of each species and of the critical density. This time-dependence is naturally taken into account when plotting Fig.~\ref{fig:energy}.} Then, the first Friedmann equation [Eq.~(\ref{eq:friedmann1})] can be written in the following form (which sometimes goes under the name of sum rule):
\begin{eqnarray}
\sum_i \Omega_i = 1\,,
\label{eq:sumrule}
\end{eqnarray}
where the sum runs over all the components of the Universe (radiation, matter, cosmological constant, curvature). For reasons which will become obvious later (highlighting the problems which led to the need for an epoch of inflation), it is also convenient to express Eq.~(\ref{eq:sumrule}) as follows:
\begin{eqnarray}
\vert \Omega(a) - 1 \vert = \frac{\vert k \vert}{a^2H^2}\,,
\label{eq:omega}
\end{eqnarray}
where $\Omega(a)$ denotes the total energy density of the Universe \textit{without including the contribution from curvature}. So far we have discussed the evolution of energy densities as a function of time $t$ or scale factor $a$. For several cosmological discussions, it is more convenient to describe the flow of time in terms of redshift $z$, defined as a function of scale factor as:
\begin{eqnarray}
\frac{a}{a_0} \equiv \frac{1}{1+z}\,.
\label{eq:redshift}
\end{eqnarray}
With this definition, $z=0$ today, while $z \to \infty$ in the very far past. The concept of redshift has a simple physical interpretation. Consider a wave emitted with a wavelength $\lambda_{\rm em}$ at redshift $z_{\rm em}$ and observed at redshift $z_{\rm obs}$. Due to the expansion of the Universe the wave is redshifted, \textit{i.e.} its wavelength is stretched, and as a result the observed wavelength $\lambda_{\rm obs}$ is given by $\lambda_{\rm obs} = \lambda_{\rm em}(1+z_{\rm em})/(1+z_{\rm obs}) = \lambda_{\rm em}a_{\rm obs}/a_{\rm em}$.

Introducing the concept of redshift we can rewrite the evolution of the Hubble parameter as follows:
\begin{eqnarray}
H(z) = H_0\sqrt{\Omega_r(1+z)^4+\Omega_m(1+z)^3+\Omega_{\Lambda}+\Omega_k(1+z)^2}\,.
\label{eq:hz}
\end{eqnarray}
When introducing massive neutrinos into the picture (and allowing for a more general DE component with constant EoS $w$, with density parameter $\Omega_{\rm DE}$), Eq.~(\ref{eq:hz}) becomes:
\begin{eqnarray}
H(z) = H_0\sqrt{\Omega_r(1+z)^4+\Omega_m(1+z)^3+\Omega_{\rm DE}(1+z)^{3(1+w)}+\Omega_k(1+z)^2+\frac{\rho_{\nu}(z)}{\rho_{\rm crit}}}\,,
\label{eq:hznu}
\end{eqnarray}
where $\rho_{\nu}(z)$ denotes the neutrino energy density as a function of redshift: we have not specified a functional form for $\rho_{\nu}(z)$ since neutrinos behave as radiation in the early Universe and matter at late times, implying that the scaling of their energy density with $z$ is non-trivial. Nonetheless, as anticipated earlier we know that in the very early Universe (when neutrinos behave as radiation), $\rho_{\nu}(z) \propto (1+z)^4$, whereas at very late times (when neutrinos behave as matter), $\rho_{\nu}(z) \propto (1+z)^3$.~\footnote{Anticipating a bit, cosmological observations tell us that $H_0 \approx 70\,{\rm km}{\rm s}^{-1}{\rm Mpc}^{-1}$, $\Omega_r \approx 5 \times 10^{-5}$, $\Omega_m \approx 0.3$, $\Omega_{\rm DE} \approx 0.7$, and $\Omega_k \approx 0$ (\textit{i.e.} the energy density of the Universe is very close to the critical energy density $\rho_{\rm crit}$, and thus the Universe is very close to being flat)~\cite{Ade:2015xua,Aghanim:2018eyx}.} In addition, we define the physical density parameter of species $i$, $\omega_i$, as $\omega_i \equiv \Omega_ih^2$, where $h$ is the \textit{reduced} Hubble parameter, defined by $h \equiv H_0/(100\,{\rm km}{\rm s}^{-1}{\rm Mpc}^{-1})$.

Later, we shall see that cosmological observables very often carry the imprint of particular length scales, in relation to specific physical effects responsible for shaping the observables themselves.~\footnote{For instance, as we will discuss in more detail later, the typical angular separation between hot and cold spots in the CMB is sensitive to the sound horizon at photon decoupling, as well as the angular diameter distance to the CMB itself. On the other hand, BAO distance measurements are sensitive to the sound horizon at the baryon drag epoch.} For this reason, it is convenient to briefly recall basic concepts pertaining to distances in cosmology. In an expanding Universe, the notion of distance can be a bit tricker than in our everyday life. Let us first define the comoving distance to an object located at redshift $z_e$ (\textit{i.e.} the distance travelled to reach us by a photon emitted by the object at time $t_e$, such a distance remaining fixed as the Universe expands), $\chi(z_e)$:
\begin{eqnarray}
\chi(z_e) = \int_{t_e}^{t_0}\frac{dt}{a(t)} = \int_{0}^{z_e}\frac{dz}{H(z)}\,,
\label{eq:chiz}
\end{eqnarray}
with $H(z)$ given by Eq.~(\ref{eq:hz}), or Eq.~(\ref{eq:hznu}) in the presence of massive neutrinos and a generic dark energy component with constant equation of state $w$.~\footnote{Other two important distances often being discussed in physical cosmology are the angular diameter distance $d_A(z)$ and luminosity distance $d_L(z)$. We will not discuss them further here, but simply note that they are related to the comoving distance $\chi$ given in Eq.~(\ref{eq:chiz}) through $d_A(z) = \chi(z)/(1+z)$ and $d_L(z) = (1+z)\chi(z)$. The angular diameter distance relates the the physical size of an object to the angle it subtends on the sky. The luminosity distance instead relates the observed flux of an object to its intrinsic luminosity.} Another important distance notion is the concept of comoving particle horizon $\chi_h$, the maximum distance a photon could travel from a very early time ($t=0$, $z=\infty$) until time $t$ (redshift $z$):
\begin{eqnarray}
\chi_h(z) = \int_{0}^{t}\frac{dt'}{a(t')} = \int_{z}^{\infty}\frac{dz'}{H(z')}\,.
\label{eq:comovingparticlehorizon}
\end{eqnarray}
With $H(z)$ given by Eq.~(\ref{eq:hz}), it is easy to show that the comoving particle horizon grows as $\chi_h(z) \propto (1+z)^{-1} \propto a$ during radiation domination, and as $\chi_h(z) \propto (1+z)^{-1/2} \propto a^{1/2}$ during matter domination. As we shall see later, in the early Universe the interplay between photon pressure and gravity (mostly provided by baryons and dark matter) set up sound waves which propagated in the tightly coupled baryon-photon plasma: these sound waves left imprints which we see today in the statistics of fluctuations in the temperature of the Cosmic Microwave Background, as well as in the large-scale distribution of galaxies. Therefore, it is convenient to define a comoving sound horizon $r_s$ as the maximum distance a sound wave could travel from the Big Bang until time $t$/redshift $z$:
\begin{eqnarray}
r_s(z) = \int_{0}^{t}dt'\,\frac{c_s(t')}{a(t')} = \int_{z}^{\infty}dz'\,\frac{c_s(z')}{H(z')}\,,
\label{eq:soundhorizon}
\end{eqnarray}
where the sound speed $c_s$ is given by:
\begin{eqnarray}
c_s = 1/\sqrt{3(1+R)}\,,
\label{eq:cs}
\end{eqnarray}
with the baryon-to-photon momentum density ratio $R$ given by:
\begin{eqnarray}
R \equiv \frac{p_b+\rho_b}{p_{\gamma}+\rho_{\gamma}}\,.
\label{eq:r}
\end{eqnarray}
In the early Universe, when photons dominate over baryons, $c_s \simeq 1/\sqrt{3}$ and hence $r_s \simeq \chi_h/\sqrt{3}$.

In its form given by Eq.~(\ref{eq:hz}), or Eq.~(\ref{eq:hznu}), the first Friedmann equation is one of the most important equations of physical cosmology. It allows us to describe the background expansion of the Universe as a function of the energy content of the Universe itself. However, this equation does not tell us how the Universe's content came to be, nor does it take into account the role of temperature in determining the content of the Universe. In fact, as the Universe expands its temperature drops and certain reactions between particles, previously maintained in equilibrium by frequent interactions, \textit{freeze-out} and lead to \textit{decoupling} of particles from each other. Moreover, temperature also determines how particles behave, depending on whether they are relativistic or not: this, as we shall see, plays a crucial role in the case of massive neutrinos. To address these issues, I will briefly review the theory of the \textit{Hot Big Bang} and describe the thermal history of the Universe.

\section{The Hot Big Bang theory}
\label{sec:hbb}

If we wind the tape of the Universe back in time, the scale factor decreases and the Universe becomes denser and denser. In such a dense Universe, reactions are generally fast enough to maintain thermodynamic equilibrium, and the Universe consisted of a hot and dense soup of particles in equilibrium at a common temperature $T$ (from now on, we will use $T = T_{\gamma}$ to denote the temperature of the photons). More generally, given a specific reaction with rate $\Gamma$, to determine whether the reaction is in equilibrium at any given time we need to compare $\Gamma$ to the expansion rate of the Universe $H$: if $\Gamma \gg H$, the reaction is in equilibrium, whereas the contrary holds if $\Gamma \ll H$.

\begin{figure}[!t]
\centering
\includegraphics[width=1.0\textwidth]{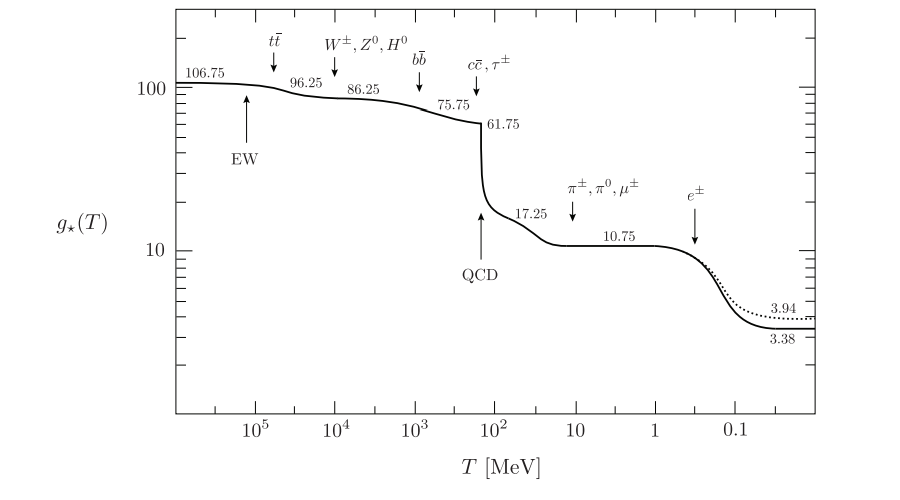}
\caption{Evolution of the effective number of relativistic degrees of freedom $g_{\star}$ (solid line) and the effective number of entropy degrees of freedom $g^s_{\star}$ (dashed line) assuming the particle content of the Standard Model, as a function of the temperature of the Universe. It is clear that both $g_{\star}$ and $g^s_{\star}$ decrease when particles annihilate or become non-relativistic. However, two events during which $g_{\star}$ and $g^s_{\star}$ decrease abruptly stand out in particular: the QCD phase transition at $T \sim 100\,{\rm MeV}$, and $e^+e^-$ annihilation at $T \sim 1\,{\rm MeV}$. Figure taken from~\cite{Baumann:2013ghw}.}
\label{fig:geff}
\end{figure}

As long as a given particle is in equilibrium, its phase space distribution $f(p,T)$, with $p \equiv \vert \mathbf{p} \vert$ the norm of the momentum and $T$ temperature,~\footnote{Because of isotropy we assume that the phase space distribution is independent of the spatial coordinate $\mathbf{x}$, and only depends on $p$ and not $\mathbf{p}$. For simplicity we also neglect the chemical potential of particles, \textit{i.e.} we set $\mu=0$. Allowing for a non-vanishing chemical potential does not change our subsequent discussion significantly.} is given by:
\begin{eqnarray}
f(\mathbf{p},T) = \frac{g}{(2\pi)^3}\frac{1}{e^{E(p)/T} \pm 1}\,,
\label{eq:distribution}
\end{eqnarray}
where $g$ is the number of internal degrees of freedom, $E(p) = \sqrt{p^2+m^2}$, and the $+$($-$) sign holds for Fermi-Dirac (Bose-Einstein) distributions respectively. From Eq.~(\ref{eq:distribution}), we can compute the number density $n(T)$, energy density $\rho(T)$, and pressure $P(T)$ of the species in question, which are given by the following~\cite{Bergstrom:1999kd,Dodelson:2003ft,Mukhanov:2005sc,Durrer:2008eom,
Weinberg:2008zzc,Lesgourgues:2018ncw}:
\begin{eqnarray}
\label{eq:nt}
n(T) &=& \int d^3\mathbf{p}\,f(\mathbf{p},T)\,,\\
\label{eq:rhot}
\rho(T) &=& \int d^3\mathbf{p}\,f(\mathbf{p},T)E(\mathbf{p})\,,\\
\label{eq:pt}
P(T) &=& \int d^3\mathbf{p}\,f(\mathbf{p},T)\frac{p^2}{3E(\mathbf{p})}\,.
\end{eqnarray}
Two limiting cases of Eqs.~(\ref{eq:nt},\ref{eq:rhot},\ref{eq:pt}) are of particular interest. The first is the \textit{relativistic} limit, where $T \gg m$ and the particle behaves as radiation. In this case, one finds:
\begin{eqnarray}
  n(T) = \begin{cases}
      \frac{3\zeta(3)}{4\pi^2}gT^3 & ({\rm FD}) \\
      \frac{\zeta(3)}{\pi^2}gT^3 & ({\rm BE})
    \end{cases}\,,\quad
  \rho(T) = \begin{cases}
      \frac{7\pi^2}{240}gT^4 & ({\rm FD}) \\
      \frac{\pi^2}{30}gT^2 & ({\rm BE})
    \end{cases}\,,\quad
  P = \frac{\rho}{3}\,,
\label{eq:relativistic}
\end{eqnarray}
where $\zeta(3) \approx 1.202$ is the Riemann zeta function of $3$, and FD/BE stand for Fermi-Dirac/Bose-Einstein respectively. From Eq.~(\ref{eq:relativistic}), it is clear that for radiation $w=1/3$. Summing over the energy densities of all relativistic species we obtain the total relativistic energy density $\rho_r$ (dominating the energy budget in the early Universe), which at any given temperature $T$ can be expressed in terms of an effective number of relativistic degrees of freedom $g_{\star}$:
\begin{eqnarray}
\rho_r = \sum_i \rho_i \equiv \frac{\pi^2}{30}g_{\star}(T)T^4\,,
\label{eq:rhorelativistic}
\end{eqnarray}
where $g_{\star}$ is given by~\cite{Bergstrom:1999kd,Dodelson:2003ft,Mukhanov:2005sc,Durrer:2008eom,
Weinberg:2008zzc,Lesgourgues:2018ncw}:
\begin{eqnarray}
g_{\star}(T) \simeq \sum_{i={\rm bosons}}\Theta(T-m_i) \left ( \frac{T_i}{T} \right )^4 + \frac{7}{8}\sum_{j={\rm fermions}}\Theta(T-m_j) \left ( \frac{T_j}{T} \right )^4\,.
\label{eq:geff}
\end{eqnarray}
In Eq.~(\ref{eq:geff}), the Heaviside step function highlights the fact that only for $T \gtrsim m_i$ or $T \gtrsim m_j$ do bosons $i$ or fermions $j$ contribute to the relativistic energy density. The evolution of $g_{\star}$ as a function of temperature is shown in Fig.~\ref{fig:geff} (solid line). As shown in Fig.~\ref{fig:geff}, $g_{\star}$ remains roughly constant except for noticeable drops during the EW phase transition, the QCD phase transition, and $e^+e^-$ annihilation, reflecting the abrupt decrease in the number of degrees of freedom in the early Universe following these events.

On the other hand, in the non-relativistic limit where $T \ll m$, the particle behaves as matter and one finds that the following holds~\cite{Bergstrom:1999kd,Dodelson:2003ft,Mukhanov:2005sc,Durrer:2008eom,
Weinberg:2008zzc,Lesgourgues:2018ncw}:
\begin{eqnarray}
n(T) = g \left ( \frac{mT}{2\pi} \right )^{\frac{3}{2}}e^{-\frac{m}{T}}\,, \quad \rho(T) = mn(T)+\frac{3}{2}nT \approx mn(T)\,, \quad P(T) = nT \ll \rho(T)\,.
\label{eq:nonrelativistic}
\end{eqnarray}
In this limit, the number density of particles is Boltzmann suppressed [due to the exponential appearing in the expression for $n(T)$ in Eq.~(\ref{eq:nonrelativistic})]: particles and antiparticles annihilate into photons, but the bath of photons does not possess enough energy to pair-create the particle-antiparticle pairs again, leading to an overall decrease in their number density. From Eq.~(\ref{eq:nonrelativistic}), it is clear that for matter $w \approx0$.

Another important concept in the Hot Big Bang theory is that of entropy density of particle species. Neglecting chemical potentials, we can define the entropy density of species $i$, $s_i$, as~\cite{Bergstrom:1999kd,Dodelson:2003ft,Mukhanov:2005sc,Durrer:2008eom,
Weinberg:2008zzc,Lesgourgues:2018ncw}:
\begin{eqnarray}
s_i \equiv \frac{\rho_i+P_i}{T_i}\,.
\label{eq:entropy}
\end{eqnarray}
As for the total relativistic energy density, we can write the total entropy density (which is dominated by relativistic species, due to Boltzmann suppression of non-relativistic ones) as follows:
\begin{eqnarray}
s = \sum_i s_i \equiv \frac{2\pi^2}{45} g^s_{\star}(T) T^3\,,
\label{eq:srelativistic}
\end{eqnarray}
where the effective number of entropy degrees of freedom $g^s_{\star}$ is defined analogously to $g_{\star}$ as~\cite{Bergstrom:1999kd,Dodelson:2003ft,Mukhanov:2005sc,Durrer:2008eom,
Weinberg:2008zzc,Lesgourgues:2018ncw}:
\begin{eqnarray}
g^s_{\star}(T) \equiv \sum_{i={\rm bosons}}\Theta(T-m_i)g_i \left ( \frac{T_i}{T} \right )^3 + \frac{7}{8}\sum_{j={\rm fermions}}\Theta(T-m_j)g_j \left ( \frac{T_j}{T} \right )^3\,.
\label{eq:gseff}
\end{eqnarray}
As $g_{\star}$, also $g^s_{\star}$ remains roughly constant except for noticeable drops during the EW phase transition, the QCD phase transition, and $e^+e^-$ annihilation, reflecting the evolution of the particle content of the primordial plasma. The evolution of $g^s_{\star}$ as a function of temperature is plotted in Fig.~\ref{fig:geff} (dashed line). For adiabatic expansion, the total entropy of the Universe is conserved, \textit{i.e.} $d(sa^3)=0$, from which $s \propto a^{-3}$. This implies that the temperature of Universe scales as~\cite{Bergstrom:1999kd,Dodelson:2003ft,Mukhanov:2005sc,Durrer:2008eom,
Weinberg:2008zzc,Lesgourgues:2018ncw}:
\begin{eqnarray}
T \propto \frac{1}{\sqrt[3]{g^{s}_{\star}}a}\,.
\label{eq:}
\end{eqnarray}
Therefore, the temperature of the primordial plasma usually scales as $1/a$, decreasing as the Universe expands adiabatically. When particle/antiparticle pairs annihilate (or phase transition occurs), entropy is released to the thermal bath (and hence to any particle coupled to photons): this results in a small sudden temperature jump, and the temperature of the plasma decreases less slowly as $1/\sqrt[3]{g^{s}_{\star}}a$, until the annihilation process/phase transition is over, at which point the cooling reverts to the previous $T \propto 1/a$ behaviour. On the other hand, decoupled particles do not enjoy the entropy injection and hence keep cooling as $T \propto 1/a$, remaining cooler than photons. As we shall see in Chapter~\ref{chap:4}, this is particularly important for neutrinos, as they decouple around the time of $e^+e^-$ annihilation and hence do not enjoy the injection of entropy from this process: as a result, today $T_{\nu} = (4/11)^{1/3}T_{\gamma}$.

So far we have discussed equilibrium thermodynamics. Equilibrium holds as long as the rate of a given reaction, $\Gamma$, is larger than the Hubble rate, $H$. When $\Gamma \sim H$, the reaction is said to freeze-out. When all the reactions keeping a given particle in equilibrium freeze-out, the particle decouples from the primordial plasma, is no longer in thermodynamic equilibrium, and free-streams. In this regime, the evolution of the particle's phase space distribution $f$ obeys the Boltzmann equation:
\begin{eqnarray}
{\cal L}[f] = {\cal C}[f]\,,
\label{eq:boltzmann}
\end{eqnarray}
where the Liouville operator ${\cal L}$ is a total derivative with respect to time and ${\cal C}$ is the collision operator. There are many excellent texts which do justice to the wonderful subject which is the Boltzmann equation and its applications to cosmology. The interested reader might want to consult e.g.~\cite{Bergstrom:1999kd,Dodelson:2003ft,Mukhanov:2005sc,Durrer:2008eom,
Weinberg:2008zzc,Lesgourgues:2018ncw}. I will not discuss this topic further here, as even a sensible discussion would basically require at least a hundred pages. Here, I just want to point out that Eq.~(\ref{eq:boltzmann}) is used to track the evolution of the phase space distributions of \textit{all} particles throughout the expansion history of the Universe. For each particle species (baryons, dark matter, dark energy, photons, neutrinos) it results in a set of coupled differential equations for the evolution of their density and velocity perturbations.

Solving the Boltzmann equations efficiently has become much an art as it is science, and is commonly done through so-called Boltzmann solvers. Two state-of-the-art examples of Boltzmann solvers, widely used in the community, are \texttt{CAMB}~\cite{Lewis:1999bs} and \texttt{CLASS}~\cite{Blas:2011rf}. The former is written in \texttt{Fortran}, whereas the latter is written in \texttt{C} (and both can be run through \texttt{Python} wrappers), and both are based on the line-of-sight approach developed in~\cite{Seljak:1996is} and for the first time applied in the \texttt{CMBFAST} code. Other Boltzmann solvers, no longer maintained or in use, are \texttt{DASh}~\cite{Kaplinghat:2002mh} and \texttt{CMBEASY}~\cite{Doran:2003sy}. There are several other important examples of Boltzmann codes, most of which are based on \texttt{CAMB} or \texttt{CLASS}: some of these are appropriately targeted for testing models of modified gravity, are \texttt{MGCAMB}~\cite{Hojjati:2011ix}, \texttt{ISITGR}~\cite{Dossett:2011tn}, \texttt{EFTCAMB}~\cite{Hu:2013twa,Raveri:2014cka,Hu:2014oga,
Hu:2014sea,Frusciante:2016xoj}, \texttt{hi\_class}~\cite{Zumalacarregui:2016pph}, whereas others are targeted for specific observables or theories~\cite{Bean:2006up,Zuntz:2008zz,Barreira:2012kk,Avilez:2013dxa,
DiDio:2013bqa,Hlozek:2014lca,Renk:2016olm,Zucca:2016iur,Stocker:2018avm,Casalino:2018mna} (see~\cite{Bellini:2017avd} for a recent comparison of Boltzmann solvers for theories beyond General Relativity and a more general overview of these codes).

\subsection{Brief thermal history of the Universe}
\label{subsec:history}

We now have the theoretical tools to understand the thermal history of the Universe. We know the phase space distribution of particles in thermal equilibrium, whereas we can track the distribution of decoupled particles through the Boltzmann equation. I will now describe the main events which occurred during the history of the Universe. Recall that early on the Universe was radiation dominated.
\begin{itemize}
\item \textbf{Baryogenesis}. Presumably at very early times baryogenesis occurred, resulting in our having significantly more matter than antimatter today. There are several viable baryogenesis models~\cite{Sakharov:1967dj,Trodden:1998ym,Riotto:1998bt}, although none of them have been experimentally verified to date, so it is unclear at what point baryogenesis took place (assuming it did). However, it is likely to have occurred above the electroweak phase transition, $T \gtrsim 125\,{\rm GeV}$.
\item \textbf{Electroweak phase transition}. At a temperature of $T \approx 125\,{\rm GeV}$, corresponding to a redshift $z \approx 10^{15}$, the Higgs field acquired a non-zero vacuum expectation value, breaking EW symmetry and providing masses to most particles~\cite{Englert:1964et,Higgs:1964ia,Higgs:1964pj,Higgs:1966ev,Guralnik:1964eu,Aad:2012tfa,Chatrchyan:2012xdj}. At this point the Universe is about $10^{-11}\,{\rm s}$ old.
\item \textbf{QCD phase transition}. At a temperature of $T \approx 100\,{\rm MeV}$, corresponding to a redshift $z \approx 10^{12}$, the QCD phase transition occurred~\cite{Gross:1973ju}. During this phase, quarks confine and form hadrons and mesons, thus substantially reducing $g_{\star}$ and $g^s_{\star}$. At this point the Universe is about $10^{-5}\,{\rm s}$ old.
\item \textbf{Neutrino decoupling}. At a temperature of $T \approx 1\,{\rm MeV}$, corresponding to a redshift $z \approx 5 \times 10^{9}$, the typical rate for weak interactions drops below the Hubble rate~\cite{Hannestad:1995rs,Mangano:2005cc,deSalas:2016ztq,Escudero:2018mvt}. As a result, weak interactions freeze out and neutrinos, previously in equilibrium with the primordial plasma, decouple and start free-streaming. At this point the Universe is about $1\,{\rm s}$ old.
\item \textbf{Electron-positron annihilation}. When the temperature of the Universe drops below the electron mass, $T \approx 0.5\,{\rm MeV}$, corresponding to a redshift $z \approx 3 \times 10^{9}$, the annihilation/pair-production process $e^++e^- \to \gamma \gamma$ can only proceed in the forward direction, as the reverse direction becomes energetically unfavourable. As a result, the electron/positron entropy is transferred to the photons (which thus cool a bit slower than $a^{-1}$, as we have seen earlier), but not to the neutrinos, since they are decoupled.~\footnote{In reality, as we shall see in Chapter~\ref{chap:4}, neutrino decoupling is not instantaneous and neutrinos were not completely decoupled by the time electron-positron annihilation occurred~\cite{Dicus:1982bz,Dodelson:1992km,Dolgov:1992qg,Fields:1992zb,
Hannestad:1995rs,Fornengo:1997wa,Dolgov:1998sf,Esposito:2000hi,Dolgov:2002wy}, meaning that they still gained some of the entropy resulting from electron-positron annihilation~\cite{Seljak:1996is,Mangano:2001iu,Mangano:2005cc}.} As we will show in Chapter~\ref{chap:4}, this results in the neutrino temperature being lower than the photon temperature by a factor of $(4/11)^{1/3}$, which can be derived by entropy conservation considerations. At this point the Universe is about $6\,{\rm s}$ old.
\item \textbf{Big Bang Nucleosynthesis}. At a temperature of $T \approx 100\,{\rm keV}$, corresponding to a redshift $z \approx 4 \times 10^8$, the synthesis of light elements (mostly $^4{\rm He}$) begins thanks to nuclear reactions binding nucleons into light nuclei, in a process known as \textit{Big Bang Nucleosynthesis} (BBN)~\cite{Tytler:2000qf,Fields:2006ga,Steigman:2012ve,Fields:2014uja}. At this point the Universe is about $3\,{\rm min}$ old. The yields of the light elements depend strongly on the energy density of baryons and radiation, and are in excellent agreement with observations.~\footnote{There is one notable exception to this statement, and it is the fact that the observed abundance of $^7{\rm Li}$ lies below the theoretical predictions of BBN. This is known as the \textit{Lithium problem} (see e.g.~\cite{Fields:2011zzb} for a review and e.g.~\cite{Poulin:2015woa,Salvati:2016jng} for proposed solutions).}
\item \textbf{Matter-radiation equality}. Matter-radiation equality is defined as the time when the contributions of matter and radiation to the right-hand side of the first Friedmann equation, Eq.~(\ref{eq:hz}), are equal. Ignoring neutrinos (we will reinsert them into the picture in Chapter~\ref{chap:4}), we see that this occurs at a redshift $z_{\rm eq} = \Omega_m/\Omega_r-1 \approx 3400$, at a temperature of $T \approx 0.75\,{\rm eV}$. At this point the Universe is about $60000\,{\rm yrs}$ old.
\item \textbf{Recombination}. At a temperature of $T \approx 0.3\,{\rm eV}$, corresponding to a redshift $z \approx 1100$, the reaction $e^-+p^+\to H+\gamma$ (with $H$ neutral Hydrogen) ceases to be in equilibrium, and becomes energetically favourable only in the forward direction. The net effect is that nuclei capture free electrons and form Hydrogen and Helium atoms. From this point on the Universe ceases to be ionized and opaque to radiation. At this point the Universe is about $370000\,{\rm yrs}$ old.
\item \textbf{Photon decoupling}. At a temperature of $T \approx 0.25\,{\rm eV}$, corresponding to a redshift $z_{\rm dec} \approx 1090$, the rate for the process of Thomson scattering $e^-+\gamma \to e^-+\gamma$ drops below the Hubble rate, mostly due to the density of free electrons dropping significantly as most of them recombine with protons to form neutral Hydrogen (see above). As a result, photons decouple and start free-streaming. They travel (almost) unimpeded until us, forming what we usually refer to as the Cosmic Microwave Background (CMB): a snapshot of the infant Universe and an incredible mine of information both on cosmology and fundamental physics. At this point the Universe is about $380000\,{\rm yrs}$ old.
\item \textbf{Drag epoch}. Even though photons have already decoupled, the small baryon-to-photon-ratio $\eta \sim 10^{-9}$ keeps the baryons coupled to the photons for a small amount of time after decoupling. The drag epoch is defined as the time when baryons stop feeling the photon drag and hence are released from the photons. This occurs at a temperature of $T \approx 0.20\,{\rm eV}$, corresponding to a redshift $z_{\rm drag} \approx 1060$. At this point the Universe is about $400000\,{\rm yrs}$ old.
\item \textbf{Dark ages}. From the drag epoch until the subsequent reionization, the Universe is transparent to radiation: this period is referred to as the ``dark ages''~\cite{MiraldaEscude:2003yt,Natarajan:2014rra,Furlanetto:2019jso}.
\item \textbf{Reionization}. When the first stars form, the ensuing UV radiation reionizes neutral Hydrogen in the intergalactic medium. As a result the Universe returns to being (partially) opaque to radiation. About $5\%$ of the CMB photons are rescattered by the ionized electrons. When exactly reionization occurred is not yet known to high accuracy, but we believe it occurred when the temperature of the Universe was about $5\,{\rm meV}$, corresponding to a redshift $z \approx 15$~\cite{Barkana:2000fd,Zaroubi:2012in}. At this point the Universe is about $200\,{\rm Myrs}$ old.
\item \textbf{Matter-dark energy equality}. Analogously to how we defined matter-radiation equality, we can define matter-dark energy equality as the time when the contributions of matter and dark energy to the right-hand side of the first Friedmann equation, Eq.~(\ref{eq:hz}), are equal. At this point dark energy takes over and the Universe starts accelerating. For a cosmological constant, we see that this occurs at a redshift $z_{\Lambda} = \Omega_\Lambda/\Omega_m-1 \approx 0.3$, at a temperature of $T \approx 0.75\,{\rm eV}$. At this point the Universe is about $9\,{\rm Gyrs}$ old.
\item \textbf{Today}. Today, the temperature of the Universe is of $T \approx 0.24\,{\rm meV}$, redshift is $z=0$ by definition, and the Universe is about $13.8\,{\rm Gyrs}$ old. As we saw in Chapter~\ref{chap:2}, the Universe today is made up for about $73\%$ by dark energy, for about $23\%$ by dark matter, and for less than $4\%$ by baryons~\cite{Ade:2015xua,Aghanim:2018eyx}.
\end{itemize}

\subsection{Inflation}
\label{subsec:inflation}

Before closing this Chapter and discussing how neutrinos fit within the picture discussed, I want to perform a brief qualitative digression to introduce the concept of \textit{inflation}, a hypothetical period of early accelerated expansion postulated to address a series of difficulties faced by the hot Big Bang model described thus far and which might have seeded the initial density fluctuations which later grow under the effect of gravity to form the structure we observe today. In some sense, inflation provides the initial conditions for the hot Big Bang, and in cosmology when we refer to Big Bang we usually really refer to inflation. Despite not being directly connected to neutrino cosmology, a qualitative understanding of inflation will be necessary to understand Paper~V, where we have studied whether our ignorance of neutrino properties affects the conclusions we draw about inflation, and hence the initial conditions of the Universe.

Notwithstanding the cosmological principle, observationally we know that the Universe is far from homogeneous. The density of the Universe features fluctuations around the mean density. We can imagine decomposing these fluctuations in terms of their (comoving) scale $\lambda$ (or equivalently Fourier modes $k$), which stays fixed as the Universe expands. As long as $\lambda>\chi_h$, with $\chi_h$ the comoving particle horizon given in Eq.~(\ref{eq:comovingparticlehorizon}), the mode is said to be \textit{super-horizon}. It remains frozen, since causal physics cannot act on it. On the other hand, as the horizon increases, more and more modes enter the horizon, and become \textit{sub-horizon}. At this point, they are no longer frozen and can be acted upon by causal physics (for instance, the effects of gravity and pressure).

We have already seen earlier that $\chi_h(z) \propto (1+z)^{-1}$ during radiation domination and $\chi(z) \propto (1+z)^{-1/2}$ during matter domination. More generally, consider a Universe dominated by a single fluid with equation of state $w$. Then, it is trivial to show that as long as $1+3w>0$, $\chi_h(z)$ grows as $(1+z)^{-(1+3w)/2}$. Notice that, from Eq.~(\ref{eq:friedmann2}), $1+3w>0$ ensures that $\ddot{a}<0$ and hence the Universe is decelerating. This implies that as long as the Universe is decelerating (which is the case for the conventional eras of radiation and matter domination), the comoving horizon is a monotonically increasing function of time (on the other hand, for dark energy domination, $\chi_h(z)$ decreases, and hence in the future we will be able to see increasingly less of the Universe). A consequence is that photons which are causally disconnected at a given redshift could never have been in causal contact before. The previous observation is particularly problematic in the case of CMB. Indeed, we observe the CMB to be remarkably uniform, to the level of $10^{-1}$, \textit{across the whole sky}. However, the particle horizon at the time of decoupling would only subtend an angle of about $1^{\circ}$ on the sky if the previous expansion history were only due to the conventional radiation and matter-dominated eras. Patches separated by more than $1^{\circ}$ on the sky have not had time to causally interact until then, thus making it surprising that they share the same temperature to such an accuracy. This is known as the \textit{horizon problem}~\cite{Lyth:1998xn,Liddle:1999mq,Riotto:2002yw,Tsujikawa:2003jp,Linde:2005ht,
Linde:2007fr,Kinney:2009vz,Baumann:2009ds,Senatore:2016aui,Vazquez:2018qdg}.

It is also worth taking a closer look at Eq.~(\ref{eq:omega}). In a Universe filled with a single fluid with equation of state $w$, it is trivial to show that $\vert \Omega - 1 \vert \propto \Omega_k \propto \vert k \vert(1+z)^{-1+3w}$. Considering again the conventional radiation and matter dominated eras, where $1+3w>0$, we see that $d\Omega_k/da>0$. Therefore, as we go back in time $\Omega=1$ is a past attractor, and the Universe gets closer and closer to flat. On the other hand, even a small amount of curvature in the early Universe gets disproportionately blown up as time increases. Observations have indicated that our Universe is remarkably close to flat. To explain this observation, it is necessary that the early Universe was flat to $\vert 1-\Omega_k \vert < 10^{-55}$! While it could well be that the initial conditions of the Universe were such that it was flat to such a degree, most people would agree that this looks like an unnatural fine-tuning, leading to what is known as the \textit{flatness problem}~\cite{Lyth:1998xn,Liddle:1999mq,Riotto:2002yw,Tsujikawa:2003jp,Linde:2005ht,
Linde:2007fr,Kinney:2009vz,Baumann:2009ds,Senatore:2016aui,Vazquez:2018qdg}.

Remarkably, both the horizon and flatness problem can be solved if we introduce an early era of accelerated expansion, \textit{i.e.} a period where the Universe was dominated by a fluid with $w<-1/3$, known as \textit{inflation}. The first inflationary models were proposed in a series of seminal papers in the early 1980s~\cite{Starobinsky:1980te,Kazanas:1980tx,Guth:1980zm,Sato:1981ds,Mukhanov:1981xt,Linde:1981mu,Albrecht:1982wi}. For a quasi-de Sitter expansion where $w \sim -1$, the Hubble rate is approximately constant, whereas the scale factor grows exponentially, implying that the physical particle horizon blows up exponentially. Therefore, we can imagine starting with a tiny patch wherein causality has been established by physical processes. Then this patch is exponentially blown and can constitute our whole observable Universe. Thinking in terms of comoving scales instead of physical scales, the comoving horizon given in Eq.~(\ref{eq:comovingparticlehorizon}) decreases during an era of exponential expansion. Therefore, as time goes on, comoving scales previously in causal contact progressively exit the horizon. They subsequently re-enter the horizon as the latter grows during the conventional radiation and matter dominated eras. Provided inflation lasted sufficiently long, scales corresponding to the particle horizon at the time of recombination were once in causal contact before they exited the horizon, explaining the remarkably uniform temperature of the CMB. In fact, if inflation lasted sufficiently long, even the largest scales we observe today (and scales which are still super-horizon today) might have been in causal contact early on. As for the flatness problem, we see that for $w<-1/3$, $d\Omega_k/da<0$ and $\Omega=1$ becomes an attractor. In other words, by exponentially blowing up physical scales, inflation ``flattens'' our Universe or can at least trick us into believing it is flat, given the very large curvature radius.~\footnote{This is the same principle which brings flat-Earthers think the Earth is flat! An increasingly large curvature radius makes it increasingly harder to detect curvature.} The duration of inflation is typically quantified in terms of e-folds $N$, where $N \equiv \log(a_{\rm end}/a_{\rm in})$, with $a_{\rm end}$ and $a_{\rm in}$ the scale factor at the end and at the beginning of inflation. To solve the horizon problem, at least $45$ e-folds of inflation are required. On the other hand, the contribution of curvature to the expansion rate is reduced by $e^{-2N}$.

Models for inflation abound in the literature, including both particle physics models and models based on modifications of gravity. For an incomplete list of works dealing with inflationary model-building (which goes far from doing justice to the wide variety of well-motivated existing inflationary models, both particle models and modified gravity ones, but which should give the reader an idea of how wide the arena of inflationary models is), see for example~\cite{Starobinsky:1980te,Kazanas:1980tx,Guth:1980zm,Sato:1981ds,Mukhanov:1981xt,Linde:1981mu,Albrecht:1982wi,
Ellis:1982ed,Steinhardt:1984jj,Abbott:1984ba,Linde:1986fc,Silk:1986vc,
Turok:1987pg,Ford:1989me,Freese:1990rb,Adams:1992bn,Borde:1993xh,Vilenkin:1994pv,Ross:1995dq,Lazarides:1995vr,Berera:1995ie,Faraoni:1996rf,
Binetruy:1996xj,Peebles:1998qn,Dvali:1998pa,ArmendarizPicon:1999rj,Maartens:1999hf,
Nojiri:2000gb,Nojiri:2003ft,Kachru:2003sx,ArkaniHamed:2003uz,BlancoPillado:2004ns,Boubekeur:2005zm,Kinney:2005vj,Anisimov:2005ne,
Nojiri:2005pu,Capozziello:2005tf,Dimopoulos:2005ac,
Savage:2006tr,Ferraro:2006jd,Bezrukov:2007ep,Cognola:2007zu,Freese:2008if,Silverstein:2008sg,Kaloper:2008fb,
Bessada:2009ns,Germani:2010hd,Maleknejad:2011jw,Kobayashi:2011nu,Visinelli:2011jy,Endlich:2012pz,Martin:2013tda,Kallosh:2013hoa,
Dong:2013swa,Sebastiani:2013eqa,Czerny:2014wza,Freese:2014nla,Marchesano:2014mla,Rinaldi:2014gua,Nojiri:2014zqa,Rinaldi:2014gha,Visinelli:2014qla,
Kannike:2015apa,Myrzakulov:2015fra,DeLaurentis:2015fea,Chakraborty:2015qga,Myrzakulov:2015qaa,Rinaldi:2015yoa,Sepehri:2015eea,
Sebastiani:2015kfa,Kappl:2015esy,Rinaldi:2015uvu,Cognola:2016gjy,Barenboim:2016mmw,Visinelli:2016teo,Visinelli:2016rhn,
Ballesteros:2016xej,Nojiri:2017ncd,Sebastiani:2017mkv,Odintsov:2017hbk,Freese:2017ace,
Odintsov:2018ggm,Kleidis:2018fdu,Achucarro:2018vey,Kehagias:2018uem,Kinney:2018nny,Haro:2019gsv,Nojiri:2019dqc,Chowdhury:2019otk,Vicentini:2019etr} and pedagogical reviews~\cite{Lyth:1998xn,Liddle:1999mq,Riotto:2002yw,Tsujikawa:2003jp,Linde:2005ht,
Linde:2007fr,Kinney:2009vz,Baumann:2009ds,Senatore:2016aui,Vazquez:2018qdg}. Most models, however, typically posit the existence of a scalar field (the \textit{inflaton} field $\phi$) rolling down along a potential. If the field moves sufficiently slowly (\textit{i.e.} its kinetic energy is sub-dominant with respect to its potential energy), its effective equation of state is close to $-1$, leading to a quasi-de Sitter expansion. However, inflation must end at some point. This typically occurs when the potential steepens and the kinetic energy of the inflaton starts to dominate. In most models, eventually the inflaton reaches the bottom of the potential, and transfers its energy to SM particles through a process known as \textit{reheating}. Presumably reheating occurred at very high temperature (\textit{i.e.} well above the electro-weak scale), but cosmological observations actually only tell us that reheating should have occurred at least $5\,{\rm MeV}$, in order not to disrupt successful BBN~\cite{Hannestad:2004px,deSalas:2015glj} (see~\cite{Mielczarek:2010ag,Dai:2014jja,Munoz:2014eqa,Domcke:2015iaa,Drewes:2015coa} for other important works dealing with cosmological constraints on reheating).

Finally, besides solving the horizon and flatness problems, inflation might also be responsible for the generation of primordial density perturbations which we observe as temperature anisotropies in the CMB, and which later grow under the effect of gravity to form the large-scale structure we observe today: galaxies, clusters, super-clusters, voids, walls, sheets, and filaments~\cite{Dodelson:2003ft}. This idea was first developed by Mukhanov and Chibisov in~\cite{Mukhanov:1981xt}, and later in~\cite{Mukhanov:1982nu,Hawking:1982cz,Starobinsky:1982ee,Guth:1982ec,Bardeen:1983qw} during the course of the 3-week Nuffield Workshop at the University of Cambridge (see also~\cite{Mukhanov:1988jd,Mukhanov:1990me,Mukhanov:2013tua} for later important work, and~\cite{Riotto:2002yw,Langlois:2010xc} for reviews). Heuristically, quantum fluctuations $\delta \phi$ naturally lead to inflation lasting slightly different amounts of time in different regions of the Universe. This leads to fluctuations in curvature perturbations ${\cal R}$, which in turn can be related to fluctuations in the density field $\delta$. One can then take the Fourier transform of these fluctuations, which are Gaussian distributed with mean zero, and uncorrelated among modes with different wavelengths. The variance of each Fourier mode can be obtained by computing their power spectrum, which quantifies the amount of fluctuations on any given scale.

It has been shown in classical papers~\cite{Mukhanov:1981xt,Mukhanov:1982nu,Hawking:1982cz,
Starobinsky:1982ee,Guth:1982ec,Bardeen:1983qw} that inflation generically predicts a nearly scale-invariant primordial power spectrum of curvature perturbations. This is typically parametrized through a dimensionless primordial power spectrum of curvature perturbations, ${\cal P}_{\cal R}(k)$, as follows:
\begin{eqnarray}
{\cal P}_{\cal R}(k) \equiv A_s \left ( \frac{k}{k_{\star}} \right )^{n_s-1}\,,
\label{eq:pr}
\end{eqnarray}
where $A_s$ quantifies the amplitude of the primordial power spectrum, $n_s$ quantifies its tilt, and $k_{\star}$ is a pivot scale (typically $k_{\star}=0.05\,{\rm Mpc}^{-1}$). A nearly scale-invariant power spectrum has $n_s \approx 1$, and most inflation models in fact predict a slightly ``red'' spectrum, with $n_s<1$. Observations indicate that $n_s \simeq 0.96$, thus strengthening the case for inflation~\cite{Martin:2013tda,Okada:2014lxa,Martin:2015dha,
Huang:2015cke,Chowdhury:2019otk}.~\footnote{Notice that the notation ${\cal P}$ really refers to the dimensionless power spectrum, which quantifies the excess of power in a bin of width $dk$ centered in $k$. The dimensionless power spectrum is related to the actual power spectrum $P$ by ${\cal P} \propto k^3P$. Therefore, a scale-invariant curvature power spectrum scales as $P(k) \propto k^{-3}$. This reflects the fact that we are taking a 3D Fourier transform, and therefore that the variance of a mode should scale as $k^{-3}$ to compensate for the fact the number of modes within a given volume scales as $k^3$. Curvature perturbations are directly related (and in fact proportional) to gravitational potential $\Phi$. From the Poisson equation (see e.g.~\cite{Dodelson:2003ft,Mukhanov:2005sc,Durrer:2008eom,
Weinberg:2008zzc,Lesgourgues:2018ncw}) we know that matter perturbations $\delta$ are related to gravitational potentials (in Fourier space) through $\delta \propto k^2\Phi$. Therefore, we expect a scale-invariant curvature perturbation power spectrum to lead to a primordial power spectrum of density fluctuations $P_{\delta}(k) \propto k^4 \times k^{-3} \propto k$. From now on, unless otherwise specified, when we say ``power spectrum'' and use the notation $P(k)$, we shall be referring to the power spectrum of density fluctuations $P_{\delta}(k)$.}

\section{The concordance $\Lambda$CDM model}
\label{sec:concordance}

We now have all the ingredients in place to discuss the concordance $\Lambda$CDM model we had already anticipated in Chapter~\ref{chap:2}. That is, the mathematical framework describing the evolution of the Universe to the best of our understanding. We have basically covered all the equations describing the $\Lambda$CDM model, so all that remains to discuss are the free parameters of the model itself. In its minimal incarnation, the $\Lambda$CDM model has six free parameters. These are:
\begin{itemize}
\item The physical baryon density $\omega_b \equiv \Omega_bh^2$, where baryons consist mostly of Hydrogen and Helium.
\item The physical cold dark matter density $\omega_c \equiv \Omega_ch^2$, where dark matter is assumed to be pressureless, stable, and non-interacting.
\item The amplitude of the primordial power spectrum, $A_s$, evaluated at the pivot scale $k_{\star}=0.05\,{\rm Mpc}^{-1}$. Note that in practice, usually one works with $\ln(10^{10}A_s)$, since $A_s$ takes values of order $\approx 10^{-9}$.
\item The tilt of the primordial power spectrum $n_s$, evaluated at the same pivot scale.
\item The angular size of the sound horizon at decoupling $\theta_s = r_s(z_{\rm dec})/\chi_{\star}$, with $z_{\rm dec}$ the redshift of decoupling and $\chi_{\star}$ the comoving distance to the CMB.
\item The optical depth to reionization $\tau$, quantifying the amount of reionization which the Universe underwent.
\end{itemize}
In the spirit of Occam's razor, these 6 parameters constitute the minimal set of parameters required to describe current cosmological observations to high precision (or at least, no one has come up with either a more satisfying model with less parameters, or a model featuring additional parameters which provides a significantly better fit to warrant the presence of these extra parameters)~\cite{Heavens:2017hkr}. The latest measurements from the 2018 reanalysis of data from the \textit{Planck} satellite has determined these six parameters to exquisite precision, with 68\% confidence regions given by $\omega_b = 0.0224 \pm 0.0001$, $\omega_c = 0.120 \pm 0.001$, $\ln(10^{10}A_s)$, $n_s = 0.965 \pm 0.004$, $100\theta_s = 1.0411 \pm 0.0003$, and $\tau = 0.054 \pm 0.007$~\cite{Aghanim:2018eyx}.~\footnote{These 6 parameters are treated as ``fundamental'': in practice, these are the parameters that are varied when analysing the data through standard by using standard Markov Chain Monte Carlo methods (see Chapter~\ref{sec:bayesianpractice} for more details). Any other parameter one can think of is either fixed (and varied only in the context of extended models), or ``derived'' from these fundamental parameters. Examples of derived parameters of common use are the Hubble constant $H_0=67.36 \pm 0.54$, the physical matter density $\omega_m = 0.1430 \pm 0.0011$, the matter density parameter $\Omega_m = 0.3153 \pm 0.0073$, the age of the Universe $t = (13.797 \pm 0.023)\,{\rm Gyr}$, the amplitude of matter fluctuations when smoothed on a scale of $8\,h^{-1}\,{\rm Mpc}$ $\sigma_8=0.8111 \pm 0.0060$, the redshift of reionization in the limit of instantaneous reionization $z_{\rm re}=7.67 \pm 0.73$, and the sound horizon at the time of baryon drag $r_{\rm drag}=(147.09 \pm 0.26)\,{\rm Mpc}$.} This model is referred to as concordance model since different observational probes of varying nature seem to point to the same values for the fundamental parameters, modulo mild tensions between high- and low-redshift probes which overall do not yet appear to undermine the consistency of the $\Lambda$CDM model.

However, this successful minimal model can be extended to include additional free parameters which are otherwise kept fixed. This approach has in fact been advocated by some, arguing that the minimal $\Lambda$CDM model does not do justice to the extremely high quality of the most recent data~\cite{DiValentino:2015ola} (see also~\cite{Giusarma:2011zq,DiValentino:2012yg,Archidiacono:2012gv,Benetti:2013wla,
Said:2013hta,Gerbino:2013ova,Gerbino:2014eqa,Cabass:2015jwe,Cabass:2016ldu,
DiValentino:2016ucb,DiValentino:2017zyq,Capparelli:2017tyx,DiValentino:2018zjj,
Yang:2018prh,Pan:2019brc} for important work on extended models). In fact, the minimal $\Lambda$CDM model was already used when analysing the 1998 data from BOOMERanG~\cite{deBernardis:2000sbo}. Rejecting extensions of the minimal $\Lambda$CDM model solely in the name of Occam's razor is not a healthy approach. For instance, fixing the sum of the neutrino masses to $0\,{\rm eV}$ or a small value is completely arbitrary and unnecessary, since cosmological data is sensitive to variations in $M_{\nu}$ of about $0.1\,{\rm eV}$. Moreover a number of well-motivated particle physics models predict a sizeable contribution of primordial gravitational waves. And finally, accepting that dark energy is a simple cosmological constant leads us to accept theoretical issues due to fine-tuning~\cite{Weinberg:1988cp,Carroll:1991mt,Carroll:2000fy,Weinberg:2000yb,
Sahni:2002kh,Peebles:2002gy,Padmanabhan:2002ji,Nobbenhuis:2004wn,Polchinski:2006gy} and the coincidence problem~\cite{Steinhardt:1999nw,Vilenkin:2001bs,Velten:2014nra}.

Some of the parameters one could consider varying in addition to the 6 base parameters include (just to mention a few) the sum of the neutrino masses $M_{\nu}$ (otherwise fixed to $M_{\nu}=0.06\,{\rm eV}$), the effective number of neutrinos $N_{\rm eff}$ (otherwise fixed to $N_{\rm eff}=3.046$; this is a parameter we will discuss in Chapter~\ref{subsec:evolutionneutrinos}), the dark energy equation of state $w$ (otherwise fixed to $w=-1$), the tensor-to-scalar ratio $r$ (otherwise fixed to $r=0$), the running of the spectral index $dn_s/d\ln k$ (otherwise fixed to $dn_s/d\ln k=0$), or the curvature density parameter $\Omega_k$ (otherwise fixed to $\Omega_k=0$). In this thesis, we will mostly be interested in $M_{\nu}$ as an additional free parameter, and hence we will mostly focus on the 7-parameter $\Lambda$CDM+$M_{\nu}$ model. Occasionally, we will consider additional extensions featuring for instance a free $w$ (Paper~I), a free $\Omega_k$ (Paper~I), a free time-varying dark energy $w(z)$ (Paper~IV), a free $r$ (Paper~V), and a free $N_{\rm eff}$ (Paper~V).

This concludes our discussion of the thermal history of the Hot Big Bang model, its problems (and how inflation solves them), and the concordance $\Lambda$CDM model. At this point, in Chapter~\ref{chap:4} we are ready to examine how neutrinos fit into the whole picture. As we shall see, the peculiar behaviour of neutrinos, a combination of their free-streaming nature and the fact that they first behave as radiation and then as matter, imprints very distinctive signatures in a set of cosmogical observables, which in turn we can use to go after neutrino properties. We will briefly review what the main observables are, and how to use them to constrain neutrino properties, especially their mass. After that, modulo a brief digression into statistical methods which we will carry out in Chapter~\ref{chap:5}, we will have all the tools in place to understand the results of this thesis.

\chapter{Massive neutrinos and how to search for them with cosmological observations}
\label{chap:4}

\begin{chapquote}{\textit{Frank Sinatra} by CAKE in \textit{Fashion Nugget} (1995)}
``We know of an ancient radiation \\
That haunts dismembered constellations \\
A faintly glimmering radio station''
\end{chapquote}

In Chapter~\ref{chap:3}, I have provided an overview of physical cosmology, in particular of the main events shaping the Universe over the course of the expansion history. If by now we are fairly confident most of these events occurred the way we imagine they occurred, we owe it to a rich suite of cosmological observations, whose precision is ever-increasing. On the other hand, there remain several open questions which near-future cosmological observations might be able to address. Some of these questions pertain to neutrinos and their unknown properties: their mass, mass ordering (also referred to as mass hierarchy), and effective number, just to mention a few. In this Chapter, I will begin by describing more in detail massive neutrinos. I will then proceed to present a selection of cosmological observations which we can use to study the Universe, focusing especially on Cosmic Microwave Background (CMB) and Large-Scale Structure (LSS) probes. Finally, I will tie everything together discussing the evolution of neutrinos during the expansion history of the Universe, how their peculiar behaviour imprints characteristic signatures in cosmological observations, and how we can use cosmological observations to learn about neutrino properties.

\section{Neutrinos and the quest for their mass}
\label{sec:massiveneutrinos}

For several years, it was widely believed throughout the community that neutrinos were massless particles. In fact, as we have seen in Chapter~\ref{sec:sm}, the Standard Model was precisely constructed in such a way as to have massless neutrinos. However, since the late 1990s, it has been widely established that neutrinos are, in fact, massive particles. We know this because of the observation of neutrino oscillations, which can only occur if at least two out of the three mass eigenstates are massive. I will now briefly sketch the standard theory of neutrino oscillations. The interested reader who wants to learn more is invited to consult more pedagogical references, e.g.~\cite{Lesgourgues:2018ncw,Maltoni:2004ei,Bilenky:1978nj,Bilenky:1987ty,
McKeown:2004yq,Visinelli:2008ds,Hernandez:2010mi,Visinelli:2014xsa,Esteban:2016qun}. Following that, I will provide an overview of the evolution of neutrinos across the expansion history of the Universe. This picture will therefore complement the thermal history of the Universe provided in Chapter~\ref{subsec:history}, by zooming in a little more detail into the role played by neutrinos.

\subsection{Neutrino oscillations}
\label{subsec:oscillations}

Neutrinos are produced by charged-current weak interactions in a definite flavour eigenstate $\nu_{\alpha}$ ($\alpha=e,\mu,\tau$), with the flavour determined by the charged lepton participating in the interaction. A flavour eigenstate $\vert \nu_{\alpha} \rangle$ is a quantum superposition of three mass eigenstates $\vert \nu_i \rangle$, with $i=1,2,3$:
\begin{eqnarray}
\vert \nu_{\alpha} \rangle = \sum_i U_{\alpha i}^{\star}\vert \nu_i \rangle\,,
\label{eq:flavour}
\end{eqnarray}
where $U$ is known as the Pontecorvo-Maki-Nakagawa-Sasaka (PMNS) matrix~\cite{Pontecorvo:1957qd,Maki:1962mu} and is the neutrino analogous of the Cabibbo-Kobayashi-Maskawa (CKM) matrix for the quarks (see e.g.~\cite{Fogli:2005cq,Giganti:2017fhf} for reviews). Let us denote the masses of the three eigenstates as $m_i$. After being produced by a source in a definite flavour eigenstate, a neutrino propagates and the different mass eigenstates pick up different phases (essentially because their phase velocities are different). The result is that at some distance away from the source there is a non-zero probability that the flavour of the arriving neutrino is different from the original one. This phenomenon is known \textit{neutrino oscillations}, and was first proposed by Bruno Pontecorvo in the late 1950s~\cite{Pontecorvo:1957cp,Pontecorvo:1967fh}, albeit through the introduction of a sterile neutrino (because only the electron neutrino was known).

The probability of a neutrino of flavour $\alpha$ turning into a neutrino of flavour $\beta$ after travelling across a distance $L$ is given by the following master formula (for a detailed derivation see e.g.~\cite{Grossman:2003eb,deGouvea:2004gd,Hernandez:2010mi,
Kayser:2012ce,Fantini:2018itu}):
\begin{eqnarray}
P(\nu_{\alpha} \to \nu_{\beta}) = \sum_{ij}U_{\alpha i}^{\star}U_{\beta i}U_{\alpha j}U_{\beta j}^{\star}e^{-i\frac{\Delta m_{ji}^2L}{2\vert \boldsymbol{p} \vert}}\,,
\label{eq:masterformula}
\end{eqnarray}
where $\Delta m_{ij}^2 \equiv m_i^2-m_j^2$ is the $i-j$ mass-squared splitting and $\boldsymbol{p}$ is the neutrino momentum. In practice, although we have three flavours and mass eigenstates, the measured values of the mixing matrix are such that for physically interesting situations it is usually only two of these eigenstates which matter at any given time. It is then instructive to consider the simplified two-family mixing case. In this case, we can parametrize the mixing matrix $U$ in terms of one mixing angle $\theta$:
\begin{eqnarray}
U = \begin{pmatrix}
    \cos \theta & \sin \theta \\
    -\sin \theta & \cos \theta
    \end{pmatrix}\,.
\label{eq:mixing}
\end{eqnarray}
In this case, the probability appearing in Eq.~(\ref{eq:masterformula}) takes a particularly simple form. Introducing convenient physical units, one finds:
\begin{eqnarray}
P(\nu_{\alpha} \to \nu_{\beta}) &=& \sin^2 \left ( 2\theta \right )\sin^2 \left [ 1.27 \left ( \frac{\Delta m^2}{{\rm eV}^2} \right ) \left ( \frac{L}{{\rm km}} \right ) \left ( \frac{E_{\nu}}{{\rm GeV}} \right )^{-1} \right ] \quad (\alpha \neq \beta) \nonumber \\
P(\nu_{\alpha} \to \nu_{\alpha}) &=& 1-P(\nu_{\alpha} \to \nu_{\beta})\,,
\label{eq:masterformulaunits}
\end{eqnarray}
where $E_{\nu}$ is the neutrino energy and $\Delta m^2$ is the mass-squared splitting between the two mass eigenstates. It is clear from Eqs.~(\ref{eq:masterformula},\ref{eq:masterformulaunits}) that the observation of neutrino oscillations requires non-zero mass-squared splittings: in the simplified two-family case, this requires at least one neutrino mass eigenstate to be massive. Notice that neutrino oscillation experiments are not sensitive to the absolute neutrino mass scale, \textit{i.e.} to the mass of the lightest eigenstate, but only to mass-squared differences. Cosmology can come to the rescue by being sensitive to the sum of the three neutrino masses $M_{\nu} \equiv \sum_i m_i$, as we shall see later.
\begin{figure}[!t]
\centering
\includegraphics[width=0.7\linewidth]{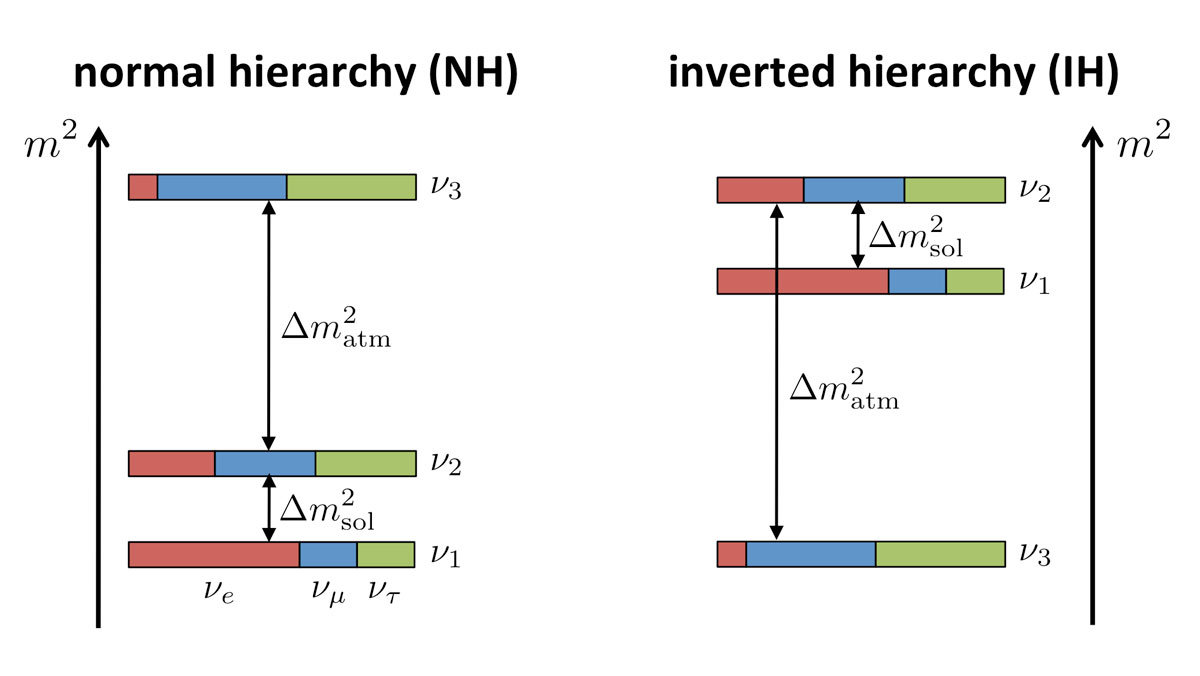}
\caption{A visual representation of the two possible neutrino mass orderings/hierarchies. On the left side, the normal ordering, where $m_1<m_2<m_3$, and the atmospheric mass-squared splitting is positive. On the right side, the inverted ordering, where $m_3<m_1<m_2$ and the atmospheric mass-squared splitting is negative. The relative proportion of red ($\nu_e$), blue ($\nu_{\mu}$), and green ($\nu_{\tau}$) in the box corresponding to the mass eigenstates quantifies the relative probability of finding the $\alpha$ flavour eigenstate in the corresponding mass eigenstate. Credits: JUNO collaboration~\cite{Juno:2017ghw}.}
\label{fig:orderings}
\end{figure}

Until recently, the only evidence of neutrino oscillations has come from solar and atmospheric neutrino oscillation experiments, which have measured to exquisite precision two non-zero mass-squared splittings: $\Delta m_{21}^2$ and $\vert \Delta m_{31}^2 \vert$, also known respectively as the solar and atmospheric mass splittings. Through thermonuclear reactions burning hydrogen into Helium, the Sun is a powerful source of MeV $\nu_e$~\cite{Robertson:2012ib,Gann:2015yta,Vissani:2017dto}. Since the time of the Homestake experiment~\cite{Davis:1968cp,Davis:1994jw}, solar neutrinos have been detected, and already then it was clear that the detected flux was lower compared to expectations (in the absence of oscillations) by about a factor of $3$~\cite{Bahcall:1976zz}. This was known as the \textit{solar neutrino problem}~\cite{Haxton:1995hv}, and it was later understood to be caused by $\nu_e \to \nu_{\mu}$ oscillations thanks to SNO~\cite{Ahmad:2002jz}.~\footnote{Interestingly, a few years later another currently unresolved problem emerged, known as the solar abundance problem~\cite{Asplund:2004eu,Asplund:2009fu,Serenelli:2009yc,Frandsen:2010yj}. This is a problem which I have worked on during my PhD~\cite{Vagnozzi:2016cmr,Vagnozzi:2017wge}.} From global fits to oscillation experiments we now know that the solar mass splitting is $\Delta m_{21}^2 \simeq 7.55 \times 10^{-5}\,{\rm eV}^2$ (see e.g.~\cite{GonzalezGarcia:2012sz,Gonzalez-Garcia:2014bfa,Gonzalez-Garcia:2015qrr,Esteban:2016qun,deSalas:2017kay,deSalas:2018bym}). Atmospheric neutrinos are instead produced when cosmic rays interact with the atoms of the atmosphere~\cite{Kajita:2012vc,Choubey:2016gps}. As solar neutrinos did, atmospheric neutrinos presented clear signs of oscillations, first observed by SuperKamiokande~\cite{Fukuda:1998mi,Kajita:2016cak}. The three experiments combining together to clarify this picture beyond any doubt were SuperKamiokande (in 1998~\cite{Fukuda:1998mi,Kajita:2016cak}), SNO (in 2001~\cite{Ahmad:2002jz,McDonald:2016ixn}), and KamLAND (in 2002~\cite{Eguchi:2002dm}). From global fits to oscillation experiments we know that the atmospheric mass splitting is larger than the solar one by about two orders of magnitude, $\vert \Delta m_{31}^2 \vert 2.5 \simeq \times 10^{-3}\,{\rm eV}^2$ (see e.g.~\cite{GonzalezGarcia:2012sz,Gonzalez-Garcia:2014bfa,Gonzalez-Garcia:2015qrr,Esteban:2016qun,deSalas:2017kay,deSalas:2018bym}).

Notice that the sign of the atmospheric mass splittings is currently unknown. This leaves open two possibilities for the neutrino mass spectrum, known as mass orderings or mass hierarchies. The first possibility, known as \textit{normal ordering} (\texttt{NO}) or \textit{normal hierarchy}, occurs when $\Delta m_{31}^2>0$, and hence $m_3 > m_1$. In this case, we have that $m_1 < m_2 < m_3$, and the sum of the three neutrino masses $M_{\nu}^{\texttt{NO}}$ (the quantity to which cosmology is sensitive) is given by:
\begin{eqnarray}
M_{\nu}^{\texttt{NO}} = m_1 + \sqrt{m_1^2+\Delta m_{21}^2} + \sqrt{m_1^2+\Delta m_{31}^2}\,.
\label{eq:mnuno}
\end{eqnarray}
The situation where $\Delta m_{31}^2<0$ and therefore $m_3 < m_1$ is instead known as the \textit{inverted ordering} (\texttt{IO}) or \textit{inverted hierarchy}. In this case $m_3 < m_1 < m_2$, and the sum of the three neutrino masses $M_{\nu}^{\texttt{IO}}$ is given by:
\begin{eqnarray}
M_{\nu}^{\texttt{IO}} = m_3 + \sqrt{m_3^2-\Delta m_{31}^2} + \sqrt{m_3^2-\Delta m_{31}^2+\Delta m_{21}^2}\,.
\label{eq:mnuio}
\end{eqnarray}
A visual representation of the two mass orderings/mass hierarchies is given in Fig.~\ref{fig:orderings}.

Let us also define $m_{\rm light}$ to be the mass of the lightest eigenstate, \textit{i.e.} $m_{\rm light}=m_1$ for \texttt{NO} and $m_{\rm light}=m_3$ for \texttt{NO}. Then, it is clear that for each of the two possible mass orderings there exists a minimal value of $M_{\nu}$, obtained by setting $m_{\rm light}=0\,{\rm eV}$. For \texttt{NO}, this minimal value is given by $M_{\nu,\min}^{\texttt{NO}} \approx 0.06\,{\rm eV}$, while for \texttt{IO} it is $M_{\nu,\min}^{\texttt{IO}} \approx 0.1\,{\rm eV}$. I suggest the reader keep the value $0.1\,{\rm eV}$ in mind as it will be a very important number in the continuation of this thesis (especially for Paper~I). In fact, since $M_{\nu} \gtrsim 0.1\,{\rm eV}$ for \texttt{IO}, it is clear that if cosmology tells us that $M_{\nu}<0.1\,{\rm eV}$ (a constraint which, as we shall see in Paper~I, is not at all far from current limits and well within the reach of cosmology in the next few years if not months!), \texttt{IO} will be to some extent excluded.

\begin{figure}[!t]
\centering
\includegraphics[width=0.7\linewidth]{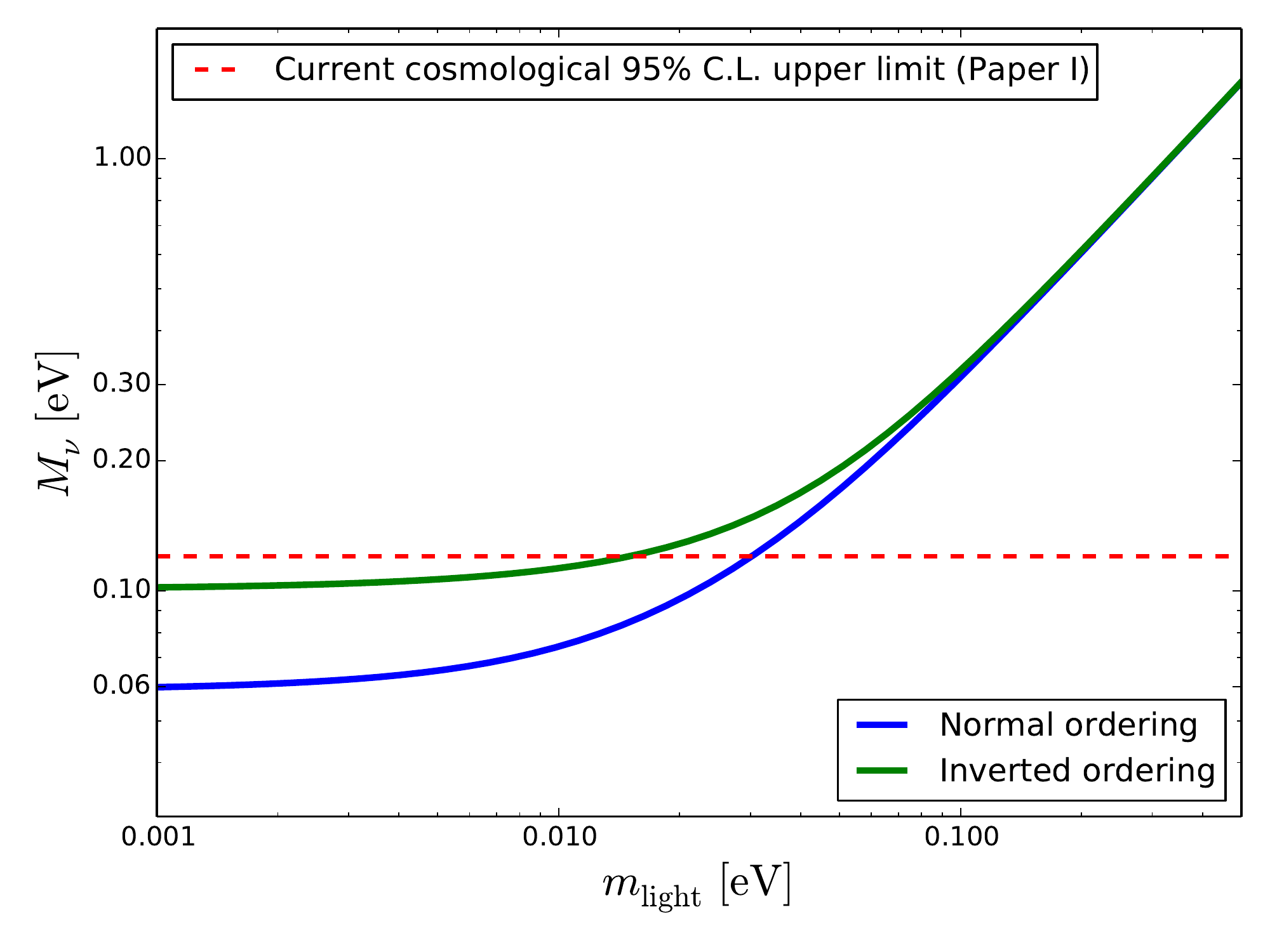}
\caption{Sum of the neutrino masses $M_{\nu}$ as a function of the mass of the lightest eigenstate $m_{\rm light}$ for \texttt{NO} (blue line) and \texttt{IO} (green line). The nearly indistinguishable width of the two lines is representative of the current $3\sigma$ uncertainties on the two mass-squared splittings. The horizontal red dashed line represents the current cosmological upper limit on the sum of the neutrino masses $M_{\nu}<0.12\,{\rm eV}$ obtained in Paper~I, \cite{Palanque-Delabrouille:2015pga}, and~\cite{Aghanim:2018eyx}.}
\label{fig:mtot_vs_mlight}
\end{figure}

Excluding \texttt{IO} would be a very important discovery, given that the mass ordering is currently unknown. Moreover, determining the mass ordering would provide more insight into the physics responsible for generating neutrino masses, and would have profound consequences in relation to the question of whether neutrinos are Dirac or Majorana~\cite{Qian:2015waa}. Plans are underway to determine the mass ordering in long-baseline experiments such as DUNE~\cite{Adams:2013qkq,Acciarri:2015uup,Acciarri:2016ooe,
Acciarri:2016crz,Strait:2016mof}, by exploiting matter effects such as the Mikheyev-Smirnov-Wolfenstein (MSW) effect~\cite{Wolfenstein:1977ue,Mikheev:1986gs,Mikheyev:1989dy}, whose result is an oscillation pattern which depends on the sign of $\Delta m_{31}^2$. Notice that these same matter effects, affecting neutrino propagation in the Sun, have allowed us to determine the sign of $\Delta m_{21}^2$. In Fig.~\ref{fig:mtot_vs_mlight} I show $M_{\nu}$ as a function of $m_{\rm light}$ for the two mass orderings: \texttt{NO} (blue) and \texttt{IO} (green). From the figure it is clear that $M_{\nu} > 0.10\,{\rm eV}$ for \texttt{IO} and $M_{\nu} > 0.06\,{\rm eV}$ for \texttt{NO}, and that current cosmological data (red) is putting \texttt{IO} under pressure. See~\cite{deSalas:2018bym} for a comprehensive overview on prospects for the determination of the mass ordering from a number of observational probes including cosmology (the discussion therein on the potential of cosmology to probe the mass ordering is partly based on our results in Paper~I).

\subsection{The history of cosmic neutrinos}
\label{subsec:evolutionneutrinos}

I will now describe in a bit more detail the evolution and peculiar behaviour of neutrinos across the expansion history of the Universe. This picture will be necessary to understand the signatures neutrinos imprint in cosmological observations (which in turn we can use to hunt these ghostly particles), which will be discussed in Chapter~\ref{sec:signaturesnu}.

In the very early Universe, neutrinos are kept in equilibrium with the primeval plasma at a temperature which is the same as that of the photons, $T_{\gamma}$, by frequent weak interactions, with typical interaction rate $\Gamma \approx G_F^2T_{\gamma}^5$, where $G_F$ is Fermi's constant. While in equilibrium, the phase-space distribution of neutrinos is given by the Fermi-Dirac distribution already seen in Eq.~(\ref{eq:distribution}):
\begin{eqnarray}
f(p,z) = \frac{g}{(2\pi)^3}\frac{1}{e^{p/T_{\nu}(z)} + 1}\,,
\label{eq:distributionfd}
\end{eqnarray}
where $g=2$ for a single neutrino species, and knowing in hindsight that neutrinos decouple when $T \approx 1\,{\rm MeV}$, we have approximated $E(p) \approx p$. The distribution depends neither on spatial coordinates, nor on the direction of the momentum, because of the assumption of homogeneity and isotropy. When the temperature of the Universe drops sufficiently, $\Gamma < H$ and weak interactions become too infrequent to keep neutrinos in equilibrium. It can be easily shown that this occurs at a temperature $T_{\nu,{\rm dec}} \approx 1\,{\rm MeV}$. Since we know from cosmology that $M_{\nu}$ is sub-eV, neutrinos decouple while ultra-relativistic. At this point, neutrinos start propagating freely. The shape of their distribution is preserved, albeit with an effective temperature $T_{\nu}(z) \propto (1+z)$. Notice that referring to $T_{\nu}(z)$ post-decoupling as a temperature is technically speaking a misnomer, since neutrinos are no longer in equilibrium. It is important to note that, because the form of the distribution is preserved, even at late times when neutrinos are non-relativistic we can neglect their mass in the distribution function.

Shortly after neutrinos decouple, electrons and positrons annihilate and release their entropy to the photon bath. However, as decoupled particles, neutrinos do not enjoy this entropy release. As a consequence, the photon temperature decreases slightly more slowly than $T \propto (1+z)$ for a reduced period of time [whereas $T_{\nu}$ continues to decrease as $(1+z)$], resulting in the photon temperature today being slightly higher than the neutrino temperature. It is a classic exercise (which I will not show here, see e.g.~\cite{Bergstrom:1999kd,Dodelson:2003ft,Mukhanov:2005sc,Durrer:2008eom,
Weinberg:2008zzc,Lesgourgues:2018ncw} for a full derivation) to use conservation of entropy as discussed in Chapter~\ref{sec:hbb}, implying that $T \propto 1/(\sqrt[3]{g^s_{\rm eff}}a)$, to show that after electron-positron annihilation the ratio between the neutrino and photon temperatures is given by $T_{\nu}/T = (4/11)^{1/3} \approx 0.71$. The current photon temperature is known to high accuracy by measuring the measurements of the CMB blackbody spectrum, and is given by $T_{\gamma,0} \simeq 2.725\,{\rm K}$. Therefore the current neutrino temperature is $T_{\nu,0} \simeq 1.95\,{\rm K}$.

At late times, when the temperature of the Universe, has dropped significantly, neutrinos become non-relativistic and start contributing to the matter budget of the Universe alongside baryons and cold dark matter. For a neutrino at temperature $T_{\nu}$, the average momentum is $\langle p \rangle \approx 3.15T_{\nu}$. Defining the non-relativistic transition redshift $z_{\rm nr}$ as the redshift when $\langle p \rangle (z) = M_{\nu}$, and using the fact that $\langle p \rangle (z) = 3.15 \times (4/11)^{(1/3)}T_{\gamma,0}(1+z)$, we find that~\cite{Lattanzi:2017ubx}:
\begin{eqnarray}
z_{\rm nr} \approx 1900 \left ( \frac{M_{\nu}}{{\rm eV}} \right )-1\,.
\label{eq:znr}
\end{eqnarray}
Therefore, neutrinos with mass $M_{\nu} \lesssim 0.6\,{\rm eV}$ become non-relativistic after recombination. Moreover, given the mass-squared splittings measured from solar and atmospheric transitions, at least two out of three neutrino mass eigenstates are non-relativistic today (whereas the lightest eigenstate could be massless and hence always relativistic).

We have already seen in Eq.~(\ref{eq:relativistic}) that the energy density of neutrinos as a function of temperature (or effective temperature) is given by:
\begin{eqnarray}
n_{\nu}(T_{\nu}) = \frac{3g\zeta(3)}{4\pi^2}T_{\nu}^3\,.
\label{eq:nnu}
\end{eqnarray}
where for a single neutrino species $g=2$ to account for particle and antiparticle. Summing over all three flavour and knowing $T_{\nu}$ today, the relic density of cosmic neutrinos is about $340\,{\rm particles}/{\rm cm}^3$. On the other hand, the energy density of neutrinos depends on whether or not they are relativistic. As a function of the effective neutrino temperature $T_{\nu}$ [see Eqs.~(\ref{eq:rhorelativistic},\ref{eq:nonrelativistic})], the neutrino energy density $\rho_{\nu}$ is given by:
\begin{eqnarray}
\rho_{\nu}(T_{\nu}) = \begin{cases}
      \frac{7\pi^2}{120}T^4 & (T_{\nu} \gg M_{\nu}) \\
      M_{\nu}n_{\nu} & (T_{\nu} \ll M_{\nu})
    \end{cases}\,,
\label{eq:rhonu}
\end{eqnarray}
consistent with the expectation that $\rho_{\nu} \propto (1+z)^4$ in the early Universe and $\rho_{\nu} \propto (1+z)^3$ in the late Universe.

It is worth taking a closer look at the neutrino energy density at early- and late-times appearing in Eq.~(\ref{eq:rhonu}). Given that the present-day neutrino number density is entirely determined by the neutrino temperature $T_{\nu}$ [see Eq.~(\ref{eq:nnu})], which in turn is entirely determined by the CMB temperature today (exquisitely measured), the energy density of neutrinos today depends only on one free parameter, $M_{\nu}$. Inserting numbers, we find that the neutrino density parameter $\Omega_{\nu}$ is given by:
\begin{eqnarray}
\Omega_{\nu} \simeq \frac{M_{\nu}}{93.14\,{\rm eV}h^2}\,,
\label{eq:omeganu}
\end{eqnarray}
While neutrinos are still relativistic (\textit{i.e.} in the early Universe), it is useful to relate their energy density to the photon energy density. Recall from Eq.~(\ref{eq:relativistic}) that the photon energy density is given by $\rho_{\gamma} = (\pi^2/15)T^4$ (since $g=2$ for photons). Given the relation between $T_{\nu}$ and $T$, we can express the total energy density in relativistic species (photons+neutrinos) in the early Universe, $\rho_r$, as:
\begin{eqnarray}
\rho_r = \rho_{\gamma} \left [ 1+\frac{7}{8} \left ( \frac{4}{11} \right )^{\frac{4}{3}}N_{\rm eff} \right ] \approx \rho_{\gamma} \left ( 1+0.2271N_{\rm eff} \right ) \,.
\label{eq:rhor}
\end{eqnarray}
The new parameter $N_{\rm eff}$ we introduced, known as the effective number of relativistic species or effective number of neutrinos, deserves a few clarifications. We would expect $N_{\rm eff}=3$, reflecting the fact that we have three neutrino species, each with $g=2$ just as the photon. In reality, $N_{\rm eff}=3$ slightly underestimates the neutrino contribution to the radiation energy density. The reason is that neutrino decoupling is not an instantaneous process, and during electron-positron annihilation neutrinos are still weakly coupled to the primeval plasma and hence receive a small part of the entropy resulting from the annihilation process~\cite{Seljak:1996is,Mangano:2001iu,Mangano:2005cc}. The net effect is to increase the total energy density of the three neutrino species, which is no longer given by $3\rho_{\nu}$ [with $\rho_{\nu}$ given by Eq.~(\ref{eq:rhonu})], but by $N_{\rm eff}\rho_{\nu}$, with $N_{\rm eff} \approx 3.046$~\cite{Mangano:2005cc} (a recent calculation revised this to $N_{\rm eff}=3.045$~\cite{deSalas:2016ztq}, but we will stick to $N_{\rm eff}=3.046$ to conform with existing literature, because near-future cosmological probes will not be sensitive to $N_{\rm eff}$ to the level where the third digit matters).

In general, many extensions of the Standard Model of Particle Physics predict the existence of additional light species in the light Universe, generically referred to ask dark radiation, which would contribute to the relativistic energy density at early times (see e.g.~\cite{Foot:1991bp,Ackerman:2008mha,Feng:2009mn,Kaplan:2009de,Archidiacono:2011gq,
Cline:2012is,Blennow:2012de,CyrRacine:2012fz,Fan:2013tia,Conlon:2013isa,
Vogel:2013raa,Fan:2013bea,Foot:2014mia,Petraki:2014uza,Marsh:2014gca,Foot:2014uba,
Foot:2014osa,Pearce:2015zca,Heikinheimo:2015kra,Chacko:2015noa,Clarke:2015gqw,
DiValentino:2016ikp,Foot:2016wvj,Agrawal:2016quu,Agrawal:2017rvu,Krall:2017xcw,
Archidiacono:2017slj,Buen-Abad:2017gxg,Baldes:2017gzu,Munoz:2018jwq} for various examples of models featuring dark radiation). It is then customary to use Eq.~(\ref{eq:rhor}) as a definition for $N_{\rm eff}$, which provides a convenient way to express the total radiation energy density. In fact, one of the simplest extensions of the $\Lambda$CDM model is the $\Lambda$CDM+$N_{\rm eff}$ model, where $N_{\rm eff}$ is a free parameter. In principle, $N_{\rm eff}$ might also be lower than the canonical value of $N_{\rm eff}=3.046$. This can occur if neutrinos have not had time to fully thermalize by decoupling, for instance if reheating occurs at very low temperature~\cite{Davidson:2000dw,Gelmini:2004ah,Ichikawa:2005vw,
Visinelli:2009kt,Freese:2017ace}. Note that cosmological data, in principle, allow for a reheating temperature as low as $5\,{\rm MeV}$~\cite{Hannestad:2004px,deSalas:2015glj}. These low-reheating models are, admittedly, more exotic. However, we will consider them in Paper~V, which is why I have considered them worth mentioning here.

I will now cover one last important point in the history of cosmic neutrinos, related to the neutrino free-streaming scale. After decoupling, neutrinos start free-streaming at  a high velocity: in other words, they move along geodesics like freely falling particles. Qualitatively, we can expect free-streaming to be extremely important, especially for structure formation. The reason is that due to their large velocities, below some scale set by the typical distance covered by a free-streaming neutrino over a Hubble time, neutrinos cannot remain confined within potential wells: this should reflect in an increased difficulty in forming structure on small scales. It is then useful to introduce a \textit{free-streaming scale} $\lambda_{\rm fs}$ (or correspondingly a free-streaming wavenumber $k_{\rm fs}$). At any given time, $\lambda_{\rm fs}$ sets the scale above which free-streaming can be neglected.~\footnote{Qualitatively, this is similar to the concept of Jeans length, which gives the scale below which pressure inhibits gravitational collapse in a fluid~\cite{Jeans:1902ghw}.} Following~\cite{Lesgourgues:2018ncw}, $\lambda_{\rm fs}$ is defined as follows:
\begin{eqnarray}
\lambda_{\rm fs}(z) \equiv \frac{2\pi}{(1+z)k_{\rm fs}} \equiv 2\pi\sqrt{\frac{2}{3}}\frac{c_{\nu}(z)}{H(z)}\,,
\label{eq:lambdafs}
\end{eqnarray}
where $c_{\nu}(z)$ is the neutrino speed as a function of redshift.~\footnote{In~\cite{Lesgourgues:2018ncw}, another quantity known as the comoving free-streaming horizon $r_{\rm fs}$ is defined as $r_{\rm fs}(z) \equiv \int_z^{\infty}dz'\,c_{\nu}(z')/H(z')$, in analogy to the particle horizon we have already seen in Eq.~(\ref{eq:comovingparticlehorizon}). It turns out that for the range of neutrino masses allowed by cosmology and oscillation experiments, \textit{i.e.} $0.06\,{\rm eV} \lesssim M_{\nu} \lesssim 0.3\,{\rm eV}$, leading to neutrinos which turned non-relativistic during matter domination, $\lambda_{\rm fs}$ and $r_{\rm fs}$ differ very little~\cite{Lesgourgues:2018ncw}. Their physical interpretation is slightly different though: whereas $\lambda_{\rm fs}$ sets the scale above which free-streaming can be neglected at any given time, $r_{\rm fs}$ sets the scale above which there is no way free-streaming could have any effect from causality arguments.}

To make progress, we need to find a convenient form for $c_{\nu}(z)$. While neutrinos are relativistic, $c_{\nu}(z) \approx 1$, while after the non-relativistic transition, introducing convenient units, $c_{\nu}(z)$ can be expressed as~\cite{Lesgourgues:2018ncw}:
\begin{eqnarray}
c_{\nu}(z) = \frac{\langle p \rangle}{M_{\nu}} \approx \frac{3.15T_{\nu}(z)}{M_{\nu}} \approx 158(1+z) \left ( \frac{M_{\nu}}{{\rm eV}} \right )^{-1}\,\frac{{\rm km}}{{\rm s}}\,.
\label{eq:cnuz}
\end{eqnarray}
As expected, $c_{\nu}(z)$ decreases as $(1+z)$ since all its redshift-dependence is encoded in the redshifting of the temperature. Inserting numbers, we see that $\lambda_{\rm fs}$ and $k_{\rm fs}$ are given by~\cite{Lesgourgues:2018ncw}:
\begin{eqnarray}
\lambda_{\rm fs} \approx 8.1(1+z)\frac{H_0}{H(z)} \left ( \frac{M_{\nu}}{{\rm eV}} \right )\,h^{-1}{\rm Mpc}\,,\nonumber \\
k_{\rm fs} \approx 0.776(1+z)^{-2}\frac{H(z)}{H_0} \left ( \frac{M_{\nu}}{{\rm eV}} \right )\,h{\rm Mpc}^{-1}\,.
\label{eq:lambdak}
\end{eqnarray}
For the range of neutrino masses allowed by cosmology and oscillation experiments, \textit{i.e.} $0.06\,{\rm eV} \lesssim M_{\nu} \lesssim 0.3\,{\rm eV}$, at least two out of three neutrinos turned non-relativistic during matter domination, when $H(z) \propto (1+z)^{3/2}$.

While during matter domination but prior to the non-relativistic transition $k_{\rm fs}(z)$ decreases as $(1+z)^{1/2}$ [see Eq.~(\ref{eq:lambdafs}) with $c=1$ and $H(z) \propto (1+z)^{3/2}$], after the non-relativistic transition $k_{\rm fs}(z)$ starts increasing $(1+z)^{-1/2}$. During the non-relativistic transition, $k_{\rm fs}$ passes through a minimum (corresponding to a \textit{maximum} free-streaming scale!). This minimum, usually denoted by $k_{\rm nr}$, sets the wavenumber above which free-streaming cannot be neglected (equivalently, the scale below which free-streaming cannot be neglected), and is found by evaluating $k_{\rm fs}$ [given by Eq.~(\ref{eq:lambdak})] at $z_{\rm nr}$ [given by Eq.~(\ref{eq:znr})]:
\begin{eqnarray}
k_{\rm nr} \approx 0.02\sqrt{\Omega_m} \left ( \frac{M_{\nu}}{{\rm eV}} \right )^{\frac{1}{2}}\,h{\rm Mpc}^{-1}\,.
\label{eq:knr}
\end{eqnarray}
It is worth noting that $k_{\rm nr}$ is numerically very similar (up to a factor of $\sqrt{3/2}$) to the wavenumber $k$ satisfying $k=a_{\rm nr}H(a_{\rm nr})$, \textit{i.e.} the wavenumber of a perturbation entering the horizon at the non-relativistic transition. In terms of physical interpretation, small-scale neutrino density fluctuations for $k>k_{\rm nr}$ are damped (and metric perturbations, \textit{i.e.} gravitational potentials, are also damped from gravitational back-reaction) and hence structure grows more slowly, because it is not possible to confine free-streaming neutrinos on small scales~\cite{Lesgourgues:2018ncw}. Modes with $k<k_{\rm nr}$ are instead never affected by free-streaming: on such scales, neutrinos behave as cold dark matter.

In summary, neutrinos exhibit a very peculiar behaviour across the expansion history of the Universe. Initially coupled to the primordial plasma through weak interactions, at $T \sim 1\,{\rm MeV}$ these interactions become too infrequent, hence neutrinos decouple and start free-streaming. At late times, at least two out of three neutrinos turn non-relativistic during matter domination, and start contributing to the matter budget of the Universe. Their free-streaming nature imprints a scale, $\lambda_{\rm nr}$ (or equivalently a wavenumber $k_{\rm nr}$), below which neutrinos cannot be kept within gravitational potentials due to their large velocities. As we shall see later, this is reflected in a suppression of structure formation on small scales, an effect increasing as we increase $M_{\nu}$, and providing one of the cleanest observational signatures of neutrino masses.

\section{Cosmological observations}
\label{sec:cosmologicalobservations}

In the following section, I will briefly describe the main cosmological observations currently being used to study the Universe. There is an extremely wide class of cosmological observations, and I cannot describe all of them in detail. For this reason, I have chosen to focus on CMB and LSS probes. Even then, my discussion will inevitably be quite limited. My aim will mostly be to endow the reader with a qualitative (and at times heuristic) understanding of the physics at play in shaping these observations, and how these observations respond to changes in the cosmological parameters. The interested reader who wants to learn more should refer to classic textbooks and reviews where such topics are covered pedagogically (e.g.~\cite{Weinberg:1972kfs,Peebles:1994xt,RowanRobinson:1996nv,Liddle:1998ew,
Peacock:1999ye,Dodelson:2003ft,Mukhanov:2005sc,Durrer:2008eom,Giovannini:2008zzb,
Weinberg:2008zzc,Bernardeau:2001qr,Percival:2013awa,Lesgourgues:2018ncw}.) For the CMB, particularly useful dedicated reviews can be found in~\cite{Hu:2001bc,Samtleben:2007zz,Bucher:2015eia,Wands:2015fua,Durrer:2015lza}.

\subsection{Cosmic Microwave Background}
\label{subsec:cmb}

As we have seen in Chapter~\ref{subsec:history}, at $z_{\rm dec} \approx 1090$, photons decouple from electrons, mostly thanks to the significantly reduced number of free electrons after recombination. From that point on, these photons (mostly) free-stream until the present time, forming what is known as the Cosmic Microwave Background (CMB). The Universe at the time of decoupling was incredibly isotropic, to about $1$ part in $10^5$. However, small anisotropies in both temperature and polarization were present: these anisotropies carry an extraordinary amount of information on the physics at $z = z_{\rm dec} \approx 1090$, and to some extent on the earlier evolution of the Universe. But there is more: since to reach us the CMB photons have had to traverse the $z<z_{\rm dec}$ Universe, they carry some (integrated) information about the post-decoupling Universe, in particular with regards to the effect of lensing from the intervening LSS, and reionization.

Let us consider the CMB temperature field as a function of angle on the sky, $T(\boldsymbol{\hat{n}})$, whose average is $T_{\rm CMB} \approx 2.725\,{\rm Kelvin}$. Let us also denote the fractional difference with respect to the mean temperature across the sky as $\Theta(\boldsymbol{\hat{n}}) \equiv (T(\boldsymbol{\hat{n}}) - T_{\rm CMB})/T_{\rm CMB}$. Since we are considering a function defined on the surface of a sphere, it makes sense to expand $\Theta(\boldsymbol{\hat{n}})$ in spherical harmonics $Y_{lm}(\boldsymbol{\hat{n}})$, as follows:
\begin{eqnarray}
\Theta(\boldsymbol{\hat{n}}) = \sum_{lm} a_{lm}Y_{lm}(\boldsymbol{\hat{n}})\,,
\label{eq:alm}
\end{eqnarray}
where the $a_{lm}$s are the expansion coefficients. For each multipole $\ell$, one has that $m=-\ell,...,\ell$. Then, assuming isotropy, we define the power spectrum of the temperature anisotropies $C_{\ell}$ (or temperature power spectrum in short) as being:
\begin{eqnarray}
C_{\ell} = \langle a_{lm}a_{lm}^{\star} \rangle\,,
\label{eq:cell}
\end{eqnarray}
where $\langle \rangle$ denotes an ensemble average. The power spectrum $C_{\ell}$ is of particular interest since it is the quantity which can be predicted by cosmological models. In other words, a given model cannot predict whether a point in the sky will be hotter and colder than the average, but it can predict the statistics of these anisotropies. The power spectrum at a given multiple $\ell$ provides information about the typical variance in temperature fluctuations at an angular scale $\theta \approx \pi/\ell$: therefore, small multipoles correspond to large angular scales, and large multipoles correspond to small angular scales.

%
%
%
%

The statistics of anisotropies in the CMB sky provide information about the physical conditions at the time the CMB was released. However, they also provide information about physical processes acting prior to decoupling (provided such processes leave a signature in the CMB - we shall see that this is the case for baryon acoustic oscillations), as well as information on the content of the Universe between decoupling and us, which can affect the propagation of CMB photons. The complete formula for the observed temperature anisotropy $\Theta(\boldsymbol{\hat{n}})\vert_{\rm obs}$ in the linear regime includes three main contributions, and was derived in a seminal paper by Sachs and Wolfe in 1967~\cite{Sachs:1967er}:
\begin{eqnarray}
\Theta(\boldsymbol{\hat{n}}) = \underbrace{\left [ \Theta_0 + \phi \right ] \vert_{\rm dec}}_{\text{Sachs-Wolfe}} + \underbrace{\boldsymbol{\hat{n}} \cdot \boldsymbol{v_b}}_{\text{Doppler}} + \underbrace{\int_{\rm decoupling}^{\rm today} \left ( \phi^{'}+\psi^{'} \right )}_{\text{Integrated Sachs-Wolfe}}\,.
\label{eq:sachs}
\end{eqnarray}
In the above, the Sachs-Wolfe contribution (first two terms) includes a contribution from the intrinsic temperature fluctuation at decoupling $\Theta_0$, and the gravitational Doppler shift due to the gravitational potential $\phi$ at the time of decoupling (in other words, photons sitting in a gravitational potential at decoupling need to climb out of it to reach us, losing energy in the process, and viceversa for photons sitting in an underdensity).~\footnote{In Eq.~(\ref{eq:sachs}), $\phi$ and $\psi$ denote the two gravitational potentials, commonly utilized to describe scalar perturbations to the FLRW metric in the Newtonian gauge. In the presence of scalar perturbations and in the Newtonian gauge, the FLRW line element Eq.~(\ref{eq:flrw}) is modified to:
\begin{eqnarray}
ds^2 = (1-2\phi)dt^2-(1-2\psi)a^2(t) \left [ \frac{dr^2}{1-kr^2}+r^2(d\theta^2 + \sin^2\theta d\phi^2) \right ]\,.
\label{eq:perturbations}
\end{eqnarray}
In the absence of anisotropic stress, $\phi = -\psi$.} The Sachs-Wolfe contribution is dominant on large scales, where knowledge of the microphysics involved is irrelevant. The second contribution (third term) is a standard Doppler shift, where $\boldsymbol{v_b}$ denotes the peculiar velocity of the photon-baryon fluid from which photons are emitted when they decouple. The third contribution is the integrated Sachs-Wolfe (ISW) effect, and is driven by the time variation of the gravitational potentials $\phi$ and $\psi$ between us and decoupling. In a purely matter-dominated Universe, the gravitational potentials $\phi$ and $\psi$ are constant and there is no ISW term~\cite{Bergstrom:1999kd,Dodelson:2003ft,Mukhanov:2005sc,Durrer:2008eom,
Weinberg:2008zzc,Lesgourgues:2018ncw}. Therefore, the ISW effect receives contributions from two epochs: just before decoupling, because the Universe was not yet completely matter-dominated (in a radiation dominated Universe potentials decay); and at late times, when dark energy comes to dominate, again causing potentials to decay. I will now discuss more in detail the physics determining the shape of the temperature power spectrum, discussing first primary anisotropies, generated by processes operating at recombination or earlier [in other words, the physics behind $\Theta_0$ in Eq.~(\ref{eq:sachs})].

\subsubsection{Primary anisotropies}
\label{subsubsec:primary}

The shape of the temperature power spectrum reflects a host of physical processes taking place before, during, and after recombination and decoupling. One process of particular importance is that of Baryon Acoustic Oscillations (BAO). Before decoupling, baryons and photons were tightly coupled in the so-called baryon-photon fluid. Inhomogeneities in this fluid were acted upon by two contrasting forces: gravity tended to make such inhomogeneities grow (making overdensities even more overdense), but such growth was hindered by the large radiation pressure of photons. Considering the overdensity field $\delta$, a cartoon version of the equation governing its evolution in Fourier space looks like:
\begin{eqnarray}
\ddot{\delta} + k^2c_s^2\delta = F\,,
\label{eq:bao}
\end{eqnarray}
where $c_s$ is the baryon-photon sound speed already seen in Eq.~(\ref{eq:cs}), whereas $F$ is a driving force which depends on the gravitational potential. The equation governing the evolution of overdensities in the Universe looks like that of a forced harmonic oscillator. As a result, acoustic waves were set in the tightly coupled baryon-photon fluid. The moment photons decouple from the plasma, the waves freeze. This leads to two important effects. Firstly, we can expect these waves to carry a typical scale, namely the sound horizon at decoupling, given by Eq.~(\ref{eq:soundhorizon}) with $z=z_{\rm dec}$. Second, we expect BAOs to imprint their signature on $C_{\ell}^{TT}$ as a sequence of peaks and troughs. Why? There is a particular oscillation mode with frequency such that at decoupling it had the time to exactly compress once (thus complete a quarter of an oscillation), so it freezes when its amplitude is maximal. We expect to observe large fluctuations/temperature anisotropies on angular scales corresponding to this mode, and thus a peak in the temperature power spectrum. A mode with an oscillation frequency twice as large instead had the time to complete half an oscillation cycle: at decoupling it is caught in phase with the background, thus with an amplitude close to zero. We expect to observe very tiny fluctuations/temperature anisotropies on angular scales corresponding to this mode, and thus a trough in the temperature power spectrum. Similarly, a mode with oscillation frequency three times that of the first peak will have gone through a compression and a rarefaction, wherein it is caught at the time of decoupling: therefore, it corresponds to a peak.

Denoting by $n=1,2,3,...$ the number of the peak, we expect the peaks to correspond to wavenumbers $k_n$ given by the following:
\begin{eqnarray}
k_n \approx n\pi/r_s(z_{\rm dec})\,,
\label{eq:kn}
\end{eqnarray}
where $r_s(z_{\rm dec})$ is the sound horizon at decoupling. Inhomogeneities corresponding to a perturbation with wavenumber $k$ contribute mostly to anisotropies at multipoles $\ell \approx k\chi_{\star}$, with $\chi_{\star} = \chi(z_{\rm dec})$ the comoving distance to the redshift of decoupling, sometimes also referred to as last-scattering. Therefore we expect the $n$th peak to appear roughly at multipoles $\ell_n$:
\begin{eqnarray}
\ell_n \approx \frac{n\pi\chi_{\star}}{r_s(z_{\rm dec})}\,,
\label{eq:lp}
\end{eqnarray}
corresponding to angular scales $\theta_n$ given by:
\begin{eqnarray}
\theta_n \approx \frac{r_s(z_{\rm dec})}{n\chi_{\star}}\,.
\end{eqnarray}
For parameters around the best-fit cosmological parameters from \textit{Planck} 2015, the first peak appears at multipoles $\ell_{\rm peak} \approx 220$, corresponding to an angular scale of approximately one degree.

Historically, the first peak of the CMB has always been regarded with great importance, and the angular scale of the first peak is usually denoted by $\theta_s$. In fact, $\theta_s$ is one of the six fundamental parameters of the $\Lambda$CDM model. It is given by the ratio between the sound horizon at decoupling and the comoving distance to decoupling:
\begin{eqnarray}
\theta_s = \frac{r_s(z_{\rm dec})}{\chi_{\star}}\,.
\label{eq:thetas}
\end{eqnarray}
It is clear that the position of the first peak provides valuable information about the geometry and energy content of the Universe, since these typically result in the first peak being projected on different angular scales $\theta_s$ (as they change the distance scales involved). Typically, modifying the late-time expansion rate affects $\chi_{\star}$, whereas modifying the early-time expansion rate affects $r_s$: in both cases, $\theta_s$ is modified. Besides its position, the amplitude of the first peak also provides valuable information on the content of the Universe. In fact, the height of the first peak is very sensitive to the integrated Sachs-Wolfe contribution to the temperature anisotropies, given by the rightmost term in Eq.~(\ref{eq:sachs}). An incomplete matter domination at the time of decoupling leads to residual time variations in the potentials $\phi$ and $\psi$, which boost the temperature anisotropies and hence the height of all peaks (but especially of the first peak). This contribution is commonly referred to as early integrated Sachs-Wolfe (EISW) effect, to distinguish it from the late integrated Sachs-Wolfe (LISW) effect due to dark energy domination at late times. The height of the first peak is therefore very sensitive to the redshift of matter-radiation equality $z_{\rm eq}$, since an earlier onset of matter domination leads to less decay of the potentials at decoupling, and hence a smaller EISW effect and a lower first peak. Conversely, a later onset of matter domination leads to a higher first peak. See~\cite{Cabass:2015xfa} for a recent work constraining the amplitude of the EISW effect.

In summary, we expect a series of peaks and troughs in the CMB temperature power spectrum $C_{\ell}^{TT}$, corresponding to oscillation modes caught at extrema of compression or rarefaction (peaks), or in phase with the background (troughs). At scales much larger than the sound horizon at decoupling, and correspondingly multipoles $\ell \lesssim \ell_{1}$ (with $\ell_1$ the multiple of the first CMB acoustic peak), perturbations are frozen to the initial conditions presumably provided by inflation. This simple picture, wherein we would expect an unending sequence of peaks of equal height, is slightly complicated by the presence of baryons. As we have already seen [Eqs.~(\ref{eq:cs},\ref{eq:r})], the amount of baryons affects the sound speed of the baryon-photon fluid, but also the amount of gravitational force felt by overdensities in the baryon-photon fluid: both quantities appear in the cartoon equation Eq.~(\ref{eq:bao}). It turns out the net effect of baryons is to enhance the compression (odd) peaks over the rarefaction (even) ones. Heuristically, this is simple to understand: increasing the amount of baryons, we are increasing the amount of gravitational pull (which drives the compression peaks), while not changing the amount of radiation pressure (which drives the rarefaction peaks). Therefore, we are enhancing the odd peaks, leading to an asymmetry between odd and even peaks.

If this were the end of the story, all odd peaks would have the same height, and so would all the even peaks. In reality, the whole peak structure is further modulated by an exponential damping envelope. This damping reflects an effect known as \textit{Silk damping}~\cite{Silk:1967kq}. Silk damping is a diffusion damping effect, due to the fact that decoupling is not an instantaneous process, but occurs over a finite but small range of redshift: CMB photons are therefore last-scattered over a shell of finite thickness (see e.g.~\cite{Pires:2004pi,Pan:2016zla,Hadzhiyska:2018mwh} for papers where limits on the duration of last-scattering are obtained). During this time, CMB photons perform a random walk through baryons, effectively erasing anisotropies on scales below their typical mean free path $r_d$, given by~\cite{Dodelson:2003ft,Lesgourgues:2018ncw}
\begin{eqnarray}
r_d = \sqrt{\pi^2\int_{0}^{a_{\rm dec}}\frac{da}{a^3\sigma_T n_e(a)H(a)} \left [ \frac{R^2+\frac{16}{15}(1+R)}{6(1+R^2)} \right ]}\,,
\label{eq:rd}
\end{eqnarray}
where $a_{\rm dec}$ is the scale factor at decoupling, $n_e$ is the number density of free electrons, and $\sigma_T$ is the Thomson cross-section. Silk damping results in a damping envelope which is particularly evident for $\ell \gtrsim 1000$, from the third peak on. In the same way the first peak contains the imprint of the angular size of the sound horizon at decoupling $\theta_s$, the damped high-multipole peaks contain the imprint of the angular size of the damping scale, $\theta_d = r_d/\chi_{\star}$.

\subsubsection{Secondary anisotropies}
\label{subsubsec:secondary}

The features of the temperature power spectrum we have discussed so far have been generated at decoupling or earlier, and are referred to as primary anisotropies. However, as CMB photons travel along the line of sight to us, new anisotropies are generated due to late-time effects. These are referred to as secondary anisotropies. See~\cite{Aghanim:2007bt} for a comprehensive review on secondary anisotropies in the CMB.

One of the most important sources of secondary anisotropies is CMB lensing: that is, the lensing of CMB photons due to the intervening matter distribution~\cite{Bartelmann:1999yn,Lewis:2006fu,Hanson:2009kr}. Lensing is mainly sensitive to low-redshift inhomogeneities, at redshifts $z \lesssim 5$. The angular scale associated to lensing is $\approx 2'$, so that lensing becomes important at multipoles $\ell \gtrsim 1000$. The effect of lensing is that of blurring the primary anisotropies, smoothing the high-multipole peaks.

The effect of lensing can be quantified through the lensing potential $\phi(\boldsymbol{\hat{n}})$, which is related to the deflection angle experienced by a CMB photon, $\boldsymbol{\alpha}(\boldsymbol{\hat{n}})$, through $\boldsymbol{\alpha}(\boldsymbol{\hat{n}}) = \boldsymbol{\nabla}\phi(\boldsymbol{\hat{n}})$. Because lensing is a non-linear effect, it creates a small amount of non-Gaussianity in the pattern of temperature anisotropies, leading to subtle correlations between temperature anisotropies on different angular scales. By using these subtle correlations, one can reconstruct the lensing potential on the sky, and from that compute the lensing potential angular power spectrum $C_{\ell}^{\phi\phi}$ (a closely related quantity, as we shall see in Chapter~\ref{sec:paper2}, is the lensing convergence $\kappa$). See e.g.~\cite{Okamoto:2003zw,Kesden:2003cc,Hanson:2010rp,
Smith:2010gu,Namikawa:2011cs,Pearson:2014qna,Sherwin:2015baa,
Namikawa:2015tjy,Larsen:2016wpa,Sehgal:2016eag,Yu:2017djs,
Horowitz:2017iql,Manzotti:2017oby,Madhavacheril:2018bxi,Schaan:2018tup} for a number of important works concerning CMB lensing reconstruction. CMB lensing is an integrated effect, sensitive to the matter distribution along the line of sight, appropriately projected. To extract the lensing signal from a specific redshift range, one can instead cross-correlate the CMB lensing effect with appropriate tracers of the LSS at that redshift~\cite{Hirata:2008cb,Sherwin:2012mr,Ade:2013hjl,Nicola:2016eua,Doux:2017tsv}.

Another important source of secondary anisotropies is reionization. Reionization drastically increases the fraction of free electrons in the late Universe, providing an extra channel for additional scattering of CMB photons which would otherwise free-stream to us. The net result is that, on scales below the horizon at reionization ($\ell \gtrsim 40$), the temperature anisotropies are exponentially suppressed by $e^{-2\tau}$, where the parameter $\tau$ is known as optical depth to reionization and quantifies the line-of-sight free-electron opacity to CMB photons. The value of $\tau$ is related to the probability that a CMB photon is rescattered due to reionization. Under the (unrealistic but nevertheless useful) assumption of instantaneous reionization, the value of $\tau$ can be related to the redshift of reionization.~\footnote{Observational signatures and strategies for probing more complex reionization models have been considered in a number of paper, see e.g.~\cite{Mellema:2012ht,Smith:2016lnt,Meyers:2017rtf,Namikawa:2017uke,
Villanueva-Domingo:2017ahx,Roy:2018gcv,Ferraro:2018izc,Giri:2018dln,
Giri:2019pxr,Roy:2019qsl}}. Details of reionization aside, a larger value of $\tau$ indicates an earlier onset of galaxy/star formation, whereas $\tau=0$ would indicate the absence of reionization.

A last important source of secondary anisotropies is known as the late integrated Sachs-Wolfe (LISW) effect, sometimes also referred to as Rees-Sciama effect~\cite{Rees:1968zza}. It is analogous to the EISW effect we discussed earlier affecting the first peak, but driven in this case by the decay of gravitational potentials as dark energy comes to dominate the late Universe. The LISW signal results in a boost of power on large scales ($\ell \lesssim 20$), corresponding to scales entering the horizon after matter-dark energy equality. However, on such scales measurements of the CMB temperature power spectrum are plagued by cosmic variance and hence not much can be said about the LISW effect from CMB measurements alone. The LISW signal can instead be extracted at a higher statistical significance by cross-correlating the CMB temperature anisotropies with tracers of the LSS such as galaxies or quasars (see e.g.~\cite{Fosalba:2003ge,Vielva:2004zg,Padmanabhan:2004fy,McEwen:2006my,Cabre:2007rv,
Ho:2008bz,Zhao:2008bn,Zhao:2010dz,Ilic:2011hh,Ilic:2013cn,
Ferraro:2014msa,Nishizawa:2014vga,Cabass:2015xfa,Renk:2017rzu,Bolis:2018vzs} for important works in this direction).

\subsubsection{A brief discussion on polarization}
\label{subsec:polarization}

Before going on to discuss how we can extract the 6 base cosmological parameters of $\Lambda$CDM from CMB measurements, I will briefly discuss polarization anisotropies. In fact, CMB photons are polarized, and polarization anisotropies carry valuable information on the physics of the tightly coupled baryon-photon plasma, CMB lensing, reionization, and possibly on primordial gravitational waves from inflation. Polarization of the CMB is an incredibly complex topic, especially from the mathematical point of view. My goal here will be to provide the reader a heuristic level of understanding, sufficient to understand why measuring polarization of the CMB is useful for extracting cosmological parameters, including parameters related to neutrinos. For a pedagogical and more complete coverage of the physics of CMB polarization I refer the reader to seminal reviews, e.g.~\cite{Hu:1997hv,Kosowsky:1998mb,Challinor:2005ye}.

\begin{figure}[!t]
\centering
\includegraphics[width=0.5\linewidth]{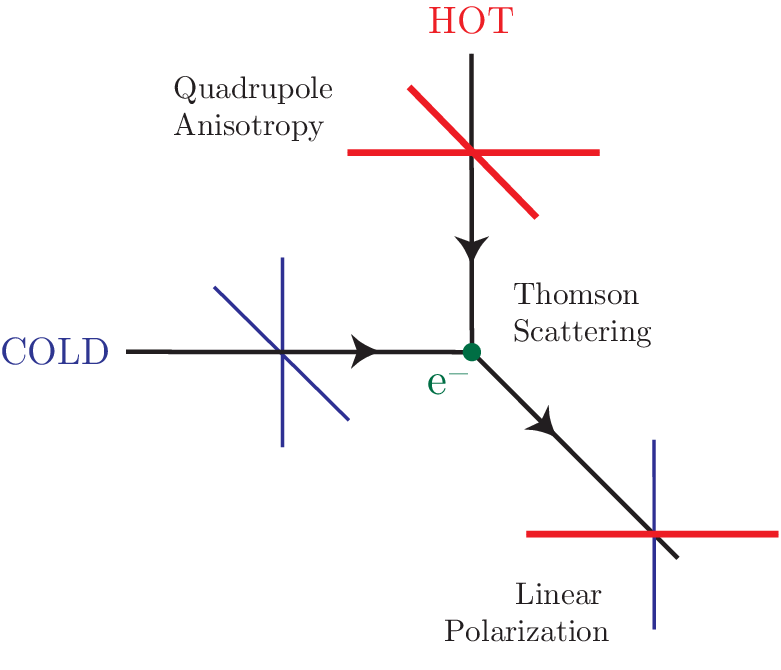}
\caption{A schematic representation of how Thomson scattering of radiation with quadrupole anisotropy generates linear polarization. Reproduced from~\cite{Hu:1997hv} with permission from Elsevier.}
\label{fig:thomson}
\end{figure}

Polarization is generated by Thomson scattering, the scattering of electromagnetic radiation off a non-relativistic electron. The differential cross-section for this scattering process $d\sigma_T/d\Omega$ is not isotropic but goes like $d\sigma_T/d\Omega \propto \vert \boldsymbol{\hat{\epsilon}'} \cdot \boldsymbol{\hat{\epsilon}} \vert$, with $\boldsymbol{\hat{\epsilon}'}$ and $\boldsymbol{\hat{\epsilon}}$ the outgoing and incoming polarization vectors (see e.g.~\cite{Chandrasekhar:1950ghw,Griffiths:1995ghw}). From a heuristic perspective, incoming radiation shakes an electron in the direction $\boldsymbol{\hat{\epsilon}}$ and causes it to radiate with intensity peaking in the direction of the incoming polarization. However, the outgoing polarization direction $\boldsymbol{\hat{\epsilon}'}$ must also be orthogonal to the direction of propagation. Therefore, incoming radiation polarized parallel to the outgoing direction does not scatter. See for instance Fig.~\ref{fig:thomson}, a cartoon version of Thomson scattering of an electron by an incoming quadrupole source, generating a net linear polarization.
In fact, it is easy to convince oneself that, in order to generate a net polarization from Thomson scattering, the incoming radiation should be anisotropic. More specifically, at least a quadrupole anisotropy is required (intensity varying at a $\pi/2$ angle), since a dipole anisotropy would lead to no net polarization.

The early pre-recombination Universe, during which the baryon-photon fluid underwent BAOs, was highly isotropic precisely due to the tight coupling between baryons and photons. For this reason, no net polarization could be generated during the time. However, towards the end of recombination, photons can start to diffuse between hot and cold regions (recall this is the process that generates Silk damping). At this point, a quadrupole moment can form, leading to net linear polarization~\cite{Kosowsky:1994cy,Seljak:1996ti,Kamionkowski:1996ks}. From these simple considerations, we can expect the size of the quadrupole to depend on the typical photon velocity (given by the dipole moment instead). It has been shown~\cite{Bond:1987ub} that the strength of the quadrupole anisotropy is suppressed with respect to the main temperature fluctuations, which is expected since the scattering generating polarization is also responsible for destroying the same information (much the same way Silk damping does), and thus we expect the polarization power spectrum to be significantly lower than the temperature one. On the other hand, we can expect the oscillating velocity field to be out of phase by $\pi/2$ with respect to the oscillating (over)density field: when the amplitude of a density mode is maximal (\textit{i.e.} it is either maximally overdense or underdense), the velocity is zero (as the oscillation is turning around), whereas when the density mode is in phase with the background, the velocity is maximal.~\footnote{Mathematically, if the density field oscillates as $\delta \propto \cos(kr_s)$, the velocity field oscillates as $v \propto \sin(kr_s)$. This is what one usually expects for the position and velocity of a harmonic oscillator, which are maximally out of phase.} Therefore, we expect the polarization power spectrum to carry the imprint of BAOs, although maximally out of phase with respect to the signature of BAOs in the temperature power spectrum: in other words, at multipoles where in temperature we have a peak, in polarization we should see a trough.

This simple picture is slightly complicated by the fact that polarization has both a strength and an orientation. A thorough description of the underlying mathematics would require us to delve into the (fascinating) realm of spin-2 fields, way beyond the scope of this thesis. For the purpose of understanding the broad features of the polarization spectra, it suffices to say that the orientation can be described by decomposing the polarization field into a curl-free $E$ and a divergence-free $B$ components~\cite{Kosowsky:1994cy,Zaldarriaga:1996xe,Kamionkowski:1996ks}. In the small-scale limit, the wavevector of a scalar perturbation $\boldsymbol{k}$ picks up a preferred direction along which to measure polarization: then, the $E$ component measures polarization aligned or orthogonal with respect to $\boldsymbol{k}$, whereas the $B$ component measures polarization crossed at $\pm \pi/4$ with respect to $\boldsymbol{k}$. Going beyond the small-scale limit does not change these qualitative features~\cite{Dodelson:2003ft}. Moreover, scalar (density) perturbations can only generate $E$-type polarization, whereas gravitational wave (tensor) perturbations generate both $E$- and $B$-type polarization (at least at the level of primary polarization anisotropies)~\cite{Seljak:1996ti,Seljak:1996gy,Kamionkowski:1997av,
Krauss:2013pha,Kamionkowski:2015yta}.

The signature of BAOs, being generate from density fluctuations, is thus only imprinted in the $E$-mode power spectrum, $C_{\ell}^{EE}$. As discussed previously, the acoustic peaks are maximally out of phase with respect to those in $C_{\ell}^{TT}$. In particular, the first peak in $E$ polarization should appear around $\ell \approx 100$. Moreover, the overall amplitude of $C_{\ell}^{EE}$ is significantly lower than that of $C_{\ell}^{TT}$, and we expect $C_{\ell}^{EE}$ to drop sharply both at large scales (small $\ell$, because polarization cannot be generated at scales which are super-horizon at recombination) and small scales (large $\ell$, because scattering erases information on small-scale anisotropies). Moreover, given the phase relation between $C_{\ell}^{TT}$ and $C_{\ell}^{EE}$, we expect a non-zero cross-correlation between temperature and $E$ polarization, with a spectrum $C_{\ell}^{TE}$ featuring oscillations at twice the frequency of the oscillations in $C_{\ell}^{TT}$ or $C_{\ell}^{EE}$. Hence, once $C_{\ell}^{TT}$ is measured, the shape of both $C_{\ell}^{EE}$ and $C_{\ell}^{TE}$ is (mostly) already determined, and can thus be used as a powerful cross-check. The primary $B$-mode power spectrum, $C_{\ell}^{BB}$, is instead generated in the presence of primordial gravitational waves, whose amplitude is quantified by the tensor-to-scalar ratio $r$. In this case, the relevant scale is the horizon at decoupling, thus we expect $C_{\ell}^{BB}$ to peak around $\ell \approx 100$ (corresponding to angular scales of about a degree), and to drop rapidly at both ends.

Secondary anisotropies, discussed earlier in Chapter~\ref{subsubsec:secondary} in the context of temperature anisotropies, affect polarization anisotropies as well. The two main sources of secondary anisotropies are lensing and reionization. As in temperature, lensing acts on small scales ($\ell \gtrsim 1000$), and results in the generation of $B$ modes from $E$ modes: heuristically, this occurs because lensing warps $E$ modes in a way that is not related to the direction of polarization, effectively generating $B$ modes, referred to as lensing $B$ modes~\cite{Zaldarriaga:1998ar,Knox:2002pe,
Dodelson:2003bv,Seljak:2003pn,Hanson:2013hsb}. With regards to reionization, on small scales the physical picture is the same as it is in temperature, leading to an $e^{-2\tau}$ suppression of the polarization power spectra. However, reionization also provides an additional source of scattering by increasing the fraction of free electrons. This leads to an enhancement of power on scales corresponding to the horizon at reionization ($\ell \approx 10$), usually referred to as the ``reionization bump''.

\subsubsection{Cosmological parameters from CMB measurements}

\begin{figure}[!t]
\centering
\includegraphics[width=1.0\textwidth]{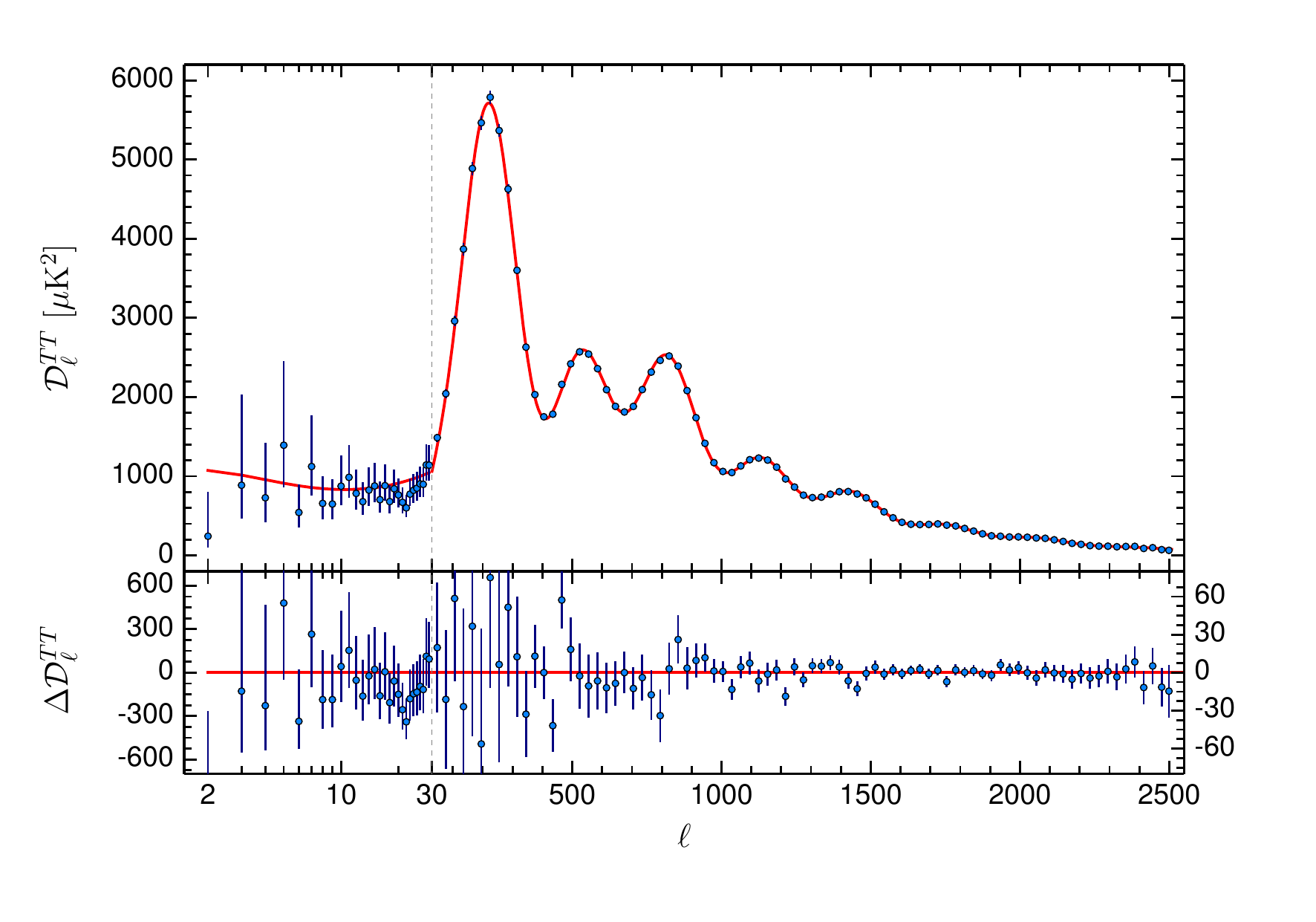}
\caption{Temperature power spectrum from the \textit{Planck} 2015 data release. \textit{Upper panel}: the blue points are the actual measurements with error bars (nearly invisible for $\ell \gg 30$), whereas the red curve is the theoretical power spectrum computed using the best-fit parameters obtained analysing temperature and large-scale polarization data.  Notice that, as per standard convention in the field, the quantity plotted on the $y$ axis is $T_{\rm CMB}^2\ell(\ell+1)C_{\ell}$, with $T_{\rm CMB} \approx 2.725\,{\rm K}$ the CMB temperature today.\textit{Lower panel}: residuals with respect to the best-fit model. Reproduced from~\cite{Ade:2015xua} with permission from EDP Sciences.}
\label{fig:tt}
\end{figure}

So far we have provided a mostly qualitative picture of the physics underlying the CMB temperature and polarization anisotropies spectra and cross-spectra. Various actors have come into play at different times and scales: prior to decoupling the interplay of gravity and pressure in the tightly coupled baryon-photon fluid set up acoustic oscillations showing up on intermediate scales in the temperature and $E$-mode polarization (albeit out of phase) power spectra, as well as in their cross-correlation. On small scales, these spectra are suppressed from Silk damping due to photons random-walking around the time of decoupling, as well as from scattering on free electrons during and after reionization. However, reionization also provides an extra source of $E$- polarization on very large scales. On intermediate scales, primordial $B$-mode polarization is generated if primordial gravitational waves (presumably from inflation) were set up in the very early Universe. On very large scales, the temperature power spectrum reflects the initial conditions presumably set by inflation, modulo additional anisotropies generated at late times when dark energy takes over, through the LISW effect. Finally, on small scales, gravitational lensing becomes important and blurs the temperature anisotropies, while generating $B$-mode polarization from $E$-mode polarization. The CMB temperature power spectrum as measured by 2015 data release of the \textit{Planck} satellite~\cite{Ade:2015xua} is shown in Fig.~\ref{fig:tt}: from the figure, we can clearly see the imprints of all effects discussed so far.

The question then is: can we use the measured spectra to pin down cosmological parameters? The answer, of course, is yes! As we anticipated in Chapter~\ref{sec:concordance}, 6 parameters appear to be sufficient in describing the CMB spectra, within the so-called concordance $\Lambda$CDM model. The parameters are: $\theta_s$, $\omega_c$, $\omega_b$, $A_s$, $n_s$, and $\tau$. Given a set of cosmological parameters, we can compute theoretical predictions for the CMB temperature, polarization, and lensing spectra, using state-of-the-art Boltzmann solvers such as \texttt{CAMB}~\cite{Lewis:1999bs} or \texttt{CLASS}~\cite{Audren:2012vy}. In the rest of this thesis, we will be concerned with a 1-parameter extension of this very successful model, the $\Lambda$CDM+$M_{\nu}$ model, where the sum of the neutrino masses $M_{\nu}$ is treated as a free parameter. For the moment, let me sketch how the 6 base parameters can be extracted from measurements of the CMB spectra. I will return in more detail to the effect of $M_{\nu}$ on the CMB spectra in Chapter~\ref{subsec:signaturesnucmb}.

The following discussion will closely follow~\cite{Lesgourgues:2018ncw}, and I recommend that the interested reader read the end of Section~5.1.6 thereof. In a simplified but overall rather complete picture, we can envisage the CMB temperature power spectrum as mostly being governed by 8 effects (referred to as C1 through to C8 in~\cite{Lesgourgues:2018ncw}):
\begin{enumerate}
\item The position of the first peak depends on $\theta_s = r_s/\chi_{\star}$. $r_s$ depends on the expansion history prior to decoupling, and is affected by changes in the photon-baryon sound speed. Hence, it is sensitive to $\omega_b$ (which controls the sound speeed) and $\omega_m$ (which controls $z_{\rm eq}$). On the other hand, $\chi_{\star}$ depends on the expansion between decoupling and us, and is affected by quantities such as $\Omega_{\Lambda}$ or $h$.
\item The relative height between odd and even peaks depends on $\omega_b/\omega_{\gamma}$ (but recall that $\omega_{\gamma}$ is basically fixed), \textit{i.e.} on the relative pressure-gravity balance.
\item The height of all peaks is controlled by the amount of expansion between equality and decoupling, during which acoustic oscillations are damped. Hence, this effect is mostly controlled by $\omega_m$ (and thus by $\omega_c$, once $\omega_b$ is known).
\item The amplitude of the high-multipole peaks is controlled by $\theta_d = r_d/\chi_{\star}$, with $r_d$ depending on the expansion history prior to decoupling and hence on $\omega_b$ and $\omega_m$ (for $\chi_{\star}$ see Point~1 above).
\item The overall amplitude of the power spectrum is controlled by $A_s$.
\item The overall tilt of the power spectrum is controlled by $n_s$.
\item The slope of the power spectrum at low-multipoles is controlled by the LISW effect and hence by $\Omega_{\Lambda}$ and $h$.
\item The amplitude at $\ell \gg 40$ versus the amplitude at $\ell < 40$ is controlled by $\tau$.
\end{enumerate}

\begin{figure}[!t]
\centering
\includegraphics[width=1.0\textwidth]{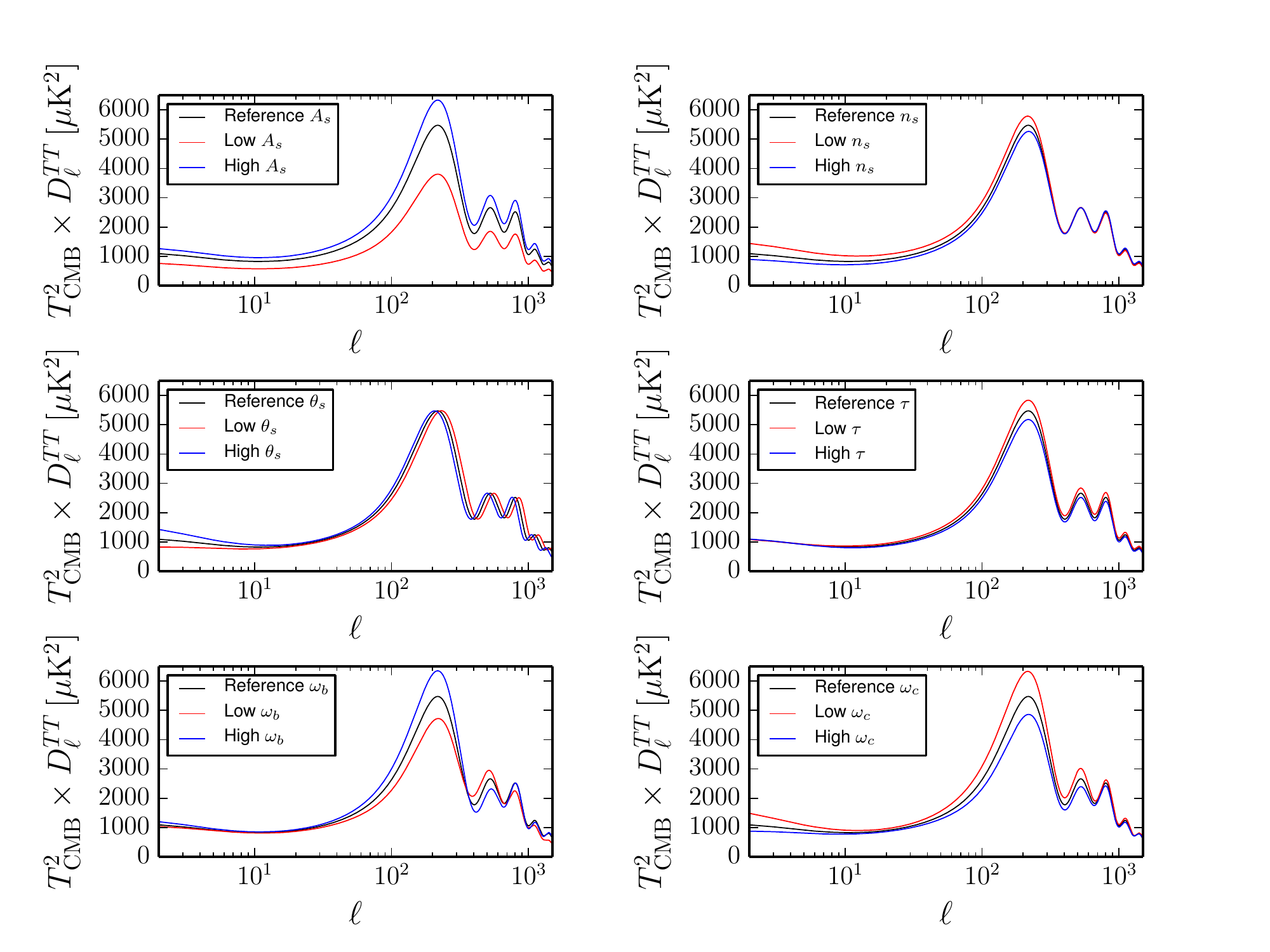}
\caption{Impact of varying the six fundamental $\Lambda$CDM parameters on the CMB temperature power spectrum. The chosen baseline model has $\omega_b=0.02$, $\omega_c=0.12$, $100\theta_s=1.054$, $\tau=0.072$, $A_s=2.16 \times 10^{-9}$, and $n_s=0.96$. Derived parameters of particular interest are $h=0.7$, $\Omega_{\Lambda}=0.713$, $z_{\rm eq}=3345.55$, and $100\theta_d=0.167$. The spectra have been produced through the Boltzmann solver \texttt{CAMB}~\cite{Lewis:1999bs}, which takes $h$ as input and not $\theta_s$. When $\omega_b$ and $\omega_c$ are varied, I manually adjust $h$ to keep $\theta_s$ fixed. Varying $\theta_s$ is accomplished by manually varying $h$. Notice that, as per standard convention in the field, the quantity plotted on the $y$ axis is $T_{\rm CMB}^2\ell(\ell+1)C_{\ell}$, with $T_{\rm CMB} \approx 2.725\,{\rm K}$ the CMB temperature today.}
\label{fig:parameterscmb}
\end{figure}

Therefore, simplifying a bit, the route towards determining cosmological parameters from the CMB power spectrum proceeds as follows: the position of the first peak directly determines $\theta_s$, which in turn depends on a certain combination of $\omega_c+\omega_b$ and $h$ (the latter a derived parameter). The height of the first peak determines $z_{\rm eq}$ and hence $\omega_c+\omega_b$, and in combination with the position of the first peak determine $h$. Comparing the amplitude of the even and odd peaks allows us to determine $\omega_b$ (and from that $\omega_c$), whose determination is improved by measuring the damping tail. The overall amplitude of the temperature power spectrum depends on the combination $A_se^{-2\tau}$, while the overall slope determines $n_s$. Measuring polarization at large scales allows one to measure $\tau$, and hence disentangle $A_s$. More generally, the acoustic peaks in polarization are sharper~\cite{Galli:2014kla}, thus allowing for a better determination of $\omega_c$ and $\omega_b$, as well as $h$. It is worth noting that the position and height of the first peak in temperature are extremely well measured, and thus $\theta_s$ and $z_{\rm eq}$ are basically fixed. In Fig.~\ref{fig:parameterscmb}, I show the effect of varying the six fundamental $\Lambda$CDM parameters on the CMB temperature power spectrum.

As we anticipated in Chapter~\ref{sec:concordance}, the base 6-parameter model can be extended by allowing other parameters, more or less physically motivated, to vary.~\footnote{Notable examples considered in the literature include the sum of the neutrino masses $M_{\nu}$, the dark energy equation of state $w$ and possibly its time derivative $w_a$, the running of the scalar spectral index $n_{\rm run} \equiv dn_s/d\ln k$, the running of the running $n_{\rm runrun} \equiv dn_{\rm run}/d\ln k$, the tensor-to-scalar ratio $r$, the primordial Helium fraction $Y_p$, the effective number of relativistic degrees of freedom $N_{\rm eff}$ (which we will discuss shortly in Chapter~\ref{sec:signaturesnu}), the curvature density parameter $\Omega_k$, the effective mass of a sterile neutrino $m_s^{\rm eff}$, as well as a phenomenological parameter controlling the amplitude of lensing, $A_L$. See, for instance, the important work~\cite{DiValentino:2015ola}, where up to 12 parameters at the same time were treated as being free.} From a purely statistical point of view (to be quantified more thoroughly in Chapter~\ref{chap:5}), it is worth noting that data does not ``favour'' any of these extensions, in the sense that the added layer of complication brought upon by introducing new parameters does not lead to an improvement in fit sufficient to justify the introduction of these parameters~\cite{Heavens:2017hkr}. Still, it is worth considering simple extensions of the $\Lambda$CDM model, since some of these extensions are particularly justified (this is particularly true in the case of the $\Lambda$CDM+$M_{\nu}$ model). The important thing to note is that freeing up additional parameters opens up degeneracies/correlations between parameters which data might not be able to resolve. In other words, different combinations of parameters might lead to the same physical effects, and hence data might not be able to disentangle them, while instead only being sensitive to a particular combination of cosmological parameters~\cite{Efstathiou:1998xx,Efstathiou:2001cv,
Howlett:2012mh,Li:2012ug,Archidiacono:2016lnv}. Effectively, we can think about this problem mathematically as that of an underdetermined system: we have more variables than constraints. Usually degeneracies can be broken by considering additional cosmological data (\textit{e.g.} large-scale structure probes) which are sensitive to ``orthogonal'' combinations of parameters. It is also worth noting that most of the degeneracies opening up in the presence of additional free parameters are related to the so-called \textit{geometrical degeneracy}: this refers to the possibility of adjusting parameters governing the background expansion in such a way as to keep the angular size of the first peak, $\theta_s$, fixed. As we shall see later, this degeneracy is particularly important when treating $M_{\nu}$ as a free parameter.

\subsection{Large-scale structure}
\label{subsec:lss}

Besides the CMB, the clustering of the large-scale structure (LSS) is another extremely powerful probe of cosmological parameters. Under the effect of gravity, the ${\cal O}(10^{-5})$ inhomogeneities present at decoupling and reflected in the anisotropies of the CMB grow and collapse to form the structures we see in the Universe today. One can therefore expect statistical probes of inhomogeneities in the matter density field to probe cosmological parameters, much as the anisotropies in the CMB do. Moreover, unlike the CMB (which is at a fixed redshift), we can observe the LSS at various redshifts and thus perform a tomographic analysis.

An interesting way of studying any given field is to examine its distribution of fluctuations over various scales/frequencies by taking its Fourier transform. Taking the inner product with its complex conjugate then gives us the field's power spectrum, which quantifies the variance of the field at any given scale. In the case of the matter overdensity field $\delta_m(\boldsymbol{k},z)$, we define its power spectrum $P_m(k,z)$ as:
\begin{eqnarray}
\left \langle \delta_m(\boldsymbol{k},z)\delta_m(\boldsymbol{k'},z) \right \rangle \equiv P_m(k,z)\delta(\boldsymbol{k}-\boldsymbol{k'})\,,
\label{eq:powerspectrum}
\end{eqnarray}
where $\delta$ denotes the Dirac delta. The power spectrum of matter density fluctuations $P_m$ contains a substantial amount of information on cosmological parameters, and is the LSS counterpart of the $C_{\ell}$s for the CMB. In fact, given a set of cosmological parameters, Boltzmann solvers can be used to make a theoretical prediction for $P_m$. The real-space counterpart of the matter power spectrum, instead, is known as the correlation function and is usually denoted by $\xi(r)$.

Mathematically speaking, the same amount of information is contained in $P(k)$ and $\xi(r)$. Historically, though, the two have always been analysed separately and used to obtain different cosmological measurements. In particular, real-space analyses are typically performed with the goal of providing a BAO distance measurement (which is essentially a background probe), while Fourier-space analyses typically measure the galaxy power spectrum $P_k$ (which depends on both the background and perturbation evolution). In the following, I will discuss the physics shaping these types of measurements: as done earlier with the CMB, my goal will be to endow the reader with an intuitive understanding of the physical processes at play and how the observables are shaped by these physical processes and respond to changes in the cosmological parameters. For more in-depth and technical treatments, I invite the reader to consult e.g.~\cite{Weinberg:1972kfs,Mukhanov:2005sc,Weinberg:2008zzc,Bernardeau:2001qr,
Dodelson:2003ft,Percival:2013awa,Lesgourgues:2018ncw}.

\subsubsection{Galaxy power spectrum}
\label{subsubsec:galaxypowerspectrum}

A large number of galaxy surveys are currently underway, measuring the clustering of matter on large scales and late times.~\footnote{A few important names among current and past galaxy surveys include (but are certainly not limited to) the Sloan Digital Sky Survey (SDSS;~\cite{York:2000gk}), the Baryon Oscillation Spectroscopic Survey (BOSS;~\cite{Dawson:2012va}), the Dark Energy Survey (DES;~\cite{Abbott:2005bi}), the extended Baryon Oscillation Spectroscopic Survey (eBOSS;~\cite{Dawson:2015wdb}), the WiggleZ Dark Energy Survey (WiggleZ;~\cite{Drinkwater:2009sd}), the 6dF Galaxy Survey (6dFGS;~\cite{Beutler:2011hx}), and the 2dF Galaxy Redshift Survey (2dFGRS~\cite{Colless:2001gk}). A few important names among upcoming surveys includes Euclid~\cite{Laureijs:2011gra}, the Dark Energy Spectroscopic Instrument (DESI;~\cite{Aghamousa:2016zmz}), the Large Synoptic Space Telescope (LSST;~\cite{Ivezic:2008fe}), the Wide Field Infrared Survey Telescope (WFIRST;~\cite{Green:2012mj}), and the Spectro-Photometer for the History of the Universe, Epoch of Reionization, and Ices Explorer (SPHEREx;~\cite{Dore:2014cca}).} Typically, these surveys provide catalogues containing a large number (usually between $100000$ and $1000000$) of galaxies. More specifically, each galaxy in these catalogues is associated to two angles and a redshift: the former two specify its position on the sky, whereas the latter can be used to determine its distance from us, assuming a fiducial cosmology. Assuming a fiducial cosmology, it is possible to convert these angles-redshift triples into a set of comoving coordinates, effectively constructing a $3D$ galaxy map. From such a map, one can construct a map of the corresponding galaxy overdensity $\delta_g$, where $\delta_g \equiv (\rho_g - \bar{\rho}_g)/\bar{\rho}_g$, with $\rho_g$ the galaxy density field and $\bar{\rho}_g$ the mean galaxy density. Finally, taking the square of the Fourier transform of $\delta_g$ (let us denote the Fourier transform of $\delta_g$ as $\delta_g(k)$, where the $k$ argument makes it clear that we are working in Fourier space), one can estimate the galaxy power spectrum $P_g(k,z)$: a practical method for doing this, used by most collaborations, is outlined in the seminal paper by Feldman, Kaiser, and Peacock~\cite{Feldman:1993ky} (such method is often referred to as FKP method from the initials of the authors). Typically, a galaxy sample from a given redshift survey lives in a narrow redshift range and can be thought of as being at a single effective redshift $z_{\rm eff}$. The galaxy power spectrum one computes then is effectively $P_g(k,z_{\rm eff})$. At this point, note a subtlety: I have been talking about \textit{galaxy} power spectrum $P_g$, whereas earlier I talked about \textit{matter} power spectrum $P_m$ (it is the latter which can be directly computed from first principles). I will return to this subtlety and its implications later.

We saw earlier in Chapter~\ref{subsec:inflation} that inflation predicts a primordial power spectrum of metric fluctuations/gravitational potentials $P_{\Phi} \propto k^{n_s-4}$ (with $n_s \approx 1$), and this translates to a primordial power spectrum of matter fluctuations $P_{\rm prim} \propto k^{n_s}$. The late-time power spectrum we observe from galaxy surveys is a ``processed'' version of the primordial power spectrum, accounting for all the physical processes occurring between inflation and today. To understand the shape of the late-time matter power spectrum, we have to understand how such processes affect perturbations in the matter field.

It is useful to make a distinction between scales which entered the horizon during radiation domination (small scales, large $k$), and scales which entered the horizon during matter domination (large scales, small $k$). The reason is that the growth of subhorizon matter perturbations is expected to be significantly different depending on whether the perturbation entered during radiation or matter domination (on the other hand, superhorizon perturbations are frozen to their initial conditions at the end of inflation). During radiation domination, the significant pressure provided by radiation prevents the growth of matter overdensities, which only grow logarithmically with the scale factor: $\delta \propto \ln a$~\cite{Bergstrom:1999kd,Dodelson:2003ft,Mukhanov:2005sc,Durrer:2008eom,
Weinberg:2008zzc,Lesgourgues:2018ncw}. On the other hand, during matter domination perturbations grow linearly with the scale factor: $\delta \propto a$~\cite{Bergstrom:1999kd,Dodelson:2003ft,Mukhanov:2005sc,Durrer:2008eom,
Weinberg:2008zzc,Lesgourgues:2018ncw}. Thus, we expect a turn-around in the late-time power spectrum, at a wavenumber $k_{\rm eq} = aH\vert_{\rm eq}$ corresponding to a scale entering the horizon at matter-radiation equality. The relation between the primordial power spectrum $P_{\rm prim}(k)$ and the late-time one $P(k)$ is quantified through the \textit{transfer function}, $T(k)$, such that $P(k) \propto P_{\rm prim}(k)T^2(k)$.

We expect the small-$k$ ($k \ll k_{\rm eq}$) part of the galaxy power spectrum to directly trace the primordial power spectrum of scalar perturbations generated by inflation: in other words, $T(k) \approx 1$ for $k \ll k_{\rm eq}$, and $P(k) \propto k^{n_s}$ (thus scaling roughly as $k^{1}$). On small scales, fits to numerical solutions show that $T(k) \propto \frac{\Omega_m}{k^2} \ln (k/k_{\rm eq})$, and we therefore expect $P(k) \propto k^{n_s-4}\ln^2(k)$ (thus scaling roughly as $k^{-3}\ln^2(k)$). Moreover, on small scales, the effect of BAOs is imprinted as a series of wiggles in the matter power spectrum. For a full numerical fit to the matter power spectrum on small scales, see Eq.~(6.51) of~\cite{Lesgourgues:2018ncw}.

\begin{figure}[!t]
\centering
\includegraphics[width=1.0\textwidth]{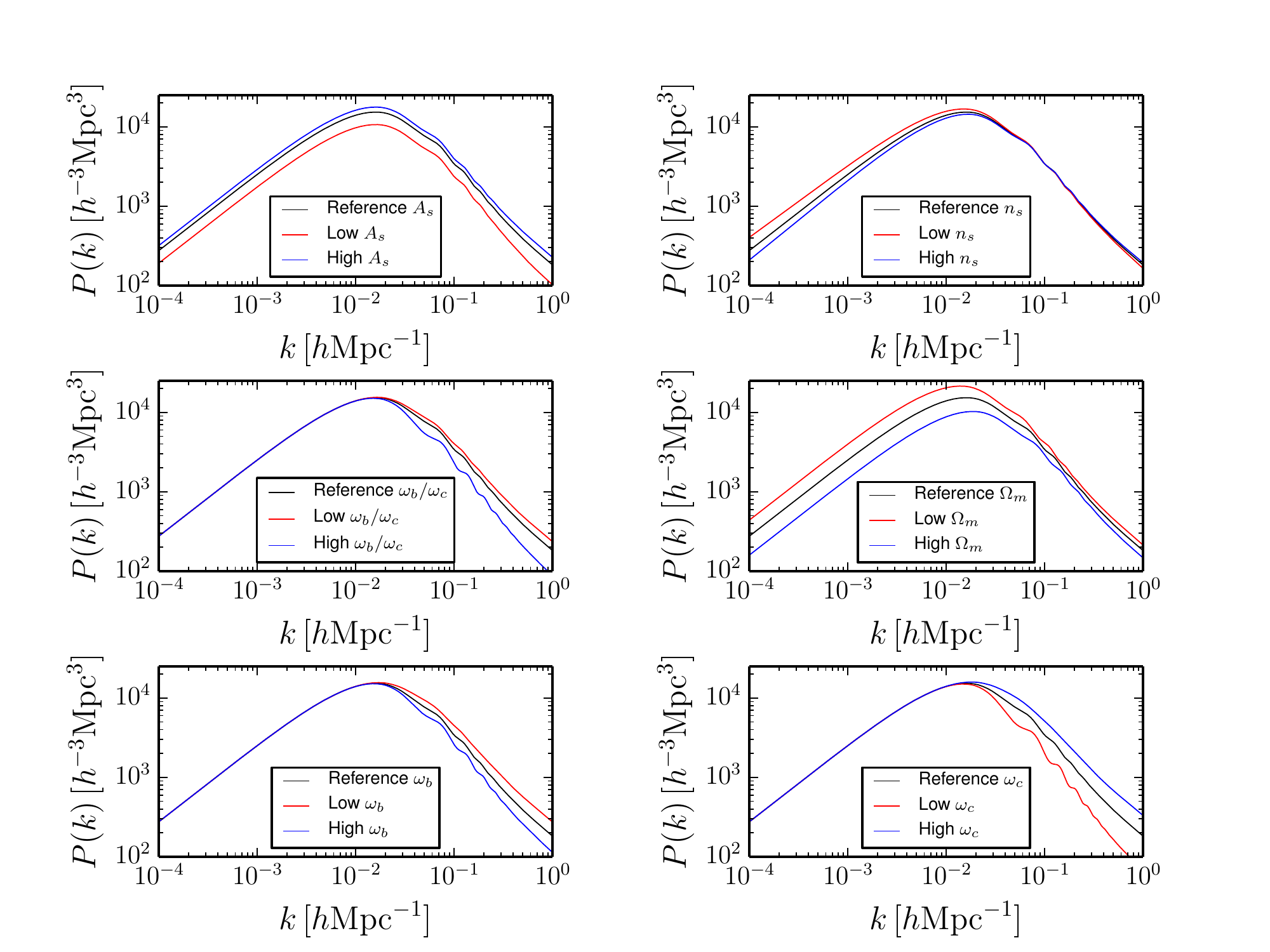}
\caption{Impact of varying the six fundamental $\Lambda$CDM parameters on the matter power spectrum. The chosen baseline model has $\omega_b=0.02$, $\omega_c=0.12$, $A_s=2.16 \times 10^{-9}$, and $n_s=0.96$. Derived parameters of particular interest are $h=0.7$, $\Omega_{\Lambda}=0.713$, $z_{\rm eq}=3345.55$, and $100\theta_d=0.167$. The spectra have been produced through the Boltzmann solver \texttt{CAMB}~\cite{Lewis:1999bs}. When $\omega_b$ and $\omega_c$, and $\omega_b/\omega_c$ are varied, I manually adjust $h$ to keep $\Omega_m$ and $z_{\rm eq}$ fixed. Varying $\Omega_m$ is accomplished by manually varying $h$.}
\label{fig:parameterslss}
\end{figure}

As we did earlier with the CMB, it is useful to identify a number of physical effects governing the shape of the matter power spectrum (in~\cite{Lesgourgues:2018ncw}, these effects are referred to as P1 through to P5):
\begin{enumerate}
\item The matter power spectrum $P(k)$ exhibits a turn-around at $k_{\rm eq} = \sqrt{2\Omega_m(1+z_{\rm eq})}$. On larger scales (smaller $k$) $P(k)$ traces the primordial power spectrum set up by inflation, whereas on smaller scales (larger $k$) it is suppressed by $k^{-4}\ln^2(k)$.
\item The amplitude of the small-scale part of the power spectrum is suppressed as $\omega_b/\omega_c$ increases, accounting for the fact that CDM perturbations grow more slowly in the presence of baryons.
\item On small scales, the power spectrum contains the imprint of BAOs in the form of wiggles, whose amplitude and phase depends on $r_d$, and hence on $\omega_b$.
\item The overall amplitude of $P(k)$ depends on $\Omega_m$ and $A_s$.
\item The overall tilt of $P(k)$ depends on $n_s$.
\end{enumerate}
In Fig.~\ref{fig:parameterslss}, I show the impact on the matter power spectrum of varying selected cosmological parameters. Clearly, of the six fundamental parameters of $\Lambda$CDM, $\theta_s$ and $\tau$ have no impact on $P(k)$ whatsoever. Instead, by looking at the five effects above, it is clear that $\omega_b/\omega_c$ and $\Omega_m$ play important roles, and therefore I consider the effect of varying these parameters as well.

Boltzmann solvers such as \texttt{CAMB}~\cite{Lewis:1999bs} or \texttt{CLASS}~\cite{Audren:2012vy} are used to compute the \textit{linear} power spectrum. In practice, above a certain redshift-dependent wavenumber $k_{\rm nl}$, typical perturbations in the matter overdensity field have grown non-linear today, and hence linear theory is no longer reliable. As a rule of thumb, $k_{\rm nl} \approx 0.2\,h{\rm Mpc}^{-1}$ today. In the non-linear regime, it is only possible to reliable study the power spectrum using N-body simulations. A discussion of N-body simulations is well beyond the scope of this thesis, and I refer the reader to dedicated articles discussing this issue, e.g.~\cite{Heitmann:2008eq,Heitmann:2013bra,Kwan:2013jva,
Banerjee:2018bxy,Rizzo:2016mdr,Liu:2017now,Springel:2005mi,Springel:2011yw,
Springel:2014ona,Dakin:2017idt,Castorina:2015bma,
Carbone:2016nzj,Ruggeri:2017dda,Dehnen:2011fj,Dolag:2008ki,Kuhlen:2012ft}.

There is one final subtle issue related to comparing the theoretical power spectrum with the observed one. Most of the matter field is made up of invisible dark matter, which we cannot observe directly (only indirectly through its effect on gravitational lensing). The only direct way to observe the matter field is through luminous tracers, such as galaxies. Therefore, what we really are observing is the galaxy power spectrum $P_g(k)$, not the matter power spectrum $P(k)$. The two quantities are only equal if the galaxy overdensity field faithfully traces the matter overdensity field. However, this is not the case, as galaxies are \textit{biased} tracers of the underlying matter overdensity field. Because galaxies form from peaks in the matter overdensity field which collapse under the effect of gravity, they preferentially trace more overdense regions and will in general be more clustered than the underlying matter field from which they originated~\cite{Kaiser:1984sw,Bardeen:1985tr,Mo:1995cs,
Fry:1996fg,Mann:1997df,Tegmark:1998wm}. It can be shown that the emergence of galaxy bias is a consequence of galaxy formation being a \textit{threshold process}, \textit{i.e.} galaxies can only form once the matter overdensity has reached a threshold level.

The statistical relation between the galaxy overdensity field and the matter overdensity field is commonly referred to as galaxy bias, see~\cite{Desjacques:2016bnm} for a recent complete review on the subject. On large, linear scales, analytical approaches to study galaxy formation (such as Press-Schechter theory~\cite{Press:1973iz,Sheth:2001dp}) suggest that the galaxy bias is a redshift-dependent constant~\cite{Dekek:1986gu,Cole:1989ghw,Fry:1992vr,Coles:1993ghw,Kauffmann:1995kx,
Mann:1997df,Sheth:1999mn,Benson:1999mva,Matsubara:1999qq,
Coles:2007be,Desjacques:2010gz,Assassi:2014fva}, and the galaxy and matter overdensities $\delta_g$ and $\delta$ are simply proportional to each other:
\begin{eqnarray}
\delta_g(k,z) = b(z)\delta(k,z)\,.
\label{eq:biasconstant}
\end{eqnarray}
The actual value of the bias depends on the LSS tracer in question (\textit{i.e.} different tracers will have a different bias), reflecting how ``difficult'' it is to create the tracer in first place: tracers which require a higher overdensity to form in first place, such as quasars~\cite{Burbidge:1967uc,Schmidt:1969vq,Antonucci:1993sg}, are more strongly biased~\cite{Slosar:2008hx,Han:2018izq}. For the same reason, the bias of a given tracer typically increases with redshift, as typical overdensities are lower as we go back in time and it is thus harder to form the tracer in question.

At the level of power spectrum, Eq.~(\ref{eq:biasconstant}) translates to:
\begin{eqnarray}
\underbrace{P_g(k,z)}_{\text{what we measure}} = b^2(z) \times \underbrace{P(k,z)}_{\text{what we would like to measure}}\,,
\label{eq:biaspkconstant}
\end{eqnarray}
where I have highlighted the fact that the true source of information on cosmological parameters is $P(k)$, but we only have access to $P_g(k)$. In practice, analyses of galaxy clustering are usually restricted to large, linear scales, where the galaxy bias can be treated as a constant nuisance parameter to be marginalized over (see Chapter~\ref{chap:5} for more details on the process of marginalization). If one wishes to move to more non-linear scales, a more careful treatment of the galaxy bias is necessary. On mildly non-linear scales, non-locality effects in galaxy formation start showing up, and complicate the simple picture wherein the galaxy bias is constant (see e.g.~\cite{Matsubara:1999qq,Coles:2007be} where heuristic examples of how different models of galaxy formation lead to a scale-dependent bias are presented). Several independent approaches to galaxy biasing have argued that the leading order correction to a constant bias in Fourier space, relevant on mildly non-linear scales, is a $k^2$ correction, \textit{i.e.} $b(k) \propto {\rm const} + k^2$ (see~\cite{Kaiser:1984sw,Bardeen:1985tr,Mo:1995cs,
Fry:1996fg,Mann:1997df,Tegmark:1998wm} for important early work, see~\cite{Smith:2006ne,Desjacques:2008jj,Desjacques:2010gz,Musso:2012ch,
Paranjape:2012ks,Schmidt:2012ys,Assassi:2014fva,Verde:2014nwa,
Biagetti:2014pha,Senatore:2014eva,Mirbabayi:2014zca} for later developments, and see~\cite{Desjacques:2016bnm} for a pedagogical explanation of why the lowest order correction scales as $k^2$). This will be relevant in Paper~II, where we study the impact of moving beyond the constant bias approximation in galaxy survey analyses.

On top of the difficulties brought upon by galaxy bias, another complication is that we do not observe galaxies in real space but in redshift space. In other words, galaxy surveys provide two angles and a redshift, and not three comoving coordinates. In order to obtain the latter, we need to assume a fiducial cosmology (which essentially is used to convert the redshift information into a $z$ coordinate), but this conversion only accounts for the Hubble flow and not for peculiar velocities. This mismatch between real and redshift space due to peculiar velocities is responsible for a phenomenon known as redshift-space distortions (RSD). RSDs manifests as elongation or flattening of structures, either due to random peculiar velocities in bound structures (Fingers of God effect)~\cite{Jackson:2008yv} or coherent motions of galaxies (Kaiser effect)~\cite{Kaiser:1987qv}. Fortunately, we have a rather good idea as for how to model these effects at the level of galaxy power spectrum in the linear regime (see e.g.~\cite{Hamilton:1997zq,Percival:2011ghw} for reviews), although the question of how to model non-linear RSD is well and truly open (see for instance~\cite{Heavens:1998es,Bharadwaj:2001zf,Pandey:2004jf,Matsubara:2007wj,
Shaw:2008aa,Samushia:2011cs,Sato:2011qr,delaTorre:2012dg,Jennings:2015lea,Zhu:2017vtj,
Hernandez-Aguayo:2018oxg,Jullo:2019lgq} for important work in this direction).

\subsubsection{Baryon Acoustic Oscillation distance measurements}
\label{subsubsec:bao}

\begin{figure}[!t]
\centering
\includegraphics[width=0.7\textwidth]{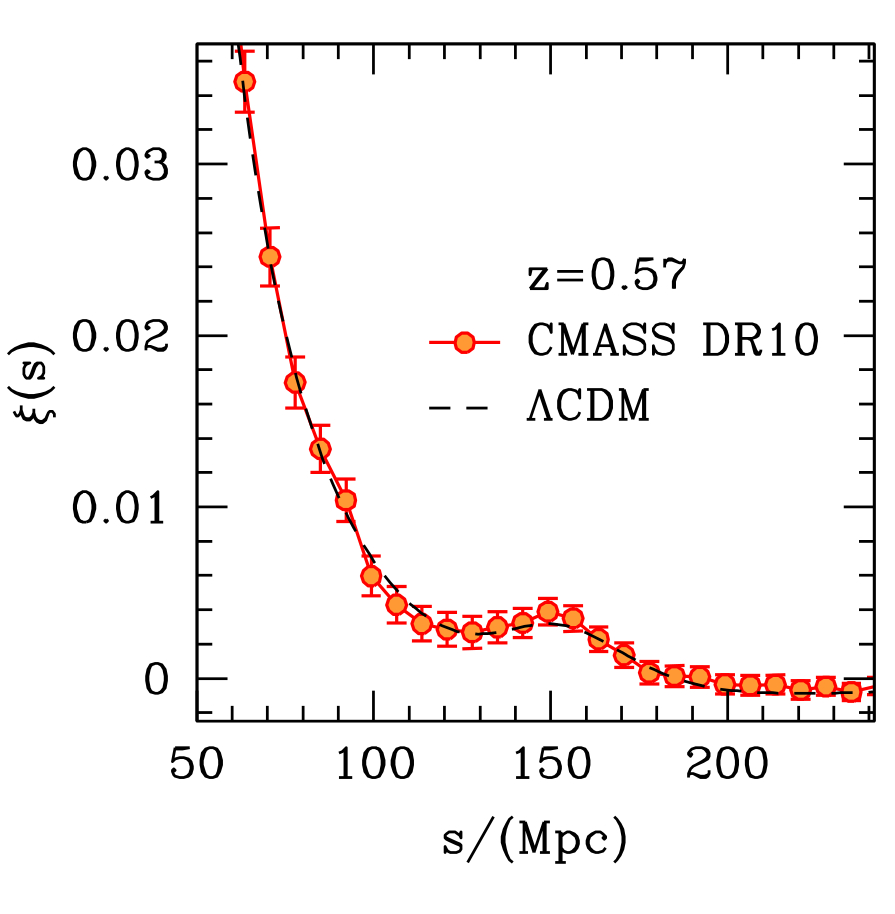}
\caption{Two point-correlation function measured from the CMASS sample of the BOSS DR10 galaxies. The ``bump'' at comoving separations of $\simeq 150\,{\rm Mpc}$ is clearly visible. Credits: BOSS collaboration~\cite{Boss:2009ghw}.}
\label{fig:xi}
\end{figure}

As we anticipated earlier, galaxy surveys can be analysed in real space or Fourier space. In the latter case, one measures the galaxy power spectrum $P_g(k)$ which we discussed in Chapter~\ref{subsubsec:galaxypowerspectrum}. In real space, one measures the 2-point correlation function $\xi(r)$, whose Fourier transform is $P_g(k)$. To get a physical understanding for the correlation function, consider a galaxy survey with mean number density $\bar{n}$, and two small regions of volume $dV_1$ and $dV_2$, separated by a distance $r$. Then, the expected number of pairs of galaxies with one galaxy in $dV_1$ and the other galaxy in $dV_2$, $\langle n_{\rm pair} \rangle$, is given by:
\begin{eqnarray}
\langle n_{\rm pair} \rangle = \bar{n}^2 \left [ 1+\xi(r) \right ]dV_1dV_2\,.
\label{eq:xir}
\end{eqnarray}
Therefore, $\xi(r)$ measures the excess clustering of galaxies at any given separation $r$. If $\xi(r)=0$, galaxies are unclustered, \textit{i.e.} randomly distributed. Conversely, $\xi(r)>0$ ($\xi(r)<0$) corresponds to stronger clustering (anti-clustering).

\begin{figure}[!t]
\centering
\includegraphics[width=1.0\textwidth]{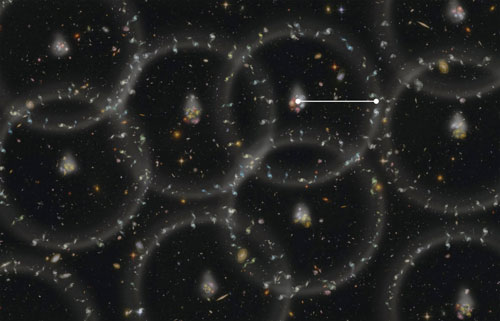}
\caption{Cartoon version of BAOs, showing spheres of baryons around initial dark matter clumps, with an excess clustering at a scale corresponding to the sound horizon at decoupling. Credits: BOSS collaboration~\cite{Boss:2009ghw}.}
\label{fig:bao}
\end{figure}

As a function of separation $r$, the 2-point correlation function $\xi(r)$ drops roughly as a power-law, $\xi(r) \propto r^{-\gamma}$ with $\gamma \sim -2$ (see e.g.~\cite{Totsujikihara,Peebles,Hauserpeebles,Peebleshauser,Peeblesnew,Watson:2011cz}). On top of the power-law, $\xi(r)$ exhibits a ``bump'' at comoving separations of about $150\,{\rm Mpc}$. This is a signature of the BAOs which were set up in the photon-baryon fluid. Heuristically, we can imagine several superimposed acoustic waves propagating simultaneously, and freezing at the time of decoupling (more precisely, at the drag epoch when baryons were released from the photon drag, see Chapter~\ref{subsec:history}). An exaggerated cartoon version of this situation is shown in Fig.~\ref{fig:bao}. The result is a slight preference for perturbations (which later grow into galaxies) separated by a distance $r_s(z_{\rm drag})$, since that is the distance travelled by sound waves at the time baryons were released from the photon drag and the waves froze.

The BAO bump in the two-point correlation function is the real-space counterpart of the BAO wiggles in the power spectrum (see e.g.~\cite{Bassett:2009mm} for a comprehensive review). The sound horizon at baryon drag $r_s(z_{\rm drag})$ is a quantity of known and fixed length. Thus, comparing its apparent size to its known size allows us to determine the distance to the galaxy survey in question, and indirectly measure the low-redshift expansion rate of the Universe. If one has sufficient sensitivity as to separate line-of-sight and transverse separations, the two can be used to constrain the combinations $r_s(z_{\rm drag})H(z)$ and $\chi(z)/r_s(z_{\rm drag})$ respectively, where $\chi(z)$ is the comoving distance to the redshift of the galaxy sample. Until recently, most galaxy surveys did not have sufficient sensitivity to do so, and instead performed an isotropic analysis, sensitive to the quantity $d_V(z)$ known as volume distance~\cite{Eisenstein:2005su,Percival:2007yw,Bassett:2009mm,Percival:2013awa}:
\begin{eqnarray}
d_V(z) = \left [ \frac{z\chi(z)^2}{H(z)} \right ]^{\frac{1}{3}}\,.
\label{eq:dv}
\end{eqnarray}
Most BAO distance measurements are reported in terms of constraints on $d_V(z)/r_s(z_{\rm drag})$, which tightly limit parameters determining the late-time expansion of the Universe. In particular, it can be shown that BAO distance measurements mostly constrain $\Omega_m$ and $H_0$~\cite{Aubourg:2014yra,Addison:2017fdm}, and are thus highly complementary to CMB measurements. In fact, BAO distance measurements are typically used in combination with CMB measurements to break degeneracies among cosmological parameters which would otherwise be present when only using the latter.

\section{Neutrino signatures in cosmological observations}
\label{sec:signaturesnu}

So far, we have provided a qualitative but rather complete picture of CMB and LSS probes. In particular, we have seen how these probes are sensitive to various cosmological parameters. Of course, these probes are also sensitive to neutrino properties, otherwise we wouldn't be here to talk about it. The natural question, then, is what are the signatures of neutrinos in CMB and LSS probes? In answering this question, I will follow closely~\cite{Lattanzi:2017ubx}, as well as the classic textbook~\cite{Lesgourgues:2018ncw}, but keeping the discussion as brief as possible: I invite the interested reader who wants to dig deeper into this interesting question to read Chapter~5.1.3, 6.1.3, and 6.1.4 of~\cite{Lesgourgues:2018ncw}. In the literature there are a number of excellent resources covering the effects of neutrinos on cosmological observations discussed here, with varying level of technicality and details: an incomplete list is given by~\cite{Dolgov:2002wy,Hannestad:2004nb,Fukugita:2005sb,Hannestad:2005ey,Lesgourgues:2006nd,Dolgov:2008hz,Hannestad:2010kz,Wong:2011ip,
Lesgourgues:2012uu,Balantekin:2013gqa,Lesgourgues:2014zoa,Verde:2015ana,Abazajian:2016hbv,Archidiacono:2017tlz,Gerbino:2018jee}. I will first discuss neutrino signatures in the CMB anisotropies, and then in the matter power spectrum.

\subsection{Signatures of neutrinos in the CMB anisotropies}
\label{subsec:signaturesnucmb}

As we have seen in Chapter~\ref{subsec:cmb}, several effects depending on various parameters or combinations of parameters mix between each other when determining the shape of the CMB temperature power spectrum. As a result, it is not simple to discuss the direct impact of neutrinos (or of any given species, for that matter), as this would require to some extent separating these effects. To make progress, it is useful to classify effects of neutrinos on the CMB anisotropies in two categories: \textit{background effects} and \textit{perturbation effects}. The former are generally considered more ``indirect'', and can usually be reabsorbed by suitably tuning the other cosmological parameters when varying neutrino parameters, whereas the latter are generally considered more ``direct'', a tell-tale of neutrinos. Background effects are related to changes in the evolution of the scale-factor and consequently to the background evolution of $H(z)$. As we have seen earlier, the CMB anisotropy spectra are sensitive to a number of characteristic scales (such as $z_{\rm eq}$, $r_s$, and $\chi_{\star}$). Varying neutrino parameters while na\"{i}vely keeping other cosmological parameters fixed will generally change these scales: however, since these are very well fixed by observations, it would be wise to instead vary other cosmological parameters at the same time to keep these scales fixed. We will later show that such a choice makes a significant difference, and allows to isolate the ``direct'' signature of neutrinos more cleanly. On the other hand, perturbation effects are related to the impact of neutrinos on metric fluctuations (gravitational potentials), which back-react on perturbations to the photon-baryon fluid. Such effects are mostly related to changes in the EISW and LISW effects, as well as in the gravitational lensing of CMB photons.
\begin{figure}[!t]
\centering
\includegraphics[width=1.0\textwidth]{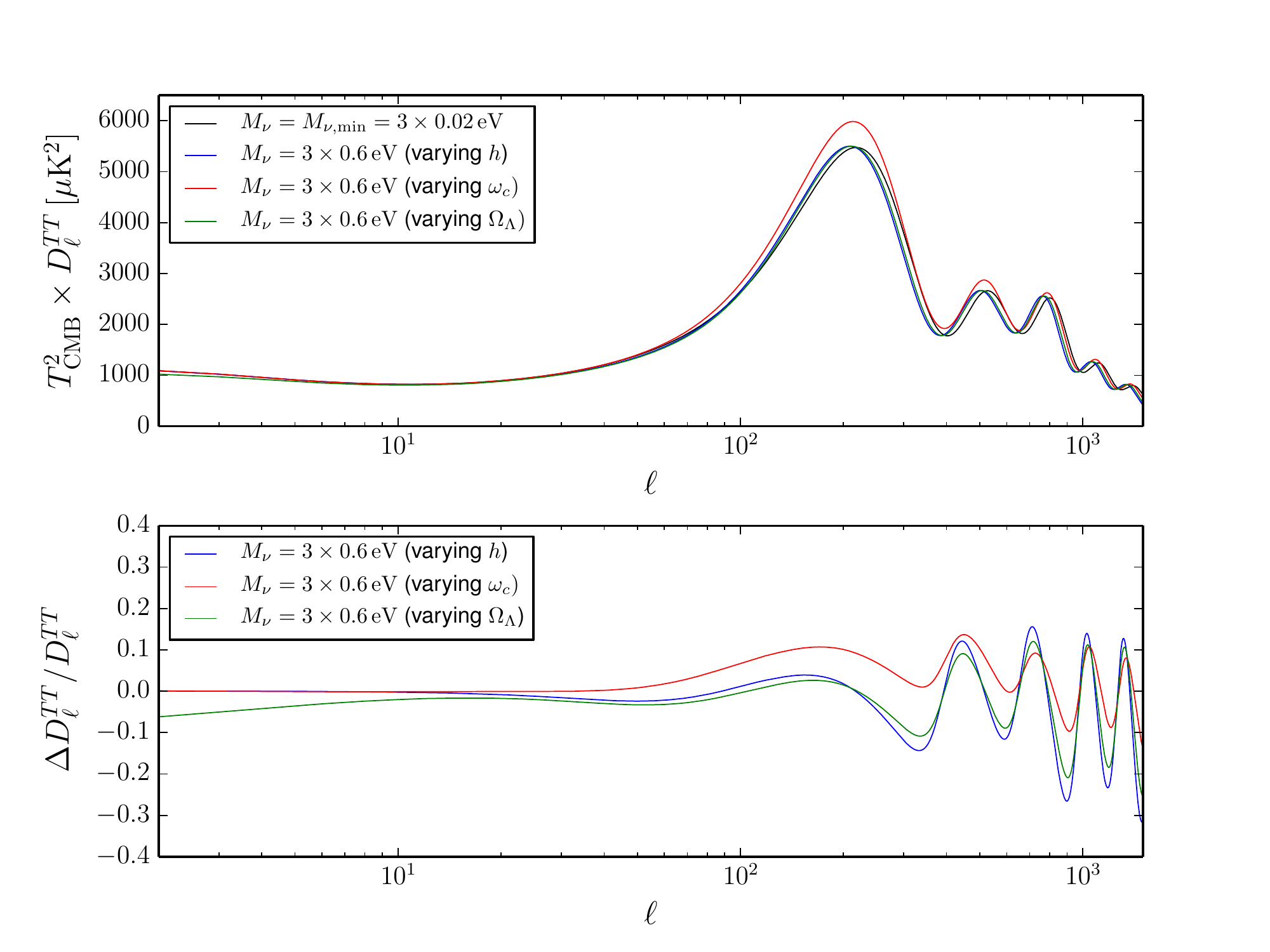}
\caption{Impact of increasing the sum of the neutrino masses $M_{\nu}$ on the CMB temperature power spectrum. \textit{Upper panel}: the black curve is the power spectrum for the baseline model where $M_{\nu}=0.06\,{\rm eV}$. In addition, we set $h=0.7$, $\omega_c=0.12$, and $\Omega_{\Lambda}=0.713$. The other three curves are obtained for $M_{\nu}=1.8\,{\rm eV}$, where the increase in $M_{\nu}$ is compensated by setting $h=74.48$ (blue curve), $\omega_c=0.10144$ (red curve), and $\Omega_{\Lambda}=0.675$ (green curve). Notice that, as per standard convention in the field, the quantity plotted on the $y$ axis is $T_{\rm CMB}^2\ell(\ell+1)C_{\ell}$, with $T_{\rm CMB} \approx 2.725\,{\rm K}$ the CMB temperature today. \textit{Lower panel}: relative change in power with respect to the baseline model, with the same color coding as above. The main changes are due to: an increase in $\theta_s$ when varying $h$ (blue curve); an increase in $\theta_s$ and an enhanced EISW effect when varying $\omega_c$ (red curve); and an increase in $\theta_s$ and a reduced LISW effect when varying $\Omega_{\Lambda}$ (green curve).}
\label{fig:neutrinoscmb}
\end{figure}

I first begin by discussing signatures of neutrino masses, in other words the impact of $M_{\nu}$ on the CMB anisotropy spectra. I will focus exclusively on the CMB temperature power spectrum, although very similar considerations apply to polarization and temperature-polarization spectra. As we have seen in Chapter~\ref{subsec:lss}, neutrinos with total mass $M_{\nu} \lesssim 1.8\,{\rm eV}$ turn non-relativistic after matter-radiation equality. Since cosmological data strongly favour $M_{\nu} \ll 1\,{\rm eV}$, in the following we will always count neutrinos as radiation at matter-radiation equality, recombination, and decoupling. In other words, $z_{\rm eq}$ is given by:
\begin{eqnarray}
z_{\rm eq} = \frac{\omega_b+\omega_c}{\omega_{\gamma} \left [ 1+\frac{7}{8} \left ( \frac{4}{11} \right )^{\frac{4}{3}}N_{\rm eff} \right ]} \equiv  \frac{\omega_b+\omega_c}{\alpha\omega_{\gamma}} \,,
\label{eq:zeqneutrinos}
\end{eqnarray}
where $\alpha \equiv [1+7/8(4/11)^{4/3}N_{\rm eff}] \approx (1+0.2271N_{\rm eff})$. I first follow the approach of~\cite{Lattanzi:2017ubx}, wherein $M_{\nu}$ is varied without attempting to keep the characteristic times and scales governing the CMB power spectrum fixed. At a later moment, I will follow the approach of~\cite{Lesgourgues:2018ncw}, where $M_{\nu}$ is varied while keeping these times and scales fixed. The approach of~\cite{Lattanzi:2017ubx} is more simple to follow especially for non-expert readers, albeit it obscures the direct neutrino signature. We have seen in Eq.~(\ref{eq:sumrule}) that the sum of all the density parameters $\Omega_i$ at present time should be equal to $1$. Defining the physical density parameters $\omega_i \equiv \Omega_ih^2$ and restricting ourselves to a minimal $\Lambda$CDM+$M_{\nu}$ model, the sum rule can be rewritten as:
\begin{eqnarray}
\omega_{\gamma} + \omega_b + \omega_c + \omega_{\Lambda} + \omega_{\nu} = h^2\,,
\label{eq:sumreduced}
\end{eqnarray}
Recall that $\omega_{\gamma}$ is accurately determined by measuring the CMB temperature, so it is for all intents and purposes fixed. On the other hand, $\omega_{\nu} \propto M_{\nu}$, so increasing $M_{\nu}$ directly increases $\omega_{\nu}$. However, Eq.~(\ref{eq:sumreduced}) must always be satisfied as $M_{\nu}$ is increased, so an increase in $M_{\nu}$ must be compensated for by varying one or more among $h$, $\omega_b$, $\omega_c$, and $\omega_{\Lambda}$. As we have seen in Chapter~\ref{subsec:cmb}, the relative height between odd and even peaks accurately fixes $\omega_b$ (and $\omega_b$ also strongly influences the abundances of light elements produced by BBN), so directly varying $\omega_b$ is not a wise choice. Following the pedagogical approach of~\cite{Lattanzi:2017ubx}, I choose $h$, $\omega_c$ and $\Omega_{\Lambda}$ as the parameters to be varied (one at a time) to compensate for the increase in $M_{\nu}$ and ensure that the sum rule remains satisfied. Notice that in all of this, $A_s$, $n_s$, $\tau$, and $\omega_b$ remain fixed. It is useful to rewrite Eq.~(\ref{eq:hznu}):
\begin{eqnarray}
H(z) = H_0\sqrt{(\Omega_b + \Omega_c)(1+z)^3 + \Omega_{\gamma}(1+z)^4 + \Omega_{\Lambda} + \frac{\rho_{\nu}(z)}{\rho_{\rm crit}}}\,.
\label{eq:hznureduced}
\end{eqnarray}
When considering the impact of varying $M_{\nu}$, we make comparisons with respect to a baseline model where $M_{\nu}=0.06\,{\rm eV}$ (distributed across 3 degenerate neutrinos of equal mass $0.02\,{\rm eV}$). The CMB temperature power spectrum for this baseline model is given by the black curve in the upper panel of Fig.~\ref{fig:neutrinoscmb}.

Let us consider a first case where we compensate for the increase in $M_{\nu}$ (and hence $\omega_{\nu}$) by increasing $h$ while keeping $\omega_c$ and $\Omega_{\Lambda}$ fixed. By inspecting Eq.~(\ref{eq:hznureduced}), it is easy to show that well before the neutrino non-relativistic transition ($z \gg z_{\rm nr}$), $M_{\nu}$ does not affect the expansion history, while for $z \lesssim z_{\rm nr}$ increasing $M_{\nu}$ increases the expansion rate. This implies that $r_s$ is left unchanged, but $\chi_{\star}$ decreases: therefore $\theta_s$ increases, and all peaks are projected to smaller multipoles. On the other hand, the height of the first peak should remain approximately unchanged, since $z_{\rm eq}$ remains unchanged [see Eq.~(\ref{eq:zeqneutrinos})] and therefore so does the EISW effect.

We now consider a second case where we compensate for the increase in $M_{\nu}$ by decreasing $\omega_c$ while keeping $h$ and $\Omega_{\Lambda}$ fixed. In this case, by inspecting Eq.~(\ref{eq:hznureduced}), we see that the expansion rate is unchanged for $z \gg z_{\rm eq}$ and for $z \ll z_{\rm nr}$, while for $z_{\rm nr} \lesssim z \lesssim z_{\rm eq}$ the expansion rate is decreased. This increases both $r_s$ (due to the decrease in $H$ between $z_{\rm eq}$ and $z_{\rm dec}$) and $\chi_{\star}$ (due to the decrease in $H$ between $z_{\rm dec}$ and $z_{\rm nr}$): numerically, we find that the former effect dominates over the latter, the net effect being again an increase in $\theta_s$ and a shift of all peaks to smaller multipoles. Moreover, from Eq.~(\ref{eq:zeqneutrinos}) we see that decreasing $\omega_c$ delays equality, the net result being an enhanced EISW effect and hence a higher first peak.

Finally, we consider the third case where we compensate for the increase in $M_{\nu}$ by decreasing $\Omega_{\Lambda}$ while keeping $h$ and $\omega_c$ fixed. Inspecting Eq.~(\ref{eq:hznureduced}) leads us to conclude that the expansion rate is unchanged for $z \gg z_{\rm nr}$, whereas numerically we find that for $z \lesssim z_{\rm nr}$, $H$ increases. As a result $r_s$ is unchanged, whereas $\chi_{\star}$ decreases, and the net effect is again that $\theta_s$ increases and all peaks are shifted to smaller multipoles. Moreover, since $z_{\rm eq}$ is unchanged, the height of the first peak remains the same. However, since decreasing $\Omega_{\Lambda}$ decreases the period of dark energy domination, we expect the LISW effect to be reduced and hence a decrease in power at very low $\ell$ (which however would be very hard to detect because of the large error bars due to cosmic variance).

The temperature power spectra in the three cases discussed above are shown in the upper panel of Fig.~\ref{fig:neutrinoscmb}, and confirm all our expectations: a shift in the peaks towards smaller $\ell$s for the case where $h$ is increased (blue curve), a similar shift with in addition an enhanced first peak for the case where $\omega_c$ is decreased (red curve), and again a similar shift with in addition a reduction in power at low $\ell$s when $\Omega_{\Lambda}$ is decreased (green curve). The lower panel of Fig.~\ref{fig:neutrinoscmb} instead shows the relative change in the power spectra with respect to the baseline case, and is helpful in making these shifts more evident.

In the three cases we just discussed, we have seen that increasing $M_{\nu}$ led to changes in the CMB power spectrum due to shifts in the background quantities $r_s$, $\chi_{\star}$, $z_{\rm eq}$, and $z_{\Lambda}$. This has the effect of concealing the ``direct'' effect of $M_{\nu}$ behind larger effects due to shifting background quantities. The more meaningful comparison between models with different $M_{\nu}$ should therefore be performed trying to keep the previous scales constant whenever possible. This is the approach advocated in~\cite{Lesgourgues:2018ncw}. It is easy to convince oneself that within the framework of the minimal $\Lambda$CDM+$M_{\nu}$ model, there isn't sufficient freedom to vary $M_{\nu}$ and keep all four the previous scales fixed. However, since the physical effects controlled by the first three are much more constrained than the LISW effect controlled by $z_{\Lambda}$, the most meaningful comparison between models with different $M_{\nu}$, at least as far as CMB data is concerned, should be performed keeping $r_s$, $\chi_{\star}$, and $z_{\rm eq}$, while allowing $z_{\Lambda}$ to vary. This can be achieved keeping $\omega_b$ and $\omega_c$ fixed, while decreasing $h$ and $\Omega_{\Lambda}$.
\begin{figure}[!t]
\centering
\includegraphics[width=1.0\textwidth]{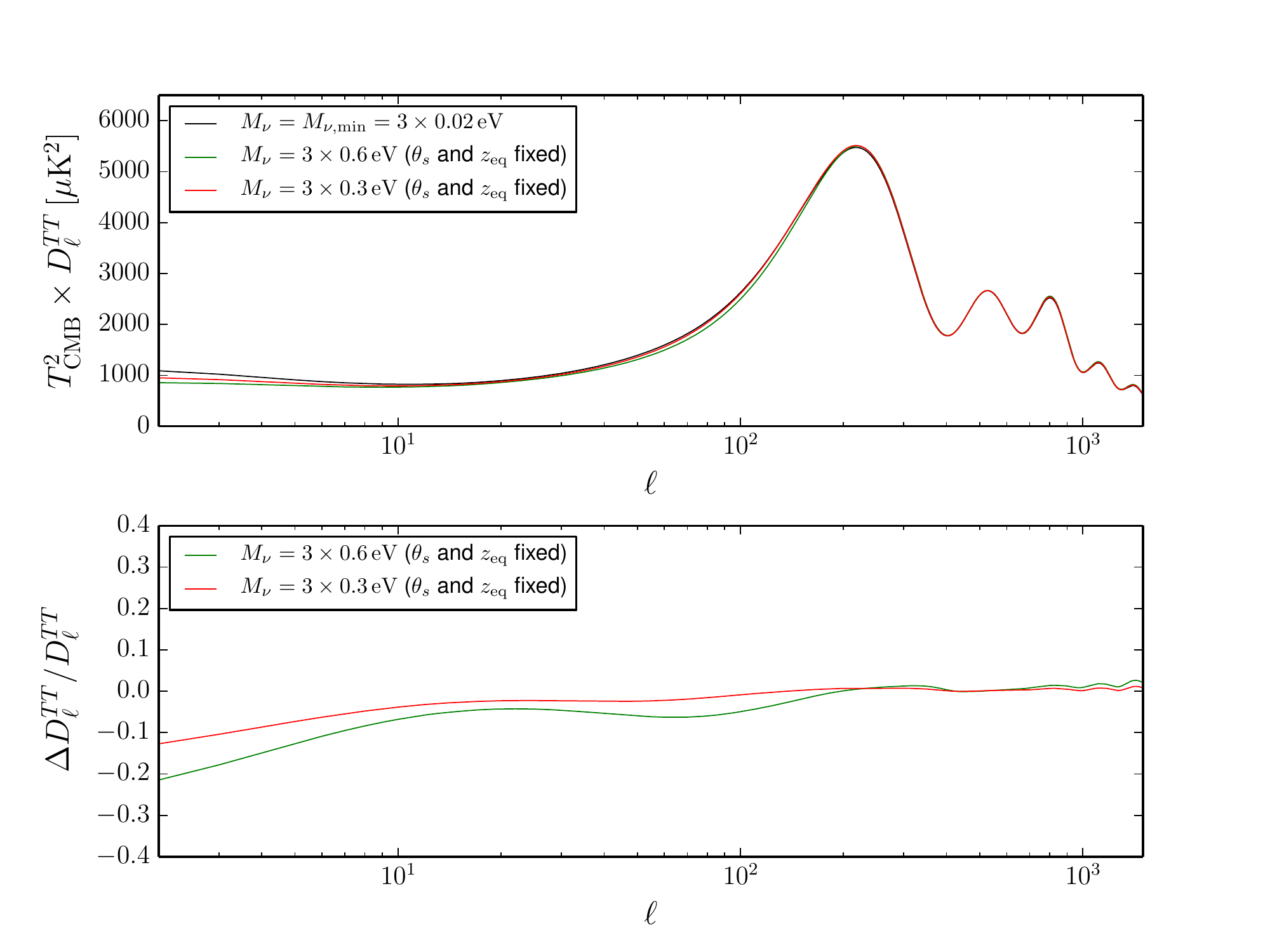}
\caption{Impact of increasing the sum of the neutrino masses $M_{\nu}$ on the CMB temperature power spectrum, adjusting $h$ and $\Omega_{\Lambda}$ to keep $\theta_s$ and $z_{\rm eq}$ fixed at the expense of a small shift in $z_{\Lambda}$. \textit{Upper panel}: the black curve is the power spectrum for the baseline model where $M_{\nu}=0.06\,{\rm eV}$, $h=0.7$, and $\Omega_{\Lambda}=0.713$. The green (red) curve is obtained for $M_{\nu}=1.8\,{\rm eV}$ ($M_{\nu}=0.9\,{\rm eV}$), where the increase in $M_{\nu}$ is compensated for by setting $h=0.569$ ($h=0.628$) and $\Omega_{\Lambda}=0.508$ ($\Omega_{\Lambda}=0.621$). Notice that, as per standard convention in the field, the quantity plotted on the $y$ axis is $T_{\rm CMB}^2\ell(\ell+1)C_{\ell}$, with $T_{\rm CMB} \approx 2.725\,{\rm K}$ the CMB temperature today. \textit{Lower panel}: relative change in power with respect to the baseline model, with the same color coding as above. The main changes are due to a reduced LISW effect, a reduced EISW effect, a minuscule change in the damping scale, and a reduction of the lensing effect.}
\label{fig:neutrinoscmbtheta}
\end{figure}

The effect on the CMB power spectrum of increasing $M_{\nu}$ while keeping $\theta_s$ and $z_{\rm eq}$ fixed is shown in Fig.~\ref{fig:neutrinoscmbtheta}. The large changes due to the shift of $\theta_s$ and the enhanced EISW effect, previously visible in Fig.~\ref{fig:neutrinoscmb}, have now basically been removed, and it is clear that the direct effect of neutrino masses turns out to be quite subtle. The largest change is the reduction in power at low-$\ell$ due to a reduced LISW effect, expected given that we chose to vary $z_{\Lambda}$ by decreasing $\Omega_{\Lambda}$ (decreasing the duration of dark energy domination). Tiny shifts at high-$\ell$ ($\ell \gtrsim 500$) are instead due to minuscule shifts in the damping scale. Moreover, at high-$\ell$, neutrinos suppress the lensing power spectrum. Because of their free-streaming nature we have discussed in Chapter~\ref{subsec:evolutionneutrinos}, and for reasons that will become clearer in Chapter~\ref{subsec:signaturesnulss}, at late times neutrinos suppress the growth of structure, resulting in less structure which lenses the CMB. The effect of lensing is to smear the high-$\ell$ peaks, and as such increasing $M_{\nu}$ sharpens the peaks. This effect, however, is small and hardly visible in Fig.~\ref{fig:neutrinoscmbtheta}. We also expect the shift in the damping scale to show up on the high-$\ell$ part of the $EE$, $TE$, and $BB$ spectra, whereas the reduction of the lensing potential will reduce the amount of lensing $B$-modes (showing up as a reduction in power in the high-$\ell$ part of the $BB$ spectrum).

The direct perturbation effects due to massive neutrinos instead show up on scales $50 \lesssim \ell \lesssim 200$, where we see that increasing $M_{\nu}$ reduces power by $\Delta D_{\ell}^{TT}/D_{\ell}^{TT} \approx -(M_{\nu}/10\,{\rm eV})$~\cite{Lesgourgues:2018ncw}. The reason is to be found in a reduced EISW effect. In fact, on large scales, neutrinos behave as a clustering component, \textit{i.e.} more like matter than radiation: this leads to less decay of the gravitational potential (recall that gravitational potentials decay in a radiation-dominated Universe and are constant in a pure-matter Universe), and hence a reduced EISW effect, since the latter is driven by time variations of the gravitational potential.~\footnote{Technically, this effect depends on the masses of the individual eigenstates, but in practice the effect of the individual masses is below sub-percent, and hence unobservable even with next-generation CMB experiments.}

\begin{figure}[!t]
\centering
\includegraphics[width=1.0\textwidth]{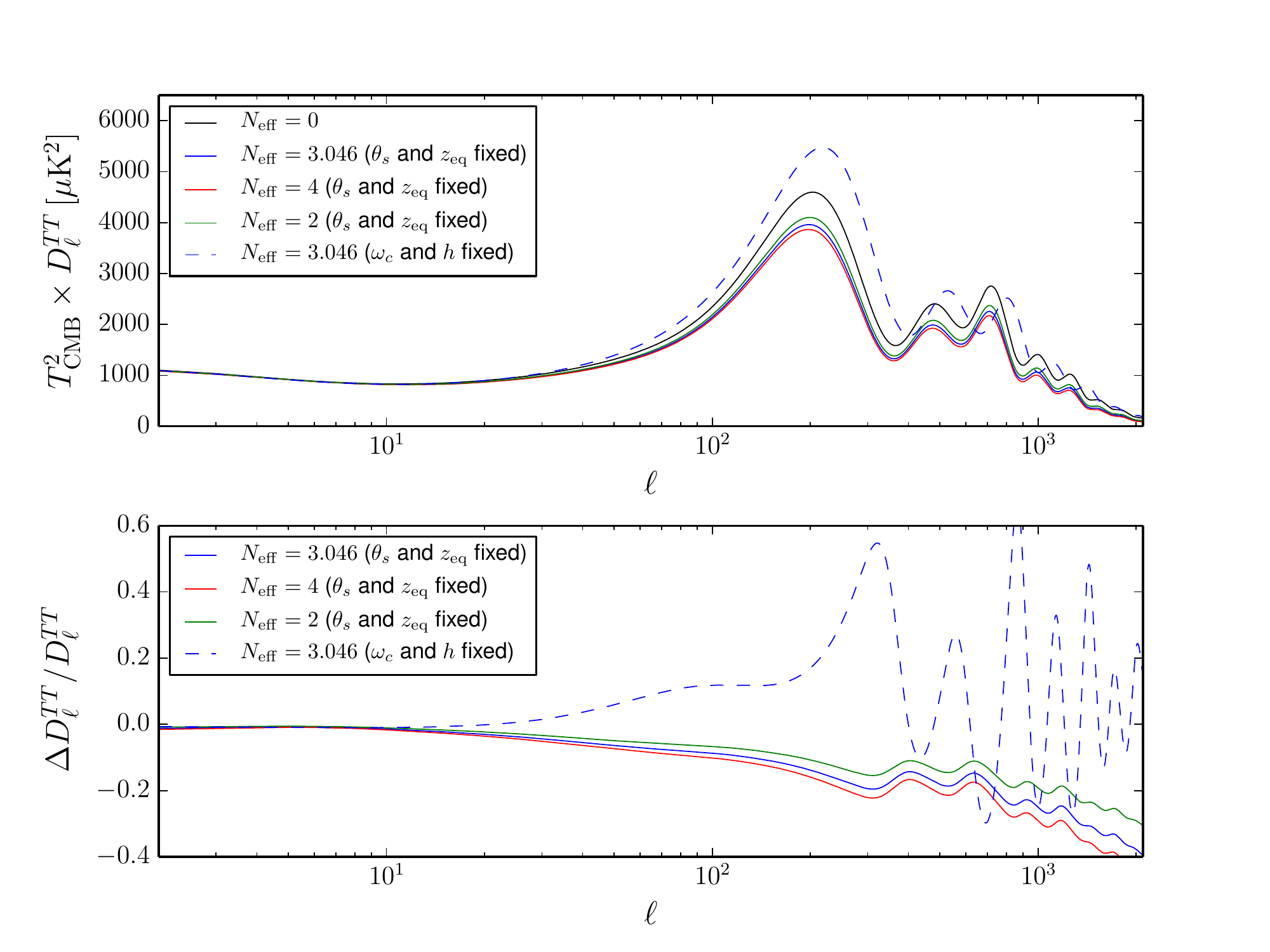}
\caption{Impact of increasing the effective number of neutrinos $N_{\rm eff}$ on the CMB temperature power spectrum. \textit{Upper panel}: the black curve is the power spectrum for the baseline model where $N_{\rm eff}=0$. In addition, we set $\omega_c=0.12$ and $h=0.7$. The dashed blue curve is obtained for $N_{\rm eff}=3.046$, keeping $\omega_c$ and $h$ fixed. The other three curves are obtained for $N_{\rm eff}=3.046$ (solid blue curve), $N_{\rm eff}=2$ (solid green curve), and $N_{\rm eff}=4$ (solid red curve), where the increase in $N_{\rm eff}$ is compensated by setting $\omega_c=0.217$, $h=0.9105$ (solid blue curve), $\omega_c=0.184$, $h=0.8441$ (solid red curve), and $\omega_c=0.247$, $h=0.9670$ (solid blue curve). Notice that, as per standard convention in the field, the quantity plotted on the $y$ axis is $T_{\rm CMB}^2\ell(\ell+1)C_{\ell}$, with $T_{\rm CMB} \approx 2.725\,{\rm K}$ the CMB temperature today. \textit{Lower panel}: relative change in power with respect to the baseline model, with the same color coding as above. The main changes are due to shifts in $\theta_s$, $z_{\rm eq}$, and $r_d$ when keeping $\omega_c$ and $h$ fixed (dashed blue curve), and shifts in $r_d$ as well as direct perturbation effects (reduced EISW effect and phase shift) for the remaining three cases.}
\label{fig:neffcmb}
\end{figure}

So far we have looked at the effect of neutrino masses on the CMB power spectrum, parametrized through $M_{\nu}$, a parameter which will interest us a lot in this Thesis (see Chapter~\ref{chap:6}). We will also be interested, albeit to a significantly lesser extent, in the effective number of relativistic species or effective number of neutrino species $N_{\rm eff}$, a parameter controlling the energy density of neutrinos while in the relativistic regime (or of any extra relativistic species for that matter). For this reason, I will now discuss the effect of $N_{\rm eff}$ on the CMB power spectrum, albeit more briefly than I did previously for $M_{\nu}$. Despite being unphysical, let me for purely instructive purposes consider a baseline model $N_{\rm eff}=0$. The power spectrum of such model is given by the black curve in Fig.~\ref{fig:neffcmb}.

The considerations made earlier for $M_{\nu}$ hold here as well: when varying $N_{\rm eff}$ it is important to try and isolate effects due to shifts in background quantities from ``direct'' perturbation effects due to $N_{\rm eff}$. When $N_{\rm eff}$ increases, na\"{i}vely $z_{\rm eq}$ decreases according to Eq.~(\ref{eq:zeqneutrinos}), leading to an enhanced EISW effect and hence an increase in the height of the first peak. In addition, the early time expansion rate is increased, leading to a decrease in the sound horizon and hence in $\theta_s$, shifting all peaks to larger multipoles. The same increase in the early expansion rate also changes the damping scale. All these effects are clearly seen in the dashed blue curve in the upper panel of Fig.~\ref{fig:neffcmb}, plotted for $N_{\rm eff}=3.046$ and keeping $\omega_c$ and $h$ fixed to the same values I used for the $N_{\rm eff}=0$ case.

As discussed in~\cite{Lesgourgues:2018ncw}, there is a way to increase $N_{\rm eff}$ while keeping $z_{\rm eq}$, $r_s$, and $\chi_{\star}$ (and hence $\theta_s$) fixed. This involves performing the transformations $h \to h\sqrt{\alpha}$ and $\omega_c \to \omega_c+(\alpha-1)\omega_m$, with $\alpha$ defined in Eq.~(\ref{eq:zeqneutrinos}). In this way, one reabsorbs the changes due to the shift in $\theta_s$ and the enhanced EISW effect. The effect on the CMB power spectrum of increasing $N_{\rm eff}$ while keeping $\theta_s$ and $z_{\rm eq}$ fixed is shown in the solid green, blue, and red curves in the upper panel of Fig.~\ref{fig:neffcmb}. Most of the remaining changes are then due to the change in the damping scale (which is still a background quantity), and to a lesser extent from direct perturbation effects. As argued in~\cite{Hu:1995en,Bashinsky:2003tk}, direct perturbation effects are related to a suppression in the EISW effect, the reason being that neutrinos cannot cluster on small scales and hence reduce time variations in the gravitational potential on those scales: this leads to a suppression of $\Delta D_{\ell}^{TT}/D_{\ell}^{TT} \approx -0.072\Delta N_{\rm eff}$. Moreover, during the BAO epoch, neutrinos travel at a speed close to the speed of light, whereas temperature fluctuations travel at the speed of sound (lower by a factor of $\simeq \sqrt{3}$): this mismatch in speed leads to neutrinos dragging out temperature fluctuations from potential wells. In the temperature power spectrum, this shows up in a phase shift, \textit{i.e.} a shift in the peaks towards smaller $\ell$ even when $\theta_s$ is kept fixed.

\begin{figure}[!t]
\centering
\includegraphics[width=1.0\textwidth]{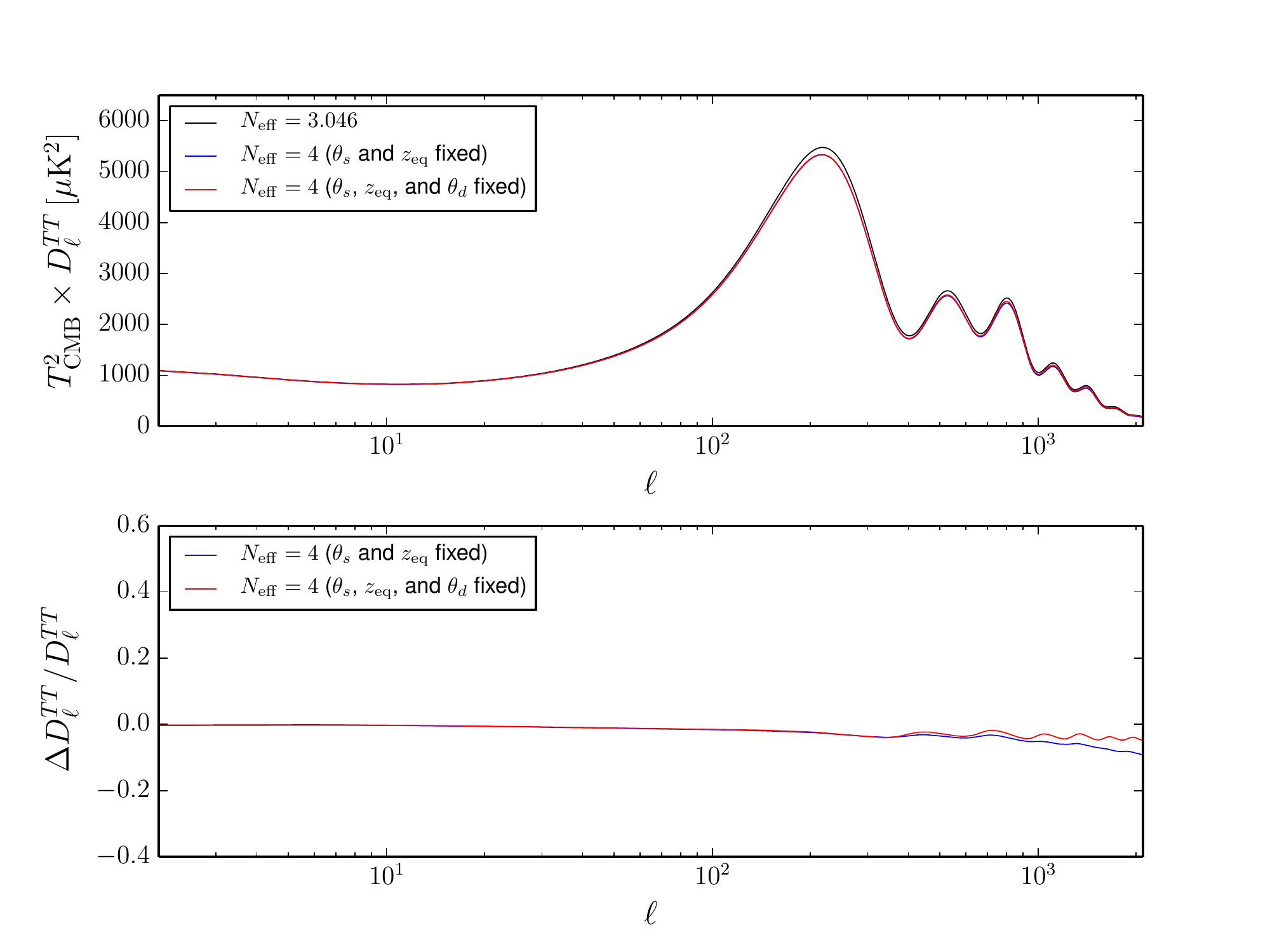}
\caption{Impact of increasing the effective number of neutrinos $N_{\rm eff}$ on the CMB temperature power spectrum while adjusting the Helium fraction $Y_p$ to keep the damping scale $r_d$ fixed. \textit{Upper panel}: the black curve is the power spectrum for a baseline model where $N_{\rm eff}=3.046$. In addition, we set $\omega_c=0.12$, $h=0.7$, and $Y_p=0.24$. The blue curve is obtained by increasing $N_{\rm eff}=4$ and compensating this increase by setting $\omega_c=0.138$ and $h=0.7435$, in order to keep $\theta_s$ and $z_{\rm eq}$, but not $r_d$ fixed. The red curve is obtained by further setting $Y_p=0.19$ to keep $r_d$ fixed. However, this is an unrealistically low value for $Y_p$, so this exercise is to be considered purely illustrative. Notice that, as per standard convention in the field, the quantity plotted on the $y$ axis is $T_{\rm CMB}^2\ell(\ell+1)C_{\ell}$, with $T_{\rm CMB} \approx 2.725\,{\rm K}$ the CMB temperature today. \textit{Lower panel}: relative change in power with respect to the baseline model, with the same color coding as above. The main changes are due to the shift in $r_d$ when not varying $Y_p$ (blue curve), and direct perturbation effects (reduced EISW effect and phase shift) when varying $Y_p$ (red curve).}
\label{fig:neffheliumcmb}
\end{figure}

As an aside, since the shift in the damping scale is still a background shift, it would be somewhat desirable to reabsorb it. Unfortunately, within the minimal $\Lambda$CDM+$N_{\rm eff}$ this is not possible while also keeping $\theta_s$ and $z_{\rm eq}$ fixed (as earlier for the $\Lambda$CDM+$M_{\nu}$ model it was not possible to keep $z_{\Lambda}$ fixed)~\cite{Lesgourgues:2018ncw}. To keep $r_d$ fixed, it is necessary to change the recombination history. One way to do so, pursued in~\cite{Bashinsky:2003tk,Hou:2011ec} is to decrease the primordial Helium fraction $Y_p$, which therefore rescales the density of free electrons $n_e$ appearing in Eq.~\ref{eq:rd}. I show the result of following this approach in Fig.~\ref{fig:neffheliumcmb}, where I compare a reference model with $N_{\rm eff}=3.046$ to a model with $N_{\rm eff}=4$, after reabsorbing the shifts in $\theta_s$ and $z_{\rm eq}$ as discussed earlier by shifting $\omega_c$ and $h$, and reabsorbing the shift in $r_d$ by decreasing $Y_p$. However, I find that $Y_p$ needs to be decreased to unrealistically low values (in practice, $Y_p$ is basically fixed to $0.24$ by BBN~\cite{Pisanti:2007hk,Iocco:2008va,Aver:2010wq,Izotov:2010ca,Consiglio:2017pot}), and hence such an exercise is to be considered purely illustrative.

In summary, I have argued that in order to understand the direct impact of neutrino parameters on the CMB spectra it is necessary to reabsorb na\"{i}vely shifts in background quantities ($r_s$, $\chi_{\star}$, $z_{\rm eq}$) as much as possible by tuning other parameters while $M_{\nu}$ and $N_{\rm eff}$ are varied. In this way, we found that the direct effect of neutrino masses on the CMB temperature spectrum shows up as a depletion of power at intermediate scales due to a reduced EISW effect, as well as a reduction of the lensing potential on small scales and a reduced LISW effect on large scales. The direct effects of varying the effective number of neutrinos are instead reflected in a reduced EISW effect and a phase shift of the acoustic peaks due to the neutrino drag effect in the early Universe (as well as a shift in the damping scale which cannot be removed if not by setting the primordial Helium fraction to unrealistically low values which are excluded by BBN).

\subsection{Signatures of neutrinos in the matter power spectrum}
\label{subsec:signaturesnulss}

\begin{figure}[!t]
\centering
\includegraphics[width=1.0\textwidth]{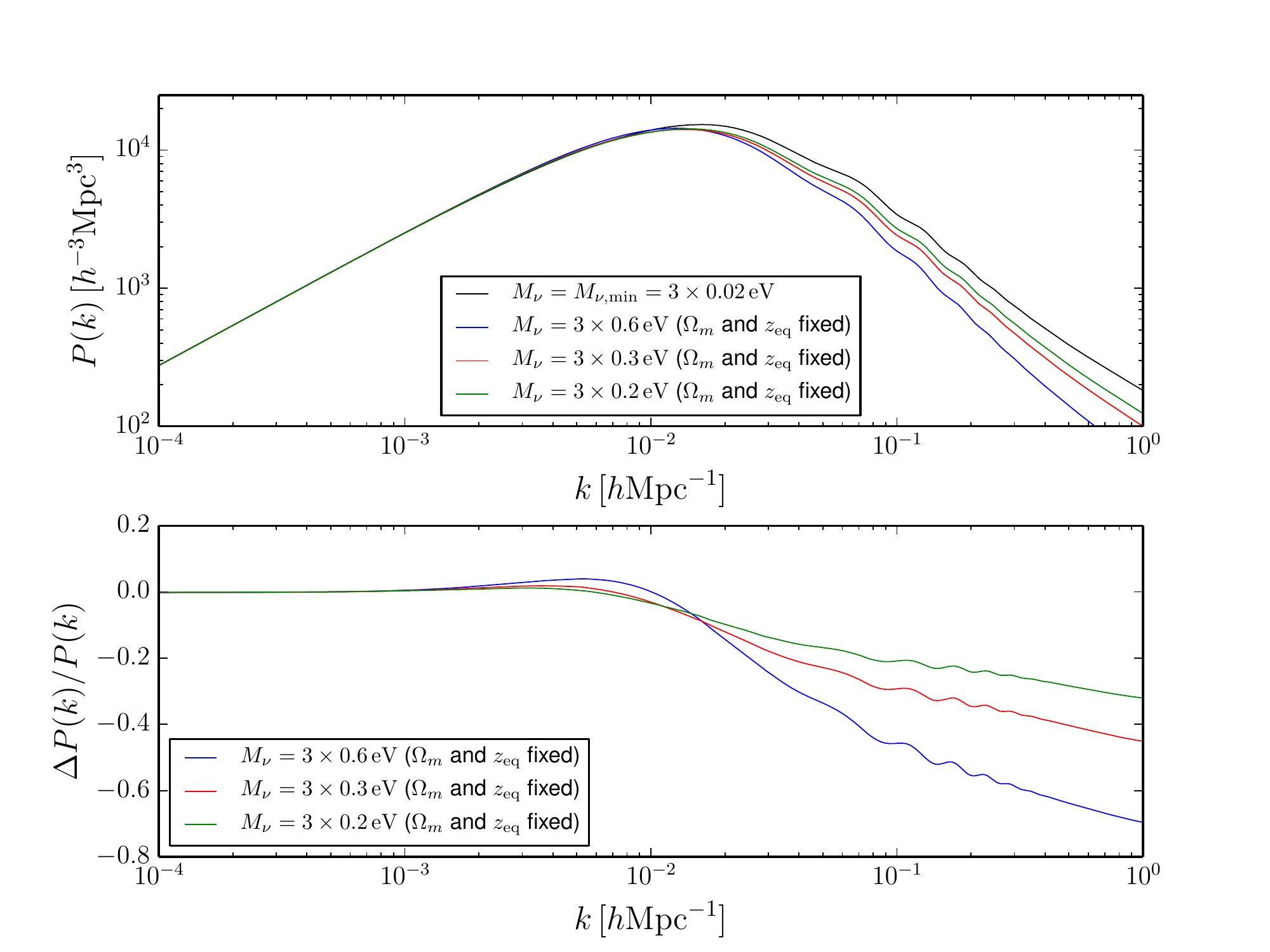}
\caption{Impact of increasing the sum of the neutrino masses $M_{\nu}$ on the linear matter power spectrum, keeping $\omega_b$ and $\omega_c$ (and hence $z_{\rm eq}$) fixed, and increasing $h$ to keep $\Omega_m$ fixed. \textit{Upper panel}: the black curve is the power spectrum for the baseline model where $M_{\nu}=0.06\,{\rm eV}$, $\omega_b=0.02$, $\omega_c=0.12$, $h=0.7$, and hence $\Omega_m=0.287$. The blue (red) [green] curves are obtained for $M_{\nu}=1.8\,{\rm eV}$ ($M_{\nu}=0.9\,{\rm eV}$) [$M_{\nu}=0.6\,{\rm eV}$], where the increase in $M_{\nu}$ is compensated for by setting $h=0.7447$ ($h=0.7218$) [$h=0.7141$]. \textit{Lower panel}: relative change in power with respect to the baseline model, with the same color coding as above. The main changes are due to the small-scale power suppression induced by neutrino free-streaming, which saturates on small scales at a value $\Delta P(k)/P(k) \approx -8f_{\nu}$, with $f_{\nu} \equiv \Omega_{\nu}/\Omega_m$.}
\label{fig:neutrinospk}
\end{figure}
To discuss the effect of neutrinos on the matter power spectrum, we will follow an approach similar to the one we carried out earlier for the CMB. In Chapter~\ref{subsec:lss}, we have already identified the scales, as well as parameters/combinations of parameters, most responsible for shaping the matter power spectrum. We have already seen that $z_{\rm eq}$ is a key quantity, as it sets the scale at which $P(k)$ turns around, reflecting the different growth experienced by modes which entered the horizon prior vs after matter-radiation equality. Moreover, the overall amplitude of $P(k)$ is governed by $\Omega_m$, whereas $\omega_b$ and $\omega_b/\omega_c$ govern the high-$k$ part of the spectrum.
\begin{figure}[!t]
\centering
\includegraphics[width=1.0\textwidth]{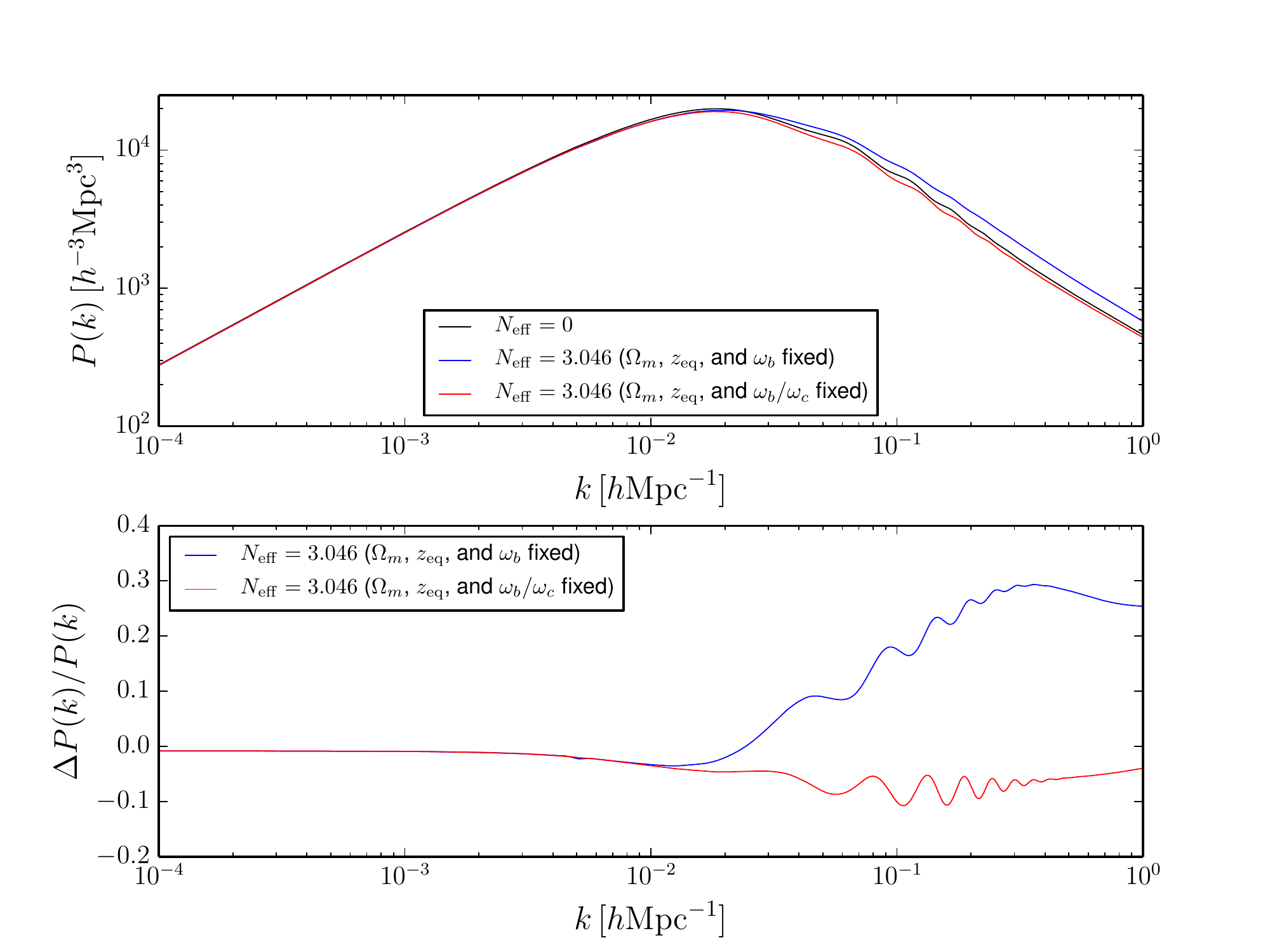}
\caption{Impact of increasing the effective number of neutrinos $N_{\rm eff}$ on the linear matter power spectrum, keeping $\omega_b$ and $\omega_c$ (and hence $z_{\rm eq}$) fixed, and increasing $h$ to keep $\Omega_m$ fixed. \textit{Upper panel}: the black curve is the power spectrum for the baseline model where $M_{\nu}=0.06\,{\rm eV}$, $\omega_b=0.02$, $\omega_c=0.12$, $h=0.7$, and hence $\Omega_m=0.287$. The blue (red) [green] curves are obtained for $M_{\nu}=1.8\,{\rm eV}$ ($M_{\nu}=0.9\,{\rm eV}$) [$M_{\nu}=0.6\,{\rm eV}$], where the increase in $M_{\nu}$ is compensated for by setting $h=0.7447$ ($h=0.7218$) [$h=0.7141$]. \textit{Lower panel}: relative change in power with respect to the baseline model, with the same color coding as above. The main changes are due to the induced changes in $\omega_b/\omega_c$ and $\omega_b$ respectively.}
\label{fig:neffpk}
\end{figure}

Therefore, a meaningful comparison of the matter power spectrum for different values of $M_{\nu}$ should be made keeping $z_{\rm eq}$, $\Omega_m$, $\omega_b$, and $\omega_b/\omega_c$ constant. Since $z_{\rm eq}$ is given by Eq.~(\ref{eq:zeqneutrinos}), increasing $M_{\nu}$ while keeping $\omega_b$ and $\omega_c$ kept fixed will result in both $z_{\rm eq}$ and $\omega_b/\omega_c$ remaining fixed (and of course, by construction, $\omega_b$ is fixed). As for $\Omega_m$, since neutrinos are non-relativistic at late times, $\Omega_m=\Omega_c+\Omega_b+\Omega_{\nu} = (\omega_c+\omega_b+\omega_{\nu})/h^2$. Since we are fixing $\omega_c$ and $\omega_b$, the only way to keep $\Omega_m$ fixed is to increase $h$ as $M_{\nu}$ is increased.

In Fig.~\ref{fig:neutrinospk}, we show the impact on $P(k)$ of increasing $M_{\nu}$ with $\Omega_m$ and $z_{\rm eq}$ fixed, thus reabsorbing any shifts in background quantities. In this way, the most prominent signature of neutrino masses is a step-like suppression in $P(k)$ on small scales (large $k$). This is a result of two effects working in the same direction. Firstly, below their free-streaming scale, neutrinos do not cluster. Secondly, subhorizon perturbations in cold dark matter and baryons grow slower in the presence of massive neutrinos. In a purely matter-dominated Universe, a perturbation $\delta$ grows as $\delta \propto a$, with $a$ the scale factor. On the other hand, in the presence of massive neutrinos, numerical solutions find that $\delta \propto a^{1-3f_{\nu}/5}$, where $f_{\nu} \equiv \Omega_{\nu}/\Omega_m$ is fraction of the matter energy density stored in neutrinos~\cite{Lesgourgues:2018ncw}. These two effects combine to result in a maximal suppression of $\delta P(k)/P(k) \approx -8f_{\nu}$ in the linear regime~\cite{Lesgourgues:2018ncw} (from numerical simulations it has been found that non-linear effects enhance this suppression to $-10f_{\nu}$)~\cite{Brandbyge:2008rv,Saito:2009ah,Viel:2010bn,Bird:2011rb,Hannestad:2011td,Mead:2016zqy}.~\footnote{Notice that, since $f_{\nu} \propto M_{\nu}$, the result that $\Delta P(k)/P(k) \propto -f_{\nu}$ is rather counterintuitive since it suggests that the suppression is larger for more massive and hence slower neutrinos, which free-stream less. This result follows because the amplitude of the suppression is a reflection of the mismatch between the fraction of matter clustering on large vs small scales. This mismatch is quantified by the energy density of neutrinos $\omega_{\nu}$, which is in fact proportional to $M_{\nu}$.}

Since the suppression in the matter power spectrum depends on the energy density stored in neutrinos, one would expect the matter power spectrum to be sensitive not only to $M_{\nu}$, but also to the masses of the individual eigenstates. In particular, one would expect there to be three ``kinks'' (or two if the lightest neutrino is massless) corresponding to the $k_{\rm nr}$ of each eigenstate. This expectation is correct, and a number of early works explored the possibility of measuring the masses of the individual eigenstates using high-precision LSS and CMB lensing data~\cite{Lesgourgues:2004ps,Lesgourgues:2005yv,Pritchard:2008wy,DeBernardis:2009di,
Jimenez:2010ev,Hall:2012kg,Jimenez:2016ckl}. However, the imprint of the individual mass eigenstates has been found to be too small to be probed by current and near-future LSS surveys. Therefore, we will not pursue this possibility further in this Thesis (although this is something I have devoted coming-and-going thoughts to, from time to time, during my PhD). For this reason, the effect of massive neutrinos on cosmological observables is parametrized in terms of $M_{\nu}$ (or equivalently $m_{\rm light}$), since that is (to zeroth order) the quantity cosmology is sensitive to.

Let us finally discuss the impact of the effective number of neutrinos $N_{\rm eff}$ on the matter power spectrum. Again, we should try to perform the comparison between different values of $N_{\rm eff}$ keeping $z_{\rm eq}$, $\Omega_m$, $\omega_b$, and $\omega_b/\omega_c$ fixed. Keeping the former two fixed is particularly important as it governs the turnaround point and the overall amplitude of the power spectrum. However, it is easy to convince oneself that within the framework of the minimal $\Lambda$CDM+$N_{\rm eff}$ model, it is impossible to keep both $\omega_b$ and $\omega_b/\omega_c$ fixed, once $N_{\rm eff}$ is varied fixing $z_{\rm eq}$ and $\Omega_m$. I will follow the approach of~\cite{Lesgourgues:2018ncw}, and first consider the case where $z_{\rm eq}$, $\Omega_m$, and $\omega_b/\omega_c$ are fixed with $\omega_b$ varying, and then the case where $z_{\rm eq}$, $\Omega_m$, and $\omega_b$ are fixed with $\omega_b/\omega_c$ varying. The latter case is more useful when CMB and LSS data are combined, since CMB data fix $\omega_b$ to high precision using the relative height of the odd/even peaks. I consider a baseline model where $N_{\rm eff}=0$, whose power spectrum is given by the black curve of the upper panel of Fig.~\ref{fig:neffpk}.

I first consider the case where $N_{\rm eff}$ is increased keeping $z_{\rm eq}$, $\Omega_m$, and $\omega_b$ fixed. It is easy to show that this can be achieved by performing the transformations $\omega_c \to \omega_c+(\alpha-1)\omega_m$ and $h \to h\sqrt{\alpha}$ we already saw when discussing the impact of $N_{\rm eff}$ on the CMB (keeping $z_{\rm eq}$ and $\theta_s$ fixed) in Chapter~\ref{subsec:signaturesnucmb}. The result is given by the blue curve in the upper panel of Fig.~\ref{fig:neffpk}. The transformation we have performed has kept $\omega_b$ fixed at the expense of decreasing $\omega_b/\omega_c$. As we have seen earlier in Chapter~\ref{subsec:lss} (see also Fig.~\ref{fig:parameterslss}), this results in more power on small scales (due to less reduction in the growth rate of dark matter perturbations), as well as damped BAO. The neutrino drag effect we have already seen in the CMB in Chapter~\ref{subsec:signaturesnucmb} is present here as well: albeit the effect is tiny, it is more evident in the lower panel.

I then consider the case where $N_{\rm eff}$ is increased keeping $z_{\rm eq}$, $\Omega_m$, and $\omega_b/\omega_c$ fixed. This can be achieved by performing the transformations $\omega_c \to \alpha \omega_c$, $\omega_b \to \alpha \omega_b$, and $h \to \sqrt{\alpha}h$. The result is given by the red curve in the upper panel of Fig.~\ref{fig:neffpk}. In this case, we have kept $\omega_b/\omega_c$ fixed at the expense of increasing $\omega_b$. The result is that of altering the phase and amplitude of the BAO, because the sound horizon $r_s$ is altered. The neutrino drag effect is present here as well, and more evident in the lower panel.

In summary, we have seen that the direct effect of neutrino masses on the matter power spectrum shows up as a suppression in power on small scales, for $k \gtrsim k_{\rm nr}$, reflecting the fact that neutrinos do not cluster on small scales, and slow down the growth rate of dark matter perturbations. The effect is proportional to $M_{\nu}$, and more precisely to $f_{\nu} \equiv \Omega_{\nu}/\Omega_m$. The direct effects of varying $N_{\rm eff}$ instead depend on whether this parameter is increased keeping $\omega_b$ or $\omega_b/\omega_c$ fixed. At any rate, it results in a change in the amplitude and phase of the BAO appearing in the matter power spectrum on intermediate and small scales.

So far I have given an overview of the main features governing the shape of the CMB and matter power spectra, and how neutrino parameters affect these spectra. The next natural step would be to actually go after these effects on real data, in order to constrain neutrino properties. Before doing so, however, a general overview of statistical methods widely used in cosmology will be necessary: this will be the topic of the next Chapter.

\chapter{A brief interlude: statistical methods in cosmology}
\label{chap:5}

\begin{chapquote}{(?) Arthur James Balfour (1982, often attributed to Mark Twain)}
``There are three kinds of lies: lies, damned lies, and statistics''
\end{chapquote}

The field of observational cosmology is inevitably intertwined with that of statistics, necessary in order to make sense of the vast amounts of data provided by the Universe. At this point in our journey, it is therefore useful to review a number of statistical and data analysis concepts widely used in cosmology, necessary in order to fully understand the remaining chapters of the thesis, as well as the included papers. In particular, the machinery of Bayesian statistics underlies most of the statistical methods adopted in cosmology. For practical reasons, I will not attempt to provide an in-depth review of these concepts. Instead, I redirect the interested reader to a number of excellent cosmology-oriented resources covering Bayesian statistics and data analysis present in the literature~\cite{Verde:2007wf,Trotta:2008qt,dataanalysis,Heavens:2009nx,Verde:2009tu,
bayesianstatistics,practicalstatistics,Trotta:2017wnx,
practicalbayesian,bayesianastrophysics} for a necessarily incomplete list.

This Chapter is organized as follows. I begin in Chapter.~\ref{sec:bayesianfrequentist} by providing a brief overview of the Bayesian school of thought, contrasting it to the main competing school of thought, namely the frequentist one, and briefly discussing possible reasons for the widespread use of Bayesian methods in cosmology. I continue in Chapter~\ref{sec:elementarynotions} by covering some of the main notions of Bayesian statistics including Bayes' theorem in Chapter~\ref{subsec:bayestheorem}, and the concepts of marginalization, credible regions, and model comparison in Chapter~\ref{subsec:marginalization}, before discussing in Chapter~\ref{sec:bayesianpractice} how these methods are applied in practice when analysing cosmological data.

\section{Bayesian vs frequentist statistics}
\label{sec:bayesianfrequentist}

It is quite remarkable that a rather simple mathematical result obtained by an obscure Presbyterian minister nearly 300 years ago and then published posthumously~\cite{Bayes:1764vd}, has come to become the cornerstone of the statistical methods underlying several disciplines, ranging from biology, to economy, and of course to cosmology. Bayesian statistics, named after Reverend Thomas Bayes, is unquestionably on the rise, for a number of very good reasons.

The Bayesian school of thought is customarily contrasted to the frequentist one. At the root, the two differ essentially in their interpretation of probability. Within the frequentist school of thought, the probability of an event is viewed as being the limit of the relative frequency of occurrence of the given event in the limit of an infinite number of equiprobable trials. In contrast, the Bayesian interpretation of probabilities views the latter as measuring the degree of belief in an event. In other words, from the Bayesian point of view, probabilities (which can be associated both to repeated or one-off events) quantify our state of knowledge (or ignorance) in the presence of partial information.

Already from this very brief discussion it is clear that there are very fundamental differences between the Bayesian and frequentist views of probability. From the frequentist point of view, model parameters and hypotheses are fixed and immutable: most importantly, they are not assigned probabilities. In Bayesian statistics, the probability or degree of belief in an event can (and will) change as new information is gathered, and depends on prior knowledge such as personal beliefs or results of earlier experiments. In fact, one of the guiding principles of Bayesian statistics is that no inference can be made without first specifying prior assumptions, forcing one to question one's assumptions and state of knowledge before even embarking into a statistical inference problem.

A question often heard is whether one of the two approaches is ``better'' then the other, and the statistics community is far from settled on this debate, with heated discussions often ensuing. I would argue that this question is irrelevant and take the more pragmatic stand of preferring the school of thought that provides me tools and results best suited to my objective. In this respect, one could argue that, at least as far as cosmology is concerned, Bayesian methods do appear to have a slight edge over frequentist ones, for a number of reasons, among which:
\begin{enumerate}
\item We only have one Universe on which we can ``experiment''. Barring ergodicity considerations, speaking about long-run results wherein we observe N Universes, necessary to embrace the frequentist picture does not really make sense in the context of cosmology. Similarly, ``replicating'' cosmological experiments is usually tricky, if not impossible.
\item Sociological effects are important as well. The widespread use of Bayesian parameter inference tools in cosmology, such as \texttt{CosmoMC}~\cite{Lewis:2002ah} and \texttt{Montepython}~\cite{Audren:2012wb} has certainly contributed to the preference for Bayesian statistics in cosmology.
\item Bayesian statistics provides a natural framework for comparing the performance of models (see Chapter~\ref{subsec:marginalization}), which is a question often of interest in cosmology.
\end{enumerate}

Let us now move on to discuss elementary notions of Bayesian statistics, and in particular the mathematical foundations thereof.

\section{Elementary notions of Bayesian statistics}
\label{sec:elementarynotions}

\subsection{Bayes' theorem}
\label{subsec:bayestheorem}

The whole machinery of Bayesian statistics rests upon a simple mathematical result known as Bayes' theorem, after Reverend Thomas Bayes, who formulated a specific case of this theorem in his most famous paper~\cite{Bayes:1764vd}, published posthumously thanks to Richard Price. Before presenting this theorem, let us first introduce our notation. With $A$ and $B$ being two propositions (to which we can assign probabilities as per the Bayesian school of thought), we will use the notation $p(A \vert B)$ to denote the probability we assign to proposition $A$ \textit{conditional} on assuming that proposition $B$ is true. Let us also denote by $p(A,B)$ the \textit{joint} probability of $A$ and $B$. Finally, let us denote by $I$ any relevant background information which is assumed to be true (for instance, if we are considering a coin toss experiment, $I$ can reflect the fact that the coin is known to be fair).

Let us recall the Kolmogorov definition of conditional probability of proposition $A$ given proposition $B$~\cite{Kolmogorov:1960ghw}:
\begin{eqnarray}
p(A \vert B,I) = \frac{p(A,B \vert I)}{p(B \vert I)} \,.
\label{eq:conditional1}
\end{eqnarray}
Obviously, the following trivially holds:
\begin{eqnarray}
p(A,B \vert I) = p(B,A \vert I) \,.
\label{eq:conditional2}
\end{eqnarray}
Combining Eqs.~(\ref{eq:conditional1},\ref{eq:conditional2}), we then trivially arrive at:
\begin{eqnarray}
p(B \vert A,I) = \frac{p(A \vert B,I)p(B \vert I)}{p(A \vert I)}
\label{eq:bayes}
\end{eqnarray}
In its simplicity, Eq.~(\ref{eq:bayes}) is known as \textit{Bayes' theorem} and lies at the heart of Bayesian methods. Notice that, as clearly discussed in~\cite{Trotta:2008qt}, Bayes' theorem is a mathematical statement, and as such it is not controversial: any controversy on the matter (especially in relation to Bayesian vs frequentist debates) is solely related to whether it should be used to perform statistical inference.

I will now change the notation of Eq.~(\ref{eq:bayes}) very slightly to make its interpretation more obvious. In doing so, I will switch from discrete events to continuous random variables. One can still convince oneself that Eq.~(\ref{eq:bayes}) will remain unchanged, with the $p$s now describing probability distribution functions rather than probabilities themselves. Let us consider a situation where we have some data/observations $\boldsymbol{d}$ and a model ${\cal M}$ described by some parameters $\boldsymbol{\theta}$. Then, I will rewrite Eq.~(\ref{eq:bayes}) performing the substitutions $A \rightarrow \boldsymbol{d}$, $B \rightarrow \boldsymbol{\theta}$, and $I \rightarrow {\cal M}$:
\begin{eqnarray}
p(\boldsymbol{\theta} \vert \boldsymbol{d},{\cal M}) = \frac{p(\boldsymbol{d} \vert \boldsymbol{\theta},{\cal M})p(\boldsymbol{\theta} \vert {\cal M})}{p(\boldsymbol{d} \vert I)} \,.
\label{eq:bayes2}
\end{eqnarray}
In the form given by Eq.~(\ref{eq:bayes2}), the utility of Bayes' theorem becomes more obvious. In cosmology, it is typically the case the one has a model ${\cal M}$ in mind, from which it is often relatively easy to compute predictions for what observations $\boldsymbol{d}$ should look like, given a set of parameters $\boldsymbol{\theta}$. Therefore, it is relatively easy to compute the $p(\boldsymbol{d} \vert \boldsymbol{\theta},{\cal M})$ term on the right-hand side of Eq.~(\ref{eq:bayes2}). However, the question one is usually more interested in is: ``\textit{given the data I just observed, what do I learn about the model parameters?}''. The answer to this question is given by $p(\boldsymbol{\theta} \vert \boldsymbol{d},{\cal M})$, the left-hand side of Eq.~(\ref{eq:bayes2}). Bayes' theorem gives us a simple route for going from quantities we know how to compute, to quantities we are interested in. In fact, one can really view Bayes' theorem as a prescription for how we learn from experience: we start from some initial belief (irrespective of the data), quantified by $p(\boldsymbol{\theta} \vert {\cal M})$, and then update our state of belief after having observed the data, to get $p(\boldsymbol{\theta} \vert \boldsymbol{d},{\cal M})$.

Let us introduce some terminology and further clear up our notation a bit. First of all, as long as we are concerned with \textit{parameter inference} (as opposed to \textit{model comparison} which will be covered later in Chapter~\ref{subsec:marginalization}, \textit{i.e.} we have one specific model in mind and are only interested in inferring the probability distribution of its parameters given the data) all probability distributions in Eq.~(\ref{eq:bayes2}) are implicitly conditioned on the same model ${\cal M}$: hence, for notation simplicity, I will drop the symbol ${\cal M}$ which will always be implicitly understood. The left-hand side of Eq.~(\ref{eq:bayes2}), $p(\boldsymbol{\theta} \vert \boldsymbol{d})$, is typically referred to as the \textit{posterior distribution} of the model parameters \textit{after} having observed the data. The quantity $p(\boldsymbol{d} \vert \boldsymbol{\theta})$ is typically referred to as the \textit{likelihood function}: I will denote it by ${\cal L}(\boldsymbol{d} \vert \boldsymbol{\theta})$. Still on the numerator of the right-hand side, $p(\boldsymbol{\theta})$ is known as the \textit{prior distribution} for the model parameters, and I will denote it by ${\cal P}(\boldsymbol{\theta})$. Finally, the denominator of the right-hand side is known as the \textit{Bayesian evidence} or \textit{marginal likelihood} (the reason why will become apparent in a while), and I will denote it by ${\cal E}(\boldsymbol{d})$. Using this notation, we can finally express Bayes' theorem as follows:
\begin{eqnarray}
\underbrace{p(\boldsymbol{\theta} \vert \boldsymbol{d})}_{\rm posterior} = \frac{\overbrace{{\cal L}(\boldsymbol{d} \vert \boldsymbol{\theta})}^{\rm likelihood}\overbrace{{\cal P}(\boldsymbol{\theta})}^{\rm prior}}{\underbrace{{\cal E}(\boldsymbol{d})}_{\rm evidence}} \,.
\label{eq:bayes3}
\end{eqnarray}

At this point three comments on Bayes' theorem, Eq.~(\ref{eq:bayes}), are in order. As a first comment, note the inevitable dependence of the result of any Bayesian inference process on the prior choice [${\cal P}(\boldsymbol{\theta})$]. This has historically been considered one of the main problems in Bayesian statistics, for two reasons. Firstly, to begin with this might be seen as undermining objectivity. Secondly, there is no indication as to how the prior should be selected besides the fact that it should reflect one's degree of belief and state of knowledge. I will not dive into discussions as to whether the dependence on the prior is actually a problem or a strength: the interested reader is invited to consult many excellent references present in the literature, and in particular Sec.~2.3 of~\cite{Trotta:2008qt}. Instead, I want to point out that there are many instances wherein including reasonable prior choices is not only desirable, but also necessary.~\footnote{For instance, a central topic in this thesis is that of inferring the sum of the neutrino masses $\mnu$ from cosmological data. As $\mnu$ is a mass, it is necessarily a positive quantity: hence, to avoid the parameter inference process producing unphysical results, one should include the information $\mnu \geq 0\,{\rm eV}$ in the prior choice.} Moreover, as long as the likelihood is large only within the support of the prior (the support being the subset of the prior domain wherein the prior is non-zero), the posterior distribution will mostly depend on the likelihood rather than the prior. In other words, the data is informative and the process of parameter inference is driven by the data rather than the prior. If the data is not informative or weakly informative, the prior plays an important role and at that point it is responsibility of whoever is performing the statistical analysis to ensure that this dependence is adequately discussed and taken into account.~\footnote{As we shall see, this is currently the situation with cosmological determinations of neutrino masses: cosmological data is currently unable to detect a non-zero $\mnu$, but only provides upper limits on the latter. Therefore, these upper limits are inevitably driven by prior choices, and in particular the choice of prior for $\mnu$. This topic will be discussed later in the thesis, as well as in the included papers, and has been the subject of much debate in the recent literature (see for instance the discussions in~\cite{Hannestad:2016fog,Gerbino:2016ehw,Vagnozzi:2017ovm,Hannestad:2017ypp,Long:2017dru,
Gariazzo:2018pei,Heavens:2018adv,Handley:2018gel,Gariazzo:2018tft}; see also~\cite{Simpson:2017qvj}, as well as the response paper~\cite{Schwetz:2017fey}).}

A second comment is that the posterior distribution \textit{viewed as a function of model parameters $\boldsymbol{\theta}$} is a probability distribution, hence it should be normalized. Demanding that the posterior be normalized in turn gives us an expression for the Bayesian evidence:
\begin{eqnarray}
\int d\boldsymbol{\theta}\,p(\boldsymbol{\theta} \vert \boldsymbol{d}) = \frac{1}{{\cal E}(\boldsymbol{d})}\int d\boldsymbol{\theta}\,{\cal L}(\boldsymbol{d} \vert \boldsymbol{\theta}){\cal P}(\boldsymbol{\theta}) = 1 \implies {\cal E}(\boldsymbol{d}) = \int d\boldsymbol{\theta}\,{\cal L}(\boldsymbol{d} \vert \boldsymbol{\theta}){\cal P}(\boldsymbol{\theta})\,.
\label{eq:evidence}
\end{eqnarray}
A third comment relates to the fact that the evidence is independent of the model parameters. In fact, as we have just seen in Eq.~(\ref{eq:evidence}), it simply acts as an overall normalization constant for the posterior distribution. However, as long as one is concerned with parameter inference as opposed to model comparison, one cares about the ratio between the values of the posterior distribution at different values of the model parameters. For this purpose, all one really needs to know is that the posterior is \textit{normalizable}, but the actual normalization [as provided in Eq.~(\ref{eq:evidence})] is in itself irrelevant. Therefore, for the purposes of parameter inference, it is actually sufficient to write Bayes' theorem in the following form:
\begin{eqnarray}
p(\boldsymbol{\theta} \vert \boldsymbol{d}) \propto {\cal L}(\boldsymbol{d} \vert \boldsymbol{\theta}){\cal P}(\boldsymbol{\theta}) \,.
\label{eq:bayes4}
\end{eqnarray}
We will return later to the subtleties of Bayesian model comparison and the complications they bring.

\subsection{Marginalization, credible regions, and model comparison}
\label{subsec:marginalization}

I will now briefly discuss a number of other important concepts in Bayesian statistics. The first is that of marginalization. In general, we will not be interested in the whole parameter vector $\boldsymbol{\theta}$. In fact, some of the parameters will be of limited physical interest, and are used to model instrumental calibration, systematics, and so on. Parameters we are not interested in are referred to as \textit{nuisance parameters}. Since we are not interested in them, an useful operation we can perform is to report the probability distribution for the parameters of interest after having integrated out the uncertainty on the nuisance parameters: this operation is known as marginalization.

Consider the simple case where we are interested in the parameter $\theta_1$, whereas $\theta_2,...,\theta_n$ are our nuisance parameters. Then, we are interested in obtaining the marginal posterior distribution for $\theta_1$, $p(\theta_1)$, rather than the joint posterior distribution on all parameters $p(\boldsymbol{\theta} \vert \boldsymbol{d})$:
\begin{eqnarray}
p(\theta_1) = \int d\theta_2...d\theta_n\,p(\boldsymbol{\theta} \vert \boldsymbol{d})\,.
\label{eq:marginalization}
\end{eqnarray}
The generalization of Eq.~(\ref{eq:marginalization}) to the case where we are interested in more than one parameter is trivial. It is customary practice in Bayesian statistics to first compute the joint posterior (including both the parameters of interest and the nuisance parameters), and then to plot one- or two-dimensional marginal posteriors for selected parameters/subsets of parameters, with all the other parameters marginalized over. For instance, in this thesis we will often be interested in the 1D marginal posterior distribution for $M_{\nu}$, where all the other parameters (including the 6 $\Lambda$CDM parameters) are treated as nuisance parameters and marginalized over. Alternatively, when exploring degeneracies/correlations between $M_{\nu}$ and any other parameter, we will be considering 2D marginal posteriors for $M_{\nu}$ and this other parameter.

Another important concept is that of credible regions. A $100 \times f\%$ credible region encloses a fraction $f$ of the posterior probability. In other words, denoting a $f\%$ credible region by ${\cal F}$, and considering a normalized posterior distribution (\textit{i.e.} such that $\int d\boldsymbol{\theta}\,p(\boldsymbol{\theta} \vert \boldsymbol{d})=1$), we have that:
\begin{eqnarray}
\int_{{\cal F}}d\boldsymbol{\theta}\,p(\boldsymbol{\theta} \vert \boldsymbol{d}) = f\,.
\label{eq:credible}
\end{eqnarray}
It is common practice to consider various nested credible regions, usually corresponding to values $f \approx 0.683$, $f \approx 0.954$, and $f \approx 0.997$, and colloquially referred to as $1\sigma$, $2\sigma$, and $3\sigma$ confidence regions. In the case of a single parameter, confidence regions are usually referred to as confidence intervals.~\footnote{Note that there is a subtle difference between Bayesian and frequentist confidence intervals. Considering for definiteness a 95\% confidence interval, in the Bayesian case a parameter falls within this interval with 95\% probability. In other words, the interval is fixed and the parameter is the random variable. In the frequentist case, the situation is in some sense reversed: the parameter is fixed, whereas it is rather the interval which is the random variable. In particular, for a large number of repeated samples, 95\% of the intervals calculated adopting this prescription include the fixed (unknown) value of the parameter.}

Note that there is generally ambiguity in the choice of a $100 \times f\%$ confidence region, as usually several regions can be constructed satisfying Eq.~(\ref{eq:credible}), but still being different between each other. The common choice is then to consider highest posterior density regions, ${\cal F}^{\star}$, such that $p(\boldsymbol{\theta} \vert \boldsymbol{d}) \geq p$ for all points in parameter space belonging to ${\cal F}^{\star}$, with $p(\boldsymbol{\theta} \vert \boldsymbol{d}) = p$ defining the boundary of the credible region. For well-behaved unimodal distributions, ${\cal F}^{\star}$ is usually uniquely defined for any given $p$.

When talking about confidence intervals, we refer to 1D 2-tail symmetric $100 \times f\%$ confidence intervals as intervals enclosing a fraction $f$ of the probability, with the remaining $(1-f)/2$ of the probability being enclosed on either side outside the confidence interval. Sometimes, it is instead more convenient to talk about a $100 \times f\%$ upper/lower limit (very often referred to, with a slight abuse of language, as $100 \times f\%$ confidence level [C.L.] upper limits), indicating the value below/above which a fraction $f$ of the probability is enclosed. In this thesis, we will almost always report $95\%$ upper limits on $M_{\nu}$. The reason is that the 1D marginal posteriors on $M_{\nu}$ will always be highly asymmetric and peaked at $M_{\nu}=0\,{\rm eV}$, which also happens to be the lower boundary of the prior we impose on $M_{\nu}$. In other words, cosmological measurements are currently only consistent with an upper limit on $M_{\nu}$ and not a detection of non-zero $M_{\nu}$.

The final important concept I want to briefly discuss is that of model comparison. So far, we have worked within the assumption of a given model ${\cal M}$, described by a parameter vector $\boldsymbol{\theta}$. Doing so, we were only interested in the posterior distribution of $\boldsymbol{\theta}$, $p(\boldsymbol{\theta} \vert \boldsymbol{d},{\cal M})$, and more specifically in the ratio between the values of the posterior distribution at different values of the model parameters. This has allowed us to neglect the overall normalization given by the evidence ${\cal E}(\boldsymbol{d})$ in  Eq.~(\ref{eq:bayes3}) [see Eq.~(\ref{eq:bayes4})].

However, in Bayesian statistics it is possible to work at a ``higher'' level and compare models themselves. In fact, one can conceive a situation where there are several competing models, and it is desirable to evaluate their relative probabilities. The ``best'' model will be the one that reaches an ideal balance between quality of fit and predictivity. In other words, it is often the case that a more complex model with more parameters will fit the data better. However, added layers of complexity should be avoided whenever a simpler model is able to provide an adequate description of the observations, in the spirit of Occam's razor. Bayesian model comparison provides a quantification of Occam's razor, evaluating whether an extra layer of complexity provided by a model is warranted by the data or is unnecessary. Note that Bayesian model comparison is a \textit{comparison} process: that is, it only makes sense insofar as there is more than one competing model.

Often, it is the case that one wishes to compare two competing models in light of data $\boldsymbol{d}$. Let us refer to the two models as ${\cal M}_0$ (described by parameter vector $\boldsymbol{\theta_0}$) and ${\cal M}_1$ (described by parameter vector $\boldsymbol{\theta_1}$). Then, we can apply Eq.~(\ref{eq:bayes4}) with ${\cal M}$ in place of $\boldsymbol{\theta}$, as follows:
\begin{eqnarray}
p({\cal M}_i \vert \boldsymbol{d}) \propto {\cal P}({\cal M}_i){\cal L}(\boldsymbol{d} \vert {\cal M}_i)\,, \quad i=0\,,1\,,
\label{eq:bayesmodels}
\end{eqnarray}
where this time ${\cal L}(\boldsymbol{d} \vert {\cal M}_i)$ is none other than the evidence ${\cal E}(\boldsymbol{d})$ we have already seen in Eq.~(\ref{eq:evidence}), where recall we had dropped the $\vert {\cal M}$ bit for simplicity since we were only considering one model. Similarly, $p({\cal M})$ is the prior probability assigned to the model itself. If no prior information is present and one has $N$ models to compare, the typical conservative choice is to set $p({\cal M}_i) = 1/N$ for $i=1,...,N$. Then, the quantity of interest when comparing two models is the \textit{odds ratio}, given by:
\begin{eqnarray}
\frac{p({\cal M}_0 \vert \boldsymbol{d})}{p({\cal M}_1 \vert \boldsymbol{d})} = \frac{{\cal L}(\boldsymbol{d} \vert {\cal M}_0)}{{\cal L}(\boldsymbol{d} \vert {\cal M}_1)}\frac{{\cal P}({\cal M}_0)}{{\cal P}({\cal M}_1)}\,.
\label{eq:odds}
\end{eqnarray}

As said previously, it is often the case that all competing models are assigned equal prior probabilities, so the second fraction on the right-hand side of Eq.~(\ref{eq:odds}) simplifies to 1. Then, one is left with the first fraction on the right-hand side of Eq.~(\ref{eq:odds}), which is usually referred to as \textit{Bayes factor}:
\begin{eqnarray}
B_{01} \equiv \frac{{\cal L}(\boldsymbol{d} \vert {\cal M}_0)}{{\cal L}(\boldsymbol{d} \vert {\cal M}_1)} = \frac{{\cal E}(\boldsymbol{d} \vert {\cal M}_0)}{{\cal E}(\boldsymbol{d} \vert {\cal M}_1)}\,.
\label{eq:factor}
\end{eqnarray}
The Bayes factor $B_{01}$ quantifies the increase/decrease (for $B_{01}>1$ and $B_{01}<1$ respectively) of the support in favour of model ${\cal M}_0$ versus model ${\cal M}_1$ after observing the data. It is given by the evidence ratio of model ${\cal M}_0$ to model ${\cal M}_1$, with the evidences computed from Eq.~(\ref{eq:evidence}). 

Traditionally, computing the Bayesian evidence in Eq.~(\ref{eq:evidence}) has always been a challenging task, due to the multi-dimensional integral over the whole parameter space. This has been one of the factors hampering a more widespread use of Bayesian model comparison (whereas efficient methods for performing parameter estimation have existed for quite some time, see Chapter~\ref{sec:bayesianpractice}). Recently, a number of efficient methods for performing the integral in Eq.~(\ref{eq:evidence}) have been devised, including nested sampling~\cite{Skilling:2006gxv}, applied in a cosmological context in e.g.~\cite{Mukherjee:2005wg,Shaw:2007jj,Feroz:2007kg,Bassett:2004wz,Bridges:2006mt}, aided by the development of the \texttt{MultiNest} software~\cite{Feroz:2008xx}. In general, if one is interested in performing a Bayesian model comparison analysis, it is always a good idea to try and simplify the evidence computation as much as possible. In this thesis, I will consider an explicit case in Paper~I, where we were interested in computing the posterior odds for normal versus inverted mass ordering.

It is customary to interpret the values one obtains for Bayes factors on empirically calibrated scales qualifying the strength of the evidence for one model with respect to the other. One widely used scale is the Jeffreys scale~\cite{Jeffreys:1939xee}, presented in Tab.~\ref{tab:jeffreysscale}. Related alternative scales are also used in the literature, for instance the Kass-Raftery scale~\cite{Kass:1995loi}.
\begin{table}[h!]
\centering
\begin{tabular}{|c|c|c|}
\hline
$\boldsymbol{B_{01}}$ & \textbf{Strength of evidence for model} $\boldsymbol{{\cal M}_0}$ \\
\hline\hline
 $<10^0$ & Negative (data supports model ${\cal M}_1$) \\
\hline
 $10^0$ to $10^{1/2}$ & Barely worth mentioning \\
\hline
 $10^{1/2}$ to $10^1$ & Substantial \\
\hline
 $10^1$ to $10^{3/2}$ & Strong \\
\hline
 $10^{3/2}$ to $10^2$ & Very strong \\
\hline
 $>10^2$ & Decisive \\
\hline
\end{tabular}
\caption{Jeffreys scale for comparing the strength of the evidence for model ${\cal M}_0$ against model ${\cal M}_1$, when the Bayes factor $B_{01}$ is known~\cite{Jeffreys:1939xee}.}
\label{tab:jeffreysscale}
\end{table}

\section{Bayesian statistics in practice: MCMC methods}
\label{sec:bayesianpractice}

At the lowest level, the way we want to apply Bayesian statistics in cosmology is to perform \textit{parameter estimation}. We have some data $\boldsymbol{d}$ and have a model ${\cal M}$ in mind, specified by parameters $\boldsymbol{\theta}$. Given the data, we want to determine the posterior distributions of the parameters. In particular, theoretical predictions for the observations enter within the likelihood. In practice, in cosmology, each evaluation of the likelihood typically involves a call to Boltzmann solvers (e.g. \texttt{CAMB}~\cite{Lewis:1999bs} or \texttt{CLASS}~\cite{Audren:2012vy}). Evaluating the posterior is, in principle, easily done using Eq.~(\ref{eq:bayes4}) to evaluate the joint posterior for the parameters. In practice, in cosmology we are usually dealing with ${\cal O}(10)$ parameters (the minimal $\Lambda$CDM model alone has 6 parameters, and each experiment carries a number of nuisance parameters to account for calibration, systematics, etc.). A na\"{i}ve grid exploration of the parameter space, which was the approach initially followed in the 1990s, clearly becomes untenable as soon as one is dealing with more than $\approx 5$ parameters. In any case, such an approach would be a waste since typically the posterior is extremely low in most of the parameter space hypervolume. Clearly, a smarter way of sampling the posterior distribution, concentrating on regions where such a distribution is highest, is needed.

Fortunately, there a number of numerical methods which come to our rescue. Nowadays, the most widely used methods is the Monte Carlo Markov Chain (MCMC) method (see e.g.~\cite{Brooks:2011ghw} for a pedagogical introduction to MCMC methods). The aim of MCMC methods is to generate a ``chain'', wherein each node of the chain consists of a point in parameter space. The distribution of points in asymptotically proportional to the \textit{target density} one wishes to sample, in this case the posterior distribution. This then makes it possible to estimate any quantity of interest from the distribution (e.g. mean, variance, and so on). An MCMC algorithm makes random draws in a Markovian way, meaning that at each step the next sample depends only on the current sample, but not on the previous ones.

The basic procedure works as follows. Say at the current step the chain has landed in the point $\boldsymbol{\theta}$. Then, a new point $\boldsymbol{\theta^{\star}}$ is proposed from a proposal distribution $q(\boldsymbol{\theta^{\star}} \vert \boldsymbol{\theta})$. One of the most popular MCMC algorithms is based on the \textit{Metropolis-Hastings} algorithm~\cite{Metropolis:1953ghw,Hastings:1970ghw}, which envisages accepting $\boldsymbol{\theta^{\star}}$ with an acceptance probability of:
\begin{eqnarray}
\alpha = \min \left ( 1,\frac{p(\boldsymbol{\theta^{\star}})}{p(\boldsymbol{\theta})}\frac{q(\boldsymbol{\theta^{\star}} \vert \boldsymbol{\theta})}{q(\boldsymbol{\theta} \vert \boldsymbol{\theta^{\star}})} \right )\,,
\label{eq:acceptancemetropolishastings}
\end{eqnarray}
where $p(\boldsymbol{\theta})$ [$p(\boldsymbol{\theta^{\star}})$] denotes the posterior probability (and more generically the target density) evaluated at $\boldsymbol{\theta}$ [$\boldsymbol{\theta^{\star}}$]. Usually the proposal distribution is chosen to be symmetric, \textit{i.e.} $q(\boldsymbol{\theta^{\star}} \vert \boldsymbol{\theta}) = q(\boldsymbol{\theta} \vert \boldsymbol{\theta^{\star}})$. In this case, we refer to the algorithm simply as Metropolis instead of Metropolis-Hastings, and Eq.~(\ref{eq:acceptancemetropolishastings}) simplifies to:
\begin{eqnarray}
\alpha = \min \left ( 1,\frac{p(\boldsymbol{\theta^{\star}})}{p(\boldsymbol{\theta})} \right )\,.
\label{eq:acceptancemetropolis}
\end{eqnarray}
In practice, this acceptance/rejection step can be performed by drawing a random number $y$ between $0$ and $1$, and accepting the the point if $y<\alpha$ (and rejecting it otherwise).~\footnote{The choice of proposal distribution is, in practice, a crucial one. We will not discuss it further here, but just note some general results suggesting that the optimal proposal distribution should lead to an acceptance rate of about $\approx 25\%$. Rather than the exact shape of the distribution, what's important is its ``scale'' (which can be the variance of the distribution if it is a Gaussian, or its half-width if it is a top-hat function). The optimal scale for the proposal distribution has been found to be $2.4/\sqrt{d}$~\cite{Gelman:1996ghw}, where $d$ is the dimensionality of the parameter space. If the scale is too small, the MCMC algorithm can be stuck locally and not explore the parameter space efficiently. If the scale is too large, the chain acceptance rate might be very low and the chain not jump very frequently, again resulting in an inefficient exploration of the parameter space. Usually, a Gaussian proposal distribution is chosen, with covariance matrix estimated from an earlier MCMC run or from an exploratory MCMC run.}

Implementing the Metropolis algorithm in practice is very simple, and can be done in a couple of lines in \texttt{Python}. There are a number of important issues (going under the name of burn-in, convergence, thinning) which I will not cover here, related to the necessity o fmaking sure the MCMC algorithm has explored the posterior distribution in an acceptable way (see~\cite{Verde:2007wf,Trotta:2008qt,dataanalysis,Heavens:2009nx,Verde:2009tu,
bayesianstatistics,practicalstatistics,Trotta:2017wnx,
practicalbayesian,bayesianastrophysics} for a complete coverage of these issues within the context of cosmology). Assuming these issues have been dealt with, an MCMC run returns us a (or more than one) chain containing $N$ elements $\boldsymbol{\theta}^{(n)}$, $n=1,...,N$. At this point, estimating Monte Carlo estimates for any function of the parameters becomes trivial. Considering a simple one-dimensional case where we have a parameter $\theta$, we can estimate the expectation value of $\theta$, $\langle \theta \rangle$, as:
\begin{eqnarray}
\langle \theta \rangle \approx \frac{1}{N}\sum_{i=1}^{N} \theta^{(i)}\,.
\label{eq:mcmctheta}
\end{eqnarray}
Similarly, we can estimate the expectation value of any function of $\theta$, as:
\begin{eqnarray}
\langle f(\theta) \approx \rangle \frac{1}{N}\sum_{i=1}^{N} f(\theta^{(i)})\,.
\label{eq:mcmcf}
\end{eqnarray}
Marginalization is also a simple business. Imagine we want the marginal distribution of parameter $\theta_1$, and want to ignore parameters $\theta_2,...,\theta_n$. Then it is sufficient to construct a histogram of the $\theta_1$ values for each point in the chain, ignoring the values of the other parameters. Higher-dimensional marginal posteriors are obtained analogously.

MCMC methods are widespread in cosmology, and several efficient MCMC samplers exist on the code market. Two names emerge above all the others: \texttt{CosmoMC}~\cite{Lewis:2002ah} is written in \texttt{Fortran} and interfaced with the Boltzmann solver \texttt{CAMB}~\cite{Lewis:1999bs}, whereas \texttt{Montepython}~\cite{Audren:2012vy} is written (you guessed it...) in \texttt{Python} and interfaced with the Boltzmann solver \texttt{CLASS}~\cite{Blas:2011rf}. In this thesis, I have performed parameter estimation and forecasts using both \texttt{CosmoMC} (in Paper~I, Paper~II, Paper~IV, and Paper~V) and \texttt{Montepython} (in Paper~III).

\chapter{Results and discussion of included papers}
\label{chap:6}

\begin{chapquote}{(?) William Edwards Deming (1978?)}
``In God we trust. All others must bring data.''
\end{chapquote}

Armed with the necessary machinery in cosmology and statistics briefly described in the previous Chapters, we are now ready to discuss the results obtained in the included papers. This Chapter will inevitably be quite succinct in nature, and I invited the interest reader to read the papers for more details. From a broad picture perspective, the included papers follow a rather coherent storyline, which broadly proceeds as follows:
\begin{itemize}
\item \textbf{Q1}: ``\textit{What does current (as of 2017) cosmological data tell us about the neutrino mass scale? How can we use this information to make statements about the neutrino mass ordering in a statistically robust way?}''
\item \textbf{A1}: Current cosmological data places rather tight constraints on the neutrino mass scale, with the most robust bound being $M_{\nu}<0.12\,{\rm eV}$ at 95\% confidence level. The use of galaxy clustering data seems especially promising. We can start to say something interesting about the mass ordering, with the normal ordering being weakly favoured due to parameter space volume effects. I can certainly tell you more in Paper~I and Chapter~\ref{sec:paper1}.
\item \textbf{Q2}: ``\textit{How can we improve from here especially in our use of galaxy clustering data?}''
\item \textbf{A2}: A better understanding of galaxy bias is crucial. It would be great to also nail down its scale-dependence better. People have been talking about doing this using CMB lensing-galaxy cross-correlations for a long time, but for the first time we got around to doing it using real data. Let me tell you more in Paper~II and Chapter~\ref{sec:paper2}.
\item \textbf{Q3}: ``\textit{I heard that when putting massive neutrinos into the picture, the galaxy bias becomes scale-dependent even on large scales. Is this true and should people worry about it?}''
\item \textbf{A3}: Yes, this is true. And yes, people should worry about it (although they haven't so far), else future determinations of cosmological parameters from galaxy clustering data will be biased (no pun intended). You can read more in Paper~III and Chapter~\ref{sec:paper3}.
\item \textbf{Q4}: ``\textit{So far we've looked at the simplest $\Lambda$CDM+$M_{\nu}$ model. But I would imagine your tight limits on $M_{\nu}$ degrade if you relax your assumptions, due to parameter degeneracies. Is it always true that neutrino mass upper limits degrade when opening up your parameter space? And if not, can this be used to learn something interesting?}''
\item \textbf{A4}: Interesting question! In fact, it isn't always true, and we found an important exception: a non-phantom dark energy component, i.e. with time-dependent equation of state $w(z) \geq -1$ throughout the expansion history, as for instance quintessence. And yes, this information can be used to potentially rule out dark energy models from laboratory measurements of the neutrino mass ordering. If you find this confusing or unexpected (and you should), I'll tell you more in Paper~IV and Chapter~\ref{sec:paper4}.
\item \textbf{Q5}: ``\textit{Still along the lines of parameter degeneracies, certainly the reverse argument is also a problem, i.e. our ignorance of neutrino properties can bias our determination of other parameters? For instance, is our knowledge about inflation (and hence in some sense the initial conditions of the Universe) affected by our ignorance of neutrino properties?}''
\item \textbf{A5}: Good point! Luckily, for the specific case of inflation, it turns out that what we learn isn't really affected by our ignorance of neutrino properties (mass, mass ordering, effective number). If you still aren't convinced, take a look at Paper~V and Chapter~\ref{sec:paper5}.
\end{itemize}

The rest of this Chapter is organized as summarized in the Q-A thread above, with Secs.~\ref{sec:paper1},~\ref{sec:paper2},~\ref{sec:paper3},~\ref{sec:paper4},~\ref{sec:paper5} briefly summarizing the results of Papers~I,~II,~III,~IV,~V respectively.

\section{Early 2017 limits on neutrino masses and mass ordering}
\label{sec:paper1}

In early 2017, we set ourselves to analyse a selection of the most recent cosmological datasets. Just a few months back, the BOSS collaboration~\cite{Dawson:2012va} had released cosmological products from their final data release, DR12, containing over a million galaxies~\cite{Alam:2015mbd,Alam:2016hwk}. This was the largest spectroscopic sample of galaxies to date and one could certainly expect great cosmological constraints from it. Our goals were twofold: to understand how far down current cosmological data could push the upper limits on $M_{\nu}$, and to address what was starting to become a hot question at the time, namely whether, and in case how, we could use these limits to make statements about the neutrino mass ordering in a statistically robust way. Our results were discussed in Paper~I~\cite{Vagnozzi:2017ovm}, which at the time of writing (early 2019) still reports the tightest upper limits on $M_{\nu}$. An incomplete list of recent related works examining cosmological constraints on neutrino masses, both in light of current and future data, can be found in~\cite{Palanque-Delabrouille:2015pga,Zhen:2015yba,Gerbino:2015ixa,DiValentino:2015wba,Allison:2015qca,DiValentino:2015sam,Cuesta:2015iho,Huang:2015wrx,Moresco:2016nqq,
Giusarma:2016phn,Oh:2016wls,Archidiacono:2016lnv,Yeche:2017upn,Capozzi:2017ipn,Couchot:2017pvz,Caldwell:2017mqu,Doux:2017tsv,Wang:2017htc,Chen:2017ayg,
Upadhye:2017hdl,Salvati:2017rsn,Nunes:2017xon,Emami:2017wqa,Boyle:2017lzt,Zennaro:2017qnp,Sprenger:2018tdb,Wang:2018lun,Mishra-Sharma:2018ykh,Choudhury:2018byy,Choudhury:2018adz,Brinckmann:2018owf,Aghanim:2018eyx,Ade:2018sbj,Yu:2018tem,Liu:2018dsw,Li:2018owg,Coulton:2018ebd,
Loureiro:2018pdz,Gariazzo:2018meg,Marques:2018ctl}. I will briefly discuss the content of Paper~I: there, we analysed several datasets and even more (28) dataset combinations. It is not my goal here to discuss all these dataset combinations, but only to focus on the essential findings, while the full details can be found in Paper~I.

\begin{figure}[!t]
\centering
\includegraphics[width=0.7\textwidth]{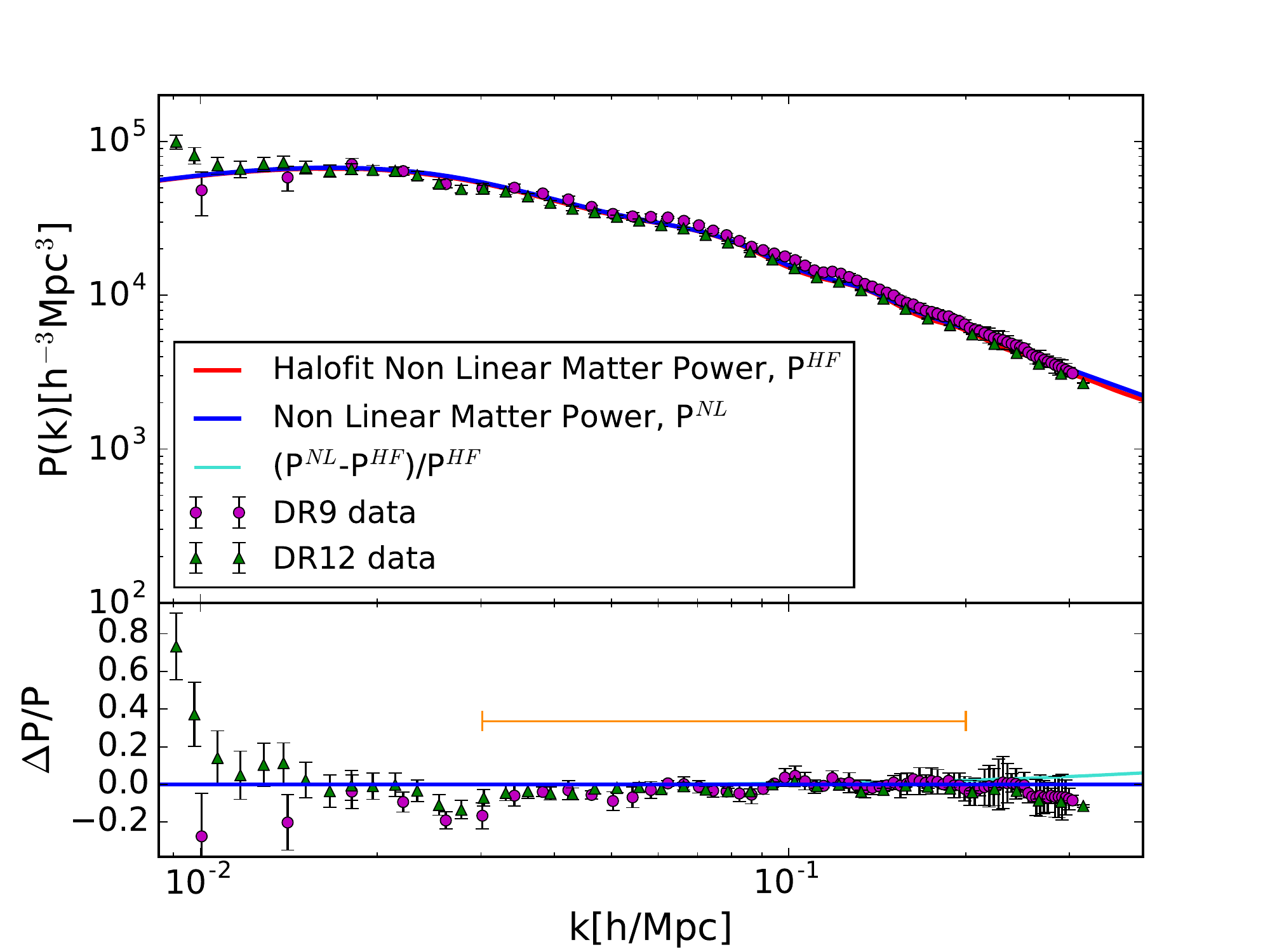}
\caption{\textit{Top panel}: nonlinear galaxy power spectrum computed using \texttt{CAMB}+\texttt{Halofit} (red curve), compared with the same quantity computed using the Coyote emulator. Both quantities are plotted assuming the \textit{Planck} 2015 best-fit parameters and $M_{\nu}=0\,{\rm eV}$ and a bias $b \approx 2$. The green triangles denote the galaxy power spectrum measured from the BOSS DR12 CMASS sample, whereas the purple circles denote the galaxy power spectrum measured from the BOSS DR9 CMASS sample. \textit{Bottom panel}: the blue line denotes the fractional difference between the power spectrum computed using the Coyote emulator vs using \texttt{CAMB}+\texttt{Halofit}. The orange line denotes the wavenumber range we use in~\cite{Vagnozzi:2017ovm}, which is safe both against systematics on large scales and nonlinear corrections on small scales. Reproduced from~\cite{Vagnozzi:2017ovm} (Paper~I) with permission from APS.}
\label{fig:paper1fig1}
\end{figure}

Given their importance in motivating our study, let me first briefly describe the BOSS DR12 product we used. We considered measurements of the spherically averaged power spectrum of galaxies from the BOSS DR12 CMASS sample~\cite{Gil-Marin:2015sqa}, containing 777,202 massive galaxies in the redshift range $0.43<z<0.7$. Later we will refer to this particular dataset as \textit{P(k)}. The low-level modelling of the power spectrum is described in Paper~I, and notably involves convolving the theoretical power spectrum with a window function accounting for mode mixing due to the finite size of the survey. Here, I will briefly discuss our treatment of galaxy bias. Denoting by $P_{\rm th}^g$ the \textit{theoretical} galaxy power spectrum (theoretical because it is what we then compare against observations, after convolution with the window function as discussed previously), we modelled this quantity as:
\begin{eqnarray}
P_{\rm th}^g(k,z) = b^2P_{{\rm HF}\nu}^m(k,z)+P^{\rm shot}\,.
\label{eq:pg}
\end{eqnarray}
In Eq.~(\ref{eq:pg}), $P_{{\rm HF}\nu}$ is the matter power spectrum computed using the Boltzmann solver \texttt{CAMB}~\cite{Lewis:1999bs} and corrected for non-linear effects using the \texttt{Halofit} method~\cite{Smith:2002dz,Takahashi:2012em}, and in particular the version of Bird, Viel, \& Haehnelt calibrated to simulations of massive neutrinos~\cite{Bird:2011rb}. $P^{\rm shot}$ is a constant shot-noise contribution included to reflect the fact that galaxies are discrete tracers of the underlying cosmic web. Finally, $b$ is a bias factor which we take to be constant (scale-independent). This is motivated by the fact that we limited our analysis to scales $0.03\,h{\rm Mpc}^{-1}<k<0.2\,h{\rm Mpc}^{-1}$, \textit{i.e.} scales which are at most mildly non-linear at the redshift in question. As we have seen in Chapter~\ref{subsec:lss}, on linear scales the galaxy bias is expected to be constant.~\footnote{Notice, however, that this approximation breaks down in the presence of massive neutrinos (see for instance~\cite{Castorina:2013wga}), although this subtle effect turns out not to be important currently, given the limited sensitivity of current data. A further investigation of large-scale galaxy bias in the presence of massive neutrinos will be the topic of Paper~III. Moreover, the role of scale-dependent galaxy bias on mildly non-linear scales will be studied in Paper~II.}. The measured BOSS DR12 CMASS power spectrum is shown in Fig.~\ref{fig:paper1fig1}, where it is compared against the theoretical nonlinear power spectrum computed using \texttt{CAMB}+\texttt{Halofit}, as well as using the Coyote emulator~\cite{Heitmann:2008eq,Heitmann:2013bra,Kwan:2013jva} calibrated onto several large N-body simulations (the figure also compares the measurements with those from the BOSS DR9 CMASS sample~\cite{Anderson:2012sa}, included for comparison with our earlier work~\cite{Giusarma:2016phn}),

We analyse cosmological data using the cosmological MCMC sampler \texttt{CosmoMC}~\cite{Lewis:2002ah}. Data-wise, we first considered measurements of the CMB temperature anisotropies, as well as large-scale polarization anisotropies, from the \textit{Planck} 2015 data release~\cite{Ade:2015xua}: we denote this dataset combination as \textit{base} (for reference, this dataset is typically referred to as ``\textit{PlanckTT+lowP}'' in the literature). Using this dataset, we recover the well-known 95\%~C.L. upper limit $\underline{\boldsymbol{M_{\nu}<0.72\,{\rm eV}}}$.~\footnote{Henceforth all our upper limits are 95\%~C.L. upper limits unless otherwise stated.} This limit is driven both by the effect of massive neutrinos on the first peak through the early ISW effect (see Chapter~\ref{subsec:signaturesnucmb}), as well as on the higher-multipole peaks through modifications to the lensing potential. When adding the \textit{P(k)} dataset to our \textit{base} dataset combination, the upper limit on $M_{\nu}$ considerably improves to $\underline{\boldsymbol{M_{\nu}<0.30\,{\rm eV}}}$. This improvement is driven by the suppression effect of massive neutrinos on the power spectrum, as well as the degeneracy-breaking ability of power spectrum measurements. We then add BAO distance measurements from the 6dFGS~\cite{Beutler:2011hx}, WiggleZ~\cite{Blake:2011en}, and BOSS DR11 LOWZ surveys~\cite{Anderson:2013zyy}, referring to these datasets \textit{BAO}. When combining the \textit{BAO} dataset with our \textit{base} and \textit{P(k)} datasets, we denote this combination by \textit{basePK} and find that the upper limit on $M_{\nu}$ improves to $\underline{\boldsymbol{M_{\nu}<0.25\,{\rm eV}}}$. BAO distance measurements help in pinning down the late-time expansion rate, and in particular $H_0$, alleviating the $M_{\nu}$-$H_0$ degeneracy and hence aiding a tighter determination of $M_{\nu}$ (see Chapter~\ref{subsec:signaturesnulss}). Finally, we included a Gaussian prior on the optical depth to reionization, $\tau = 0.055 \pm 0.009$, which we denoted by $\tau 0p055$: this prior is intended to mimic, to the best of our knowledge, the new large-scale polarization measurements to be delivered by the \textit{Planck} collaboration in 2019.~\footnote{The value of $\tau$ we included was obtained from the \textit{Planck} collaboration after identifying, modelling, and removing previously unidentified systematics in large-scale polarization data from the High Frequency Instrument (HFI), resulting in an improved determination of the optical depth to reionization, and a shift of the latter towards lower values~\cite{Aghanim:2016yuo,Adam:2016hgk}. While the full HFI low-$\ell$ likelihood was not available at the time, an important part of the cosmological information contained in large-scale polarization measurements resides in the value of the optical depth to reionization, which determines the shape and location of the reionization bump in the polarization power spectra (see Chapter~\ref{subsec:cmb}): for this reason, we made the conservative choice of not including large-scale polarization data in order to avoid double-counting information, while retaining large-scale temperature data.} We found that including the $\tau 0p055$ dataset improved our upper limit to $\underline{\boldsymbol{M_{\nu}<0.20\,{\rm eV}}}$, because of the mutual degeneracies between $M_{\nu}$, $A_s$, and $\tau$ (discussed in more detail in Paper~I).

We then considered the impact of including small-scale CMB polarization data. As discussed in detail in Paper~I, due to the presence of tiny residual systematics in this dataset, the resulting limits should be interpreted with more caution. We denoted by \textit{basepol} the dataset resulting from combining small-scale polarization with our \textit{base} dataset (for reference, this dataset combination is typically referred to as ``\textit{PlanckTTTEEE+lowP}'' in the literature). We find that the upper limit on $M_{\nu}$ improves from the $0.72\,{\rm eV}$ found for the \textit{base} dataset to $\underline{\boldsymbol{M_{\nu}<0.49\,{\rm eV}}}$. The fact that small-scale CMB polarization measurements can considerably improve cosmological parameter estimation, including the determination of the neutrino mass, is well known and was recently emphasized in~\cite{Galli:2014kla}.~\footnote{There are several reasons why this is the case, even though the signal-to-noise ratio is lower in polarization than it is in temperature. As shown in~\cite{Galli:2014kla}, the change of the spectra under a variation of cosmological parameters compared to the noise is larger in polarization than it is in temperature. This is particularly true for the TE cross-correlation spectrum. In other words, the response of the polarization spectra to changes in cosmological parameters is substantially greater than that of the temperature spectrum. Besides this, the acoustic peaks are also sharper in polarization than in temperature, and the small-scale polarization spectrum is less sensitive to astrophysical foregrounds than the temperature spectrum at the same scales, which is affected by contamination from unresolved radio and infrared galaxies~\cite{Tucci:2004zy,Seiffert:2006vh}.} We now gradually add additional LSS datasets to the \textit{basepol} dataset combination, in order to improve the determination of $M_{\nu}$. When we add the \textit{P(k)} dataset the limit improves to $\underline{\boldsymbol{M_{\nu}<0.27\,{\rm eV}}}$, while further adding the \textit{BAO} dataset (resulting combination referred to as \textit{basepolPK}, in analogy to the previous \textit{basePK} dataset) our limit improves to $\underline{\boldsymbol{M_{\nu}<0.21\,{\rm eV}}}$. Finally, including the $\tau 0p055$ prior on the optical depth to reionization, our limit improves to $\underline{\boldsymbol{M_{\nu}<0.18\,{\rm eV}}}$. In Paper~I we tested the inclusion of other datasets (for instance, direct measurements of the Hubble parameter~\cite{Efstathiou:2013via,Riess:2016jrr}, or SZ cluster counts~\cite{Ade:2015fva,Ade:2015gva}): these lead to even tighter limits (up to $M_{\nu}<0.11\,{\rm eV}$), at the price of being less robust. For this reason I will not discuss the corresponding results here, but invited the interested reader to read Paper~I for more details.

The careful reader will have noticed that, whenever we included \textit{P(k)} measurements from the BOSS DR12 CMASS sample, we did not include BAO distance measurements from the BOSS DR11 CMASS sample (despite these being readily available and widely used). The reason is that there is a substantial overlap in volume between the two samples, so using both measurements would lead to double-counting of data. This naturally raises the question: which of these two datasets would be more constraining? A na\"{i}ve guess would be that a \textit{P(k)} measurement is more constraining than a BAO distance measurement: if anything, \textit{P(k)} technically already contain the BAO information (see Chapter~\ref{subsec:lss}), so loosely speaking the BAO measurement extracted from a given survey should be a ``subset'' of the power spectrum measurement extracted from the same survey. Our guess was that replacing the \textit{P(k)} dataset with the BAO distance measurement from the BOSS DR11 CMASS sample at $z=0.57$ should have resulted in looser constraints on $M_{\nu}$.

Of course, in Paper~I we checked our guess. We first removed the \textit{P(k)} dataset, while augmenting the \textit{BAO} dataset with the BAO distance measurement from BOSS DR11 CMASS: the resulting combination of four BAO measurements was referred to as \textit{BAOFULL}. We denoted the combination of the \textit{base} and \textit{BAOFULL} datasets as \textit{baseBAO}. For this dataset combination, we found $\underline{\boldsymbol{M_{\nu}<0.19\,{\rm eV}}}$ (compare with $M_{\nu}<0.25\,{\rm eV}$ found for the \textit{basePK} dataset). Surprisingly, replacing power spectrum measurements with BAO distance measurements resulted in a tighter limit on $M_{\nu}$! This trend was confirmed for other dataset combinations we tested. When adding the $\tau 0p055$ prior to the \textit{baseBAO} dataset combination, we found $\underline{\boldsymbol{M_{\nu}<0.15\,{\rm eV}}}$ (compare with $M_{\nu}<0.20\,{\rm eV}$ found for the \textit{basePK}+$\tau 0p055$ dataset). Analogous results were obtained when using small-scale polarization data. Combining the \textit{basepol} and \textit{BAOFULL} datasets (combination denoted by \textit{basepolBAO}), we found $\underline{\boldsymbol{M_{\nu}<0.15\,{\rm eV}}}$ (compare with $M_{\nu}<0.21\,{\rm eV}$ found for the \textit{basePK} dataset); finally, adding the $\tau 0p055$ prior to the \textit{basepolBAO} combination, we found $\underline{\boldsymbol{M_{\nu}<0.12\,{\rm eV}}}$ (compare with $M_{\nu}<0.18\,{\rm eV}$ found for the \textit{basePK} dataset). For the reader's convenience, the content of the datasets/dataset combinations adopted is briefly summarized in Tab.~\ref{tab:tab1paper1}, while the limits from the 12 dataset combinations we discussed are summarized in Tab.~\ref{tab:tab2paper1}.
\begin{table}[h!]
\centering
\begin{tabular}{|c|c|}
\hline
Dataset & Content \\
\hline\hline
\textit{\textbf{base}} & \textit{Planck} CMB temperature and large-scale polarization \\
\hline
\textit{\textbf{basepol}} & \textit{base}+small-scale polarization \\
\hline
\textit{\textbf{P(k)}} & BOSS DR12 CMASS spherically averaged power spectrum \\
\hline
\textit{\textbf{BAO}} & BAO from 6dFGS BAO, WiggleZ, BOSS DR11 LOWZ \\
\hline
\textit{\textbf{BAOFULL}} & BAO from 6dFGS, WiggleZ, BOSS DR11 LOWZ \& CMASS \\
\hline
\textit{\textbf{basePK}} & \textit{base}+\textit{P(k)}+\textit{BAO} \\
\hline
\textit{\textbf{basepolPK}} & \textit{basepol}+\textit{P(k)}+\textit{BAO} \\
\hline
\textit{\textbf{baseBAO}} & \textit{base}+\textit{BAOFULL} \\
\hline
\textit{\textbf{basepolBAO}} & \textit{basepol}+\textit{BAOFULL} \\
\hline
\end{tabular}
\caption{Content of datasets and/or dataset combinations used in Paper~I.}
\label{tab:tab1paper1}
\end{table}
\begin{table}[h!]
\centering
\begin{tabular}{|c|c|}
\hline
Dataset	& Upper limit on $M_\nu$ (95\%~C.L.)\\
\hline\hline
\textit{base} & $0.72\,{\rm eV}$ \\	
\textit{base}+\textit{P(k)} & $0.30\,{\rm eV}$ \\
\textit{basePK} & $0.25\,{\rm eV}$ \\
\textit{basePK}+$\tau0p055$ & $0.20\,{\rm eV}$ \\
\textit{basepol} & $0.49\,{\rm eV}$ \\
\textit{basepol}+\textit{P(k)} & $0.27\,{\rm eV}$ \\
\textit{basepolPK} & $0.21\,{\rm eV}$ \\
\textit{basepolPK}+$\tau 0p055$ & $0.18\,{\rm eV}$ \\
\textit{baseBAO} & $0.19\,{\rm eV}$ \\
\textit{baseBAO}+$\tau 0p055$ & $0.15\,{\rm eV}$ \\
\textit{basepolBAO} & $0.15\,{\rm eV}$ \\
\textit{basepolBAO}+$\tau 0p055$ & $0.12\,{\rm eV}$ \\
\hline
\end{tabular}
\caption{95\%~C.L. upper bounds on the sum of the three active neutrino masses $M_\nu$ (in eV). The left column shows the combination of cosmological datasets adopted (see Tab.~\ref{tab:tab1paper1} for further details on these datasets), while the right column shows the 95\%~C.L. upper limits obtained for the specific combinations.}
\label{tab:tab2paper1}
\end{table}

\begin{figure}[!t]
\centering
\includegraphics[width=0.8\textwidth]{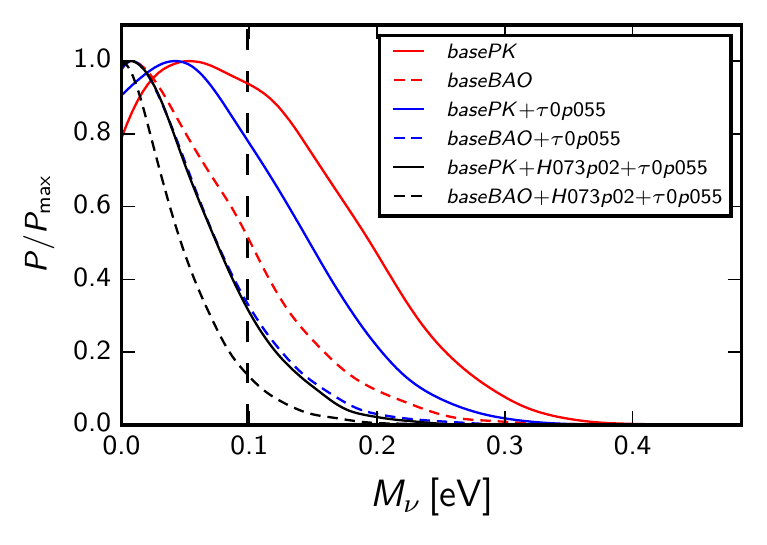}
\caption{Posteriors on $M_{\nu}$ (normalized to their maximum values) obtained using different dataset combinations. The figure should be read as follows: to make the BAO vs \textit{P(k)} comparison, choose a given color and compare the solid curve [\textit{P(k)}] against the dashed curve [BAO]. It is clear that BAO (dashed curves) leads to tighter constraints. Notice that the black curves are obtained including a prior on $H_0$ based on the locally measured value, not discussed in this Chapter (see Paper~I for more details). Reproduced from~\cite{Vagnozzi:2017ovm} (Paper~I) with permission from APS.}
\label{fig:paper1fig2}
\end{figure}

The results of Paper~I, surprisingly, indicated that BAO distance measurements appear to be more constraining than \textit{P(k)} measurements, despite the latter carrying more information than the former. Note that our results are confirmed by related earlier findings of~\cite{Hamann:2010pw,Giusarma:2012ph}. The only sensible explanation must be that, somehow, we are not analysing \textit{P(k)} data in a wise way, and this is preventing us from fully retrieving the information therein contained. One limitation in the modelling of \textit{P(k)} data is the need to introduce several nuisance parameters. In our case, we introduced two extra parameters (a constant bias and a shot-noise term): marginalizing over these extra parameters, especially on the bias, loosens the constraints on $M_{\nu}$. Clearly, a better handle on the bias (and eventually its scale-dependence) is highly desirable. A long-standing idea in this direction has been to use cross-correlations between CMB lensing and galaxies~\cite{Pen:2004rm,More:2014uva,Amendola:2015pha,Giannantonio:2015ahz,Pujol:2016lfe,
Singh:2016xey,Singh:2016edo,Joudaki:2017zdt,Simon:2017osp,Singh:2018kmr}: for the first time, we realized this idea on real data, but the reader will have to wait until Chapter~\ref{sec:paper2} (and Paper~II) to read more. Recall also that we set a hard cutoff at $k=0.2\,h{\rm Mpc}^{-1}$ to avoid delving into the non-linear regime. While the BAO feature in the 2-point correlation function appears on rather linear scales, the extraction of distance measurements benefits from what is known as the reconstruction procedure~\cite{Eisenstein:2006nk,Padmanabhan:2012hf,White:2015eaa}, which sharpens the BAO peak but introduces some amount of non-linear information (which we are instead conservatively choosing not to use when analysing \textit{P(k)} data). Our conclusion in Paper~I was that, albeit \textit{prima facie} BAO information counterintuitively leads to tighter limits than \textit{P(k)} measurements, this results reflects not so much a limitation of \textit{P(k)} data, but rather a limitation of \textit{the way we analyse} \textit{P(k)} data, and that improvements in that direction are certainly warranted (see e.g. Paper~II). A visual representation of the BAO vs \textit{P(k)} comparison is shown in Fig.~\ref{fig:paper1fig2} (notice that the figure also contains results obtained using a prior on $H_0$ based on the locally measured value by Riess \textit{et al.}~\cite{Riess:2016jrr}, not discussed in this Chapter but extensively discussed in Paper~I, see Sec.~IIID of Paper~I for more details).

Let me now cover the final major point discussed in Paper~I, namely how to robustly quantify the preference for one of the two mass orderings: normal ordering (\texttt{\texttt{NO}}) and inverted ordering (\texttt{IO}).~\footnote{The following discussion will deviate slightly from that in Sec.~IIB of Paper~I. Both qualitatively and quantitatively, the results are basically unchanged. The approach I will choose here allows one of the main conclusions of this part of Paper~I, the fact that cosmology will always prefer the \texttt{NO} due to volume effects, to be more easily grasped and understood. Over the 2 years between writing Paper~I and writing this thesis, by giving a number of talks on the subject I have realized that the chosen approach is more effective at conveying the message. In any case, the reader might want to be aware of these changes.} Intuitively, since the \texttt{IO} requires $M_{\nu}>0.1\,{\rm eV}$, the reader might expect that the closer our upper limits get to $0.1\,{\rm eV}$, the more the \texttt{IO} is under pressure. This expectation is certainly correct. However, the na\"{i}ve guess that a 95\%~C.L. upper limit of $M_{\nu}<0.1\,{\rm eV}$ would exclude the \texttt{IO} at 95\%~C.L. would be incorrect. In reality, as pointed out in the important earlier paper~\cite{Hannestad:2016fog}, the problem one has to solve here is a Bayesian model comparison problem (see Chapter~\ref{subsec:marginalization}) between two competing models: \texttt{NO} and \texttt{IO}. Or, in other words, to determine whether the \texttt{IO} hypothesis can be rejected in favour of the \texttt{NO} hypothesis at some confidence. The goal is then to compute the Bayesian evidence for \texttt{NO} and \texttt{IO}, and hence the Bayes factor of \texttt{NO} vs \texttt{IO}. As we discussed in Chapter~\ref{subsec:marginalization}, computing Bayesian evidences and Bayes factors is usually computationally expensive. However, in this case the situation is rather simplified: we have two very similar competing models sharing the same parameter space. The only difference is that one of the two models has access to a larger region of parameter space for a specific parameter: the \texttt{NO} can access the region $0.06\,{\rm eV}<M_{\nu}<0.1\,{\rm eV}$, whereas the \texttt{IO} cannot.

Our highly simplified situation makes it easy to write down a simple and illuminating expression for the Bayes factor of \texttt{NO} vs \texttt{IO}, $B_{\rm \texttt{NO},\texttt{IO}}$. Under the valid assumptions that the prior on $M_{\nu}$ is factorizable from the priors on the other cosmological parameters and that the likelihood does not depend on the chosen mass ordering (\textit{i.e.} that all difference between the two mass orderings resides in the different volume of parameter space accessible, a reasonable assumption given that cosmological data cares about $M_{\nu}$ and not about the individual masses), we find that $B_{\rm \texttt{NO},\texttt{IO}}$ can be written as follows:
\begin{eqnarray}
B_{\rm \texttt{NO},\texttt{IO}} = \frac{\int_{0.06\,{\rm eV}}^{\infty}{\cal P}(M_{\nu})p(M_{\nu} \vert \boldsymbol{d})}{\int_{0.10\,{\rm eV}}^{\infty}{\cal P}(M_{\nu})p(M_{\nu} \vert \boldsymbol{d})}\,,
\label{eq:bayesnoio}
\end{eqnarray}
where ${\cal P}(M_{\nu})$ denotes the prior on $M_{\nu}$ (flat in our case), and $p(M_{\nu} \vert \boldsymbol{d})$ denotes the posterior of $M_{\nu}$ given data $\boldsymbol{d}$. The confidence level at which we can exclude the inverted ordering (or equivalently, the posterior odds for the normal ordering), is given by $B_{\rm \texttt{NO},\texttt{IO}}/(1+B_{\rm \texttt{NO},\texttt{IO}})$. Generalizing Eq.~(\ref{eq:bayesnoio}) to the case where \texttt{NO} and \texttt{IO} are not taken to be equally likely a priori is trivial. Three comments on Eq.~(\ref{eq:bayesnoio}) are useful:
\begin{enumerate}
\item The integrand is the same in the numerator and the denominator: the only difference is the range of integration, which is wider for the numerator.
\item Combining the above with the fact that the integrand is a strictly positive quantity (it is a product of two probability distributions), it will always be the case that $B_{\rm \texttt{NO},\texttt{IO}}>1$!
\item The inevitable appearance of ${\cal P}(M_{\nu})$ implies that the result is sensitive, to a greater or less extent, to how one chooses to weigh one's prior volume.
\end{enumerate}
Using Eq.~(\ref{eq:bayesnoio}), I computed the Bayes factor for \texttt{NO} vs \texttt{IO} for the 12 different dataset combinations discussed above and summarized in Tab.~\ref{tab:tab2paper1}. I found that the Bayes factor remains rather low for all combinations, and in any case always below the threshold value of $\sqrt{10}$ necessary for claiming a substantial preference for the \texttt{NO} according to the Jeffreys scale presented in Tab.~\ref{tab:jeffreysscale}: according to the same scale, the preference for the \texttt{NO} remains always barely worth mentioning. The highest value of the Bayes factor is achieved for the \textit{basepolBAO}+$\tau 0p055$ dataset combination, which gives $B_{\rm \texttt{NO},\texttt{IO}} \approx 2.4$ ($M_{\nu}<0.12\,{\rm eV}$), a figure which excludes the \texttt{IO} at only 71\%~C.L.! As mentioned previously, in Paper~I we tested other less robust dataset combinations which led to tighter limits, but in any case the highest value of $B_{\rm \texttt{NO},\texttt{IO}}$ we obtained was $3.3$ (excluding \texttt{IO} at 77\%~C.L.), even though for the same dataset we found the extremely tight limit $M_{\nu}<0.093\,{\rm eV}$ (see Paper~I for more details).

Our findings in Paper~I highlighted the fact that cosmology will \textit{always} prefer the \texttt{NO} over the \texttt{IO} [which is self-evident from Eq.~(\ref{eq:bayesnoio}), since $B_{\rm \texttt{NO},\texttt{IO}}>1$ will always hold]. This preference arises entirely due to \textit{volume effects}, \textit{i.e.} the fact that the \texttt{NO} has access to a larger region of parameter space, and not due to physical effects (since besides these volume considerations, the data is not sensitive to differences between the two orderings). Notice that a corollary of these findings is that cosmology will only be able to determine the mass ordering if Nature has chosen the \texttt{NO} and a value of $M_{\nu}$ substantially lower than $0.1\,{\rm eV}$. As a back of the envelope estimate, in the best event a sensitivity $\sigma_{M_{\nu}} \sim 0.02\,{\rm eV}$ would be needed for a $2\sigma$ discrimination of the mass ordering (confirmed quantitatively in~\cite{Hannestad:2016fog}). These conclusions hold insofar as cosmological data remains mostly sensitive to $M_{\nu}$ rather than the masses of the individual eigenstates, which is expected to remain the case for the foreseeable future. The cosmological preference for the normal ordering being due to volume effects also warrants a careful investigation into the choice of prior on $M_{\nu}$. This is a fascinating discussion which however is well beyond the scope of Paper~I: I invited the interested reader to consult a number of papers which appeared around the same time, or later than Paper~I, see for instance~\cite{Hannestad:2016fog,Gerbino:2016ehw,Vagnozzi:2017ovm,Simpson:2017qvj,Schwetz:2017fey,Hannestad:2017ypp,Long:2017dru,
Gariazzo:2018pei,Heavens:2018adv,Handley:2018gel}.

\subsection{Executive summary of Paper~I}
\label{subsec:paper1}

Let me finally wrap up and summarize our results in Paper~I. We analysed a suite of state-of-the-art cosmological datasets (including the galaxy power spectrum from the CMASS sample of the BOSS final data release). The tightest upper limit on $M_{\nu}$ we found and deemed sufficiently robust was $M_{\nu}<0.12\,{\rm eV}$ 95\%~C.L., which at the time of writing remains the tightest upper limit on $M_{\nu}$ ever reported (matched by~\cite{Palanque-Delabrouille:2015pga,Aghanim:2018eyx}). This upper limit is tantalizingly close to $0.10\,{\rm eV}$, the minimum allowed value of $M_{\nu}$ within the inverted ordering, suggesting that cosmological data might be putting the inverted ordering under pressure. We devised a simple method for quantifying the preference for the normal ordering in a statistically robust way [Eq.~(\ref{eq:bayesnoio})], based on Bayesian model comparison. In doing so, we clarified that cosmological data, insofar as only sensitive to $M_{\nu}$ and not the masses of the individual eigenstates, will always prefer the normal ordering due to parameter space volume effects, thus emphasizing the role of choice of prior. Applying our method we found that the dataset combination leading to $M_{\nu}<0.12\,{\rm eV}$ indicates a $2.4$:$1$ preference for the normal ordering, barely worth mentioning according to the Jeffreys scale of Tab.~\ref{tab:jeffreysscale}. Finally, we analysed the relative constraining power of power spectrum versus BAO distance measurements, finding the counterintuitive result that BAO distance measurements appear to be more constraining. We argued that this finding indicates the necessity of devising wiser ways of analysing power spectrum data, and in particular improving the determination of the galaxy bias. The natural continuation of this work is therefore in Paper~II (to be discussed in Chapter~\ref{sec:paper2}), where we devise a method representing a first step in this direction. 

\section{Scale-dependent galaxy bias and CMB lensing-galaxy cross-correlations}
\label{sec:paper2}

As we have argued in Chapter~\ref{sec:paper1} and Paper~I, galaxy clustering (\textit{i.e.} power spectrum) data represents a powerful probe of massive neutrinos (and more generally of free-streaming species). However, na\"{i}vely comparing the constraining power of \textit{P(k)} vs BAO measurements revealed that improvements are needed in order to fully harness the constraining power of the former, especially in terms of getting a better handle on the galaxy bias. For quite some time, a long-standing idea in this direction has been that of using cross-correlations between CMB lensing and galaxy maps to calibrate the galaxy bias and possibly its scale-dependence. While some steps had been taken in this direction (e.g.~\cite{Pen:2004rm,More:2014uva,Amendola:2015pha,Giannantonio:2015ahz,Pujol:2016lfe,
Singh:2016xey,Singh:2016edo,Joudaki:2017zdt,Simon:2017osp,Singh:2018kmr}), nobody had ever tried to fully apply this idea on real data. In early 2017, we decided the time was ripe to try our this idea on real data, understand what the practical difficulties (both theoretical and observational) were, and see whether we could use this to improve our limits on neutrino masses: the results of our work were described are described in Paper~II~\cite{Giusarma:2018jei}, and will be summarized in this Chapter.

Before starting, I want to heuristically argue that using cross-correlations between CMB lensing and galaxy maps, in combination with galaxy clustering [\textit{i.e.} \textit{P(k)}] measurements, is a good idea. Let us for the moment just consider a constant linear bias $b$. We have already seen [e.g. Eq.~(\ref{eq:pg})] that galaxy power spectrum measurements are proportional to $b^2$, with $b$ treated as a nuisance parameter which is marginalized over. Being somehow able to at the same time measure another quantity which scales like a different power of $b$ (e.g. $b^1$) would help us nail down $b$ even better, which would reflect in marginalized parameter constraints being less loose than they would otherwise be. How do we construct a quantity proportional to $b^1$? We can try cross-correlating the galaxy overdensity field (which carries one power of $b$) with another field carrying no dependence on $b$ (and thus directly tracing the underlying matter overdensity field). CMB lensing cares about the (projected) matter overdensity field, and hence appears as an excellent candidate for the latter field. Cross-correlations between the CMB lensing field and galaxy maps should thus be proportional to one power of $b$.

Lensing acts to remap the direction of photons reaching us from the CMB, see e.g.~\cite{Bartelmann:1999yn,Lewis:2006fu,Hanson:2009kr} for seminal reviews. Assume we receive a photon coming from direction $\hat{\boldsymbol{n}}$. We define the deflection field $\boldsymbol{d}$ to point from $\hat{\boldsymbol{n}}$ to the direction from which the photon was originally emitted: this field can be measured from CMB maps due to the subtle effects lensing imprints on the statistics of CMB fluctuations (see e.g.~\cite{Hu:2001kj,Hanson:2010rp} for more details). From $\boldsymbol{d}$ one can determine the \textit{lensing convergence} $\kappa \equiv -\boldsymbol{\nabla} \cdot \boldsymbol{d}/2$. In a direction $\hat{\boldsymbol{n}}$, $\kappa$ is given by a weighted projection of the matter overdensity $\delta$~\cite{Lewis:2006fu}:
\begin{eqnarray}
\kappa(\hat{\boldsymbol{n}}) = \int_{0}^{z_{\rm dec}} dz\,W^{\kappa}(z)\delta(\chi(z)\hat{\boldsymbol{n}},z)\,,
\label{eq:kappaproj}
\end{eqnarray}
where the lensing kernel $W^{\kappa}(z)$, in a flat Universe, is given by:
\begin{eqnarray}
W^{\kappa}(z) = \frac{3}{2H(z)}\Omega_m H_0^2(1+z)\chi(z)\frac{\chi_{\star}(z)-\chi(z)}{\chi_{\star}(z)}\,.
\end{eqnarray}

The meaning of the often heard statement that CMB lensing cares about the projected matter overdensity field between us and last-scattering is reflected in Eq.~(\ref{eq:kappaproj}). Given a galaxy survey, we can now consider the fractional galaxy overdensity field and cross-correlate that with the CMB lensing convergence field. The lensing convergence-galaxy overdensity cross-power spectrum, which we will refer to as $C_{\ell}^{\kappa g}$, is given by (see e.g.~\cite{Peiris:2000kb,Hirata:2008cb,Bleem:2012gm,Sherwin:2012mr,Pearson:2013iha,
Bianchini:2014dla,Bianchini:2015yly,Singh:2016xey,Schmittfull:2017ffw,Bianchini:2018mwv}):
\begin{eqnarray}
C_\ell^{\kappa g} = \int_{z_0}^{z_1} dz\frac{H(z)}{\chi^{2}(z)}W^{\kappa}(z)f_g(z)P_{mg}\left(k=\frac{\ell}{\chi(z)},z\right)\,,
\label{eq:clkg}
\end{eqnarray}
where the galaxy sample is assumed to reside in the redshift range between $z_0$ and $z_1$ and to have a normalized redshift distribution given by $f_g(z)$. On the other hand, $P_{mg}(k)$ is the matter-galaxy cross-power spectrum. Since it results from correlating the galaxy overdensity field (carrying one power of bias) with the underlying matter field (independent of bias), it will carry only one power of bias $b$.
\begin{figure}[!h]
\centering
\includegraphics[width=0.7\textwidth]{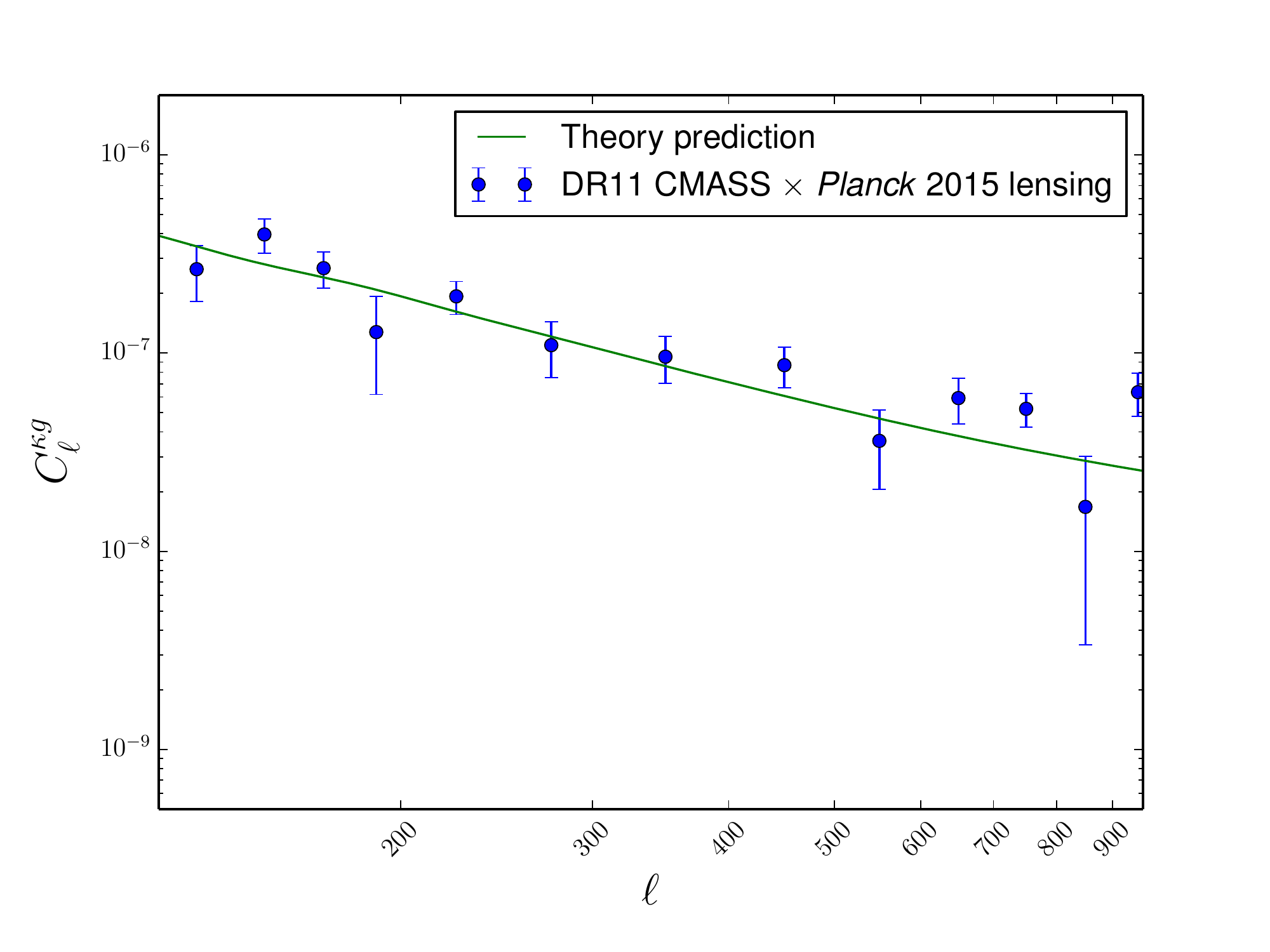}
\caption{Measured CMB lensing convergence-galaxy overdensity cross-power spectrum from cross-correlating \textit{Planck} 2015 lensing maps with galaxies from the BOSS DR11 CMASS sample (blue points), compared against the theory predictions (green curve). Theory predictions are made assuming a scale-dependent bias $b_{\rm cross}(k)$ with parameters $a$ and $c$ fixed to their central values inferred from the \textit{PlanckTT}+\textit{lowP}+$C_{\ell}^{\kappa g}$+\textit{P(k)} dataset combination, $a=1.95$ and $c=0.48\,h^{-2}{\rm Mpc}^2$ (see Tab.~I in Paper~II).}
\label{fig:paper2fig1}
\end{figure}

Let us discuss in more detail $P_{mg}$, and its relation to galaxy bias and the galaxy power spectrum $P_g$, which is less trivial than one might imagine. Let us also reinstate into the picture the leading-order scale-dependence of galaxy bias due to complexities inherent in the processes leading to galaxy formation (see Chapter~\ref{subsec:lss}): $b(k) = b_0+b_1k^2$. One might na\"{i}vely guess that $P_{mg}(k)$ and $P_g(k)$ are related to the underlying matter power spectrum $P(k)$ as follows:
\begin{eqnarray}
P_{mg}(k) = b(k)P(k)\,,\quad P_g(k) = b^2(k)P(k)\,.
\label{eq:pmpg}
\end{eqnarray}
The implicit assumption in Eq.~(\ref{eq:pmpg}) is that the bias appearing in cross-correlation and auto-correlation measurements is the same quantity. This assumption turns out to be not entirely correct. In fact, it is more correct to rewrite Eq.~(\ref{eq:pmpg}) as follows:
\begin{eqnarray}
P_{mg}(k) = b_{\rm cross}(k)P(k)\,,\quad P_g(k) = b_{\rm auto}^2(k)P(k)\,,
\label{eq:pmpgcorrect}
\end{eqnarray}
where $b_{\rm cross}$ and $b_{\rm auto}$ share the same functional forms (\textit{i.e.} a constant plus a $k^2$ correction) and same large-scale value (\textit{i.e.} the constant term is the same in both), but have different coefficients in front of the $k^2$ correction. In Paper~II we therefore chose to parametrize these two biases as follows:
\begin{eqnarray}
b_{\rm cross}(k) = a+ck^2\,,\quad b_{\rm auto}(k) = a+dk^2\,,
\label{eq:crossauto}
\end{eqnarray}
with $a$, $c$, and $d$ being free nuisance parameters which we will eventually marginalize over. From simulations and theoretical considerations, one expects $db_{\rm cross}/dk>0$ and $db_{\rm auto}/dk<0$: in other words, after the large-scale plateau where both biases are constant and equal to each other, the biases in cross- and auto-correlation increase and decrease with decreasing scale (increasing wavenumber) respectively. This behaviour is clearly seen in the simulations of~\cite{Okumura:2012xh}: see the short-dashed ($b_{\rm cross}$) and long-dashed ($b_{\rm auto}$) curves in the top row panels of Fig.~2 in~\cite{Okumura:2012xh}. We therefore expect $c>0$ and $d<0$. It would of course be highly desirable if a relation between $c$ and $d$ existed (perhaps calibrated to simulations), but to the best of our knowledge no such relation exists: therefore, in the following, we will treat them as independent (nuisance) parameters. The origin of the differences between $b_{\rm cross}(k)$ and $b_{\rm auto}(k)$, and in particular their different behaviour on small scales, are discussed in much more detail towards the end of Section~II of Paper~II. These differences can be traced back to the discrete nature of galaxies as tracers of the matter density field, as well as the principle of halo exclusion, and I invite the interested reader to read Paper~II for more details (see also e.g.~\cite{CasasMiranda:2001ym,Smith:2006ne,Baldauf:2013hka}).

Our idea in Paper~II was to combine clustering [\textit{i.e.} \textit{P(k)}] and CMB lensing-galaxy cross-correlation [\textit{i.e.} $C_{\ell}^{\kappa g}$] measurements, to interpret them within a theoretically motivated scale-dependent bias model [Eq.~(\ref{eq:crossauto})], and to see whether this would lead to substantial improvements in the upper limits on $M_{\nu}$. Recall that in Paper~I, we found $M_{\nu}<0.30\,{\rm eV}$ for our \textit{base}+\textit{P(k)} dataset combination. This limit will be our yardstick for quantifying improvements in the limits on $M_{\nu}$ brought upon our work. Our modelling of the data is discussed in more detail in Paper~II. We place flat priors on the bias parameters $a$, $b$, and $c$ appearing in Eq.~(\ref{eq:crossauto}). Although from the discussion in the previous paragraph we expect $c>0$ and $d<0$, we place flat priors on these quantities which still allow for $c<0$ and $d>0$ as well: we decided to leave it up to data to choose the sign of $c$ and $d$, in an attempt to be as conservative as possible.
\begin{figure}[!h]
\centering
\includegraphics[width=0.7\textwidth]{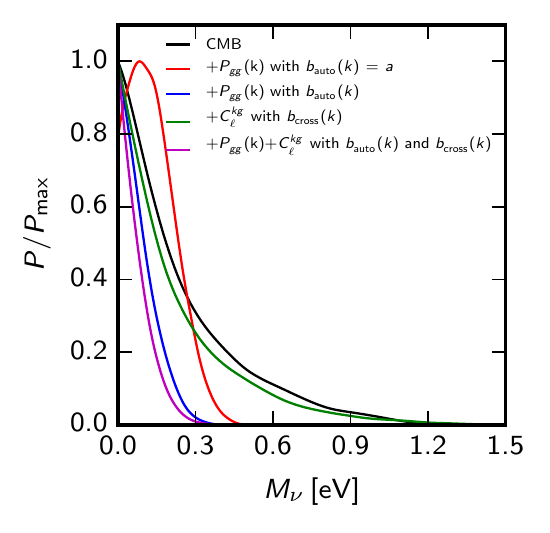}
\caption{Posterior distributions for $M_{\nu}$ (normalized to their maximum values) obtained using different datasets and making different assumptions on the galaxy bias: CMB (\textit{PlanckTT}+\textit{lowP}; black curve), CMB+\textit{P(k)} (BOSS DR12 CMASS) with constant bias (from Paper~I~\cite{Vagnozzi:2017ovm}; red curve), CMB+$C_{\ell}^{\kappa g}$ (BOSS DR11 CMASS $\times$ \textit{Planck} 2015 lensing) using scale-dependent $b_{\rm cross}(k)$ (from Eq.~(\ref{eq:crossauto}); green curve), CMB+\textit{P(k)} using scale-dependent $b_{\rm auto}(k)$ (from Eq.~(\ref{eq:crossauto}); blue curve), and CMB+$C_{\ell}^{\kappa g}$+\textit{P(k)} with scale-dependent $b_{\rm cross}(k)$ and $b_{\rm auto}(k)$ (purple curve). Reproduced from~\cite{Giusarma:2018jei} (Paper~II) with permission from APS.}
\label{fig:paper2fig2}
\end{figure}

We combined CMB temperature and large-scale polarization data from the \textit{Planck} 2015 data release with galaxy power spectrum data from the BOSS DR12 CMASS sample (already discussed in Paper~I), and the cross-correlation between CMB lensing convergence maps from the \textit{Planck} 2015 data release and galaxy maps from the BOSS DR11 CMASS sample~\cite{Anderson:2013zyy,Pullen:2015vtb}. The measured cross-correlation is shown in Fig.~\ref{fig:paper2fig1}. Using this dataset combination and parametrizing the scale-dependent biases appearing in \textit{P(k)} and $C_{\ell}^{\kappa g}$ with $b_{\rm auto}$ and $b_{\rm cross}$ as in Eq.~(\ref{eq:crossauto}) respectively, we find that the upper limit on $M_{\nu}$ improves to $\underline{\boldsymbol{M_{\nu}<0.19\,{\rm eV}}}$: this represents a substantial improvement over the previous $M_{\nu}<0.30\,{\rm eV}$ upper limit. In Fig.~\ref{fig:paper2fig2} we show the posterior distributions for various dataset combinations (including the earlier result of~\cite{Vagnozzi:2017ovm}, red curve): note that the posterior distribution for the $M_{\nu}<0.19\,{\rm eV}$ limit previously quoted is given by the purple curve.

As for the bias parameters, for the scale-independent parameter we find $a=1.95 \pm 0.07$ (consistent with expectations~\cite{Alam:2016hwk}), while for the scale-dependent parameters we find $c=0.48 \pm 0.90\,h^{-2}{\rm Mpc}^2$ and $d=-14.13 \pm 4.02\,h^{-2}{\rm Mpc}^2$. This is quite remarkable: despite not imposing that $c>0$ and $d<0$ at the level of priors, we find that the sign of these quantities is consistent with theoretical expectations! Notice also that we ``detect'' a scale-dependence in the galaxy power spectrum at over $3\sigma$ (\textit{i.e.} $d=0$ is more than $3\sigma$ away from the measured value). The measured value of $d$ naturally defines a scale $k_{sd}$ at which the complexities of galaxy formation lead to strong scale-dependence in the bias: $k_{sd} \equiv 1/\sqrt{d} \approx 0.27\,h{\rm Mpc}^{-1}$. This is consistent with the expectation that the $k^2$ correction we have considered in Paper~II should become prominent somewhere between $0.2\,h{\rm Mpc}^{-1}$ and $0.3\,h{\rm Mpc}^{-1}$. Clearly, our analysis shows that even at mildly non-linear scales (we set $k_{\max} = 0.2\,h{\rm Mpc}^{-1}$) scale-dependent galaxy bias should no longer be ignored. As future data becomes more precise, so should the theoretical modelling of the bias, considering even terms beyond $k^2$, and possibly relying on a perturbation theory-based approach (see e.g.~\cite{Carlson:2012bu,Modi:2016dah,Modi:2017wds}).

\subsection{Executive summary of Paper~II}
\label{subsec:paper2}

In summary, in Paper~II we have realized on real data the long-standing idea of using CMB lensing-galaxy cross-correlations to help nail down the (scale-dependent) bias in clustering measurements. In doing so, we have clarified an issue, far from widely known, pertaining to the different behaviour of the bias parameter in auto-correlation and cross-correlation measurements. We demonstrated that our method improves the constraining power of galaxy clustering measurements by finding substantial improvements in our upper limits on the sum of the neutrino masses, which improved from $M_{\nu}<0.30\,{\rm eV}$ to $M_{\nu}<0.19\,{\rm eV}$. We detected scale-dependence in the auto-correlation bias at moderate significance, with sign and magnitude consistent with expectations from simulations and theory. Our results suggested that, even in the mildly non-linear regime, it is time to start worrying about higher-order corrections to the usually adopted approach of a constant galaxy bias. As a natural continuation of this work, I asked myself whether our assumption of a constant bias on large scales was justified? This question had been nagging me for a while, so I set myself to find a definitive answer, the quest towards which is described in Paper~III (to be discussed in Chapter~\ref{sec:paper3}).

\section{Scale-dependent galaxy bias induced by massive neutrinos}
\label{sec:paper3}

So far we have assumed that we could safely treat the galaxy bias as being scale-independent (\textit{i.e.} constant) on large scales (small wavenumber $k$). In the absence of massive neutrinos, this is a simple and well-known result known at least since~\cite{Mann:1997df} (see also the review~\cite{Desjacques:2016bnm}), following from simple Press-Schechter theory~\cite{Press:1973iz}. However, once massive neutrinos are introduced into the picture, the situation is no longer so simple. To see why, recall so far we have defined the bias as the factor relating the galaxy and matter overdensities:
\begin{eqnarray}
\delta_g = b_m \delta_m \,,
\label{eq:bm}
\end{eqnarray}
In Eq.~(\ref{eq:bm}) I have introduced a subscript $_m$ to reflect the fact that we are defining the bias \textit{with respect to the matter field}. Heuristically, this means we are implicitly assuming that the tracers on the left-hand side (in this case, galaxies) form from the field on the right-hand side (in this case, matter).

Is the previous assumption still true when one introduces massive neutrinos into the picture? At late times, \textit{i.e.} those relevant for the formation of galaxies, neutrinos are non-relativistic and hence contribute to the matter field.~\footnote{This is true for at least two out of three neutrinos. However even if the lightest eigenstate were massless the energy density of the two non-relativistic species would completely dominate over the energy density of the massless one.} However, the wavenumbers relevant for galaxy formation are $k \gg k_{\rm fs}$, with $k_{\rm fs}$ the neutrino free-streaming scale introduced in Chapter~\ref{subsec:signaturesnulss}. In other words, on the scales relevant for galaxy formation, neutrinos are free-streaming and cannot be kept within the potential wells from which galaxies will form.

From the above discussions, it becomes clear that the previous assumption of galaxies forming from the total matter field (where by total I mean including CDM, baryons, and non-relativistic neutrinos), implicitly entering into the definition of Eq.~(\ref{eq:bm}), is no longer valid. Instead, galaxies can only form from the CDM+baryons field, and a meaningful definition of galaxy bias should reflect this simple observation. We therefore define a different galaxy bias, $b_{cb}$ (where the subscript $_{cb}$ refers obviously to CDM+baryons), as follows:
\begin{eqnarray}
\delta_g = b_{cb} \delta_{cb} \,,
\label{eq:bcb}
\end{eqnarray}
where as usual $\delta_{cb}$ denotes the CDM+baryons overdensity field. At the level of power spectra, Eq.~(\ref{eq:bcb}) translates to:
\begin{eqnarray}
P_g(k,z) = b_{cb}^2(k,z)P_{cb}(k,z) \,,
\label{eq:pcb}
\end{eqnarray}
with $P_{cb}$ the CDM+baryons power spectrum.

In the presence of massive neutrinos, the bias $b_m$ as defined in Eq.~(\ref{eq:bm}) becomes scale-dependent \textit{even on large scales}! The reason is that on large scales ($k \ll k_{\rm fs}$) neutrino free-streaming is irrelevant and neutrinos behave as CDM, and therefore galaxies trace the total matter field (including massive neutrinos). On small scales ($k \gg k_{\rm fs}$) galaxies instead only trace the CDM+baryons field. The transition between the two regimes (non-free-streaming and free-streaming) marks a change in behaviour in the clustering of galaxies, and will be reflected in a scale-dependence of the bias. This scale-dependence will depend on the value of $M_{\nu}$ (governing the free-streaming scale), hence the bias will also depend on $M_{\nu}$.

On the other hand we can expect $b_{cb}$ to be a more ``meaningful'' definition of galaxy bias in the presence of massive neutrinos, where by ``meaningful'' I mean a definition which preserves the properties one would expect hold for galaxy bias: namely, a quantity which is scale-independent on large scales, and independent of $M_{\nu}$.~\footnote{Notice that both $b_m$ and $b_{cb}$ are anyway scale-dependent on small scales, with the leading-order correction in Fourier space being a $k^2$ correction, as already discussed in Chapter~\ref{sec:paper2}.} The above expectation has been verified by dedicated simulations carried out by Castorina \textit{et al.} in~\cite{Castorina:2013wga} (see also~\cite{Villaescusa-Navarro:2013pva,Costanzi:2013bha} for other two papers in the same series exploring cosmology with massive neutrinos through state-of-the-art simulations, and the later~\cite{Villaescusa-Navarro:2017mfx}). These simulations verified that the bias defined with respect to the CDM+baryons field as in Eq.~(\ref{eq:bcb}), $b_{cb}$ is to very good approximation scale-independent on large scales, as well as \textit{universal} (\textit{i.e.} independent of $M_{\nu}$). On the other hand, the bias defined with respect to the total matter field as in~\ref{eq:bm}, $b_m$, is scale-dependent on large scales, and the scale-dependence depends on the value of $M_{\nu}$: we refer to this effect as \textit{neutrino-induced scale-dependent bias} (NISDB).~\footnote{Although in reality neutrinos are not actually inducing any scale-dependence in the bias, but rather it is the definition of bias which needs to be revised.}

Notice that, in principle, there is nothing wrong in using the bias $b_m$ as defined in Eq.~(\ref{eq:bm}): the only practical obstacle is that it is much more difficult to model $b_m$ than it is to model $b_{cb}$! As long as one is consistent and careful in one's treatment of bias, one is free to use either of the two biases. The problem, however, is the following: most analyses of neutrino masses from galaxy clustering data have been using the bias $b_m$, while treating it as being scale-independent on large scales, \textit{i.e.} as if it were actually $b_{cb}$, in other words mixing the two. This is clearly inconsistent, and begs the question: ``\textit{Is this inconsistency in our treatment of galaxy bias in the presence of massive neutrinos a problem for current and future analyses?}'' This was a very important open question at the time I started my PhD, and a question we set ourselves to answer in Paper~III. The answer, as it turns out, is yes! In the following, I will very briefly summarize the results obtained in Paper~III. The interested reader is invited to read through Paper~III for more details.

When accounting for RSD effects and dropping all $z$-dependences, Eq.~(\ref{eq:pcb}) becomes:
\begin{eqnarray}
P_g(k,M_{\nu}) = \left ( b_{cb}(k) + f_{cb}(k,M_{\nu}) \right )^2P_{cb}(k,z) \,,
\label{eq:pcbpieno}
\end{eqnarray}
where the growth rate of the CDM+baryons perturbations $f_{cb}$ is defined as:
\begin{eqnarray}
f_{cb}(k,M_{\nu}) \equiv \frac{d\ln \left ( \sqrt{P_{cb}(k,z,M_{\nu})} \right )}{d\ln a} \,.
\label{eq:fcb}
\end{eqnarray}
The validity of Eq.~(\ref{eq:pcbpieno}) has been checked explicitly using simulations in~\cite{Villaescusa-Navarro:2017mfx}, and the appearance of the $f_{cb}$ factor implies that it is solely the CDM+baryon component which is driving RSD effects. In summary, the name of the game here is to compute the CDM+baryons power spectrum $P_{cb}$, as well as the CDM+baryons growth rate $f_{cb}$, in order to then model the tracer power spectrum as in Eq.~(\ref{eq:pcbpieno}). We modified the \texttt{CLASS} Boltzmann solver~\cite{Lesgourgues:2011re,Blas:2011rf,Lesgourgues:2011rg,Lesgourgues:2011rh} to compute both quantities: this patch was made public in \texttt{v2.7} of the code.

Our goal in the rest of Paper~III was to then check whether the heretofore inconsistent treatment of galaxy bias in cosmologies with massive neutrinos (\textit{i.e.} treating $b_m$ as if it were $b_{cb}$) will affect analyses of future galaxy clustering data, and if so to what extent. We chose to focus on future galaxy clustering data from the \textit{Euclid} satellite~\cite{Laureijs:2011gra,Amendola:2012ys,Amendola:2016saw}. \textit{Euclid} is a space telescope scheduled to launch in 2022, which will measure spectra and shapes of galaxies up to redshift 2, with the aim of unveiling the nature of cosmic acceleration through Baryon Acoustic Oscillation and weak lensing measurements. Since of course \textit{Euclid} data is not yet available, we perform an MCMC sensitivity forecast (see e.g. the seminal~\cite{Perotto:2006rj}, as well as relevant follow-up  papers such as~\cite{Audren:2012vy,Sprenger:2018tdb,Brinckmann:2018owf}), proceeding through the following five basic steps:
\begin{enumerate}
\item Choose a fiducial model. The fiducial values of the cosmological parameters are given by Tab.~I in Paper~III.
\item Generate mock power spectrum data consistent with \textit{Euclid}'s sensitivity.~\footnote{The reader is invited to consult Paper~III for a full discussion of our modelling of the galaxy spectrum, accounting for effects such as redshift-space distortions, Fingers of God, limited instrumental resolution, Alcock-Paczy\'{n}ski effect, uncertainties in the bias model, and other approximations.}
\item Analyse the mock data with standard MCMC techniques \textit{with the NISDB properly taken into account}, \textit{i.e.} with the galaxy power spectrum modelled as in Eq.~(\ref{eq:pcbpieno}) (notice that this analysis presumably should recover the input fiducial parameters).
\item Analyse the mock data with standard MCMC techniques \textit{with the NISDB not taken into account}. In other words, we model the galaxy power as in Eq.~(\ref{eq:pcbpieno}), but with $b_m$ and $f_m$ in place of $b_{cb}$ and $f_{cb}$ [with $f_m$ defined analogously to $f_{cb}$ in Eq.~(\ref{eq:fcb})].
\item Compare the cosmological parameters extracted for the two cases.
\end{enumerate}
Notice from Tab.~I of Paper~III that we pessimistically $M_{\nu}=0.06\,{\rm eV}$, \textit{i.e.} the minimal value allowed within the normal ordering. Our motivation is twofold: first, this value would be the hardest to detect. Second, as argued in~\cite{Castorina:2013wga}, the size of the NISDB effect is $\approx f_{\nu} \propto M_{\nu}$ (where by ``size'' I mean the difference between $P_{cb}$ and $P_m$ on the scales under consideration). Therefore, if we find that the NISDB effect is important for the minimal allowed value of $M_{\nu}$, the same conclusion will hold to an even great extent for any other value of $M_{\nu}$!
\begin{figure*}[!t]
\centering
\begin{tabular}{cc}
\includegraphics[width=0.5\textwidth]{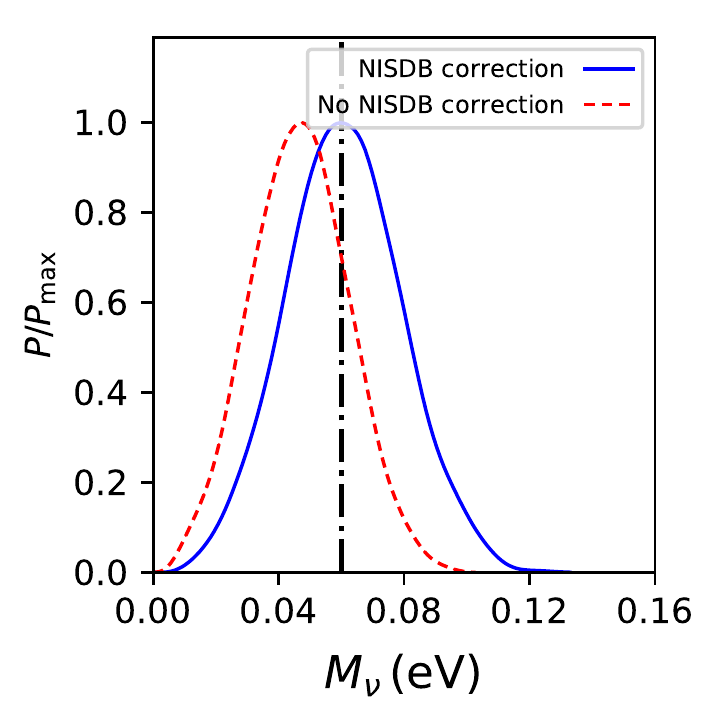}&\includegraphics[width=0.5\textwidth]{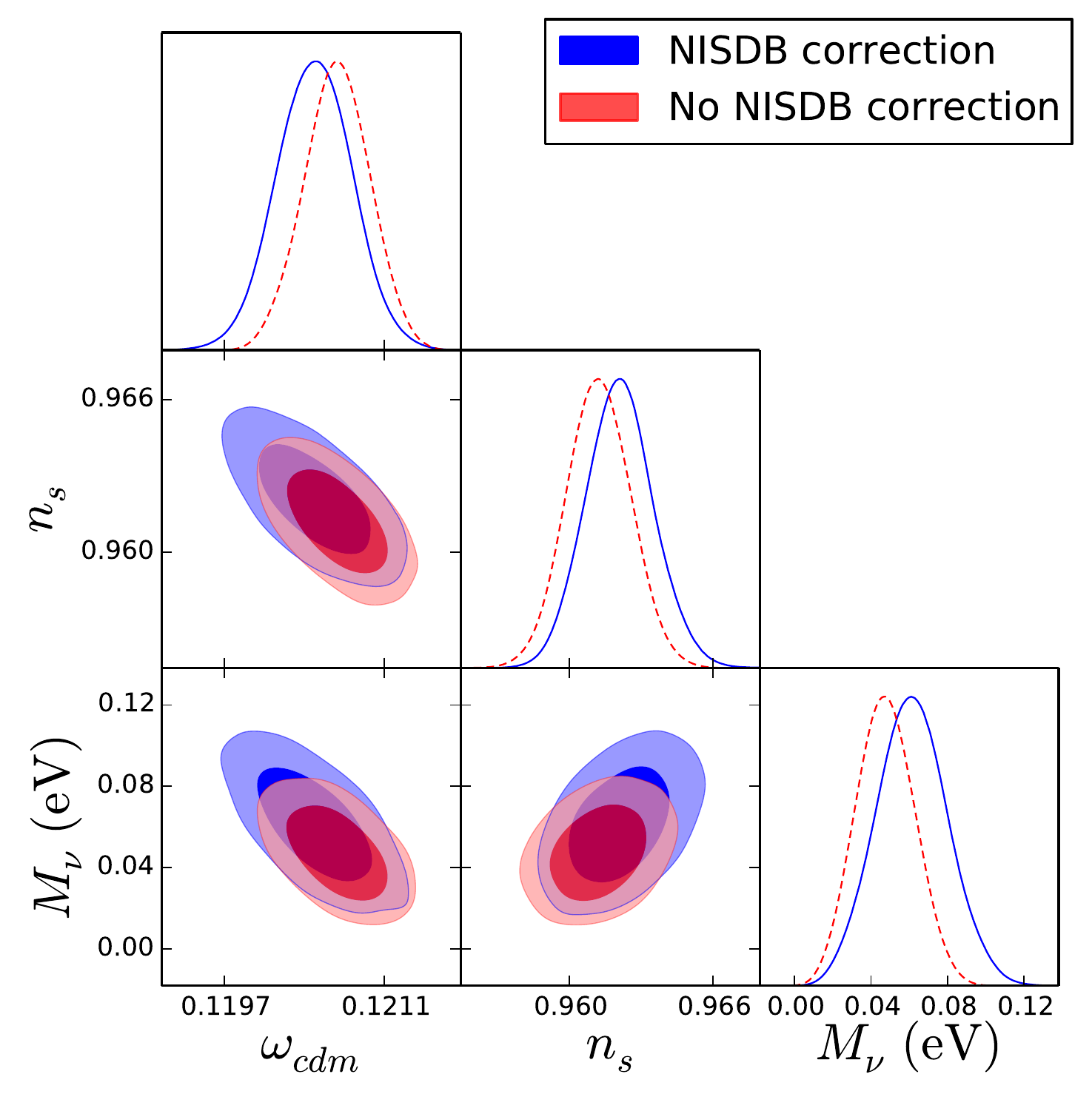}\\
\end{tabular}
\caption{The impact of not correctly accounting for the NISDB effect when analyzing mock galaxy clustering data from \textit{Euclid}. \textit{Left panel}: one-dimensional posterior distributions for $M_{\nu}$ normalized to their maximum values, when the NISDB effect is correctly accounted for (blue solid), or not accounted for (red dashed). The dot-dashed vertical line denotes the input fiducial value $M_{\nu}=0.06\,{\rm eV}$. \textit{Right panel}: triangular plot showing joint and one-dimensional marginalized posterior distributions for $M_{\nu}$, $\omega_{\rm cdm} \equiv \omega_c$, and $n_s$, when the NISDB is correctly accounted for (blue contours/solid curves) and when it is not accounted for (red contours/dashed curves). Reproduced from~\cite{Vagnozzi:2018pwo} (Paper~III) with permission from IoP.}
\label{fig:paper3fig1}
\end{figure*}

Our result is conveniently summarized in Fig.~\ref{fig:paper3fig1}, where we plot (left panel) the posterior distributions of $M_{\nu}$ we obtain when correctly accounting for the NISDB effect (blue curve) and when we fail to do so (red curve). The vertical dot-dashed line denotes the input fiducial value $M_{\nu}=0.06\,{\rm eV}$, which is perfectly recovered when the NISDB effect is correctly accounted for [$M_{\nu} = (0.061 \pm 0.019)\,{\rm eV}$]. On the other hand, when not accounting for the NISDB effect, our determination of $M_{\nu}$ is biased: we find $M_{\nu} = (0.046 \pm 0.015)\,{\rm eV}$, a shift of about $0.6\sigma$ from the ``true'' fiducial value. We also get a spurious increase in sensitivity, since the error bar obtained when not accounting for the NISDB effect is about $25\%$ smaller than the one obtained when correctly accounting for the effect. The magnitude of such shifts are consistent with theoretical expectations, as we explain in Paper~III. Notice of course that the shifts in the recovered value of $M_{\nu}$ would be proportionally larger if the fiducial value of $M_{\nu}$ were larger. For instance, for $M_{\nu}=0.18\,{\rm eV}$ (still marginally allowed by cosmological limits) such shifts would be three times as large (\textit{i.e.} almost $2\sigma$). Moreover, shifts in $M_{\nu}$ would naturally propagate to other parameters correlated with $M_{\nu}$. As an example, the triangular plot in the right panel of Fig.~\ref{fig:paper3fig1} shows the induced shifts in $\omega_c$ and $n_s$, two of the parameters most strongly correlated with $M_{\nu}$. As we see from the figure, the shifts in $\omega_c$ and $n_s$ are comparable in size to those in $M_{\nu}$. Our result confirms and supports earlier findings of~\cite{Raccanelli:2017kht} (see also the later work~\cite{Valcin:2019fxe}), who performed a similar analysis but using a Fisher matrix formalism.

There is one final caveat I want to briefly discuss, which we did not consider in~\cite{Vagnozzi:2018pwo}. Namely, we have assumed that $b_{cb}$ is a constant on large scales. In reality, it is known that a small residual scale-dependence, due to the effect of massive neutrinos on the process of halo collapse, should be imprinted in $b_{cb}$ on large scales~\cite{Parfrey:2010uy,LoVerde:2014pxa,LoVerde:2016ahu,Munoz:2018ajr,
Chiang:2018laa,Fidler:2018dcy}. This effect was recently seen in simulations for the first time in~\cite{Chiang:2018laa}, and is smaller than the NISDB effect we have studied in~\cite{Vagnozzi:2018pwo}. The question of whether this residual scale-dependence is important for parameter estimation is still unclear~\cite{Munoz:2018ajr}, and detailed studies (mirroring what we have done in~\cite{Vagnozzi:2018pwo}) are underway.

\subsection{Executive summary of Paper~III}
\label{subsec:paper3}

In conclusion, in Paper~III we found that an incorrect treatment of galaxy bias in the presence of massive neutrinos leads to ${\cal O}(\sigma)$ shifts in the determined cosmological parameters: this affects both $M_{\nu}$ as well as other parameters correlated with $M_{\nu}$ (for instance $n_s$ or $\omega_c$). In the era of precision sub-percent cosmology, systematic shifts of such magnitude are clearly unacceptable. We therefore encourage the cosmology community to carefully take the neutrino-induced scale-dependent bias effect into account, especially when analysing future galaxy clustering data.

\section{Massive neutrinos meet (non-phantom) dark energy}
\label{sec:paper4}

The greatest weakness of cosmological limits on neutrino masses is their (in)stability against a larger parameter space: typically, limits degrade considerably when relaxing assumptions on the underlying cosmological model and allowing for an extended parameter space. An example is discussed in Sec.~IVC of Paper~I (not discussed in Chapter~\ref{sec:paper1}), where we treated the dark energy equation of state as a free parameter (this quantity is fixed to $w=-1$ in $\Lambda$CDM): for a particular dataset combination, this broadened our upper limit from $M_{\nu}<0.19\,{\rm eV}$ to $M_{\nu}<0.31\,{\rm eV}$.~\footnote{For an incomplete list of other recent works examining neutrino mass constraints in extended cosmological models, see e.g.~\cite{Hannestad:2005gj,Joudaki:2012fx,Archidiacono:2013lva,Zhang:2015uhk,
Wang:2016tsz,Zhao:2016ecj,Kumar:2016zpg,Xu:2016ddc,Guo:2017hea,Zhang:2017rbg,
Li:2017iur,Yang:2017amu,Lorenz:2017fgo,Sutherland:2018ghu,Guo:2018gyo,Choudhury:2018byy,
Choudhury:2018adz,Zhao:2018fjj,Hagstotz:2019gsv,Feng:2019mym}.} The reason is that marginalizing over additional parameters strongly correlated with $M_{\nu}$ (for instance $w$) significantly broadens the $M_{\nu}$ distribution.~\footnote{The reason there is a strong correlation between $M_{\nu}$ and $w$ is that one can vary one parameter and then adjust the other to keep the observables fixed. In this case, one can increase $M_{\nu}$ and correspondingly decrease $w$ to keep the angular size of the first peak of the CMB $\theta_s$ roughly fixed: thus, we expect there to be an inverse correlation between $M_{\nu}$ and $w$ (see Fig.~4 of Paper~I).} This observation, however, begs the question: ``\textit{Will moving to an extended parameter space always broaden the $M_{\nu}$ distribution?}'' The answer, as we found in Paper~IV, is no! In the remainder of this Chapter, I will briefly summarize the results of Paper~IV, providing an explanation for this unexpected result.

In Paper~IV we relaxed the assumption, implicit in $\Lambda$CDM, wherein DE consists of a cosmological constant with constant EoS $w=-1$. Instead we allowed for a \textit{dynamical dark energy} (DDE) component with EoS varying with redshift, $w(z)$. Several parametrizations of the EoS of DDE components exist in the literature, some more phenomenological in nature and others more closely rooted to specific models. Aiming for a rather model-independent approach, we considered a simple two-parameter description of a time-varying EoS which usually goes under the name of Chevallier-Polarski-Linder (CPL) parametrization, where the evolution of the EoS with redshift is given by the following~\cite{Chevallier:2000qy,Linder:2002et}:
\begin{eqnarray}
w(z) = w_0+w_a\frac{z}{1+z}\,.
\label{eq:cpl}
\end{eqnarray}
Rewriting Eq.~(\ref{eq:cpl}) in terms of scale factor rather than redshift, we arrive at the expression:
\begin{eqnarray}
w(a) = w_0+w_a(1-a)\,,
\label{eq:cpla}
\end{eqnarray}
which one immediately recognizes as a Taylor expansion of the DE EoS as a function of the scale factor $a = (1+z)^{-1}$ around the present time ($a_0=1$), truncated to first order. Physically speaking, $w_0$ corresponds to the EoS today, whereas $w_a$ corresponds to the derivative of the EoS with respect to the scale factor, up to a minus sign. The energy density of a dark energy component whose EoS is of the CPL form, $\rho_{\rm DDE}(z)$, is given by:
\begin{eqnarray}
\rho_{\rm DDE}(z) = \rho_{\rm DE\,,0}(1+z)^{3(1+w_0+w_a)}\exp \left ( -3w_a\frac{z}{1+z} \right )\,,
\label{eq:rhodde}
\end{eqnarray}
where $\rho_{\rm DE\,,0}$ is the DDE energy density at the present time.

The CPL parametrization is probably the most widely used DDE parametrization, for several reasons (e.g. discussed in~\cite{Linder:2002et}): besides being highly manageable due to its 2-dimensional nature, this parametrization is bounded at high redshift (unlike the previously used linear-in-redshift parametrization), and has a simple physical interpretation. Most importantly, it has a direct connection to several physical dark energy models, notably quintessence dark energy. First proposed by Ratra and Peebles in 1988~\cite{Peebles:1987ek} (see e.g.~\cite{Caldwell:1997ii,Carroll:1998zi,Zlatev:1998tr,
Steinhardt:1999nw,Amendola:1999er} for other seminal papers), in its simplest incarnation quintessence consists of a class of dark energy models wherein the role of dark energy is played by a rolling scalar field, $\phi$. It has been shown that Eq.~(\ref{eq:cpl}) is accurate to sub-percent level in recovering observables for quintessence models~\cite{Linder:2002et,Linder:2002wx,
Linder:2006sv,Linder:2007wa,Linder:2008pp}.~\footnote{By ``observables'', I mean quantities to which the main cosmological observations (CMB, BAO, Supernovae, weak lensing) are sensitive, such as Hubble parameter and/or distance measurements. Notice that the EoS itself is not directly observable, thus there is fundamentally no strong case for obtaining parametrizations which provide sub-percent accuracy in the EoS. Notice also that in the whole literature there exist only two simple physical parametrizations of the EoS of scalar field DE, in the sense of being tested against exact solutions of the Klein-Gordon equation. Besides the CPL parametrization, the other physical parametrization is the 4-parameter Copeland-Corasaniti-Linder-Huterer parametrization~\cite{Corasaniti:2002vg,Linder:2005ne}.}

Even when adopting a parametrized framework, such as in Eq.~(\ref{eq:cpl}), it is always of paramount importance to make contact with known and physically viable theories. In Paper~IV, our initial goal in using Eq.~(\ref{eq:cpl}) was to make contact with a model we can refer to as \textit{standard quintessence}, where DE consists of a single, minimally coupled scalar field, with a canonical kinetic term. In other words, we are considering the following simple Lagrangian for the quintessence field $\phi$:
\begin{eqnarray}
{\cal L} = \frac{1}{2}\partial_{\mu}\phi\partial^{\mu}\phi - V(\phi)\,.
\label{eq:lagrangian}
\end{eqnarray}
It can be shown that the EoS of a standard quintessence field will in general be time-dependent, but will always satisfy $w(z) \geq -1$~\footnote{To show this, it is sufficient to compute the stress-energy tensor of the quintessence field from Eq.~(\ref{eq:lagrangian}). From that one can read off the pressure and energy density of the scalar field, the ratio of which gives the EoS. The fact that $w(z) \geq -1$ follows if one neglects spatial derivatives in the EoS. This is justified since late-time acceleration requires a very light scalar field, whose Compton wavelength will typically be larger than the Hubble scale. Therefore, the quintessence field will typically be smooth within the Hubble scale.}. The value $w=-1$ is usually referred to as \textit{phantom divide} (and, correspondingly, a DE component with EoS $w<-1$ is referred to as phantom DE), and cannot be crossed by standard quintessence models.~\footnote{Crossing the phantom divide requires either a wrong-sign kinetic term (e.g.~\cite{Carroll:2003st,Vikman:2004dc,Hu:2004kh,Caldwell:2005ai,Creminelli:2008wc}), using multiple fields (e.g.~\cite{Hu:2004kh,Guo:2004fq,Saridakis:2010mf}), non-minimally coupling the scalar field to gravity (e.g.~\cite{Carroll:2004hc}), including higher derivative operators (e.g.~\cite{Creminelli:2008wc}), or mixing the metric and scalar kinetic terms through kinetic braiding (e.g.~\cite{Deffayet:2010qz,Easson:2016klq}). Notice also that several modified gravity models can feature an effective phantom behaviour (see e.g.~\cite{Gannouji:2006jm,Nesseris:2006er}).} Standard quintessence is arguably one of the simplest models of DDE: hereafter, whenever we refer to ``quintessence'', it will be understood that we are referring to standard quintessence. In phantom dark energy models, the dark energy density keeps growing with time and correspondingly the acceleration of the Universe's expansion increases: in most phantom dark energy models, this results in the end of the Universe through the dissociation of any bound structure (including atoms), a rather tragic prospect known as ``Big Rip''~\cite{Caldwell:2003vq}.~\footnote{However, the Big Rip does not occur if $w \to -1$ asymptotically in the future, which occurs frequently in certain modified gravity models, see \textit{e.g.}~\cite{BouhmadiLopez:2004me,Wei:2005fq,Zhang:2009xj,Frampton:2011sp,
vonStrauss:2011mq,Astashenok:2012tv,Saridakis:2012jy,Myrzakulov:2013mja,Akrami:2015qga,
Odintsov:2015zza,Oikonomou:2015qha,Schmidt-May:2015vnx,Odintsov:2015ynk,Mortsell:2017fog,Oikonomou:2018qsc}.}

Upon choosing to parametrize the EoS of quintessence through the CPL parametrization in Eq.~(\ref{eq:cpl}), it is important to keep the \textit{non-phantom} nature of quintessence in mind. That is, we better make sure that $w(z) \geq -1$ for all $z$: this can easily be satisfied by imposing the following two conditions:
\begin{eqnarray}
w_0 \geq -1 \,, \quad w_0+w_a \geq -1\,.
\label{eq:npdde}
\end{eqnarray}
Eq.~(\ref{eq:npdde}) forces the DE component to be non-phantom both today ($w_0 \geq -1$) as well as in the far past ($z \to \infty$, $w_0+w_a \geq -1$). The monotonic nature of the CPL parametrization will then ensure that the DE component remains non-phantom throughout the expansion history. We refer to the model parametrized by the combination of CPL equation of state [Eq.~(\ref{eq:cpl})], restricted by the conditions in Eq.~(\ref{eq:npdde}), as \textit{non-phantom dynamical dark energy} (NPDDE in short). On the other hand, we refer to the model parametrized by the CPL equation of state without further restrictions on the values of $w_0$ and $w_a$ as $w_0w_a$CDM. Notice that, during the period of DE domination, the energy density of a NPDDE component [Eq.~(\ref{eq:rhodde})] is always greater than that of a cosmological constant with the same $\rho_{\rm DE\,,0}$.

In Paper~IV, we compared the upper limits on $M_{\nu}$ obtained assuming the standard $\Lambda$CDM scenario, against those obtained assuming the NPDDE model (which contains two extra parameters). For completeness, we also considered how these upper limits change when assuming the $w_0w_a$CDM model. We considered two different combinations of datasets. The first combination, which we refer to as \textit{base}, contains measurements of the CMB temperature anisotropies from the \textit{Planck} 2015 data release, a Gaussian prior on the optical depth to reionization $\tau = 0.055 \pm 0.009$ (intended to mimic large-scale polarization measurements from the \textit{Planck} 2019 data release), SNeIa distance measurements from the JLA catalogue, and finally BAO distance measurements from the BOSS DR11 CMASS and LOWZ samples~\cite{Anderson:2013zyy}, the SDSS DR7 MGS~\cite{Ross:2014qpa}, and the 6dFGS survey~\cite{Beutler:2011hx}. The second combination, which we refer to as \textit{pol}, contains small-scale polarization and temperature-polarization cross-correlation spectra from the \textit{Planck} 2015 data release in addition to the aforementioned datasets.

\begin{figure*}[!h]
\centering
\begin{tabular}{cc}
\includegraphics[width=0.5\textwidth]{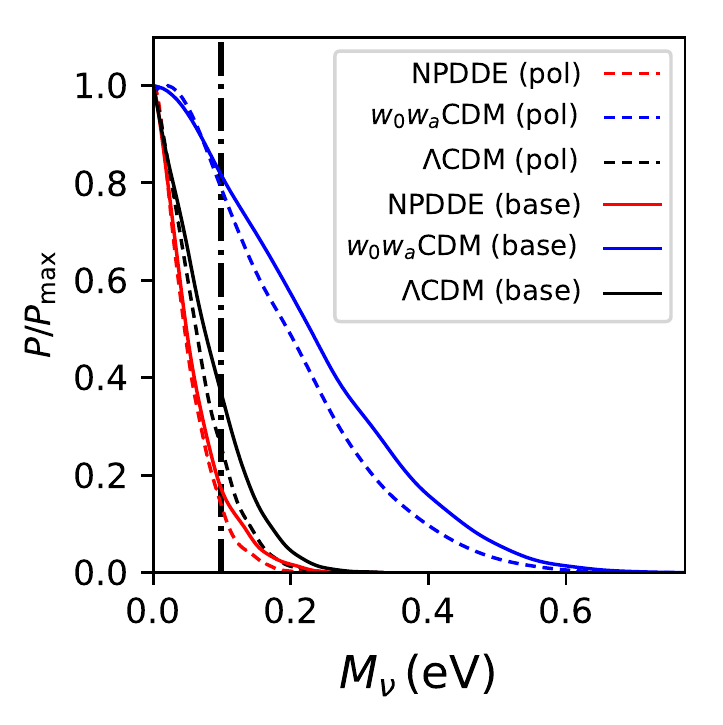}&\includegraphics[width=0.5\textwidth]{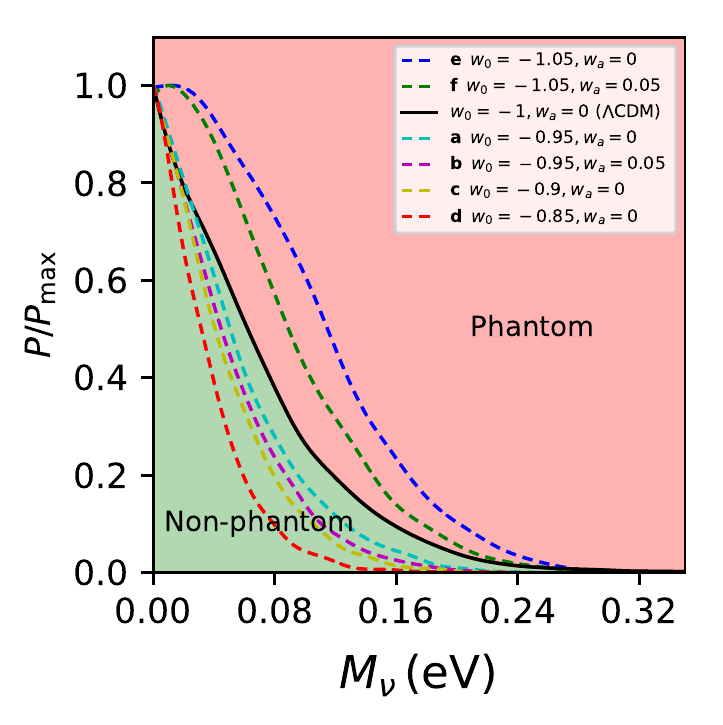}\\
\end{tabular}
\caption{\textit{Left panel}: one-dimensional posterior distributions for $M_{\nu}$ normalized to their maximum values, assuming $\Lambda$CDM (black), the $w_0w_a$CDM model (blue), and the NPDDE model (red), and using the \textit{base} (solid) or \textit{pol} (dashed) dataset. The dot-dashed vertical line denotes $M_{\nu}=0.1\,{\rm eV}$, the minimum value of the sum of the neutrino masses allowed for the inverted ordering. \textit{Right panel}: one-dimensional posterior distributions for $M_{\nu}$ for a selection of cosmological models where $w_0$ and $w_a$ are \textit{fixed}. The $\Lambda$CDM posterior is given by the solid black curve. The posteriors to the left/right of the $\Lambda$CDM posterior, lying in the ``non-phantom''/``phantom'' region, are obtained fixing $w_0$ and $w_a$ fixed to values satisfying/not satisfying the NPDDE condition [Eq.~(\ref{eq:npdde})]. Reproduced from~\cite{Vagnozzi:2018jhn} (Paper~IV) with permission from APS.}
\label{fig:paper4fig1}
\end{figure*}

For the $\Lambda$CDM case we find the 95\%~C.L. upper limit $M_{\nu}<0.16\,{\rm eV}$ for the \textit{base} dataset combination. When considering the $w_0w_a$CDM model, unsurprisingly we found that the upper limit degrades significantly to $M_{\nu}<0.41\,{\rm eV}$. When considering the NPDDE model, we found that the upper limit tightened by about 20\% to $M_{\nu}<0.13\,{\rm eV}$. This is very surprising especially considering that $\Lambda$CDM is a special case of the NPDDE model, given that it is recovered when we set $w_0=-1$ and $w_a=0$. We find similar values for the \textit{pol} dataset combination, namely $M_{\nu}<0.13\,{\rm eV}$ ($\Lambda$CDM), $M_{\nu}<0.37\,{\rm eV}$ ($w_0w_a$CDM), and $M_{\nu}<0.11\,{\rm eV}$ (NPDDE). The posterior distributions for $M_{\nu}$ obtained in the six cases just discussed are plotted in the left panel of Fig.~\ref{fig:paper4fig1}.

The explanation for this result relies on the observation that, during DE domination, the energy density of a NPDDE component [Eq.~(\ref{eq:rhodde}) with $w_0$ and $w_a$ satisfying Eq.~(\ref{eq:npdde})] is always greater than the energy density of a cosmological constant with the same $\rho_{\rm DE\,,0}$. Let us consider the normalized expansion rate $E(z)$, defined as follows:
\begin{eqnarray}
E(z) \equiv \frac{H(z)}{H_0} \approx \sqrt{(\Omega_c+\Omega_b)(1+z)^3+\Omega_{\rm DDE}(z)+\Omega_{\nu}(z)}\,,
\label{eq:ez}
\end{eqnarray}
where in the last approximation we have neglected the radiation energy density, which is negligible during DE domination. During the same period, $\Omega_{\nu}(z)$ is proportional to $M_{\nu}$, given that at least two out of three neutrino species are non-relativistic. Keeping $\Omega_c$, $\Omega_b$, $M_{\nu}$, and $\rho_{\rm DE\,,0}$ fixed, it is clear that the late-time normalized expansion rate is higher in a NPDDE model than in $\Lambda$CDM. As we already saw in Chapter~\ref{subsec:cmb}, CMB data accurately constrains $\theta_s$, the ratio between the comoving sound horizon at decoupling $r_s$ and the comoving distance to the CMB $\chi_{\star}$. Late-time physics cannot change $r_s$ (which is fixed by pre-recombination physics), so whatever change in the dark energy sector better keep $\chi_{\star}$ (approximately) fixed in order not to change $\theta_s$. Up to proportionality factors, $\chi_{\star}$ can be written as [see Eq.~(\ref{eq:chiz})]:
\begin{eqnarray}
\chi_{\star} \propto \frac{1}{H_0}\int_0^{z_{\rm dec}} \frac{dz}{E(z)}\,,
\label{eq:da}
\end{eqnarray}
where $E(z)$ is the normalized expansion rate we saw in Eq.~(\ref{eq:ez}). Combining Eq.~(\ref{eq:da}) and Eq.~(\ref{eq:ez}), it is clear that to keep $\chi_{\star}$ fixed when introducing a NPDDE model in place of the cosmological constant, both $H_0$ and $M_{\nu}$ need to decrease (decreasing $\Omega_c$ and $\Omega_b$ is not a valid option as it would change the redshift of matter-radiation equality, which is also strongly constrained by the CMB). Indeed, this is precisely what we find: $M_{\nu}$ decreases (more precisely, the upper limits on $M_{\nu}$ become tighter), but so does $H_0$ (see Fig.~2 in Paper~IV).

In Paper~IV, we have provided a more intuitive explanation for the fact that the limits on $M_{\nu}$ are tighter for the NPDDE model compared to $\Lambda$CDM, building upon the (Bayesian) statistical method adopted, and the role of the marginalization process. If we imagine \textit{fixing} (instead of varying) $w_0$ and $w_a$ to values satisfying Eq.~(\ref{eq:npdde}), the resulting limits on $M_{\nu}$ are always tighter than the $\Lambda$CDM limit (obtained with $w_0=-1$, $w_a=0$). This is clearly shown in the right panel of Fig.~\ref{fig:paper4fig1} (see the four example curves lying in the region labelled ``non-phantom'' to the left of the solid black curve, the latter representing the posterior obtained assuming $\Lambda$CDM). In reality, however, we vary $w_0$ and $w_a$ and then marginalize over them. Heuristically, marginalizing over $w_0$ and $w_a$ for the NPDDE model results in a $M_{\nu}$ posterior which is a weighted average of the posteriors lying in the ``non-phantom'' region in the right panel of Fig.~\ref{fig:paper4fig1}: since all of these posteriors result in limits tighter than the $\Lambda$CDM limit, the same is going to be true for their weighted average.

I want to conclude this Chapter arguing that our findings can be very interesting in the event of a non-cosmological measurement of the neutrino mass ordering. From earlier discussions in Chapter~\ref{sec:paper1} and Paper~I, it is clear that NPDDE models prefer the normal ordering more strongly than $\Lambda$CDM does: in other words NPDDE models such as quintessence more strongly prefer lighter neutrinos, which cannot be reconciled with the inverted ordering. An extensive program of long-baseline oscillation experiments (such as T2K~\cite{Itow:2001ee,Abe:2011ks}, NO$\nu$A~\cite{Ayres:2004js,Ayres:2007tu,Patterson:2012zs}, and DUNE~\cite{Adams:2013qkq,Acciarri:2015uup,Acciarri:2016ooe,
Acciarri:2016crz,Strait:2016mof}), completely independent from cosmology, are aiming to determine the neutrino mass ordering within the next 5-10 years. If these experiments were to determine that the neutrino mass ordering is inverted (recall that to zeroth order cosmology can instead only determine the mass ordering if it is normal!), non-phantom DDE models would be under strong pressure. In other words DE, if dynamical, would likely have to have crossed the phantom divide at some point. Of course, this conclusion excludes non-standard exotic physics in the neutrino sector, such as models with a vanishing neutrino energy density (due perhaps to annihilation into light bosons at late times, e.g.~\cite{Beacom:2004yd}), mass-varying neutrinos (e.g.~\cite{Fardon:2003eh,Cirelli:2005sg,Horvat:2005ua,Barger:2005mh,
Brookfield:2005bz,Franca:2009xp,Wetterich:2013jsa,Geng:2015haa,Lorenz:2018fzb}), non-standard neutrino interactions~\cite{Escudero:2015yka,Stadler:2018dsa,Kreisch:2019yzn,
Stadler:2019dii,Park:2019ibn}, and so on. In our view, the findings of Paper~IV constituted a rather interesting result, providing unexpected connections between two fields one would normally not relate: neutrino oscillation experiments and the nature of dark energy.

\subsection{Executive summary of Paper~IV}
\label{subsec:paper4}

To conclude, in Paper~IV, we found that it is not always true that the upper limits on $M_{\nu}$ degrade when moving to an extended parameter space. We demonstrated this explicitly by considering a non-phantom dark energy model [NPDDE; $w(z) \geq -1$] containing two extra parameters with respect to $\Lambda$CDM, but which recovers $\Lambda$CDM for a particular choice of these two parameters. We showed that the upper limits on $M_{\nu}$, in fact, become tighter than in the $\Lambda$CDM case. This implies that the preference for the normal ordering is even stronger in NPDDE models: on the other hand, should near-future long-baseline neutrino oscillation experiments determine that the neutrino ordering is inverted, the viability of such models (which include quintessence) would be put in jeopardy.

\section{Massive neutrinos meet inflation}
\label{sec:paper5}

As briefly discussed at the start of Chapter~\ref{sec:paper4}, the main weakness of cosmological limits on neutrino masses is their (in)stability against a larger parameter space, particularly when the extended parameters are strongly correlated/degenerate with $M_{\nu}$. So far, we focused on the effect other parameters have on the limits on $M_{\nu}$ (e.g. Paper~IV). Of course, this problem can be in some sense reversed: if a particular parameter $X$ is strongly correlated with $M_{\nu}$, the values inferred for $X$ might be sensitive to the assumptions I make when introducing $M_{\nu}$ into the picture, or to the very fact that I introduced $M_{\nu}$ in first place. Recall, in fact, that in the baseline $\Lambda$CDM model, $M_{\nu}$ is not a free parameter, but is fixed to $M_{\nu}=0.06\,{\rm eV}$, the minimum value allowed by oscillations data. Of the six base $\Lambda$CDM parameters, one whose correlation with $M_{\nu}$ is particularly strong is the scalar spectral index $n_s$ (e.g.~\cite{Takada:2005si,Carbone:2010ik,Oyama:2015gma,Canac:2016smv,Archidiacono:2016lnv}). The scalar spectral index plays a particularly important role when observationally discriminating between competing inflationary models~\cite{Dodelson:1997hr}. Current cosmological data can already differentiate between inflationary models, and has ruled some out (see e.g.~\cite{Martin:2013tda,Planck:2013jfk,
Martin:2013nzq,Ade:2015lrj,Escudero:2015wba,Huang:2015cke}). Given the correlation between $n_s$ and $M_{\nu}$, we asked ourselves the following question: ``\textit{are our conclusions about inflationary models strongly affected by our assumptions about unknowns in the neutrino sector?}'' The answer, fortunately, turns out to be no. In the following, I will briefly discuss our investigation of this question, which is reported in Paper~V~\cite{Gerbino:2016sgw}.

It is worth reminding the reader of the three main assumptions/approximations usually made with regards to the neutrino sector when analysing cosmological data. In the baseline $\Lambda$CDM model, $M_{\nu}$ is not a free parameter, and is fixed to the minimum value allowed by oscillations data if the normal ordering (\texttt{NO}) is realized, $M_{\nu}=0.06\,{\rm eV}$. When fixed to this value, usually one follows the \textit{1mass} approximation: here, the neutrino mass spectrum is approximated as consisting of two massless and one massive neutrino (\textbf{approximation} $\mathbf{\# 1}$): clearly this is an approximation because even in the \texttt{NO} minimal mass case, in reality one has two massive eigenstates beyond the lightest massless one. When $M_{\nu}$ is not fixed but varying, the \textit{3deg} approximation is adopted, where neutrino mass spectrum consists of three degenerate massive neutrinos each carrying mass $m_i=M_{\nu}/3$ (\textbf{approximation} $\mathbf{\# 2}$): this is also an approximation, since it neglects the mass splittings between the three eigenstates. Finally, in the standard $\Lambda$CDM model, the effective number of relativistic species at recombination (also referred to as effective number of neutrinos) $N_{\rm eff}$ is not varied but fixed to its standard value of $N_{\rm eff}=3.046$~\cite{Mangano:2005cc} (recently re-evaluated to be $N_{\rm eff}=3.045$~\cite{deSalas:2016ztq}) (\textbf{approximation} $\mathbf{\# 3}$): this is an approximation in a ``broader'' sense, but still one worth checking. Broadly speaking these three assumptions, related to the neutrino mass, mass ordering, and effective number, are those we decided to check in Paper~V.

We considered CMB temperature and large-scale polarization data from the \textit{Planck} 2015 data release (referred to as \textit{PlanckTT}+\textit{lowP}), BAO distance measurements from the 6dFGS~\cite{Beutler:2011hx}, SDSS-MGS~\cite{Ross:2014qpa}, and BOSS DR11 surveys~\cite{Anderson:2013zyy} (referred to as \textit{BAO}); when also varying the tensor-to-scalar ratio $r$ (results only briefly discussed here, refer to Paper~V for more details), we also include degree-scale measurements of the B-mode power spectrum from the BICEP/Keck collaboration~\cite{Array:2015xqh} (referred to as \textit{BK14}).

We first investigated the impact on the estimation of $n_s$ of our assumptions on the neutrino mass ordering and total mass $M_{\nu}$. Considering only the \textit{PlanckTT}+\textit{lowP} dataset, for the baseline $\Lambda$CDM model where $M_{\nu}$ is fixed to $0.06\,{\rm eV}$ and the neutrino mass spectrum is treated following the \textit{1mass} approximation, we find $\underline{\boldsymbol{n_s=0.9656 \pm 0.0063}}$. When instead still fixing $M_{\nu}=0.06\,{\rm eV}$ but modelling the mass spectrum following the exact \texttt{NO} with mass-squared splittings given by oscillations global fits~\cite{GonzalezGarcia:2012sz,Gonzalez-Garcia:2014bfa,Gonzalez-Garcia:2015qrr,Esteban:2016qun,deSalas:2017kay,deSalas:2018bym}, we find $\underline{\boldsymbol{n_s=0.9655 \pm 0.0063}}$. The shift in moving from the \textit{1mass} approximation to the exact \texttt{NO} is negligible and consistent with statistical fluctuations from the MCMC algorithm. We then move on to test assumptions on the neutrino mass (more specifically, assumptions on the cosmological model adopted, $\Lambda$CDM+$M_{\nu}$ vs $\Lambda$CDM), and allow $M_{\nu}$ to vary while adopting the \textit{3deg} approximation. We find a larger shift this time, with $\underline{\boldsymbol{n_s=0.9636 \pm 0.0071}}$. The error bar broadening is consistent with the expectation from having introduced an additional parameter to marginalize over. Within the $\Lambda$CDM+$M_{\nu}$ model, we re-test the assumptions on the mass splittings by abandoning the \textit{3deg} approximation in favour of an exact \texttt{NO} modelling, and find a modest shift to $\underline{\boldsymbol{n_s=0.9629 \pm 0.0069}}$. The shift in $n_s$ when moving from the \textit{3deg} approximation to the exact \texttt{NO} for the $\Lambda$CDM+$M_{\nu}$ model, while small, is about 7 times larger than the shift obtained when moving from the \textit{1mass} approximation to the exact \texttt{NO} for the $\Lambda$CDM model ($M_{\nu}$ fixed), na\"{i}vely suggesting that under the $\Lambda$CDM+$M_{\nu}$ model cosmological data might be more sensitive to the exact mass splittings than it is under the $\Lambda$CDM model. Later I will argue that this is not the case: on the contrary, the shift is entirely a consequence of volume effects and can be removed by a suitable choice of prior on $M_{\nu}$ in the \textit{3deg} case.

Overall, we noticed a (small but non-negligible) shift of $n_s$ to smaller values when marginalizing over $M_{\nu}$ and using only CMB data, compared to the case when $M_{\nu}$ is fixed to $0.06\,{\rm eV}$. As far as CMB data is concerned, there exists a rather strong degeneracy between $M_{\nu}$ and $n_s$ due to their competing effects both on the damping tail ($\ell \gtrsim 500$) and on larger scales ($\ell \lesssim 500$). Increasing $M_{\nu}$ suppresses structure formation and hence suppresses the lensing potential, which reduces the smearing effect of lensing on small scales (adding power to the damping tail). This can be compensated by having a redder primordial power spectrum and hence decreasing $n_s$, in other words tilting the spectrum to give less power to small scales. Moreover, as we have seen in~\ref{subsec:signaturesnucmb}, increasing $M_{\nu}$ while decreasing $\Omega_{\Lambda}$ to keep $\theta_s$ fixed reduces the amplitude of the early and late ISW effects, resulting in an overall depletion of power for $\ell \lesssim 500$ (see e.g. Fig.~6 of~\cite{Hou:2012xq} and Fig.~\ref{fig:neutrinoscmbtheta} in this Thesis). This effect too can be compensated by decreasing $n_s$ to give more power to large scales. In summary, when using CMB data alone we expect a rather strong inverse correlation between $M_{\nu}$ and $n_s$: this is clearly visible from the red contours in Fig.~3 in Paper~V. When we marginalize over $M_{\nu}$ instead of keeping it fixed, we open up the $M_{\nu}$-$n_s$ degeneracy, which results overall in a lower value for $n_s$ (as well as a slightly larger error bar).

Coming back to the shift when moving from the \textit{3deg} approximation to the exact \texttt{NO} for the $\Lambda$CDM+$M_{\nu}$ model, the key point to note is that when adopting the \textit{3deg} approximation we are allowing values of $M_{\nu}$ as low as $0\,{\rm eV}$ (\textit{i.e.} the prior we set is $M_{\nu} \geq 0\,{\rm eV}$), whereas when modelling the exact \texttt{NO}, values in the range $0\,{\rm eV}<M_{\nu}<0.06\,{\rm eV}$ are by construction no longer explored by the MCMC algorithm. The astute reader might have understood that we are once more getting into the land of volume effects already discussed in Chapter~\ref{sec:paper1}: within the same model, the \textit{3deg} approximation has access to a larger region of parameter space than when modelling the exact \texttt{NO}. An ``apples to apples'' comparison between \textit{3deg} and \texttt{NO} for the $\Lambda$CDM+$M_{\nu}$ should somehow take this into account. We therefore tried using the \textit{3deg} approximation but this time applying a prior $M_{\nu} \geq 0.06\,{\rm eV}$: in this case we found $n_s=0.9630 \pm 0.0070$, a completely negligible shift with respect to the value found when using the exact \texttt{NO}. Our conclusion therefore was that the shifts when moving from \textit{3deg} to \texttt{NO} were entirely due to parameter space volume effects and not a sign that data is mildly sensitive to the exact modelling of the neutrino mass spectrum. The exact \texttt{NO} cuts the region of low $M_{\nu}$: given the inverse correlation between $M_{\nu}$ and $n_s$, this implies cutting the region of high $n_s$, which explains why we find a lower value of $n_s$ when using the exact \texttt{NO}. When including also BAO data, the correlation between $M_{\nu}$ and $n_s$ changes sign, for reasons discussed in detail in Paper~V. Therefore, most of the shifts we had seen earlier for the CMB-only case change direction (e.g. $n_s$ increases when $M_{\nu}$ is marginalized over, instead of decreasing), but our main conclusions are totally unchanged: our determination of $n_s$ is basically insensitive to assumptions/approximations on the neutrino mass spectrum if not through parameter space volume effects, and only mildly sensitive to the choice of cosmological model (\textit{i.e.} the choice of whether or not to include $M_{\nu}$ as a free parameter). In Paper~V, we also tested the impact of further marginalizing over the tensor-to-scalar ratio $r$ (in that case also including the \textit{BK14} dataset), finding that the previous conclusions are qualitatively unchanged.

\begin{figure}[!h]
\centering
\includegraphics[width=0.7\linewidth]{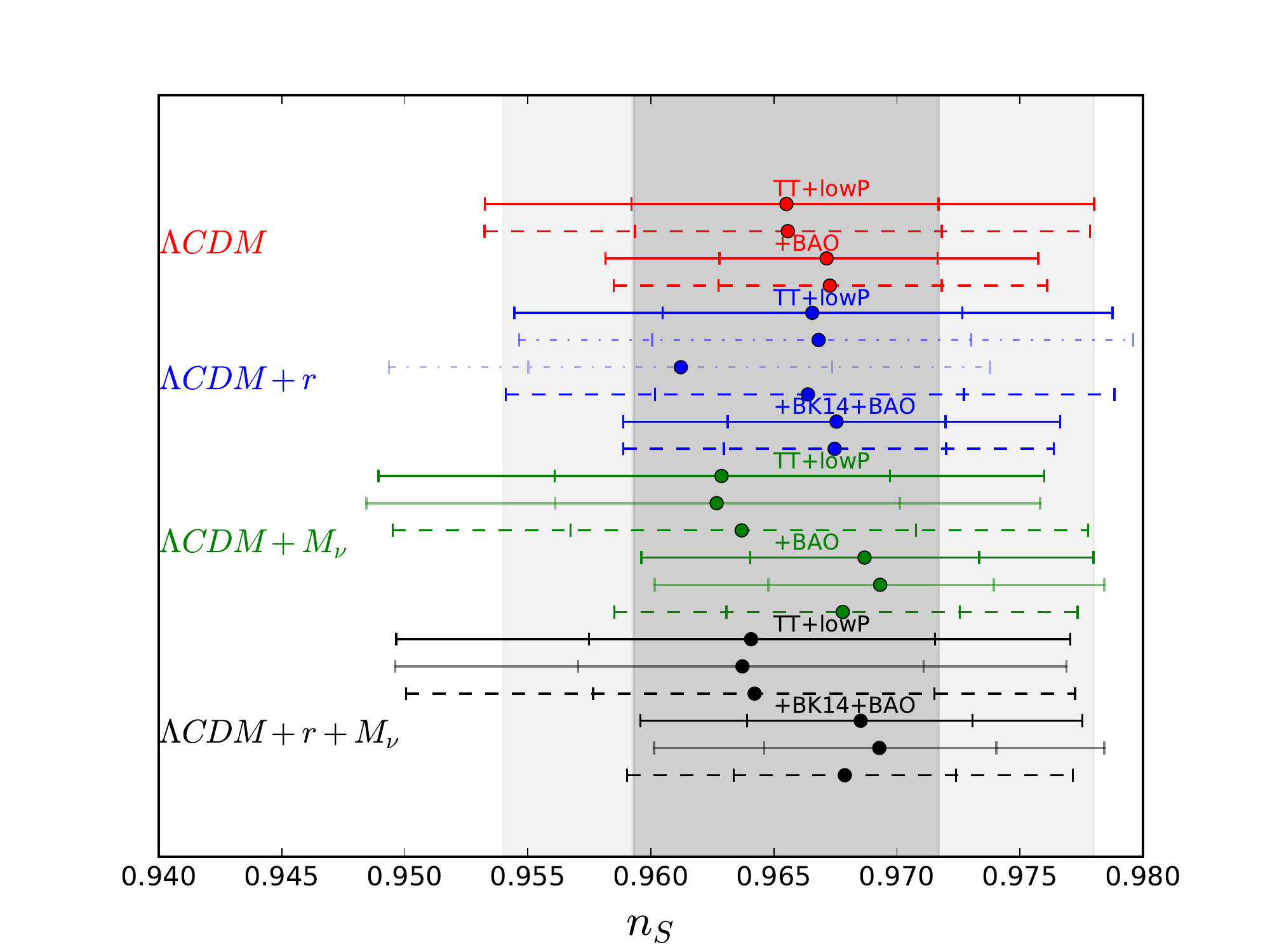}
\caption{Marginalized 68\% and 95\% confidence intervals for $n_s$ for different choices of cosmological models ($\Lambda$CDM, $\Lambda$CDM+$r$, $\Lambda$CDM+$M_{\nu}$, and $\Lambda$CDM+$r$+$M_{\nu}$), cosmological datasets (combinations of \textit{PlanckTT}+\textit{lowP}, \textit{BAO}, and \textit{BK14}), and approximations on the neutrino mass spectrum (\texttt{NO} or \textit{1mass}/\textit{3deg} approximations). The solid bold lines are obtained using the exact \texttt{NO} modelling, solid light lines using the exact \texttt{IO} modelling, and dashed lines for the approximations: \textit{1mass} approximation when $M_{\nu}$ is fixed ($\Lambda$CDM and $\Lambda$CDM+$r$ models), and \textit{3deg} approximation when $M_{\nu}$ is varied ($\Lambda$CDM+$M_{\nu}$ and $\Lambda$CDM+$r$+$M_{\nu}$ models). Only for the case of the $\Lambda$CDM+$r$ model, we considered two additional cases where $M_{\nu}$ is fixed to values higher than the standard $M_{\nu}=0.06\,{\rm eV}$, to enlarge the impact of $M_{\nu}$ on $n_s$: the results are the two dashed-dotted blue lines, where the top line has $M_{\nu}=0.07\,{\rm eV}$ and the bottom line has $M_{\nu}=0.5\,{\rm eV}$. The vertical grey bands are the 68\% and 95\% confidence intervals limits obtained by the Planck collaboration for the baseline $\Lambda$CDM model for the \textit{PlanckTT}+\textit{lowP} dataset (which of course basically reproduce our topmost solid red interval). Reproduced from~\cite{Gerbino:2016sgw} (Paper~V) with permission from APS.}
\label{fig:paper5fig1}
\end{figure}

Our results so far are conveniently summarized in Tab.~\ref{tab:tab1paper5}. The table should be roughly read as follows: for a given dataset, shifts brought upon by marginalizing over $M_{\nu}$ (\textit{i.e.} due to the assumption on the cosmological model), which are the largest ones, can be seen by remaining on a given row and moving from the left to the right. On the other hand, for a given dataset, shifts brought upon by assumptions on the neutrino mass spectrum (exact \texttt{NO} vs approximations) can be seen by remaining on a given column and moving downwards by one row. When doing so for the $\Lambda$CDM+$M_{\nu}$ model (second column), it should be kept in mind that the shift is due to volume effects and can be reabsorbed by adopting the prior $M_{\nu} \geq 0.06\,{\rm eV}$ when using the \textit{3deg} approximation. A visual representation of the shifts in $n_s$ is given in Fig.~\ref{fig:paper5fig1}, including in this case also the results obtained when marginalizing over $r$ ($\Lambda$CDM+$r$ model), not discussed here (see Sec.~IIIB of Paper~V for more details).
\begin{table}[!h]
\centering
\begin{tabular}{|cc||c|c|}
\hline \hline
&&$\Lambda$CDM	&$\Lambda$CDM+$M_{\nu}$\\
\hline
\multirow{2}{*}{\textit{PlanckTT}+\textit{lowP}} & \texttt{NO} &$ 0.9655\pm 0.0063$	& $0.9629\pm 0.0069$\\	
                                                    & approx & $ 0.9656\pm 0.0063$		& $0.9636\pm 0.0071$\\
\hline
\multirow{2}{*}{\textit{PlanckTT}+\textit{lowP}+\textit{BAO}} & \texttt{NO} & $0.9671\pm 0.0045$	& $0.9686\pm 0.0047$\\	
                                    & approx & $ 0.9673\pm 0.0045$		& $0.9678\pm 0.0048$\\
\hline
\end{tabular}
\caption{Marginalized 68\% confidence intervals for $n_s$ for different choices of cosmological models, cosmological datasets, and approximations on the neutrino mass spectrum (\texttt{NO} or approx). Rows labelled ``approx'' refer to the \textit{1mass} approximation (first column, $\Lambda$CDM model with $M_{\nu}$ fixed to $0.06\,{\rm eV}$) or the \textit{3deg} approximation (second column, $\Lambda$CDM+$M_{\nu}$, $M_{\nu}$ marginalized over).}
\label{tab:tab1paper5}
\end{table}

Afterwards, we moved on to test assumptions on the neutrino effective number $N_{\rm eff}$. We can expect a direct correlation between $N_{\rm eff}$ and $n_s$. As discussed in~\cite{Hou:2011ec}, increasing $N_{\rm eff}$ while adjusting other parameters in such a way as to keep $\theta_s$ fixed, leads to increased Silk damping, and less power in the damping tail ($\ell \gtrsim 500$): this is clearly shown in Fig.~1 of~\cite{Hou:2011ec}. This effect can be compensated by increasing $n_s$ to give more power to the damping tail: therefore, we can expect that in general adding $N_{\rm eff}$ as a free parameter should shift $n_s$ to higher values. We also investigated the impact of marginalizing over $M_{\nu}$ in addition to $N_{\rm eff}$: that is, we compare the values of $n_s$ obtained for the $\Lambda$CDM+$N_{\rm eff}$ model ($M_{\nu}$ fixed to $0.06\,{\rm eV}$) and the $\Lambda$CDM+$N_{\rm eff}$+$M_{\nu}$ model. Since we previously found that the exact modelling of the mass splittings played essentially no role in determining $n_s$, we choose for simplicity to model the neutrino mass spectrum following the \textit{1mass} approximation when $M_{\nu}$ is fixed, and the \textit{3deg} approximation when $M_{\nu}$ is varying. As far as $N_{\rm eff}$ is concerned, we test two different possible scenarios. In a first case, we apply a ``broad'' flat prior on $N_{\rm eff}$ between $0$ and $10$. In a second case, we apply a ``hard'' prior $N_{\rm eff} \leq 3.046$: this prior is a proxy for low-reheating scenarios~\cite{Davidson:2000dw,Gelmini:2004ah,Ichikawa:2005vw,
Visinelli:2009kt,Freese:2017ace}, where thermalization is incomplete by the time of neutrino decoupling, effectively leading to a value of $N_{\rm eff}$ lower compared to the usual expectations. In this case, given the direction of the $N_{\rm eff}$-$n_s$ correlation previously discussed, we expect that $n_s$ should instead shift to lower values, as we are artificially excluding the region of high $N_{\rm eff}$ which would pull $n_s$ to higher values (this is again a volume effect argument).

\begin{figure}[!h]
\centering
\includegraphics[width=0.7\linewidth]{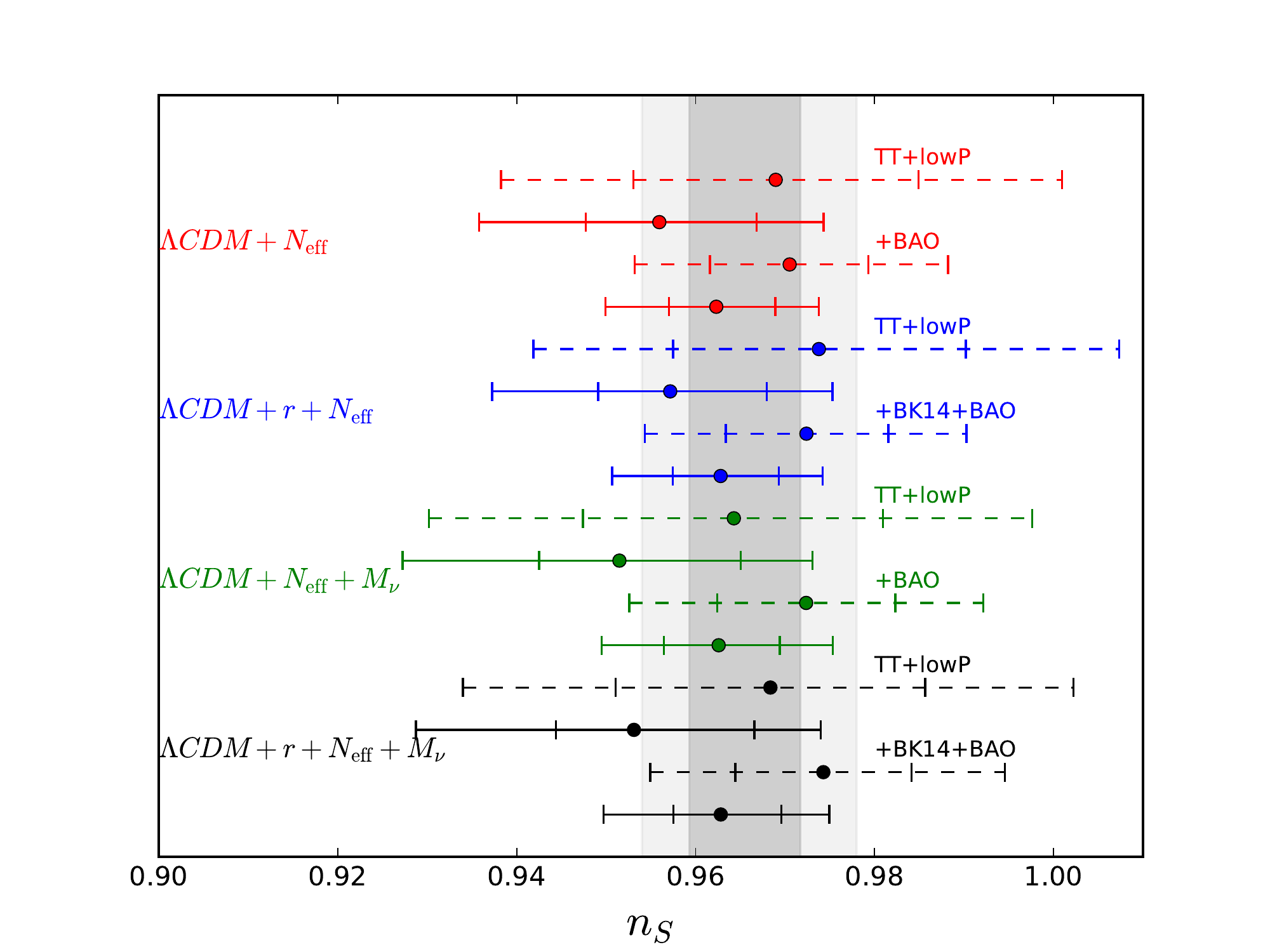}
\caption{Marginalized 68\% and 95\% confidence intervals for $n_s$ for different choices of cosmological models ($\Lambda$CDM+$N_{\rm eff}$, $\Lambda$CDM+$r$+$N_{\rm eff}$, $\Lambda$CDM+$N_{\rm eff}$+$M_{\nu}$, and $\Lambda$CDM+$r$+$N_{\rm eff}$+$M_{\nu}$), cosmological datasets (combinations of \textit{PlanckTT}+\textit{lowP}, \textit{BAO}, and \textit{BK14}), and assumptions about the neutrino effective number (``broad'' $0 \leq N_{\rm eff} \leq 10$ prior or ``hard'' $N_{\rm eff} \leq 3.046$ prior). Solid lines are for the ``broad'' prior while dashed lines are for the ``hard'' prior. Vertical grey bands as in Fig.~\ref{fig:paper5fig1}. Reproduced from~\cite{Gerbino:2016sgw} (Paper~V) with permission from APS.}
\label{fig:paper5fig2}
\end{figure}

When using only \textit{PlanckTT}+\textit{lowP} data, the baseline value of $n_s$ to compare against is $n_s=0.9656 \pm 0.0063$ when adopting the $\Lambda$CDM+$N_{\rm eff}$ model, and $n_s=0.9636 \pm 0.0071$ when adopting the $\Lambda$CDM+$N_{\rm eff}$+$M_{\nu}$ model (see Tab.~\ref{tab:tab1paper5}). For the ``broad'' $\Lambda$CDM+$N_{\rm eff}$ case, we find as expected a shift of $n_s$ towards larger values: $\underline{\boldsymbol{n_s=0.969 \pm 0.016}}$. For the ``hard'' $\Lambda$CDM+$N_{\rm eff}$ case, again as expected we found a shift of $n_s$ towards smaller values: $\underline{\boldsymbol{n_s=0.956^{+0.011}_{-0.008}}}$. We then move to the $\Lambda$CDM+$N_{\rm eff}$+$M_{\nu}$ model, where for the ``broad'' case we find $\underline{\boldsymbol{n_s=0.964 \pm 0.0017}}$ and for the ``hard'' case we find $\underline{\boldsymbol{n_s = 0.951^{+0.014}_{-0.009}}}$. From these shifts we have drawn two conclusions. Firstly, we have two degeneracies at play which pull in opposite directions: the $N_{\rm eff}$-$n_s$ degeneracy and the $M_{\nu}$-$n_s$ degeneracy. Our results suggest that the former is more relevant than the latter, since even when $M_{\nu}$ is marginalized over for the ``broad'' case the net effect is still an increase in $n_s$, which indicates that the ``pull'' due to the $N_{\rm eff}$-$n_s$ degeneracy is stronger than the ``pull'' due to the $M_{\nu}$-$n_s$ one. The second conclusion is that the freedom induced by changing our assumptions on $N_{\rm eff}$ has a rather non-negligible impact on $n_s$. For instance, assuming low-reheating scenarios (``hard'' prior) lowered the value of $n_s$ by almost $1\sigma$. Our results concerning shifts in $n_s$ as we change our assumptions on $N_{\rm eff}$ are summarized in Tab.~\ref{tab:tab2paper5} and  Fig.~\ref{fig:paper5fig2} (again in this case including in also the results obtained when marginalizing over $r$ not discussed here: see Sec.~IIID of Paper~V for more details).
\begin{table}[!h]
\centering
\begin{tabular}{|cc||c|c|}
\hline \hline
&&$\Lambda$CDM+$N_{\rm eff}$	&$\Lambda$CDM+$N_{\rm eff}$+$M_{\nu}$\\
\hline
\multirow{2}{*}{\textit{PlanckTT}+\textit{lowP}} & broad ($0 \leq N_{\rm eff} \leq 10$) &$ 0.969 \pm 0.016$	& $0.964 \pm 0.017$\\	
                                                    & hard ($N_{\rm eff} \leq 3.046$) & $ 0.956^{+0.011}_{-0.008}$		& $0.951^{+0.014}_{-0.009}$\\
\hline
\multirow{2}{*}{\textit{PlanckTT}+\textit{lowP}+\textit{BAO}} & broad ($0 \leq N_{\rm eff} \leq 10$) & $0.971 \pm 0.009$	& $0.973 \pm 0.010$\\	
                                    & hard ($N_{\rm eff} \leq 3.046$) & $ 0.962^{+0.007}_{-0.005}$		& $0.962^{+0.007}_{-0.006}$\\
\hline
\end{tabular}
\caption{Marginalized 68\% confidence intervals for $n_s$ for different choices of cosmological models, cosmological datasets, and approximations on the neutrino effective number (``broad'' or ``hard'' prior on $N_{\rm eff}$, described in the table). Note that we adopt the \textit{1mass} approximation when $M_{\nu}$ is fixed ($\Lambda$CDM+$N_{\rm eff}$ model) and the \textit{3deg} approximation when $M_{\nu}$ is varying ($\Lambda$CDM+$N_{\rm eff}$+$M_{\nu}$ model), given our earlier findings that modelling the exact mass splittings leads to negligible shifts in $n_s$.}
\label{tab:tab2paper5}
\end{table}

\begin{figure*}[!h]
\centering
\begin{tabular}{cc}
\includegraphics[width=0.5\textwidth]{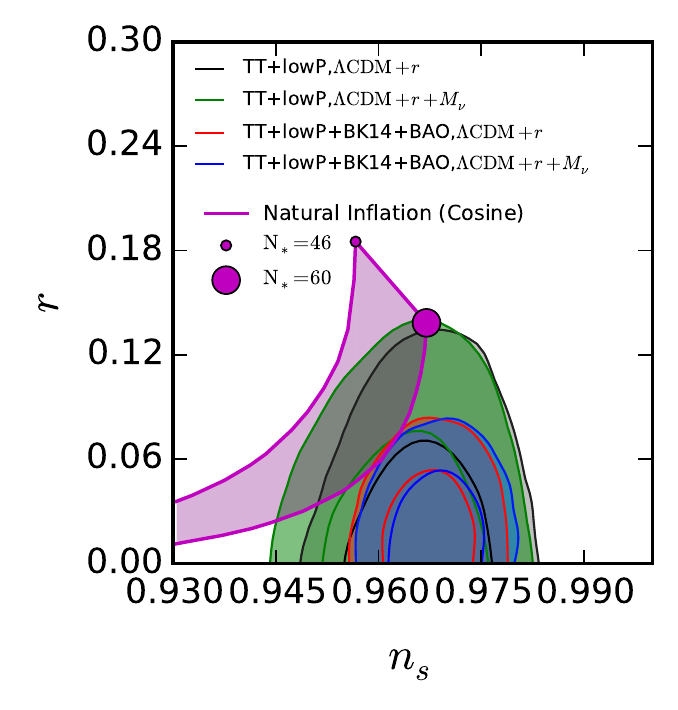}&\includegraphics[width=0.5\textwidth]{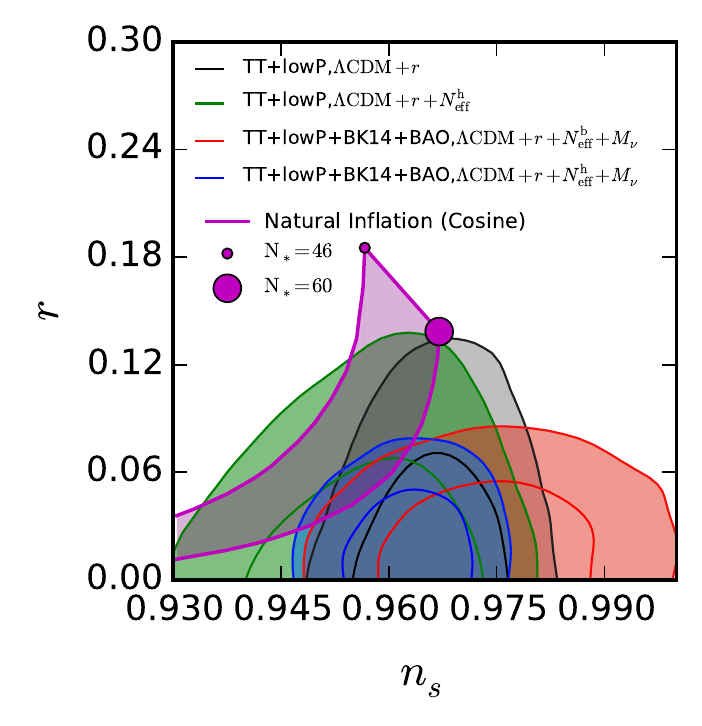}\\
\end{tabular}
\caption{68\% and 95\%~C.L. joint probability contours in the $n_s$-$r$ plane for the datasets and models indicated. The predictions for the cosine natural inflation model are shown in purple for $46 \leq N_{*} \leq 60$, with $N_{*}$ is the number of e-folds of inflation. \textit{Left panel}: contours computed assuming \texttt{NO}. \textit{Right panel}: ``h'' and ``b'' stand for the hard ($N_{\rm eff} \leq 3.046$) and broad ($0 \leq N_{\rm eff} \leq 10$) priors imposed on $N_{\rm eff}$, contours computed assuming the \textit{1mass} approximation when $M_{\nu}$ is fixed ($\Lambda$CDM+$r$ and $\Lambda$CDM+$r$+$N_{\rm eff}$ models), and the \textit{3deg} approximation when $M_{\nu}$ is varying ($\Lambda$CDM+$r$+$N_{\rm eff}$+$M_{\nu}$ model). Reproduced from~\cite{Gerbino:2016sgw} (Paper~V) with permission from APS.}
\label{fig:paper5fig3}
\end{figure*}

Our findings can be important when assessing the validity of inflationary models in light of precision cosmological data. Usually, inflationary models are compared against observations by plotting their predictions in the $n_s$-$r$ plane, assuming a minimal $\Lambda$CDM+$r$ model. As a concrete example, in Fig.~\ref{fig:paper5fig3} we compare the predictions of the original cosine natural inflation model of Freese \textit{et al.}~\cite{Freese:1990rb} (see e.g.~\cite{Adams:1992bn,ArkaniHamed:2003wu,Freese:2004un,Dimopoulos:2005ac,
Savage:2006tr,Freese:2008if,Kaloper:2008fb,Czerny:2014wza,Freese:2014nla,Kappl:2015esy} for other important works) against observational constraints in the $n_s$-$r$ plane, within the different cosmological models we have considered in Paper~V. Within the minimal $\Lambda$CDM+$r$ model and including \textit{BK14} data, cosine natural inflation is excluded at more than $2\sigma$ (see left panel of Fig.~\ref{fig:paper5fig3}): it can however be ``rescued'' by relaxing the assumptions on the neutrino effective number, particularly when considering low-reheating scenarios (see right panel of Fig.~\ref{fig:paper5fig3}), although these scenarios are admittedly a bit more exotic. In Paper~V we have also provided forecasts (using the methodology we described in Chapter~\ref{sec:paper3}) for future CMB experiments such as COrE~\cite{Bouchet:2011ck} and CMB-S4~\cite{Abazajian:2016yjj}, and shown that our conclusions are still relevant for future data: in other words, even with future data marginalizing over $N_{\rm eff}$ can lead to shifts of order $1\sigma$ of $n_s$. The reader is invited to read Sec.~IV of Paper~V for more details on our forecasts.

\subsection{Executive summary of Paper~V}
\label{subsec:paper5}

To conclude, in Paper~V we have studied how our assumptions about the neutrino unknowns (mass, mass ordering, effective number) impact the inferred values of inflationary parameters, focusing on the scalar spectral index $n_s$. We have found that modelling the exact mass ordering leads to negligible shifts in $n_s$, modulo shifts due to volume effects which can be reabsorbed by an appropriate prior. To put it differently, when allowing $M_{\nu}$ to vary, adopting the \textit{3deg} approximation of 3 degenerate eigenstates is for all intents and purposes a good enough approximation, and results obtained modelling the exact \texttt{NO} (\texttt{IO}) are basically equivalent to those obtained assuming \textit{3deg} approximation and assuming a prior $M_{\nu} \geq 0.06\,{\rm eV}$ ($M_{\nu} \geq 0.10\,{\rm eV}$). The biggest shifts in $n_s$ occur when relaxing the assumptions on the effective neutrino number, particularly when allowing for more exotic low-reheating scenarios where $N_{\rm eff}$ can be lower than the canonical value $3.046$. Despite these shifts are at most of order $1\sigma$, a complete assessment of the impact of neutrino properties on the estimation of inflationary parameters is important as certain inflationary models which are currently marginally excluded (e.g. cosine natural inflation) are observationally viable once we allow for more freedom in the neutrino sector (see also~\cite{Tram:2016rcw,Barenboim:2019tux}).

\chapter{Summary and outlook}
\label{chap:7}

\begin{chapquote}{Ramsay Snow to Theon Greyjoy in \textit{Game of Thrones}, Season 3, Episode 6: ``The Climb'' (2013)}
``If you think this has a happy ending, you haven’t been paying attention.''
\end{chapquote}

We have come to the end of this journey into the realm of neutrino cosmology, and at this point I will briefly summarize the results described in more detail in Chapter~\ref{chap:6} and the included papers, and provide an outlook into future research directions which could build upon these results. If you, reader, have managed to follow me until here, I believe there is no need to convince you that neutrino cosmology is an extremely fascinating subject, and one which promises to be ripe with discoveries in the coming years as more data, and especially more precise data, pours in. At the time I started my PhD, a number of open questions in the field of cosmology begged for answers (see Chapter~\ref{sec:cosmology} for an outline of these questions), and I believe this thesis contributed to answering them.

In Paper~I (Chapter~\ref{sec:paper1}), I have shown that already current cosmological data provides a great deal of information about massive neutrinos. In particular, I have shown that a combination of current CMB and clustering data sets the limit $M_{\nu}<0.12\,{\rm eV}$, currently the tightest upper limit on the sum of the neutrino masses. Moreover, in Paper~I I have devised a simple method to quantify the preference for the normal neutrino mass ordering from cosmology (a slightly different, but conceptually identical, method is discussed in Chapter~\ref{sec:paper1}). In fact, a byproduct of such method has been showing that cosmology will always prefer, even if only slightly, the normal neutrino mass ordering, due not to physical effects but parameter space volume effects. In Paper~I, I have found that current data shows at most a weak preference for the normal ordering, with odds of about $3:1$.

One of the side results of Paper~I was that galaxy clustering data appears to be less constraining than BAO distance measurements, despite in principle containing more information than the latter. In Paper~I I argued that this reflects a limit in our our analysis methodology, warranting a wiser treatment of galaxy bias. This was the path followed in Paper~II (Chapter~\ref{sec:paper2}): we took an old idea of using cross-correlations between CMB lensing and galaxy maps, in combination with galaxy clustering measurements, to provide a better handle on the scale-dependent galaxy bias, and for the first time realized this idea on real data. In doing so, we clarified a number of subtleties having to do with scale-dependent bias in auto- and cross-correlation measurements.

Another important issue in the use of galaxy clustering data to study neutrino properties is that of properly defining the galaxy bias in first place. Virtually all analyses so far have defined the bias with respect to the total matter field, whereas it is known from simulations that a meaningful definition of bias is with respect to the cold dark matter+baryons field. In Paper~III (Chapter~\ref{sec:paper3}), we have checked whether this mismatch could be a problem for the analysis of future clustering data. We have found that an incorrect definition of bias can lead to misestimated parameters, among which the sum of the neutrino masses. In Paper~III we have also provided public tools for accounting for this effect in a simple and efficient way.

In the two remaining papers, we have instead examined correlations between the sum of the neutrino masses and other cosmological parameters, and on the consequences of such correlations. In Paper~IV (Chapter~\ref{sec:paper4}), we have shown that in dynamical dark energy cosmologies where dark energy is forced to be non-phantom (\textit{i.e.} $w(z) \geq -1$), the upper limits on $M_{\nu}$ become counterintuitively tighter than the $\Lambda$CDM upper limits. As a consequence, non-phantom dark energy models (which include standard quintessence models) prefer the normal neutrino ordering more strongly than $\Lambda$CDM does. Their viability could therefore be jeopardized should upcoming laboratory experiments should determine that the neutrino mass ordering is inverted. The result of Paper~IV provides an unexpected window, that of neutrino laboratory experiments, into the physics of what is driving cosmic acceleration.

Finally, in Paper~V (Chapter~\ref{sec:paper5}) we have checked whether our ignorance of neutrino properties can bias our determination of inflationary parameters, and hence of the initial conditions of the Universe. We have found that, fortunately, this is not a concern. The only case where important shifts in inflationary parameters are obtained is when low-reheating scenarios, which are quite exotic, are considered. Therefore, in Paper~V we have concluded that our uncertainties about the physics in the neutrino sector do not affect our determination of inflationary parameters to a significant extent, neither with current nor with future data.

Building upon the results in the included papers, there are several directions which could be pursued in future works (some of which I am already pursuing). One interesting direction building upon Paper~I could be that of robustly combining cosmology and laboratory ($\beta$ decay, double $\beta$ decay, and oscillation) experiments, along the lines of~\cite{Gerbino:2015ixa}. More interestingly, such an approach could be used to study sterile neutrinos, including sterile neutrinos at the ${\rm eV}$ scale which have been suggested as possible solutions to a series of anomalies~\cite{Melchiorri:2008gq,Archidiacono:2012ri,Archidiacono:2013xxa,Mirizzi:2013gnd,Gariazzo:2013gua,Archidiacono:2014apa,Gariazzo:2015rra,
Gariazzo:2016ehl,Archidiacono:2016kkh,Bridle:2016isd,Abazajian:2017tcc,Dentler:2018sju,Giunti:2019aiy}.

Beyond neutrinos, another intriguing study related to Paper~I could involve using the same galaxy clustering data to constrain light relics. That is, species which decoupled while relativistic like neutrinos. If the species are heavy enough, they essentially behave as cold dark matter at decoupling, and there is no hope of constraining them from the CMB. However, their free-streaming would result in a suppression of power on small scales, exactly as with neutrinos. Moreover, if heavy enough, the light relic would become non-relativistic during radiation domination. As shown in~\cite{Viel:2005qj,Viel:2006kd,Boyarsky:2008xj}, this has the effect of enhancing the maximum suppression, making it $14f_x$ (with $f_x \equiv \Omega_x/\Omega_m$ the fraction of the energy density in the relic $x$) instead of $8f_{\nu}$ as in the neutrino case. This suggests that light relics should be a promising target for large-scale structure probes.

Paper~II also warrants several follow-up directions. In light of the precision of future data, it is important to try and model the non-linear galaxy power spectrum as precisely as possible, and this includes understanding non-linear bias. It would be interesting to explore whether combining CMB lensing and galaxy clustering, and possibly higher order correlators of the lensing and/or galaxy fields could help constraining non-linear bias.

As for Paper~III, the most immediate follow-up work would be to make sure that current and upcoming LSS surveys follow our recommendations, eventually updating their pipelines if necessary. Besides that, an interesting follow-up would be to explore the issue of proper definition of bias in cosmologies beyond those with massive neutrinos (for instance mixed dark matter cosmologies).

Concerning instead Paper~IV, it could be worth going ``non-parametric'', \textit{i.e.} to see how much our results change if we do not adopt parametric form for $w(z)$. Possible approaches include for instance adopting a principal component analysis (PCA) approach using PCA components from future surveys (along the lines of the work in~\cite{Said:2013jxa,DiazRivero:2019ukx}), or using Gaussian Processes reconstruction binning the equation of state in time (along the lines of the work in~\cite{Gerardi:2019obr}). Another interesting follow-up would instead be to repeat the analysis for specific well-motivated quintessence models, or factoring more general theoretical considerations concerning the theoretical health and mathematical soundness of the theoretical models one is parametrizing (along the lines of the work in~\cite{Peirone:2017lgi}).

Finally, there are two very ambitious research directions I plan to at least keep thinking about moving forward. One is related to the possibility of going beyond the sum of the neutrino masses $M_{\nu}$ and measuring the masses of the individual eigenstates from cosmological data. A series of earlier works had examined this problem about a decade ago, concluding that it will be infeasible in the foreseeable future, due to insufficient sensitivity in LSS data~\cite{Pritchard:2008wy,DeBernardis:2009di,Jimenez:2010ev,Hall:2012kg,Jimenez:2016ckl}. However, it might be worth re-examining the issue as we get closer to the launch date for a number of important LSS surveys, and as we understand the performance of such surveys better. Should this be possible, our conclusions in Paper~I about cosmology only being sensitive to the mass ordering through volume effects would be surpassed, and it might be possible to determine the mass ordering even if it is inverted.

The second ambitious direction is related to the cosmic neutrino background (CNB)~\cite{Audren:2014lsa}, which currently remains undetected. Experimental efforts through capture of cosmic relic neutrinos on tritium such as \textit{Ptolemy}~\cite{Baracchini:2018wwj,Betti:2019ouf} are underway to try and detect the CNB, and it is far from clear whether these will succeed.~\footnote{Other methods to detect the CNB and with it the sum of the neutrino masses have been proposed, e.g.~\cite{Millar:2018hkv}, but do not appear promising.} If, however, they should succeed and we were to detect the CNB, a whole new field of cosmology could open up by studying anisotropies in the CNB. The same way anisotropies in the CMB have provided, and are still providing, a mine of information, the same would definitely hold for anisotropies in the CNB. It might be worth, in the meanwhile, thinking about how best to exploit these anisotropies, should we one day manage to detect them~\cite{Michney:2006mk,Hannestad:2009xu}.

Technicalities aside, I hope I have convinced the reader that neutrino cosmology is an extremely exciting and active area of research. The next 5 to 10 years will be extremely crucial in this direction, as we expect a first detection of non-zero neutrino masses from a combination of future CMB and LSS probes~\cite{Hannestad:2002cn,Joudaki:2011nw,Carbone:2011by,Hamann:2012fe,
Allison:2015qca,Archidiacono:2016lnv,Boyle:2017lzt,Sprenger:2018tdb,
Mishra-Sharma:2018ykh,Brinckmann:2018owf,Kreisch:2018var,Ade:2018sbj,Yu:2018tem,
Boyle:2018rva,Hanany:2019lle}. In this thesis, I have contributed to addressing a number of critical issues whose resolution is crucial if we want to make sure that a robust detection is reached. Detecting the neutrino mass scale and possibly the mass ordering would open the door towards new physics beyond the Standard Model, possibly shedding light onto processes operating at energy scales we will likely never be able to reach down on Earth. There is all the reason to believe that cosmological data will provide the first glimpse onto this realm, and hence all the reason to be excited and stay tuned.

\cleardoublepage
\addtocounter{chapter}{1}
\renewcommand{\bibname}{References}
\renewcommand{\ThumbBoxColor}{red}

\newcommand{\RefsBox}{}
\fancypagestyle{plain}
{
   \fancyhead[CE]{\RefsBox}
   \fancyhead[CO]{\RefsBox}
}
\fancyhead[CE]{\RefsBox}
\fancyhead[CO]{\RefsBox}

\bibliographystyle{JHEP}
{\footnotesize \bibliography{PhD_thesis}}

\providecommand{\href}[2]{#2}\begingroup\raggedright\begin{thebibliography}{1000}

\bibitem{Tanabashi:2018oca}
{\scshape Particle Data Group} collaboration, M.~Tanabashi et~al.,
  \emph{{Review of Particle Physics}},
  \href{https://doi.org/10.1103/PhysRevD.98.030001}{\emph{Phys. Rev.}
  {\bfseries D98} (2018) 030001}.

\bibitem{GonzalezGarcia:2012sz}
M.~C. González-García, M.~Maltoni, J.~Salvadó and T.~Schwetz, \emph{{Global
  fit to three neutrino mixing: critical look at present precision}},
  \href{https://doi.org/10.1007/JHEP12(2012)123}{\emph{JHEP} {\bfseries 12}
  (2012) 123}, [\href{https://arxiv.org/abs/1209.3023}{{\ttfamily 1209.3023}}].

\bibitem{Gonzalez-Garcia:2014bfa}
M.~C. González-García, M.~Maltoni and T.~Schwetz, \emph{{Updated fit to three
  neutrino mixing: status of leptonic CP violation}},
  \href{https://doi.org/10.1007/JHEP11(2014)052}{\emph{JHEP} {\bfseries 11}
  (2014) 052}, [\href{https://arxiv.org/abs/1409.5439}{{\ttfamily 1409.5439}}].

\bibitem{Gonzalez-Garcia:2015qrr}
M.~C. González-García, M.~Maltoni and T.~Schwetz, \emph{{Global Analyses of
  Neutrino Oscillation Experiments}},
  \href{https://doi.org/10.1016/j.nuclphysb.2016.02.033}{\emph{Nucl. Phys.}
  {\bfseries B908} (2016) 199--217},
  [\href{https://arxiv.org/abs/1512.06856}{{\ttfamily 1512.06856}}].

\bibitem{Esteban:2016qun}
I.~Esteban, M.~C. González-García, M.~Maltoni, I.~Martínez-Soler and
  T.~Schwetz, \emph{{Updated fit to three neutrino mixing: exploring the
  accelerator-reactor complementarity}},
  \href{https://doi.org/10.1007/JHEP01(2017)087}{\emph{JHEP} {\bfseries 01}
  (2017) 087}, [\href{https://arxiv.org/abs/1611.01514}{{\ttfamily
  1611.01514}}].

\bibitem{deSalas:2017kay}
P.~F. de~Salas, D.~V. Forero, C.~A. Ternes, M.~Tórtola and J.~W.~F. Valle,
  \emph{{Status of neutrino oscillations 2018: 3$\sigma$ hint for normal mass
  ordering and improved CP sensitivity}},
  \href{https://doi.org/10.1016/j.physletb.2018.06.019}{\emph{Phys. Lett.}
  {\bfseries B782} (2018) 633--640},
  [\href{https://arxiv.org/abs/1708.01186}{{\ttfamily 1708.01186}}].

\bibitem{deSalas:2018bym}
P.~F. de~Salas, S.~Gariazzo, O.~Mena, C.~A. Ternes and M.~Tórtola,
  \emph{{Neutrino Mass Ordering from Oscillations and Beyond: 2018 Status and
  Future Prospects}},
  \href{https://doi.org/10.3389/fspas.2018.00036}{\emph{Front. Astron. Space
  Sci.} {\bfseries 5} (2018) 36},
  [\href{https://arxiv.org/abs/1806.11051}{{\ttfamily 1806.11051}}].

\bibitem{Hannestad:2002cn}
S.~Hannestad, \emph{{Can cosmology detect hierarchical neutrino masses?}},
  \href{https://doi.org/10.1103/PhysRevD.67.085017}{\emph{Phys. Rev.}
  {\bfseries D67} (2003) 085017},
  [\href{https://arxiv.org/abs/astro-ph/0211106}{{\ttfamily
  astro-ph/0211106}}].

\bibitem{Joudaki:2011nw}
S.~Joudaki and M.~Kaplinghat, \emph{{Dark Energy and Neutrino Masses from
  Future Measurements of the Expansion History and Growth of Structure}},
  \href{https://doi.org/10.1103/PhysRevD.86.023526}{\emph{Phys. Rev.}
  {\bfseries D86} (2012) 023526},
  [\href{https://arxiv.org/abs/1106.0299}{{\ttfamily 1106.0299}}].

\bibitem{Carbone:2011by}
C.~Carbone, C.~Fedeli, L.~Moscardini and A.~Cimatti, \emph{{Measuring the
  neutrino mass from future wide galaxy cluster catalogues}},
  \href{https://doi.org/10.1088/1475-7516/2012/03/023}{\emph{JCAP} {\bfseries
  1203} (2012) 023}, [\href{https://arxiv.org/abs/1112.4810}{{\ttfamily
  1112.4810}}].

\bibitem{Hamann:2012fe}
J.~Hamann, S.~Hannestad and Y.~Y.~Y. Wong, \emph{{Measuring neutrino masses
  with a future galaxy survey}},
  \href{https://doi.org/10.1088/1475-7516/2012/11/052}{\emph{JCAP} {\bfseries
  1211} (2012) 052}, [\href{https://arxiv.org/abs/1209.1043}{{\ttfamily
  1209.1043}}].

\bibitem{Allison:2015qca}
R.~Allison, P.~Caucal, E.~Calabrese, J.~Dunkley and T.~Louis, \emph{{Towards a
  cosmological neutrino mass detection}},
  \href{https://doi.org/10.1103/PhysRevD.92.123535}{\emph{Phys. Rev.}
  {\bfseries D92} (2015) 123535},
  [\href{https://arxiv.org/abs/1509.07471}{{\ttfamily 1509.07471}}].

\bibitem{Archidiacono:2016lnv}
M.~Archidiacono, T.~Brinckmann, J.~Lesgourgues and V.~Poulin, \emph{{Physical
  effects involved in the measurements of neutrino masses with future
  cosmological data}},
  \href{https://doi.org/10.1088/1475-7516/2017/02/052}{\emph{JCAP} {\bfseries
  1702} (2017) 052}, [\href{https://arxiv.org/abs/1610.09852}{{\ttfamily
  1610.09852}}].

\bibitem{Boyle:2017lzt}
A.~Boyle and E.~Komatsu, \emph{{Deconstructing the neutrino mass constraint
  from galaxy redshift surveys}},
  \href{https://doi.org/10.1088/1475-7516/2018/03/035}{\emph{JCAP} {\bfseries
  1803} (2018) 035}, [\href{https://arxiv.org/abs/1712.01857}{{\ttfamily
  1712.01857}}].

\bibitem{Sprenger:2018tdb}
T.~Sprenger, M.~Archidiacono, T.~Brinckmann, S.~Clesse and J.~Lesgourgues,
  \emph{{Cosmology in the era of Euclid and the Square Kilometre Array}},
  \href{https://doi.org/10.1088/1475-7516/2019/02/047}{\emph{JCAP} {\bfseries
  1902} (2019) 047}, [\href{https://arxiv.org/abs/1801.08331}{{\ttfamily
  1801.08331}}].

\bibitem{Mishra-Sharma:2018ykh}
S.~Mishra-Sharma, D.~Alonso and J.~Dunkley, \emph{{Neutrino masses and
  beyond-$\Lambda$CDM cosmology with LSST and future CMB experiments}},
  \href{https://doi.org/10.1103/PhysRevD.97.123544}{\emph{Phys. Rev.}
  {\bfseries D97} (2018) 123544},
  [\href{https://arxiv.org/abs/1803.07561}{{\ttfamily 1803.07561}}].

\bibitem{Brinckmann:2018owf}
T.~Brinckmann, D.~C. Hooper, M.~Archidiacono, J.~Lesgourgues and T.~Sprenger,
  \emph{{The promising future of a robust cosmological neutrino mass
  measurement}},
  \href{https://doi.org/10.1088/1475-7516/2019/01/059}{\emph{JCAP} {\bfseries
  1901} (2019) 059}, [\href{https://arxiv.org/abs/1808.05955}{{\ttfamily
  1808.05955}}].

\bibitem{Kreisch:2018var}
C.~D. Kreisch, A.~Pisani, C.~Carbone, J.~Liu, A.~J. Hawken, E.~Massara et~al.,
  \emph{{Massive Neutrinos Leave Fingerprints on Cosmic Voids}},
  \href{https://arxiv.org/abs/1808.07464}{{\ttfamily 1808.07464}}.

\bibitem{Ade:2018sbj}
{\scshape Simons Observatory} collaboration, P.~Ade et~al., \emph{{The Simons
  Observatory: Science goals and forecasts}},
  \href{https://doi.org/10.1088/1475-7516/2019/02/056}{\emph{JCAP} {\bfseries
  1902} (2019) 056}, [\href{https://arxiv.org/abs/1808.07445}{{\ttfamily
  1808.07445}}].

\bibitem{Yu:2018tem}
B.~Yu, R.~Z. Knight, B.~D. Sherwin, S.~Ferraro, L.~Knox and M.~Schmittfull,
  \emph{{Towards Neutrino Mass from Cosmology without Optical Depth
  Information}},  \href{https://arxiv.org/abs/1809.02120}{{\ttfamily
  1809.02120}}.

\bibitem{Boyle:2018rva}
A.~Boyle, \emph{{Understanding the neutrino mass constraints achievable by
  combining CMB lensing and spectroscopic galaxy surveys}},
  \href{https://arxiv.org/abs/1811.07636}{{\ttfamily 1811.07636}}.

\bibitem{Hanany:2019lle}
{\scshape NASA PICO} collaboration, S.~Hanany et~al., \emph{{PICO: Probe of
  Inflation and Cosmic Origins}},
  \href{https://arxiv.org/abs/1902.10541}{{\ttfamily 1902.10541}}.

\bibitem{Bhattacharyya:2008ez}
G.~Bhattacharyya, \emph{{Electroweak Symmetry Breaking and BSM Physics (A
  Review)}}, \href{https://doi.org/10.1007/s12043-009-0004-0}{\emph{Pramana}
  {\bfseries 72} (2009) 37--54},
  [\href{https://arxiv.org/abs/0807.3883}{{\ttfamily 0807.3883}}].

\bibitem{Lykken:2010mc}
J.~D. Lykken, \emph{{Beyond the Standard Model}},  in \emph{{CERN Yellow Report
  CERN-2010-002, 101-109}}, 2010,
  \href{https://arxiv.org/abs/1005.1676}{{\ttfamily 1005.1676}}.

\bibitem{Allanach:2016yth}
B.~C. Allanach, \emph{{Beyond the Standard Model Lectures for the 2016 European
  School of High-Energy Physics}},  in \emph{{Proceedings, 2016 European School
  of High-Energy Physics (ESHEP2016): Skeikampen, Norway, June 15-28 2016}},
  pp.~123--152, 2017, \href{https://arxiv.org/abs/1609.02015}{{\ttfamily
  1609.02015}}.

\bibitem{Rosenfeld:2017ksi}
R.~Rosenfeld, \emph{{Physics Beyond the Standard Model}},  in
  \emph{{Proceedings, 8th CERN–Latin-American School of High-Energy Physics
  (CLASHEP2015): Ibarra, Ecuador, March 05-17, 2015}}, pp.~159--164, 2016,
  \href{https://arxiv.org/abs/1708.00800}{{\ttfamily 1708.00800}}.

\bibitem{Graverini:2018riw}
{\scshape ATLAS, CMS, LHCb} collaboration, E.~Graverini, \emph{{Flavour
  anomalies: a review}},
  \href{https://doi.org/10.1088/1742-6596/1137/1/012025}{\emph{J. Phys. Conf.
  Ser.} {\bfseries 1137} (2019) 012025},
  [\href{https://arxiv.org/abs/1807.11373}{{\ttfamily 1807.11373}}].

\bibitem{Joyce:2014kja}
A.~Joyce, B.~Jain, J.~Khoury and M.~Trodden, \emph{{Beyond the Cosmological
  Standard Model}},
  \href{https://doi.org/10.1016/j.physrep.2014.12.002}{\emph{Phys. Rept.}
  {\bfseries 568} (2015) 1--98},
  [\href{https://arxiv.org/abs/1407.0059}{{\ttfamily 1407.0059}}].

\bibitem{Bull:2015stt}
P.~Bull et~al., \emph{{Beyond $\Lambda$CDM: Problems, solutions, and the road
  ahead}}, \href{https://doi.org/10.1016/j.dark.2016.02.001}{\emph{Phys. Dark
  Univ.} {\bfseries 12} (2016) 56--99},
  [\href{https://arxiv.org/abs/1512.05356}{{\ttfamily 1512.05356}}].

\bibitem{Freedman:2017yms}
W.~L. Freedman, \emph{{Cosmology at a Crossroads}},
  \href{https://doi.org/10.1038/s41550-017-0121}{\emph{Nat. Astron.} {\bfseries
  1} (2017) 0121}, [\href{https://arxiv.org/abs/1706.02739}{{\ttfamily
  1706.02739}}].

\bibitem{DiValentino:2017gzb}
E.~Di~Valentino, \emph{{Crack in the cosmological paradigm}},
  \href{https://doi.org/10.1038/s41550-017-0236-8}{\emph{Nat. Astron.}
  {\bfseries 1} (2017) 569--570},
  [\href{https://arxiv.org/abs/1709.04046}{{\ttfamily 1709.04046}}].

\bibitem{Douspis:2018xlj}
M.~Douspis, L.~Salvati and N.~Aghanim, \emph{{On the Tension between Large
  Scale Structures and Cosmic Microwave Background}},
  \href{https://doi.org/10.22323/1.335.0037}{\emph{PoS} {\bfseries EDSU2018}
  (2018) 037}, [\href{https://arxiv.org/abs/1901.05289}{{\ttfamily
  1901.05289}}].

\bibitem{Mandl:1985bg}
F.~Mandl and G.~Shaw, \emph{{Quantum Field Theory}}.
\newblock 1985.

\bibitem{Cheng:1985bj}
T.~P. Cheng and L.~F. Li, \emph{{Gauge Theory of Elementary particle physics}}.
\newblock 1984.

\bibitem{Kane:1987gb}
G.~L. Kane, \emph{{Modern Elementary Particle Physics}}.
\newblock Cambridge University Press, 2017.

\bibitem{Aitchison:1989bs}
I.~J.~R. Aitchison and A.~J.~G. Hey, \emph{{Gauge Theories in Particle Physics:
  A Practical Introduction}}.
\newblock 1989.

\bibitem{Collins:1989kn}
P.~D.~B. Collins, A.~D. Martin and E.~J. Squires, \emph{{Particle Physics and
  Cosmology}}.
\newblock 1989.

\bibitem{Donoghue:1992dd}
J.~F. Donoghue, E.~Golowich and B.~R. Holstein, \emph{{Dynamics of the standard
  model}}, \href{https://doi.org/10.1017/CBO9780511524370}{\emph{Camb. Monogr.
  Part. Phys. Nucl. Phys. Cosmol.} {\bfseries 2} (1992) 1--540}.

\bibitem{Cottingham:2007zz}
W.~N. Cottingham and D.~A. Greenwood, \emph{{An introduction to the standard
  model of particle physics}}.
\newblock Cambridge University Press, 2007.

\bibitem{Griffiths:2008zz}
D.~Griffiths, \emph{{Introduction to elementary particles}}.
\newblock 2008.

\bibitem{Langacker:2010zza}
P.~Langacker, \emph{{The standard model and beyond}}.
\newblock 2010.

\bibitem{Schwartz:2013pla}
M.~D. Schwartz, \emph{{Quantum Field Theory and the Standard Model}}.
\newblock Cambridge University Press, 2014.

\bibitem{Fukuda:1998mi}
{\scshape Super-Kamiokande} collaboration, Y.~Fukuda et~al., \emph{{Evidence
  for oscillation of atmospheric neutrinos}},
  \href{https://doi.org/10.1103/PhysRevLett.81.1562}{\emph{Phys. Rev. Lett.}
  {\bfseries 81} (1998) 1562--1567},
  [\href{https://arxiv.org/abs/hep-ex/9807003}{{\ttfamily hep-ex/9807003}}].

\bibitem{Minkowski:1977sc}
P.~Minkowski, \emph{{$\mu \to e\gamma$ at a Rate of One Out of $10^{9}$ Muon
  Decays?}}, \href{https://doi.org/10.1016/0370-2693(77)90435-X}{\emph{Phys.
  Lett.} {\bfseries 67B} (1977) 421--428}.

\bibitem{Yanagida:1979as}
T.~Yanagida, \emph{{Horizontal gauge symmetry and masses of neutrinos}},
  {\emph{Conf. Proc.} {\bfseries C7902131} (1979) 95--99}.

\bibitem{GellMann:1980vs}
M.~Gell-Mann, P.~Ramond and R.~Slansky, \emph{{Complex Spinors and Unified
  Theories}}, {\emph{Conf. Proc.} {\bfseries C790927} (1979) 315--321},
  [\href{https://arxiv.org/abs/1306.4669}{{\ttfamily 1306.4669}}].

\bibitem{Mohapatra:1979ia}
R.~N. Mohapatra and G.~Senjanović, \emph{{Neutrino Mass and Spontaneous Parity
  Nonconservation}},
  \href{https://doi.org/10.1103/PhysRevLett.44.912}{\emph{Phys. Rev. Lett.}
  {\bfseries 44} (1980) 912}.

\bibitem{Schechter:1980gr}
J.~Schechter and J.~W.~F. Valle, \emph{{Neutrino Masses in SU(2) x U(1)
  Theories}}, \href{https://doi.org/10.1103/PhysRevD.22.2227}{\emph{Phys. Rev.}
  {\bfseries D22} (1980) 2227}.

\bibitem{Magg:1980ut}
M.~Magg and C.~Wetterich, \emph{{Neutrino Mass Problem and Gauge Hierarchy}},
  \href{https://doi.org/10.1016/0370-2693(80)90825-4}{\emph{Phys. Lett.}
  {\bfseries 94B} (1980) 61--64}.

\bibitem{Lazarides:1980nt}
G.~Lazarides, Q.~Shafi and C.~Wetterich, \emph{{Proton Lifetime and Fermion
  Masses in an SO(10) Model}},
  \href{https://doi.org/10.1016/0550-3213(81)90354-0}{\emph{Nucl. Phys.}
  {\bfseries B181} (1981) 287--300}.

\bibitem{Mohapatra:1980yp}
R.~N. Mohapatra and G.~Senjanović, \emph{{Neutrino Masses and Mixings in Gauge
  Models with Spontaneous Parity Violation}},
  \href{https://doi.org/10.1103/PhysRevD.23.165}{\emph{Phys. Rev.} {\bfseries
  D23} (1981) 165}.

\bibitem{Wetterich:1981bx}
C.~Wetterich, \emph{{Neutrino Masses and the Scale of B-L Violation}},
  \href{https://doi.org/10.1016/0550-3213(81)90279-0}{\emph{Nucl. Phys.}
  {\bfseries B187} (1981) 343--375}.

\bibitem{Foot:1988aq}
R.~Foot, H.~Lew, X.~G. He and G.~C. Joshi, \emph{{Seesaw Neutrino Masses
  Induced by a Triplet of Leptons}},
  \href{https://doi.org/10.1007/BF01415558}{\emph{Z. Phys.} {\bfseries C44}
  (1989) 441}.

\bibitem{Ma:1998dn}
E.~Ma, \emph{{Pathways to naturally small neutrino masses}},
  \href{https://doi.org/10.1103/PhysRevLett.81.1171}{\emph{Phys. Rev. Lett.}
  {\bfseries 81} (1998) 1171--1174},
  [\href{https://arxiv.org/abs/hep-ph/9805219}{{\ttfamily hep-ph/9805219}}].

\bibitem{Ma:2002pf}
E.~Ma and D.~P. Roy, \emph{{Heavy triplet leptons and new gauge boson}},
  \href{https://doi.org/10.1016/S0550-3213(02)00815-5}{\emph{Nucl. Phys.}
  {\bfseries B644} (2002) 290--302},
  [\href{https://arxiv.org/abs/hep-ph/0206150}{{\ttfamily hep-ph/0206150}}].

\bibitem{Zee:1985rj}
A.~Zee, \emph{{Charged Scalar Field and Quantum Number Violations}},
  \href{https://doi.org/10.1016/0370-2693(85)90625-2}{\emph{Phys. Lett.}
  {\bfseries 161B} (1985) 141--145}.

\bibitem{Zee:1985id}
A.~Zee, \emph{{Quantum Numbers of Majorana Neutrino Masses}},
  \href{https://doi.org/10.1016/0550-3213(86)90475-X}{\emph{Nucl. Phys.}
  {\bfseries B264} (1986) 99--110}.

\bibitem{Babu:1988ki}
K.~S. Babu, \emph{{Model of 'Calculable' Majorana Neutrino Masses}},
  \href{https://doi.org/10.1016/0370-2693(88)91584-5}{\emph{Phys. Lett.}
  {\bfseries B203} (1988) 132--136}.

\bibitem{Asaka:2005an}
T.~Asaka, S.~Blanchet and M.~Shaposhnikov, \emph{{The nuMSM, dark matter and
  neutrino masses}},
  \href{https://doi.org/10.1016/j.physletb.2005.09.070}{\emph{Phys. Lett.}
  {\bfseries B631} (2005) 151--156},
  [\href{https://arxiv.org/abs/hep-ph/0503065}{{\ttfamily hep-ph/0503065}}].

\bibitem{Asaka:2005pn}
T.~Asaka and M.~Shaposhnikov, \emph{{The nuMSM, dark matter and baryon
  asymmetry of the universe}},
  \href{https://doi.org/10.1016/j.physletb.2005.06.020}{\emph{Phys. Lett.}
  {\bfseries B620} (2005) 17--26},
  [\href{https://arxiv.org/abs/hep-ph/0505013}{{\ttfamily hep-ph/0505013}}].

\bibitem{Boyarsky:2009ix}
A.~Boyarsky, O.~Ruchayskiy and M.~Shaposhnikov, \emph{{The Role of sterile
  neutrinos in cosmology and astrophysics}},
  \href{https://doi.org/10.1146/annurev.nucl.010909.083654}{\emph{Ann. Rev.
  Nucl. Part. Sci.} {\bfseries 59} (2009) 191--214},
  [\href{https://arxiv.org/abs/0901.0011}{{\ttfamily 0901.0011}}].

\bibitem{Boucenna:2014zba}
S.~M. Boucenna, S.~Morisi and J.~W.~F. Valle, \emph{{The low-scale approach to
  neutrino masses}}, \href{https://doi.org/10.1155/2014/831598}{\emph{Adv. High
  Energy Phys.} {\bfseries 2014} (2014) 831598},
  [\href{https://arxiv.org/abs/1404.3751}{{\ttfamily 1404.3751}}].

\bibitem{King:2015aea}
S.~F. King, \emph{{Models of Neutrino Mass, Mixing and CP Violation}},
  \href{https://doi.org/10.1088/0954-3899/42/12/123001}{\emph{J. Phys.}
  {\bfseries G42} (2015) 123001},
  [\href{https://arxiv.org/abs/1510.02091}{{\ttfamily 1510.02091}}].

\bibitem{King:2003jb}
S.~F. King, \emph{{Neutrino mass models}},
  \href{https://doi.org/10.1088/0034-4885/67/2/R01}{\emph{Rept. Prog. Phys.}
  {\bfseries 67} (2004) 107--158},
  [\href{https://arxiv.org/abs/hep-ph/0310204}{{\ttfamily hep-ph/0310204}}].

\bibitem{Cai:2017jrq}
Y.~Cai, J.~Herrero-García, M.~A. Schmidt, A.~Vicente and R.~R. Volkas,
  \emph{{From the trees to the forest: a review of radiative neutrino mass
  models}}, \href{https://doi.org/10.3389/fphy.2017.00063}{\emph{Front.in
  Phys.} {\bfseries 5} (2017) 63},
  [\href{https://arxiv.org/abs/1706.08524}{{\ttfamily 1706.08524}}].

\bibitem{Mohapatra:2006gs}
R.~N. Mohapatra and A.~Y. Smirnov, \emph{{Neutrino Mass and New Physics}},
  \href{https://doi.org/10.1146/annurev.nucl.56.080805.140534}{\emph{Ann. Rev.
  Nucl. Part. Sci.} {\bfseries 56} (2006) 569--628},
  [\href{https://arxiv.org/abs/hep-ph/0603118}{{\ttfamily hep-ph/0603118}}].

\bibitem{Altarelli:2010gt}
G.~Altarelli and F.~Feruglio, \emph{{Discrete Flavor Symmetries and Models of
  Neutrino Mixing}},
  \href{https://doi.org/10.1103/RevModPhys.82.2701}{\emph{Rev. Mod. Phys.}
  {\bfseries 82} (2010) 2701--2729},
  [\href{https://arxiv.org/abs/1002.0211}{{\ttfamily 1002.0211}}].

\bibitem{Meloni:2017cig}
D.~Meloni, \emph{{GUT and flavor models for neutrino masses and mixing}},
  \href{https://doi.org/10.3389/fphy.2017.00043}{\emph{Front.in Phys.}
  {\bfseries 5} (2017) 43}, [\href{https://arxiv.org/abs/1709.02662}{{\ttfamily
  1709.02662}}].

\bibitem{Capozzi:2018ubv}
F.~Capozzi, E.~Lisi, A.~Marrone and A.~Palazzo, \emph{{Current unknowns in the
  three neutrino framework}},
  \href{https://doi.org/10.1016/j.ppnp.2018.05.005}{\emph{Prog. Part. Nucl.
  Phys.} {\bfseries 102} (2018) 48--72},
  [\href{https://arxiv.org/abs/1804.09678}{{\ttfamily 1804.09678}}].

\bibitem{Hubble:1929ig}
E.~Hubble, \emph{{A relation between distance and radial velocity among
  extra-galactic nebulae}},
  \href{https://doi.org/10.1073/pnas.15.3.168}{\emph{Proc. Nat. Acad. Sci.}
  {\bfseries 15} (1929) 168--173}.

\bibitem{Lemaitre:1927zz}
G.~Lemaître, \emph{{A Homogeneous Universe of Constant Mass and Growing Radius
  Accounting for the Radial Velocity of Extragalactic Nebulae}},
  \href{https://doi.org/10.1007/s10714-013-1548-3}{\emph{Annales Soc. Sci.
  Bruxelles A} {\bfseries 47} (1927) 49--59}.

\bibitem{Zwicky:1933gu}
F.~Zwicky, \emph{{Die Rotverschiebung von extragalaktischen Nebeln}},
  \href{https://doi.org/10.1007/s10714-008-0707-4}{\emph{Helv. Phys. Acta}
  {\bfseries 6} (1933) 110--127}.

\bibitem{Alpher:1948ve}
R.~A. Alpher, H.~Bethe and G.~Gamow, \emph{{The origin of chemical elements}},
  \href{https://doi.org/10.1103/PhysRev.73.803}{\emph{Phys. Rev.} {\bfseries
  73} (1948) 803--804}.

\bibitem{Alpher:1948we}
R.~A. Alpher and R.~C. Herman, \emph{{On the Relative Abundance of the
  Elements}}, \href{https://doi.org/10.1103/PhysRev.74.1737}{\emph{Phys. Rev.}
  {\bfseries 74} (1948) 1737--1742}.

\bibitem{Penzias:1965wn}
A.~A. Penzias and R.~W. Wilson, \emph{{A Measurement of excess antenna
  temperature at 4080-Mc/s}},
  \href{https://doi.org/10.1086/148307}{\emph{Astrophys. J.} {\bfseries 142}
  (1965) 419--421}.

\bibitem{Dicke:1965zz}
R.~H. Dicke, P.~J.~E. Peebles, P.~G. Roll and D.~T. Wilkinson, \emph{{Cosmic
  Black-Body Radiation}},
  \href{https://doi.org/10.1086/148306}{\emph{Astrophys. J.} {\bfseries 142}
  (1965) 414--419}.

\bibitem{Fixsen:1996nj}
D.~J. Fixsen, E.~S. Cheng, J.~M. Gales, J.~C. Mather, R.~A. Shafer and E.~L.
  Wright, \emph{{The Cosmic Microwave Background spectrum from the full COBE
  FIRAS data set}}, \href{https://doi.org/10.1086/178173}{\emph{Astrophys. J.}
  {\bfseries 473} (1996) 576},
  [\href{https://arxiv.org/abs/astro-ph/9605054}{{\ttfamily
  astro-ph/9605054}}].

\bibitem{Smoot:1992td}
{\scshape COBE} collaboration, G.~F. Smoot et~al., \emph{{Structure in the COBE
  differential microwave radiometer first year maps}},
  \href{https://doi.org/10.1086/186504}{\emph{Astrophys. J.} {\bfseries 396}
  (1992) L1--L5}.

\bibitem{Bennett:2003ba}
{\scshape WMAP} collaboration, C.~L. Bennett et~al., \emph{{The Microwave
  Anisotropy Probe (MAP) mission}},
  \href{https://doi.org/10.1086/345346}{\emph{Astrophys. J.} {\bfseries 583}
  (2003) 1--23}, [\href{https://arxiv.org/abs/astro-ph/0301158}{{\ttfamily
  astro-ph/0301158}}].

\bibitem{Spergel:2006hy}
{\scshape WMAP} collaboration, D.~N. Spergel et~al., \emph{{Wilkinson Microwave
  Anisotropy Probe (WMAP) three year results: implications for cosmology}},
  \href{https://doi.org/10.1086/513700}{\emph{Astrophys. J. Suppl.} {\bfseries
  170} (2007) 377}, [\href{https://arxiv.org/abs/astro-ph/0603449}{{\ttfamily
  astro-ph/0603449}}].

\bibitem{Bennett:2012zja}
{\scshape WMAP} collaboration, C.~L. Bennett et~al., \emph{{Nine-Year Wilkinson
  Microwave Anisotropy Probe (WMAP) Observations: Final Maps and Results}},
  \href{https://doi.org/10.1088/0067-0049/208/2/20}{\emph{Astrophys. J. Suppl.}
  {\bfseries 208} (2013) 20},
  [\href{https://arxiv.org/abs/1212.5225}{{\ttfamily 1212.5225}}].

\bibitem{Planck:2006aa}
{\scshape Planck} collaboration, J.~Tauber, M.~Bersanelli, J.~M. Lamarre,
  G.~Efstathiou, C.~Lawrence, F.~Bouchet et~al., \emph{{The Scientific
  programme of Planck}},
  \href{https://arxiv.org/abs/astro-ph/0604069}{{\ttfamily astro-ph/0604069}}.

\bibitem{Ade:2013sjv}
{\scshape Planck} collaboration, P.~A.~R. Ade et~al., \emph{{Planck 2013
  results. I. Overview of products and scientific results}},
  \href{https://doi.org/10.1051/0004-6361/201321529}{\emph{Astron. Astrophys.}
  {\bfseries 571} (2014) A1},
  [\href{https://arxiv.org/abs/1303.5062}{{\ttfamily 1303.5062}}].

\bibitem{Ade:2013kta}
{\scshape Planck} collaboration, P.~A.~R. Ade et~al., \emph{{Planck 2013
  results. XV. CMB power spectra and likelihood}},
  \href{https://doi.org/10.1051/0004-6361/201321573}{\emph{Astron. Astrophys.}
  {\bfseries 571} (2014) A15},
  [\href{https://arxiv.org/abs/1303.5075}{{\ttfamily 1303.5075}}].

\bibitem{Ade:2013zuv}
{\scshape Planck} collaboration, P.~A.~R. Ade et~al., \emph{{Planck 2013
  results. XVI. Cosmological parameters}},
  \href{https://doi.org/10.1051/0004-6361/201321591}{\emph{Astron. Astrophys.}
  {\bfseries 571} (2014) A16},
  [\href{https://arxiv.org/abs/1303.5076}{{\ttfamily 1303.5076}}].

\bibitem{Adam:2015rua}
{\scshape Planck} collaboration, R.~Adam et~al., \emph{{Planck 2015 results. I.
  Overview of products and scientific results}},
  \href{https://doi.org/10.1051/0004-6361/201527101}{\emph{Astron. Astrophys.}
  {\bfseries 594} (2016) A1},
  [\href{https://arxiv.org/abs/1502.01582}{{\ttfamily 1502.01582}}].

\bibitem{Ade:2015xua}
{\scshape Planck} collaboration, P.~A.~R. Ade et~al., \emph{{Planck 2015
  results. XIII. Cosmological parameters}},
  \href{https://doi.org/10.1051/0004-6361/201525830}{\emph{Astron. Astrophys.}
  {\bfseries 594} (2016) A13},
  [\href{https://arxiv.org/abs/1502.01589}{{\ttfamily 1502.01589}}].

\bibitem{Aghanim:2015xee}
{\scshape Planck} collaboration, N.~Aghanim et~al., \emph{{Planck 2015 results.
  XI. CMB power spectra, likelihoods, and robustness of parameters}},
  \href{https://doi.org/10.1051/0004-6361/201526926}{\emph{Astron. Astrophys.}
  {\bfseries 594} (2016) A11},
  [\href{https://arxiv.org/abs/1507.02704}{{\ttfamily 1507.02704}}].

\bibitem{Akrami:2018vks}
{\scshape Planck} collaboration, Y.~Akrami et~al., \emph{{Planck 2018 results.
  I. Overview and the cosmological legacy of Planck}},
  \href{https://arxiv.org/abs/1807.06205}{{\ttfamily 1807.06205}}.

\bibitem{Aghanim:2018eyx}
{\scshape Planck} collaboration, N.~Aghanim et~al., \emph{{Planck 2018 results.
  VI. Cosmological parameters}},
  \href{https://arxiv.org/abs/1807.06209}{{\ttfamily 1807.06209}}.

\bibitem{Rubin:1970zza}
V.~C. Rubin and W.~K. Ford, Jr., \emph{{Rotation of the Andromeda Nebula from a
  Spectroscopic Survey of Emission Regions}},
  \href{https://doi.org/10.1086/150317}{\emph{Astrophys. J.} {\bfseries 159}
  (1970) 379--403}.

\bibitem{Rubin:1978kmz}
V.~C. Rubin, W.~K. Ford, Jr. and N.~Thonnard, \emph{{Extended rotation curves
  of high-luminosity spiral galaxies. IV. Systematic dynamical properties, Sa
  through Sc}}, \href{https://doi.org/10.1086/182804}{\emph{Astrophys. J.}
  {\bfseries 225} (1978) L107--L111}.

\bibitem{Rubin:1980zd}
V.~C. Rubin, N.~Thonnard and W.~K. Ford, Jr., \emph{{Rotational properties of
  21 SC galaxies with a large range of luminosities and radii, from NGC 4605 /R
  = 4kpc/ to UGC 2885 /R = 122 kpc/}},
  \href{https://doi.org/10.1086/158003}{\emph{Astrophys. J.} {\bfseries 238}
  (1980) 471}.

\bibitem{Rubin:1982kyu}
V.~C. Rubin, W.~K. Ford, Jr., N.~Thonnard and D.~Burstein, \emph{{Rotational
  properties of 23 SB galaxies}},
  \href{https://doi.org/10.1086/160355}{\emph{Astrophys. J.} {\bfseries 261}
  (1982) 439}.

\bibitem{Rubin:1985ze}
V.~C. Rubin, D.~Burstein, W.~K. Ford, Jr. and N.~Thonnard, \emph{{Rotation
  velocities of 16 SA galaxies and a comparison of Sa, Sb, and SC rotation
  properties}}, \href{https://doi.org/10.1086/162866}{\emph{Astrophys. J.}
  {\bfseries 289} (1985) 81}.

\bibitem{Riess:1998cb}
{\scshape Supernova Search Team} collaboration, A.~G. Riess et~al.,
  \emph{{Observational evidence from supernovae for an accelerating universe
  and a cosmological constant}},
  \href{https://doi.org/10.1086/300499}{\emph{Astron. J.} {\bfseries 116}
  (1998) 1009--1038}, [\href{https://arxiv.org/abs/astro-ph/9805201}{{\ttfamily
  astro-ph/9805201}}].

\bibitem{Perlmutter:1998np}
{\scshape Supernova Cosmology Project} collaboration, S.~Perlmutter et~al.,
  \emph{{Measurements of Omega and Lambda from 42 high redshift supernovae}},
  \href{https://doi.org/10.1086/307221}{\emph{Astrophys. J.} {\bfseries 517}
  (1999) 565--586}, [\href{https://arxiv.org/abs/astro-ph/9812133}{{\ttfamily
  astro-ph/9812133}}].

\bibitem{Peebles:2002gy}
P.~J.~E. Peebles and B.~Ratra, \emph{{The Cosmological constant and dark
  energy}}, \href{https://doi.org/10.1103/RevModPhys.75.559}{\emph{Rev. Mod.
  Phys.} {\bfseries 75} (2003) 559--606},
  [\href{https://arxiv.org/abs/astro-ph/0207347}{{\ttfamily
  astro-ph/0207347}}].

\bibitem{Huterer:2017buf}
D.~Huterer and D.~L. Shafer, \emph{{Dark energy two decades after: Observables,
  probes, consistency tests}},
  \href{https://doi.org/10.1088/1361-6633/aa997e}{\emph{Rept. Prog. Phys.}
  {\bfseries 81} (2018) 016901},
  [\href{https://arxiv.org/abs/1709.01091}{{\ttfamily 1709.01091}}].

\bibitem{Nojiri:2006ri}
S.~Nojiri and S.~D. Odintsov, \emph{{Introduction to modified gravity and
  gravitational alternative for dark energy}},
  \href{https://doi.org/10.1142/S0219887807001928}{\emph{Int. J. Geom. Meth.
  Mod. Phys.} {\bfseries 4} (2007) 115},
  [\href{https://arxiv.org/abs/hep-th/0601213}{{\ttfamily hep-th/0601213}}].

\bibitem{Silvestri:2009hh}
A.~Silvestri and M.~Trodden, \emph{{Approaches to Understanding Cosmic
  Acceleration}},
  \href{https://doi.org/10.1088/0034-4885/72/9/096901}{\emph{Rept. Prog. Phys.}
  {\bfseries 72} (2009) 096901},
  [\href{https://arxiv.org/abs/0904.0024}{{\ttfamily 0904.0024}}].

\bibitem{Tsujikawa:2010zza}
S.~Tsujikawa, \emph{{Modified gravity models of dark energy}},
  \href{https://doi.org/10.1007/978-3-642-10598-2_3}{\emph{Lect. Notes Phys.}
  {\bfseries 800} (2010) 99--145},
  [\href{https://arxiv.org/abs/1101.0191}{{\ttfamily 1101.0191}}].

\bibitem{Clifton:2011jh}
T.~Clifton, P.~G. Ferreira, A.~Padilla and C.~Skordis, \emph{{Modified Gravity
  and Cosmology}},
  \href{https://doi.org/10.1016/j.physrep.2012.01.001}{\emph{Phys. Rept.}
  {\bfseries 513} (2012) 1--189},
  [\href{https://arxiv.org/abs/1106.2476}{{\ttfamily 1106.2476}}].

\bibitem{Joyce:2016vqv}
A.~Joyce, L.~Lombriser and F.~Schmidt, \emph{{Dark Energy Versus Modified
  Gravity}},
  \href{https://doi.org/10.1146/annurev-nucl-102115-044553}{\emph{Ann. Rev.
  Nucl. Part. Sci.} {\bfseries 66} (2016) 95--122},
  [\href{https://arxiv.org/abs/1601.06133}{{\ttfamily 1601.06133}}].

\bibitem{York:2000gk}
{\scshape SDSS} collaboration, D.~G. York et~al., \emph{{The Sloan Digital Sky
  Survey: Technical Summary}},
  \href{https://doi.org/10.1086/301513}{\emph{Astron. J.} {\bfseries 120}
  (2000) 1579--1587}, [\href{https://arxiv.org/abs/astro-ph/0006396}{{\ttfamily
  astro-ph/0006396}}].

\bibitem{Eisenstein:2005su}
{\scshape SDSS} collaboration, D.~J. Eisenstein et~al., \emph{{Detection of the
  Baryon Acoustic Peak in the Large-Scale Correlation Function of SDSS Luminous
  Red Galaxies}}, \href{https://doi.org/10.1086/466512}{\emph{Astrophys. J.}
  {\bfseries 633} (2005) 560--574},
  [\href{https://arxiv.org/abs/astro-ph/0501171}{{\ttfamily
  astro-ph/0501171}}].

\bibitem{Abbott:2016blz}
{\scshape LIGO Scientific, Virgo} collaboration, B.~P. Abbott et~al.,
  \emph{{Observation of Gravitational Waves from a Binary Black Hole Merger}},
  \href{https://doi.org/10.1103/PhysRevLett.116.061102}{\emph{Phys. Rev. Lett.}
  {\bfseries 116} (2016) 061102},
  [\href{https://arxiv.org/abs/1602.03837}{{\ttfamily 1602.03837}}].

\bibitem{Abbott:2016nmj}
{\scshape LIGO Scientific, Virgo} collaboration, B.~P. Abbott et~al.,
  \emph{{GW151226: Observation of Gravitational Waves from a 22-Solar-Mass
  Binary Black Hole Coalescence}},
  \href{https://doi.org/10.1103/PhysRevLett.116.241103}{\emph{Phys. Rev. Lett.}
  {\bfseries 116} (2016) 241103},
  [\href{https://arxiv.org/abs/1606.04855}{{\ttfamily 1606.04855}}].

\bibitem{Abramovici:1992ah}
A.~Abramovici et~al., \emph{{LIGO: The Laser interferometer gravitational wave
  observatory}},
  \href{https://doi.org/10.1126/science.256.5055.325}{\emph{Science} {\bfseries
  256} (1992) 325--333}.

\bibitem{Abbott:2007kv}
{\scshape LIGO Scientific} collaboration, B.~P. Abbott et~al., \emph{{LIGO: The
  Laser interferometer gravitational-wave observatory}},
  \href{https://doi.org/10.1088/0034-4885/72/7/076901}{\emph{Rept. Prog. Phys.}
  {\bfseries 72} (2009) 076901},
  [\href{https://arxiv.org/abs/0711.3041}{{\ttfamily 0711.3041}}].

\bibitem{Evans:2016mbw}
{\scshape LIGO Scientific} collaboration, B.~P. Abbott et~al., \emph{{Exploring
  the Sensitivity of Next Generation Gravitational Wave Detectors}},
  \href{https://doi.org/10.1088/1361-6382/aa51f4}{\emph{Class. Quant. Grav.}
  {\bfseries 34} (2017) 044001},
  [\href{https://arxiv.org/abs/1607.08697}{{\ttfamily 1607.08697}}].

\bibitem{TheLIGOScientific:2017qsa}
{\scshape LIGO Scientific, Virgo} collaboration, B.~P. Abbott et~al.,
  \emph{{GW170817: Observation of Gravitational Waves from a Binary Neutron
  Star Inspiral}},
  \href{https://doi.org/10.1103/PhysRevLett.119.161101}{\emph{Phys. Rev. Lett.}
  {\bfseries 119} (2017) 161101},
  [\href{https://arxiv.org/abs/1710.05832}{{\ttfamily 1710.05832}}].

\bibitem{GBM:2017lvd}
{\scshape LIGO Scientific, Virgo, Fermi GBM, INTEGRAL, IceCube, AstroSat
  Cadmium Zinc Telluride Imager Team, IPN, Insight-Hxmt, ANTARES, Swift, AGILE
  Team, 1M2H Team, Dark Energy Camera GW-EM, DES, DLT40, GRAWITA, Fermi-LAT,
  ATCA, ASKAP, Las Cumbres Observatory Group, OzGrav, DWF (Deeper Wider Faster
  Program), AST3, CAASTRO, VINROUGE, MASTER, J-GEM, GROWTH, JAGWAR,
  CaltechNRAO, TTU-NRAO, NuSTAR, Pan-STARRS, MAXI Team, TZAC Consortium, KU,
  Nordic Optical Telescope, ePESSTO, GROND, Texas Tech University, SALT Group,
  TOROS, BOOTES, MWA, CALET, IKI-GW Follow-up, H.E.S.S., LOFAR, LWA, HAWC,
  Pierre Auger, ALMA, Euro VLBI Team, Pi of Sky, Chandra Team at McGill
  University, DFN, ATLAS Telescopes, High Time Resolution Universe Survey,
  RIMAS, RATIR, SKA South Africa/MeerKAT} collaboration, B.~P. Abbott et~al.,
  \emph{{Multi-messenger Observations of a Binary Neutron Star Merger}},
  \href{https://doi.org/10.3847/2041-8213/aa91c9}{\emph{Astrophys. J.}
  {\bfseries 848} (2017) L12},
  [\href{https://arxiv.org/abs/1710.05833}{{\ttfamily 1710.05833}}].

\bibitem{Monitor:2017mdv}
{\scshape LIGO Scientific, Virgo, Fermi-GBM, INTEGRAL} collaboration, B.~P.
  Abbott et~al., \emph{{Gravitational Waves and Gamma-rays from a Binary
  Neutron Star Merger: GW170817 and GRB 170817A}},
  \href{https://doi.org/10.3847/2041-8213/aa920c}{\emph{Astrophys. J.}
  {\bfseries 848} (2017) L13},
  [\href{https://arxiv.org/abs/1710.05834}{{\ttfamily 1710.05834}}].

\bibitem{TheLIGOScientific:2016src}
{\scshape LIGO Scientific, Virgo} collaboration, B.~P. Abbott et~al.,
  \emph{{Tests of general relativity with GW150914}},
  \href{https://doi.org/10.1103/PhysRevLett.116.221101,
  10.1103/PhysRevLett.121.129902}{\emph{Phys. Rev. Lett.} {\bfseries 116}
  (2016) 221101}, [\href{https://arxiv.org/abs/1602.03841}{{\ttfamily
  1602.03841}}]. [Erratum: Phys. Rev. Lett.121,no.12,129902(2018)].

\bibitem{Blas:2016qmn}
D.~Blas, M.~M. Ivanov, I.~Sawicki and S.~Sibiryakov, \emph{{On constraining the
  speed of gravitational waves following GW150914}},
  \href{https://doi.org/10.1134/S0021364016100040,
  10.7868/S0370274X16100039}{\emph{JETP Lett.} {\bfseries 103} (2016)
  624--626}, [\href{https://arxiv.org/abs/1602.04188}{{\ttfamily 1602.04188}}].

\bibitem{Creminelli:2017sry}
P.~Creminelli and F.~Vernizzi, \emph{{Dark Energy after GW170817 and
  GRB170817A}},
  \href{https://doi.org/10.1103/PhysRevLett.119.251302}{\emph{Phys. Rev. Lett.}
  {\bfseries 119} (2017) 251302},
  [\href{https://arxiv.org/abs/1710.05877}{{\ttfamily 1710.05877}}].

\bibitem{Sakstein:2017xjx}
J.~Sakstein and B.~Jain, \emph{{Implications of the Neutron Star Merger
  GW170817 for Cosmological Scalar-Tensor Theories}},
  \href{https://doi.org/10.1103/PhysRevLett.119.251303}{\emph{Phys. Rev. Lett.}
  {\bfseries 119} (2017) 251303},
  [\href{https://arxiv.org/abs/1710.05893}{{\ttfamily 1710.05893}}].

\bibitem{Ezquiaga:2017ekz}
J.~M. Ezquiaga and M.~Zumalacárregui, \emph{{Dark Energy After GW170817: Dead
  Ends and the Road Ahead}},
  \href{https://doi.org/10.1103/PhysRevLett.119.251304}{\emph{Phys. Rev. Lett.}
  {\bfseries 119} (2017) 251304},
  [\href{https://arxiv.org/abs/1710.05901}{{\ttfamily 1710.05901}}].

\bibitem{Baker:2017hug}
T.~Baker, E.~Bellini, P.~G. Ferreira, M.~Lagos, J.~Noller and I.~Sawicki,
  \emph{{Strong constraints on cosmological gravity from GW170817 and GRB
  170817A}}, \href{https://doi.org/10.1103/PhysRevLett.119.251301}{\emph{Phys.
  Rev. Lett.} {\bfseries 119} (2017) 251301},
  [\href{https://arxiv.org/abs/1710.06394}{{\ttfamily 1710.06394}}].

\bibitem{Boran:2017rdn}
S.~Boran, S.~Desai, E.~O. Kahya and R.~P. Woodard, \emph{{GW170817 Falsifies
  Dark Matter Emulators}},
  \href{https://doi.org/10.1103/PhysRevD.97.041501}{\emph{Phys. Rev.}
  {\bfseries D97} (2018) 041501},
  [\href{https://arxiv.org/abs/1710.06168}{{\ttfamily 1710.06168}}].

\bibitem{Green:2017qcv}
M.~A. Green, J.~W. Moffat and V.~T. Toth, \emph{{Modified Gravity (MOG), the
  speed of gravitational radiation and the event GW170817/GRB170817A}},
  \href{https://doi.org/10.1016/j.physletb.2018.03.015}{\emph{Phys. Lett.}
  {\bfseries B780} (2018) 300--302},
  [\href{https://arxiv.org/abs/1710.11177}{{\ttfamily 1710.11177}}].

\bibitem{Nojiri:2017hai}
S.~Nojiri and S.~D. Odintsov, \emph{{Cosmological Bound from the Neutron Star
  Merger GW170817 in scalar-tensor and $F(R)$ gravity theories}},
  \href{https://doi.org/10.1016/j.physletb.2018.01.078}{\emph{Phys. Lett.}
  {\bfseries B779} (2018) 425--429},
  [\href{https://arxiv.org/abs/1711.00492}{{\ttfamily 1711.00492}}].

\bibitem{Arai:2017hxj}
S.~Arai and A.~Nishizawa, \emph{{Generalized framework for testing gravity with
  gravitational-wave propagation. II. Constraints on Horndeski theory}},
  \href{https://doi.org/10.1103/PhysRevD.97.104038}{\emph{Phys. Rev.}
  {\bfseries D97} (2018) 104038},
  [\href{https://arxiv.org/abs/1711.03776}{{\ttfamily 1711.03776}}].

\bibitem{Jana:2017ost}
S.~Jana, G.~K. Chakravarty and S.~Mohanty, \emph{{Constraints on Born-Infeld
  gravity from the speed of gravitational waves after GW170817 and GRB
  170817A}}, \href{https://doi.org/10.1103/PhysRevD.97.084011}{\emph{Phys.
  Rev.} {\bfseries D97} (2018) 084011},
  [\href{https://arxiv.org/abs/1711.04137}{{\ttfamily 1711.04137}}].

\bibitem{Amendola:2017orw}
L.~Amendola, M.~Kunz, I.~D. Saltas and I.~Sawicki, \emph{{Fate of Large-Scale
  Structure in Modified Gravity After GW170817 and GRB170817A}},
  \href{https://doi.org/10.1103/PhysRevLett.120.131101}{\emph{Phys. Rev. Lett.}
  {\bfseries 120} (2018) 131101},
  [\href{https://arxiv.org/abs/1711.04825}{{\ttfamily 1711.04825}}].

\bibitem{Visinelli:2017bny}
L.~Visinelli, N.~Bolis and S.~Vagnozzi, \emph{{Brane-world extra dimensions in
  light of GW170817}},
  \href{https://doi.org/10.1103/PhysRevD.97.064039}{\emph{Phys. Rev.}
  {\bfseries D97} (2018) 064039},
  [\href{https://arxiv.org/abs/1711.06628}{{\ttfamily 1711.06628}}].

\bibitem{Crisostomi:2017lbg}
M.~Crisostomi and K.~Koyama, \emph{{Vainshtein mechanism after GW170817}},
  \href{https://doi.org/10.1103/PhysRevD.97.021301}{\emph{Phys. Rev.}
  {\bfseries D97} (2018) 021301},
  [\href{https://arxiv.org/abs/1711.06661}{{\ttfamily 1711.06661}}].

\bibitem{Langlois:2017dyl}
D.~Langlois, R.~Saito, D.~Yamauchi and K.~Noui, \emph{{Scalar-tensor theories
  and modified gravity in the wake of GW170817}},
  \href{https://doi.org/10.1103/PhysRevD.97.061501}{\emph{Phys. Rev.}
  {\bfseries D97} (2018) 061501},
  [\href{https://arxiv.org/abs/1711.07403}{{\ttfamily 1711.07403}}].

\bibitem{Gumrukcuoglu:2017ijh}
A.~E. Gümrükçüoğlu, M.~Saravani and T.~P. Sotiriou, \emph{{Hořava gravity
  after GW170817}},
  \href{https://doi.org/10.1103/PhysRevD.97.024032}{\emph{Phys. Rev.}
  {\bfseries D97} (2018) 024032},
  [\href{https://arxiv.org/abs/1711.08845}{{\ttfamily 1711.08845}}].

\bibitem{Kreisch:2017uet}
C.~D. Kreisch and E.~Komatsu, \emph{{Cosmological Constraints on Horndeski
  Gravity in Light of GW170817}},
  \href{https://doi.org/10.1088/1475-7516/2018/12/030}{\emph{JCAP} {\bfseries
  1812} (2018) 030}, [\href{https://arxiv.org/abs/1712.02710}{{\ttfamily
  1712.02710}}].

\bibitem{Bartolo:2017ibw}
N.~Bartolo, P.~Karmakar, S.~Matarrese and M.~Scomparin, \emph{{Cosmic
  structures and gravitational waves in ghost-free scalar-tensor theories of
  gravity}}, \href{https://doi.org/10.1088/1475-7516/2018/05/048}{\emph{JCAP}
  {\bfseries 1805} (2018) 048},
  [\href{https://arxiv.org/abs/1712.04002}{{\ttfamily 1712.04002}}].

\bibitem{Dima:2017pwp}
A.~Dima and F.~Vernizzi, \emph{{Vainshtein Screening in Scalar-Tensor Theories
  before and after GW170817: Constraints on Theories beyond Horndeski}},
  \href{https://doi.org/10.1103/PhysRevD.97.101302}{\emph{Phys. Rev.}
  {\bfseries D97} (2018) 101302},
  [\href{https://arxiv.org/abs/1712.04731}{{\ttfamily 1712.04731}}].

\bibitem{Pardo:2018ipy}
K.~Pardo, M.~Fishbach, D.~E. Holz and D.~N. Spergel, \emph{{Limits on the
  number of spacetime dimensions from GW170817}},
  \href{https://doi.org/10.1088/1475-7516/2018/07/048}{\emph{JCAP} {\bfseries
  1807} (2018) 048}, [\href{https://arxiv.org/abs/1801.08160}{{\ttfamily
  1801.08160}}].

\bibitem{Casalino:2018tcd}
A.~Casalino, M.~Rinaldi, L.~Sebastiani and S.~Vagnozzi, \emph{{Mimicking dark
  matter and dark energy in a mimetic model compatible with GW170817}},
  \href{https://doi.org/10.1016/j.dark.2018.10.001}{\emph{Phys. Dark Univ.}
  {\bfseries 22} (2018) 108},
  [\href{https://arxiv.org/abs/1803.02620}{{\ttfamily 1803.02620}}].

\bibitem{Jana:2018djs}
S.~Jana and S.~Mohanty, \emph{{Constraints on $f(R)$ theories of gravity from
  GW170817}}, \href{https://doi.org/10.1103/PhysRevD.99.044056}{\emph{Phys.
  Rev.} {\bfseries D99} (2019) 044056},
  [\href{https://arxiv.org/abs/1807.04060}{{\ttfamily 1807.04060}}].

\bibitem{Ezquiaga:2018btd}
J.~M. Ezquiaga and M.~Zumalacárregui, \emph{{Dark Energy in light of
  Multi-Messenger Gravitational-Wave astronomy}},
  \href{https://doi.org/10.3389/fspas.2018.00044}{\emph{Front. Astron. Space
  Sci.} {\bfseries 5} (2018) 44},
  [\href{https://arxiv.org/abs/1807.09241}{{\ttfamily 1807.09241}}].

\bibitem{Abbott:2018lct}
{\scshape LIGO Scientific, Virgo} collaboration, B.~P. Abbott et~al.,
  \emph{{Tests of General Relativity with GW170817}},
  \href{https://arxiv.org/abs/1811.00364}{{\ttfamily 1811.00364}}.

\bibitem{Casalino:2018wnc}
A.~Casalino, M.~Rinaldi, L.~Sebastiani and S.~Vagnozzi, \emph{{Alive and well:
  mimetic gravity and a higher-order extension in light of GW170817}},
  \href{https://doi.org/10.1088/1361-6382/aaf1fd}{\emph{Class. Quant. Grav.}
  {\bfseries 36} (2019) 017001},
  [\href{https://arxiv.org/abs/1811.06830}{{\ttfamily 1811.06830}}].

\bibitem{Raveri:2014eea}
M.~Raveri, C.~Baccigalupi, A.~Silvestri and S.-Y. Zhou, \emph{{Measuring the
  speed of cosmological gravitational waves}},
  \href{https://doi.org/10.1103/PhysRevD.91.061501}{\emph{Phys. Rev.}
  {\bfseries D91} (2015) 061501},
  [\href{https://arxiv.org/abs/1405.7974}{{\ttfamily 1405.7974}}].

\bibitem{Lombriser:2015sxa}
L.~Lombriser and A.~Taylor, \emph{{Breaking a Dark Degeneracy with
  Gravitational Waves}},
  \href{https://doi.org/10.1088/1475-7516/2016/03/031}{\emph{JCAP} {\bfseries
  1603} (2016) 031}, [\href{https://arxiv.org/abs/1509.08458}{{\ttfamily
  1509.08458}}].

\bibitem{Lombriser:2016yzn}
L.~Lombriser and N.~A. Lima, \emph{{Challenges to Self-Acceleration in Modified
  Gravity from Gravitational Waves and Large-Scale Structure}},
  \href{https://doi.org/10.1016/j.physletb.2016.12.048}{\emph{Phys. Lett.}
  {\bfseries B765} (2017) 382--385},
  [\href{https://arxiv.org/abs/1602.07670}{{\ttfamily 1602.07670}}].

\bibitem{Bettoni:2016mij}
D.~Bettoni, J.~M. Ezquiaga, K.~Hinterbichler and M.~Zumalacárregui,
  \emph{{Speed of Gravitational Waves and the Fate of Scalar-Tensor Gravity}},
  \href{https://doi.org/10.1103/PhysRevD.95.084029}{\emph{Phys. Rev.}
  {\bfseries D95} (2017) 084029},
  [\href{https://arxiv.org/abs/1608.01982}{{\ttfamily 1608.01982}}].

\bibitem{Alonso:2016suf}
D.~Alonso, E.~Bellini, P.~G. Ferreira and M.~Zumalacárregui,
  \emph{{Observational future of cosmological scalar-tensor theories}},
  \href{https://doi.org/10.1103/PhysRevD.95.063502}{\emph{Phys. Rev.}
  {\bfseries D95} (2017) 063502},
  [\href{https://arxiv.org/abs/1610.09290}{{\ttfamily 1610.09290}}].

\bibitem{Schutz:1986gp}
B.~F. Schutz, \emph{{Determining the Hubble Constant from Gravitational Wave
  Observations}}, \href{https://doi.org/10.1038/323310a0}{\emph{Nature}
  {\bfseries 323} (1986) 310--311}.

\bibitem{Nissanke:2009kt}
S.~Nissanke, D.~E. Holz, S.~A. Hughes, N.~Dalal and J.~L. Sievers,
  \emph{{Exploring short gamma-ray bursts as gravitational-wave standard
  sirens}}, \href{https://doi.org/10.1088/0004-637X/725/1/496}{\emph{Astrophys.
  J.} {\bfseries 725} (2010) 496--514},
  [\href{https://arxiv.org/abs/0904.1017}{{\ttfamily 0904.1017}}].

\bibitem{Tamanini:2016zlh}
N.~Tamanini, C.~Caprini, E.~Barausse, A.~Sesana, A.~Klein and A.~Petiteau,
  \emph{{Science with the space-based interferometer eLISA. III: Probing the
  expansion of the Universe using gravitational wave standard sirens}},
  \href{https://doi.org/10.1088/1475-7516/2016/04/002}{\emph{JCAP} {\bfseries
  1604} (2016) 002}, [\href{https://arxiv.org/abs/1601.07112}{{\ttfamily
  1601.07112}}].

\bibitem{Caprini:2016qxs}
C.~Caprini and N.~Tamanini, \emph{{Constraining early and interacting dark
  energy with gravitational wave standard sirens: the potential of the eLISA
  mission}}, \href{https://doi.org/10.1088/1475-7516/2016/10/006}{\emph{JCAP}
  {\bfseries 1610} (2016) 006},
  [\href{https://arxiv.org/abs/1607.08755}{{\ttfamily 1607.08755}}].

\bibitem{Cai:2016sby}
R.-G. Cai and T.~Yang, \emph{{Estimating cosmological parameters by the
  simulated data of gravitational waves from the Einstein Telescope}},
  \href{https://doi.org/10.1103/PhysRevD.95.044024}{\emph{Phys. Rev.}
  {\bfseries D95} (2017) 044024},
  [\href{https://arxiv.org/abs/1608.08008}{{\ttfamily 1608.08008}}].

\bibitem{Cai:2017yww}
R.-G. Cai, N.~Tamanini and T.~Yang, \emph{{Reconstructing the dark sector
  interaction with LISA}},
  \href{https://doi.org/10.1088/1475-7516/2017/05/031}{\emph{JCAP} {\bfseries
  1705} (2017) 031}, [\href{https://arxiv.org/abs/1703.07323}{{\ttfamily
  1703.07323}}].

\bibitem{Chen:2017rfc}
H.-Y. Chen, M.~Fishbach and D.~E. Holz, \emph{{A two per cent Hubble constant
  measurement from standard sirens within five years}},
  \href{https://doi.org/10.1038/s41586-018-0606-0}{\emph{Nature} {\bfseries
  562} (2018) 545--547}, [\href{https://arxiv.org/abs/1712.06531}{{\ttfamily
  1712.06531}}].

\bibitem{Feeney:2018mkj}
S.~M. Feeney, H.~V. Peiris, A.~R. Williamson, S.~M. Nissanke, D.~J. Mortlock,
  J.~Alsing et~al., \emph{{Prospects for resolving the Hubble constant tension
  with standard sirens}},
  \href{https://doi.org/10.1103/PhysRevLett.122.061105}{\emph{Phys. Rev. Lett.}
  {\bfseries 122} (2019) 061105},
  [\href{https://arxiv.org/abs/1802.03404}{{\ttfamily 1802.03404}}].

\bibitem{Wang:2018lun}
L.-F. Wang, X.-N. Zhang, J.-F. Zhang and X.~Zhang, \emph{{Impacts of
  gravitational-wave standard siren observation of the Einstein Telescope on
  weighing neutrinos in cosmology}},
  \href{https://doi.org/10.1016/j.physletb.2018.05.027}{\emph{Phys. Lett.}
  {\bfseries B782} (2018) 87--93},
  [\href{https://arxiv.org/abs/1802.04720}{{\ttfamily 1802.04720}}].

\bibitem{Belgacem:2018lbp}
E.~Belgacem, Y.~Dirian, S.~Foffa and M.~Maggiore, \emph{{Modified
  gravitational-wave propagation and standard sirens}},
  \href{https://doi.org/10.1103/PhysRevD.98.023510}{\emph{Phys. Rev.}
  {\bfseries D98} (2018) 023510},
  [\href{https://arxiv.org/abs/1805.08731}{{\ttfamily 1805.08731}}].

\bibitem{DiValentino:2018jbh}
E.~Di~Valentino, D.~E. Holz, A.~Melchiorri and F.~Renzi, \emph{{The
  cosmological impact of future constraints on $H_0$ from gravitational-wave
  standard sirens}},
  \href{https://doi.org/10.1103/PhysRevD.98.083523}{\emph{Phys. Rev.}
  {\bfseries D98} (2018) 083523},
  [\href{https://arxiv.org/abs/1806.07463}{{\ttfamily 1806.07463}}].

\bibitem{Nunes:2018evm}
R.~C. Nunes, S.~Pan and E.~N. Saridakis, \emph{{New observational constraints
  on $f(T)$ gravity through gravitational-wave astronomy}},
  \href{https://doi.org/10.1103/PhysRevD.98.104055}{\emph{Phys. Rev.}
  {\bfseries D98} (2018) 104055},
  [\href{https://arxiv.org/abs/1810.03942}{{\ttfamily 1810.03942}}].

\bibitem{Du:2018tia}
M.~Du, W.~Yang, L.~Xu, S.~Pan and D.~F. Mota, \emph{{Future Constraints on
  Dynamical Dark-Energy using Gravitational-Wave Standard Sirens}},
  \href{https://arxiv.org/abs/1812.01440}{{\ttfamily 1812.01440}}.

\bibitem{Calcagni:2019kzo}
G.~Calcagni, S.~Kuroyanagi, S.~Marsat, M.~Sakellariadou, N.~Tamanini and
  G.~Tasinato, \emph{{Gravitational-wave luminosity distance in quantum
  gravity}},  \href{https://arxiv.org/abs/1904.00384}{{\ttfamily 1904.00384}}.

\bibitem{Yang:2019bpr}
W.~Yang, S.~Pan, E.~Di~Valentino, B.~Wang and A.~Wang, \emph{{Forecasting
  Interacting Vacuum-Energy Models using Gravitational Waves}},
  \href{https://arxiv.org/abs/1904.11980}{{\ttfamily 1904.11980}}.

\bibitem{Furlanetto:2006jb}
S.~Furlanetto, S.~P. Oh and F.~Briggs, \emph{{Cosmology at Low Frequencies: The
  21 cm Transition and the High-Redshift Universe}},
  \href{https://doi.org/10.1016/j.physrep.2006.08.002}{\emph{Phys. Rept.}
  {\bfseries 433} (2006) 181--301},
  [\href{https://arxiv.org/abs/astro-ph/0608032}{{\ttfamily
  astro-ph/0608032}}].

\bibitem{Morales:2009gs}
M.~F. Morales and J.~S.~B. Wyithe, \emph{{Reionization and Cosmology with 21 cm
  Fluctuations}},
  \href{https://doi.org/10.1146/annurev-astro-081309-130936}{\emph{Ann. Rev.
  Astron. Astrophys.} {\bfseries 48} (2010) 127--171},
  [\href{https://arxiv.org/abs/0910.3010}{{\ttfamily 0910.3010}}].

\bibitem{Pritchard:2011xb}
J.~R. Pritchard and A.~Loeb, \emph{{21-cm cosmology}},
  \href{https://doi.org/10.1088/0034-4885/75/8/086901}{\emph{Rept. Prog. Phys.}
  {\bfseries 75} (2012) 086901},
  [\href{https://arxiv.org/abs/1109.6012}{{\ttfamily 1109.6012}}].

\bibitem{Munoz:2015eqa}
J.~B. Muñoz, Y.~Ali-Haïmoud and M.~Kamionkowski, \emph{{Primordial
  non-gaussianity from the bispectrum of 21-cm fluctuations in the dark ages}},
  \href{https://doi.org/10.1103/PhysRevD.92.083508}{\emph{Phys. Rev.}
  {\bfseries D92} (2015) 083508},
  [\href{https://arxiv.org/abs/1506.04152}{{\ttfamily 1506.04152}}].

\bibitem{Villaescusa-Navarro:2016kbz}
F.~Villaescusa-Navarro, D.~Alonso and M.~Viel, \emph{{Baryonic acoustic
  oscillations from 21 cm intensity mapping: the Square Kilometre Array case}},
  \href{https://doi.org/10.1093/mnras/stw3224}{\emph{Mon. Not. Roy. Astron.
  Soc.} {\bfseries 466} (2017) 2736--2751},
  [\href{https://arxiv.org/abs/1609.00019}{{\ttfamily 1609.00019}}].

\bibitem{Obuljen:2017jiy}
A.~Obuljen, E.~Castorina, F.~Villaescusa-Navarro and M.~Viel,
  \emph{{High-redshift post-reionization cosmology with 21cm intensity
  mapping}}, \href{https://doi.org/10.1088/1475-7516/2018/05/004}{\emph{JCAP}
  {\bfseries 1805} (2018) 004},
  [\href{https://arxiv.org/abs/1709.07893}{{\ttfamily 1709.07893}}].

\bibitem{Barkana:2018lgd}
R.~Barkana, \emph{{Possible interaction between baryons and dark-matter
  particles revealed by the first stars}},
  \href{https://doi.org/10.1038/nature25791}{\emph{Nature} {\bfseries 555}
  (2018) 71--74}, [\href{https://arxiv.org/abs/1803.06698}{{\ttfamily
  1803.06698}}].

\bibitem{Barkana:2018cct}
R.~Barkana, N.~J. Outmezguine, D.~Redigolo and T.~Volansky, \emph{{Strong
  constraints on light dark matter interpretation of the EDGES signal}},
  \href{https://doi.org/10.1103/PhysRevD.98.103005}{\emph{Phys. Rev.}
  {\bfseries D98} (2018) 103005},
  [\href{https://arxiv.org/abs/1803.03091}{{\ttfamily 1803.03091}}].

\bibitem{Fraser:2018acy}
S.~Fraser et~al., \emph{{The EDGES 21 cm Anomaly and Properties of Dark
  Matter}}, \href{https://doi.org/10.1016/j.physletb.2018.08.035}{\emph{Phys.
  Lett.} {\bfseries B785} (2018) 159--164},
  [\href{https://arxiv.org/abs/1803.03245}{{\ttfamily 1803.03245}}].

\bibitem{Munoz:2018jwq}
J.~B. Muñoz, C.~Dvorkin and A.~Loeb, \emph{{21-cm Fluctuations from Charged
  Dark Matter}},
  \href{https://doi.org/10.1103/PhysRevLett.121.121301}{\emph{Phys. Rev. Lett.}
  {\bfseries 121} (2018) 121301},
  [\href{https://arxiv.org/abs/1804.01092}{{\ttfamily 1804.01092}}].

\bibitem{Kovetz:2018zan}
E.~D. Kovetz, V.~Poulin, V.~Gluscevic, K.~K. Boddy, R.~Barkana and
  M.~Kamionkowski, \emph{{Tighter limits on dark matter explanations of the
  anomalous EDGES 21 cm signal}},
  \href{https://doi.org/10.1103/PhysRevD.98.103529}{\emph{Phys. Rev.}
  {\bfseries D98} (2018) 103529},
  [\href{https://arxiv.org/abs/1807.11482}{{\ttfamily 1807.11482}}].

\bibitem{Bowman:2018yin}
J.~D. Bowman, A.~E.~E. Rogers, R.~A. Monsalve, T.~J. Mozdzen and N.~Mahesh,
  \emph{{An absorption profile centred at 78 megahertz in the sky-averaged
  spectrum}}, \href{https://doi.org/10.1038/nature25792}{\emph{Nature}
  {\bfseries 555} (2018) 67--70},
  [\href{https://arxiv.org/abs/1810.05912}{{\ttfamily 1810.05912}}].

\bibitem{Nebrin:2018vqt}
O.~Nebrin, R.~Ghara and G.~Mellema, \emph{{Fuzzy dark matter at cosmic dawn:
  new 21-cm constraints}},
  \href{https://doi.org/10.1088/1475-7516/2019/04/051}{\emph{JCAP} {\bfseries
  1904} (2019) 051}, [\href{https://arxiv.org/abs/1812.09760}{{\ttfamily
  1812.09760}}].

\bibitem{Munoz:2019fkt}
J.~B. Muñoz, \emph{{A Standard Ruler at Cosmic Dawn}},
  \href{https://arxiv.org/abs/1904.07868}{{\ttfamily 1904.07868}}.

\bibitem{Munoz:2019rhi}
J.~B. Muñoz, \emph{{Velocity-induced Acoustic Oscillations at Cosmic Dawn}},
  \href{https://arxiv.org/abs/1904.07881}{{\ttfamily 1904.07881}}.

\bibitem{Einstein:1916vd}
A.~Einstein, \emph{{The Foundation of the General Theory of Relativity}},
  \href{https://doi.org/10.1002/andp.200590044,
  10.1002/andp.19163540702}{\emph{Annalen Phys.} {\bfseries 49} (1916)
  769--822}.

\bibitem{Bergstrom:1999kd}
L.~Bergström and A.~Goobar, \emph{{Cosmology and particle astrophysics}}.
\newblock 1999.

\bibitem{Dodelson:2003ft}
S.~Dodelson, \emph{{Modern Cosmology}}.
\newblock Academic Press, Amsterdam, 2003.

\bibitem{Mukhanov:2005sc}
V.~Mukhanov, \emph{{Physical Foundations of Cosmology}}.
\newblock Cambridge University Press, Oxford, 2005.

\bibitem{Durrer:2008eom}
R.~Durrer, \emph{{The Cosmic Microwave Background}}.
\newblock Cambridge University Press, Cambridge, 2008.

\bibitem{Weinberg:2008zzc}
S.~Weinberg, \emph{{Cosmology}}.
\newblock 2008.

\bibitem{Lesgourgues:2018ncw}
J.~Lesgourgues, G.~Mangano, G.~Miele and S.~Pastor, \emph{{Neutrino
  Cosmology}}.
\newblock Cambridge University Press, 2018.

\bibitem{Carroll:2004st}
S.~M. Carroll, \emph{{Spacetime and geometry: An introduction to general
  relativity}}.
\newblock 2004.

\bibitem{Friedmann:1924bb}
A.~Friedmann, \emph{{On the Possibility of a world with constant negative
  curvature of space}}, \href{https://doi.org/10.1007/BF01328280}{\emph{Z.
  Phys.} {\bfseries 21} (1924) 326--332}.

\bibitem{Lemaitre:1931zza}
G.~Lemaître, \emph{{A homogeneous universe of constant mass and increasing
  radius accounting for the radial velocity of extra-galactic nebulae}},
  {\emph{Mon. Not. Roy. Astron. Soc.} {\bfseries 91} (1931) 483--490}.

\bibitem{Robertson:1935zz}
H.~P. Robertson, \emph{{Kinematics and World-Structure}},
  \href{https://doi.org/10.1086/143681}{\emph{Astrophys. J.} {\bfseries 82}
  (1935) 284--301}.

\bibitem{Walker}
A.~G. {Walker}, \emph{{On Milne's Theory of World-Structure}},
  \href{https://doi.org/10.1112/plms/s2-42.1.90}{\emph{Proceedings of the
  London Mathematical Society, (Series 2) volume 42, p.~90-127} {\bfseries 42}
  (1937) 90--127}.

\bibitem{TheConversation:2017ghw}
{The Conversation}, \emph{https://theconversation.com},  2017.

\bibitem{Einstein:1917ce}
A.~Einstein, \emph{{Cosmological Considerations in the General Theory of
  Relativity}}, {\emph{Sitzungsber. Preuss. Akad. Wiss. Berlin (Math. Phys.)}
  {\bfseries 1917} (1917) 142--152}.

\bibitem{Weinberg:1988cp}
S.~Weinberg, \emph{{The Cosmological Constant Problem}},
  \href{https://doi.org/10.1103/RevModPhys.61.1}{\emph{Rev. Mod. Phys.}
  {\bfseries 61} (1989) 1--23}.

\bibitem{Carroll:1991mt}
S.~M. Carroll, W.~H. Press and E.~L. Turner, \emph{{The Cosmological
  constant}},
  \href{https://doi.org/10.1146/annurev.aa.30.090192.002435}{\emph{Ann. Rev.
  Astron. Astrophys.} {\bfseries 30} (1992) 499--542}.

\bibitem{Carroll:2000fy}
S.~M. Carroll, \emph{{The Cosmological constant}},
  \href{https://doi.org/10.12942/lrr-2001-1}{\emph{Living Rev. Rel.} {\bfseries
  4} (2001) 1}, [\href{https://arxiv.org/abs/astro-ph/0004075}{{\ttfamily
  astro-ph/0004075}}].

\bibitem{Weinberg:2000yb}
S.~Weinberg, \emph{{The Cosmological constant problems}},  in \emph{{Sources
  and detection of dark matter and dark energy in the universe. Proceedings,
  4th International Symposium, DM 2000, Marina del Rey, USA, February 23-25,
  2000}}, pp.~18--26, 2000,
  \href{https://arxiv.org/abs/astro-ph/0005265}{{\ttfamily astro-ph/0005265}}.

\bibitem{Sahni:2002kh}
V.~Sahni, \emph{{The Cosmological constant problem and quintessence}},
  \href{https://doi.org/10.1088/0264-9381/19/13/304}{\emph{Class. Quant. Grav.}
  {\bfseries 19} (2002) 3435--3448},
  [\href{https://arxiv.org/abs/astro-ph/0202076}{{\ttfamily
  astro-ph/0202076}}].

\bibitem{Padmanabhan:2002ji}
T.~Padmanabhan, \emph{{Cosmological constant: The Weight of the vacuum}},
  \href{https://doi.org/10.1016/S0370-1573(03)00120-0}{\emph{Phys. Rept.}
  {\bfseries 380} (2003) 235--320},
  [\href{https://arxiv.org/abs/hep-th/0212290}{{\ttfamily hep-th/0212290}}].

\bibitem{Nobbenhuis:2004wn}
S.~Nobbenhuis, \emph{{Categorizing different approaches to the cosmological
  constant problem}},
  \href{https://doi.org/10.1007/s10701-005-9042-8}{\emph{Found. Phys.}
  {\bfseries 36} (2006) 613--680},
  [\href{https://arxiv.org/abs/gr-qc/0411093}{{\ttfamily gr-qc/0411093}}].

\bibitem{Polchinski:2006gy}
J.~Polchinski, \emph{{The Cosmological Constant and the String Landscape}},  in
  \emph{{The Quantum Structure of Space and Time: Proceedings of the 23rd
  Solvay Conference on Physics. Brussels, Belgium. 1 - 3 December 2005}},
  pp.~216--236, 2006, \href{https://arxiv.org/abs/hep-th/0603249}{{\ttfamily
  hep-th/0603249}}.

\bibitem{Peccei:1977hh}
R.~D. Peccei and H.~R. Quinn, \emph{{CP Conservation in the Presence of
  Instantons}}, \href{https://doi.org/10.1103/PhysRevLett.38.1440}{\emph{Phys.
  Rev. Lett.} {\bfseries 38} (1977) 1440--1443}.

\bibitem{Wilczek:1977pj}
F.~Wilczek, \emph{{Problem of Strong $P$ and $T$ Invariance in the Presence of
  Instantons}}, \href{https://doi.org/10.1103/PhysRevLett.40.279}{\emph{Phys.
  Rev. Lett.} {\bfseries 40} (1978) 279--282}.

\bibitem{Hu:2000ke}
W.~Hu, R.~Barkana and A.~Gruzinov, \emph{{Cold and fuzzy dark matter}},
  \href{https://doi.org/10.1103/PhysRevLett.85.1158}{\emph{Phys. Rev. Lett.}
  {\bfseries 85} (2000) 1158--1161},
  [\href{https://arxiv.org/abs/astro-ph/0003365}{{\ttfamily
  astro-ph/0003365}}].

\bibitem{Dodelson:1993je}
S.~Dodelson and L.~M. Widrow, \emph{{Sterile-neutrinos as dark matter}},
  \href{https://doi.org/10.1103/PhysRevLett.72.17}{\emph{Phys. Rev. Lett.}
  {\bfseries 72} (1994) 17--20},
  [\href{https://arxiv.org/abs/hep-ph/9303287}{{\ttfamily hep-ph/9303287}}].

\bibitem{McDonald:1993ex}
J.~McDonald, \emph{{Gauge singlet scalars as cold dark matter}},
  \href{https://doi.org/10.1103/PhysRevD.50.3637}{\emph{Phys. Rev.} {\bfseries
  D50} (1994) 3637--3649},
  [\href{https://arxiv.org/abs/hep-ph/0702143}{{\ttfamily hep-ph/0702143}}].

\bibitem{Jungman:1995df}
G.~Jungman, M.~Kamionkowski and K.~Griest, \emph{{Supersymmetric dark matter}},
  \href{https://doi.org/10.1016/0370-1573(95)00058-5}{\emph{Phys. Rept.}
  {\bfseries 267} (1996) 195--373},
  [\href{https://arxiv.org/abs/hep-ph/9506380}{{\ttfamily hep-ph/9506380}}].

\bibitem{Kusenko:1997si}
A.~Kusenko and M.~E. Shaposhnikov, \emph{{Supersymmetric Q balls as dark
  matter}}, \href{https://doi.org/10.1016/S0370-2693(97)01375-0}{\emph{Phys.
  Lett.} {\bfseries B418} (1998) 46--54},
  [\href{https://arxiv.org/abs/hep-ph/9709492}{{\ttfamily hep-ph/9709492}}].

\bibitem{Moroi:1999zb}
T.~Moroi and L.~Randall, \emph{{Wino cold dark matter from anomaly mediated
  SUSY breaking}},
  \href{https://doi.org/10.1016/S0550-3213(99)00748-8}{\emph{Nucl. Phys.}
  {\bfseries B570} (2000) 455--472},
  [\href{https://arxiv.org/abs/hep-ph/9906527}{{\ttfamily hep-ph/9906527}}].

\bibitem{Burgess:2000yq}
C.~P. Burgess, M.~Pospelov and T.~ter Veldhuis, \emph{{The Minimal model of
  nonbaryonic dark matter: A Singlet scalar}},
  \href{https://doi.org/10.1016/S0550-3213(01)00513-2}{\emph{Nucl. Phys.}
  {\bfseries B619} (2001) 709--728},
  [\href{https://arxiv.org/abs/hep-ph/0011335}{{\ttfamily hep-ph/0011335}}].

\bibitem{Covi:2001nw}
L.~Covi, H.-B. Kim, J.~E. Kim and L.~Roszkowski, \emph{{Axinos as dark
  matter}}, \href{https://doi.org/10.1088/1126-6708/2001/05/033}{\emph{JHEP}
  {\bfseries 05} (2001) 033},
  [\href{https://arxiv.org/abs/hep-ph/0101009}{{\ttfamily hep-ph/0101009}}].

\bibitem{TuckerSmith:2001hy}
D.~Tucker-Smith and N.~Weiner, \emph{{Inelastic dark matter}},
  \href{https://doi.org/10.1103/PhysRevD.64.043502}{\emph{Phys. Rev.}
  {\bfseries D64} (2001) 043502},
  [\href{https://arxiv.org/abs/hep-ph/0101138}{{\ttfamily hep-ph/0101138}}].

\bibitem{Servant:2002aq}
G.~Servant and T.~M.~P. Tait, \emph{{Is the lightest Kaluza-Klein particle a
  viable dark matter candidate?}},
  \href{https://doi.org/10.1016/S0550-3213(02)01012-X}{\emph{Nucl. Phys.}
  {\bfseries B650} (2003) 391--419},
  [\href{https://arxiv.org/abs/hep-ph/0206071}{{\ttfamily hep-ph/0206071}}].

\bibitem{Cheng:2002ej}
H.-C. Cheng, J.~L. Feng and K.~T. Matchev, \emph{{Kaluza-Klein dark matter}},
  \href{https://doi.org/10.1103/PhysRevLett.89.211301}{\emph{Phys. Rev. Lett.}
  {\bfseries 89} (2002) 211301},
  [\href{https://arxiv.org/abs/hep-ph/0207125}{{\ttfamily hep-ph/0207125}}].

\bibitem{Foot:2004wz}
R.~Foot and R.~R. Volkas, \emph{{Spheroidal galactic halos and mirror dark
  matter}}, \href{https://doi.org/10.1103/PhysRevD.70.123508}{\emph{Phys. Rev.}
  {\bfseries D70} (2004) 123508},
  [\href{https://arxiv.org/abs/astro-ph/0407522}{{\ttfamily
  astro-ph/0407522}}].

\bibitem{Cirelli:2005uq}
M.~Cirelli, N.~Fornengo and A.~Strumia, \emph{{Minimal dark matter}},
  \href{https://doi.org/10.1016/j.nuclphysb.2006.07.012}{\emph{Nucl. Phys.}
  {\bfseries B753} (2006) 178--194},
  [\href{https://arxiv.org/abs/hep-ph/0512090}{{\ttfamily hep-ph/0512090}}].

\bibitem{Pospelov:2007mp}
M.~Pospelov, A.~Ritz and M.~B. Voloshin, \emph{{Secluded WIMP Dark Matter}},
  \href{https://doi.org/10.1016/j.physletb.2008.02.052}{\emph{Phys. Lett.}
  {\bfseries B662} (2008) 53--61},
  [\href{https://arxiv.org/abs/0711.4866}{{\ttfamily 0711.4866}}].

\bibitem{ArkaniHamed:2008qn}
N.~Arkani-Hamed, D.~P. Finkbeiner, T.~R. Slatyer and N.~Weiner, \emph{{A Theory
  of Dark Matter}},
  \href{https://doi.org/10.1103/PhysRevD.79.015014}{\emph{Phys. Rev.}
  {\bfseries D79} (2009) 015014},
  [\href{https://arxiv.org/abs/0810.0713}{{\ttfamily 0810.0713}}].

\bibitem{Sikivie:2009qn}
P.~Sikivie and Q.~Yang, \emph{{Bose-Einstein Condensation of Dark Matter
  Axions}}, \href{https://doi.org/10.1103/PhysRevLett.103.111301}{\emph{Phys.
  Rev. Lett.} {\bfseries 103} (2009) 111301},
  [\href{https://arxiv.org/abs/0901.1106}{{\ttfamily 0901.1106}}].

\bibitem{Kaplan:2009ag}
D.~E. Kaplan, M.~A. Luty and K.~M. Zurek, \emph{{Asymmetric Dark Matter}},
  \href{https://doi.org/10.1103/PhysRevD.79.115016}{\emph{Phys. Rev.}
  {\bfseries D79} (2009) 115016},
  [\href{https://arxiv.org/abs/0901.4117}{{\ttfamily 0901.4117}}].

\bibitem{Visinelli:2009zm}
L.~Visinelli and P.~Gondolo, \emph{{Dark Matter Axions Revisited}},
  \href{https://doi.org/10.1103/PhysRevD.80.035024}{\emph{Phys. Rev.}
  {\bfseries D80} (2009) 035024},
  [\href{https://arxiv.org/abs/0903.4377}{{\ttfamily 0903.4377}}].

\bibitem{Visinelli:2009bg}
L.~Visinelli and P.~Gondolo, \emph{{Axion cold dark matter revisited}},
  \href{https://doi.org/10.1088/1742-6596/203/1/012035}{\emph{J. Phys. Conf.
  Ser.} {\bfseries 203} (2010) 012035},
  [\href{https://arxiv.org/abs/0910.3941}{{\ttfamily 0910.3941}}].

\bibitem{Beltran:2010ww}
M.~Beltrán, D.~Hooper, E.~W. Kolb, Z.~A.~C. Krusberg and T.~M.~P. Tait,
  \emph{{Maverick dark matter at colliders}},
  \href{https://doi.org/10.1007/JHEP09(2010)037}{\emph{JHEP} {\bfseries 09}
  (2010) 037}, [\href{https://arxiv.org/abs/1002.4137}{{\ttfamily 1002.4137}}].

\bibitem{Feng:2011vu}
J.~L. Feng, J.~Kumar, D.~Marfatia and D.~Sanford, \emph{{Isospin-Violating Dark
  Matter}}, \href{https://doi.org/10.1016/j.physletb.2011.07.083}{\emph{Phys.
  Lett.} {\bfseries B703} (2011) 124--127},
  [\href{https://arxiv.org/abs/1102.4331}{{\ttfamily 1102.4331}}].

\bibitem{Arias:2012az}
P.~Arias, D.~Cadamuro, M.~Goodsell, J.~Jäckel, J.~Redondo and A.~Ringwald,
  \emph{{WISPy Cold Dark Matter}},
  \href{https://doi.org/10.1088/1475-7516/2012/06/013}{\emph{JCAP} {\bfseries
  1206} (2012) 013}, [\href{https://arxiv.org/abs/1201.5902}{{\ttfamily
  1201.5902}}].

\bibitem{Tulin:2013teo}
S.~Tulin, H.-B. Yu and K.~M. Zurek, \emph{{Beyond Collisionless Dark Matter:
  Particle Physics Dynamics for Dark Matter Halo Structure}},
  \href{https://doi.org/10.1103/PhysRevD.87.115007}{\emph{Phys. Rev.}
  {\bfseries D87} (2013) 115007},
  [\href{https://arxiv.org/abs/1302.3898}{{\ttfamily 1302.3898}}].

\bibitem{Petraki:2013wwa}
K.~Petraki and R.~R. Volkas, \emph{{Review of asymmetric dark matter}},
  \href{https://doi.org/10.1142/S0217751X13300287}{\emph{Int. J. Mod. Phys.}
  {\bfseries A28} (2013) 1330028},
  [\href{https://arxiv.org/abs/1305.4939}{{\ttfamily 1305.4939}}].

\bibitem{Zurek:2013wia}
K.~M. Zurek, \emph{{Asymmetric Dark Matter: Theories, Signatures, and
  Constraints}},
  \href{https://doi.org/10.1016/j.physrep.2013.12.001}{\emph{Phys. Rept.}
  {\bfseries 537} (2014) 91--121},
  [\href{https://arxiv.org/abs/1308.0338}{{\ttfamily 1308.0338}}].

\bibitem{Visinelli:2014twa}
L.~Visinelli and P.~Gondolo, \emph{{Axion cold dark matter in view of BICEP2
  results}}, \href{https://doi.org/10.1103/PhysRevLett.113.011802}{\emph{Phys.
  Rev. Lett.} {\bfseries 113} (2014) 011802},
  [\href{https://arxiv.org/abs/1403.4594}{{\ttfamily 1403.4594}}].

\bibitem{Baldes:2015lka}
I.~Baldes, N.~F. Bell, A.~J. Millar and R.~R. Volkas, \emph{{Asymmetric Dark
  Matter and CP Violating Scatterings in a UV Complete Model}},
  \href{https://doi.org/10.1088/1475-7516/2015/10/048}{\emph{JCAP} {\bfseries
  1510} (2015) 048}, [\href{https://arxiv.org/abs/1506.07521}{{\ttfamily
  1506.07521}}].

\bibitem{Visinelli:2015wha}
L.~Visinelli, \emph{{Condensation of Galactic Cold Dark Matter}},
  \href{https://doi.org/10.1088/1475-7516/2016/07/009}{\emph{JCAP} {\bfseries
  1607} (2016) 009}, [\href{https://arxiv.org/abs/1509.05871}{{\ttfamily
  1509.05871}}].

\bibitem{Blennow:2015hzp}
M.~Blennow, S.~Clementz and J.~Herrero-García, \emph{{Pinning down inelastic
  dark matter in the Sun and in direct detection}},
  \href{https://doi.org/10.1088/1475-7516/2016/04/004}{\emph{JCAP} {\bfseries
  1604} (2016) 004}, [\href{https://arxiv.org/abs/1512.03317}{{\ttfamily
  1512.03317}}].

\bibitem{Visinelli:2016hzy}
L.~Visinelli, \emph{{Analytic expressions for the kinetic decoupling of
  WIMPs}}, \href{https://doi.org/10.1088/1742-6596/718/4/042059}{\emph{J. Phys.
  Conf. Ser.} {\bfseries 718} (2016) 042059},
  [\href{https://arxiv.org/abs/1601.00817}{{\ttfamily 1601.00817}}].

\bibitem{Escudero:2016tzx}
M.~Escudero, N.~Rius and V.~Sanz, \emph{{Sterile neutrino portal to Dark Matter
  I: The $U(1)_{B-L}$ case}},
  \href{https://doi.org/10.1007/JHEP02(2017)045}{\emph{JHEP} {\bfseries 02}
  (2017) 045}, [\href{https://arxiv.org/abs/1606.01258}{{\ttfamily
  1606.01258}}].

\bibitem{Escudero:2016ksa}
M.~Escudero, N.~Rius and V.~Sanz, \emph{{Sterile Neutrino portal to Dark Matter
  II: Exact Dark symmetry}},
  \href{https://doi.org/10.1140/epjc/s10052-017-4963-x}{\emph{Eur. Phys. J.}
  {\bfseries C77} (2017) 397},
  [\href{https://arxiv.org/abs/1607.02373}{{\ttfamily 1607.02373}}].

\bibitem{Baum:2016oow}
S.~Baum, L.~Visinelli, K.~Freese and P.~Stengel, \emph{{Dark matter capture,
  subdominant WIMPs, and neutrino observatories}},
  \href{https://doi.org/10.1103/PhysRevD.95.043007}{\emph{Phys. Rev.}
  {\bfseries D95} (2017) 043007},
  [\href{https://arxiv.org/abs/1611.09665}{{\ttfamily 1611.09665}}].

\bibitem{Blennow:2016gde}
M.~Blennow, S.~Clementz and J.~Herrero-García, \emph{{Self-interacting
  inelastic dark matter: A viable solution to the small scale structure
  problems}}, \href{https://doi.org/10.1088/1475-7516/2017/03/048}{\emph{JCAP}
  {\bfseries 1703} (2017) 048},
  [\href{https://arxiv.org/abs/1612.06681}{{\ttfamily 1612.06681}}].

\bibitem{Visinelli:2017imh}
L.~Visinelli, \emph{{Light axion-like dark matter must be present during
  inflation}}, \href{https://doi.org/10.1103/PhysRevD.96.023013}{\emph{Phys.
  Rev.} {\bfseries D96} (2017) 023013},
  [\href{https://arxiv.org/abs/1703.08798}{{\ttfamily 1703.08798}}].

\bibitem{Brinckmann:2017uve}
T.~Brinckmann, J.~Zavala, D.~Rapetti, S.~H. Hansen and M.~Vogelsberger,
  \emph{{The structure and assembly history of cluster-sized haloes in
  self-interacting dark matter}},
  \href{https://doi.org/10.1093/mnras/stx2782}{\emph{Mon. Not. Roy. Astron.
  Soc.} {\bfseries 474} (2018) 746--759},
  [\href{https://arxiv.org/abs/1705.00623}{{\ttfamily 1705.00623}}].

\bibitem{Foot:2017dgx}
R.~Foot, \emph{{Dissipative dark matter halos: The steady state solution}},
  \href{https://doi.org/10.1103/PhysRevD.97.043012}{\emph{Phys. Rev.}
  {\bfseries D97} (2018) 043012},
  [\href{https://arxiv.org/abs/1707.02528}{{\ttfamily 1707.02528}}].

\bibitem{Boucenna:2017ghj}
S.~M. Boucenna, F.~Kühnel, T.~Ohlsson and L.~Visinelli, \emph{{Novel
  Constraints on Mixed Dark-Matter Scenarios of Primordial Black Holes and
  WIMPs}}, \href{https://doi.org/10.1088/1475-7516/2018/07/003}{\emph{JCAP}
  {\bfseries 1807} (2018) 003},
  [\href{https://arxiv.org/abs/1712.06383}{{\ttfamily 1712.06383}}].

\bibitem{Hui:2016ltb}
L.~Hui, J.~P. Ostriker, S.~Tremaine and E.~Witten, \emph{{Ultralight scalars as
  cosmological dark matter}},
  \href{https://doi.org/10.1103/PhysRevD.95.043541}{\emph{Phys. Rev.}
  {\bfseries D95} (2017) 043541},
  [\href{https://arxiv.org/abs/1610.08297}{{\ttfamily 1610.08297}}].

\bibitem{Baum:2017enm}
S.~Baum, M.~Carena, N.~R. Shah and C.~E.~M. Wagner, \emph{{Higgs portals for
  thermal Dark Matter. EFT perspectives and the NMSSM}},
  \href{https://doi.org/10.1007/JHEP04(2018)069}{\emph{JHEP} {\bfseries 04}
  (2018) 069}, [\href{https://arxiv.org/abs/1712.09873}{{\ttfamily
  1712.09873}}].

\bibitem{Foot:2018qpw}
R.~Foot, \emph{{Dissipative dark matter halos: The steady state solution II}},
  \href{https://doi.org/10.1103/PhysRevD.97.103006}{\emph{Phys. Rev.}
  {\bfseries D97} (2018) 103006},
  [\href{https://arxiv.org/abs/1801.09359}{{\ttfamily 1801.09359}}].

\bibitem{Escudero:2018thh}
M.~Escudero, L.~Lopez-Honorez, O.~Mena, S.~Palomares-Ruiz and
  P.~Villanueva-Domingo, \emph{{A fresh look into the interacting dark matter
  scenario}}, \href{https://doi.org/10.1088/1475-7516/2018/06/007}{\emph{JCAP}
  {\bfseries 1806} (2018) 007},
  [\href{https://arxiv.org/abs/1803.08427}{{\ttfamily 1803.08427}}].

\bibitem{Escudero:2018fwn}
M.~Escudero, S.~J. Witte and N.~Rius, \emph{{The dispirited case of gauged
  U(1)$_{B-L}$ dark matter}},
  \href{https://doi.org/10.1007/JHEP08(2018)190}{\emph{JHEP} {\bfseries 08}
  (2018) 190}, [\href{https://arxiv.org/abs/1806.02823}{{\ttfamily
  1806.02823}}].

\bibitem{Sokolenko:2018noz}
A.~Sokolenko, K.~Bondarenko, T.~Brinckmann, J.~Zavala, M.~Vogelsberger,
  T.~Bringmann et~al., \emph{{Towards an improved model of self-interacting
  dark matter haloes}},
  \href{https://doi.org/10.1088/1475-7516/2018/12/038}{\emph{JCAP} {\bfseries
  1812} (2018) 038}, [\href{https://arxiv.org/abs/1806.11539}{{\ttfamily
  1806.11539}}].

\bibitem{Akarsu:2018aro}
O.~Akarsu, N.~Katırci, S.~Kumar, R.~C. Nunes and M.~Sami, \emph{{Cosmological
  implications of scale-independent energy-momentum squared gravity: Pseudo
  nonminimal interactions in dark matter and relativistic relics}},
  \href{https://doi.org/10.1103/PhysRevD.98.063522}{\emph{Phys. Rev.}
  {\bfseries D98} (2018) 063522},
  [\href{https://arxiv.org/abs/1807.01588}{{\ttfamily 1807.01588}}].

\bibitem{Elor:2018twp}
G.~Elor, M.~Escudero and A.~Nelson, \emph{{Baryogenesis and Dark Matter from
  $B$ Mesons}}, \href{https://doi.org/10.1103/PhysRevD.99.035031}{\emph{Phys.
  Rev.} {\bfseries D99} (2019) 035031},
  [\href{https://arxiv.org/abs/1810.00880}{{\ttfamily 1810.00880}}].

\bibitem{Kumar:2019gfl}
S.~Kumar, R.~C. Nunes and S.~K. Yadav, \emph{{Testing warmness of dark
  matter}},  \href{https://arxiv.org/abs/1901.07549}{{\ttfamily 1901.07549}}.

\bibitem{Dvorkin:2019zdi}
C.~Dvorkin, T.~Lin and K.~Schutz, \emph{{Making dark matter out of light:
  freeze-in from plasma effects}},
  \href{https://arxiv.org/abs/1902.08623}{{\ttfamily 1902.08623}}.

\bibitem{Drukier:1986tm}
A.~K. Drukier, K.~Freese and D.~N. Spergel, \emph{{Detecting Cold Dark Matter
  Candidates}}, \href{https://doi.org/10.1103/PhysRevD.33.3495}{\emph{Phys.
  Rev.} {\bfseries D33} (1986) 3495--3508}.

\bibitem{Spergel:1999mh}
D.~N. Spergel and P.~J. Steinhardt, \emph{{Observational evidence for
  selfinteracting cold dark matter}},
  \href{https://doi.org/10.1103/PhysRevLett.84.3760}{\emph{Phys. Rev. Lett.}
  {\bfseries 84} (2000) 3760--3763},
  [\href{https://arxiv.org/abs/astro-ph/9909386}{{\ttfamily
  astro-ph/9909386}}].

\bibitem{Oikonomou:2006mh}
V.~K. Oikonomou, J.~D. Vergados and C.~C. Moustakidis, \emph{{Direct Detection
  of Dark Matter-Rates for Various Wimps}},
  \href{https://doi.org/10.1016/j.nuclphysb.2007.03.014}{\emph{Nucl. Phys.}
  {\bfseries B773} (2007) 19--42},
  [\href{https://arxiv.org/abs/hep-ph/0612293}{{\ttfamily hep-ph/0612293}}].

\bibitem{Goodman:2010ku}
J.~Goodman, M.~Ibe, A.~Rajaraman, W.~Shepherd, T.~M.~P. Tait and H.-B. Yu,
  \emph{{Constraints on Dark Matter from Colliders}},
  \href{https://doi.org/10.1103/PhysRevD.82.116010}{\emph{Phys. Rev.}
  {\bfseries D82} (2010) 116010},
  [\href{https://arxiv.org/abs/1008.1783}{{\ttfamily 1008.1783}}].

\bibitem{Aartsen:2012kia}
{\scshape IceCube} collaboration, M.~G. Aartsen et~al., \emph{{Search for dark
  matter annihilations in the Sun with the 79-string IceCube detector}},
  \href{https://doi.org/10.1103/PhysRevLett.110.131302}{\emph{Phys. Rev. Lett.}
  {\bfseries 110} (2013) 131302},
  [\href{https://arxiv.org/abs/1212.4097}{{\ttfamily 1212.4097}}].

\bibitem{Agnese:2013rvf}
{\scshape CDMS} collaboration, R.~Agnese et~al., \emph{{Silicon Detector Dark
  Matter Results from the Final Exposure of CDMS II}},
  \href{https://doi.org/10.1103/PhysRevLett.111.251301}{\emph{Phys. Rev. Lett.}
  {\bfseries 111} (2013) 251301},
  [\href{https://arxiv.org/abs/1304.4279}{{\ttfamily 1304.4279}}].

\bibitem{Viel:2013apy}
M.~Viel, G.~D. Becker, J.~S. Bolton and M.~G. Haehnelt, \emph{{Warm dark matter
  as a solution to the small scale crisis: New constraints from high redshift
  Lyman-$\alpha$ forest data}},
  \href{https://doi.org/10.1103/PhysRevD.88.043502}{\emph{Phys. Rev.}
  {\bfseries D88} (2013) 043502},
  [\href{https://arxiv.org/abs/1306.2314}{{\ttfamily 1306.2314}}].

\bibitem{Bernabei:2013xsa}
R.~Bernabei et~al., \emph{{Final model independent result of
  DAMA/LIBRA-phase1}},
  \href{https://doi.org/10.1140/epjc/s10052-013-2648-7}{\emph{Eur. Phys. J.}
  {\bfseries C73} (2013) 2648},
  [\href{https://arxiv.org/abs/1308.5109}{{\ttfamily 1308.5109}}].

\bibitem{Visinelli:2013fia}
L.~Visinelli, \emph{{Axion-Electromagnetic Waves}},
  \href{https://doi.org/10.1142/S0217732313501629}{\emph{Mod. Phys. Lett.}
  {\bfseries A28} (2013) 1350162},
  [\href{https://arxiv.org/abs/1401.0709}{{\ttfamily 1401.0709}}].

\bibitem{Daylan:2014rsa}
T.~Daylan, D.~P. Finkbeiner, D.~Hooper, T.~Linden, S.~K.~N. Portillo, N.~L.
  Rodd et~al., \emph{{The characterization of the gamma-ray signal from the
  central Milky Way: A case for annihilating dark matter}},
  \href{https://doi.org/10.1016/j.dark.2015.12.005}{\emph{Phys. Dark Univ.}
  {\bfseries 12} (2016) 1--23},
  [\href{https://arxiv.org/abs/1402.6703}{{\ttfamily 1402.6703}}].

\bibitem{Visinelli:2015eka}
L.~Visinelli and P.~Gondolo, \emph{{Kinetic decoupling of WIMPs: analytic
  expressions}}, \href{https://doi.org/10.1103/PhysRevD.91.083526}{\emph{Phys.
  Rev.} {\bfseries D91} (2015) 083526},
  [\href{https://arxiv.org/abs/1501.02233}{{\ttfamily 1501.02233}}].

\bibitem{Freese:2015mta}
K.~Freese, T.~Rindler-Daller, D.~Spolyar and M.~Valluri, \emph{{Dark Stars: A
  Review}}, \href{https://doi.org/10.1088/0034-4885/79/6/066902}{\emph{Rept.
  Prog. Phys.} {\bfseries 79} (2016) 066902},
  [\href{https://arxiv.org/abs/1501.02394}{{\ttfamily 1501.02394}}].

\bibitem{Blennow:2015oea}
M.~Blennow, J.~Herrero-García and T.~Schwetz, \emph{{A halo-independent lower
  bound on the dark matter capture rate in the Sun from a direct detection
  signal}}, \href{https://doi.org/10.1088/1475-7516/2015/05/036}{\emph{JCAP}
  {\bfseries 1505} (2015) 036},
  [\href{https://arxiv.org/abs/1502.03342}{{\ttfamily 1502.03342}}].

\bibitem{Ackermann:2015zua}
{\scshape Fermi-LAT} collaboration, M.~Ackermann et~al., \emph{{Searching for
  Dark Matter Annihilation from Milky Way Dwarf Spheroidal Galaxies with Six
  Years of Fermi Large Area Telescope Data}},
  \href{https://doi.org/10.1103/PhysRevLett.115.231301}{\emph{Phys. Rev. Lett.}
  {\bfseries 115} (2015) 231301},
  [\href{https://arxiv.org/abs/1503.02641}{{\ttfamily 1503.02641}}].

\bibitem{Giesen:2015ufa}
G.~Giesen, M.~Boudaud, Y.~Génolini, V.~Poulin, M.~Cirelli, P.~Salati et~al.,
  \emph{{AMS-02 antiprotons, at last! Secondary astrophysical component and
  immediate implications for Dark Matter}},
  \href{https://doi.org/10.1088/1475-7516/2015/09/023,
  10.1088/1475-7516/2015/9/023}{\emph{JCAP} {\bfseries 1509} (2015) 023},
  [\href{https://arxiv.org/abs/1504.04276}{{\ttfamily 1504.04276}}].

\bibitem{Blennow:2015gta}
M.~Blennow, J.~Herrero-García, T.~Schwetz and S.~Vogl, \emph{{Halo-independent
  tests of dark matter direct detection signals: local DM density, LHC, and
  thermal freeze-out}},
  \href{https://doi.org/10.1088/1475-7516/2015/08/039}{\emph{JCAP} {\bfseries
  1508} (2015) 039}, [\href{https://arxiv.org/abs/1505.05710}{{\ttfamily
  1505.05710}}].

\bibitem{Poulin:2015pna}
V.~Poulin, P.~D. Serpico and J.~Lesgourgues, \emph{{Dark Matter annihilations
  in halos and high-redshift sources of reionization of the universe}},
  \href{https://doi.org/10.1088/1475-7516/2015/12/041}{\emph{JCAP} {\bfseries
  1512} (2015) 041}, [\href{https://arxiv.org/abs/1508.01370}{{\ttfamily
  1508.01370}}].

\bibitem{Blennow:2015yca}
M.~Blennow, P.~Coloma, E.~Fernández-Martínez, P.~A.~N. Machado and
  B.~Zaldívar, \emph{{Global constraints on vector-like WIMP effective
  interactions}},
  \href{https://doi.org/10.1088/1475-7516/2016/04/015}{\emph{JCAP} {\bfseries
  1604} (2016) 015}, [\href{https://arxiv.org/abs/1509.01587}{{\ttfamily
  1509.01587}}].

\bibitem{Bird:2016dcv}
S.~Bird, I.~Cholis, J.~B. Muñoz, Y.~Ali-Haïmoud, M.~Kamionkowski, E.~D.
  Kovetz et~al., \emph{{Did LIGO detect dark matter?}},
  \href{https://doi.org/10.1103/PhysRevLett.116.201301}{\emph{Phys. Rev. Lett.}
  {\bfseries 116} (2016) 201301},
  [\href{https://arxiv.org/abs/1603.00464}{{\ttfamily 1603.00464}}].

\bibitem{Munoz:2016tmg}
J.~B. Muñoz, E.~D. Kovetz, L.~Dai and M.~Kamionkowski, \emph{{Lensing of Fast
  Radio Bursts as a Probe of Compact Dark Matter}},
  \href{https://doi.org/10.1103/PhysRevLett.117.091301}{\emph{Phys. Rev. Lett.}
  {\bfseries 117} (2016) 091301},
  [\href{https://arxiv.org/abs/1605.00008}{{\ttfamily 1605.00008}}].

\bibitem{Carr:2016drx}
B.~Carr, F.~Kühnel and M.~Sandstad, \emph{{Primordial Black Holes as Dark
  Matter}}, \href{https://doi.org/10.1103/PhysRevD.94.083504}{\emph{Phys. Rev.}
  {\bfseries D94} (2016) 083504},
  [\href{https://arxiv.org/abs/1607.06077}{{\ttfamily 1607.06077}}].

\bibitem{Akerib:2016vxi}
{\scshape LUX} collaboration, D.~S. Akerib et~al., \emph{{Results from a search
  for dark matter in the complete LUX exposure}},
  \href{https://doi.org/10.1103/PhysRevLett.118.021303}{\emph{Phys. Rev. Lett.}
  {\bfseries 118} (2017) 021303},
  [\href{https://arxiv.org/abs/1608.07648}{{\ttfamily 1608.07648}}].

\bibitem{Escudero:2016gzx}
M.~Escudero, A.~Berlin, D.~Hooper and M.-X. Lin, \emph{{Toward (Finally!)
  Ruling Out Z and Higgs Mediated Dark Matter Models}},
  \href{https://doi.org/10.1088/1475-7516/2016/12/029}{\emph{JCAP} {\bfseries
  1612} (2016) 029}, [\href{https://arxiv.org/abs/1609.09079}{{\ttfamily
  1609.09079}}].

\bibitem{TheMADMAXWorkingGroup:2016hpc}
{\scshape MADMAX Working Group} collaboration, A.~Caldwell, G.~Dvali,
  B.~Majorovits, A.~Millar, G.~Raffelt, J.~Redondo et~al., \emph{{Dielectric
  Haloscopes: A New Way to Detect Axion Dark Matter}},
  \href{https://doi.org/10.1103/PhysRevLett.118.091801}{\emph{Phys. Rev. Lett.}
  {\bfseries 118} (2017) 091801},
  [\href{https://arxiv.org/abs/1611.05865}{{\ttfamily 1611.05865}}].

\bibitem{Escudero:2016kpw}
M.~Escudero, D.~Hooper and S.~J. Witte, \emph{{Updated Collider and Direct
  Detection Constraints on Dark Matter Models for the Galactic Center Gamma-Ray
  Excess}}, \href{https://doi.org/10.1088/1475-7516/2017/02/038}{\emph{JCAP}
  {\bfseries 1702} (2017) 038},
  [\href{https://arxiv.org/abs/1612.06462}{{\ttfamily 1612.06462}}].

\bibitem{Millar:2016cjp}
A.~J. Millar, G.~G. Raffelt, J.~Redondo and F.~D. Steffen, \emph{{Dielectric
  Haloscopes to Search for Axion Dark Matter: Theoretical Foundations}},
  \href{https://doi.org/10.1088/1475-7516/2017/01/061}{\emph{JCAP} {\bfseries
  1701} (2017) 061}, [\href{https://arxiv.org/abs/1612.07057}{{\ttfamily
  1612.07057}}].

\bibitem{Gariazzo:2017pzb}
S.~Gariazzo, M.~Escudero, R.~Diamanti and O.~Mena, \emph{{Cosmological searches
  for a noncold dark matter component}},
  \href{https://doi.org/10.1103/PhysRevD.96.043501}{\emph{Phys. Rev.}
  {\bfseries D96} (2017) 043501},
  [\href{https://arxiv.org/abs/1704.02991}{{\ttfamily 1704.02991}}].

\bibitem{Aprile:2017iyp}
{\scshape XENON} collaboration, E.~Aprile et~al., \emph{{First Dark Matter
  Search Results from the XENON1T Experiment}},
  \href{https://doi.org/10.1103/PhysRevLett.119.181301}{\emph{Phys. Rev. Lett.}
  {\bfseries 119} (2017) 181301},
  [\href{https://arxiv.org/abs/1705.06655}{{\ttfamily 1705.06655}}].

\bibitem{Poulin:2017bwe}
V.~Poulin, P.~D. Serpico, F.~Calore, S.~Clesse and K.~Kohri, \emph{{CMB bounds
  on disk-accreting massive primordial black holes}},
  \href{https://doi.org/10.1103/PhysRevD.96.083524}{\emph{Phys. Rev.}
  {\bfseries D96} (2017) 083524},
  [\href{https://arxiv.org/abs/1707.04206}{{\ttfamily 1707.04206}}].

\bibitem{Millar:2017eoc}
A.~J. Millar, J.~Redondo and F.~D. Steffen, \emph{{Dielectric haloscopes:
  sensitivity to the axion dark matter velocity}},
  \href{https://doi.org/10.1088/1475-7516/2017/10/006,
  10.1088/1475-7516/2018/05/E02}{\emph{JCAP} {\bfseries 1710} (2017) 006},
  [\href{https://arxiv.org/abs/1707.04266}{{\ttfamily 1707.04266}}]. [Erratum:
  JCAP1805,no.05,E02(2018)].

\bibitem{Kumar:2017bpv}
S.~Kumar and R.~C. Nunes, \emph{{Observational constraints on dark
  matter–dark energy scattering cross section}},
  \href{https://doi.org/10.1140/epjc/s10052-017-5334-3}{\emph{Eur. Phys. J.}
  {\bfseries C77} (2017) 734},
  [\href{https://arxiv.org/abs/1709.02384}{{\ttfamily 1709.02384}}].

\bibitem{Baum:2017kfa}
S.~Baum, R.~Catena, J.~Conrad, K.~Freese and M.~B. Krauss, \emph{{Determining
  dark matter properties with a XENONnT/LZ signal and LHC Run 3 monojet
  searches}}, \href{https://doi.org/10.1103/PhysRevD.97.083002}{\emph{Phys.
  Rev.} {\bfseries D97} (2018) 083002},
  [\href{https://arxiv.org/abs/1709.06051}{{\ttfamily 1709.06051}}].

\bibitem{Escudero:2017yia}
M.~Escudero, S.~J. Witte and D.~Hooper, \emph{{Hidden Sector Dark Matter and
  the Galactic Center Gamma-Ray Excess: A Closer Look}},
  \href{https://doi.org/10.1088/1475-7516/2017/11/042}{\emph{JCAP} {\bfseries
  1711} (2017) 042}, [\href{https://arxiv.org/abs/1709.07002}{{\ttfamily
  1709.07002}}].

\bibitem{Visinelli:2017ooc}
L.~Visinelli, S.~Baum, J.~Redondo, K.~Freese and F.~Wilczek, \emph{{Dilute and
  dense axion stars}},
  \href{https://doi.org/10.1016/j.physletb.2017.12.010}{\emph{Phys. Lett.}
  {\bfseries B777} (2018) 64--72},
  [\href{https://arxiv.org/abs/1710.08910}{{\ttfamily 1710.08910}}].

\bibitem{Visinelli:2017qga}
L.~Visinelli, \emph{{(Non-)thermal production of WIMPs during kination}},
  \href{https://doi.org/10.3390/sym10110546}{\emph{Symmetry} {\bfseries 10}
  (2018) 546}, [\href{https://arxiv.org/abs/1710.11006}{{\ttfamily
  1710.11006}}].

\bibitem{Zumalacarregui:2017qqd}
M.~Zumalacárregui and U.~Seljak, \emph{{Limits on stellar-mass compact objects
  as dark matter from gravitational lensing of type Ia supernovae}},
  \href{https://doi.org/10.1103/PhysRevLett.121.141101}{\emph{Phys. Rev. Lett.}
  {\bfseries 121} (2018) 141101},
  [\href{https://arxiv.org/abs/1712.02240}{{\ttfamily 1712.02240}}].

\bibitem{Munoz:2018pzp}
J.~B. Muñoz and A.~Loeb, \emph{{A small amount of mini-charged dark matter
  could cool the baryons in the early Universe}},
  \href{https://doi.org/10.1038/s41586-018-0151-x}{\emph{Nature} {\bfseries
  557} (2018) 684}, [\href{https://arxiv.org/abs/1802.10094}{{\ttfamily
  1802.10094}}].

\bibitem{Kumar:2018yhh}
S.~Kumar, R.~C. Nunes and S.~K. Yadav, \emph{{Cosmological bounds on dark
  matter-photon coupling}},
  \href{https://doi.org/10.1103/PhysRevD.98.043521}{\emph{Phys. Rev.}
  {\bfseries D98} (2018) 043521},
  [\href{https://arxiv.org/abs/1803.10229}{{\ttfamily 1803.10229}}].

\bibitem{Baum:2018ekm}
S.~Baum, K.~Freese and C.~Kelso, \emph{{Dark Matter implications of
  DAMA/LIBRA-phase2 results}},
  \href{https://doi.org/10.1016/j.physletb.2018.12.036}{\emph{Phys. Lett.}
  {\bfseries B789} (2019) 262--269},
  [\href{https://arxiv.org/abs/1804.01231}{{\ttfamily 1804.01231}}].

\bibitem{Knirck:2018knd}
S.~Knirck, A.~J. Millar, C.~A.~J. O'Hare, J.~Redondo and F.~D. Steffen,
  \emph{{Directional axion detection}},
  \href{https://doi.org/10.1088/1475-7516/2018/11/051}{\emph{JCAP} {\bfseries
  1811} (2018) 051}, [\href{https://arxiv.org/abs/1806.05927}{{\ttfamily
  1806.05927}}].

\bibitem{Baum:2018tfw}
S.~Baum, A.~K. Drukier, K.~Freese, M.~Górski and P.~Stengel, \emph{{Searching
  for Dark Matter with Paleo-Detectors}},
  \href{https://arxiv.org/abs/1806.05991}{{\ttfamily 1806.05991}}.

\bibitem{Visinelli:2018zif}
L.~Visinelli and H.~Terças, \emph{{A Kinetic Theory of Axions in Magnetized
  Plasmas: the Axionon}},  \href{https://arxiv.org/abs/1807.06828}{{\ttfamily
  1807.06828}}.

\bibitem{Visinelli:2018wza}
L.~Visinelli and J.~Redondo, \emph{{Axion Miniclusters in Modified Cosmological
  Histories}},  \href{https://arxiv.org/abs/1808.01879}{{\ttfamily
  1808.01879}}.

\bibitem{Drukier:2018pdy}
A.~K. Drukier, S.~Baum, K.~Freese, M.~Górski and P.~Stengel,
  \emph{{Paleo-detectors: Searching for Dark Matter with Ancient Minerals}},
  \href{https://doi.org/10.1103/PhysRevD.99.043014}{\emph{Phys. Rev.}
  {\bfseries D99} (2019) 043014},
  [\href{https://arxiv.org/abs/1811.06844}{{\ttfamily 1811.06844}}].

\bibitem{Edwards:2019puy}
T.~D.~P. Edwards, B.~J. Kavanagh, C.~Weniger, S.~Baum, A.~K. Drukier, K.~Freese
  et~al., \emph{{Digging for dark matter: Spectral analysis and discovery
  potential of paleo-detectors}},
  \href{https://doi.org/10.1103/PhysRevD.99.043541}{\emph{Phys. Rev.}
  {\bfseries D99} (2019) 043541},
  [\href{https://arxiv.org/abs/1811.10549}{{\ttfamily 1811.10549}}].

\bibitem{Baum:2018sxd}
S.~Baum, R.~Catena and M.~B. Krauss, \emph{{Constraints on Simplified Models
  for Dark Matter from LHC Dijet Searches}},
  \href{https://arxiv.org/abs/1812.01585}{{\ttfamily 1812.01585}}.

\bibitem{Baum:2018lua}
S.~Baum, R.~Catena and M.~B. Krauss, \emph{{Impact of a XENONnT Signal on LHC
  Dijet Searches}},  \href{https://arxiv.org/abs/1812.01594}{{\ttfamily
  1812.01594}}.

\bibitem{Wu:2019nhd}
Y.~Wu, K.~Freese, C.~Kelso and P.~Stengel, \emph{{Uncertainties in Direct Dark
  Matter Detection in Light of GAIA}},
  \href{https://arxiv.org/abs/1904.04781}{{\ttfamily 1904.04781}}.

\bibitem{Ramberg:2019dgi}
N.~Ramberg and L.~Visinelli, \emph{{Probing the Early Universe with Axion
  Physics and Gravitational Waves}},
  \href{https://arxiv.org/abs/1904.05707}{{\ttfamily 1904.05707}}.

\bibitem{Lawson:2019brd}
M.~Lawson, A.~J. Millar, M.~Pancaldi, E.~Vitagliano and F.~Wilczek,
  \emph{{Tunable axion plasma haloscopes}},
  \href{https://arxiv.org/abs/1904.11872}{{\ttfamily 1904.11872}}.

\bibitem{Milgrom:1983ca}
M.~Milgrom, \emph{{A Modification of the Newtonian dynamics as a possible
  alternative to the hidden mass hypothesis}},
  \href{https://doi.org/10.1086/161130}{\emph{Astrophys. J.} {\bfseries 270}
  (1983) 365--370}.

\bibitem{Mannheim:1988dj}
P.~D. Mannheim and D.~Kazanas, \emph{{Exact Vacuum Solution to Conformal Weyl
  Gravity and Galactic Rotation Curves}},
  \href{https://doi.org/10.1086/167623}{\emph{Astrophys. J.} {\bfseries 342}
  (1989) 635--638}.

\bibitem{Mannheim:1992vj}
P.~D. Mannheim, \emph{{Linear potentials and galactic rotation curves}},
  \href{https://doi.org/10.1086/173468}{\emph{Astrophys. J.} {\bfseries 419}
  (1993) 150--154}, [\href{https://arxiv.org/abs/hep-ph/9212304}{{\ttfamily
  hep-ph/9212304}}].

\bibitem{Bento:2003dj}
M.~C. Bento, O.~Bertolami and A.~A. Sen, \emph{{Generalized Chaplygin gas
  model: Dark energy - dark matter unification and CMBR constraints}},
  \href{https://doi.org/10.1023/A:1026207312105}{\emph{Gen. Rel. Grav.}
  {\bfseries 35} (2003) 2063--2069},
  [\href{https://arxiv.org/abs/gr-qc/0305086}{{\ttfamily gr-qc/0305086}}].

\bibitem{Zhang:2004gc}
X.~Zhang, F.-Q. Wu and J.~Zhang, \emph{{A New generalized Chaplygin gas as a
  scheme for unification of dark energy and dark matter}},
  \href{https://doi.org/10.1088/1475-7516/2006/01/003}{\emph{JCAP} {\bfseries
  0601} (2006) 003}, [\href{https://arxiv.org/abs/astro-ph/0411221}{{\ttfamily
  astro-ph/0411221}}].

\bibitem{Moffat:2004bm}
J.~W. Moffat, \emph{{Gravitational theory, galaxy rotation curves and cosmology
  without dark matter}},
  \href{https://doi.org/10.1088/1475-7516/2005/05/003}{\emph{JCAP} {\bfseries
  0505} (2005) 003}, [\href{https://arxiv.org/abs/astro-ph/0412195}{{\ttfamily
  astro-ph/0412195}}].

\bibitem{Arik:2005ir}
M.~Arık and M.~C. Çalık, \emph{{Can Brans-Dicke scalar field account for
  dark energy and dark matter?}},
  \href{https://doi.org/10.1142/S021773230602055X}{\emph{Mod. Phys. Lett.}
  {\bfseries A21} (2006) 1241--1248},
  [\href{https://arxiv.org/abs/gr-qc/0505035}{{\ttfamily gr-qc/0505035}}].

\bibitem{Moffat:2005si}
J.~W. Moffat, \emph{{Scalar-tensor-vector gravity theory}},
  \href{https://doi.org/10.1088/1475-7516/2006/03/004}{\emph{JCAP} {\bfseries
  0603} (2006) 004}, [\href{https://arxiv.org/abs/gr-qc/0506021}{{\ttfamily
  gr-qc/0506021}}].

\bibitem{Capozziello:2006ph}
S.~Capozziello, V.~F. Cardone and A.~Troisi, \emph{{Low surface brightness
  galaxies rotation curves in the low energy limit of $R^n$ gravity: no need
  for dark matter?}},
  \href{https://doi.org/10.1111/j.1365-2966.2007.11401.x}{\emph{Mon. Not. Roy.
  Astron. Soc.} {\bfseries 375} (2007) 1423--1440},
  [\href{https://arxiv.org/abs/astro-ph/0603522}{{\ttfamily
  astro-ph/0603522}}].

\bibitem{Sereno:2006mw}
M.~Sereno and P.~Jetzer, \emph{{Dark matter vs. modifications of the
  gravitational inverse-square law. Results from planetary motion in the solar
  system}}, \href{https://doi.org/10.1111/j.1365-2966.2006.10670.x}{\emph{Mon.
  Not. Roy. Astron. Soc.} {\bfseries 371} (2006) 626--632},
  [\href{https://arxiv.org/abs/astro-ph/0606197}{{\ttfamily
  astro-ph/0606197}}].

\bibitem{Zlosnik:2006zu}
T.~G. Złośnik, P.~G. Ferreira and G.~D. Starkman, \emph{{Modifying gravity
  with the Aether: An alternative to Dark Matter}},
  \href{https://doi.org/10.1103/PhysRevD.75.044017}{\emph{Phys. Rev.}
  {\bfseries D75} (2007) 044017},
  [\href{https://arxiv.org/abs/astro-ph/0607411}{{\ttfamily
  astro-ph/0607411}}].

\bibitem{Nojiri:2006gh}
S.~Nojiri and S.~D. Odintsov, \emph{{Modified f(R) gravity consistent with
  realistic cosmology: From matter dominated epoch to dark energy universe}},
  \href{https://doi.org/10.1103/PhysRevD.74.086005}{\emph{Phys. Rev.}
  {\bfseries D74} (2006) 086005},
  [\href{https://arxiv.org/abs/hep-th/0608008}{{\ttfamily hep-th/0608008}}].

\bibitem{Cognola:2006sp}
G.~Cognola, E.~Elizalde, S.~Nojiri, S.~Odintsov and S.~Zerbini,
  \emph{{String-inspired Gauss-Bonnet gravity reconstructed from the universe
  expansion history and yielding the transition from matter dominance to dark
  energy}}, \href{https://doi.org/10.1103/PhysRevD.75.086002}{\emph{Phys. Rev.}
  {\bfseries D75} (2007) 086002},
  [\href{https://arxiv.org/abs/hep-th/0611198}{{\ttfamily hep-th/0611198}}].

\bibitem{Brownstein:2007sr}
J.~R. Brownstein and J.~W. Moffat, \emph{{The Bullet Cluster 1E0657-558
  evidence shows Modified Gravity in the absence of Dark Matter}},
  \href{https://doi.org/10.1111/j.1365-2966.2007.12275.x}{\emph{Mon. Not. Roy.
  Astron. Soc.} {\bfseries 382} (2007) 29--47},
  [\href{https://arxiv.org/abs/astro-ph/0702146}{{\ttfamily
  astro-ph/0702146}}].

\bibitem{Saffari:2007xc}
Y.~Sobouti, \emph{{An f(r) gravitation for galactic environments}},
  \href{https://doi.org/10.1051/0004-6361:20077452,
  10.1051/0004-6361:20065188}{\emph{Astron. Astrophys.} {\bfseries 464} (2007)
  921}, [\href{https://arxiv.org/abs/0704.3345}{{\ttfamily 0704.3345}}].
  [Erratum: Astron. Astrophys.472,833(2007)].

\bibitem{Kahya:2007zy}
E.~O. Kahya and R.~P. Woodard, \emph{{A Generic Test of Modified Gravity Models
  which Emulate Dark Matter}},
  \href{https://doi.org/10.1016/j.physletb.2007.07.029}{\emph{Phys. Lett.}
  {\bfseries B652} (2007) 213--216},
  [\href{https://arxiv.org/abs/0705.0153}{{\ttfamily 0705.0153}}].

\bibitem{Boehmer:2007kx}
C.~G. Böhmer, T.~Harko and F.~S.~N. Lobo, \emph{{Dark matter as a geometric
  effect in f(R) gravity}},
  \href{https://doi.org/10.1016/j.astropartphys.2008.04.003}{\emph{Astropart.
  Phys.} {\bfseries 29} (2008) 386--392},
  [\href{https://arxiv.org/abs/0709.0046}{{\ttfamily 0709.0046}}].

\bibitem{Nojiri:2008nt}
S.~Nojiri and S.~D. Odintsov, \emph{{Dark energy, inflation and dark matter
  from modified F(R) gravity}}, {\emph{TSPU Bulletin} {\bfseries N8(110)}
  (2011) 7--19}, [\href{https://arxiv.org/abs/0807.0685}{{\ttfamily
  0807.0685}}].

\bibitem{Mukohyama:2009mz}
S.~Mukohyama, \emph{{Dark matter as integration constant in Hořava-Lifshitz
  gravity}}, \href{https://doi.org/10.1103/PhysRevD.80.064005}{\emph{Phys.
  Rev.} {\bfseries D80} (2009) 064005},
  [\href{https://arxiv.org/abs/0905.3563}{{\ttfamily 0905.3563}}].

\bibitem{Lim:2010yk}
E.~A. Lim, I.~Sawicki and A.~Vikman, \emph{{Dust of Dark Energy}},
  \href{https://doi.org/10.1088/1475-7516/2010/05/012}{\emph{JCAP} {\bfseries
  1005} (2010) 012}, [\href{https://arxiv.org/abs/1003.5751}{{\ttfamily
  1003.5751}}].

\bibitem{Sebastiani:2010ct}
L.~Sebastiani, \emph{{Dark Viscous Fluid coupled with Dark Matter and future
  singularity}},
  \href{https://doi.org/10.1140/epjc/s10052-010-1398-z}{\emph{Eur. Phys. J.}
  {\bfseries C69} (2010) 547--553},
  [\href{https://arxiv.org/abs/1006.1610}{{\ttfamily 1006.1610}}].

\bibitem{Capozziello:2012qt}
S.~Capozziello, T.~Harko, T.~S. Koivisto, F.~S.~N. Lobo and G.~J. Olmo,
  \emph{{The virial theorem and the dark matter problem in hybrid
  metric-Palatini gravity}},
  \href{https://doi.org/10.1088/1475-7516/2013/07/024}{\emph{JCAP} {\bfseries
  1307} (2013) 024}, [\href{https://arxiv.org/abs/1212.5817}{{\ttfamily
  1212.5817}}].

\bibitem{Chamseddine:2013kea}
A.~H. Chamseddine and V.~Mukhanov, \emph{{Mimetic Dark Matter}},
  \href{https://doi.org/10.1007/JHEP11(2013)135}{\emph{JHEP} {\bfseries 11}
  (2013) 135}, [\href{https://arxiv.org/abs/1308.5410}{{\ttfamily 1308.5410}}].

\bibitem{Chamseddine:2014vna}
A.~H. Chamseddine, V.~Mukhanov and A.~Vikman, \emph{{Cosmology with Mimetic
  Matter}}, \href{https://doi.org/10.1088/1475-7516/2014/06/017}{\emph{JCAP}
  {\bfseries 1406} (2014) 017},
  [\href{https://arxiv.org/abs/1403.3961}{{\ttfamily 1403.3961}}].

\bibitem{Boehmer:2014ipa}
C.~G. Böhmer, N.~Tamanini and M.~Wright, \emph{{On galaxy rotation curves from
  a continuum mechanics approach to modified gravity}},
  \href{https://doi.org/10.1142/S0218271818500074}{\emph{Int. J. Mod. Phys.}
  {\bfseries D27} (2017) 1850007},
  [\href{https://arxiv.org/abs/1403.4110}{{\ttfamily 1403.4110}}].

\bibitem{Mirzagholi:2014ifa}
L.~Mirzagholi and A.~Vikman, \emph{{Imperfect Dark Matter}},
  \href{https://doi.org/10.1088/1475-7516/2015/06/028}{\emph{JCAP} {\bfseries
  1506} (2015) 028}, [\href{https://arxiv.org/abs/1412.7136}{{\ttfamily
  1412.7136}}].

\bibitem{Tamanini:2015iia}
N.~Tamanini, \emph{{Phenomenological models of dark energy interacting with
  dark matter}}, \href{https://doi.org/10.1103/PhysRevD.92.043524}{\emph{Phys.
  Rev.} {\bfseries D92} (2015) 043524},
  [\href{https://arxiv.org/abs/1504.07397}{{\ttfamily 1504.07397}}].

\bibitem{Myrzakulov:2015nqa}
R.~Myrzakulov, L.~Sebastiani, S.~Vagnozzi and S.~Zerbini, \emph{{Mimetic
  covariant renormalizable gravity}}, {\emph{Fund. J. Mod. Phys.} {\bfseries 8}
  (2015) 119--124}, [\href{https://arxiv.org/abs/1505.03115}{{\ttfamily
  1505.03115}}].

\bibitem{Ramazanov:2015pha}
S.~Ramazanov, \emph{{Initial Conditions for Imperfect Dark Matter}},
  \href{https://doi.org/10.1088/1475-7516/2015/12/007}{\emph{JCAP} {\bfseries
  1512} (2015) 007}, [\href{https://arxiv.org/abs/1507.00291}{{\ttfamily
  1507.00291}}].

\bibitem{Guendelman:2015rea}
E.~Guendelman, E.~Nissimov and S.~Pacheva, \emph{{Dark Energy and Dark Matter
  From Hidden Symmetry of Gravity Model with a Non-Riemannian Volume Form}},
  \href{https://doi.org/10.1140/epjc/s10052-015-3699-8}{\emph{Eur. Phys. J.}
  {\bfseries C75} (2015) 472},
  [\href{https://arxiv.org/abs/1508.02008}{{\ttfamily 1508.02008}}].

\bibitem{Myrzakulov:2015kda}
R.~Myrzakulov, L.~Sebastiani, S.~Vagnozzi and S.~Zerbini, \emph{{Static
  spherically symmetric solutions in mimetic gravity: rotation curves and
  wormholes}},
  \href{https://doi.org/10.1088/0264-9381/33/12/125005}{\emph{Class. Quant.
  Grav.} {\bfseries 33} (2016) 125005},
  [\href{https://arxiv.org/abs/1510.02284}{{\ttfamily 1510.02284}}].

\bibitem{Salzano:2016udu}
V.~Salzano, D.~F. Mota, M.~P. Dąbrowski and S.~Capozziello, \emph{{No need for
  dark matter in galaxy clusters within Galileon theory}},
  \href{https://doi.org/10.1088/1475-7516/2016/10/033}{\emph{JCAP} {\bfseries
  1610} (2016) 033}, [\href{https://arxiv.org/abs/1607.02606}{{\ttfamily
  1607.02606}}].

\bibitem{Babichev:2016bxi}
E.~Babichev, L.~Marzola, M.~Raidal, A.~Schmidt-May, F.~Urban, H.~Veermäe
  et~al., \emph{{Heavy spin-2 Dark Matter}},
  \href{https://doi.org/10.1088/1475-7516/2016/09/016}{\emph{JCAP} {\bfseries
  1609} (2016) 016}, [\href{https://arxiv.org/abs/1607.03497}{{\ttfamily
  1607.03497}}].

\bibitem{Zaregonbadi:2016xna}
R.~Zaregonbadi, M.~Farhoudi and N.~Riazi, \emph{{Dark Matter From f(R,T)
  Gravity}}, \href{https://doi.org/10.1103/PhysRevD.94.084052}{\emph{Phys.
  Rev.} {\bfseries D94} (2016) 084052},
  [\href{https://arxiv.org/abs/1608.00469}{{\ttfamily 1608.00469}}].

\bibitem{Rinaldi:2016oqp}
M.~Rinaldi, \emph{{Mimicking dark matter in Horndeski gravity}},
  \href{https://doi.org/10.1016/j.dark.2017.02.003}{\emph{Phys. Dark Univ.}
  {\bfseries 16} (2017) 14--21},
  [\href{https://arxiv.org/abs/1608.03839}{{\ttfamily 1608.03839}}].

\bibitem{Verlinde:2016toy}
E.~P. Verlinde, \emph{{Emergent Gravity and the Dark Universe}},
  \href{https://doi.org/10.21468/SciPostPhys.2.3.016}{\emph{SciPost Phys.}
  {\bfseries 2} (2017) 016},
  [\href{https://arxiv.org/abs/1611.02269}{{\ttfamily 1611.02269}}].

\bibitem{Sebastiani:2016ras}
L.~Sebastiani, S.~Vagnozzi and R.~Myrzakulov, \emph{{Mimetic gravity: a review
  of recent developments and applications to cosmology and astrophysics}},
  \href{https://doi.org/10.1155/2017/3156915}{\emph{Adv. High Energy Phys.}
  {\bfseries 2017} (2017) 3156915},
  [\href{https://arxiv.org/abs/1612.08661}{{\ttfamily 1612.08661}}].

\bibitem{Marttens:2017njo}
R.~F. von Marttens, L.~Casarini, W.~Zimdahl, W.~S. Hipólito-Ricaldi and D.~F.
  Mota, \emph{{Does a generalized Chaplygin gas correctly describe the
  cosmological dark sector?}},
  \href{https://doi.org/10.1016/j.dark.2017.02.001}{\emph{Phys. Dark Univ.}
  {\bfseries 15} (2017) 114--124},
  [\href{https://arxiv.org/abs/1702.00651}{{\ttfamily 1702.00651}}].

\bibitem{Calmet:2017voc}
X.~Calmet and I.~Kuntz, \emph{{What is modified gravity and how to
  differentiate it from particle dark matter?}},
  \href{https://doi.org/10.1140/epjc/s10052-017-4695-y}{\emph{Eur. Phys. J.}
  {\bfseries C77} (2017) 132},
  [\href{https://arxiv.org/abs/1702.03832}{{\ttfamily 1702.03832}}].

\bibitem{Hirano:2017zox}
S.~Hirano, S.~Nishi and T.~Kobayashi, \emph{{Healthy imperfect dark matter from
  effective theory of mimetic cosmological perturbations}},
  \href{https://doi.org/10.1088/1475-7516/2017/07/009}{\emph{JCAP} {\bfseries
  1707} (2017) 009}, [\href{https://arxiv.org/abs/1704.06031}{{\ttfamily
  1704.06031}}].

\bibitem{Koutsoumbas:2017fxp}
G.~Koutsoumbas, K.~Ntrekis, E.~Papantonopoulos and E.~N. Saridakis,
  \emph{{Unification of Dark Matter - Dark Energy in Generalized Galileon
  Theories}}, \href{https://doi.org/10.1088/1475-7516/2018/02/003}{\emph{JCAP}
  {\bfseries 1802} (2018) 003},
  [\href{https://arxiv.org/abs/1704.08640}{{\ttfamily 1704.08640}}].

\bibitem{Vagnozzi:2017ilo}
S.~Vagnozzi, \emph{{Recovering a MOND-like acceleration law in mimetic
  gravity}}, \href{https://doi.org/10.1088/1361-6382/aa838b}{\emph{Class.
  Quant. Grav.} {\bfseries 34} (2017) 185006},
  [\href{https://arxiv.org/abs/1708.00603}{{\ttfamily 1708.00603}}].

\bibitem{Marzola:2017lbt}
L.~Marzola, M.~Raidal and F.~R. Urban, \emph{{Oscillating Spin-2 Dark Matter}},
  \href{https://doi.org/10.1103/PhysRevD.97.024010}{\emph{Phys. Rev.}
  {\bfseries D97} (2018) 024010},
  [\href{https://arxiv.org/abs/1708.04253}{{\ttfamily 1708.04253}}].

\bibitem{Dutta:2017fjw}
J.~Dutta, W.~Khyllep, E.~N. Saridakis, N.~Tamanini and S.~Vagnozzi,
  \emph{{Cosmological dynamics of mimetic gravity}},
  \href{https://doi.org/10.1088/1475-7516/2018/02/041}{\emph{JCAP} {\bfseries
  1802} (2018) 041}, [\href{https://arxiv.org/abs/1711.07290}{{\ttfamily
  1711.07290}}].

\bibitem{Frandsen:2018ftj}
M.~T. Frandsen and J.~Petersen, \emph{{Investigating Dark Matter and MOND
  Models with Galactic Rotation Curve Data}},
  \href{https://arxiv.org/abs/1805.10706}{{\ttfamily 1805.10706}}.

\bibitem{Zlosnik:2018qvg}
T.~Złośnik, F.~Urban, L.~Marzola and T.~Koivisto, \emph{{Spacetime and dark
  matter from spontaneous breaking of Lorentz symmetry}},
  \href{https://doi.org/10.1088/1361-6382/aaea96}{\emph{Class. Quant. Grav.}
  {\bfseries 35} (2018) 235003},
  [\href{https://arxiv.org/abs/1807.01100}{{\ttfamily 1807.01100}}].

\bibitem{Barrientos:2018giw}
E.~Barrientos and S.~Mendoza, \emph{{MOND as the weak field limit of an
  extended metric theory of gravity with a matter-curvature coupling}},
  \href{https://doi.org/10.1103/PhysRevD.98.084033}{\emph{Phys. Rev.}
  {\bfseries D98} (2018) 084033},
  [\href{https://arxiv.org/abs/1808.01386}{{\ttfamily 1808.01386}}].

\bibitem{Burrage:2018zuj}
C.~Burrage, E.~J. Copeland, C.~Käding and P.~Millington, \emph{{Symmetron
  scalar fields: Modified gravity, dark matter, or both?}},
  \href{https://doi.org/10.1103/PhysRevD.99.043539}{\emph{Phys. Rev.}
  {\bfseries D99} (2019) 043539},
  [\href{https://arxiv.org/abs/1811.12301}{{\ttfamily 1811.12301}}].

\bibitem{Lisanti:2018qam}
M.~Lisanti, M.~Moschella, N.~J. Outmezguine and O.~Slone, \emph{{Testing Dark
  Matter and Modifications to Gravity using Local Milky Way Observables}},
  \href{https://arxiv.org/abs/1812.08169}{{\ttfamily 1812.08169}}.

\bibitem{Odintsov:2019mlf}
S.~D. Odintsov and V.~K. Oikonomou, \emph{{$f(R)$ Gravity Inflation with
  String-Corrected Axion Dark Matter}},
  \href{https://doi.org/10.1103/PhysRevD.99.064049}{\emph{Phys. Rev.}
  {\bfseries D99} (2019) 064049},
  [\href{https://arxiv.org/abs/1901.05363}{{\ttfamily 1901.05363}}].

\bibitem{Horndeski:1974wa}
G.~W. Horndeski, \emph{{Second-order scalar-tensor field equations in a
  four-dimensional space}},
  \href{https://doi.org/10.1007/BF01807638}{\emph{Int. J. Theor. Phys.}
  {\bfseries 10} (1974) 363--384}.

\bibitem{Bilic:2001cg}
N.~Bilić, G.~B. Tupper and R.~D. Viollier, \emph{{Unification of dark matter
  and dark energy: The Inhomogeneous Chaplygin gas}},
  \href{https://doi.org/10.1016/S0370-2693(02)01716-1}{\emph{Phys. Lett.}
  {\bfseries B535} (2002) 17--21},
  [\href{https://arxiv.org/abs/astro-ph/0111325}{{\ttfamily
  astro-ph/0111325}}].

\bibitem{Freese:2002sq}
K.~Freese and M.~Lewis, \emph{{Cardassian expansion: A Model in which the
  universe is flat, matter dominated, and accelerating}},
  \href{https://doi.org/10.1016/S0370-2693(02)02122-6}{\emph{Phys. Lett.}
  {\bfseries B540} (2002) 1--8},
  [\href{https://arxiv.org/abs/astro-ph/0201229}{{\ttfamily
  astro-ph/0201229}}].

\bibitem{Bento:2002ps}
M.~C. Bento, O.~Bertolami and A.~A. Sen, \emph{{Generalized Chaplygin gas,
  accelerated expansion and dark energy matter unification}},
  \href{https://doi.org/10.1103/PhysRevD.66.043507}{\emph{Phys. Rev.}
  {\bfseries D66} (2002) 043507},
  [\href{https://arxiv.org/abs/gr-qc/0202064}{{\ttfamily gr-qc/0202064}}].

\bibitem{Sahni:2002dx}
V.~Sahni and Y.~Shtanov, \emph{{Brane world models of dark energy}},
  \href{https://doi.org/10.1088/1475-7516/2003/11/014}{\emph{JCAP} {\bfseries
  0311} (2003) 014}, [\href{https://arxiv.org/abs/astro-ph/0202346}{{\ttfamily
  astro-ph/0202346}}].

\bibitem{Li:2004rb}
M.~Li, \emph{{A Model of holographic dark energy}},
  \href{https://doi.org/10.1016/j.physletb.2004.10.014}{\emph{Phys. Lett.}
  {\bfseries B603} (2004) 1},
  [\href{https://arxiv.org/abs/hep-th/0403127}{{\ttfamily hep-th/0403127}}].

\bibitem{Elizalde:2004mq}
E.~Elizalde, S.~Nojiri and S.~D. Odintsov, \emph{{Late-time cosmology in
  (phantom) scalar-tensor theory: Dark energy and the cosmic speed-up}},
  \href{https://doi.org/10.1103/PhysRevD.70.043539}{\emph{Phys. Rev.}
  {\bfseries D70} (2004) 043539},
  [\href{https://arxiv.org/abs/hep-th/0405034}{{\ttfamily hep-th/0405034}}].

\bibitem{Nojiri:2005vv}
S.~Nojiri, S.~D. Odintsov and M.~Sasaki, \emph{{Gauss-Bonnet dark energy}},
  \href{https://doi.org/10.1103/PhysRevD.71.123509}{\emph{Phys. Rev.}
  {\bfseries D71} (2005) 123509},
  [\href{https://arxiv.org/abs/hep-th/0504052}{{\ttfamily hep-th/0504052}}].

\bibitem{Kolb:2005da}
E.~W. Kolb, S.~Matarrese and A.~Riotto, \emph{{On cosmic acceleration without
  dark energy}}, \href{https://doi.org/10.1088/1367-2630/8/12/322}{\emph{New J.
  Phys.} {\bfseries 8} (2006) 322},
  [\href{https://arxiv.org/abs/astro-ph/0506534}{{\ttfamily
  astro-ph/0506534}}].

\bibitem{Nojiri:2005jg}
S.~Nojiri and S.~D. Odintsov, \emph{{Modified Gauss-Bonnet theory as
  gravitational alternative for dark energy}},
  \href{https://doi.org/10.1016/j.physletb.2005.10.010}{\emph{Phys. Lett.}
  {\bfseries B631} (2005) 1--6},
  [\href{https://arxiv.org/abs/hep-th/0508049}{{\ttfamily hep-th/0508049}}].

\bibitem{Alnes:2005rw}
H.~Alnes, M.~Amarzguioui and O.~Grøn, \emph{{An inhomogeneous alternative to
  dark energy?}}, \href{https://doi.org/10.1103/PhysRevD.73.083519}{\emph{Phys.
  Rev.} {\bfseries D73} (2006) 083519},
  [\href{https://arxiv.org/abs/astro-ph/0512006}{{\ttfamily
  astro-ph/0512006}}].

\bibitem{Cognola:2006eg}
G.~Cognola, E.~Elizalde, S.~Nojiri, S.~D. Odintsov and S.~Zerbini, \emph{{Dark
  energy in modified Gauss-Bonnet gravity: Late-time acceleration and the
  hierarchy problem}},
  \href{https://doi.org/10.1103/PhysRevD.73.084007}{\emph{Phys. Rev.}
  {\bfseries D73} (2006) 084007},
  [\href{https://arxiv.org/abs/hep-th/0601008}{{\ttfamily hep-th/0601008}}].

\bibitem{Copeland:2006wr}
E.~J. Copeland, M.~Sami and S.~Tsujikawa, \emph{{Dynamics of dark energy}},
  \href{https://doi.org/10.1142/S021827180600942X}{\emph{Int. J. Mod. Phys.}
  {\bfseries D15} (2006) 1753--1936},
  [\href{https://arxiv.org/abs/hep-th/0603057}{{\ttfamily hep-th/0603057}}].

\bibitem{Amendola:2006kh}
L.~Amendola, D.~Polarski and S.~Tsujikawa, \emph{{Are f(R) dark energy models
  cosmologically viable ?}},
  \href{https://doi.org/10.1103/PhysRevLett.98.131302}{\emph{Phys. Rev. Lett.}
  {\bfseries 98} (2007) 131302},
  [\href{https://arxiv.org/abs/astro-ph/0603703}{{\ttfamily
  astro-ph/0603703}}].

\bibitem{Carroll:2006jn}
S.~M. Carroll, I.~Sawicki, A.~Silvestri and M.~Trodden, \emph{{Modified-Source
  Gravity and Cosmological Structure Formation}},
  \href{https://doi.org/10.1088/1367-2630/8/12/323}{\emph{New J. Phys.}
  {\bfseries 8} (2006) 323},
  [\href{https://arxiv.org/abs/astro-ph/0607458}{{\ttfamily
  astro-ph/0607458}}].

\bibitem{Amendola:2006we}
L.~Amendola, R.~Gannouji, D.~Polarski and S.~Tsujikawa, \emph{{Conditions for
  the cosmological viability of f(R) dark energy models}},
  \href{https://doi.org/10.1103/PhysRevD.75.083504}{\emph{Phys. Rev.}
  {\bfseries D75} (2007) 083504},
  [\href{https://arxiv.org/abs/gr-qc/0612180}{{\ttfamily gr-qc/0612180}}].

\bibitem{Hu:2007nk}
W.~Hu and I.~Sawicki, \emph{{Models of f(R) Cosmic Acceleration that Evade
  Solar-System Tests}},
  \href{https://doi.org/10.1103/PhysRevD.76.064004}{\emph{Phys. Rev.}
  {\bfseries D76} (2007) 064004},
  [\href{https://arxiv.org/abs/0705.1158}{{\ttfamily 0705.1158}}].

\bibitem{Pogosian:2007sw}
L.~Pogosian and A.~Silvestri, \emph{{The pattern of growth in viable f(R)
  cosmologies}}, \href{https://doi.org/10.1103/PhysRevD.77.023503,
  10.1103/PhysRevD.81.049901}{\emph{Phys. Rev.} {\bfseries D77} (2008) 023503},
  [\href{https://arxiv.org/abs/0709.0296}{{\ttfamily 0709.0296}}]. [Erratum:
  Phys. Rev.D81,049901(2010)].

\bibitem{Tsujikawa:2007xu}
S.~Tsujikawa, \emph{{Observational signatures of f(R) dark energy models that
  satisfy cosmological and local gravity constraints}},
  \href{https://doi.org/10.1103/PhysRevD.77.023507}{\emph{Phys. Rev.}
  {\bfseries D77} (2008) 023507},
  [\href{https://arxiv.org/abs/0709.1391}{{\ttfamily 0709.1391}}].

\bibitem{Saridakis:2007wx}
E.~N. Saridakis, \emph{{Holographic Dark Energy in Braneworld Models with a
  Gauss-Bonnet Term in the Bulk. Interacting Behavior and the w =-1 Crossing}},
  \href{https://doi.org/10.1016/j.physletb.2008.02.032}{\emph{Phys. Lett.}
  {\bfseries B661} (2008) 335--341},
  [\href{https://arxiv.org/abs/0712.3806}{{\ttfamily 0712.3806}}].

\bibitem{Jhingan:2008ym}
S.~Jhingan, S.~Nojiri, S.~D. Odintsov, M.~Sami, I.~Thongkool and S.~Zerbini,
  \emph{{Phantom and non-phantom dark energy: The Cosmological relevance of
  non-locally corrected gravity}},
  \href{https://doi.org/10.1016/j.physletb.2008.04.054}{\emph{Phys. Lett.}
  {\bfseries B663} (2008) 424--428},
  [\href{https://arxiv.org/abs/0803.2613}{{\ttfamily 0803.2613}}].

\bibitem{Gavela:2009cy}
M.~B. Gavela, D.~Hernández, L.~Lopez~Honorez, O.~Mena and S.~Rigolin,
  \emph{{Dark coupling}}, \href{https://doi.org/10.1088/1475-7516/2010/05/E01,
  10.1088/1475-7516/2009/07/034}{\emph{JCAP} {\bfseries 0907} (2009) 034},
  [\href{https://arxiv.org/abs/0901.1611}{{\ttfamily 0901.1611}}]. [Erratum:
  JCAP1005,E01(2010)].

\bibitem{Saridakis:2009bv}
E.~N. Saridakis, \emph{{Hořava-Lifshitz Dark Energy}},
  \href{https://doi.org/10.1140/epjc/s10052-010-1294-6}{\emph{Eur. Phys. J.}
  {\bfseries C67} (2010) 229--235},
  [\href{https://arxiv.org/abs/0905.3532}{{\ttfamily 0905.3532}}].

\bibitem{Zumalacarregui:2010wj}
M.~Zumalacárregui, T.~S. Koivisto, D.~F. Mota and P.~Ruíz-Lapuente,
  \emph{{Disformal Scalar Fields and the Dark Sector of the Universe}},
  \href{https://doi.org/10.1088/1475-7516/2010/05/038}{\emph{JCAP} {\bfseries
  1005} (2010) 038}, [\href{https://arxiv.org/abs/1004.2684}{{\ttfamily
  1004.2684}}].

\bibitem{Dent:2011zz}
J.~B. Dent, S.~Dutta and E.~N. Saridakis, \emph{{f(T) gravity mimicking
  dynamical dark energy. Background and perturbation analysis}},
  \href{https://doi.org/10.1088/1475-7516/2011/01/009}{\emph{JCAP} {\bfseries
  1101} (2011) 009}, [\href{https://arxiv.org/abs/1010.2215}{{\ttfamily
  1010.2215}}].

\bibitem{Elizalde:2011ds}
E.~Elizalde, S.~D. Odintsov, L.~Sebastiani and S.~Zerbini, \emph{{Oscillations
  of the F(R) dark energy in the accelerating universe}},
  \href{https://doi.org/10.1140/epjc/s10052-011-1843-7}{\emph{Eur. Phys. J.}
  {\bfseries C72} (2012) 1843},
  [\href{https://arxiv.org/abs/1108.6184}{{\ttfamily 1108.6184}}].

\bibitem{Geng:2011aj}
C.-Q. Geng, C.-C. Lee, E.~N. Saridakis and Y.-P. Wu, \emph{{“Teleparallel”
  dark energy}},
  \href{https://doi.org/10.1016/j.physletb.2011.09.082}{\emph{Phys. Lett.}
  {\bfseries B704} (2011) 384--387},
  [\href{https://arxiv.org/abs/1109.1092}{{\ttfamily 1109.1092}}].

\bibitem{Bamba:2012cp}
K.~Bamba, S.~Capozziello, S.~Nojiri and S.~D. Odintsov, \emph{{Dark energy
  cosmology: the equivalent description via different theoretical models and
  cosmography tests}},
  \href{https://doi.org/10.1007/s10509-012-1181-8}{\emph{Astrophys. Space Sci.}
  {\bfseries 342} (2012) 155--228},
  [\href{https://arxiv.org/abs/1205.3421}{{\ttfamily 1205.3421}}].

\bibitem{Bamba:2012qi}
K.~Bamba, A.~López-Revelles, R.~Myrzakulov, S.~D. Odintsov and L.~Sebastiani,
  \emph{{Cosmic history of viable exponential gravity: Equation of state
  oscillations and growth index from inflation to dark energy era}},
  \href{https://doi.org/10.1088/0264-9381/30/1/015008}{\emph{Class. Quant.
  Grav.} {\bfseries 30} (2013) 015008},
  [\href{https://arxiv.org/abs/1207.1009}{{\ttfamily 1207.1009}}].

\bibitem{Pan:2012ki}
S.~Pan, S.~Bhattacharya and S.~Chakraborty, \emph{{An analytic model for
  interacting dark energy and its observational constraints}},
  \href{https://doi.org/10.1093/mnras/stv1495}{\emph{Mon. Not. Roy. Astron.
  Soc.} {\bfseries 452} (2015) 3038--3046},
  [\href{https://arxiv.org/abs/1210.0396}{{\ttfamily 1210.0396}}].

\bibitem{Myrzakulov:2013hca}
R.~Myrzakulov, L.~Sebastiani and S.~Zerbini, \emph{{Some aspects of generalized
  modified gravity models}},
  \href{https://doi.org/10.1142/S0218271813300176}{\emph{Int. J. Mod. Phys.}
  {\bfseries D22} (2013) 1330017},
  [\href{https://arxiv.org/abs/1302.4646}{{\ttfamily 1302.4646}}].

\bibitem{Pan:2013rha}
S.~Pan and S.~Chakraborty, \emph{{Will there be again a transition from
  acceleration to deceleration in course of the dark energy evolution of the
  universe?}}, \href{https://doi.org/10.1140/epjc/s10052-013-2575-7}{\emph{Eur.
  Phys. J.} {\bfseries C73} (2013) 2575},
  [\href{https://arxiv.org/abs/1303.5602}{{\ttfamily 1303.5602}}].

\bibitem{Gleyzes:2013ooa}
J.~Gleyzes, D.~Langlois, F.~Piazza and F.~Vernizzi, \emph{{Essential Building
  Blocks of Dark Energy}},
  \href{https://doi.org/10.1088/1475-7516/2013/08/025}{\emph{JCAP} {\bfseries
  1308} (2013) 025}, [\href{https://arxiv.org/abs/1304.4840}{{\ttfamily
  1304.4840}}].

\bibitem{Cognola:2013fva}
G.~Cognola, R.~Myrzakulov, L.~Sebastiani and S.~Zerbini, \emph{{Einstein
  gravity with Gauss-Bonnet entropic corrections}},
  \href{https://doi.org/10.1103/PhysRevD.88.024006}{\emph{Phys. Rev.}
  {\bfseries D88} (2013) 024006},
  [\href{https://arxiv.org/abs/1304.1878}{{\ttfamily 1304.1878}}].

\bibitem{Maggiore:2014sia}
M.~Maggiore and M.~Mancarella, \emph{{Nonlocal gravity and dark energy}},
  \href{https://doi.org/10.1103/PhysRevD.90.023005}{\emph{Phys. Rev.}
  {\bfseries D90} (2014) 023005},
  [\href{https://arxiv.org/abs/1402.0448}{{\ttfamily 1402.0448}}].

\bibitem{Rinaldi:2014yta}
M.~Rinaldi, \emph{{Higgs Dark Energy}},
  \href{https://doi.org/10.1088/0264-9381/32/4/045002}{\emph{Class. Quant.
  Grav.} {\bfseries 32} (2015) 045002},
  [\href{https://arxiv.org/abs/1404.0532}{{\ttfamily 1404.0532}}].

\bibitem{Rinaldi:2015iza}
M.~Rinaldi, \emph{{Dark energy as a fixed point of the Einstein Yang-Mills
  Higgs Equations}},
  \href{https://doi.org/10.1088/1475-7516/2015/10/023}{\emph{JCAP} {\bfseries
  1510} (2015) 023}, [\href{https://arxiv.org/abs/1508.04576}{{\ttfamily
  1508.04576}}].

\bibitem{Rabochaya:2015haa}
Y.~Rabochaya and S.~Zerbini, \emph{{A note on a mimetic scalar–tensor
  cosmological model}},
  \href{https://doi.org/10.1140/epjc/s10052-016-3926-y}{\emph{Eur. Phys. J.}
  {\bfseries C76} (2016) 85},
  [\href{https://arxiv.org/abs/1509.03720}{{\ttfamily 1509.03720}}].

\bibitem{Wang:2016lxa}
B.~Wang, E.~Abdalla, F.~Atrio-Barandela and D.~Pavón, \emph{{Dark Matter and
  Dark Energy Interactions: Theoretical Challenges, Cosmological Implications
  and Observational Signatures}},
  \href{https://doi.org/10.1088/0034-4885/79/9/096901}{\emph{Rept. Prog. Phys.}
  {\bfseries 79} (2016) 096901},
  [\href{https://arxiv.org/abs/1603.08299}{{\ttfamily 1603.08299}}].

\bibitem{Elizalde:2017mrn}
E.~Elizalde, S.~D. Odintsov, L.~Sebastiani and R.~Myrzakulov,
  \emph{{Beyond-one-loop quantum gravity action yielding both inflation and
  late-time acceleration}},
  \href{https://doi.org/10.1016/j.nuclphysb.2017.06.003}{\emph{Nucl. Phys.}
  {\bfseries B921} (2017) 411--435},
  [\href{https://arxiv.org/abs/1706.01879}{{\ttfamily 1706.01879}}].

\bibitem{Saridakis:2017rdo}
E.~N. Saridakis, \emph{{Ricci-Gauss-Bonnet holographic dark energy}},
  \href{https://doi.org/10.1103/PhysRevD.97.064035}{\emph{Phys. Rev.}
  {\bfseries D97} (2018) 064035},
  [\href{https://arxiv.org/abs/1707.09331}{{\ttfamily 1707.09331}}].

\bibitem{Bahamonde:2017ize}
S.~Bahamonde, C.~G. Böhmer, S.~Carloni, E.~J. Copeland, W.~Fang and
  N.~Tamanini, \emph{{Dynamical systems applied to cosmology: dark energy and
  modified gravity}},
  \href{https://doi.org/10.1016/j.physrep.2018.09.001}{\emph{Phys. Rept.}
  {\bfseries 775-777} (2018) 1--122},
  [\href{https://arxiv.org/abs/1712.03107}{{\ttfamily 1712.03107}}].

\bibitem{Bahamonde:2018miw}
S.~Bahamonde, M.~Marciu and P.~Rudra, \emph{{Generalised teleparallel quintom
  dark energy non-minimally coupled with the scalar torsion and a boundary
  term}}, \href{https://doi.org/10.1088/1475-7516/2018/04/056}{\emph{JCAP}
  {\bfseries 1804} (2018) 056},
  [\href{https://arxiv.org/abs/1802.09155}{{\ttfamily 1802.09155}}].

\bibitem{Saridakis:2018unr}
E.~N. Saridakis, K.~Bamba, R.~Myrzakulov and F.~K. Anagnostopoulos,
  \emph{{Holographic dark energy through Tsallis entropy}},
  \href{https://doi.org/10.1088/1475-7516/2018/12/012}{\emph{JCAP} {\bfseries
  1812} (2018) 012}, [\href{https://arxiv.org/abs/1806.01301}{{\ttfamily
  1806.01301}}].

\bibitem{Odintsov:2018nch}
S.~D. Odintsov, V.~K. Oikonomou and S.~Banerjee, \emph{{Dynamics of inflation
  and dark energy from $F(R,G)$ gravity}},
  \href{https://doi.org/10.1016/j.nuclphysb.2018.07.013}{\emph{Nucl. Phys.}
  {\bfseries B938} (2019) 935--956},
  [\href{https://arxiv.org/abs/1807.00335}{{\ttfamily 1807.00335}}].

\bibitem{Marsh:2018kub}
M.~C. David~Marsh, \emph{{The Swampland, Quintessence and the Vacuum Energy}},
  \href{https://doi.org/10.1016/j.physletb.2018.11.001}{\emph{Phys. Lett.}
  {\bfseries B789} (2019) 639--642},
  [\href{https://arxiv.org/abs/1809.00726}{{\ttfamily 1809.00726}}].

\bibitem{Odintsov:2018zai}
S.~D. Odintsov and V.~K. Oikonomou, \emph{{Finite-time Singularities in
  Swampland-related Dark Energy Models}},
  \href{https://arxiv.org/abs/1810.03575}{{\ttfamily 1810.03575}}.

\bibitem{Langlois:2018dxi}
D.~Langlois, \emph{{Dark Energy and Modified Gravity in Degenerate Higher-Order
  Scalar-Tensor (DHOST) theories: a review}},
  \href{https://doi.org/10.1142/S0218271819420069}{\emph{Int. J. Mod. Phys.}
  {\bfseries D28} (2019) 1942006},
  [\href{https://arxiv.org/abs/1811.06271}{{\ttfamily 1811.06271}}].

\bibitem{Barros:2019rdv}
B.~J. Barros, \emph{{Kinetically coupled dark energy}},
  \href{https://doi.org/10.1103/PhysRevD.99.064051}{\emph{Phys. Rev.}
  {\bfseries D99} (2019) 064051},
  [\href{https://arxiv.org/abs/1901.03972}{{\ttfamily 1901.03972}}].

\bibitem{Paliathanasis:2019hbi}
A.~Paliathanasis, S.~Pan and W.~Yang, \emph{{Dynamics of nonlinear interacting
  dark energy models}},  \href{https://arxiv.org/abs/1903.02370}{{\ttfamily
  1903.02370}}.

\bibitem{Huterer:1998qv}
D.~Huterer and M.~S. Turner, \emph{{Prospects for probing the dark energy via
  supernova distance measurements}},
  \href{https://doi.org/10.1103/PhysRevD.60.081301}{\emph{Phys. Rev.}
  {\bfseries D60} (1999) 081301},
  [\href{https://arxiv.org/abs/astro-ph/9808133}{{\ttfamily
  astro-ph/9808133}}].

\bibitem{Perlmutter:1999jt}
S.~Perlmutter, M.~S. Turner and M.~J. White, \emph{{Constraining dark energy
  with SNe Ia and large scale structure}},
  \href{https://doi.org/10.1103/PhysRevLett.83.670}{\emph{Phys. Rev. Lett.}
  {\bfseries 83} (1999) 670--673},
  [\href{https://arxiv.org/abs/astro-ph/9901052}{{\ttfamily
  astro-ph/9901052}}].

\bibitem{Huterer:2000mj}
D.~Huterer and M.~S. Turner, \emph{{Probing the dark energy: Methods and
  strategies}}, \href{https://doi.org/10.1103/PhysRevD.64.123527}{\emph{Phys.
  Rev.} {\bfseries D64} (2001) 123527},
  [\href{https://arxiv.org/abs/astro-ph/0012510}{{\ttfamily
  astro-ph/0012510}}].

\bibitem{Hannestad:2002ur}
S.~Hannestad and E.~Mörtsell, \emph{{Probing the dark side: Constraints on the
  dark energy equation of state from CMB, large scale structure and Type Ia
  supernovae}}, \href{https://doi.org/10.1103/PhysRevD.66.063508}{\emph{Phys.
  Rev.} {\bfseries D66} (2002) 063508},
  [\href{https://arxiv.org/abs/astro-ph/0205096}{{\ttfamily
  astro-ph/0205096}}].

\bibitem{Melchiorri:2002ux}
A.~Melchiorri, L.~Mersini-Houghton, C.~J. Odman and M.~Trodden, \emph{{The
  State of the dark energy equation of state}},
  \href{https://doi.org/10.1103/PhysRevD.68.043509}{\emph{Phys. Rev.}
  {\bfseries D68} (2003) 043509},
  [\href{https://arxiv.org/abs/astro-ph/0211522}{{\ttfamily
  astro-ph/0211522}}].

\bibitem{Blake:2003rh}
C.~Blake and K.~Glazebrook, \emph{{Probing dark energy using baryonic
  oscillations in the galaxy power spectrum as a cosmological ruler}},
  \href{https://doi.org/10.1086/376983}{\emph{Astrophys. J.} {\bfseries 594}
  (2003) 665--673}, [\href{https://arxiv.org/abs/astro-ph/0301632}{{\ttfamily
  astro-ph/0301632}}].

\bibitem{Wang:2004ru}
Y.~Wang and K.~Freese, \emph{{Probing dark energy using its density instead of
  its equation of state}},
  \href{https://doi.org/10.1016/j.physletb.2005.10.083}{\emph{Phys. Lett.}
  {\bfseries B632} (2006) 449--452},
  [\href{https://arxiv.org/abs/astro-ph/0402208}{{\ttfamily
  astro-ph/0402208}}].

\bibitem{Hannestad:2004cb}
S.~Hannestad and E.~Mörtsell, \emph{{Cosmological constraints on the dark
  energy equation of state and its evolution}},
  \href{https://doi.org/10.1088/1475-7516/2004/09/001}{\emph{JCAP} {\bfseries
  0409} (2004) 001}, [\href{https://arxiv.org/abs/astro-ph/0407259}{{\ttfamily
  astro-ph/0407259}}].

\bibitem{Blomqvist:2008ud}
M.~Blomqvist, E.~Mörtsell and S.~Nobili, \emph{{Probing Dark Energy
  Inhomogeneities with Supernovae}},
  \href{https://doi.org/10.1088/1475-7516/2008/06/027}{\emph{JCAP} {\bfseries
  0806} (2008) 027}, [\href{https://arxiv.org/abs/0806.0496}{{\ttfamily
  0806.0496}}].

\bibitem{Zhao:2009fn}
G.-B. Zhao, L.~Pogosian, A.~Silvestri and J.~Zylberberg, \emph{{Cosmological
  Tests of General Relativity with Future Tomographic Surveys}},
  \href{https://doi.org/10.1103/PhysRevLett.103.241301}{\emph{Phys. Rev. Lett.}
  {\bfseries 103} (2009) 241301},
  [\href{https://arxiv.org/abs/0905.1326}{{\ttfamily 0905.1326}}].

\bibitem{Giannantonio:2009gi}
T.~Giannantonio, M.~Martinelli, A.~Silvestri and A.~Melchiorri, \emph{{New
  constraints on parametrised modified gravity from correlations of the CMB
  with large scale structure}},
  \href{https://doi.org/10.1088/1475-7516/2010/04/030}{\emph{JCAP} {\bfseries
  1004} (2010) 030}, [\href{https://arxiv.org/abs/0909.2045}{{\ttfamily
  0909.2045}}].

\bibitem{Sherwin:2011gv}
B.~D. Sherwin et~al., \emph{{Evidence for dark energy from the cosmic microwave
  background alone using the Atacama Cosmology Telescope lensing
  measurements}},
  \href{https://doi.org/10.1103/PhysRevLett.107.021302}{\emph{Phys. Rev. Lett.}
  {\bfseries 107} (2011) 021302},
  [\href{https://arxiv.org/abs/1105.0419}{{\ttfamily 1105.0419}}].

\bibitem{Hojjati:2011ix}
A.~Hojjati, L.~Pogosian and G.-B. Zhao, \emph{{Testing gravity with CAMB and
  CosmoMC}}, \href{https://doi.org/10.1088/1475-7516/2011/08/005}{\emph{JCAP}
  {\bfseries 1108} (2011) 005},
  [\href{https://arxiv.org/abs/1106.4543}{{\ttfamily 1106.4543}}].

\bibitem{Zumalacarregui:2012pq}
M.~Zumalacárregui, J.~García-Bellido and P.~Ruíz-Lapuente, \emph{{Tension in
  the Void: Cosmic Rulers Strain Inhomogeneous Cosmologies}},
  \href{https://doi.org/10.1088/1475-7516/2012/10/009}{\emph{JCAP} {\bfseries
  1210} (2012) 009}, [\href{https://arxiv.org/abs/1201.2790}{{\ttfamily
  1201.2790}}].

\bibitem{Bellini:2012qn}
E.~Bellini, N.~Bartolo and S.~Matarrese, \emph{{Spherical Collapse in covariant
  Galileon theory}},
  \href{https://doi.org/10.1088/1475-7516/2012/06/019}{\emph{JCAP} {\bfseries
  1206} (2012) 019}, [\href{https://arxiv.org/abs/1202.2712}{{\ttfamily
  1202.2712}}].

\bibitem{Zumalacarregui:2012us}
M.~Zumalacárregui, T.~S. Koivisto and D.~F. Mota, \emph{{DBI Galileons in the
  Einstein Frame: Local Gravity and Cosmology}},
  \href{https://doi.org/10.1103/PhysRevD.87.083010}{\emph{Phys. Rev.}
  {\bfseries D87} (2013) 083010},
  [\href{https://arxiv.org/abs/1210.8016}{{\ttfamily 1210.8016}}].

\bibitem{Bartolo:2013ws}
N.~Bartolo, E.~Bellini, D.~Bertacca and S.~Matarrese, \emph{{Matter bispectrum
  in cubic Galileon cosmologies}},
  \href{https://doi.org/10.1088/1475-7516/2013/03/034}{\emph{JCAP} {\bfseries
  1303} (2013) 034}, [\href{https://arxiv.org/abs/1301.4831}{{\ttfamily
  1301.4831}}].

\bibitem{Salvatelli:2013wra}
V.~Salvatelli, A.~Marchini, L.~Lopez-Honorez and O.~Mena, \emph{{New
  constraints on Coupled Dark Energy from the Planck satellite experiment}},
  \href{https://doi.org/10.1103/PhysRevD.88.023531}{\emph{Phys. Rev.}
  {\bfseries D88} (2013) 023531},
  [\href{https://arxiv.org/abs/1304.7119}{{\ttfamily 1304.7119}}].

\bibitem{Bellini:2014fua}
E.~Bellini and I.~Sawicki, \emph{{Maximal freedom at minimum cost: linear
  large-scale structure in general modifications of gravity}},
  \href{https://doi.org/10.1088/1475-7516/2014/07/050}{\emph{JCAP} {\bfseries
  1407} (2014) 050}, [\href{https://arxiv.org/abs/1404.3713}{{\ttfamily
  1404.3713}}].

\bibitem{Archidiacono:2014msa}
M.~Archidiacono, L.~Lopez-Honorez and O.~Mena, \emph{{Current constraints on
  early and stressed dark energy models and future 21 cm perspectives}},
  \href{https://doi.org/10.1103/PhysRevD.90.123016}{\emph{Phys. Rev.}
  {\bfseries D90} (2014) 123016},
  [\href{https://arxiv.org/abs/1409.1802}{{\ttfamily 1409.1802}}].

\bibitem{Baum:2014rka}
S.~Baum, G.~Cantatore, D.~H.~H. Hoffmann, M.~Karuza, Y.~K. Semertzidis,
  A.~Upadhye et~al., \emph{{Detecting solar chameleons through radiation
  pressure}}, \href{https://doi.org/10.1016/j.physletb.2014.10.055}{\emph{Phys.
  Lett.} {\bfseries B739} (2014) 167--173},
  [\href{https://arxiv.org/abs/1409.3852}{{\ttfamily 1409.3852}}].

\bibitem{Ade:2015rim}
{\scshape Planck} collaboration, P.~A.~R. Ade et~al., \emph{{Planck 2015
  results. XIV. Dark energy and modified gravity}},
  \href{https://doi.org/10.1051/0004-6361/201525814}{\emph{Astron. Astrophys.}
  {\bfseries 594} (2016) A14},
  [\href{https://arxiv.org/abs/1502.01590}{{\ttfamily 1502.01590}}].

\bibitem{Bellini:2015oua}
E.~Bellini and M.~Zumalacárregui, \emph{{Nonlinear evolution of the baryon
  acoustic oscillation scale in alternative theories of gravity}},
  \href{https://doi.org/10.1103/PhysRevD.92.063522}{\emph{Phys. Rev.}
  {\bfseries D92} (2015) 063522},
  [\href{https://arxiv.org/abs/1505.03839}{{\ttfamily 1505.03839}}].

\bibitem{Frusciante:2015maa}
N.~Frusciante, M.~Raveri, D.~Vernieri, B.~Hu and A.~Silvestri, \emph{{Hořava
  Gravity in the Effective Field Theory formalism: From cosmology to
  observational constraints}},
  \href{https://doi.org/10.1016/j.dark.2016.03.002}{\emph{Phys. Dark Univ.}
  {\bfseries 13} (2016) 7--24},
  [\href{https://arxiv.org/abs/1508.01787}{{\ttfamily 1508.01787}}].

\bibitem{Alam:2015rsa}
S.~Alam, S.~Ho and A.~Silvestri, \emph{{Testing deviations from $\Lambda$CDM
  with growth rate measurements from six large-scale structure surveys at $z =
  $0.06–1}}, \href{https://doi.org/10.1093/mnras/stv2935}{\emph{Mon. Not.
  Roy. Astron. Soc.} {\bfseries 456} (2016) 3743--3756},
  [\href{https://arxiv.org/abs/1509.05034}{{\ttfamily 1509.05034}}].

\bibitem{Bellini:2015xja}
E.~Bellini, A.~J. Cuesta, R.~Jiménez and L.~Verde, \emph{{Constraints on
  deviations from $\Lambda$CDM within Horndeski gravity}},
  \href{https://doi.org/10.1088/1475-7516/2016/06/E01,
  10.1088/1475-7516/2016/02/053}{\emph{JCAP} {\bfseries 1602} (2016) 053},
  [\href{https://arxiv.org/abs/1509.07816}{{\ttfamily 1509.07816}}]. [Erratum:
  JCAP1606,no.06,E01(2016)].

\bibitem{Bianchini:2015iaa}
F.~Bianchini and A.~Silvestri, \emph{{Kinetic Sunyaev-Zel’dovich effect in
  modified gravity}},
  \href{https://doi.org/10.1103/PhysRevD.93.064026}{\emph{Phys. Rev.}
  {\bfseries D93} (2016) 064026},
  [\href{https://arxiv.org/abs/1510.08844}{{\ttfamily 1510.08844}}].

\bibitem{Hu:2016zrh}
B.~Hu, M.~Raveri, M.~Rizzato and A.~Silvestri, \emph{{Testing Hu–Sawicki f(R)
  gravity with the effective field theory approach}},
  \href{https://doi.org/10.1093/mnras/stw775}{\emph{Mon. Not. Roy. Astron.
  Soc.} {\bfseries 459} (2016) 3880--3889},
  [\href{https://arxiv.org/abs/1601.07536}{{\ttfamily 1601.07536}}].

\bibitem{Nunes:2016dlj}
R.~C. Nunes, S.~Pan and E.~N. Saridakis, \emph{{New constraints on interacting
  dark energy from cosmic chronometers}},
  \href{https://doi.org/10.1103/PhysRevD.94.023508}{\emph{Phys. Rev.}
  {\bfseries D94} (2016) 023508},
  [\href{https://arxiv.org/abs/1605.01712}{{\ttfamily 1605.01712}}].

\bibitem{Pogosian:2016pwr}
L.~Pogosian and A.~Silvestri, \emph{{What can cosmology tell us about gravity?
  Constraining Horndeski gravity with $\Sigma$ and $\mu$}},
  \href{https://doi.org/10.1103/PhysRevD.94.104014}{\emph{Phys. Rev.}
  {\bfseries D94} (2016) 104014},
  [\href{https://arxiv.org/abs/1606.05339}{{\ttfamily 1606.05339}}].

\bibitem{Pan:2016ngu}
S.~Pan and G.~S. Sharov, \emph{{A model with interaction of dark components and
  recent observational data}},
  \href{https://doi.org/10.1093/mnras/stx2278}{\emph{Mon. Not. Roy. Astron.
  Soc.} {\bfseries 472} (2017) 4736--4749},
  [\href{https://arxiv.org/abs/1609.02287}{{\ttfamily 1609.02287}}].

\bibitem{Raveri:2017qvt}
M.~Raveri, P.~Bull, A.~Silvestri and L.~Pogosian, \emph{{Priors on the
  effective Dark Energy equation of state in scalar-tensor theories}},
  \href{https://doi.org/10.1103/PhysRevD.96.083509}{\emph{Phys. Rev.}
  {\bfseries D96} (2017) 083509},
  [\href{https://arxiv.org/abs/1703.05297}{{\ttfamily 1703.05297}}].

\bibitem{Dhawan:2017leu}
S.~Dhawan, A.~Goobar, E.~Mörtsell, R.~Amanullah and U.~Feindt,
  \emph{{Narrowing down the possible explanations of cosmic acceleration with
  geometric probes}},
  \href{https://doi.org/10.1088/1475-7516/2017/07/040}{\emph{JCAP} {\bfseries
  1707} (2017) 040}, [\href{https://arxiv.org/abs/1705.05768}{{\ttfamily
  1705.05768}}].

\bibitem{Yang:2017zjs}
W.~Yang, S.~Pan and J.~D. Barrow, \emph{{Large-scale Stability and Astronomical
  Constraints for Coupled Dark-Energy Models}},
  \href{https://doi.org/10.1103/PhysRevD.97.043529}{\emph{Phys. Rev.}
  {\bfseries D97} (2018) 043529},
  [\href{https://arxiv.org/abs/1706.04953}{{\ttfamily 1706.04953}}].

\bibitem{Yang:2017alx}
W.~Yang, S.~Pan and A.~Paliathanasis, \emph{{Latest astronomical constraints on
  some non-linear parametric dark energy models}},
  \href{https://doi.org/10.1093/mnras/sty019}{\emph{Mon. Not. Roy. Astron.
  Soc.} {\bfseries 475} (2018) 2605--2613},
  [\href{https://arxiv.org/abs/1708.01717}{{\ttfamily 1708.01717}}].

\bibitem{Dhawan:2017kft}
S.~Dhawan, A.~Goobar and E.~Mörtsell, \emph{{The effect of inhomogeneities on
  dark energy constraints}},
  \href{https://doi.org/10.1088/1475-7516/2018/07/024}{\emph{JCAP} {\bfseries
  1807} (2018) 024}, [\href{https://arxiv.org/abs/1710.02374}{{\ttfamily
  1710.02374}}].

\bibitem{Guo:2017deu}
J.-J. Guo, J.-F. Zhang, Y.-H. Li, D.-Z. He and X.~Zhang, \emph{{Probing the
  sign-changeable interaction between dark energy and dark matter with current
  observations}}, \href{https://doi.org/10.1007/s11433-017-9131-9}{\emph{Sci.
  China Phys. Mech. Astron.} {\bfseries 61} (2018) 030011},
  [\href{https://arxiv.org/abs/1710.03068}{{\ttfamily 1710.03068}}].

\bibitem{Peirone:2017vcq}
S.~Peirone, N.~Frusciante, B.~Hu, M.~Raveri and A.~Silvestri, \emph{{Do current
  cosmological observations rule out all Covariant Galileons?}},
  \href{https://doi.org/10.1103/PhysRevD.97.063518}{\emph{Phys. Rev.}
  {\bfseries D97} (2018) 063518},
  [\href{https://arxiv.org/abs/1711.04760}{{\ttfamily 1711.04760}}].

\bibitem{Peirone:2017ywi}
S.~Peirone, K.~Koyama, L.~Pogosian, M.~Raveri and A.~Silvestri,
  \emph{{Large-scale structure phenomenology of viable Horndeski theories}},
  \href{https://doi.org/10.1103/PhysRevD.97.043519}{\emph{Phys. Rev.}
  {\bfseries D97} (2018) 043519},
  [\href{https://arxiv.org/abs/1712.00444}{{\ttfamily 1712.00444}}].

\bibitem{Pan:2017zoh}
S.~Pan, E.~N. Saridakis and W.~Yang, \emph{{Observational Constraints on
  Oscillating Dark-Energy Parametrizations}},
  \href{https://doi.org/10.1103/PhysRevD.98.063510}{\emph{Phys. Rev.}
  {\bfseries D98} (2018) 063510},
  [\href{https://arxiv.org/abs/1712.05746}{{\ttfamily 1712.05746}}].

\bibitem{Garcia-Garcia:2018hlc}
C.~García-García, E.~V. Linder, P.~Ruíz-Lapuente and M.~Zumalacárregui,
  \emph{{Dark energy from $\alpha$-attractors: phenomenology and observational
  constraints}},
  \href{https://doi.org/10.1088/1475-7516/2018/08/022}{\emph{JCAP} {\bfseries
  1808} (2018) 022}, [\href{https://arxiv.org/abs/1803.00661}{{\ttfamily
  1803.00661}}].

\bibitem{Poulin:2018dzj}
V.~Poulin, T.~L. Smith, D.~Grin, T.~Karwal and M.~Kamionkowski,
  \emph{{Cosmological implications of ultralight axionlike fields}},
  \href{https://doi.org/10.1103/PhysRevD.98.083525}{\emph{Phys. Rev.}
  {\bfseries D98} (2018) 083525},
  [\href{https://arxiv.org/abs/1806.10608}{{\ttfamily 1806.10608}}].

\bibitem{Espejo:2018hxa}
J.~Espejo, S.~Peirone, M.~Raveri, K.~Koyama, L.~Pogosian and A.~Silvestri,
  \emph{{Phenomenology of Large Scale Structure in scalar-tensor theories:
  joint prior covariance of $w_{\textrm{DE}}$, $\Sigma$ and $\mu$ in
  Horndeski}}, \href{https://doi.org/10.1103/PhysRevD.99.023512}{\emph{Phys.
  Rev.} {\bfseries D99} (2019) 023512},
  [\href{https://arxiv.org/abs/1809.01121}{{\ttfamily 1809.01121}}].

\bibitem{Visinelli:2018utg}
L.~Visinelli and S.~Vagnozzi, \emph{{Cosmological window onto the string
  axiverse and the supersymmetry breaking scale}},
  \href{https://doi.org/10.1103/PhysRevD.99.063517}{\emph{Phys. Rev.}
  {\bfseries D99} (2019) 063517},
  [\href{https://arxiv.org/abs/1809.06382}{{\ttfamily 1809.06382}}].

\bibitem{Abbott:2018xao}
{\scshape DES} collaboration, T.~M.~C. Abbott et~al., \emph{{Dark Energy Survey
  Year 1 Results: Constraints on Extended Cosmological Models from Galaxy
  Clustering and Weak Lensing}},
  \href{https://arxiv.org/abs/1810.02499}{{\ttfamily 1810.02499}}.

\bibitem{Brush:2018dhg}
M.~Brush, E.~V. Linder and M.~Zumalacárregui, \emph{{No Slip CMB}},
  \href{https://doi.org/10.1088/1475-7516/2019/01/029}{\emph{JCAP} {\bfseries
  1901} (2019) 029}, [\href{https://arxiv.org/abs/1810.12337}{{\ttfamily
  1810.12337}}].

\bibitem{Contigiani:2018hbn}
O.~Contigiani, V.~Vardanyan and A.~Silvestri, \emph{{Splashback radius in
  symmetron gravity}},
  \href{https://doi.org/10.1103/PhysRevD.99.064030}{\emph{Phys. Rev.}
  {\bfseries D99} (2019) 064030},
  [\href{https://arxiv.org/abs/1812.05568}{{\ttfamily 1812.05568}}].

\bibitem{Yang:2018qec}
W.~Yang, N.~Banerjee, A.~Paliathanasis and S.~Pan, \emph{{Reconstructing the
  dark matter and dark energy interaction scenarios from observations}},
  \href{https://arxiv.org/abs/1812.06854}{{\ttfamily 1812.06854}}.

\bibitem{Zucca:2019xhg}
A.~Zucca, L.~Pogosian, A.~Silvestri and G.-B. Zhao, \emph{{MGCAMB with massive
  neutrinos and dynamical dark energy}},
  \href{https://arxiv.org/abs/1901.05956}{{\ttfamily 1901.05956}}.

\bibitem{Bambi:2019tjh}
C.~Bambi, K.~Freese, S.~Vagnozzi and L.~Visinelli, \emph{{Testing the
  rotational nature of the supermassive object M87* from the circularity and
  size of its first image}},
  \href{https://arxiv.org/abs/1904.12983}{{\ttfamily 1904.12983}}.

\bibitem{Efstathiou:2013via}
G.~Efstathiou, \emph{{$H_0$ Revisited}},
  \href{https://doi.org/10.1093/mnras/stu278}{\emph{Mon. Not. Roy. Astron.
  Soc.} {\bfseries 440} (2014) 1138--1152},
  [\href{https://arxiv.org/abs/1311.3461}{{\ttfamily 1311.3461}}].

\bibitem{Raveri:2015maa}
M.~Raveri, \emph{{Are cosmological data sets consistent with each other within
  the $\Lambda$ cold dark matter model?}},
  \href{https://doi.org/10.1103/PhysRevD.93.043522}{\emph{Phys. Rev.}
  {\bfseries D93} (2016) 043522},
  [\href{https://arxiv.org/abs/1510.00688}{{\ttfamily 1510.00688}}].

\bibitem{Joudaki:2016mvz}
S.~Joudaki et~al., \emph{{CFHTLenS revisited: assessing concordance with Planck
  including astrophysical systematics}},
  \href{https://doi.org/10.1093/mnras/stw2665}{\emph{Mon. Not. Roy. Astron.
  Soc.} {\bfseries 465} (2017) 2033--2052},
  [\href{https://arxiv.org/abs/1601.05786}{{\ttfamily 1601.05786}}].

\bibitem{Kitching:2016hvn}
T.~D. Kitching, L.~Verde, A.~F. Heavens and R.~Jiménez, \emph{{Discrepancies
  between CFHTLenS cosmic shear and Planck: new physics or systematic
  effects?}}, \href{https://doi.org/10.1093/mnras/stw707}{\emph{Mon. Not. Roy.
  Astron. Soc.} {\bfseries 459} (2016) 971--981},
  [\href{https://arxiv.org/abs/1602.02960}{{\ttfamily 1602.02960}}].

\bibitem{Riess:2016jrr}
A.~G. Riess et~al., \emph{{A 2.4\% Determination of the Local Value of the
  Hubble Constant}},
  \href{https://doi.org/10.3847/0004-637X/826/1/56}{\emph{Astrophys. J.}
  {\bfseries 826} (2016) 56},
  [\href{https://arxiv.org/abs/1604.01424}{{\ttfamily 1604.01424}}].

\bibitem{Grandis:2016fwl}
S.~Grandis, D.~Rapetti, A.~Saro, J.~J. Mohr and J.~P. Dietrich,
  \emph{{Quantifying tensions between CMB and distance data sets in models with
  free curvature or lensing amplitude}},
  \href{https://doi.org/10.1093/mnras/stw2028}{\emph{Mon. Not. Roy. Astron.
  Soc.} {\bfseries 463} (2016) 1416--1430},
  [\href{https://arxiv.org/abs/1604.06463}{{\ttfamily 1604.06463}}].

\bibitem{DiValentino:2016hlg}
E.~Di~Valentino, A.~Melchiorri and J.~Silk, \emph{{Reconciling Planck with the
  local value of $H_0$ in extended parameter space}},
  \href{https://doi.org/10.1016/j.physletb.2016.08.043}{\emph{Phys. Lett.}
  {\bfseries B761} (2016) 242--246},
  [\href{https://arxiv.org/abs/1606.00634}{{\ttfamily 1606.00634}}].

\bibitem{Poulin:2016nat}
V.~Poulin, P.~D. Serpico and J.~Lesgourgues, \emph{{A fresh look at linear
  cosmological constraints on a decaying dark matter component}},
  \href{https://doi.org/10.1088/1475-7516/2016/08/036}{\emph{JCAP} {\bfseries
  1608} (2016) 036}, [\href{https://arxiv.org/abs/1606.02073}{{\ttfamily
  1606.02073}}].

\bibitem{Qing-Guo:2016ykt}
Q.-G. Huang and K.~Wang, \emph{{How the dark energy can reconcile Planck with
  local determination of the Hubble constant}},
  \href{https://doi.org/10.1140/epjc/s10052-016-4352-x}{\emph{Eur. Phys. J.}
  {\bfseries C76} (2016) 506},
  [\href{https://arxiv.org/abs/1606.05965}{{\ttfamily 1606.05965}}].

\bibitem{Bernal:2016gxb}
J.~L. Bernal, L.~Verde and A.~G. Riess, \emph{{The trouble with $H_0$}},
  \href{https://doi.org/10.1088/1475-7516/2016/10/019}{\emph{JCAP} {\bfseries
  1610} (2016) 019}, [\href{https://arxiv.org/abs/1607.05617}{{\ttfamily
  1607.05617}}].

\bibitem{Ko:2016uft}
P.~Ko and Y.~Tang, \emph{{Light dark photon and fermionic dark radiation for
  the Hubble constant and the structure formation}},
  \href{https://doi.org/10.1016/j.physletb.2016.10.001}{\emph{Phys. Lett.}
  {\bfseries B762} (2016) 462--466},
  [\href{https://arxiv.org/abs/1608.01083}{{\ttfamily 1608.01083}}].

\bibitem{Joudaki:2016kym}
S.~Joudaki et~al., \emph{{KiDS-450: Testing extensions to the standard
  cosmological model}}, \href{https://doi.org/10.1093/mnras/stx998}{\emph{Mon.
  Not. Roy. Astron. Soc.} {\bfseries 471} (2017) 1259--1279},
  [\href{https://arxiv.org/abs/1610.04606}{{\ttfamily 1610.04606}}].

\bibitem{Zhao:2017cud}
G.-B. Zhao et~al., \emph{{Dynamical dark energy in light of the latest
  observations}}, \href{https://doi.org/10.1038/s41550-017-0216-z}{\emph{Nat.
  Astron.} {\bfseries 1} (2017) 627--632},
  [\href{https://arxiv.org/abs/1701.08165}{{\ttfamily 1701.08165}}].

\bibitem{Kumar:2017dnp}
S.~Kumar and R.~C. Nunes, \emph{{Echo of interactions in the dark sector}},
  \href{https://doi.org/10.1103/PhysRevD.96.103511}{\emph{Phys. Rev.}
  {\bfseries D96} (2017) 103511},
  [\href{https://arxiv.org/abs/1702.02143}{{\ttfamily 1702.02143}}].

\bibitem{Zhao:2017urm}
M.-M. Zhao, D.-Z. He, J.-F. Zhang and X.~Zhang, \emph{{Search for sterile
  neutrinos in holographic dark energy cosmology: Reconciling Planck
  observation with the local measurement of the Hubble constant}},
  \href{https://doi.org/10.1103/PhysRevD.96.043520}{\emph{Phys. Rev.}
  {\bfseries D96} (2017) 043520},
  [\href{https://arxiv.org/abs/1703.08456}{{\ttfamily 1703.08456}}].

\bibitem{Camera:2019vbp}
S.~Camera, M.~Martinelli and D.~Bertacca, \emph{{Does quartessence ease cosmic
  tensions?}}, \href{https://doi.org/10.1016/j.dark.2018.11.008}{\emph{Phys.
  Dark Univ.} {\bfseries 23} (2019) 100247},
  [\href{https://arxiv.org/abs/1704.06277}{{\ttfamily 1704.06277}}].

\bibitem{DiValentino:2017iww}
E.~Di~Valentino, A.~Melchiorri and O.~Mena, \emph{{Can interacting dark energy
  solve the $H_0$ tension?}},
  \href{https://doi.org/10.1103/PhysRevD.96.043503}{\emph{Phys. Rev.}
  {\bfseries D96} (2017) 043503},
  [\href{https://arxiv.org/abs/1704.08342}{{\ttfamily 1704.08342}}].

\bibitem{Sola:2017znb}
J.~Solà, A.~Gómez-Valent and J.~de~Cruz~Pérez, \emph{{The $H_0$ tension in
  light of vacuum dynamics in the Universe}},
  \href{https://doi.org/10.1016/j.physletb.2017.09.073}{\emph{Phys. Lett.}
  {\bfseries B774} (2017) 317--324},
  [\href{https://arxiv.org/abs/1705.06723}{{\ttfamily 1705.06723}}].

\bibitem{Feeney:2017sgx}
S.~M. Feeney, D.~J. Mortlock and N.~Dalmasso, \emph{{Clarifying the Hubble
  constant tension with a Bayesian hierarchical model of the local distance
  ladder}}, \href{https://doi.org/10.1093/mnras/sty418}{\emph{Mon. Not. Roy.
  Astron. Soc.} {\bfseries 476} (2018) 3861--3882},
  [\href{https://arxiv.org/abs/1707.00007}{{\ttfamily 1707.00007}}].

\bibitem{Efstathiou:2017rgv}
G.~Efstathiou and P.~Lemos, \emph{{Statistical inconsistencies in the KiDS-450
  data set}}, \href{https://doi.org/10.1093/mnras/sty099}{\emph{Mon. Not. Roy.
  Astron. Soc.} {\bfseries 476} (2018) 151--157},
  [\href{https://arxiv.org/abs/1707.00483}{{\ttfamily 1707.00483}}].

\bibitem{Yang:2017ccc}
W.~Yang, S.~Pan and D.~F. Mota, \emph{{Novel approach toward the large-scale
  stable interacting dark-energy models and their astronomical bounds}},
  \href{https://doi.org/10.1103/PhysRevD.96.123508}{\emph{Phys. Rev.}
  {\bfseries D96} (2017) 123508},
  [\href{https://arxiv.org/abs/1709.00006}{{\ttfamily 1709.00006}}].

\bibitem{DiValentino:2017rcr}
E.~Di~Valentino, E.~V. Linder and A.~Melchiorri, \emph{{Vacuum phase transition
  solves the $H_0$ tension}},
  \href{https://doi.org/10.1103/PhysRevD.97.043528}{\emph{Phys. Rev.}
  {\bfseries D97} (2018) 043528},
  [\href{https://arxiv.org/abs/1710.02153}{{\ttfamily 1710.02153}}].

\bibitem{DiValentino:2017oaw}
E.~Di~Valentino, C.~Bœhm, E.~Hivon and F.~R. Bouchet, \emph{{Reducing the
  $H_0$ and $\sigma_8$ tensions with Dark Matter-neutrino interactions}},
  \href{https://doi.org/10.1103/PhysRevD.97.043513}{\emph{Phys. Rev.}
  {\bfseries D97} (2018) 043513},
  [\href{https://arxiv.org/abs/1710.02559}{{\ttfamily 1710.02559}}].

\bibitem{Pan:2017ent}
S.~Pan, A.~Mukherjee and N.~Banerjee, \emph{{Astronomical bounds on a
  cosmological model allowing a general interaction in the dark sector}},
  \href{https://doi.org/10.1093/mnras/sty755}{\emph{Mon. Not. Roy. Astron.
  Soc.} {\bfseries 477} (2018) 1189--1205},
  [\href{https://arxiv.org/abs/1710.03725}{{\ttfamily 1710.03725}}].

\bibitem{Abbott:2017smn}
{\scshape DES} collaboration, T.~M.~C. Abbott et~al., \emph{{Dark Energy Survey
  Year 1 Results: A Precise $H_0$ Measurement from DES Y1, BAO, and D/H Data}},
  \href{https://doi.org/10.1093/mnras/sty1939}{\emph{Mon. Not. Roy. Astron.
  Soc.} {\bfseries 480} (2018) 3879},
  [\href{https://arxiv.org/abs/1711.00403}{{\ttfamily 1711.00403}}].

\bibitem{An:2017crg}
R.~An, C.~Feng and B.~Wang, \emph{{Relieving the Tension between Weak Lensing
  and Cosmic Microwave Background with Interacting Dark Matter and Dark Energy
  Models}}, \href{https://doi.org/10.1088/1475-7516/2018/02/038}{\emph{JCAP}
  {\bfseries 1802} (2018) 038},
  [\href{https://arxiv.org/abs/1711.06799}{{\ttfamily 1711.06799}}].

\bibitem{Benetti:2017juy}
M.~Benetti, L.~L. Graef and J.~S. Alcaniz, \emph{{The $H_0$ and $\sigma_8$
  tensions and the scale invariant spectrum}},
  \href{https://doi.org/10.1088/1475-7516/2018/07/066}{\emph{JCAP} {\bfseries
  1807} (2018) 066}, [\href{https://arxiv.org/abs/1712.00677}{{\ttfamily
  1712.00677}}].

\bibitem{Feng:2017usu}
L.~Feng, J.-F. Zhang and X.~Zhang, \emph{{Search for sterile neutrinos in a
  universe of vacuum energy interacting with cold dark matter}},
  \href{https://doi.org/10.1016/j.dark.2018.100261}{\emph{Phys. Dark Univ.}
  (2017) 100261}, [\href{https://arxiv.org/abs/1712.03148}{{\ttfamily
  1712.03148}}].

\bibitem{Renzi:2017cbg}
F.~Renzi, E.~Di~Valentino and A.~Melchiorri, \emph{{Cornering the Planck
  $A_{lens}$ anomaly with future CMB data}},
  \href{https://doi.org/10.1103/PhysRevD.97.123534}{\emph{Phys. Rev.}
  {\bfseries D97} (2018) 123534},
  [\href{https://arxiv.org/abs/1712.08758}{{\ttfamily 1712.08758}}].

\bibitem{Riess:2018uxu}
A.~G. Riess et~al., \emph{{New Parallaxes of Galactic Cepheids from Spatially
  Scanning the Hubble Space Telescope: Implications for the Hubble Constant}},
  \href{https://doi.org/10.3847/1538-4357/aaadb7}{\emph{Astrophys. J.}
  {\bfseries 855} (2018) 136},
  [\href{https://arxiv.org/abs/1801.01120}{{\ttfamily 1801.01120}}].

\bibitem{Mortsell:2018mfj}
E.~Mörtsell and S.~Dhawan, \emph{{Does the Hubble constant tension call for
  new physics?}},
  \href{https://doi.org/10.1088/1475-7516/2018/09/025}{\emph{JCAP} {\bfseries
  1809} (2018) 025}, [\href{https://arxiv.org/abs/1801.07260}{{\ttfamily
  1801.07260}}].

\bibitem{Nunes:2018xbm}
R.~C. Nunes, \emph{{Structure formation in $f(T)$ gravity and a solution for
  $H_0$ tension}},
  \href{https://doi.org/10.1088/1475-7516/2018/05/052}{\emph{JCAP} {\bfseries
  1805} (2018) 052}, [\href{https://arxiv.org/abs/1802.02281}{{\ttfamily
  1802.02281}}].

\bibitem{Poulin:2018zxs}
V.~Poulin, K.~K. Boddy, S.~Bird and M.~Kamionkowski, \emph{{Implications of an
  extended dark energy cosmology with massive neutrinos for cosmological
  tensions}}, \href{https://doi.org/10.1103/PhysRevD.97.123504}{\emph{Phys.
  Rev.} {\bfseries D97} (2018) 123504},
  [\href{https://arxiv.org/abs/1803.02474}{{\ttfamily 1803.02474}}].

\bibitem{Yang:2018ubt}
W.~Yang, S.~Pan, L.~Xu and D.~F. Mota, \emph{{Effects of anisotropic stress in
  interacting dark matter – dark energy scenarios}},
  \href{https://doi.org/10.1093/mnras/sty2789}{\emph{Mon. Not. Roy. Astron.
  Soc.} {\bfseries 482} (2019) 1858--1871},
  [\href{https://arxiv.org/abs/1804.08455}{{\ttfamily 1804.08455}}].

\bibitem{Yang:2018euj}
W.~Yang, S.~Pan, E.~Di~Valentino, R.~C. Nunes, S.~Vagnozzi and D.~F. Mota,
  \emph{{Tale of stable interacting dark energy, observational signatures, and
  the $H_0$ tension}},
  \href{https://doi.org/10.1088/1475-7516/2018/09/019}{\emph{JCAP} {\bfseries
  1809} (2018) 019}, [\href{https://arxiv.org/abs/1805.08252}{{\ttfamily
  1805.08252}}].

\bibitem{Adhikari:2018wnk}
S.~Adhikari and D.~Huterer, \emph{{A new measure of tension between
  experiments}},
  \href{https://doi.org/10.1088/1475-7516/2019/01/036}{\emph{JCAP} {\bfseries
  1901} (2019) 036}, [\href{https://arxiv.org/abs/1806.04292}{{\ttfamily
  1806.04292}}].

\bibitem{Raveri:2018wln}
M.~Raveri and W.~Hu, \emph{{Concordance and Discordance in Cosmology}},
  \href{https://doi.org/10.1103/PhysRevD.99.043506}{\emph{Phys. Rev.}
  {\bfseries D99} (2019) 043506},
  [\href{https://arxiv.org/abs/1806.04649}{{\ttfamily 1806.04649}}].

\bibitem{DiValentino:2018wum}
E.~Di~Valentino and L.~Mersini-Houghton, \emph{{Testing Predictions of the
  Quantum Landscape Multiverse 3: The Hilltop Inflationary Potential}},
  \href{https://doi.org/10.3390/sym11040520}{\emph{Symmetry} {\bfseries 11}
  (2019) 520}, [\href{https://arxiv.org/abs/1807.10833}{{\ttfamily
  1807.10833}}].

\bibitem{Yang:2018xlt}
W.~Yang, S.~Pan, R.~Herrera and S.~Chakraborty, \emph{{Large-scale (in)
  stability analysis of an exactly solved coupled dark-energy model}},
  \href{https://doi.org/10.1103/PhysRevD.98.043517}{\emph{Phys. Rev.}
  {\bfseries D98} (2018) 043517},
  [\href{https://arxiv.org/abs/1808.01669}{{\ttfamily 1808.01669}}].

\bibitem{DEramo:2018vss}
F.~D'Eramo, R.~Z. Ferreira, A.~Notari and J.~L. Bernal, \emph{{Hot Axions and
  the $H_0$ tension}},
  \href{https://doi.org/10.1088/1475-7516/2018/11/014}{\emph{JCAP} {\bfseries
  1811} (2018) 014}, [\href{https://arxiv.org/abs/1808.07430}{{\ttfamily
  1808.07430}}].

\bibitem{Guo:2018ans}
R.-Y. Guo, J.-F. Zhang and X.~Zhang, \emph{{Can the $H_0$ tension be resolved
  in extensions to $\Lambda$CDM cosmology?}},
  \href{https://doi.org/10.1088/1475-7516/2019/02/054}{\emph{JCAP} {\bfseries
  1902} (2019) 054}, [\href{https://arxiv.org/abs/1809.02340}{{\ttfamily
  1809.02340}}].

\bibitem{Yang:2018uae}
W.~Yang, A.~Mukherjee, E.~Di~Valentino and S.~Pan, \emph{{Interacting dark
  energy with time varying equation of state and the $H_0$ tension}},
  \href{https://doi.org/10.1103/PhysRevD.98.123527}{\emph{Phys. Rev.}
  {\bfseries D98} (2018) 123527},
  [\href{https://arxiv.org/abs/1809.06883}{{\ttfamily 1809.06883}}].

\bibitem{Yang:2018qmz}
W.~Yang, S.~Pan, E.~Di~Valentino, E.~N. Saridakis and S.~Chakraborty,
  \emph{{Observational constraints on one-parameter dynamical dark-energy
  parametrizations and the $H_0$ tension}},
  \href{https://doi.org/10.1103/PhysRevD.99.043543}{\emph{Phys. Rev.}
  {\bfseries D99} (2019) 043543},
  [\href{https://arxiv.org/abs/1810.05141}{{\ttfamily 1810.05141}}].

\bibitem{DiValentino:2018gcu}
E.~Di~Valentino and S.~Bridle, \emph{{Exploring the Tension between Current
  Cosmic Microwave Background and Cosmic Shear Data}},
  \href{https://doi.org/10.3390/sym10110585}{\emph{Symmetry} {\bfseries 10}
  (2018) 585}.

\bibitem{Poulin:2018cxd}
V.~Poulin, T.~L. Smith, T.~Karwal and M.~Kamionkowski, \emph{{Early Dark Energy
  Can Resolve The Hubble Tension}},
  \href{https://arxiv.org/abs/1811.04083}{{\ttfamily 1811.04083}}.

\bibitem{Kumar:2019wfs}
S.~Kumar, R.~C. Nunes and S.~K. Yadav, \emph{{Dark sector interaction: a remedy
  of the tensions between CMB and LSS data}},
  \href{https://arxiv.org/abs/1903.04865}{{\ttfamily 1903.04865}}.

\bibitem{Vattis:2019efj}
K.~Vattis, S.~M. Koushiappas and A.~Loeb, \emph{{Late universe decaying dark
  matter can relieve the $H_0$ tension}},
  \href{https://arxiv.org/abs/1903.06220}{{\ttfamily 1903.06220}}.

\bibitem{Pan:2019jqh}
S.~Pan, W.~Yang, C.~Singha and E.~N. Saridakis, \emph{{Observational
  constraints on sign-changeable interaction models and alleviation of the
  $H_0$ tension}},  \href{https://arxiv.org/abs/1903.10969}{{\ttfamily
  1903.10969}}.

\bibitem{Agrawal:2019lmo}
P.~Agrawal, F.-Y. Cyr-Racine, D.~Pinner and L.~Randall, \emph{{Rock 'n' Roll
  Solutions to the Hubble Tension}},
  \href{https://arxiv.org/abs/1904.01016}{{\ttfamily 1904.01016}}.

\bibitem{Yang:2019jwn}
W.~Yang, S.~Pan, A.~Paliathanasis, S.~Ghosh and Y.~Wu, \emph{{Observational
  constraints of a new unified dark fluid and the $H_0$ tension}},
  \href{https://arxiv.org/abs/1904.10436}{{\ttfamily 1904.10436}}.

\bibitem{Baumann:2013ghw}
{Daniel Baumann}, \emph{Cosmology},  2013.

\bibitem{Lewis:1999bs}
A.~Lewis, A.~Challinor and A.~Lasenby, \emph{{Efficient computation of CMB
  anisotropies in closed FRW models}},
  \href{https://doi.org/10.1086/309179}{\emph{Astrophys. J.} {\bfseries 538}
  (2000) 473--476}, [\href{https://arxiv.org/abs/astro-ph/9911177}{{\ttfamily
  astro-ph/9911177}}].

\bibitem{Blas:2011rf}
D.~Blas, J.~Lesgourgues and T.~Tram, \emph{{The Cosmic Linear Anisotropy
  Solving System (CLASS) II: Approximation schemes}},
  \href{https://doi.org/10.1088/1475-7516/2011/07/034}{\emph{JCAP} {\bfseries
  1107} (2011) 034}, [\href{https://arxiv.org/abs/1104.2933}{{\ttfamily
  1104.2933}}].

\bibitem{Seljak:1996is}
U.~Seljak and M.~Zaldarriaga, \emph{{A Line of sight integration approach to
  cosmic microwave background anisotropies}},
  \href{https://doi.org/10.1086/177793}{\emph{Astrophys. J.} {\bfseries 469}
  (1996) 437--444}, [\href{https://arxiv.org/abs/astro-ph/9603033}{{\ttfamily
  astro-ph/9603033}}].

\bibitem{Kaplinghat:2002mh}
M.~Kaplinghat, L.~Knox and C.~Skordis, \emph{{Rapid calculation of theoretical
  cmb angular power spectra}},
  \href{https://doi.org/10.1086/342656}{\emph{Astrophys. J.} {\bfseries 578}
  (2002) 665}, [\href{https://arxiv.org/abs/astro-ph/0203413}{{\ttfamily
  astro-ph/0203413}}].

\bibitem{Doran:2003sy}
M.~Doran, \emph{{CMBEASY: an object oriented code for the cosmic microwave
  background}},
  \href{https://doi.org/10.1088/1475-7516/2005/10/011}{\emph{JCAP} {\bfseries
  0510} (2005) 011}, [\href{https://arxiv.org/abs/astro-ph/0302138}{{\ttfamily
  astro-ph/0302138}}].

\bibitem{Dossett:2011tn}
J.~N. Dossett, M.~Ishak and J.~Moldenhauer, \emph{{Testing General Relativity
  at Cosmological Scales: Implementation and Parameter Correlations}},
  \href{https://doi.org/10.1103/PhysRevD.84.123001}{\emph{Phys. Rev.}
  {\bfseries D84} (2011) 123001},
  [\href{https://arxiv.org/abs/1109.4583}{{\ttfamily 1109.4583}}].

\bibitem{Hu:2013twa}
B.~Hu, M.~Raveri, N.~Frusciante and A.~Silvestri, \emph{{Effective Field Theory
  of Cosmic Acceleration: an implementation in CAMB}},
  \href{https://doi.org/10.1103/PhysRevD.89.103530}{\emph{Phys. Rev.}
  {\bfseries D89} (2014) 103530},
  [\href{https://arxiv.org/abs/1312.5742}{{\ttfamily 1312.5742}}].

\bibitem{Raveri:2014cka}
M.~Raveri, B.~Hu, N.~Frusciante and A.~Silvestri, \emph{{Effective Field Theory
  of Cosmic Acceleration: constraining dark energy with CMB data}},
  \href{https://doi.org/10.1103/PhysRevD.90.043513}{\emph{Phys. Rev.}
  {\bfseries D90} (2014) 043513},
  [\href{https://arxiv.org/abs/1405.1022}{{\ttfamily 1405.1022}}].

\bibitem{Hu:2014oga}
B.~Hu, M.~Raveri, N.~Frusciante and A.~Silvestri, \emph{{EFTCAMB/EFTCosmoMC:
  Numerical Notes v3.0}},  \href{https://arxiv.org/abs/1405.3590}{{\ttfamily
  1405.3590}}.

\bibitem{Hu:2014sea}
B.~Hu, M.~Raveri, A.~Silvestri and N.~Frusciante, \emph{{Exploring massive
  neutrinos in dark cosmologies with $\scriptsize{EFTCAMB}$/ EFTCosmoMC}},
  \href{https://doi.org/10.1103/PhysRevD.91.063524}{\emph{Phys. Rev.}
  {\bfseries D91} (2015) 063524},
  [\href{https://arxiv.org/abs/1410.5807}{{\ttfamily 1410.5807}}].

\bibitem{Frusciante:2016xoj}
N.~Frusciante, G.~Papadomanolakis and A.~Silvestri, \emph{{An Extended action
  for the effective field theory of dark energy: a stability analysis and a
  complete guide to the mapping at the basis of EFTCAMB}},
  \href{https://doi.org/10.1088/1475-7516/2016/07/018}{\emph{JCAP} {\bfseries
  1607} (2016) 018}, [\href{https://arxiv.org/abs/1601.04064}{{\ttfamily
  1601.04064}}].

\bibitem{Zumalacarregui:2016pph}
M.~Zumalacárregui, E.~Bellini, I.~Sawicki, J.~Lesgourgues and P.~G. Ferreira,
  \emph{{\texttt{hi\_class}: Horndeski in the Cosmic Linear Anisotropy Solving
  System}}, \href{https://doi.org/10.1088/1475-7516/2017/08/019}{\emph{JCAP}
  {\bfseries 1708} (2017) 019},
  [\href{https://arxiv.org/abs/1605.06102}{{\ttfamily 1605.06102}}].

\bibitem{Bean:2006up}
R.~Bean, D.~Bernat, L.~Pogosian, A.~Silvestri and M.~Trodden, \emph{{Dynamics
  of Linear Perturbations in f(R) Gravity}},
  \href{https://doi.org/10.1103/PhysRevD.75.064020}{\emph{Phys. Rev.}
  {\bfseries D75} (2007) 064020},
  [\href{https://arxiv.org/abs/astro-ph/0611321}{{\ttfamily
  astro-ph/0611321}}].

\bibitem{Zuntz:2008zz}
J.~A. Zuntz, P.~G. Ferreira and T.~G. Złośnik, \emph{{Constraining Lorentz
  violation with cosmology}},
  \href{https://doi.org/10.1103/PhysRevLett.101.261102}{\emph{Phys. Rev. Lett.}
  {\bfseries 101} (2008) 261102},
  [\href{https://arxiv.org/abs/0808.1824}{{\ttfamily 0808.1824}}].

\bibitem{Barreira:2012kk}
A.~Barreira, B.~Li, C.~M. Baugh and S.~Pascoli, \emph{{Linear perturbations in
  Galileon gravity models}},
  \href{https://doi.org/10.1103/PhysRevD.86.124016}{\emph{Phys. Rev.}
  {\bfseries D86} (2012) 124016},
  [\href{https://arxiv.org/abs/1208.0600}{{\ttfamily 1208.0600}}].

\bibitem{Avilez:2013dxa}
A.~Avilez and C.~Skordis, \emph{{Cosmological constraints on Brans-Dicke
  theory}}, \href{https://doi.org/10.1103/PhysRevLett.113.011101}{\emph{Phys.
  Rev. Lett.} {\bfseries 113} (2014) 011101},
  [\href{https://arxiv.org/abs/1303.4330}{{\ttfamily 1303.4330}}].

\bibitem{DiDio:2013bqa}
E.~Di~Dio, F.~Montanari, J.~Lesgourgues and R.~Durrer, \emph{{The CLASSgal code
  for Relativistic Cosmological Large Scale Structure}},
  \href{https://doi.org/10.1088/1475-7516/2013/11/044}{\emph{JCAP} {\bfseries
  1311} (2013) 044}, [\href{https://arxiv.org/abs/1307.1459}{{\ttfamily
  1307.1459}}].

\bibitem{Hlozek:2014lca}
R.~Hložek, D.~Grin, D.~J.~E. Marsh and P.~G. Ferreira, \emph{{A search for
  ultralight axions using precision cosmological data}},
  \href{https://doi.org/10.1103/PhysRevD.91.103512}{\emph{Phys. Rev.}
  {\bfseries D91} (2015) 103512},
  [\href{https://arxiv.org/abs/1410.2896}{{\ttfamily 1410.2896}}].

\bibitem{Renk:2016olm}
J.~Renk, M.~Zumalacárregui and F.~Montanari, \emph{{Gravity at the horizon: on
  relativistic effects, CMB-LSS correlations and ultra-large scales in
  Horndeski's theory}},
  \href{https://doi.org/10.1088/1475-7516/2016/07/040}{\emph{JCAP} {\bfseries
  1607} (2016) 040}, [\href{https://arxiv.org/abs/1604.03487}{{\ttfamily
  1604.03487}}].

\bibitem{Zucca:2016iur}
A.~Zucca, Y.~Li and L.~Pogosian, \emph{{Constraints on Primordial Magnetic
  Fields from Planck combined with the South Pole Telescope CMB B-mode
  polarization measurements}},
  \href{https://doi.org/10.1103/PhysRevD.95.063506}{\emph{Phys. Rev.}
  {\bfseries D95} (2017) 063506},
  [\href{https://arxiv.org/abs/1611.00757}{{\ttfamily 1611.00757}}].

\bibitem{Stocker:2018avm}
P.~Stöcker, M.~Krämer, J.~Lesgourgues and V.~Poulin, \emph{{Exotic energy
  injection with ExoCLASS: Application to the Higgs portal model and
  evaporating black holes}},
  \href{https://doi.org/10.1088/1475-7516/2018/03/018}{\emph{JCAP} {\bfseries
  1803} (2018) 018}, [\href{https://arxiv.org/abs/1801.01871}{{\ttfamily
  1801.01871}}].

\bibitem{Casalino:2018mna}
A.~Casalino and M.~Rinaldi, \emph{{Testing Horndeski gravity as dark matter
  with hi\_class}},
  \href{https://doi.org/10.1016/j.dark.2018.11.004}{\emph{Phys. Dark Univ.}
  {\bfseries 23} (2019) 100243},
  [\href{https://arxiv.org/abs/1807.01995}{{\ttfamily 1807.01995}}].

\bibitem{Bellini:2017avd}
E.~Bellini et~al., \emph{{Comparison of Einstein-Boltzmann solvers for testing
  general relativity}},
  \href{https://doi.org/10.1103/PhysRevD.97.023520}{\emph{Phys. Rev.}
  {\bfseries D97} (2018) 023520},
  [\href{https://arxiv.org/abs/1709.09135}{{\ttfamily 1709.09135}}].

\bibitem{Sakharov:1967dj}
A.~D. Sakharov, \emph{{Violation of CP Invariance, C asymmetry, and baryon
  asymmetry of the universe}},
  \href{https://doi.org/10.1070/PU1991v034n05ABEH002497}{\emph{Pisma Zh. Eksp.
  Teor. Fiz.} {\bfseries 5} (1967) 32--35}.

\bibitem{Trodden:1998ym}
M.~Trodden, \emph{{Electroweak baryogenesis}},
  \href{https://doi.org/10.1103/RevModPhys.71.1463}{\emph{Rev. Mod. Phys.}
  {\bfseries 71} (1999) 1463--1500},
  [\href{https://arxiv.org/abs/hep-ph/9803479}{{\ttfamily hep-ph/9803479}}].

\bibitem{Riotto:1998bt}
A.~Riotto, \emph{{Theories of baryogenesis}},  in \emph{{Proceedings, Summer
  School in High-energy physics and cosmology: Trieste, Italy, June 29-July 17,
  1998}}, pp.~326--436, 1998,
  \href{https://arxiv.org/abs/hep-ph/9807454}{{\ttfamily hep-ph/9807454}}.

\bibitem{Englert:1964et}
F.~Englert and R.~Brout, \emph{{Broken Symmetry and the Mass of Gauge Vector
  Mesons}}, \href{https://doi.org/10.1103/PhysRevLett.13.321}{\emph{Phys. Rev.
  Lett.} {\bfseries 13} (1964) 321--323}.

\bibitem{Higgs:1964ia}
P.~W. Higgs, \emph{{Broken symmetries, massless particles and gauge fields}},
  \href{https://doi.org/10.1016/0031-9163(64)91136-9}{\emph{Phys. Lett.}
  {\bfseries 12} (1964) 132--133}.

\bibitem{Higgs:1964pj}
P.~W. Higgs, \emph{{Broken Symmetries and the Masses of Gauge Bosons}},
  \href{https://doi.org/10.1103/PhysRevLett.13.508}{\emph{Phys. Rev. Lett.}
  {\bfseries 13} (1964) 508--509}.

\bibitem{Higgs:1966ev}
P.~W. Higgs, \emph{{Spontaneous Symmetry Breakdown without Massless Bosons}},
  \href{https://doi.org/10.1103/PhysRev.145.1156}{\emph{Phys. Rev.} {\bfseries
  145} (1966) 1156--1163}.

\bibitem{Guralnik:1964eu}
G.~S. Guralnik, C.~R. Hagen and T.~W.~B. Kibble, \emph{{Global Conservation
  Laws and Massless Particles}},
  \href{https://doi.org/10.1103/PhysRevLett.13.585}{\emph{Phys. Rev. Lett.}
  {\bfseries 13} (1964) 585--587}.

\bibitem{Aad:2012tfa}
{\scshape ATLAS} collaboration, G.~Aad et~al., \emph{{Observation of a new
  particle in the search for the Standard Model Higgs boson with the ATLAS
  detector at the LHC}},
  \href{https://doi.org/10.1016/j.physletb.2012.08.020}{\emph{Phys. Lett.}
  {\bfseries B716} (2012) 1--29},
  [\href{https://arxiv.org/abs/1207.7214}{{\ttfamily 1207.7214}}].

\bibitem{Chatrchyan:2012xdj}
{\scshape CMS} collaboration, S.~Chatrchyan et~al., \emph{{Observation of a new
  boson at a mass of 125 GeV with the CMS experiment at the LHC}},
  \href{https://doi.org/10.1016/j.physletb.2012.08.021}{\emph{Phys. Lett.}
  {\bfseries B716} (2012) 30--61},
  [\href{https://arxiv.org/abs/1207.7235}{{\ttfamily 1207.7235}}].

\bibitem{Gross:1973ju}
D.~J. Gross and F.~Wilczek, \emph{{Asymptotically Free Gauge Theories - I}},
  \href{https://doi.org/10.1103/PhysRevD.8.3633}{\emph{Phys. Rev.} {\bfseries
  D8} (1973) 3633--3652}.

\bibitem{Hannestad:1995rs}
S.~Hannestad and J.~Madsen, \emph{{Neutrino decoupling in the early universe}},
  \href{https://doi.org/10.1103/PhysRevD.52.1764}{\emph{Phys. Rev.} {\bfseries
  D52} (1995) 1764--1769},
  [\href{https://arxiv.org/abs/astro-ph/9506015}{{\ttfamily
  astro-ph/9506015}}].

\bibitem{Mangano:2005cc}
G.~Mangano, G.~Miele, S.~Pastor, T.~Pinto, O.~Pisanti and P.~D. Serpico,
  \emph{{Relic neutrino decoupling including flavor oscillations}},
  \href{https://doi.org/10.1016/j.nuclphysb.2005.09.041}{\emph{Nucl. Phys.}
  {\bfseries B729} (2005) 221--234},
  [\href{https://arxiv.org/abs/hep-ph/0506164}{{\ttfamily hep-ph/0506164}}].

\bibitem{deSalas:2016ztq}
P.~F. de~Salas and S.~Pastor, \emph{{Relic neutrino decoupling with flavour
  oscillations revisited}},
  \href{https://doi.org/10.1088/1475-7516/2016/07/051}{\emph{JCAP} {\bfseries
  1607} (2016) 051}, [\href{https://arxiv.org/abs/1606.06986}{{\ttfamily
  1606.06986}}].

\bibitem{Escudero:2018mvt}
M.~Escudero, \emph{{Neutrino decoupling beyond the Standard Model: CMB
  constraints on the Dark Matter mass with a fast and precise $N_{\rm eff}$
  evaluation}},
  \href{https://doi.org/10.1088/1475-7516/2019/02/007}{\emph{JCAP} {\bfseries
  1902} (2019) 007}, [\href{https://arxiv.org/abs/1812.05605}{{\ttfamily
  1812.05605}}].

\bibitem{Dicus:1982bz}
D.~A. Dicus, E.~W. Kolb, A.~M. Gleeson, E.~C.~G. Sudarshan, V.~L. Teplitz and
  M.~S. Turner, \emph{{Primordial Nucleosynthesis Including Radiative, Coulomb,
  and Finite Temperature Corrections to Weak Rates}},
  \href{https://doi.org/10.1103/PhysRevD.26.2694}{\emph{Phys. Rev.} {\bfseries
  D26} (1982) 2694}.

\bibitem{Dodelson:1992km}
S.~Dodelson and M.~S. Turner, \emph{{Nonequilibrium neutrino statistical
  mechanics in the expanding universe}},
  \href{https://doi.org/10.1103/PhysRevD.46.3372}{\emph{Phys. Rev.} {\bfseries
  D46} (1992) 3372--3387}.

\bibitem{Dolgov:1992qg}
A.~D. Dolgov and M.~Fukugita, \emph{{Nonequilibrium effect of the neutrino
  distribution on primordial helium synthesis}},
  \href{https://doi.org/10.1103/PhysRevD.46.5378}{\emph{Phys. Rev.} {\bfseries
  D46} (1992) 5378--5382}.

\bibitem{Fields:1992zb}
B.~D. Fields, S.~Dodelson and M.~S. Turner, \emph{{Effect of neutrino heating
  on primordial nucleosynthesis}},
  \href{https://doi.org/10.1103/PhysRevD.47.4309}{\emph{Phys. Rev.} {\bfseries
  D47} (1993) 4309--4314},
  [\href{https://arxiv.org/abs/astro-ph/9210007}{{\ttfamily
  astro-ph/9210007}}].

\bibitem{Fornengo:1997wa}
N.~Fornengo, C.~W. Kim and J.~Song, \emph{{Finite temperature effects on the
  neutrino decoupling in the early universe}},
  \href{https://doi.org/10.1103/PhysRevD.56.5123}{\emph{Phys. Rev.} {\bfseries
  D56} (1997) 5123--5134},
  [\href{https://arxiv.org/abs/hep-ph/9702324}{{\ttfamily hep-ph/9702324}}].

\bibitem{Dolgov:1998sf}
A.~D. Dolgov, S.~H. Hansen and D.~V. Semikoz, \emph{{Nonequilibrium corrections
  to the spectra of massless neutrinos in the early universe: Addendum}},
  \href{https://doi.org/10.1016/S0550-3213(98)00818-9}{\emph{Nucl. Phys.}
  {\bfseries B543} (1999) 269--274},
  [\href{https://arxiv.org/abs/hep-ph/9805467}{{\ttfamily hep-ph/9805467}}].

\bibitem{Esposito:2000hi}
S.~Esposito, G.~Miele, S.~Pastor, M.~Peloso and O.~Pisanti,
  \emph{{Nonequilibrium spectra of degenerate relic neutrinos}},
  \href{https://doi.org/10.1016/S0550-3213(00)00554-X}{\emph{Nucl. Phys.}
  {\bfseries B590} (2000) 539--561},
  [\href{https://arxiv.org/abs/astro-ph/0005573}{{\ttfamily
  astro-ph/0005573}}].

\bibitem{Dolgov:2002wy}
A.~D. Dolgov, \emph{{Neutrinos in cosmology}},
  \href{https://doi.org/10.1016/S0370-1573(02)00139-4}{\emph{Phys. Rept.}
  {\bfseries 370} (2002) 333--535},
  [\href{https://arxiv.org/abs/hep-ph/0202122}{{\ttfamily hep-ph/0202122}}].

\bibitem{Mangano:2001iu}
G.~Mangano, G.~Miele, S.~Pastor and M.~Peloso, \emph{{A Precision calculation
  of the effective number of cosmological neutrinos}},
  \href{https://doi.org/10.1016/S0370-2693(02)01622-2}{\emph{Phys. Lett.}
  {\bfseries B534} (2002) 8--16},
  [\href{https://arxiv.org/abs/astro-ph/0111408}{{\ttfamily
  astro-ph/0111408}}].

\bibitem{Tytler:2000qf}
D.~Tytler, J.~M. O'Meara, N.~Suzuki and D.~Lubin, \emph{{Review of Big Bang
  nucleosynthesis and primordial abundances}},
  \href{https://doi.org/10.1238/Physica.Topical.085a00012}{\emph{Phys. Scripta}
  {\bfseries T85} (2000) 12},
  [\href{https://arxiv.org/abs/astro-ph/0001318}{{\ttfamily
  astro-ph/0001318}}].

\bibitem{Fields:2006ga}
B.~Fields and S.~Sarkar, \emph{{Big-Bang nucleosynthesis (2006 Particle Data
  Group mini-review)}},
  \href{https://arxiv.org/abs/astro-ph/0601514}{{\ttfamily astro-ph/0601514}}.

\bibitem{Steigman:2012ve}
G.~Steigman, \emph{{Neutrinos And Big Bang Nucleosynthesis}},
  \href{https://doi.org/10.1155/2012/268321}{\emph{Adv. High Energy Phys.}
  {\bfseries 2012} (2012) 268321},
  [\href{https://arxiv.org/abs/1208.0032}{{\ttfamily 1208.0032}}].

\bibitem{Fields:2014uja}
B.~D. Fields, P.~Molaro and S.~Sarkar, \emph{{Big-Bang Nucleosynthesis}},
  {\emph{Chin. Phys.} {\bfseries C38} (2014) 339--344},
  [\href{https://arxiv.org/abs/1412.1408}{{\ttfamily 1412.1408}}].

\bibitem{Fields:2011zzb}
B.~D. Fields, \emph{{The primordial lithium problem}},
  \href{https://doi.org/10.1146/annurev-nucl-102010-130445}{\emph{Ann. Rev.
  Nucl. Part. Sci.} {\bfseries 61} (2011) 47--68},
  [\href{https://arxiv.org/abs/1203.3551}{{\ttfamily 1203.3551}}].

\bibitem{Poulin:2015woa}
V.~Poulin and P.~D. Serpico, \emph{{Loophole to the Universal Photon Spectrum
  in Electromagnetic Cascades and Application to the Cosmological Lithium
  Problem}}, \href{https://doi.org/10.1103/PhysRevLett.114.091101}{\emph{Phys.
  Rev. Lett.} {\bfseries 114} (2015) 091101},
  [\href{https://arxiv.org/abs/1502.01250}{{\ttfamily 1502.01250}}].

\bibitem{Salvati:2016jng}
L.~Salvati, L.~Pagano, M.~Lattanzi, M.~Gerbino and A.~Melchiorri,
  \emph{{Breaking Be: a sterile neutrino solution to the cosmological lithium
  problem}}, \href{https://doi.org/10.1088/1475-7516/2016/08/022}{\emph{JCAP}
  {\bfseries 1608} (2016) 022},
  [\href{https://arxiv.org/abs/1606.06968}{{\ttfamily 1606.06968}}].

\bibitem{MiraldaEscude:2003yt}
J.~Miralda-Escudé, \emph{{The dark age of the universe}},
  \href{https://doi.org/10.1126/science.1085325}{\emph{Science} {\bfseries 300}
  (2003) 1904--1909}, [\href{https://arxiv.org/abs/astro-ph/0307396}{{\ttfamily
  astro-ph/0307396}}].

\bibitem{Natarajan:2014rra}
A.~Natarajan and N.~Yoshida, \emph{{The Dark Ages of the Universe and Hydrogen
  Reionization}}, \href{https://doi.org/10.1093/ptep/ptu067}{\emph{PTEP}
  {\bfseries 2014} (2014) 06B112},
  [\href{https://arxiv.org/abs/1404.7146}{{\ttfamily 1404.7146}}].

\bibitem{Furlanetto:2019jso}
S.~Furlanetto et~al., \emph{{Astro 2020 Science White Paper: Fundamental
  Cosmology in the Dark Ages with 21-cm Line Fluctuations}},
  \href{https://arxiv.org/abs/1903.06212}{{\ttfamily 1903.06212}}.

\bibitem{Barkana:2000fd}
R.~Barkana and A.~Loeb, \emph{{In the beginning: The First sources of light and
  the reionization of the Universe}},
  \href{https://doi.org/10.1016/S0370-1573(01)00019-9}{\emph{Phys. Rept.}
  {\bfseries 349} (2001) 125--238},
  [\href{https://arxiv.org/abs/astro-ph/0010468}{{\ttfamily
  astro-ph/0010468}}].

\bibitem{Zaroubi:2012in}
S.~{Zaroubi}, \emph{{The Epoch of Reionization}},  in \emph{The First Galaxies}
  (T.~{Wiklind}, B.~{Mobasher} and V.~{Bromm}, eds.), vol.~396 of
  \emph{Astrophysics and Space Science Library}, p.~45, Jan, 2013,
  \href{https://arxiv.org/abs/1206.0267}{{\ttfamily 1206.0267}}.

\bibitem{Lyth:1998xn}
D.~H. Lyth and A.~Riotto, \emph{{Particle physics models of inflation and the
  cosmological density perturbation}},
  \href{https://doi.org/10.1016/S0370-1573(98)00128-8}{\emph{Phys. Rept.}
  {\bfseries 314} (1999) 1--146},
  [\href{https://arxiv.org/abs/hep-ph/9807278}{{\ttfamily hep-ph/9807278}}].

\bibitem{Liddle:1999mq}
A.~R. Liddle, \emph{{An Introduction to cosmological inflation}},  in
  \emph{{Proceedings, Summer School in High-energy physics and cosmology:
  Trieste, Italy, June 29-July 17, 1998}}, pp.~260--295, 1999,
  \href{https://arxiv.org/abs/astro-ph/9901124}{{\ttfamily astro-ph/9901124}}.

\bibitem{Riotto:2002yw}
A.~Riotto, \emph{{Inflation and the theory of cosmological perturbations}},
  {\emph{ICTP Lect. Notes Ser.} {\bfseries 14} (2003) 317--413},
  [\href{https://arxiv.org/abs/hep-ph/0210162}{{\ttfamily hep-ph/0210162}}].

\bibitem{Tsujikawa:2003jp}
S.~Tsujikawa, \emph{{Introductory review of cosmic inflation}},  in \emph{{2nd
  Tah Poe School on Cosmology: Modern Cosmology Phitsanulok, Thailand, April
  17-25, 2003}}, 2003, \href{https://arxiv.org/abs/hep-ph/0304257}{{\ttfamily
  hep-ph/0304257}}.

\bibitem{Linde:2005ht}
A.~D. Linde, \emph{{Particle physics and inflationary cosmology}},
  {\emph{Contemp. Concepts Phys.} {\bfseries 5} (1990) 1--362},
  [\href{https://arxiv.org/abs/hep-th/0503203}{{\ttfamily hep-th/0503203}}].

\bibitem{Linde:2007fr}
A.~D. Linde, \emph{{Inflationary Cosmology}},
  \href{https://doi.org/10.1007/978-3-540-74353-8_1}{\emph{Lect. Notes Phys.}
  {\bfseries 738} (2008) 1--54},
  [\href{https://arxiv.org/abs/0705.0164}{{\ttfamily 0705.0164}}].

\bibitem{Kinney:2009vz}
W.~H. Kinney, \emph{{TASI Lectures on Inflation}},
  \href{https://arxiv.org/abs/0902.1529}{{\ttfamily 0902.1529}}.

\bibitem{Baumann:2009ds}
D.~Baumann, \emph{{Inflation}},  in \emph{{Physics of the large and the small,
  TASI 09, proceedings of the Theoretical Advanced Study Institute in
  Elementary Particle Physics, Boulder, Colorado, USA, 1-26 June 2009}},
  pp.~523--686, 2011, \href{https://arxiv.org/abs/0907.5424}{{\ttfamily
  0907.5424}}.

\bibitem{Senatore:2016aui}
L.~Senatore, \emph{{Lectures on Inflation}},  in \emph{{Proceedings,
  Theoretical Advanced Study Institute in Elementary Particle Physics: New
  Frontiers in Fields and Strings (TASI 2015): Boulder, CO, USA, June 1-26,
  2015}}, pp.~447--543, 2017,
  \href{https://arxiv.org/abs/1609.00716}{{\ttfamily 1609.00716}}.

\bibitem{Vazquez:2018qdg}
J.~A. Vázquez, L.~E. Padilla and T.~Matos, \emph{{Inflationary Cosmology: From
  Theory to Observations}},  \href{https://arxiv.org/abs/1810.09934}{{\ttfamily
  1810.09934}}.

\bibitem{Starobinsky:1980te}
A.~A. Starobinsky, \emph{{A New Type of Isotropic Cosmological Models Without
  Singularity}},
  \href{https://doi.org/10.1016/0370-2693(80)90670-X}{\emph{Phys. Lett.}
  {\bfseries B91} (1980) 99--102}.

\bibitem{Kazanas:1980tx}
D.~Kazanas, \emph{{Dynamics of the Universe and Spontaneous Symmetry
  Breaking}}, \href{https://doi.org/10.1086/183361}{\emph{Astrophys. J.}
  {\bfseries 241} (1980) L59--L63}.

\bibitem{Guth:1980zm}
A.~H. Guth, \emph{{The Inflationary Universe: A Possible Solution to the
  Horizon and Flatness Problems}},
  \href{https://doi.org/10.1103/PhysRevD.23.347}{\emph{Phys. Rev.} {\bfseries
  D23} (1981) 347--356}.

\bibitem{Sato:1981ds}
K.~Sato, \emph{{Cosmological Baryon Number Domain Structure and the First Order
  Phase Transition of a Vacuum}},
  \href{https://doi.org/10.1016/0370-2693(81)90805-4}{\emph{Phys. Lett.}
  {\bfseries 99B} (1981) 66--70}.

\bibitem{Mukhanov:1981xt}
V.~F. Mukhanov and G.~V. Chibisov, \emph{{Quantum Fluctuations and a
  Nonsingular Universe}}, {\emph{JETP Lett.} {\bfseries 33} (1981) 532--535}.

\bibitem{Linde:1981mu}
A.~D. Linde, \emph{{A New Inflationary Universe Scenario: A Possible Solution
  of the Horizon, Flatness, Homogeneity, Isotropy and Primordial Monopole
  Problems}}, \href{https://doi.org/10.1016/0370-2693(82)91219-9}{\emph{Phys.
  Lett.} {\bfseries 108B} (1982) 389--393}.

\bibitem{Albrecht:1982wi}
A.~Albrecht and P.~J. Steinhardt, \emph{{Cosmology for Grand Unified Theories
  with Radiatively Induced Symmetry Breaking}},
  \href{https://doi.org/10.1103/PhysRevLett.48.1220}{\emph{Phys. Rev. Lett.}
  {\bfseries 48} (1982) 1220--1223}.

\bibitem{Ellis:1982ed}
J.~R. Ellis, D.~V. Nanopoulos, K.~A. Olive and K.~Tamvakis, \emph{{Cosmological
  Inflation Cries Out for Supersymmetry}},
  \href{https://doi.org/10.1016/0370-2693(82)90198-8}{\emph{Phys. Lett.}
  {\bfseries 118B} (1982) 335}.

\bibitem{Steinhardt:1984jj}
P.~J. Steinhardt and M.~S. Turner, \emph{{A Prescription for Successful New
  Inflation}}, \href{https://doi.org/10.1103/PhysRevD.29.2162}{\emph{Phys.
  Rev.} {\bfseries D29} (1984) 2162--2171}.

\bibitem{Abbott:1984ba}
R.~B. Abbott, S.~M. Barr and S.~D. Ellis, \emph{{Kaluza-Klein Cosmologies and
  Inflation}}, \href{https://doi.org/10.1103/PhysRevD.30.720}{\emph{Phys. Rev.}
  {\bfseries D30} (1984) 720}.

\bibitem{Linde:1986fc}
A.~D. Linde, \emph{{Eternal Chaotic Inflation}},
  \href{https://doi.org/10.1142/S0217732386000129}{\emph{Mod. Phys. Lett.}
  {\bfseries A1} (1986) 81}.

\bibitem{Silk:1986vc}
J.~Silk and M.~S. Turner, \emph{{Double Inflation}},
  \href{https://doi.org/10.1103/PhysRevD.35.419}{\emph{Phys. Rev.} {\bfseries
  D35} (1987) 419}.

\bibitem{Turok:1987pg}
N.~Turok, \emph{{String Driven Inflation}},
  \href{https://doi.org/10.1103/PhysRevLett.60.549}{\emph{Phys. Rev. Lett.}
  {\bfseries 60} (1988) 549}.

\bibitem{Ford:1989me}
L.~H. Ford, \emph{{Inflation driven by a vector field}},
  \href{https://doi.org/10.1103/PhysRevD.40.967}{\emph{Phys. Rev.} {\bfseries
  D40} (1989) 967}.

\bibitem{Freese:1990rb}
K.~Freese, J.~A. Frieman and A.~V. Olinto, \emph{{Natural inflation with pseudo
  - Nambu-Goldstone bosons}},
  \href{https://doi.org/10.1103/PhysRevLett.65.3233}{\emph{Phys. Rev. Lett.}
  {\bfseries 65} (1990) 3233--3236}.

\bibitem{Adams:1992bn}
F.~C. Adams, J.~R. Bond, K.~Freese, J.~A. Frieman and A.~V. Olinto,
  \emph{{Natural inflation: Particle physics models, power law spectra for
  large scale structure, and constraints from COBE}},
  \href{https://doi.org/10.1103/PhysRevD.47.426}{\emph{Phys. Rev.} {\bfseries
  D47} (1993) 426--455},
  [\href{https://arxiv.org/abs/hep-ph/9207245}{{\ttfamily hep-ph/9207245}}].

\bibitem{Borde:1993xh}
A.~Borde and A.~Vilenkin, \emph{{Eternal inflation and the initial
  singularity}}, \href{https://doi.org/10.1103/PhysRevLett.72.3305}{\emph{Phys.
  Rev. Lett.} {\bfseries 72} (1994) 3305--3309},
  [\href{https://arxiv.org/abs/gr-qc/9312022}{{\ttfamily gr-qc/9312022}}].

\bibitem{Vilenkin:1994pv}
A.~Vilenkin, \emph{{Topological inflation}},
  \href{https://doi.org/10.1103/PhysRevLett.72.3137}{\emph{Phys. Rev. Lett.}
  {\bfseries 72} (1994) 3137--3140},
  [\href{https://arxiv.org/abs/hep-th/9402085}{{\ttfamily hep-th/9402085}}].

\bibitem{Ross:1995dq}
G.~G. Ross and S.~Sarkar, \emph{{Successful supersymmetric inflation}},
  \href{https://doi.org/10.1016/0550-3213(96)00013-2}{\emph{Nucl. Phys.}
  {\bfseries B461} (1996) 597--624},
  [\href{https://arxiv.org/abs/hep-ph/9506283}{{\ttfamily hep-ph/9506283}}].

\bibitem{Lazarides:1995vr}
G.~Lazarides and C.~Panagiotakopoulos, \emph{{Smooth hybrid inflation}},
  \href{https://doi.org/10.1103/PhysRevD.52.R559}{\emph{Phys. Rev.} {\bfseries
  D52} (1995) R559--R563},
  [\href{https://arxiv.org/abs/hep-ph/9506325}{{\ttfamily hep-ph/9506325}}].

\bibitem{Berera:1995ie}
A.~Berera, \emph{{Warm inflation}},
  \href{https://doi.org/10.1103/PhysRevLett.75.3218}{\emph{Phys. Rev. Lett.}
  {\bfseries 75} (1995) 3218--3221},
  [\href{https://arxiv.org/abs/astro-ph/9509049}{{\ttfamily
  astro-ph/9509049}}].

\bibitem{Faraoni:1996rf}
V.~Faraoni, \emph{{Nonminimal coupling of the scalar field and inflation}},
  \href{https://doi.org/10.1103/PhysRevD.53.6813}{\emph{Phys. Rev.} {\bfseries
  D53} (1996) 6813--6821},
  [\href{https://arxiv.org/abs/astro-ph/9602111}{{\ttfamily
  astro-ph/9602111}}].

\bibitem{Binetruy:1996xj}
P.~Binetruy and G.~R. Dvali, \emph{{D term inflation}},
  \href{https://doi.org/10.1016/S0370-2693(96)01083-0}{\emph{Phys. Lett.}
  {\bfseries B388} (1996) 241--246},
  [\href{https://arxiv.org/abs/hep-ph/9606342}{{\ttfamily hep-ph/9606342}}].

\bibitem{Peebles:1998qn}
P.~J.~E. Peebles and A.~Vilenkin, \emph{{Quintessential inflation}},
  \href{https://doi.org/10.1103/PhysRevD.59.063505}{\emph{Phys. Rev.}
  {\bfseries D59} (1999) 063505},
  [\href{https://arxiv.org/abs/astro-ph/9810509}{{\ttfamily
  astro-ph/9810509}}].

\bibitem{Dvali:1998pa}
G.~R. Dvali and S.~H.~H. Tye, \emph{{Brane inflation}},
  \href{https://doi.org/10.1016/S0370-2693(99)00132-X}{\emph{Phys. Lett.}
  {\bfseries B450} (1999) 72--82},
  [\href{https://arxiv.org/abs/hep-ph/9812483}{{\ttfamily hep-ph/9812483}}].

\bibitem{ArmendarizPicon:1999rj}
C.~Armendariz-Picon, T.~Damour and V.~F. Mukhanov, \emph{{k - inflation}},
  \href{https://doi.org/10.1016/S0370-2693(99)00603-6}{\emph{Phys. Lett.}
  {\bfseries B458} (1999) 209--218},
  [\href{https://arxiv.org/abs/hep-th/9904075}{{\ttfamily hep-th/9904075}}].

\bibitem{Maartens:1999hf}
R.~Maartens, D.~Wands, B.~A. Bassett and I.~Heard, \emph{{Chaotic inflation on
  the brane}}, \href{https://doi.org/10.1103/PhysRevD.62.041301}{\emph{Phys.
  Rev.} {\bfseries D62} (2000) 041301},
  [\href{https://arxiv.org/abs/hep-ph/9912464}{{\ttfamily hep-ph/9912464}}].

\bibitem{Nojiri:2000gb}
S.~Nojiri and S.~D. Odintsov, \emph{{Brane world inflation induced by quantum
  effects}}, \href{https://doi.org/10.1016/S0370-2693(00)00629-8}{\emph{Phys.
  Lett.} {\bfseries B484} (2000) 119--123},
  [\href{https://arxiv.org/abs/hep-th/0004097}{{\ttfamily hep-th/0004097}}].

\bibitem{Nojiri:2003ft}
S.~Nojiri and S.~D. Odintsov, \emph{{Modified gravity with negative and
  positive powers of the curvature: Unification of the inflation and of the
  cosmic acceleration}},
  \href{https://doi.org/10.1103/PhysRevD.68.123512}{\emph{Phys. Rev.}
  {\bfseries D68} (2003) 123512},
  [\href{https://arxiv.org/abs/hep-th/0307288}{{\ttfamily hep-th/0307288}}].

\bibitem{Kachru:2003sx}
S.~Kachru, R.~Kallosh, A.~D. Linde, J.~M. Maldacena, L.~P. McAllister and S.~P.
  Trivedi, \emph{{Towards inflation in string theory}},
  \href{https://doi.org/10.1088/1475-7516/2003/10/013}{\emph{JCAP} {\bfseries
  0310} (2003) 013}, [\href{https://arxiv.org/abs/hep-th/0308055}{{\ttfamily
  hep-th/0308055}}].

\bibitem{ArkaniHamed:2003uz}
N.~Arkani-Hamed, P.~Creminelli, S.~Mukohyama and M.~Zaldarriaga, \emph{{Ghost
  inflation}}, \href{https://doi.org/10.1088/1475-7516/2004/04/001}{\emph{JCAP}
  {\bfseries 0404} (2004) 001},
  [\href{https://arxiv.org/abs/hep-th/0312100}{{\ttfamily hep-th/0312100}}].

\bibitem{BlancoPillado:2004ns}
J.~J. Blanco-Pillado, C.~P. Burgess, J.~M. Cline, C.~Escoda, M.~Gómez-Reino,
  R.~Kallosh et~al., \emph{{Racetrack inflation}},
  \href{https://doi.org/10.1088/1126-6708/2004/11/063}{\emph{JHEP} {\bfseries
  11} (2004) 063}, [\href{https://arxiv.org/abs/hep-th/0406230}{{\ttfamily
  hep-th/0406230}}].

\bibitem{Boubekeur:2005zm}
L.~Boubekeur and D.~H. Lyth, \emph{{Hilltop inflation}},
  \href{https://doi.org/10.1088/1475-7516/2005/07/010}{\emph{JCAP} {\bfseries
  0507} (2005) 010}, [\href{https://arxiv.org/abs/hep-ph/0502047}{{\ttfamily
  hep-ph/0502047}}].

\bibitem{Kinney:2005vj}
W.~H. Kinney, \emph{{Horizon crossing and inflation with large eta}},
  \href{https://doi.org/10.1103/PhysRevD.72.023515}{\emph{Phys. Rev.}
  {\bfseries D72} (2005) 023515},
  [\href{https://arxiv.org/abs/gr-qc/0503017}{{\ttfamily gr-qc/0503017}}].

\bibitem{Anisimov:2005ne}
A.~Anisimov, E.~Babichev and A.~Vikman, \emph{{B-inflation}},
  \href{https://doi.org/10.1088/1475-7516/2005/06/006}{\emph{JCAP} {\bfseries
  0506} (2005) 006}, [\href{https://arxiv.org/abs/astro-ph/0504560}{{\ttfamily
  astro-ph/0504560}}].

\bibitem{Nojiri:2005pu}
S.~Nojiri and S.~D. Odintsov, \emph{{Unifying phantom inflation with late-time
  acceleration: Scalar phantom-non-phantom transition model and generalized
  holographic dark energy}},
  \href{https://doi.org/10.1007/s10714-006-0301-6}{\emph{Gen. Rel. Grav.}
  {\bfseries 38} (2006) 1285--1304},
  [\href{https://arxiv.org/abs/hep-th/0506212}{{\ttfamily hep-th/0506212}}].

\bibitem{Capozziello:2005tf}
S.~Capozziello, S.~Nojiri and S.~D. Odintsov, \emph{{Unified phantom cosmology:
  Inflation, dark energy and dark matter under the same standard}},
  \href{https://doi.org/10.1016/j.physletb.2005.11.012}{\emph{Phys. Lett.}
  {\bfseries B632} (2006) 597--604},
  [\href{https://arxiv.org/abs/hep-th/0507182}{{\ttfamily hep-th/0507182}}].

\bibitem{Dimopoulos:2005ac}
S.~Dimopoulos, S.~Kachru, J.~McGreevy and J.~G. Wacker, \emph{{N-flation}},
  \href{https://doi.org/10.1088/1475-7516/2008/08/003}{\emph{JCAP} {\bfseries
  0808} (2008) 003}, [\href{https://arxiv.org/abs/hep-th/0507205}{{\ttfamily
  hep-th/0507205}}].

\bibitem{Savage:2006tr}
C.~Savage, K.~Freese and W.~H. Kinney, \emph{{Natural Inflation: Status after
  WMAP 3-year data}},
  \href{https://doi.org/10.1103/PhysRevD.74.123511}{\emph{Phys. Rev.}
  {\bfseries D74} (2006) 123511},
  [\href{https://arxiv.org/abs/hep-ph/0609144}{{\ttfamily hep-ph/0609144}}].

\bibitem{Ferraro:2006jd}
R.~Ferraro and F.~Fiorini, \emph{{Modified teleparallel gravity: Inflation
  without inflaton}},
  \href{https://doi.org/10.1103/PhysRevD.75.084031}{\emph{Phys. Rev.}
  {\bfseries D75} (2007) 084031},
  [\href{https://arxiv.org/abs/gr-qc/0610067}{{\ttfamily gr-qc/0610067}}].

\bibitem{Bezrukov:2007ep}
F.~L. Bezrukov and M.~Shaposhnikov, \emph{{The Standard Model Higgs boson as
  the inflaton}},
  \href{https://doi.org/10.1016/j.physletb.2007.11.072}{\emph{Phys. Lett.}
  {\bfseries B659} (2008) 703--706},
  [\href{https://arxiv.org/abs/0710.3755}{{\ttfamily 0710.3755}}].

\bibitem{Cognola:2007zu}
G.~Cognola, E.~Elizalde, S.~Nojiri, S.~D. Odintsov, L.~Sebastiani and
  S.~Zerbini, \emph{{A Class of viable modified f(R) gravities describing
  inflation and the onset of accelerated expansion}},
  \href{https://doi.org/10.1103/PhysRevD.77.046009}{\emph{Phys. Rev.}
  {\bfseries D77} (2008) 046009},
  [\href{https://arxiv.org/abs/0712.4017}{{\ttfamily 0712.4017}}].

\bibitem{Freese:2008if}
K.~Freese, C.~Savage and W.~H. Kinney, \emph{{Natural Inflation: The Status
  after WMAP 3-year data}},
  \href{https://doi.org/10.1142/S0218271807011371}{\emph{Int. J. Mod. Phys.}
  {\bfseries D16} (2008) 2573--2585},
  [\href{https://arxiv.org/abs/0802.0227}{{\ttfamily 0802.0227}}].

\bibitem{Silverstein:2008sg}
E.~Silverstein and A.~Westphal, \emph{{Monodromy in the CMB: Gravity Waves and
  String Inflation}},
  \href{https://doi.org/10.1103/PhysRevD.78.106003}{\emph{Phys. Rev.}
  {\bfseries D78} (2008) 106003},
  [\href{https://arxiv.org/abs/0803.3085}{{\ttfamily 0803.3085}}].

\bibitem{Kaloper:2008fb}
N.~Kaloper and L.~Sorbo, \emph{{A Natural Framework for Chaotic Inflation}},
  \href{https://doi.org/10.1103/PhysRevLett.102.121301}{\emph{Phys. Rev. Lett.}
  {\bfseries 102} (2009) 121301},
  [\href{https://arxiv.org/abs/0811.1989}{{\ttfamily 0811.1989}}].

\bibitem{Bessada:2009ns}
D.~Bessada, W.~H. Kinney, D.~Stojkovic and J.~Wang, \emph{{Tachyacoustic
  Cosmology: An Alternative to Inflation}},
  \href{https://doi.org/10.1103/PhysRevD.81.043510}{\emph{Phys. Rev.}
  {\bfseries D81} (2010) 043510},
  [\href{https://arxiv.org/abs/0908.3898}{{\ttfamily 0908.3898}}].

\bibitem{Germani:2010hd}
C.~Germani and A.~Kehagias, \emph{{UV-Protected Inflation}},
  \href{https://doi.org/10.1103/PhysRevLett.106.161302}{\emph{Phys. Rev. Lett.}
  {\bfseries 106} (2011) 161302},
  [\href{https://arxiv.org/abs/1012.0853}{{\ttfamily 1012.0853}}].

\bibitem{Maleknejad:2011jw}
A.~Maleknejad and M.~M. Sheikh-Jabbari, \emph{{Gauge-flation: Inflation From
  Non-Abelian Gauge Fields}},
  \href{https://doi.org/10.1016/j.physletb.2013.05.001}{\emph{Phys. Lett.}
  {\bfseries B723} (2013) 224--228},
  [\href{https://arxiv.org/abs/1102.1513}{{\ttfamily 1102.1513}}].

\bibitem{Kobayashi:2011nu}
T.~Kobayashi, M.~Yamaguchi and J.~Yokoyama, \emph{{Generalized G-inflation:
  Inflation with the most general second-order field equations}},
  \href{https://doi.org/10.1143/PTP.126.511}{\emph{Prog. Theor. Phys.}
  {\bfseries 126} (2011) 511--529},
  [\href{https://arxiv.org/abs/1105.5723}{{\ttfamily 1105.5723}}].

\bibitem{Visinelli:2011jy}
L.~Visinelli, \emph{{Natural Warm Inflation}},
  \href{https://doi.org/10.1088/1475-7516/2011/09/013}{\emph{JCAP} {\bfseries
  1109} (2011) 013}, [\href{https://arxiv.org/abs/1107.3523}{{\ttfamily
  1107.3523}}].

\bibitem{Endlich:2012pz}
S.~Endlich, A.~Nicolis and J.~Wang, \emph{{Solid Inflation}},
  \href{https://doi.org/10.1088/1475-7516/2013/10/011}{\emph{JCAP} {\bfseries
  1310} (2013) 011}, [\href{https://arxiv.org/abs/1210.0569}{{\ttfamily
  1210.0569}}].

\bibitem{Martin:2013tda}
J.~Martin, C.~Ringeval and V.~Vennin, \emph{{Encyclopædia Inflationaris}},
  \href{https://doi.org/10.1016/j.dark.2014.01.003}{\emph{Phys. Dark Univ.}
  {\bfseries 5-6} (2014) 75--235},
  [\href{https://arxiv.org/abs/1303.3787}{{\ttfamily 1303.3787}}].

\bibitem{Kallosh:2013hoa}
R.~Kallosh and A.~Linde, \emph{{Universality Class in Conformal Inflation}},
  \href{https://doi.org/10.1088/1475-7516/2013/07/002}{\emph{JCAP} {\bfseries
  1307} (2013) 002}, [\href{https://arxiv.org/abs/1306.5220}{{\ttfamily
  1306.5220}}].

\bibitem{Dong:2013swa}
R.~Dong, W.~H. Kinney and D.~Stojkovic, \emph{{Symmetron Inflation}},
  \href{https://doi.org/10.1088/1475-7516/2014/01/021}{\emph{JCAP} {\bfseries
  1401} (2014) 021}, [\href{https://arxiv.org/abs/1307.4451}{{\ttfamily
  1307.4451}}].

\bibitem{Sebastiani:2013eqa}
L.~Sebastiani, G.~Cognola, R.~Myrzakulov, S.~D. Odintsov and S.~Zerbini,
  \emph{{Nearly Starobinsky inflation from modified gravity}},
  \href{https://doi.org/10.1103/PhysRevD.89.023518}{\emph{Phys. Rev.}
  {\bfseries D89} (2014) 023518},
  [\href{https://arxiv.org/abs/1311.0744}{{\ttfamily 1311.0744}}].

\bibitem{Czerny:2014wza}
M.~Czerny and F.~Takahashi, \emph{{Multi-Natural Inflation}},
  \href{https://doi.org/10.1016/j.physletb.2014.04.039}{\emph{Phys. Lett.}
  {\bfseries B733} (2014) 241--246},
  [\href{https://arxiv.org/abs/1401.5212}{{\ttfamily 1401.5212}}].

\bibitem{Freese:2014nla}
K.~Freese and W.~H. Kinney, \emph{{Natural Inflation: Consistency with Cosmic
  Microwave Background Observations of Planck and BICEP2}},
  \href{https://doi.org/10.1088/1475-7516/2015/03/044}{\emph{JCAP} {\bfseries
  1503} (2015) 044}, [\href{https://arxiv.org/abs/1403.5277}{{\ttfamily
  1403.5277}}].

\bibitem{Marchesano:2014mla}
F.~Marchesano, G.~Shiu and A.~M. Uranga, \emph{{F-term Axion Monodromy
  Inflation}}, \href{https://doi.org/10.1007/JHEP09(2014)184}{\emph{JHEP}
  {\bfseries 09} (2014) 184},
  [\href{https://arxiv.org/abs/1404.3040}{{\ttfamily 1404.3040}}].

\bibitem{Rinaldi:2014gua}
M.~Rinaldi, G.~Cognola, L.~Vanzo and S.~Zerbini, \emph{{Reconstructing the
  inflationary $f(R)$ from observations}},
  \href{https://doi.org/10.1088/1475-7516/2014/08/015}{\emph{JCAP} {\bfseries
  1408} (2014) 015}, [\href{https://arxiv.org/abs/1406.1096}{{\ttfamily
  1406.1096}}].

\bibitem{Nojiri:2014zqa}
S.~Nojiri and S.~D. Odintsov, \emph{{Mimetic $F(R)$ gravity: inflation, dark
  energy and bounce}},
  \href{https://doi.org/10.1142/S0217732314502113}{\emph{Mod. Phys. Lett.}
  {\bfseries A29} (2014) 1450211},
  [\href{https://arxiv.org/abs/1408.3561}{{\ttfamily 1408.3561}}].

\bibitem{Rinaldi:2014gha}
M.~Rinaldi, G.~Cognola, L.~Vanzo and S.~Zerbini, \emph{{Inflation in
  scale-invariant theories of gravity}},
  \href{https://doi.org/10.1103/PhysRevD.91.123527}{\emph{Phys. Rev.}
  {\bfseries D91} (2015) 123527},
  [\href{https://arxiv.org/abs/1410.0631}{{\ttfamily 1410.0631}}].

\bibitem{Visinelli:2014qla}
L.~Visinelli, \emph{{Cosmological perturbations for an inflaton field coupled
  to radiation}},
  \href{https://doi.org/10.1088/1475-7516/2015/01/005}{\emph{JCAP} {\bfseries
  1501} (2015) 005}, [\href{https://arxiv.org/abs/1410.1187}{{\ttfamily
  1410.1187}}].

\bibitem{Kannike:2015apa}
K.~Kannike, G.~Hütsi, L.~Pizza, A.~Racioppi, M.~Raidal, A.~Salvio et~al.,
  \emph{{Dynamically Induced Planck Scale and Inflation}},
  \href{https://doi.org/10.1007/JHEP05(2015)065}{\emph{JHEP} {\bfseries 05}
  (2015) 065}, [\href{https://arxiv.org/abs/1502.01334}{{\ttfamily
  1502.01334}}].

\bibitem{Myrzakulov:2015fra}
R.~Myrzakulov, L.~Sebastiani and S.~Zerbini, \emph{{Reconstruction of Inflation
  Models}}, \href{https://doi.org/10.1140/epjc/s10052-015-3443-4}{\emph{Eur.
  Phys. J.} {\bfseries C75} (2015) 215},
  [\href{https://arxiv.org/abs/1502.04432}{{\ttfamily 1502.04432}}].

\bibitem{DeLaurentis:2015fea}
M.~De~Laurentis, M.~Paolella and S.~Capozziello, \emph{{Cosmological inflation
  in $F(R,\mathcal{G})$ gravity}},
  \href{https://doi.org/10.1103/PhysRevD.91.083531}{\emph{Phys. Rev.}
  {\bfseries D91} (2015) 083531},
  [\href{https://arxiv.org/abs/1503.04659}{{\ttfamily 1503.04659}}].

\bibitem{Chakraborty:2015qga}
S.~Chakraborty, S.~Pan and S.~Saha, \emph{{A unified cosmic evolution:
  Inflation to late time acceleration}},
  \href{https://arxiv.org/abs/1503.05552}{{\ttfamily 1503.05552}}.

\bibitem{Myrzakulov:2015qaa}
R.~Myrzakulov, L.~Sebastiani and S.~Vagnozzi, \emph{{Inflation in $f(R,\phi )$
  -theories and mimetic gravity scenario}},
  \href{https://doi.org/10.1140/epjc/s10052-015-3672-6}{\emph{Eur. Phys. J.}
  {\bfseries C75} (2015) 444},
  [\href{https://arxiv.org/abs/1504.07984}{{\ttfamily 1504.07984}}].

\bibitem{Rinaldi:2015yoa}
M.~Rinaldi, L.~Vanzo, S.~Zerbini and G.~Venturi, \emph{{Inflationary
  quasiscale-invariant attractors}},
  \href{https://doi.org/10.1103/PhysRevD.93.024040}{\emph{Phys. Rev.}
  {\bfseries D93} (2016) 024040},
  [\href{https://arxiv.org/abs/1505.03386}{{\ttfamily 1505.03386}}].

\bibitem{Sepehri:2015eea}
A.~Sepehri, F.~Rahaman, M.~R. Setare, A.~Pradhan, S.~Capozziello and I.~H.
  Sardar, \emph{{Unifying inflation with late-time acceleration by a BIonic
  system}}, \href{https://doi.org/10.1016/j.physletb.2015.05.042}{\emph{Phys.
  Lett.} {\bfseries B747} (2015) 1--8},
  [\href{https://arxiv.org/abs/1505.05105}{{\ttfamily 1505.05105}}].

\bibitem{Sebastiani:2015kfa}
L.~Sebastiani and R.~Myrzakulov, \emph{{F(R) gravity and inflation}},
  \href{https://doi.org/10.1142/S0219887815300032}{\emph{Int. J. Geom. Meth.
  Mod. Phys.} {\bfseries 12} (2015) 1530003},
  [\href{https://arxiv.org/abs/1506.05330}{{\ttfamily 1506.05330}}].

\bibitem{Kappl:2015esy}
R.~Kappl, H.~P. Nilles and M.~W. Winkler, \emph{{Modulated Natural Inflation}},
  \href{https://doi.org/10.1016/j.physletb.2015.12.073}{\emph{Phys. Lett.}
  {\bfseries B753} (2016) 653--659},
  [\href{https://arxiv.org/abs/1511.05560}{{\ttfamily 1511.05560}}].

\bibitem{Rinaldi:2015uvu}
M.~Rinaldi and L.~Vanzo, \emph{{Inflation and reheating in theories with
  spontaneous scale invariance symmetry breaking}},
  \href{https://doi.org/10.1103/PhysRevD.94.024009}{\emph{Phys. Rev.}
  {\bfseries D94} (2016) 024009},
  [\href{https://arxiv.org/abs/1512.07186}{{\ttfamily 1512.07186}}].

\bibitem{Cognola:2016gjy}
G.~Cognola, R.~Myrzakulov, L.~Sebastiani, S.~Vagnozzi and S.~Zerbini,
  \emph{{Covariant Hořava-like and mimetic Horndeski gravity: cosmological
  solutions and perturbations}},
  \href{https://doi.org/10.1088/0264-9381/33/22/225014}{\emph{Class. Quant.
  Grav.} {\bfseries 33} (2016) 225014},
  [\href{https://arxiv.org/abs/1601.00102}{{\ttfamily 1601.00102}}].

\bibitem{Barenboim:2016mmw}
G.~Barenboim, W.-I. Park and W.~H. Kinney, \emph{{Eternal Hilltop Inflation}},
  \href{https://doi.org/10.1088/1475-7516/2016/05/030}{\emph{JCAP} {\bfseries
  1605} (2016) 030}, [\href{https://arxiv.org/abs/1601.08140}{{\ttfamily
  1601.08140}}].

\bibitem{Visinelli:2016teo}
L.~Visinelli, \emph{{Inflating without a flat potential: Viscous inflation}},
  \href{https://arxiv.org/abs/1604.03873}{{\ttfamily 1604.03873}}.

\bibitem{Visinelli:2016rhn}
L.~Visinelli, \emph{{Observational Constraints on Monomial Warm Inflation}},
  \href{https://doi.org/10.1088/1475-7516/2016/07/054}{\emph{JCAP} {\bfseries
  1607} (2016) 054}, [\href{https://arxiv.org/abs/1605.06449}{{\ttfamily
  1605.06449}}].

\bibitem{Ballesteros:2016xej}
G.~Ballesteros, J.~Redondo, A.~Ringwald and C.~Tamarit, \emph{{Standard
  Model—axion—seesaw—Higgs portal inflation. Five problems of particle
  physics and cosmology solved in one stroke}},
  \href{https://doi.org/10.1088/1475-7516/2017/08/001}{\emph{JCAP} {\bfseries
  1708} (2017) 001}, [\href{https://arxiv.org/abs/1610.01639}{{\ttfamily
  1610.01639}}].

\bibitem{Nojiri:2017ncd}
S.~Nojiri, S.~D. Odintsov and V.~K. Oikonomou, \emph{{Modified Gravity Theories
  on a Nutshell: Inflation, Bounce and Late-time Evolution}},
  \href{https://doi.org/10.1016/j.physrep.2017.06.001}{\emph{Phys. Rept.}
  {\bfseries 692} (2017) 1--104},
  [\href{https://arxiv.org/abs/1705.11098}{{\ttfamily 1705.11098}}].

\bibitem{Sebastiani:2017mkv}
L.~Sebastiani, S.~Myrzakul and R.~Myrzakulov, \emph{{Warm inflation in
  Horndeski gravity}},
  \href{https://doi.org/10.1007/s10714-017-2257-0}{\emph{Gen. Rel. Grav.}
  {\bfseries 49} (2017) 90},
  [\href{https://arxiv.org/abs/1707.03702}{{\ttfamily 1707.03702}}].

\bibitem{Odintsov:2017hbk}
S.~D. Odintsov, V.~K. Oikonomou and L.~Sebastiani, \emph{{Unification of
  Constant-roll Inflation and Dark Energy with Logarithmic $R^2$-corrected and
  Exponential $F(R)$ Gravity}},
  \href{https://doi.org/10.1016/j.nuclphysb.2017.08.018}{\emph{Nucl. Phys.}
  {\bfseries B923} (2017) 608--632},
  [\href{https://arxiv.org/abs/1708.08346}{{\ttfamily 1708.08346}}].

\bibitem{Freese:2017ace}
K.~Freese, E.~I. Sfakianakis, P.~Stengel and L.~Visinelli, \emph{{The Higgs
  Boson can delay Reheating after Inflation}},
  \href{https://doi.org/10.1088/1475-7516/2018/05/067}{\emph{JCAP} {\bfseries
  1805} (2018) 067}, [\href{https://arxiv.org/abs/1712.03791}{{\ttfamily
  1712.03791}}].

\bibitem{Odintsov:2018ggm}
S.~D. Odintsov and V.~K. Oikonomou, \emph{{The reconstruction of $f(\phi)R$ and
  mimetic gravity from viable slow-roll inflation}},
  \href{https://doi.org/10.1016/j.nuclphysb.2018.01.027}{\emph{Nucl. Phys.}
  {\bfseries B929} (2018) 79--112},
  [\href{https://arxiv.org/abs/1801.10529}{{\ttfamily 1801.10529}}].

\bibitem{Kleidis:2018fdu}
K.~Kleidis and V.~K. Oikonomou, \emph{{Scalar Field Assisted $f(R)$ Gravity
  Inflation}}, \href{https://doi.org/10.1142/S0219887818501372}{\emph{Int. J.
  Geom. Meth. Mod. Phys.} {\bfseries 15} (2018) 1850137},
  [\href{https://arxiv.org/abs/1803.10748}{{\ttfamily 1803.10748}}].

\bibitem{Achucarro:2018vey}
A.~Achúcarro and G.~A. Palma, \emph{{The string swampland constraints require
  multi-field inflation}},
  \href{https://doi.org/10.1088/1475-7516/2019/02/041}{\emph{JCAP} {\bfseries
  1902} (2019) 041}, [\href{https://arxiv.org/abs/1807.04390}{{\ttfamily
  1807.04390}}].

\bibitem{Kehagias:2018uem}
A.~Kehagias and A.~Riotto, \emph{{A note on Inflation and the Swampland}},
  \href{https://doi.org/10.1002/prop.201800052}{\emph{Fortsch. Phys.}
  {\bfseries 66} (2018) 1800052},
  [\href{https://arxiv.org/abs/1807.05445}{{\ttfamily 1807.05445}}].

\bibitem{Kinney:2018nny}
W.~H. Kinney, S.~Vagnozzi and L.~Visinelli, \emph{{The Zoo Plot Meets the
  Swampland: Mutual (In)Consistency of Single-Field Inflation, String
  Conjectures, and Cosmological Data}},
  \href{https://arxiv.org/abs/1808.06424}{{\ttfamily 1808.06424}}.

\bibitem{Haro:2019gsv}
J.~Haro, J.~Amorós and S.~Pan, \emph{{The Peebles - Vilenkin quintessential
  inflation model resivited}},
  \href{https://arxiv.org/abs/1901.00167}{{\ttfamily 1901.00167}}.

\bibitem{Nojiri:2019dqc}
S.~Nojiri, S.~D. Odintsov and V.~K. Oikonomou, \emph{{$k$-essence $f(R)$
  gravity inflation}},
  \href{https://doi.org/10.1016/j.nuclphysb.2019.02.008}{\emph{Nucl. Phys.}
  {\bfseries B941} (2019) 11--27},
  [\href{https://arxiv.org/abs/1902.03669}{{\ttfamily 1902.03669}}].

\bibitem{Chowdhury:2019otk}
D.~Chowdhury, J.~Martin, C.~Ringeval and V.~Vennin, \emph{{Inflation after
  Planck: Judgment Day}},  \href{https://arxiv.org/abs/1902.03951}{{\ttfamily
  1902.03951}}.

\bibitem{Vicentini:2019etr}
S.~Vicentini, L.~Vanzo and M.~Rinaldi, \emph{{Scale-invariant inflation with
  1-loop quantum corrections}},
  \href{https://arxiv.org/abs/1902.04434}{{\ttfamily 1902.04434}}.

\bibitem{Hannestad:2004px}
S.~Hannestad, \emph{{What is the lowest possible reheating temperature?}},
  \href{https://doi.org/10.1103/PhysRevD.70.043506}{\emph{Phys. Rev.}
  {\bfseries D70} (2004) 043506},
  [\href{https://arxiv.org/abs/astro-ph/0403291}{{\ttfamily
  astro-ph/0403291}}].

\bibitem{deSalas:2015glj}
P.~F. de~Salas, M.~Lattanzi, G.~Mangano, G.~Miele, S.~Pastor and O.~Pisanti,
  \emph{{Bounds on very low reheating scenarios after Planck}},
  \href{https://doi.org/10.1103/PhysRevD.92.123534}{\emph{Phys. Rev.}
  {\bfseries D92} (2015) 123534},
  [\href{https://arxiv.org/abs/1511.00672}{{\ttfamily 1511.00672}}].

\bibitem{Mielczarek:2010ag}
J.~Mielczarek, \emph{{Reheating temperature from the CMB}},
  \href{https://doi.org/10.1103/PhysRevD.83.023502}{\emph{Phys. Rev.}
  {\bfseries D83} (2011) 023502},
  [\href{https://arxiv.org/abs/1009.2359}{{\ttfamily 1009.2359}}].

\bibitem{Dai:2014jja}
L.~Dai, M.~Kamionkowski and J.~Wang, \emph{{Reheating constraints to
  inflationary models}},
  \href{https://doi.org/10.1103/PhysRevLett.113.041302}{\emph{Phys. Rev. Lett.}
  {\bfseries 113} (2014) 041302},
  [\href{https://arxiv.org/abs/1404.6704}{{\ttfamily 1404.6704}}].

\bibitem{Munoz:2014eqa}
J.~B. Muñoz and M.~Kamionkowski, \emph{{Equation-of-State Parameter for
  Reheating}}, \href{https://doi.org/10.1103/PhysRevD.91.043521}{\emph{Phys.
  Rev.} {\bfseries D91} (2015) 043521},
  [\href{https://arxiv.org/abs/1412.0656}{{\ttfamily 1412.0656}}].

\bibitem{Domcke:2015iaa}
V.~Domcke and J.~Heisig, \emph{{Constraints on the reheating temperature from
  sizable tensor modes}},
  \href{https://doi.org/10.1103/PhysRevD.92.103515}{\emph{Phys. Rev.}
  {\bfseries D92} (2015) 103515},
  [\href{https://arxiv.org/abs/1504.00345}{{\ttfamily 1504.00345}}].

\bibitem{Drewes:2015coa}
M.~Drewes, \emph{{What can the CMB tell about the microphysics of cosmic
  reheating?}},
  \href{https://doi.org/10.1088/1475-7516/2016/03/013}{\emph{JCAP} {\bfseries
  1603} (2016) 013}, [\href{https://arxiv.org/abs/1511.03280}{{\ttfamily
  1511.03280}}].

\bibitem{Mukhanov:1982nu}
V.~F. Mukhanov and G.~V. Chibisov, \emph{{The Vacuum energy and large scale
  structure of the universe}}, {\emph{Sov. Phys. JETP} {\bfseries 56} (1982)
  258--265}.

\bibitem{Hawking:1982cz}
S.~W. Hawking, \emph{{The Development of Irregularities in a Single Bubble
  Inflationary Universe}},
  \href{https://doi.org/10.1016/0370-2693(82)90373-2}{\emph{Phys. Lett.}
  {\bfseries 115B} (1982) 295}.

\bibitem{Starobinsky:1982ee}
A.~A. Starobinsky, \emph{{Dynamics of Phase Transition in the New Inflationary
  Universe Scenario and Generation of Perturbations}},
  \href{https://doi.org/10.1016/0370-2693(82)90541-X}{\emph{Phys. Lett.}
  {\bfseries 117B} (1982) 175--178}.

\bibitem{Guth:1982ec}
A.~H. Guth and S.~Y. Pi, \emph{{Fluctuations in the New Inflationary
  Universe}}, \href{https://doi.org/10.1103/PhysRevLett.49.1110}{\emph{Phys.
  Rev. Lett.} {\bfseries 49} (1982) 1110--1113}.

\bibitem{Bardeen:1983qw}
J.~M. Bardeen, P.~J. Steinhardt and M.~S. Turner, \emph{{Spontaneous Creation
  of Almost Scale - Free Density Perturbations in an Inflationary Universe}},
  \href{https://doi.org/10.1103/PhysRevD.28.679}{\emph{Phys. Rev.} {\bfseries
  D28} (1983) 679}.

\bibitem{Mukhanov:1988jd}
V.~F. Mukhanov, \emph{{Quantum Theory of Gauge Invariant Cosmological
  Perturbations}}, {\emph{Sov. Phys. JETP} {\bfseries 67} (1988) 1297--1302}.

\bibitem{Mukhanov:1990me}
V.~F. Mukhanov, H.~A. Feldman and R.~H. Brandenberger, \emph{{Theory of
  cosmological perturbations. Part 1. Classical perturbations. Part 2. Quantum
  theory of perturbations. Part 3. Extensions}},
  \href{https://doi.org/10.1016/0370-1573(92)90044-Z}{\emph{Phys. Rept.}
  {\bfseries 215} (1992) 203--333}.

\bibitem{Mukhanov:2013tua}
V.~Mukhanov, \emph{{Quantum Cosmological Perturbations: Predictions and
  Observations}},
  \href{https://doi.org/10.1140/epjc/s10052-013-2486-7}{\emph{Eur. Phys. J.}
  {\bfseries C73} (2013) 2486},
  [\href{https://arxiv.org/abs/1303.3925}{{\ttfamily 1303.3925}}].

\bibitem{Langlois:2010xc}
D.~Langlois, \emph{{Lectures on inflation and cosmological perturbations}},
  \href{https://doi.org/10.1007/978-3-642-10598-2_1}{\emph{Lect. Notes Phys.}
  {\bfseries 800} (2010) 1--57},
  [\href{https://arxiv.org/abs/1001.5259}{{\ttfamily 1001.5259}}].

\bibitem{Okada:2014lxa}
N.~Okada, V.~N. Şenoğuz and Q.~Shafi, \emph{{The Observational Status of
  Simple Inflationary Models: an Update}},
  \href{https://doi.org/10.3906/fiz-1505-7}{\emph{Turk. J. Phys.} {\bfseries
  40} (2016) 150--162}, [\href{https://arxiv.org/abs/1403.6403}{{\ttfamily
  1403.6403}}].

\bibitem{Martin:2015dha}
J.~Martin, \emph{{The Observational Status of Cosmic Inflation after Planck}},
  \href{https://doi.org/10.1007/978-3-319-44769-8_2}{\emph{Astrophys. Space
  Sci. Proc.} {\bfseries 45} (2016) 41--134},
  [\href{https://arxiv.org/abs/1502.05733}{{\ttfamily 1502.05733}}].

\bibitem{Huang:2015cke}
Q.-G. Huang, K.~Wang and S.~Wang, \emph{{Inflation model constraints from data
  released in 2015}},
  \href{https://doi.org/10.1103/PhysRevD.93.103516}{\emph{Phys. Rev.}
  {\bfseries D93} (2016) 103516},
  [\href{https://arxiv.org/abs/1512.07769}{{\ttfamily 1512.07769}}].

\bibitem{Heavens:2017hkr}
A.~Heavens, Y.~Fantaye, E.~Sellentin, H.~Eggers, Z.~Hosenie, S.~Kroon et~al.,
  \emph{{No evidence for extensions to the standard cosmological model}},
  \href{https://doi.org/10.1103/PhysRevLett.119.101301}{\emph{Phys. Rev. Lett.}
  {\bfseries 119} (2017) 101301},
  [\href{https://arxiv.org/abs/1704.03467}{{\ttfamily 1704.03467}}].

\bibitem{DiValentino:2015ola}
E.~Di~Valentino, A.~Melchiorri and J.~Silk, \emph{{Beyond six parameters:
  extending $\Lambda$CDM}},
  \href{https://doi.org/10.1103/PhysRevD.92.121302}{\emph{Phys. Rev.}
  {\bfseries D92} (2015) 121302},
  [\href{https://arxiv.org/abs/1507.06646}{{\ttfamily 1507.06646}}].

\bibitem{Giusarma:2011zq}
E.~Giusarma, M.~Archidiacono, R.~de~Putter, A.~Melchiorri and O.~Mena,
  \emph{{Sterile neutrino models and nonminimal cosmologies}},
  \href{https://doi.org/10.1103/PhysRevD.85.083522}{\emph{Phys. Rev.}
  {\bfseries D85} (2012) 083522},
  [\href{https://arxiv.org/abs/1112.4661}{{\ttfamily 1112.4661}}].

\bibitem{DiValentino:2012yg}
E.~Di~Valentino, A.~Melchiorri, V.~Salvatelli and A.~Silvestri,
  \emph{{Parametrised modified gravity and the CMB Bispectrum}},
  \href{https://doi.org/10.1103/PhysRevD.86.063517}{\emph{Phys. Rev.}
  {\bfseries D86} (2012) 063517},
  [\href{https://arxiv.org/abs/1204.5352}{{\ttfamily 1204.5352}}].

\bibitem{Archidiacono:2012gv}
M.~Archidiacono, E.~Giusarma, A.~Melchiorri and O.~Mena, \emph{{Dark Radiation
  in extended cosmological scenarios}},
  \href{https://doi.org/10.1103/PhysRevD.86.043509}{\emph{Phys. Rev.}
  {\bfseries D86} (2012) 043509},
  [\href{https://arxiv.org/abs/1206.0109}{{\ttfamily 1206.0109}}].

\bibitem{Benetti:2013wla}
M.~Benetti, M.~Gerbino, W.~H. Kinney, E.~W. Kolb, M.~Lattanzi, A.~Melchiorri
  et~al., \emph{{Cosmological data and indications for new physics}},
  \href{https://doi.org/10.1088/1475-7516/2013/10/030}{\emph{JCAP} {\bfseries
  1310} (2013) 030}, [\href{https://arxiv.org/abs/1303.4317}{{\ttfamily
  1303.4317}}].

\bibitem{Said:2013hta}
N.~Said, E.~Di~Valentino and M.~Gerbino, \emph{{Planck constraints on the
  effective neutrino number and the CMB power spectrum lensing amplitude}},
  \href{https://doi.org/10.1103/PhysRevD.88.023513}{\emph{Phys. Rev.}
  {\bfseries D88} (2013) 023513},
  [\href{https://arxiv.org/abs/1304.6217}{{\ttfamily 1304.6217}}].

\bibitem{Gerbino:2013ova}
M.~Gerbino, E.~Di~Valentino and N.~Said, \emph{{Neutrino Anisotropies after
  Planck}}, \href{https://doi.org/10.1103/PhysRevD.88.063538}{\emph{Phys. Rev.}
  {\bfseries D88} (2013) 063538},
  [\href{https://arxiv.org/abs/1304.7400}{{\ttfamily 1304.7400}}].

\bibitem{Gerbino:2014eqa}
M.~Gerbino, A.~Marchini, L.~Pagano, L.~Salvati, E.~Di~Valentino and
  A.~Melchiorri, \emph{{Blue gravity waves from BICEP2?}},
  \href{https://doi.org/10.1103/PhysRevD.90.047301}{\emph{Phys. Rev.}
  {\bfseries D90} (2014) 047301},
  [\href{https://arxiv.org/abs/1403.5732}{{\ttfamily 1403.5732}}].

\bibitem{Cabass:2015jwe}
G.~Cabass, L.~Pagano, L.~Salvati, M.~Gerbino, E.~Giusarma and A.~Melchiorri,
  \emph{{Updated Constraints and Forecasts on Primordial Tensor Modes}},
  \href{https://doi.org/10.1103/PhysRevD.93.063508}{\emph{Phys. Rev.}
  {\bfseries D93} (2016) 063508},
  [\href{https://arxiv.org/abs/1511.05146}{{\ttfamily 1511.05146}}].

\bibitem{Cabass:2016ldu}
G.~Cabass, E.~Di~Valentino, A.~Melchiorri, E.~Pajer and J.~Silk,
  \emph{{Constraints on the running of the running of the scalar tilt from CMB
  anisotropies and spectral distortions}},
  \href{https://doi.org/10.1103/PhysRevD.94.023523}{\emph{Phys. Rev.}
  {\bfseries D94} (2016) 023523},
  [\href{https://arxiv.org/abs/1605.00209}{{\ttfamily 1605.00209}}].

\bibitem{DiValentino:2016ucb}
E.~Di~Valentino and F.~R. Bouchet, \emph{{A comment on power-law inflation with
  a dark radiation component}},
  \href{https://doi.org/10.1088/1475-7516/2016/10/011}{\emph{JCAP} {\bfseries
  1610} (2016) 011}, [\href{https://arxiv.org/abs/1609.00328}{{\ttfamily
  1609.00328}}].

\bibitem{DiValentino:2017zyq}
E.~Di~Valentino, A.~Melchiorri, E.~V. Linder and J.~Silk, \emph{{Constraining
  Dark Energy Dynamics in Extended Parameter Space}},
  \href{https://doi.org/10.1103/PhysRevD.96.023523}{\emph{Phys. Rev.}
  {\bfseries D96} (2017) 023523},
  [\href{https://arxiv.org/abs/1704.00762}{{\ttfamily 1704.00762}}].

\bibitem{Capparelli:2017tyx}
L.~Capparelli, E.~Di~Valentino, A.~Melchiorri and J.~Chluba, \emph{{Impact of
  theoretical assumptions in the determination of the neutrino effective number
  from future CMB measurements}},
  \href{https://doi.org/10.1103/PhysRevD.97.063519}{\emph{Phys. Rev.}
  {\bfseries D97} (2018) 063519},
  [\href{https://arxiv.org/abs/1712.06965}{{\ttfamily 1712.06965}}].

\bibitem{DiValentino:2018zjj}
E.~Di~Valentino, A.~Melchiorri, Y.~Fantaye and A.~Heavens, \emph{{Bayesian
  evidence against the Harrison-Zel’dovich spectrum in tensions with
  cosmological data sets}},
  \href{https://doi.org/10.1103/PhysRevD.98.063508}{\emph{Phys. Rev.}
  {\bfseries D98} (2018) 063508},
  [\href{https://arxiv.org/abs/1808.09201}{{\ttfamily 1808.09201}}].

\bibitem{Yang:2018prh}
W.~Yang, S.~Pan, E.~Di~Valentino and E.~N. Saridakis, \emph{{Observational
  constraints on dynamical dark energy with pivoting redshift}},
  \href{https://arxiv.org/abs/1811.06932}{{\ttfamily 1811.06932}}.

\bibitem{Pan:2019brc}
S.~Pan, W.~Yang and A.~Paliathanasis, \emph{{Imprints of an extended
  Chevallier-Polarski-Linder parametrization on the large scales}},
  \href{https://arxiv.org/abs/1902.07108}{{\ttfamily 1902.07108}}.

\bibitem{deBernardis:2000sbo}
{\scshape Boomerang} collaboration, P.~de~Bernardis et~al., \emph{{A Flat
  universe from high resolution maps of the cosmic microwave background
  radiation}}, \href{https://doi.org/10.1038/35010035}{\emph{Nature} {\bfseries
  404} (2000) 955--959},
  [\href{https://arxiv.org/abs/astro-ph/0004404}{{\ttfamily
  astro-ph/0004404}}].

\bibitem{Steinhardt:1999nw}
P.~J. Steinhardt, L.-M. Wang and I.~Zlatev, \emph{{Cosmological tracking
  solutions}}, \href{https://doi.org/10.1103/PhysRevD.59.123504}{\emph{Phys.
  Rev.} {\bfseries D59} (1999) 123504},
  [\href{https://arxiv.org/abs/astro-ph/9812313}{{\ttfamily
  astro-ph/9812313}}].

\bibitem{Vilenkin:2001bs}
A.~Vilenkin, \emph{{Cosmological constant problems and their solutions}},  in
  \emph{{8th International Symposium on Particles Strings and Cosmology (PASCOS
  2001) Chapel Hill, North Carolina, April 10-15, 2001}}, pp.~173--182, 2001,
  \href{https://arxiv.org/abs/hep-th/0106083}{{\ttfamily hep-th/0106083}}.

\bibitem{Velten:2014nra}
H.~E.~S. Velten, R.~F. von Marttens and W.~Zimdahl, \emph{{Aspects of the
  cosmological ``coincidence problem''}},
  \href{https://doi.org/10.1140/epjc/s10052-014-3160-4}{\emph{Eur. Phys. J.}
  {\bfseries C74} (2014) 3160},
  [\href{https://arxiv.org/abs/1410.2509}{{\ttfamily 1410.2509}}].

\bibitem{Maltoni:2004ei}
M.~Maltoni, T.~Schwetz, M.~A. Tórtola and J.~W.~F. Valle, \emph{{Status of
  global fits to neutrino oscillations}},
  \href{https://doi.org/10.1088/1367-2630/6/1/122}{\emph{New J. Phys.}
  {\bfseries 6} (2004) 122},
  [\href{https://arxiv.org/abs/hep-ph/0405172}{{\ttfamily hep-ph/0405172}}].

\bibitem{Bilenky:1978nj}
S.~M. Bilenky and B.~Pontecorvo, \emph{{Lepton Mixing and Neutrino
  Oscillations}},
  \href{https://doi.org/10.1016/0370-1573(78)90095-9}{\emph{Phys. Rept.}
  {\bfseries 41} (1978) 225--261}.

\bibitem{Bilenky:1987ty}
S.~M. Bilenky and S.~T. Petcov, \emph{{Massive Neutrinos and Neutrino
  Oscillations}}, \href{https://doi.org/10.1103/RevModPhys.59.671}{\emph{Rev.
  Mod. Phys.} {\bfseries 59} (1987) 671}. [Erratum: Rev. Mod.
  Phys.60,575(1988)].

\bibitem{McKeown:2004yq}
R.~D. McKeown and P.~Vogel, \emph{{Neutrino masses and oscillations: Triumphs
  and challenges}},
  \href{https://doi.org/10.1016/j.physrep.2004.01.003}{\emph{Phys. Rept.}
  {\bfseries 394} (2004) 315--356},
  [\href{https://arxiv.org/abs/hep-ph/0402025}{{\ttfamily hep-ph/0402025}}].

\bibitem{Visinelli:2008ds}
L.~Visinelli and P.~Gondolo, \emph{{Neutrino Oscillations and Decoherence}},
  \href{https://arxiv.org/abs/0810.4132}{{\ttfamily 0810.4132}}.

\bibitem{Hernandez:2010mi}
P.~Hernández, \emph{{Neutrino physics}},  in \emph{{High-energy physics.
  Proceedings, 5th CERN-Latin-American School, Recinto Quirama, Colombia, March
  15-28, 2009}}, 2010, \href{https://arxiv.org/abs/1010.4131}{{\ttfamily
  1010.4131}}.

\bibitem{Visinelli:2014xsa}
L.~Visinelli, \emph{{Neutrino flavor oscillations in a curved space-time}},
  \href{https://doi.org/10.1007/s10714-015-1899-z}{\emph{Gen. Rel. Grav.}
  {\bfseries 47} (2015) 62}, [\href{https://arxiv.org/abs/1410.1523}{{\ttfamily
  1410.1523}}].

\bibitem{Pontecorvo:1957qd}
B.~Pontecorvo, \emph{{Inverse beta processes and nonconservation of lepton
  charge}}, {\emph{Sov. Phys. JETP} {\bfseries 7} (1958) 172--173}.

\bibitem{Maki:1962mu}
Z.~Maki, M.~Nakagawa and S.~Sakata, \emph{{Remarks on the unified model of
  elementary particles}}, \href{https://doi.org/10.1143/PTP.28.870}{\emph{Prog.
  Theor. Phys.} {\bfseries 28} (1962) 870--880}.

\bibitem{Fogli:2005cq}
G.~L. Fogli, E.~Lisi, A.~Marrone and A.~Palazzo, \emph{{Global analysis of
  three-flavor neutrino masses and mixings}},
  \href{https://doi.org/10.1016/j.ppnp.2005.08.002}{\emph{Prog. Part. Nucl.
  Phys.} {\bfseries 57} (2006) 742--795},
  [\href{https://arxiv.org/abs/hep-ph/0506083}{{\ttfamily hep-ph/0506083}}].

\bibitem{Giganti:2017fhf}
C.~Giganti, S.~Lavignac and M.~Zito, \emph{{Neutrino oscillations: the rise of
  the PMNS paradigm}},
  \href{https://doi.org/10.1016/j.ppnp.2017.10.001}{\emph{Prog. Part. Nucl.
  Phys.} {\bfseries 98} (2018) 1--54},
  [\href{https://arxiv.org/abs/1710.00715}{{\ttfamily 1710.00715}}].

\bibitem{Pontecorvo:1957cp}
B.~Pontecorvo, \emph{{Mesonium and anti-mesonium}}, {\emph{Sov. Phys. JETP}
  {\bfseries 6} (1957) 429}.

\bibitem{Pontecorvo:1967fh}
B.~Pontecorvo, \emph{{Neutrino Experiments and the Problem of Conservation of
  Leptonic Charge}}, {\emph{Sov. Phys. JETP} {\bfseries 26} (1968) 984--988}.

\bibitem{Grossman:2003eb}
Y.~Grossman, \emph{{TASI 2002 lectures on neutrinos}},  in \emph{{Particle
  physics and cosmology: The quest for physics beyond the standard model(s).
  Proceedings, Theoretical Advanced Study Institute, TASI 2002, Boulder, USA,
  June 3-28, 2002}}, pp.~5--48, 2003,
  \href{https://arxiv.org/abs/hep-ph/0305245}{{\ttfamily hep-ph/0305245}}.

\bibitem{deGouvea:2004gd}
A.~de~Gouvêa, \emph{{TASI lectures on neutrino physics}},  in \emph{{Physics
  in D >= 4. Proceedings, Theoretical Advanced Study Institute in elementary
  particle physics, TASI 2004, Boulder, USA, June 6-July 2, 2004}},
  pp.~197--258, 2004, \href{https://arxiv.org/abs/hep-ph/0411274}{{\ttfamily
  hep-ph/0411274}}.

\bibitem{Kayser:2012ce}
B.~Kayser, \emph{{Neutrino Oscillation Physics}},  in \emph{{Proceedings, 2011
  European School of High-Energy Physics (ESHEP 2011): Cheile Gradistei,
  Romania, September 7-20, 2011}}, pp.~107--117, 2014,
  \href{https://arxiv.org/abs/1206.4325}{{\ttfamily 1206.4325}}.

\bibitem{Fantini:2018itu}
G.~Fantini, A.~Gallo~Rosso, F.~Vissani and V.~Zema, \emph{{Introduction to the
  Formalism of Neutrino Oscillations}},
  \href{https://doi.org/10.1142/9789813226098_0002}{\emph{Adv. Ser. Direct.
  High Energy Phys.} {\bfseries 28} (2018) 37--119},
  [\href{https://arxiv.org/abs/1802.05781}{{\ttfamily 1802.05781}}].

\bibitem{Juno:2017ghw}
{JUNO Collaboration}, \emph{http://www.staff.uni-mainz.de/wurmm/juno.html},
  2017.

\bibitem{Robertson:2012ib}
W.~C. Haxton, R.~G. Hamish~Robertson and A.~M. Serenelli, \emph{{Solar
  Neutrinos: Status and Prospects}},
  \href{https://doi.org/10.1146/annurev-astro-081811-125539}{\emph{Ann. Rev.
  Astron. Astrophys.} {\bfseries 51} (2013) 21--61},
  [\href{https://arxiv.org/abs/1208.5723}{{\ttfamily 1208.5723}}].

\bibitem{Gann:2015yta}
G.~D. Orebi~Gann, \emph{{Everything Under the Sun: A Review of Solar
  Neutrinos}},  \href{https://arxiv.org/abs/1504.02154}{{\ttfamily
  1504.02154}}.

\bibitem{Vissani:2017dto}
F.~Vissani, \emph{{Solar neutrino physics on the beginning of 2017}},
  \href{https://doi.org/10.15407/jnpae2017.01.005}{\emph{Nucl. Phys. Atom.
  Energy} {\bfseries 18} (2017) 5--12},
  [\href{https://arxiv.org/abs/1706.05435}{{\ttfamily 1706.05435}}].

\bibitem{Davis:1968cp}
R.~Davis, Jr., D.~S. Harmer and K.~C. Hoffman, \emph{{Search for neutrinos from
  the sun}}, \href{https://doi.org/10.1103/PhysRevLett.20.1205}{\emph{Phys.
  Rev. Lett.} {\bfseries 20} (1968) 1205--1209}.

\bibitem{Davis:1994jw}
R.~Davis, \emph{{A review of the Homestake solar neutrino experiment}},
  \href{https://doi.org/10.1016/0146-6410(94)90004-3}{\emph{Prog. Part. Nucl.
  Phys.} {\bfseries 32} (1994) 13--32}.

\bibitem{Bahcall:1976zz}
J.~N. Bahcall and R.~Davis, \emph{{Solar Neutrinos - a Scientific Puzzle}},
  \href{https://doi.org/10.1126/science.191.4224.264}{\emph{Science} {\bfseries
  191} (1976) 264--267}.

\bibitem{Haxton:1995hv}
W.~C. Haxton, \emph{{The solar neutrino problem}},
  \href{https://doi.org/10.1146/annurev.aa.33.090195.002331}{\emph{Ann. Rev.
  Astron. Astrophys.} {\bfseries 33} (1995) 459--503},
  [\href{https://arxiv.org/abs/hep-ph/9503430}{{\ttfamily hep-ph/9503430}}].

\bibitem{Ahmad:2002jz}
{\scshape SNO} collaboration, Q.~R. Ahmad et~al., \emph{{Direct evidence for
  neutrino flavor transformation from neutral current interactions in the
  Sudbury Neutrino Observatory}},
  \href{https://doi.org/10.1103/PhysRevLett.89.011301}{\emph{Phys. Rev. Lett.}
  {\bfseries 89} (2002) 011301},
  [\href{https://arxiv.org/abs/nucl-ex/0204008}{{\ttfamily nucl-ex/0204008}}].

\bibitem{Asplund:2004eu}
M.~Asplund, N.~Grevesse and J.~Sauval, \emph{{The Solar chemical composition}},
  \href{https://doi.org/10.1016/j.nuclphysa.2005.06.010}{\emph{Nucl. Phys.}
  {\bfseries A777} (2006) 1--4},
  [\href{https://arxiv.org/abs/astro-ph/0410214}{{\ttfamily
  astro-ph/0410214}}].

\bibitem{Asplund:2009fu}
M.~Asplund, N.~Grevesse, A.~J. Sauval and P.~Scott, \emph{{The chemical
  composition of the Sun}},
  \href{https://doi.org/10.1146/annurev.astro.46.060407.145222}{\emph{Ann. Rev.
  Astron. Astrophys.} {\bfseries 47} (2009) 481--522},
  [\href{https://arxiv.org/abs/0909.0948}{{\ttfamily 0909.0948}}].

\bibitem{Serenelli:2009yc}
A.~Serenelli, S.~Basu, J.~W. Ferguson and M.~Asplund, \emph{{New Solar
  Composition: The Problem With Solar Models Revisited}},
  \href{https://doi.org/10.1088/0004-637X/705/2/L123}{\emph{Astrophys. J.}
  {\bfseries 705} (2009) L123--L127},
  [\href{https://arxiv.org/abs/0909.2668}{{\ttfamily 0909.2668}}].

\bibitem{Frandsen:2010yj}
M.~T. Frandsen and S.~Sarkar, \emph{{Asymmetric dark matter and the Sun}},
  \href{https://doi.org/10.1103/PhysRevLett.105.011301}{\emph{Phys. Rev. Lett.}
  {\bfseries 105} (2010) 011301},
  [\href{https://arxiv.org/abs/1003.4505}{{\ttfamily 1003.4505}}].

\bibitem{Vagnozzi:2016cmr}
S.~Vagnozzi, K.~Freese and T.~H. Zurbuchen, \emph{{Solar models in light of new
  high metallicity measurements from solar wind data}},
  \href{https://doi.org/10.3847/1538-4357/aa6931}{\emph{Astrophys. J.}
  {\bfseries 839} (2017) 55},
  [\href{https://arxiv.org/abs/1603.05960}{{\ttfamily 1603.05960}}].

\bibitem{Vagnozzi:2017wge}
S.~Vagnozzi, \emph{{New solar metallicity measurements}},
  \href{https://doi.org/10.3390/atoms7020041}{\emph{Atoms} {\bfseries 7} (2019)
  41}, [\href{https://arxiv.org/abs/1703.10834}{{\ttfamily 1703.10834}}].

\bibitem{Kajita:2012vc}
T.~Kajita, \emph{{Atmospheric neutrinos}},
  \href{https://doi.org/10.1155/2012/504715}{\emph{Adv. High Energy Phys.}
  {\bfseries 2012} (2012) 504715}.

\bibitem{Choubey:2016gps}
S.~Choubey, \emph{{Atmospheric Neutrinos: Status and Prospects}},
  \href{https://doi.org/10.1016/j.nuclphysb.2016.03.026}{\emph{Nucl. Phys.}
  {\bfseries B908} (2016) 235--249},
  [\href{https://arxiv.org/abs/1603.06841}{{\ttfamily 1603.06841}}].

\bibitem{Kajita:2016cak}
T.~Kajita, \emph{{Nobel Lecture: Discovery of atmospheric neutrino
  oscillations}},
  \href{https://doi.org/10.1103/RevModPhys.88.030501}{\emph{Rev. Mod. Phys.}
  {\bfseries 88} (2016) 030501}.

\bibitem{McDonald:2016ixn}
A.~B. McDonald, \emph{{Nobel Lecture: The Sudbury Neutrino Observatory:
  Observation of flavor change for solar neutrinos}},
  \href{https://doi.org/10.1103/RevModPhys.88.030502}{\emph{Rev. Mod. Phys.}
  {\bfseries 88} (2016) 030502}.

\bibitem{Eguchi:2002dm}
{\scshape KamLAND} collaboration, K.~Eguchi et~al., \emph{{First results from
  KamLAND: Evidence for reactor anti-neutrino disappearance}},
  \href{https://doi.org/10.1103/PhysRevLett.90.021802}{\emph{Phys. Rev. Lett.}
  {\bfseries 90} (2003) 021802},
  [\href{https://arxiv.org/abs/hep-ex/0212021}{{\ttfamily hep-ex/0212021}}].

\bibitem{Qian:2015waa}
X.~Qian and P.~Vogel, \emph{{Neutrino Mass Hierarchy}},
  \href{https://doi.org/10.1016/j.ppnp.2015.05.002}{\emph{Prog. Part. Nucl.
  Phys.} {\bfseries 83} (2015) 1--30},
  [\href{https://arxiv.org/abs/1505.01891}{{\ttfamily 1505.01891}}].

\bibitem{Adams:2013qkq}
{\scshape LBNE} collaboration, C.~Adams et~al., \emph{{The Long-Baseline
  Neutrino Experiment: Exploring Fundamental Symmetries of the Universe}},  in
  \emph{{Snowmass 2013: Workshop on Energy Frontier Seattle, USA, June 30-July
  3, 2013}}, 2013, \href{https://arxiv.org/abs/1307.7335}{{\ttfamily
  1307.7335}}.

\bibitem{Acciarri:2015uup}
{\scshape DUNE} collaboration, R.~Acciarri et~al., \emph{{Long-Baseline
  Neutrino Facility (LBNF) and Deep Underground Neutrino Experiment (DUNE)}},
  \href{https://arxiv.org/abs/1512.06148}{{\ttfamily 1512.06148}}.

\bibitem{Acciarri:2016ooe}
{\scshape DUNE} collaboration, R.~Acciarri et~al., \emph{{Long-Baseline
  Neutrino Facility (LBNF) and Deep Underground Neutrino Experiment (DUNE)}},
  \href{https://arxiv.org/abs/1601.02984}{{\ttfamily 1601.02984}}.

\bibitem{Acciarri:2016crz}
{\scshape DUNE} collaboration, R.~Acciarri et~al., \emph{{Long-Baseline
  Neutrino Facility (LBNF) and Deep Underground Neutrino Experiment (DUNE)}},
  \href{https://arxiv.org/abs/1601.05471}{{\ttfamily 1601.05471}}.

\bibitem{Strait:2016mof}
{\scshape DUNE} collaboration, J.~Strait et~al., \emph{{Long-Baseline Neutrino
  Facility (LBNF) and Deep Underground Neutrino Experiment (DUNE)}},
  \href{https://arxiv.org/abs/1601.05823}{{\ttfamily 1601.05823}}.

\bibitem{Wolfenstein:1977ue}
L.~Wolfenstein, \emph{{Neutrino Oscillations in Matter}},
  \href{https://doi.org/10.1103/PhysRevD.17.2369}{\emph{Phys. Rev.} {\bfseries
  D17} (1978) 2369--2374}.

\bibitem{Mikheev:1986gs}
S.~P. Mikheyev and A.~{\relax Yu}. Smirnov, \emph{{Resonance Amplification of
  Oscillations in Matter and Spectroscopy of Solar Neutrinos}}, {\emph{Sov. J.
  Nucl. Phys.} {\bfseries 42} (1985) 913--917}.

\bibitem{Mikheyev:1989dy}
S.~P. Mikheyev and A.~{\relax Yu}. Smirnov, \emph{{Resonant neutrino
  oscillations in matter}},
  \href{https://doi.org/10.1016/0146-6410(89)90008-2}{\emph{Prog. Part. Nucl.
  Phys.} {\bfseries 23} (1989) 41--136}.

\bibitem{Palanque-Delabrouille:2015pga}
N.~Palanque-Delabrouille et~al., \emph{{Neutrino masses and cosmology with
  Lyman-alpha forest power spectrum}},
  \href{https://doi.org/10.1088/1475-7516/2015/11/011}{\emph{JCAP} {\bfseries
  1511} (2015) 011}, [\href{https://arxiv.org/abs/1506.05976}{{\ttfamily
  1506.05976}}].

\bibitem{Lattanzi:2017ubx}
M.~Lattanzi and M.~Gerbino, \emph{{Status of neutrino properties and future
  prospects - Cosmological and astrophysical constraints}},
  \href{https://doi.org/10.3389/fphy.2017.00070}{\emph{Front.in Phys.}
  {\bfseries 5} (2018) 70}, [\href{https://arxiv.org/abs/1712.07109}{{\ttfamily
  1712.07109}}].

\bibitem{Foot:1991bp}
R.~Foot, H.~Lew and R.~R. Volkas, \emph{{A Model with fundamental improper
  space-time symmetries}},
  \href{https://doi.org/10.1016/0370-2693(91)91013-L}{\emph{Phys. Lett.}
  {\bfseries B272} (1991) 67--70}.

\bibitem{Ackerman:2008mha}
L.~Ackerman, M.~R. Buckley, S.~M. Carroll and M.~Kamionkowski, \emph{{Dark
  Matter and Dark Radiation}},
  \href{https://doi.org/10.1103/PhysRevD.79.023519,
  10.1142/9789814293792_0021}{\emph{Phys. Rev.} {\bfseries D79} (2009) 023519},
  [\href{https://arxiv.org/abs/0810.5126}{{\ttfamily 0810.5126}}].

\bibitem{Feng:2009mn}
J.~L. Feng, M.~Kaplinghat, H.~Tu and H.-B. Yu, \emph{{Hidden Charged Dark
  Matter}}, \href{https://doi.org/10.1088/1475-7516/2009/07/004}{\emph{JCAP}
  {\bfseries 0907} (2009) 004},
  [\href{https://arxiv.org/abs/0905.3039}{{\ttfamily 0905.3039}}].

\bibitem{Kaplan:2009de}
D.~E. Kaplan, G.~Z. Krnjaic, K.~R. Rehermann and C.~M. Wells, \emph{{Atomic
  Dark Matter}},
  \href{https://doi.org/10.1088/1475-7516/2010/05/021}{\emph{JCAP} {\bfseries
  1005} (2010) 021}, [\href{https://arxiv.org/abs/0909.0753}{{\ttfamily
  0909.0753}}].

\bibitem{Archidiacono:2011gq}
M.~Archidiacono, E.~Calabrese and A.~Melchiorri, \emph{{The Case for Dark
  Radiation}}, \href{https://doi.org/10.1103/PhysRevD.84.123008}{\emph{Phys.
  Rev.} {\bfseries D84} (2011) 123008},
  [\href{https://arxiv.org/abs/1109.2767}{{\ttfamily 1109.2767}}].

\bibitem{Cline:2012is}
J.~M. Cline, Z.~Liu and W.~Xue, \emph{{Millicharged Atomic Dark Matter}},
  \href{https://doi.org/10.1103/PhysRevD.85.101302}{\emph{Phys. Rev.}
  {\bfseries D85} (2012) 101302},
  [\href{https://arxiv.org/abs/1201.4858}{{\ttfamily 1201.4858}}].

\bibitem{Blennow:2012de}
M.~Blennow, E.~Fernández-Martínez, O.~Mena, J.~Redondo and P.~Serra,
  \emph{{Asymmetric Dark Matter and Dark Radiation}},
  \href{https://doi.org/10.1088/1475-7516/2012/07/022}{\emph{JCAP} {\bfseries
  1207} (2012) 022}, [\href{https://arxiv.org/abs/1203.5803}{{\ttfamily
  1203.5803}}].

\bibitem{CyrRacine:2012fz}
F.-Y. Cyr-Racine and K.~Sigurdson, \emph{{Cosmology of atomic dark matter}},
  \href{https://doi.org/10.1103/PhysRevD.87.103515}{\emph{Phys. Rev.}
  {\bfseries D87} (2013) 103515},
  [\href{https://arxiv.org/abs/1209.5752}{{\ttfamily 1209.5752}}].

\bibitem{Fan:2013tia}
J.~Fan, A.~Katz, L.~Randall and M.~Reece, \emph{{Dark-Disk Universe}},
  \href{https://doi.org/10.1103/PhysRevLett.110.211302}{\emph{Phys. Rev. Lett.}
  {\bfseries 110} (2013) 211302},
  [\href{https://arxiv.org/abs/1303.3271}{{\ttfamily 1303.3271}}].

\bibitem{Conlon:2013isa}
J.~P. Conlon and M.~C.~D. Marsh, \emph{{The Cosmophenomenology of Axionic Dark
  Radiation}}, \href{https://doi.org/10.1007/JHEP10(2013)214}{\emph{JHEP}
  {\bfseries 10} (2013) 214},
  [\href{https://arxiv.org/abs/1304.1804}{{\ttfamily 1304.1804}}].

\bibitem{Vogel:2013raa}
H.~Vogel and J.~Redondo, \emph{{Dark Radiation constraints on minicharged
  particles in models with a hidden photon}},
  \href{https://doi.org/10.1088/1475-7516/2014/02/029}{\emph{JCAP} {\bfseries
  1402} (2014) 029}, [\href{https://arxiv.org/abs/1311.2600}{{\ttfamily
  1311.2600}}].

\bibitem{Fan:2013bea}
J.~Fan, A.~Katz and J.~Shelton, \emph{{Direct and indirect detection of
  dissipative dark matter}},
  \href{https://doi.org/10.1088/1475-7516/2014/06/059}{\emph{JCAP} {\bfseries
  1406} (2014) 059}, [\href{https://arxiv.org/abs/1312.1336}{{\ttfamily
  1312.1336}}].

\bibitem{Foot:2014mia}
R.~Foot, \emph{{Mirror dark matter: Cosmology, galaxy structure and direct
  detection}}, \href{https://doi.org/10.1142/S0217751X14300130}{\emph{Int. J.
  Mod. Phys.} {\bfseries A29} (2014) 1430013},
  [\href{https://arxiv.org/abs/1401.3965}{{\ttfamily 1401.3965}}].

\bibitem{Petraki:2014uza}
K.~Petraki, L.~Pearce and A.~Kusenko, \emph{{Self-interacting asymmetric dark
  matter coupled to a light massive dark photon}},
  \href{https://doi.org/10.1088/1475-7516/2014/07/039}{\emph{JCAP} {\bfseries
  1407} (2014) 039}, [\href{https://arxiv.org/abs/1403.1077}{{\ttfamily
  1403.1077}}].

\bibitem{Marsh:2014gca}
M.~C.~D. Marsh, \emph{{The Darkness of Spin-0 Dark Radiation}},
  \href{https://doi.org/10.1088/1475-7516/2015/01/017}{\emph{JCAP} {\bfseries
  1501} (2015) 017}, [\href{https://arxiv.org/abs/1407.2501}{{\ttfamily
  1407.2501}}].

\bibitem{Foot:2014uba}
R.~Foot and S.~Vagnozzi, \emph{{Dissipative hidden sector dark matter}},
  \href{https://doi.org/10.1103/PhysRevD.91.023512}{\emph{Phys. Rev.}
  {\bfseries D91} (2015) 023512},
  [\href{https://arxiv.org/abs/1409.7174}{{\ttfamily 1409.7174}}].

\bibitem{Foot:2014osa}
R.~Foot and S.~Vagnozzi, \emph{{Diurnal modulation signal from dissipative
  hidden sector dark matter}},
  \href{https://doi.org/10.1016/j.physletb.2015.06.063}{\emph{Phys. Lett.}
  {\bfseries B748} (2015) 61--66},
  [\href{https://arxiv.org/abs/1412.0762}{{\ttfamily 1412.0762}}].

\bibitem{Pearce:2015zca}
L.~Pearce, K.~Petraki and A.~Kusenko, \emph{{Signals from dark atom formation
  in halos}}, \href{https://doi.org/10.1103/PhysRevD.91.083532}{\emph{Phys.
  Rev.} {\bfseries D91} (2015) 083532},
  [\href{https://arxiv.org/abs/1502.01755}{{\ttfamily 1502.01755}}].

\bibitem{Heikinheimo:2015kra}
M.~Heikinheimo, M.~Raidal, C.~Spethmann and H.~Veermäe, \emph{{Dark matter
  self-interactions via collisionless shocks in cluster mergers}},
  \href{https://doi.org/10.1016/j.physletb.2015.08.012}{\emph{Phys. Lett.}
  {\bfseries B749} (2015) 236--241},
  [\href{https://arxiv.org/abs/1504.04371}{{\ttfamily 1504.04371}}].

\bibitem{Chacko:2015noa}
Z.~Chacko, Y.~Cui, S.~Hong and T.~Okui, \emph{{Hidden dark matter sector, dark
  radiation, and the CMB}},
  \href{https://doi.org/10.1103/PhysRevD.92.055033}{\emph{Phys. Rev.}
  {\bfseries D92} (2015) 055033},
  [\href{https://arxiv.org/abs/1505.04192}{{\ttfamily 1505.04192}}].

\bibitem{Clarke:2015gqw}
J.~D. Clarke and R.~Foot, \emph{{Plasma dark matter direct detection}},
  \href{https://doi.org/10.1088/1475-7516/2016/01/029}{\emph{JCAP} {\bfseries
  1601} (2016) 029}, [\href{https://arxiv.org/abs/1512.06471}{{\ttfamily
  1512.06471}}].

\bibitem{DiValentino:2016ikp}
E.~Di~Valentino, S.~Gariazzo, M.~Gerbino, E.~Giusarma and O.~Mena, \emph{{Dark
  Radiation and Inflationary Freedom after Planck 2015}},
  \href{https://doi.org/10.1103/PhysRevD.93.083523}{\emph{Phys. Rev.}
  {\bfseries D93} (2016) 083523},
  [\href{https://arxiv.org/abs/1601.07557}{{\ttfamily 1601.07557}}].

\bibitem{Foot:2016wvj}
R.~Foot and S.~Vagnozzi, \emph{{Solving the small-scale structure puzzles with
  dissipative dark matter}},
  \href{https://doi.org/10.1088/1475-7516/2016/07/013}{\emph{JCAP} {\bfseries
  1607} (2016) 013}, [\href{https://arxiv.org/abs/1602.02467}{{\ttfamily
  1602.02467}}].

\bibitem{Agrawal:2016quu}
P.~Agrawal, F.-Y. Cyr-Racine, L.~Randall and J.~Scholtz, \emph{{Make Dark
  Matter Charged Again}},
  \href{https://doi.org/10.1088/1475-7516/2017/05/022}{\emph{JCAP} {\bfseries
  1705} (2017) 022}, [\href{https://arxiv.org/abs/1610.04611}{{\ttfamily
  1610.04611}}].

\bibitem{Agrawal:2017rvu}
P.~Agrawal, F.-Y. Cyr-Racine, L.~Randall and J.~Scholtz, \emph{{Dark
  Catalysis}}, \href{https://doi.org/10.1088/1475-7516/2017/08/021}{\emph{JCAP}
  {\bfseries 1708} (2017) 021},
  [\href{https://arxiv.org/abs/1702.05482}{{\ttfamily 1702.05482}}].

\bibitem{Krall:2017xcw}
R.~Krall, F.-Y. Cyr-Racine and C.~Dvorkin, \emph{{Wandering in the Lyman-alpha
  Forest: A Study of Dark Matter-Dark Radiation Interactions}},
  \href{https://doi.org/10.1088/1475-7516/2017/09/003}{\emph{JCAP} {\bfseries
  1709} (2017) 003}, [\href{https://arxiv.org/abs/1705.08894}{{\ttfamily
  1705.08894}}].

\bibitem{Archidiacono:2017slj}
M.~Archidiacono, S.~Bohr, S.~Hannestad, J.~H. Jørgensen and J.~Lesgourgues,
  \emph{{Linear scale bounds on dark matter--dark radiation interactions and
  connection with the small scale crisis of cold dark matter}},
  \href{https://doi.org/10.1088/1475-7516/2017/11/010}{\emph{JCAP} {\bfseries
  1711} (2017) 010}, [\href{https://arxiv.org/abs/1706.06870}{{\ttfamily
  1706.06870}}].

\bibitem{Buen-Abad:2017gxg}
M.~A. Buen-Abad, M.~Schmaltz, J.~Lesgourgues and T.~Brinckmann,
  \emph{{Interacting Dark Sector and Precision Cosmology}},
  \href{https://doi.org/10.1088/1475-7516/2018/01/008}{\emph{JCAP} {\bfseries
  1801} (2018) 008}, [\href{https://arxiv.org/abs/1708.09406}{{\ttfamily
  1708.09406}}].

\bibitem{Baldes:2017gzu}
I.~Baldes, M.~Cirelli, P.~Panci, K.~Petraki, F.~Sala and M.~Taoso,
  \emph{{Asymmetric dark matter: residual annihilations and
  self-interactions}},
  \href{https://doi.org/10.21468/SciPostPhys.4.6.041}{\emph{SciPost Phys.}
  {\bfseries 4} (2018) 041},
  [\href{https://arxiv.org/abs/1712.07489}{{\ttfamily 1712.07489}}].

\bibitem{Davidson:2000dw}
S.~Davidson, M.~Losada and A.~Riotto, \emph{{A New perspective on
  baryogenesis}},
  \href{https://doi.org/10.1103/PhysRevLett.84.4284}{\emph{Phys. Rev. Lett.}
  {\bfseries 84} (2000) 4284--4287},
  [\href{https://arxiv.org/abs/hep-ph/0001301}{{\ttfamily hep-ph/0001301}}].

\bibitem{Gelmini:2004ah}
G.~Gelmini, S.~Palomares-Ruiz and S.~Pascoli, \emph{{Low reheating temperature
  and the visible sterile neutrino}},
  \href{https://doi.org/10.1103/PhysRevLett.93.081302}{\emph{Phys. Rev. Lett.}
  {\bfseries 93} (2004) 081302},
  [\href{https://arxiv.org/abs/astro-ph/0403323}{{\ttfamily
  astro-ph/0403323}}].

\bibitem{Ichikawa:2005vw}
K.~Ichikawa, M.~Kawasaki and F.~Takahashi, \emph{{The Oscillation effects on
  thermalization of the neutrinos in the Universe with low reheating
  temperature}}, \href{https://doi.org/10.1103/PhysRevD.72.043522}{\emph{Phys.
  Rev.} {\bfseries D72} (2005) 043522},
  [\href{https://arxiv.org/abs/astro-ph/0505395}{{\ttfamily
  astro-ph/0505395}}].

\bibitem{Visinelli:2009kt}
L.~Visinelli and P.~Gondolo, \emph{{Axion cold dark matter in non-standard
  cosmologies}}, \href{https://doi.org/10.1103/PhysRevD.81.063508}{\emph{Phys.
  Rev.} {\bfseries D81} (2010) 063508},
  [\href{https://arxiv.org/abs/0912.0015}{{\ttfamily 0912.0015}}].

\bibitem{Jeans:1902ghw}
J.~H. {Jeans}, \emph{{The Stability of a Spherical Nebula}},
  \href{https://doi.org/10.1098/rsta.1902.0012}{\emph{Philosophical
  Transactions of the Royal Society of London Series A} {\bfseries 199} (1902)
  1--53}.

\bibitem{Weinberg:1972kfs}
S.~Weinberg, \emph{{Gravitation and Cosmology}}.
\newblock John Wiley and Sons, New York, 1972.

\bibitem{Peebles:1994xt}
P.~J.~E. Peebles, \emph{{Principles of physical cosmology}}.
\newblock 1994.

\bibitem{RowanRobinson:1996nv}
M.~Rowan-Robinson, \emph{{Cosmology}}.
\newblock 1996.

\bibitem{Liddle:1998ew}
A.~R. Liddle, \emph{{An introduction to modern cosmology}}.
\newblock 1998.

\bibitem{Peacock:1999ye}
J.~A. Peacock, \emph{{Cosmological physics}}.
\newblock 1999.

\bibitem{Giovannini:2008zzb}
M.~Giovannini, \emph{{A primer on the physics of the cosmic microwave
  background}}.
\newblock 2008.

\bibitem{Bernardeau:2001qr}
F.~Bernardeau, S.~Colombi, E.~Gaztañaga and R.~Scoccimarro, \emph{{Large scale
  structure of the universe and cosmological perturbation theory}},
  \href{https://doi.org/10.1016/S0370-1573(02)00135-7}{\emph{Phys. Rept.}
  {\bfseries 367} (2002) 1--248},
  [\href{https://arxiv.org/abs/astro-ph/0112551}{{\ttfamily
  astro-ph/0112551}}].

\bibitem{Percival:2013awa}
W.~J. Percival, \emph{{Large Scale Structure Observations}},  in
  \emph{{Proceedings, International School of Physics 'Enrico Fermi': New
  Horizons for Observational Cosmology: Rome, Italy, June 30-July 6, 2013}},
  vol.~186, pp.~101--135, 2015,
  \href{https://arxiv.org/abs/1312.5490}{{\ttfamily 1312.5490}}.

\bibitem{Hu:2001bc}
W.~Hu and S.~Dodelson, \emph{{Cosmic microwave background anisotropies}},
  \href{https://doi.org/10.1146/annurev.astro.40.060401.093926}{\emph{Ann. Rev.
  Astron. Astrophys.} {\bfseries 40} (2002) 171--216},
  [\href{https://arxiv.org/abs/astro-ph/0110414}{{\ttfamily
  astro-ph/0110414}}].

\bibitem{Samtleben:2007zz}
D.~Samtleben, S.~Staggs and B.~Winstein, \emph{{The Cosmic microwave background
  for pedestrians: A Review for particle and nuclear physicists}},
  \href{https://doi.org/10.1146/annurev.nucl.54.070103.181232}{\emph{Ann. Rev.
  Nucl. Part. Sci.} {\bfseries 57} (2007) 245--283},
  [\href{https://arxiv.org/abs/0803.0834}{{\ttfamily 0803.0834}}].

\bibitem{Bucher:2015eia}
M.~Bucher, \emph{{Physics of the cosmic microwave background anisotropy}},
  \href{https://doi.org/10.1142/S0218271815300049}{\emph{Int. J. Mod. Phys.}
  {\bfseries D24} (2015) 1530004},
  [\href{https://arxiv.org/abs/1501.04288}{{\ttfamily 1501.04288}}].

\bibitem{Wands:2015fua}
D.~Wands, O.~F. Piattella and L.~Casarini, \emph{{Physics of the Cosmic
  Microwave Background Radiation}},
  \href{https://doi.org/10.1007/978-3-319-44769-8_1}{\emph{Astrophys. Space
  Sci. Proc.} {\bfseries 45} (2016) 3--39},
  [\href{https://arxiv.org/abs/1504.06335}{{\ttfamily 1504.06335}}].

\bibitem{Durrer:2015lza}
R.~Durrer, \emph{{The cosmic microwave background: the history of its
  experimental investigation and its significance for cosmology}},
  \href{https://doi.org/10.1088/0264-9381/32/12/124007}{\emph{Class. Quant.
  Grav.} {\bfseries 32} (2015) 124007},
  [\href{https://arxiv.org/abs/1506.01907}{{\ttfamily 1506.01907}}].

\bibitem{Sachs:1967er}
R.~K. Sachs and A.~M. Wolfe, \emph{{Perturbations of a cosmological model and
  angular variations of the microwave background}},
  \href{https://doi.org/10.1007/s10714-007-0448-9}{\emph{Astrophys. J.}
  {\bfseries 147} (1967) 73--90}.

\bibitem{Cabass:2015xfa}
G.~Cabass, M.~Gerbino, E.~Giusarma, A.~Melchiorri, L.~Pagano and L.~Salvati,
  \emph{{Constraints on the early and late integrated Sachs-Wolfe effects from
  the Planck 2015 cosmic microwave background anisotropies in the angular power
  spectra}}, \href{https://doi.org/10.1103/PhysRevD.92.063534}{\emph{Phys.
  Rev.} {\bfseries D92} (2015) 063534},
  [\href{https://arxiv.org/abs/1507.07586}{{\ttfamily 1507.07586}}].

\bibitem{Silk:1967kq}
J.~Silk, \emph{{Cosmic black body radiation and galaxy formation}},
  \href{https://doi.org/10.1086/149449}{\emph{Astrophys. J.} {\bfseries 151}
  (1968) 459--471}.

\bibitem{Pires:2004pi}
N.~Pires, M.~A.~S. Nobre and J.~A.~S. Lima, \emph{{On the width of the last
  scattering surface}},
  \href{https://doi.org/10.1142/S0218271804005638}{\emph{Int. J. Mod. Phys.}
  {\bfseries D13} (2004) 1425--1429},
  [\href{https://arxiv.org/abs/astro-ph/0411657}{{\ttfamily
  astro-ph/0411657}}].

\bibitem{Pan:2016zla}
Z.~Pan, L.~Knox, B.~Mulroe and A.~Narimani, \emph{{Cosmic Microwave Background
  Acoustic Peak Locations}},
  \href{https://doi.org/10.1093/mnras/stw833}{\emph{Mon. Not. Roy. Astron.
  Soc.} {\bfseries 459} (2016) 2513--2524},
  [\href{https://arxiv.org/abs/1603.03091}{{\ttfamily 1603.03091}}].

\bibitem{Hadzhiyska:2018mwh}
B.~Hadzhiyska and D.~N. Spergel, \emph{{Measuring the Duration of Last
  Scattering}}, \href{https://doi.org/10.1103/PhysRevD.99.043537}{\emph{Phys.
  Rev.} {\bfseries D99} (2019) 043537},
  [\href{https://arxiv.org/abs/1808.04083}{{\ttfamily 1808.04083}}].

\bibitem{Aghanim:2007bt}
N.~Aghanim, S.~Majumdar and J.~Silk, \emph{{Secondary anisotropies of the
  CMB}}, \href{https://doi.org/10.1088/0034-4885/71/6/066902}{\emph{Rept. Prog.
  Phys.} {\bfseries 71} (2008) 066902},
  [\href{https://arxiv.org/abs/0711.0518}{{\ttfamily 0711.0518}}].

\bibitem{Bartelmann:1999yn}
M.~Bartelmann and P.~Schneider, \emph{{Weak gravitational lensing}},
  \href{https://doi.org/10.1016/S0370-1573(00)00082-X}{\emph{Phys. Rept.}
  {\bfseries 340} (2001) 291--472},
  [\href{https://arxiv.org/abs/astro-ph/9912508}{{\ttfamily
  astro-ph/9912508}}].

\bibitem{Lewis:2006fu}
A.~Lewis and A.~Challinor, \emph{{Weak gravitational lensing of the CMB}},
  \href{https://doi.org/10.1016/j.physrep.2006.03.002}{\emph{Phys. Rept.}
  {\bfseries 429} (2006) 1--65},
  [\href{https://arxiv.org/abs/astro-ph/0601594}{{\ttfamily
  astro-ph/0601594}}].

\bibitem{Hanson:2009kr}
D.~Hanson, A.~Challinor and A.~Lewis, \emph{{Weak lensing of the CMB}},
  \href{https://doi.org/10.1007/s10714-010-1036-y}{\emph{Gen. Rel. Grav.}
  {\bfseries 42} (2010) 2197--2218},
  [\href{https://arxiv.org/abs/0911.0612}{{\ttfamily 0911.0612}}].

\bibitem{Okamoto:2003zw}
T.~Okamoto and W.~Hu, \emph{{CMB lensing reconstruction on the full sky}},
  \href{https://doi.org/10.1103/PhysRevD.67.083002}{\emph{Phys. Rev.}
  {\bfseries D67} (2003) 083002},
  [\href{https://arxiv.org/abs/astro-ph/0301031}{{\ttfamily
  astro-ph/0301031}}].

\bibitem{Kesden:2003cc}
M.~H. Kesden, A.~Cooray and M.~Kamionkowski, \emph{{Lensing reconstruction with
  CMB temperature and polarization}},
  \href{https://doi.org/10.1103/PhysRevD.67.123507}{\emph{Phys. Rev.}
  {\bfseries D67} (2003) 123507},
  [\href{https://arxiv.org/abs/astro-ph/0302536}{{\ttfamily
  astro-ph/0302536}}].

\bibitem{Hanson:2010rp}
D.~Hanson, A.~Challinor, G.~Efstathiou and P.~Bielewicz, \emph{{CMB temperature
  lensing power reconstruction}},
  \href{https://doi.org/10.1103/PhysRevD.83.043005}{\emph{Phys. Rev.}
  {\bfseries D83} (2011) 043005},
  [\href{https://arxiv.org/abs/1008.4403}{{\ttfamily 1008.4403}}].

\bibitem{Smith:2010gu}
K.~M. Smith, D.~Hanson, M.~LoVerde, C.~M. Hirata and O.~Zahn, \emph{{Delensing
  CMB Polarization with External Datasets}},
  \href{https://doi.org/10.1088/1475-7516/2012/06/014}{\emph{JCAP} {\bfseries
  1206} (2012) 014}, [\href{https://arxiv.org/abs/1010.0048}{{\ttfamily
  1010.0048}}].

\bibitem{Namikawa:2011cs}
T.~Namikawa, D.~Yamauchi and A.~Taruya, \emph{{Full-sky lensing reconstruction
  of gradient and curl modes from CMB maps}},
  \href{https://doi.org/10.1088/1475-7516/2012/01/007}{\emph{JCAP} {\bfseries
  1201} (2012) 007}, [\href{https://arxiv.org/abs/1110.1718}{{\ttfamily
  1110.1718}}].

\bibitem{Pearson:2014qna}
R.~Pearson, B.~Sherwin and A.~Lewis, \emph{{CMB lensing reconstruction using
  cut sky polarization maps and pure-$B$ modes}},
  \href{https://doi.org/10.1103/PhysRevD.90.023539}{\emph{Phys. Rev.}
  {\bfseries D90} (2014) 023539},
  [\href{https://arxiv.org/abs/1403.3911}{{\ttfamily 1403.3911}}].

\bibitem{Sherwin:2015baa}
B.~D. Sherwin and M.~Schmittfull, \emph{{Delensing the CMB with the Cosmic
  Infrared Background}},
  \href{https://doi.org/10.1103/PhysRevD.92.043005}{\emph{Phys. Rev.}
  {\bfseries D92} (2015) 043005},
  [\href{https://arxiv.org/abs/1502.05356}{{\ttfamily 1502.05356}}].

\bibitem{Namikawa:2015tjy}
T.~Namikawa, D.~Yamauchi, B.~Sherwin and R.~Nagata, \emph{{Delensing Cosmic
  Microwave Background B-modes with the Square Kilometre Array Radio Continuum
  Survey}}, \href{https://doi.org/10.1103/PhysRevD.93.043527}{\emph{Phys. Rev.}
  {\bfseries D93} (2016) 043527},
  [\href{https://arxiv.org/abs/1511.04653}{{\ttfamily 1511.04653}}].

\bibitem{Larsen:2016wpa}
P.~Larsen, A.~Challinor, B.~D. Sherwin and D.~Mak, \emph{{Demonstration of
  cosmic microwave background delensing using the cosmic infrared background}},
  \href{https://doi.org/10.1103/PhysRevLett.117.151102}{\emph{Phys. Rev. Lett.}
  {\bfseries 117} (2016) 151102},
  [\href{https://arxiv.org/abs/1607.05733}{{\ttfamily 1607.05733}}].

\bibitem{Sehgal:2016eag}
N.~Sehgal, M.~S. Madhavacheril, B.~Sherwin and A.~van Engelen, \emph{{Internal
  Delensing of Cosmic Microwave Background Acoustic Peaks}},
  \href{https://doi.org/10.1103/PhysRevD.95.103512}{\emph{Phys. Rev.}
  {\bfseries D95} (2017) 103512},
  [\href{https://arxiv.org/abs/1612.03898}{{\ttfamily 1612.03898}}].

\bibitem{Yu:2017djs}
B.~Yu, J.~C. Hill and B.~D. Sherwin, \emph{{Multitracer CMB delensing maps from
  Planck and WISE data}},
  \href{https://doi.org/10.1103/PhysRevD.96.123511}{\emph{Phys. Rev.}
  {\bfseries D96} (2017) 123511},
  [\href{https://arxiv.org/abs/1705.02332}{{\ttfamily 1705.02332}}].

\bibitem{Horowitz:2017iql}
B.~Horowitz, S.~Ferraro and B.~D. Sherwin, \emph{{Reconstructing Small Scale
  Lenses from the Cosmic Microwave Background Temperature Fluctuations}},
  \href{https://doi.org/10.1093/mnras/stz566}{\emph{Mon. Not. Roy. Astron.
  Soc.} {\bfseries 469} (2019) 394--400},
  [\href{https://arxiv.org/abs/1710.10236}{{\ttfamily 1710.10236}}].

\bibitem{Manzotti:2017oby}
A.~Manzotti, \emph{{Future cosmic microwave background delensing with galaxy
  surveys}}, \href{https://doi.org/10.1103/PhysRevD.97.043527}{\emph{Phys.
  Rev.} {\bfseries D97} (2018) 043527},
  [\href{https://arxiv.org/abs/1710.11038}{{\ttfamily 1710.11038}}].

\bibitem{Madhavacheril:2018bxi}
M.~S. Madhavacheril and J.~C. Hill, \emph{{Mitigating Foreground Biases in CMB
  Lensing Reconstruction Using Cleaned Gradients}},
  \href{https://doi.org/10.1103/PhysRevD.98.023534}{\emph{Phys. Rev.}
  {\bfseries D98} (2018) 023534},
  [\href{https://arxiv.org/abs/1802.08230}{{\ttfamily 1802.08230}}].

\bibitem{Schaan:2018tup}
E.~Schaan and S.~Ferraro, \emph{{Foreground-immune CMB lensing with shear-only
  reconstruction}},  \href{https://arxiv.org/abs/1804.06403}{{\ttfamily
  1804.06403}}.

\bibitem{Hirata:2008cb}
C.~M. Hirata, S.~Ho, N.~Padmanabhan, U.~Seljak and N.~A. Bahcall,
  \emph{{Correlation of CMB with large-scale structure: II. Weak lensing}},
  \href{https://doi.org/10.1103/PhysRevD.78.043520}{\emph{Phys. Rev.}
  {\bfseries D78} (2008) 043520},
  [\href{https://arxiv.org/abs/0801.0644}{{\ttfamily 0801.0644}}].

\bibitem{Sherwin:2012mr}
B.~D. Sherwin et~al., \emph{{The Atacama Cosmology Telescope: Cross-Correlation
  of CMB Lensing and Quasars}},
  \href{https://doi.org/10.1103/PhysRevD.86.083006}{\emph{Phys. Rev.}
  {\bfseries D86} (2012) 083006},
  [\href{https://arxiv.org/abs/1207.4543}{{\ttfamily 1207.4543}}].

\bibitem{Ade:2013hjl}
{\scshape POLARBEAR} collaboration, P.~A.~R. Ade et~al., \emph{{Evidence for
  Gravitational Lensing of the Cosmic Microwave Background Polarization from
  Cross-correlation with the Cosmic Infrared Background}},
  \href{https://doi.org/10.1103/PhysRevLett.112.131302}{\emph{Phys. Rev. Lett.}
  {\bfseries 112} (2014) 131302},
  [\href{https://arxiv.org/abs/1312.6645}{{\ttfamily 1312.6645}}].

\bibitem{Nicola:2016eua}
A.~Nicola, A.~Refregier and A.~Amara, \emph{{Integrated approach to cosmology:
  Combining CMB, large-scale structure and weak lensing}},
  \href{https://doi.org/10.1103/PhysRevD.94.083517}{\emph{Phys. Rev.}
  {\bfseries D94} (2016) 083517},
  [\href{https://arxiv.org/abs/1607.01014}{{\ttfamily 1607.01014}}].

\bibitem{Doux:2017tsv}
C.~Doux, M.~Penna-Lima, S.~D.~P. Vitenti, J.~Tréguer, E.~Aubourg and K.~Ganga,
  \emph{{Cosmological constraints from a joint analysis of cosmic microwave
  background and spectroscopic tracers of the large-scale structure}},
  \href{https://doi.org/10.1093/mnras/sty2160}{\emph{Mon. Not. Roy. Astron.
  Soc.} {\bfseries 480} (2018) 5386--5411},
  [\href{https://arxiv.org/abs/1706.04583}{{\ttfamily 1706.04583}}].

\bibitem{Mellema:2012ht}
G.~Mellema et~al., \emph{{Reionization and the Cosmic Dawn with the Square
  Kilometre Array}},
  \href{https://doi.org/10.1007/s10686-013-9334-5}{\emph{Exper. Astron.}
  {\bfseries 36} (2013) 235--318},
  [\href{https://arxiv.org/abs/1210.0197}{{\ttfamily 1210.0197}}].

\bibitem{Smith:2016lnt}
K.~M. Smith and S.~Ferraro, \emph{{Detecting Patchy Reionization in the Cosmic
  Microwave Background}},
  \href{https://doi.org/10.1103/PhysRevLett.119.021301}{\emph{Phys. Rev. Lett.}
  {\bfseries 119} (2017) 021301},
  [\href{https://arxiv.org/abs/1607.01769}{{\ttfamily 1607.01769}}].

\bibitem{Meyers:2017rtf}
J.~Meyers, P.~D. Meerburg, A.~van Engelen and N.~Battaglia, \emph{{Beyond CMB
  cosmic variance limits on reionization with the polarized
  Sunyaev-Zel’dovich effect}},
  \href{https://doi.org/10.1103/PhysRevD.97.103505}{\emph{Phys. Rev.}
  {\bfseries D97} (2018) 103505},
  [\href{https://arxiv.org/abs/1710.01708}{{\ttfamily 1710.01708}}].

\bibitem{Namikawa:2017uke}
T.~Namikawa, \emph{{Constraints on Patchy Reionization from Planck CMB
  Temperature Trispectrum}},
  \href{https://doi.org/10.1103/PhysRevD.97.063505}{\emph{Phys. Rev.}
  {\bfseries D97} (2018) 063505},
  [\href{https://arxiv.org/abs/1711.00058}{{\ttfamily 1711.00058}}].

\bibitem{Villanueva-Domingo:2017ahx}
P.~Villanueva-Domingo, S.~Gariazzo, N.~Y. Gnedin and O.~Mena, \emph{{Was there
  an early reionization component in our universe?}},
  \href{https://doi.org/10.1088/1475-7516/2018/04/024}{\emph{JCAP} {\bfseries
  1804} (2018) 024}, [\href{https://arxiv.org/abs/1712.02807}{{\ttfamily
  1712.02807}}].

\bibitem{Roy:2018gcv}
A.~Roy, A.~Lapi, D.~Spergel and C.~Baccigalupi, \emph{{Observing patchy
  reionization with future CMB polarization experiments}},
  \href{https://doi.org/10.1088/1475-7516/2018/05/014}{\emph{JCAP} {\bfseries
  1805} (2018) 014}, [\href{https://arxiv.org/abs/1801.02393}{{\ttfamily
  1801.02393}}].

\bibitem{Ferraro:2018izc}
S.~Ferraro and K.~M. Smith, \emph{{Characterizing the epoch of reionization
  with the small-scale CMB: Constraints on the optical depth and duration}},
  \href{https://doi.org/10.1103/PhysRevD.98.123519}{\emph{Phys. Rev.}
  {\bfseries D98} (2018) 123519},
  [\href{https://arxiv.org/abs/1803.07036}{{\ttfamily 1803.07036}}].

\bibitem{Giri:2018dln}
S.~K. Giri, A.~D'Aloisio, G.~Mellema, E.~Komatsu, R.~Ghara and S.~Majumdar,
  \emph{{Position-dependent power spectra of the 21-cm signal from the epoch of
  reionization}},
  \href{https://doi.org/10.1088/1475-7516/2019/02/058}{\emph{JCAP} {\bfseries
  1902} (2019) 058}, [\href{https://arxiv.org/abs/1811.09633}{{\ttfamily
  1811.09633}}].

\bibitem{Giri:2019pxr}
S.~K. Giri, G.~Mellema, T.~Aldheimer, K.~L. Dixon and I.~T. Iliev,
  \emph{{Neutral island statistics during reionization from 21-cm tomography}},
   \href{https://arxiv.org/abs/1903.01294}{{\ttfamily 1903.01294}}.

\bibitem{Roy:2019qsl}
A.~Roy, A.~Lapi, D.~Spergel, S.~Basak and C.~Baccigalupi, \emph{{Detectability
  of the $\tau$-21cm cross-correlation: a tomographic probe of patchy
  reionization}},  \href{https://arxiv.org/abs/1904.02637}{{\ttfamily
  1904.02637}}.

\bibitem{Rees:1968zza}
M.~J. Rees and D.~W. Sciama, \emph{{Large scale Density Inhomogeneities in the
  Universe}}, \href{https://doi.org/10.1038/217511a0}{\emph{Nature} {\bfseries
  217} (1968) 511--516}.

\bibitem{Fosalba:2003ge}
P.~Fosalba, E.~Gaztañaga and F.~Castander, \emph{{Detection of the ISW and SZ
  effects from the CMB-galaxy correlation}},
  \href{https://doi.org/10.1086/379848}{\emph{Astrophys. J.} {\bfseries 597}
  (2003) L89--92}, [\href{https://arxiv.org/abs/astro-ph/0307249}{{\ttfamily
  astro-ph/0307249}}].

\bibitem{Vielva:2004zg}
P.~Vielva, E.~Martínez-González and M.~Tucci, \emph{{WMAP and NVSS
  cross-correlation in wavelet space: ISW detection and dark energy
  constraints}},
  \href{https://doi.org/10.1111/j.1365-2966.2005.09764.x}{\emph{Mon. Not. Roy.
  Astron. Soc.} {\bfseries 365} (2006) 891},
  [\href{https://arxiv.org/abs/astro-ph/0408252}{{\ttfamily
  astro-ph/0408252}}].

\bibitem{Padmanabhan:2004fy}
N.~Padmanabhan, C.~M. Hirata, U.~Seljak, D.~Schlegel, J.~Brinkmann and D.~P.
  Schneider, \emph{{Correlating the CMB with luminous red galaxies: The
  Integrated Sachs-Wolfe effect}},
  \href{https://doi.org/10.1103/PhysRevD.72.043525}{\emph{Phys. Rev.}
  {\bfseries D72} (2005) 043525},
  [\href{https://arxiv.org/abs/astro-ph/0410360}{{\ttfamily
  astro-ph/0410360}}].

\bibitem{McEwen:2006my}
J.~D. McEwen, P.~Vielva, M.~P. Hobson, E.~Martínez-González and A.~N.
  Lasenby, \emph{{Detection of the integrated Sachs–Wolfe effect and
  corresponding dark energy constraints made with directional spherical
  wavelets}},
  \href{https://doi.org/10.1111/j.1365-2966.2007.11505.x}{\emph{Mon. Not. Roy.
  Astron. Soc.} {\bfseries 376} (2007) 1211--1226},
  [\href{https://arxiv.org/abs/astro-ph/0602398}{{\ttfamily
  astro-ph/0602398}}].

\bibitem{Cabre:2007rv}
A.~Cabré, P.~Fosalba, E.~Gaztañaga and M.~Manera, \emph{{Error analysis in
  cross-correlation of sky maps: Application to the ISW detection}},
  \href{https://doi.org/10.1111/j.1365-2966.2007.12280.x}{\emph{Mon. Not. Roy.
  Astron. Soc.} {\bfseries 381} (2007) 1347},
  [\href{https://arxiv.org/abs/astro-ph/0701393}{{\ttfamily
  astro-ph/0701393}}].

\bibitem{Ho:2008bz}
S.~Ho, C.~Hirata, N.~Padmanabhan, U.~Seljak and N.~Bahcall, \emph{{Correlation
  of CMB with large-scale structure: I. ISW Tomography and Cosmological
  Implications}}, \href{https://doi.org/10.1103/PhysRevD.78.043519}{\emph{Phys.
  Rev.} {\bfseries D78} (2008) 043519},
  [\href{https://arxiv.org/abs/0801.0642}{{\ttfamily 0801.0642}}].

\bibitem{Zhao:2008bn}
G.-B. Zhao, L.~Pogosian, A.~Silvestri and J.~Zylberberg, \emph{{Searching for
  modified growth patterns with tomographic surveys}},
  \href{https://doi.org/10.1103/PhysRevD.79.083513}{\emph{Phys. Rev.}
  {\bfseries D79} (2009) 083513},
  [\href{https://arxiv.org/abs/0809.3791}{{\ttfamily 0809.3791}}].

\bibitem{Zhao:2010dz}
G.-B. Zhao, T.~Giannantonio, L.~Pogosian, A.~Silvestri, D.~J. Bacon, K.~Koyama
  et~al., \emph{{Probing modifications of General Relativity using current
  cosmological observations}},
  \href{https://doi.org/10.1103/PhysRevD.81.103510}{\emph{Phys. Rev.}
  {\bfseries D81} (2010) 103510},
  [\href{https://arxiv.org/abs/1003.0001}{{\ttfamily 1003.0001}}].

\bibitem{Ilic:2011hh}
S.~Ilić, M.~Douspis, M.~Langer, A.~Pénin and G.~Lagache,
  \emph{{Cross-correlation between the cosmic microwave and infrared
  backgrounds for integrated Sachs-Wolfe detection}},
  \href{https://doi.org/10.1111/j.1365-2966.2011.19221.x}{\emph{Mon. Not. Roy.
  Astron. Soc.} {\bfseries 416} (2011) 2688},
  [\href{https://arxiv.org/abs/1106.2328}{{\ttfamily 1106.2328}}].

\bibitem{Ilic:2013cn}
S.~Ilić, M.~Langer and M.~Douspis, \emph{{On the detection of the integrated
  Sachs-Wolfe effect with stacked voids}},
  \href{https://doi.org/10.1051/0004-6361/201321150}{\emph{Astron. Astrophys.}
  {\bfseries 556} (2013) A51},
  [\href{https://arxiv.org/abs/1301.5849}{{\ttfamily 1301.5849}}].

\bibitem{Ferraro:2014msa}
S.~Ferraro, B.~D. Sherwin and D.~N. Spergel, \emph{{WISE measurement of the
  integrated Sachs-Wolfe effect}},
  \href{https://doi.org/10.1103/PhysRevD.91.083533}{\emph{Phys. Rev.}
  {\bfseries D91} (2015) 083533},
  [\href{https://arxiv.org/abs/1401.1193}{{\ttfamily 1401.1193}}].

\bibitem{Nishizawa:2014vga}
A.~J. Nishizawa, \emph{{The integrated Sachs–Wolfe effect and the
  Rees–Sciama effect}},
  \href{https://doi.org/10.1093/ptep/ptu062}{\emph{PTEP} {\bfseries 2014}
  (2014) 06B110}, [\href{https://arxiv.org/abs/1404.5102}{{\ttfamily
  1404.5102}}].

\bibitem{Renk:2017rzu}
J.~Renk, M.~Zumalacárregui, F.~Montanari and A.~Barreira, \emph{{Galileon
  gravity in light of ISW, CMB, BAO and H$_0$ data}},
  \href{https://doi.org/10.1088/1475-7516/2017/10/020}{\emph{JCAP} {\bfseries
  1710} (2017) 020}, [\href{https://arxiv.org/abs/1707.02263}{{\ttfamily
  1707.02263}}].

\bibitem{Bolis:2018vzs}
N.~Bolis, A.~De~Felice and S.~Mukohyama, \emph{{Integrated Sachs-Wolfe-galaxy
  cross-correlation bounds on the two branches of the minimal theory of massive
  gravity}}, \href{https://doi.org/10.1103/PhysRevD.98.024010}{\emph{Phys.
  Rev.} {\bfseries D98} (2018) 024010},
  [\href{https://arxiv.org/abs/1804.01790}{{\ttfamily 1804.01790}}].

\bibitem{Hu:1997hv}
W.~Hu and M.~J. White, \emph{{A CMB polarization primer}},
  \href{https://doi.org/10.1016/S1384-1076(97)00022-5}{\emph{New Astron.}
  {\bfseries 2} (1997) 323},
  [\href{https://arxiv.org/abs/astro-ph/9706147}{{\ttfamily
  astro-ph/9706147}}].

\bibitem{Kosowsky:1998mb}
A.~Kosowsky, \emph{{Introduction to microwave background polarization}},
  \href{https://doi.org/10.1016/S1387-6473(99)00009-3}{\emph{New Astron. Rev.}
  {\bfseries 43} (1999) 157},
  [\href{https://arxiv.org/abs/astro-ph/9904102}{{\ttfamily
  astro-ph/9904102}}].

\bibitem{Challinor:2005ye}
A.~Challinor, \emph{{Cosmic microwave background polarization analysis}},
  {\emph{Lect. Notes Phys.} {\bfseries 653} (2004) 71--104},
  [\href{https://arxiv.org/abs/astro-ph/0502093}{{\ttfamily
  astro-ph/0502093}}].

\bibitem{Chandrasekhar:1950ghw}
S.~{Chandrasekhar}, \emph{{Radiative transfer.}}
\newblock 1950.

\bibitem{Griffiths:1995ghw}
D.~J. {Griffiths}, \emph{{Introduction to Quantum Mechanics}}.
\newblock 1995.

\bibitem{Kosowsky:1994cy}
A.~Kosowsky, \emph{{Cosmic microwave background polarization}},
  \href{https://doi.org/10.1006/aphy.1996.0020}{\emph{Annals Phys.} {\bfseries
  246} (1996) 49--85},
  [\href{https://arxiv.org/abs/astro-ph/9501045}{{\ttfamily
  astro-ph/9501045}}].

\bibitem{Seljak:1996ti}
U.~Seljak, \emph{{Measuring polarization in cosmic microwave background}},
  \href{https://doi.org/10.1086/304123}{\emph{Astrophys. J.} {\bfseries 482}
  (1997) 6}, [\href{https://arxiv.org/abs/astro-ph/9608131}{{\ttfamily
  astro-ph/9608131}}].

\bibitem{Kamionkowski:1996ks}
M.~Kamionkowski, A.~Kosowsky and A.~Stebbins, \emph{{Statistics of cosmic
  microwave background polarization}},
  \href{https://doi.org/10.1103/PhysRevD.55.7368}{\emph{Phys. Rev.} {\bfseries
  D55} (1997) 7368--7388},
  [\href{https://arxiv.org/abs/astro-ph/9611125}{{\ttfamily
  astro-ph/9611125}}].

\bibitem{Bond:1987ub}
J.~R. Bond and G.~Efstathiou, \emph{{The statistics of cosmic background
  radiation fluctuations}}, {\emph{Mon. Not. Roy. Astron. Soc.} {\bfseries 226}
  (1987) 655--687}.

\bibitem{Zaldarriaga:1996xe}
M.~Zaldarriaga and U.~Seljak, \emph{{An all sky analysis of polarization in the
  microwave background}},
  \href{https://doi.org/10.1103/PhysRevD.55.1830}{\emph{Phys. Rev.} {\bfseries
  D55} (1997) 1830--1840},
  [\href{https://arxiv.org/abs/astro-ph/9609170}{{\ttfamily
  astro-ph/9609170}}].

\bibitem{Seljak:1996gy}
U.~Seljak and M.~Zaldarriaga, \emph{{Signature of gravity waves in polarization
  of the microwave background}},
  \href{https://doi.org/10.1103/PhysRevLett.78.2054}{\emph{Phys. Rev. Lett.}
  {\bfseries 78} (1997) 2054--2057},
  [\href{https://arxiv.org/abs/astro-ph/9609169}{{\ttfamily
  astro-ph/9609169}}].

\bibitem{Kamionkowski:1997av}
M.~Kamionkowski and A.~Kosowsky, \emph{{Detectability of inflationary
  gravitational waves with microwave background polarization}},
  \href{https://doi.org/10.1103/PhysRevD.57.685}{\emph{Phys. Rev.} {\bfseries
  D57} (1998) 685--691},
  [\href{https://arxiv.org/abs/astro-ph/9705219}{{\ttfamily
  astro-ph/9705219}}].

\bibitem{Krauss:2013pha}
L.~M. Krauss and F.~Wilczek, \emph{{Using Cosmology to Establish the
  Quantization of Gravity}},
  \href{https://doi.org/10.1103/PhysRevD.89.047501}{\emph{Phys. Rev.}
  {\bfseries D89} (2014) 047501},
  [\href{https://arxiv.org/abs/1309.5343}{{\ttfamily 1309.5343}}].

\bibitem{Kamionkowski:2015yta}
M.~Kamionkowski and E.~D. Kovetz, \emph{{The Quest for B Modes from
  Inflationary Gravitational Waves}},
  \href{https://doi.org/10.1146/annurev-astro-081915-023433}{\emph{Ann. Rev.
  Astron. Astrophys.} {\bfseries 54} (2016) 227--269},
  [\href{https://arxiv.org/abs/1510.06042}{{\ttfamily 1510.06042}}].

\bibitem{Zaldarriaga:1998ar}
M.~Zaldarriaga and U.~Seljak, \emph{{Gravitational lensing effect on cosmic
  microwave background polarization}},
  \href{https://doi.org/10.1103/PhysRevD.58.023003}{\emph{Phys. Rev.}
  {\bfseries D58} (1998) 023003},
  [\href{https://arxiv.org/abs/astro-ph/9803150}{{\ttfamily
  astro-ph/9803150}}].

\bibitem{Knox:2002pe}
L.~Knox and Y.-S. Song, \emph{{A Limit on the detectability of the energy scale
  of inflation}},
  \href{https://doi.org/10.1103/PhysRevLett.89.011303}{\emph{Phys. Rev. Lett.}
  {\bfseries 89} (2002) 011303},
  [\href{https://arxiv.org/abs/astro-ph/0202286}{{\ttfamily
  astro-ph/0202286}}].

\bibitem{Dodelson:2003bv}
S.~Dodelson, E.~Rozo and A.~Stebbins, \emph{{Primordial gravity waves and weak
  lensing}}, \href{https://doi.org/10.1103/PhysRevLett.91.021301}{\emph{Phys.
  Rev. Lett.} {\bfseries 91} (2003) 021301},
  [\href{https://arxiv.org/abs/astro-ph/0301177}{{\ttfamily
  astro-ph/0301177}}].

\bibitem{Seljak:2003pn}
U.~Seljak and C.~M. Hirata, \emph{{Gravitational lensing as a contaminant of
  the gravity wave signal in CMB}},
  \href{https://doi.org/10.1103/PhysRevD.69.043005}{\emph{Phys. Rev.}
  {\bfseries D69} (2004) 043005},
  [\href{https://arxiv.org/abs/astro-ph/0310163}{{\ttfamily
  astro-ph/0310163}}].

\bibitem{Hanson:2013hsb}
{\scshape SPTpol} collaboration, D.~Hanson et~al., \emph{{Detection of B-mode
  Polarization in the Cosmic Microwave Background with Data from the South Pole
  Telescope}},
  \href{https://doi.org/10.1103/PhysRevLett.111.141301}{\emph{Phys. Rev. Lett.}
  {\bfseries 111} (2013) 141301},
  [\href{https://arxiv.org/abs/1307.5830}{{\ttfamily 1307.5830}}].

\bibitem{Audren:2012vy}
B.~Audren, J.~Lesgourgues, S.~Bird, M.~G. Haehnelt and M.~Viel, \emph{{Neutrino
  masses and cosmological parameters from a Euclid-like survey: Markov Chain
  Monte Carlo forecasts including theoretical errors}},
  \href{https://doi.org/10.1088/1475-7516/2013/01/026}{\emph{JCAP} {\bfseries
  1301} (2013) 026}, [\href{https://arxiv.org/abs/1210.2194}{{\ttfamily
  1210.2194}}].

\bibitem{Galli:2014kla}
S.~Galli, K.~Benabed, F.~Bouchet, J.-F. Cardoso, F.~Elsner, E.~Hivon et~al.,
  \emph{{CMB Polarization can constrain cosmology better than CMB
  temperature}}, \href{https://doi.org/10.1103/PhysRevD.90.063504}{\emph{Phys.
  Rev.} {\bfseries D90} (2014) 063504},
  [\href{https://arxiv.org/abs/1403.5271}{{\ttfamily 1403.5271}}].

\bibitem{Efstathiou:1998xx}
G.~Efstathiou and J.~R. Bond, \emph{{Cosmic confusion: Degeneracies among
  cosmological parameters derived from measurements of microwave background
  anisotropies}},
  \href{https://doi.org/10.1046/j.1365-8711.1999.02274.x}{\emph{Mon. Not. Roy.
  Astron. Soc.} {\bfseries 304} (1999) 75--97},
  [\href{https://arxiv.org/abs/astro-ph/9807103}{{\ttfamily
  astro-ph/9807103}}].

\bibitem{Efstathiou:2001cv}
G.~Efstathiou, \emph{{Principal component analysis of the cosmic microwave
  background anisotropies: revealing the tensor degeneracy}},
  \href{https://doi.org/10.1046/j.1365-8711.2002.05315.x}{\emph{Mon. Not. Roy.
  Astron. Soc.} {\bfseries 332} (2002) 193},
  [\href{https://arxiv.org/abs/astro-ph/0109151}{{\ttfamily
  astro-ph/0109151}}].

\bibitem{Howlett:2012mh}
C.~Howlett, A.~Lewis, A.~Hall and A.~Challinor, \emph{{CMB power spectrum
  parameter degeneracies in the era of precision cosmology}},
  \href{https://doi.org/10.1088/1475-7516/2012/04/027}{\emph{JCAP} {\bfseries
  1204} (2012) 027}, [\href{https://arxiv.org/abs/1201.3654}{{\ttfamily
  1201.3654}}].

\bibitem{Li:2012ug}
H.~Li and J.-Q. Xia, \emph{{Impacts on Cosmological Constraints from
  Degeneracies}},
  \href{https://doi.org/10.1088/1475-7516/2012/11/039}{\emph{JCAP} {\bfseries
  1211} (2012) 039}, [\href{https://arxiv.org/abs/1210.2037}{{\ttfamily
  1210.2037}}].

\bibitem{Dawson:2012va}
{\scshape BOSS} collaboration, K.~S. Dawson et~al., \emph{{The Baryon
  Oscillation Spectroscopic Survey of SDSS-III}},
  \href{https://doi.org/10.1088/0004-6256/145/1/10}{\emph{Astron. J.}
  {\bfseries 145} (2013) 10},
  [\href{https://arxiv.org/abs/1208.0022}{{\ttfamily 1208.0022}}].

\bibitem{Abbott:2005bi}
{\scshape DES} collaboration, T.~Abbott et~al., \emph{{The dark energy
  survey}},  \href{https://arxiv.org/abs/astro-ph/0510346}{{\ttfamily
  astro-ph/0510346}}.

\bibitem{Dawson:2015wdb}
K.~S. Dawson et~al., \emph{{The SDSS-IV extended Baryon Oscillation
  Spectroscopic Survey: Overview and Early Data}},
  \href{https://doi.org/10.3847/0004-6256/151/2/44}{\emph{Astron. J.}
  {\bfseries 151} (2016) 44},
  [\href{https://arxiv.org/abs/1508.04473}{{\ttfamily 1508.04473}}].

\bibitem{Drinkwater:2009sd}
M.~J. Drinkwater et~al., \emph{{The WiggleZ Dark Energy Survey: Survey Design
  and First Data Release}},
  \href{https://doi.org/10.1111/j.1365-2966.2009.15754.x}{\emph{Mon. Not. Roy.
  Astron. Soc.} {\bfseries 401} (2010) 1429--1452},
  [\href{https://arxiv.org/abs/0911.4246}{{\ttfamily 0911.4246}}].

\bibitem{Beutler:2011hx}
F.~Beutler, C.~Blake, M.~Colless, D.~H. Jones, L.~Staveley-Smith, L.~Campbell
  et~al., \emph{{The 6dF Galaxy Survey: Baryon Acoustic Oscillations and the
  Local Hubble Constant}},
  \href{https://doi.org/10.1111/j.1365-2966.2011.19250.x}{\emph{Mon. Not. Roy.
  Astron. Soc.} {\bfseries 416} (2011) 3017--3032},
  [\href{https://arxiv.org/abs/1106.3366}{{\ttfamily 1106.3366}}].

\bibitem{Colless:2001gk}
{\scshape 2DFGRS} collaboration, M.~Colless et~al., \emph{{The 2dF Galaxy
  Redshift Survey: Spectra and redshifts}},
  \href{https://doi.org/10.1046/j.1365-8711.2001.04902.x}{\emph{Mon. Not. Roy.
  Astron. Soc.} {\bfseries 328} (2001) 1039},
  [\href{https://arxiv.org/abs/astro-ph/0106498}{{\ttfamily
  astro-ph/0106498}}].

\bibitem{Laureijs:2011gra}
{\scshape EUCLID} collaboration, R.~Laureijs et~al., \emph{{Euclid Definition
  Study Report}},  \href{https://arxiv.org/abs/1110.3193}{{\ttfamily
  1110.3193}}.

\bibitem{Aghamousa:2016zmz}
{\scshape DESI} collaboration, A.~Aghamousa et~al., \emph{{The DESI Experiment
  Part I: Science,Targeting, and Survey Design}},
  \href{https://arxiv.org/abs/1611.00036}{{\ttfamily 1611.00036}}.

\bibitem{Ivezic:2008fe}
{\scshape LSST} collaboration, Z.~Ivezić et~al., \emph{{LSST: from Science
  Drivers to Reference Design and Anticipated Data Products}},
  \href{https://doi.org/10.3847/1538-4357/ab042c}{\emph{Astrophys. J.}
  {\bfseries 873} (2019) 111},
  [\href{https://arxiv.org/abs/0805.2366}{{\ttfamily 0805.2366}}].

\bibitem{Green:2012mj}
J.~Green et~al., \emph{{Wide-Field InfraRed Survey Telescope (WFIRST) Final
  Report}},  \href{https://arxiv.org/abs/1208.4012}{{\ttfamily 1208.4012}}.

\bibitem{Dore:2014cca}
O.~Doré et~al., \emph{{Cosmology with the SPHEREX All-Sky Spectral Survey}},
  \href{https://arxiv.org/abs/1412.4872}{{\ttfamily 1412.4872}}.

\bibitem{Feldman:1993ky}
H.~A. Feldman, N.~Kaiser and J.~A. Peacock, \emph{{Power spectrum analysis of
  three-dimensional redshift surveys}},
  \href{https://doi.org/10.1086/174036}{\emph{Astrophys. J.} {\bfseries 426}
  (1994) 23--37}, [\href{https://arxiv.org/abs/astro-ph/9304022}{{\ttfamily
  astro-ph/9304022}}].

\bibitem{Heitmann:2008eq}
K.~Heitmann, M.~White, C.~Wagner, S.~Habib and D.~Higdon, \emph{{The Coyote
  Universe I: Precision Determination of the Nonlinear Matter Power Spectrum}},
  \href{https://doi.org/10.1088/0004-637X/715/1/104}{\emph{Astrophys. J.}
  {\bfseries 715} (2010) 104--121},
  [\href{https://arxiv.org/abs/0812.1052}{{\ttfamily 0812.1052}}].

\bibitem{Heitmann:2013bra}
K.~Heitmann, E.~Lawrence, J.~Kwan, S.~Habib and D.~Higdon, \emph{{The Coyote
  Universe Extended: Precision Emulation of the Matter Power Spectrum}},
  \href{https://doi.org/10.1088/0004-637X/780/1/111}{\emph{Astrophys. J.}
  {\bfseries 780} (2014) 111},
  [\href{https://arxiv.org/abs/1304.7849}{{\ttfamily 1304.7849}}].

\bibitem{Kwan:2013jva}
J.~Kwan, K.~Heitmann, S.~Habib, N.~Padmanabhan, H.~Finkel, E.~Lawrence et~al.,
  \emph{{Cosmic Emulation: Fast Predictions for the Galaxy Power Spectrum}},
  \href{https://doi.org/10.1088/0004-637X/810/1/35}{\emph{Astrophys. J.}
  {\bfseries 810} (2015) 35},
  [\href{https://arxiv.org/abs/1311.6444}{{\ttfamily 1311.6444}}].

\bibitem{Banerjee:2018bxy}
A.~Banerjee, D.~Powell, T.~Abel and F.~Villaescusa-Navarro, \emph{{Reducing
  Noise in Cosmological N-body Simulations with Neutrinos}},
  \href{https://doi.org/10.1088/1475-7516/2018/09/028}{\emph{JCAP} {\bfseries
  1809} (2018) 028}, [\href{https://arxiv.org/abs/1801.03906}{{\ttfamily
  1801.03906}}].

\bibitem{Rizzo:2016mdr}
L.~A. Rizzo, F.~Villaescusa-Navarro, P.~Monaco, E.~Munari, S.~Borgani,
  E.~Castorina et~al., \emph{{Simulating cosmologies beyond $\Lambda$CDM with
  PINOCCHIO}}, \href{https://doi.org/10.1088/1475-7516/2017/01/008}{\emph{JCAP}
  {\bfseries 1701} (2017) 008},
  [\href{https://arxiv.org/abs/1610.07624}{{\ttfamily 1610.07624}}].

\bibitem{Liu:2017now}
J.~Liu, S.~Bird, J.~M.~Z. Matilla, J.~C. Hill, Z.~Haiman, M.~S. Madhavacheril
  et~al., \emph{{MassiveNuS: Cosmological Massive Neutrino Simulations}},
  \href{https://doi.org/10.1088/1475-7516/2018/03/049}{\emph{JCAP} {\bfseries
  1803} (2018) 049}, [\href{https://arxiv.org/abs/1711.10524}{{\ttfamily
  1711.10524}}].

\bibitem{Springel:2005mi}
V.~Springel, \emph{{The Cosmological simulation code GADGET-2}},
  \href{https://doi.org/10.1111/j.1365-2966.2005.09655.x}{\emph{Mon. Not. Roy.
  Astron. Soc.} {\bfseries 364} (2005) 1105--1134},
  [\href{https://arxiv.org/abs/astro-ph/0505010}{{\ttfamily
  astro-ph/0505010}}].

\bibitem{Springel:2011yw}
V.~Springel, \emph{{Smoothed Particle Hydrodynamics in Astrophysics}},
  \href{https://doi.org/10.1146/annurev-astro-081309-130914}{\emph{Ann. Rev.
  Astron. Astrophys.} {\bfseries 48} (2010) 391--430},
  [\href{https://arxiv.org/abs/1109.2219}{{\ttfamily 1109.2219}}].

\bibitem{Springel:2014ona}
V.~{Springel}, \emph{{High Performance Computing and Numerical Modelling}},
  \href{https://doi.org/10.1007/978-3-662-47890-5_3}{\emph{Saas-Fee Advanced
  Course} {\bfseries 43} (Jan, 2016) 251},
  [\href{https://arxiv.org/abs/1412.5187}{{\ttfamily 1412.5187}}].

\bibitem{Dakin:2017idt}
J.~Dakin, J.~Brandbyge, S.~Hannestad, T.~Haugbølle and T.~Tram,
  \emph{{$\nu$CO$N$CEPT: Cosmological neutrino simulations from the non-linear
  Boltzmann hierarchy}},
  \href{https://doi.org/10.1088/1475-7516/2019/02/052}{\emph{JCAP} {\bfseries
  1902} (2019) 052}, [\href{https://arxiv.org/abs/1712.03944}{{\ttfamily
  1712.03944}}].

\bibitem{Castorina:2015bma}
E.~Castorina, C.~Carbone, J.~Bel, E.~Sefusatti and K.~Dolag, \emph{{DEMNUni:
  The clustering of large-scale structures in the presence of massive
  neutrinos}}, \href{https://doi.org/10.1088/1475-7516/2015/07/043}{\emph{JCAP}
  {\bfseries 1507} (2015) 043},
  [\href{https://arxiv.org/abs/1505.07148}{{\ttfamily 1505.07148}}].

\bibitem{Carbone:2016nzj}
C.~Carbone, M.~Petkova and K.~Dolag, \emph{{DEMNUni: ISW, Rees-Sciama, and
  weak-lensing in the presence of massive neutrinos}},
  \href{https://doi.org/10.1088/1475-7516/2016/07/034}{\emph{JCAP} {\bfseries
  1607} (2016) 034}, [\href{https://arxiv.org/abs/1605.02024}{{\ttfamily
  1605.02024}}].

\bibitem{Ruggeri:2017dda}
R.~Ruggeri, E.~Castorina, C.~Carbone and E.~Sefusatti, \emph{{DEMNUni: Massive
  neutrinos and the bispectrum of large scale structures}},
  \href{https://doi.org/10.1088/1475-7516/2018/03/003}{\emph{JCAP} {\bfseries
  1803} (2018) 003}, [\href{https://arxiv.org/abs/1712.02334}{{\ttfamily
  1712.02334}}].

\bibitem{Dehnen:2011fj}
W.~Dehnen and J.~Read, \emph{{N-body simulations of gravitational dynamics}},
  \href{https://doi.org/10.1140/epjp/i2011-11055-3}{\emph{Eur. Phys. J. Plus}
  {\bfseries 126} (2011) 55},
  [\href{https://arxiv.org/abs/1105.1082}{{\ttfamily 1105.1082}}].

\bibitem{Dolag:2008ki}
K.~Dolag, S.~Borgani, S.~Schindler, A.~Diaferio and A.~M. Bykov,
  \emph{{Simulation techniques for cosmological simulations}},
  \href{https://doi.org/10.1007/s11214-008-9316-5}{\emph{Space Sci. Rev.}
  {\bfseries 134} (2008) 229},
  [\href{https://arxiv.org/abs/0801.1023}{{\ttfamily 0801.1023}}].

\bibitem{Kuhlen:2012ft}
M.~Kuhlen, M.~Vogelsberger and R.~Angulo, \emph{{Numerical Simulations of the
  Dark Universe: State of the Art and the Next Decade}},
  \href{https://doi.org/10.1016/j.dark.2012.10.002}{\emph{Phys. Dark Univ.}
  {\bfseries 1} (2012) 50--93},
  [\href{https://arxiv.org/abs/1209.5745}{{\ttfamily 1209.5745}}].

\bibitem{Kaiser:1984sw}
N.~Kaiser, \emph{{On the Spatial correlations of Abell clusters}},
  \href{https://doi.org/10.1086/184341}{\emph{Astrophys. J.} {\bfseries 284}
  (1984) L9--L12}.

\bibitem{Bardeen:1985tr}
J.~M. Bardeen, J.~R. Bond, N.~Kaiser and A.~S. Szalay, \emph{{The Statistics of
  Peaks of Gaussian Random Fields}},
  \href{https://doi.org/10.1086/164143}{\emph{Astrophys. J.} {\bfseries 304}
  (1986) 15--61}.

\bibitem{Mo:1995cs}
H.~J. Mo and S.~D.~M. White, \emph{{An Analytic model for the spatial
  clustering of dark matter halos}},
  \href{https://doi.org/10.1093/mnras/282.2.347}{\emph{Mon. Not. Roy. Astron.
  Soc.} {\bfseries 282} (1996) 347},
  [\href{https://arxiv.org/abs/astro-ph/9512127}{{\ttfamily
  astro-ph/9512127}}].

\bibitem{Fry:1996fg}
J.~N. Fry, \emph{{The Evolution of Bias}},
  \href{https://doi.org/10.1086/310006}{\emph{Astrophys. J.} {\bfseries 461}
  (1996) L65}.

\bibitem{Mann:1997df}
B.~Mann, J.~Peacock and A.~Heavens, \emph{{Eulerian bias and the galaxy density
  field}}, \href{https://doi.org/10.1046/j.1365-8711.1998.01053.x}{\emph{Mon.
  Not. Roy. Astron. Soc.} {\bfseries 293} (1998) 209--221},
  [\href{https://arxiv.org/abs/astro-ph/9708031}{{\ttfamily
  astro-ph/9708031}}].

\bibitem{Tegmark:1998wm}
M.~Tegmark and P.~J.~E. Peebles, \emph{{The Time evolution of bias}},
  \href{https://doi.org/10.1086/311426}{\emph{Astrophys. J.} {\bfseries 500}
  (1998) L79}, [\href{https://arxiv.org/abs/astro-ph/9804067}{{\ttfamily
  astro-ph/9804067}}].

\bibitem{Desjacques:2016bnm}
V.~Desjacques, D.~Jeong and F.~Schmidt, \emph{{Large-Scale Galaxy Bias}},
  \href{https://doi.org/10.1016/j.physrep.2017.12.002}{\emph{Phys. Rept.}
  {\bfseries 733} (2018) 1--193},
  [\href{https://arxiv.org/abs/1611.09787}{{\ttfamily 1611.09787}}].

\bibitem{Press:1973iz}
W.~H. Press and P.~Schechter, \emph{{Formation of galaxies and clusters of
  galaxies by selfsimilar gravitational condensation}},
  \href{https://doi.org/10.1086/152650}{\emph{Astrophys. J.} {\bfseries 187}
  (1974) 425--438}.

\bibitem{Sheth:2001dp}
R.~K. Sheth and G.~Tormen, \emph{{An Excursion Set Model of Hierarchical
  Clustering : Ellipsoidal Collapse and the Moving Barrier}},
  \href{https://doi.org/10.1046/j.1365-8711.2002.04950.x}{\emph{Mon. Not. Roy.
  Astron. Soc.} {\bfseries 329} (2002) 61},
  [\href{https://arxiv.org/abs/astro-ph/0105113}{{\ttfamily
  astro-ph/0105113}}].

\bibitem{Dekek:1986gu}
A.~Dekel and J.~Silk, \emph{{The origin of dwarf galaxies, cold dark matter,
  and biased galaxy formation}},
  \href{https://doi.org/10.1086/164050}{\emph{Astrophys. J.} {\bfseries 303}
  (1986) 39--55}.

\bibitem{Cole:1989ghw}
S.~{Cole} and N.~{Kaiser}, \emph{{Biased clustering in the cold dark matter
  cosmogony}}, \href{https://doi.org/10.1093/mnras/237.4.1127}{\emph{Mon. Not.
  Roy. Astron. Soc.} {\bfseries 237} (Apr., 1989) 1127--1146}.

\bibitem{Fry:1992vr}
J.~N. Fry and E.~Gaztañaga, \emph{{Biasing and hierarchical statistics in
  large scale structure}},
  \href{https://doi.org/10.1086/173015}{\emph{Astrophys. J.} {\bfseries 413}
  (1993) 447--452}, [\href{https://arxiv.org/abs/astro-ph/9302009}{{\ttfamily
  astro-ph/9302009}}].

\bibitem{Coles:1993ghw}
P.~{Coles}, \emph{{Galaxy formation with a local bias}},
  \href{https://doi.org/10.1093/mnras/262.4.1065}{\emph{Mon. Not. Roy. Astron.
  Soc.} {\bfseries 262} (June, 1993) 1065--1075}.

\bibitem{Kauffmann:1995kx}
G.~Kauffmann, A.~Nusser and M.~Steinmetz, \emph{{Galaxy formation and large
  scale bias}}, \href{https://doi.org/10.1093/mnras/286.4.795}{\emph{Mon. Not.
  Roy. Astron. Soc.} {\bfseries 286} (1997) 795--811},
  [\href{https://arxiv.org/abs/astro-ph/9512009}{{\ttfamily
  astro-ph/9512009}}].

\bibitem{Sheth:1999mn}
R.~K. Sheth and G.~Tormen, \emph{{Large scale bias and the peak background
  split}}, \href{https://doi.org/10.1046/j.1365-8711.1999.02692.x}{\emph{Mon.
  Not. Roy. Astron. Soc.} {\bfseries 308} (1999) 119},
  [\href{https://arxiv.org/abs/astro-ph/9901122}{{\ttfamily
  astro-ph/9901122}}].

\bibitem{Benson:1999mva}
A.~J. Benson, S.~Cole, C.~S. Frenk, C.~M. Baugh and C.~G. Lacey, \emph{{The
  Nature of galaxy bias and clustering}},
  \href{https://doi.org/10.1046/j.1365-8711.2000.03101.x}{\emph{Mon. Not. Roy.
  Astron. Soc.} {\bfseries 311} (2000) 793--808},
  [\href{https://arxiv.org/abs/astro-ph/9903343}{{\ttfamily
  astro-ph/9903343}}].

\bibitem{Matsubara:1999qq}
T.~Matsubara, \emph{{Stochasticity of bias and nonlocality of galaxy formation:
  Linear scales}}, \href{https://doi.org/10.1086/307931}{\emph{Astrophys. J.}
  {\bfseries 525} (1999) 543--553},
  [\href{https://arxiv.org/abs/astro-ph/9906029}{{\ttfamily
  astro-ph/9906029}}].

\bibitem{Coles:2007be}
P.~Coles and P.~Erdoğdu, \emph{{Scale-dependent Galaxy Bias}},
  \href{https://doi.org/10.1088/1475-7516/2007/10/007}{\emph{JCAP} {\bfseries
  0710} (2007) 007}, [\href{https://arxiv.org/abs/0706.0412}{{\ttfamily
  0706.0412}}].

\bibitem{Desjacques:2010gz}
V.~Desjacques, M.~Crocce, R.~Scoccimarro and R.~K. Sheth, \emph{{Modeling
  scale-dependent bias on the baryonic acoustic scale with the statistics of
  peaks of Gaussian random fields}},
  \href{https://doi.org/10.1103/PhysRevD.82.103529}{\emph{Phys. Rev.}
  {\bfseries D82} (2010) 103529},
  [\href{https://arxiv.org/abs/1009.3449}{{\ttfamily 1009.3449}}].

\bibitem{Assassi:2014fva}
V.~Assassi, D.~Baumann, D.~Green and M.~Zaldarriaga, \emph{{Renormalized Halo
  Bias}}, \href{https://doi.org/10.1088/1475-7516/2014/08/056}{\emph{JCAP}
  {\bfseries 1408} (2014) 056},
  [\href{https://arxiv.org/abs/1402.5916}{{\ttfamily 1402.5916}}].

\bibitem{Burbidge:1967uc}
E.~M. Burbidge, \emph{{Quasi-stellar objects}},
  \href{https://doi.org/10.1146/annurev.aa.05.090167.002151}{\emph{Ann. Rev.
  Astron. Astrophys.} {\bfseries 5} (1967) 399--452}.

\bibitem{Schmidt:1969vq}
M.~Schmidt, \emph{{Quasistellar objects}},
  \href{https://doi.org/10.1146/annurev.aa.07.090169.002523}{\emph{Ann. Rev.
  Astron. Astrophys.} {\bfseries 7} (1969) 527--552}.

\bibitem{Antonucci:1993sg}
R.~Antonucci, \emph{{Unified models for active galactic nuclei and quasars}},
  \href{https://doi.org/10.1146/annurev.aa.31.090193.002353}{\emph{Ann. Rev.
  Astron. Astrophys.} {\bfseries 31} (1993) 473--521}.

\bibitem{Slosar:2008hx}
A.~Slosar, C.~Hirata, U.~Seljak, S.~Ho and N.~Padmanabhan, \emph{{Constraints
  on local primordial non-Gaussianity from large scale structure}},
  \href{https://doi.org/10.1088/1475-7516/2008/08/031}{\emph{JCAP} {\bfseries
  0808} (2008) 031}, [\href{https://arxiv.org/abs/0805.3580}{{\ttfamily
  0805.3580}}].

\bibitem{Han:2018izq}
J.~Han, S.~Ferraro, E.~Giusarma and S.~Ho, \emph{{Probing Gravitational Lensing
  of the CMB with SDSS-IV Quasars}},
  \href{https://doi.org/10.1093/mnras/stz528}{\emph{Mon. Not. Roy. Astron.
  Soc.} {\bfseries 485} (2019) 1720--1726},
  [\href{https://arxiv.org/abs/1809.04196}{{\ttfamily 1809.04196}}].

\bibitem{Smith:2006ne}
R.~E. Smith, R.~Scoccimarro and R.~K. Sheth, \emph{{The Scale Dependence of
  Halo and Galaxy Bias: Effects in Real Space}},
  \href{https://doi.org/10.1103/PhysRevD.75.063512}{\emph{Phys. Rev.}
  {\bfseries D75} (2007) 063512},
  [\href{https://arxiv.org/abs/astro-ph/0609547}{{\ttfamily
  astro-ph/0609547}}].

\bibitem{Desjacques:2008jj}
V.~Desjacques, \emph{{Baryon acoustic signature in the clustering of density
  maxima}}, \href{https://doi.org/10.1103/PhysRevD.78.103503}{\emph{Phys. Rev.}
  {\bfseries D78} (2008) 103503},
  [\href{https://arxiv.org/abs/0806.0007}{{\ttfamily 0806.0007}}].

\bibitem{Musso:2012ch}
M.~Musso, A.~Paranjape and R.~K. Sheth, \emph{{Scale dependent halo bias in the
  excursion set approach}},
  \href{https://doi.org/10.1111/j.1365-2966.2012.21903.x}{\emph{Mon. Not. Roy.
  Astron. Soc.} {\bfseries 427} (2012) 3145--3158},
  [\href{https://arxiv.org/abs/1205.3401}{{\ttfamily 1205.3401}}].

\bibitem{Paranjape:2012ks}
A.~Paranjape and R.~K. Sheth, \emph{{Peaks theory and the excursion set
  approach}},
  \href{https://doi.org/10.1111/j.1365-2966.2012.21911.x}{\emph{Mon. Not. Roy.
  Astron. Soc.} {\bfseries 426} (2012) 2789--2796},
  [\href{https://arxiv.org/abs/1206.3506}{{\ttfamily 1206.3506}}].

\bibitem{Schmidt:2012ys}
F.~Schmidt, D.~Jeong and V.~Desjacques, \emph{{Peak-Background Split,
  Renormalization, and Galaxy Clustering}},
  \href{https://doi.org/10.1103/PhysRevD.88.023515}{\emph{Phys. Rev.}
  {\bfseries D88} (2013) 023515},
  [\href{https://arxiv.org/abs/1212.0868}{{\ttfamily 1212.0868}}].

\bibitem{Verde:2014nwa}
L.~Verde, R.~Jiménez, F.~Simpson, L.~Álvarez Gaumé, A.~Heavens and
  S.~Matarrese, \emph{{The bias of weighted dark matter haloes from peak
  theory}}, \href{https://doi.org/10.1093/mnras/stu1164}{\emph{Mon. Not. Roy.
  Astron. Soc.} {\bfseries 443} (2014) 122--137},
  [\href{https://arxiv.org/abs/1404.2241}{{\ttfamily 1404.2241}}].

\bibitem{Biagetti:2014pha}
M.~Biagetti, V.~Desjacques, A.~Kehagias and A.~Riotto, \emph{{Nonlocal halo
  bias with and without massive neutrinos}},
  \href{https://doi.org/10.1103/PhysRevD.90.045022}{\emph{Phys. Rev.}
  {\bfseries D90} (2014) 045022},
  [\href{https://arxiv.org/abs/1405.1435}{{\ttfamily 1405.1435}}].

\bibitem{Senatore:2014eva}
L.~Senatore, \emph{{Bias in the Effective Field Theory of Large Scale
  Structures}},
  \href{https://doi.org/10.1088/1475-7516/2015/11/007}{\emph{JCAP} {\bfseries
  1511} (2015) 007}, [\href{https://arxiv.org/abs/1406.7843}{{\ttfamily
  1406.7843}}].

\bibitem{Mirbabayi:2014zca}
M.~Mirbabayi, F.~Schmidt and M.~Zaldarriaga, \emph{{Biased Tracers and Time
  Evolution}}, \href{https://doi.org/10.1088/1475-7516/2015/07/030}{\emph{JCAP}
  {\bfseries 1507} (2015) 030},
  [\href{https://arxiv.org/abs/1412.5169}{{\ttfamily 1412.5169}}].

\bibitem{Jackson:2008yv}
J.~C. Jackson, \emph{{Fingers of God: A critique of Rees' theory of primoridal
  gravitational radiation}},
  \href{https://doi.org/10.1093/mnras/156.1.1P}{\emph{Mon. Not. Roy. Astron.
  Soc.} {\bfseries 156} (1972) 1P--5P},
  [\href{https://arxiv.org/abs/0810.3908}{{\ttfamily 0810.3908}}].

\bibitem{Kaiser:1987qv}
N.~Kaiser, \emph{{Clustering in real space and in redshift space}}, {\emph{Mon.
  Not. Roy. Astron. Soc.} {\bfseries 227} (1987) 1--27}.

\bibitem{Hamilton:1997zq}
A.~J.~S. Hamilton, \emph{{Linear redshift distortions: A Review}},  in
  \emph{{Ringberg Workshop on Large Scale Structure Ringberg, Germany,
  September 23-28, 1996}}, 1997,
  \href{https://arxiv.org/abs/astro-ph/9708102}{{\ttfamily astro-ph/9708102}}.

\bibitem{Percival:2011ghw}
W.~J. {Percival}, L.~{Samushia}, A.~J. {Ross}, C.~{Shapiro} and
  A.~{Raccanelli}, \emph{{Redshift-space distortions}}, {\emph{Philosophical
  Transactions of the Royal Society of London Series A} {\bfseries 369} (2011)
  5058--5067}.

\bibitem{Heavens:1998es}
A.~F. Heavens, S.~Matarrese and L.~Verde, \emph{{The Nonlinear redshift-space
  power spectrum of galaxies}},
  \href{https://doi.org/10.1046/j.1365-8711.1998.02052.x}{\emph{Mon. Not. Roy.
  Astron. Soc.} {\bfseries 301} (1998) 797--808},
  [\href{https://arxiv.org/abs/astro-ph/9808016}{{\ttfamily
  astro-ph/9808016}}].

\bibitem{Bharadwaj:2001zf}
S.~Bharadwaj, \emph{{Nonlinear redshift distortions: The Two point correlation
  function}},
  \href{https://doi.org/10.1046/j.1365-8711.2001.04738.x}{\emph{Mon. Not. Roy.
  Astron. Soc.} {\bfseries 327} (2001) 577--587},
  [\href{https://arxiv.org/abs/astro-ph/0105320}{{\ttfamily
  astro-ph/0105320}}].

\bibitem{Pandey:2004jf}
B.~Pandey and S.~Bharadwaj, \emph{{Modeling non-linear effects in the redshift
  space two - point correlation function and its implications for the pairwise
  velocity dispersion}},
  \href{https://doi.org/10.1111/j.1365-2966.2005.08835.x}{\emph{Mon. Not. Roy.
  Astron. Soc.} {\bfseries 358} (2005) 939--948},
  [\href{https://arxiv.org/abs/astro-ph/0403670}{{\ttfamily
  astro-ph/0403670}}].

\bibitem{Matsubara:2007wj}
T.~Matsubara, \emph{{Resumming Cosmological Perturbations via the Lagrangian
  Picture: One-loop Results in Real Space and in Redshift Space}},
  \href{https://doi.org/10.1103/PhysRevD.77.063530}{\emph{Phys. Rev.}
  {\bfseries D77} (2008) 063530},
  [\href{https://arxiv.org/abs/0711.2521}{{\ttfamily 0711.2521}}].

\bibitem{Shaw:2008aa}
J.~R. Shaw and A.~Lewis, \emph{{Non-linear Redshift-Space Power Spectra}},
  \href{https://doi.org/10.1103/PhysRevD.78.103512}{\emph{Phys. Rev.}
  {\bfseries D78} (2008) 103512},
  [\href{https://arxiv.org/abs/0808.1724}{{\ttfamily 0808.1724}}].

\bibitem{Samushia:2011cs}
L.~Samushia, W.~J. Percival and A.~Raccanelli, \emph{{Interpreting large-scale
  redshift-space distortion measurements}},
  \href{https://doi.org/10.1111/j.1365-2966.2011.20169.x}{\emph{Mon. Not. Roy.
  Astron. Soc.} {\bfseries 420} (2012) 2102--2119},
  [\href{https://arxiv.org/abs/1102.1014}{{\ttfamily 1102.1014}}].

\bibitem{Sato:2011qr}
M.~Sato and T.~Matsubara, \emph{{Nonlinear Biasing and Redshift-Space
  Distortions in Lagrangian Resummation Theory and N-body Simulations}},
  \href{https://doi.org/10.1103/PhysRevD.84.043501}{\emph{Phys. Rev.}
  {\bfseries D84} (2011) 043501},
  [\href{https://arxiv.org/abs/1105.5007}{{\ttfamily 1105.5007}}].

\bibitem{delaTorre:2012dg}
S.~de~la Torre and L.~Guzzo, \emph{{Modelling non-linear redshift-space
  distortions in the galaxy clustering pattern: systematic errors on the growth
  rate parameter}},
  \href{https://doi.org/10.1111/j.1365-2966.2012.21824.x}{\emph{Mon. Not. Roy.
  Astron. Soc.} {\bfseries 427} (2012) 327},
  [\href{https://arxiv.org/abs/1202.5559}{{\ttfamily 1202.5559}}].

\bibitem{Jennings:2015lea}
E.~Jennings, R.~H. Wechsler, S.~W. Skillman and M.~S. Warren,
  \emph{{Disentangling redshift-space distortions and non-linear bias using the
  2D power spectrum}}, \href{https://doi.org/10.1093/mnras/stv2989}{\emph{Mon.
  Not. Roy. Astron. Soc.} {\bfseries 457} (2016) 1076--1088},
  [\href{https://arxiv.org/abs/1508.01803}{{\ttfamily 1508.01803}}].

\bibitem{Zhu:2017vtj}
H.-M. Zhu, Y.~Yu and U.-L. Pen, \emph{{Nonlinear reconstruction of redshift
  space distortions}},
  \href{https://doi.org/10.1103/PhysRevD.97.043502}{\emph{Phys. Rev.}
  {\bfseries D97} (2018) 043502},
  [\href{https://arxiv.org/abs/1711.03218}{{\ttfamily 1711.03218}}].

\bibitem{Hernandez-Aguayo:2018oxg}
C.~Hernández-Aguayo, J.~Hou, B.~Li, C.~M. Baugh and A.~G. Sánchez,
  \emph{{Large-scale redshift space distortions in modified gravity theories}},
  \href{https://doi.org/10.1093/mnras/stz516}{\emph{Mon. Not. Roy. Astron.
  Soc.} {\bfseries 485} (2019) 2194--2213},
  [\href{https://arxiv.org/abs/1811.09197}{{\ttfamily 1811.09197}}].

\bibitem{Jullo:2019lgq}
E.~Jullo et~al., \emph{{Testing gravity with galaxy-galaxy lensing and
  redshift-space distortions using CFHT-Stripe 82, CFHTLenS and BOSS CMASS
  datasets}},  \href{https://arxiv.org/abs/1903.07160}{{\ttfamily 1903.07160}}.

\bibitem{Boss:2009ghw}
{BOSS collaboration}, \emph{http://www.sdss3.org/surveys/boss.php},  2009.

\bibitem{Totsujikihara}
H.~{Totsuji} and T.~{Kihara}, \emph{{The Correlation Function for the
  Distribution of Galaxies}}, {\emph{Publ.Astron.Soc.Jap.} {\bfseries 21}
  (1969) 221}.

\bibitem{Peebles}
P.~J.~E. {Peebles}, \emph{{Statistical Analysis of Catalogs of Extragalactic
  Objects. I. Theory}}, \href{https://doi.org/10.1086/152431}{\emph{Astrophys.
  J.} {\bfseries 185} (Oct., 1973) 413--440}.

\bibitem{Hauserpeebles}
M.~G. {Hauser} and P.~J.~E. {Peebles}, \emph{{Statistical Analysis of Catalogs
  of Extragalactic Objects. II. the Abell Catalog of Rich Clusters}},
  \href{https://doi.org/10.1086/152453}{\emph{Astrophys. J.} {\bfseries 185}
  (Nov., 1973) 757--786}.

\bibitem{Peebleshauser}
P.~J.~E. {Peebles} and M.~G. {Hauser}, \emph{{Statistical Analysis of Catalogs
  of Extragalactic Objects. III. The Shane-Wirtanen and Zwicky Catalogs}},
  \href{https://doi.org/10.1086/190308}{\emph{Astrophys.J.Suppl.} {\bfseries
  28} (Nov., 1974) 19}.

\bibitem{Peeblesnew}
P.~J.~E. {Peebles}, \emph{{The Nature of the Distribution of Galaxies}},
  {\emph{Astron.Astrophys.} {\bfseries 32} (May, 1974) 197}.

\bibitem{Watson:2011cz}
D.~F. Watson, A.~A. Berlind and A.~R. Zentner, \emph{{A Cosmic Coincidence: The
  Power-Law Galaxy Correlation Function}},
  \href{https://doi.org/10.1088/0004-637X/738/1/22}{\emph{Astrophys. J.}
  {\bfseries 738} (2011) 22},
  [\href{https://arxiv.org/abs/1101.5155}{{\ttfamily 1101.5155}}].

\bibitem{Bassett:2009mm}
B.~A. Bassett and R.~Hložek, \emph{{Baryon Acoustic Oscillations}},
  \href{https://arxiv.org/abs/0910.5224}{{\ttfamily 0910.5224}}.

\bibitem{Percival:2007yw}
W.~J. Percival, S.~Cole, D.~J. Eisenstein, R.~C. Nichol, J.~A. Peacock, A.~C.
  Pope et~al., \emph{{Measuring the Baryon Acoustic Oscillation scale using the
  SDSS and 2dFGRS}},
  \href{https://doi.org/10.1111/j.1365-2966.2007.12268.x}{\emph{Mon. Not. Roy.
  Astron. Soc.} {\bfseries 381} (2007) 1053--1066},
  [\href{https://arxiv.org/abs/0705.3323}{{\ttfamily 0705.3323}}].

\bibitem{Aubourg:2014yra}
E.~Aubourg et~al., \emph{{Cosmological implications of baryon acoustic
  oscillation measurements}},
  \href{https://doi.org/10.1103/PhysRevD.92.123516}{\emph{Phys. Rev.}
  {\bfseries D92} (2015) 123516},
  [\href{https://arxiv.org/abs/1411.1074}{{\ttfamily 1411.1074}}].

\bibitem{Addison:2017fdm}
G.~E. Addison, D.~J. Watts, C.~L. Bennett, M.~Halpern, G.~Hinshaw and J.~L.
  Weiland, \emph{{Elucidating $\Lambda$CDM: Impact of Baryon Acoustic
  Oscillation Measurements on the Hubble Constant Discrepancy}},
  \href{https://doi.org/10.3847/1538-4357/aaa1ed}{\emph{Astrophys. J.}
  {\bfseries 853} (2018) 119},
  [\href{https://arxiv.org/abs/1707.06547}{{\ttfamily 1707.06547}}].

\bibitem{Hannestad:2004nb}
S.~Hannestad, \emph{{Neutrinos in cosmology}},
  \href{https://doi.org/10.1088/1367-2630/6/1/108}{\emph{New J. Phys.}
  {\bfseries 6} (2004) 108},
  [\href{https://arxiv.org/abs/hep-ph/0404239}{{\ttfamily hep-ph/0404239}}].

\bibitem{Fukugita:2005sb}
M.~Fukugita, \emph{{Massive neutrinos in cosmology}},
  \href{https://doi.org/10.1016/j.nuclphysbps.2006.02.002}{\emph{Nucl. Phys.
  Proc. Suppl.} {\bfseries 155} (2006) 10--17},
  [\href{https://arxiv.org/abs/hep-ph/0511068}{{\ttfamily hep-ph/0511068}}].

\bibitem{Hannestad:2005ey}
S.~Hannestad, \emph{{Introduction to neutrino cosmology}},
  \href{https://doi.org/10.1016/j.ppnp.2005.11.028}{\emph{Prog. Part. Nucl.
  Phys.} {\bfseries 57} (2006) 309--323},
  [\href{https://arxiv.org/abs/astro-ph/0511595}{{\ttfamily
  astro-ph/0511595}}].

\bibitem{Lesgourgues:2006nd}
J.~Lesgourgues and S.~Pastor, \emph{{Massive neutrinos and cosmology}},
  \href{https://doi.org/10.1016/j.physrep.2006.04.001}{\emph{Phys. Rept.}
  {\bfseries 429} (2006) 307--379},
  [\href{https://arxiv.org/abs/astro-ph/0603494}{{\ttfamily
  astro-ph/0603494}}].

\bibitem{Dolgov:2008hz}
A.~D. Dolgov, \emph{{Cosmology and Neutrino Properties}},
  \href{https://doi.org/10.1134/S1063778808120181}{\emph{Phys. Atom. Nucl.}
  {\bfseries 71} (2008) 2152--2164},
  [\href{https://arxiv.org/abs/0803.3887}{{\ttfamily 0803.3887}}].

\bibitem{Hannestad:2010kz}
S.~Hannestad, \emph{{Neutrino physics from precision cosmology}},
  \href{https://doi.org/10.1016/j.ppnp.2010.07.001}{\emph{Prog. Part. Nucl.
  Phys.} {\bfseries 65} (2010) 185--208},
  [\href{https://arxiv.org/abs/1007.0658}{{\ttfamily 1007.0658}}].

\bibitem{Wong:2011ip}
Y.~Y.~Y. Wong, \emph{{Neutrino mass in cosmology: status and prospects}},
  \href{https://doi.org/10.1146/annurev-nucl-102010-130252}{\emph{Ann. Rev.
  Nucl. Part. Sci.} {\bfseries 61} (2011) 69--98},
  [\href{https://arxiv.org/abs/1111.1436}{{\ttfamily 1111.1436}}].

\bibitem{Lesgourgues:2012uu}
J.~Lesgourgues and S.~Pastor, \emph{{Neutrino mass from Cosmology}},
  \href{https://doi.org/10.1155/2012/608515}{\emph{Adv. High Energy Phys.}
  {\bfseries 2012} (2012) 608515},
  [\href{https://arxiv.org/abs/1212.6154}{{\ttfamily 1212.6154}}].

\bibitem{Balantekin:2013gqa}
A.~B. Balantekin and G.~M. Fuller, \emph{{Neutrinos in Cosmology and
  Astrophysics}}, \href{https://doi.org/10.1016/j.ppnp.2013.03.008}{\emph{Prog.
  Part. Nucl. Phys.} {\bfseries 71} (2013) 162--166},
  [\href{https://arxiv.org/abs/1303.3874}{{\ttfamily 1303.3874}}].

\bibitem{Lesgourgues:2014zoa}
J.~Lesgourgues and S.~Pastor, \emph{{Neutrino cosmology and Planck}},
  \href{https://doi.org/10.1088/1367-2630/16/6/065002}{\emph{New J. Phys.}
  {\bfseries 16} (2014) 065002},
  [\href{https://arxiv.org/abs/1404.1740}{{\ttfamily 1404.1740}}].

\bibitem{Verde:2015ana}
L.~Verde, \emph{{Neutrino properties from cosmology}},
  \href{https://doi.org/10.1088/1742-6596/598/1/012010}{\emph{J. Phys. Conf.
  Ser.} {\bfseries 598} (2015) 012010}.

\bibitem{Abazajian:2016hbv}
K.~N. Abazajian and M.~Kaplinghat, \emph{{Neutrino Physics from the Cosmic
  Microwave Background and Large-Scale Structure}},
  \href{https://doi.org/10.1146/annurev-nucl-102014-021908}{\emph{Ann. Rev.
  Nucl. Part. Sci.} {\bfseries 66} (2016) 401--420}.

\bibitem{Archidiacono:2017tlz}
M.~Archidiacono, T.~Brinckmann, J.~Lesgourgues and V.~Poulin, \emph{{Neutrino
  properties from cosmology}},  in \emph{{Proceedings, Prospects in Neutrino
  Physics (NuPhys2016): London, UK, December 12-14, 2016}}, 2017,
  \href{https://arxiv.org/abs/1705.00496}{{\ttfamily 1705.00496}}.

\bibitem{Gerbino:2018jee}
M.~Gerbino, \emph{{Neutrino properties from cosmology}},  in
  \emph{{Proceedings, Prospects in Neutrino Physics (NuPhys2017): London, UK,
  December 20-22, 2017}}, pp.~52--52, 2018,
  \href{https://arxiv.org/abs/1803.11545}{{\ttfamily 1803.11545}}.

\bibitem{Hu:1995en}
W.~Hu and N.~Sugiyama, \emph{{Small scale cosmological perturbations: An
  Analytic approach}}, \href{https://doi.org/10.1086/177989}{\emph{Astrophys.
  J.} {\bfseries 471} (1996) 542--570},
  [\href{https://arxiv.org/abs/astro-ph/9510117}{{\ttfamily
  astro-ph/9510117}}].

\bibitem{Bashinsky:2003tk}
S.~Bashinsky and U.~Seljak, \emph{{Neutrino perturbations in CMB anisotropy and
  matter clustering}},
  \href{https://doi.org/10.1103/PhysRevD.69.083002}{\emph{Phys. Rev.}
  {\bfseries D69} (2004) 083002},
  [\href{https://arxiv.org/abs/astro-ph/0310198}{{\ttfamily
  astro-ph/0310198}}].

\bibitem{Hou:2011ec}
Z.~Hou, R.~Keisler, L.~Knox, M.~Millea and C.~Reichardt, \emph{{How Massless
  Neutrinos Affect the Cosmic Microwave Background Damping Tail}},
  \href{https://doi.org/10.1103/PhysRevD.87.083008}{\emph{Phys. Rev.}
  {\bfseries D87} (2013) 083008},
  [\href{https://arxiv.org/abs/1104.2333}{{\ttfamily 1104.2333}}].

\bibitem{Pisanti:2007hk}
O.~Pisanti, A.~Cirillo, S.~Esposito, F.~Iocco, G.~Mangano, G.~Miele et~al.,
  \emph{{PArthENoPE: Public Algorithm Evaluating the Nucleosynthesis of
  Primordial Elements}},
  \href{https://doi.org/10.1016/j.cpc.2008.02.015}{\emph{Comput. Phys. Commun.}
  {\bfseries 178} (2008) 956--971},
  [\href{https://arxiv.org/abs/0705.0290}{{\ttfamily 0705.0290}}].

\bibitem{Iocco:2008va}
F.~Iocco, G.~Mangano, G.~Miele, O.~Pisanti and P.~D. Serpico, \emph{{Primordial
  Nucleosynthesis: from precision cosmology to fundamental physics}},
  \href{https://doi.org/10.1016/j.physrep.2009.02.002}{\emph{Phys. Rept.}
  {\bfseries 472} (2009) 1--76},
  [\href{https://arxiv.org/abs/0809.0631}{{\ttfamily 0809.0631}}].

\bibitem{Aver:2010wq}
E.~Aver, K.~A. Olive and E.~D. Skillman, \emph{{A New Approach to Systematic
  Uncertainties and Self-Consistency in Helium Abundance Determinations}},
  \href{https://doi.org/10.1088/1475-7516/2010/05/003}{\emph{JCAP} {\bfseries
  1005} (2010) 003}, [\href{https://arxiv.org/abs/1001.5218}{{\ttfamily
  1001.5218}}].

\bibitem{Izotov:2010ca}
Y.~I. Izotov and T.~X. Thuan, \emph{{The primordial abundance of 4He: evidence
  for non-standard big bang nucleosynthesis}},
  \href{https://doi.org/10.1088/2041-8205/710/1/L67}{\emph{Astrophys. J.}
  {\bfseries 710} (2010) L67--L71},
  [\href{https://arxiv.org/abs/1001.4440}{{\ttfamily 1001.4440}}].

\bibitem{Consiglio:2017pot}
R.~Consiglio, P.~F. de~Salas, G.~Mangano, G.~Miele, S.~Pastor and O.~Pisanti,
  \emph{{PArthENoPE reloaded}},
  \href{https://doi.org/10.1016/j.cpc.2018.06.022}{\emph{Comput. Phys. Commun.}
  {\bfseries 233} (2018) 237--242},
  [\href{https://arxiv.org/abs/1712.04378}{{\ttfamily 1712.04378}}].

\bibitem{Brandbyge:2008rv}
J.~Brandbyge, S.~Hannestad, T.~Haugbølle and B.~Thomsen, \emph{{The Effect of
  Thermal Neutrino Motion on the Non-linear Cosmological Matter Power
  Spectrum}}, \href{https://doi.org/10.1088/1475-7516/2008/08/020}{\emph{JCAP}
  {\bfseries 0808} (2008) 020},
  [\href{https://arxiv.org/abs/0802.3700}{{\ttfamily 0802.3700}}].

\bibitem{Saito:2009ah}
S.~Saito, M.~Takada and A.~Taruya, \emph{{Nonlinear power spectrum in the
  presence of massive neutrinos: perturbation theory approach, galaxy bias and
  parameter forecasts}},
  \href{https://doi.org/10.1103/PhysRevD.80.083528}{\emph{Phys. Rev.}
  {\bfseries D80} (2009) 083528},
  [\href{https://arxiv.org/abs/0907.2922}{{\ttfamily 0907.2922}}].

\bibitem{Viel:2010bn}
M.~Viel, M.~G. Haehnelt and V.~Springel, \emph{{The effect of neutrinos on the
  matter distribution as probed by the Intergalactic Medium}},
  \href{https://doi.org/10.1088/1475-7516/2010/06/015}{\emph{JCAP} {\bfseries
  1006} (2010) 015}, [\href{https://arxiv.org/abs/1003.2422}{{\ttfamily
  1003.2422}}].

\bibitem{Bird:2011rb}
S.~Bird, M.~Viel and M.~G. Haehnelt, \emph{{Massive Neutrinos and the
  Non-linear Matter Power Spectrum}},
  \href{https://doi.org/10.1111/j.1365-2966.2011.20222.x}{\emph{Mon. Not. Roy.
  Astron. Soc.} {\bfseries 420} (2012) 2551--2561},
  [\href{https://arxiv.org/abs/1109.4416}{{\ttfamily 1109.4416}}].

\bibitem{Hannestad:2011td}
S.~Hannestad, T.~Haugbølle and C.~Schultz, \emph{{Neutrinos in Non-linear
  Structure Formation - a Simple SPH Approach}},
  \href{https://doi.org/10.1088/1475-7516/2012/02/045}{\emph{JCAP} {\bfseries
  1202} (2012) 045}, [\href{https://arxiv.org/abs/1110.1257}{{\ttfamily
  1110.1257}}].

\bibitem{Mead:2016zqy}
A.~Mead, C.~Heymans, L.~Lombriser, J.~Peacock, O.~Steele and H.~Winther,
  \emph{{Accurate halo-model matter power spectra with dark energy, massive
  neutrinos and modified gravitational forces}},
  \href{https://doi.org/10.1093/mnras/stw681}{\emph{Mon. Not. Roy. Astron.
  Soc.} {\bfseries 459} (2016) 1468--1488},
  [\href{https://arxiv.org/abs/1602.02154}{{\ttfamily 1602.02154}}].

\bibitem{Lesgourgues:2004ps}
J.~Lesgourgues, S.~Pastor and L.~Perotto, \emph{{Probing neutrino masses with
  future galaxy redshift surveys}},
  \href{https://doi.org/10.1103/PhysRevD.70.045016}{\emph{Phys. Rev.}
  {\bfseries D70} (2004) 045016},
  [\href{https://arxiv.org/abs/hep-ph/0403296}{{\ttfamily hep-ph/0403296}}].

\bibitem{Lesgourgues:2005yv}
J.~Lesgourgues, L.~Perotto, S.~Pastor and M.~Piat, \emph{{Probing neutrino
  masses with cmb lensing extraction}},
  \href{https://doi.org/10.1103/PhysRevD.73.045021}{\emph{Phys. Rev.}
  {\bfseries D73} (2006) 045021},
  [\href{https://arxiv.org/abs/astro-ph/0511735}{{\ttfamily
  astro-ph/0511735}}].

\bibitem{Pritchard:2008wy}
J.~R. Pritchard and E.~Pierpaoli, \emph{{Constraining massive neutrinos using
  cosmological 21 cm observations}},
  \href{https://doi.org/10.1103/PhysRevD.78.065009}{\emph{Phys. Rev.}
  {\bfseries D78} (2008) 065009},
  [\href{https://arxiv.org/abs/0805.1920}{{\ttfamily 0805.1920}}].

\bibitem{DeBernardis:2009di}
F.~De~Bernardis, T.~D. Kitching, A.~Heavens and A.~Melchiorri,
  \emph{{Determining the Neutrino Mass Hierarchy with Cosmology}},
  \href{https://doi.org/10.1103/PhysRevD.80.123509}{\emph{Phys. Rev.}
  {\bfseries D80} (2009) 123509},
  [\href{https://arxiv.org/abs/0907.1917}{{\ttfamily 0907.1917}}].

\bibitem{Jimenez:2010ev}
R.~Jiménez, T.~Kitching, C.~Peña-Garay and L.~Verde, \emph{{Can we measure
  the neutrino mass hierarchy in the sky?}},
  \href{https://doi.org/10.1088/1475-7516/2010/05/035}{\emph{JCAP} {\bfseries
  1005} (2010) 035}, [\href{https://arxiv.org/abs/1003.5918}{{\ttfamily
  1003.5918}}].

\bibitem{Hall:2012kg}
A.~C. Hall and A.~Challinor, \emph{{Probing the neutrino mass hierarchy with
  CMB weak lensing}},
  \href{https://doi.org/10.1111/j.1365-2966.2012.21493.x}{\emph{Mon. Not. Roy.
  Astron. Soc.} {\bfseries 425} (2012) 1170--1184},
  [\href{https://arxiv.org/abs/1205.6172}{{\ttfamily 1205.6172}}].

\bibitem{Jimenez:2016ckl}
C.~Peña-Garay, L.~Verde and R.~Jiménez, \emph{{Neutrino footprint in Large
  Scale Structure}},
  \href{https://doi.org/10.1016/j.dark.2016.11.004}{\emph{Phys. Dark Univ.}
  {\bfseries 15} (2017) 31--34},
  [\href{https://arxiv.org/abs/1602.08430}{{\ttfamily 1602.08430}}].

\bibitem{Verde:2007wf}
L.~Verde, \emph{{A practical guide to Basic Statistical Techniques for Data
  Analysis in Cosmology}},  \href{https://arxiv.org/abs/0712.3028}{{\ttfamily
  0712.3028}}.

\bibitem{Trotta:2008qt}
R.~Trotta, \emph{{Bayes in the sky: Bayesian inference and model selection in
  cosmology}}, \href{https://doi.org/10.1080/00107510802066753}{\emph{Contemp.
  Phys.} {\bfseries 49} (2008) 71--104},
  [\href{https://arxiv.org/abs/0803.4089}{{\ttfamily 0803.4089}}].

\bibitem{dataanalysis}
V.~J. {Mart{\'{\i}}nez}, E.~{Saar}, E.~{Mart{\'{\i}}nez-Gonz{\'a}lez} and M.-J.
  {Pons-Border{\'{\i}}a}, eds., \emph{{Data Analysis in Cosmology}}, vol.~665
  of \emph{Lecture Notes in Physics, Berlin Springer Verlag}, 2009.

\bibitem{Heavens:2009nx}
A.~Heavens, \emph{{Statistical techniques in cosmology}},
  \href{https://arxiv.org/abs/0906.0664}{{\ttfamily 0906.0664}}.

\bibitem{Verde:2009tu}
L.~Verde, \emph{{Statistical methods in cosmology}},
  \href{https://doi.org/10.1007/978-3-642-10598-2_4}{\emph{Lect. Notes Phys.}
  {\bfseries 800} (2010) 147--177},
  [\href{https://arxiv.org/abs/0911.3105}{{\ttfamily 0911.3105}}].

\bibitem{bayesianstatistics}
M.~P. {Hobson}, A.~H. {Jaffe}, A.~R. {Liddle}, P.~{Mukherjee} and
  D.~{Parkinson}, \emph{{Bayesian Methods in Cosmology}}.
\newblock Dec., 2009.

\bibitem{practicalstatistics}
J.~V. {Wall} and C.~R. {Jenkins}, \emph{{Practical Statistics for
  Astronomers}}.
\newblock Apr., 2012.

\bibitem{Trotta:2017wnx}
R.~Trotta, \emph{{Bayesian Methods in Cosmology}},  2017,
  \href{https://arxiv.org/abs/1701.01467}{{\ttfamily 1701.01467}}.

\bibitem{practicalbayesian}
C.~A.~L. {Bailer-Jones}, \emph{{Practical Bayesian Inference}}.
\newblock Apr., 2017.

\bibitem{bayesianastrophysics}
J.~M. {Hilbe}, R.~S. {de Souza} and E.~E.~O. {Ishida}, \emph{{Bayesian Models
  for Astrophysical Data Using R, JAGS, Python, and Stan}}.
\newblock May, 2017.

\bibitem{Bayes:1764vd}
R.~Bayes, \emph{{An essay toward solving a problem in the doctrine of
  chances}}, \href{https://doi.org/10.1098/rstl.1763.0053}{\emph{Phil. Trans.
  Roy. Soc. Lond.} {\bfseries 53} (1764) 370--418}.

\bibitem{Lewis:2002ah}
A.~Lewis and S.~Bridle, \emph{{Cosmological parameters from CMB and other data:
  A Monte Carlo approach}},
  \href{https://doi.org/10.1103/PhysRevD.66.103511}{\emph{Phys. Rev.}
  {\bfseries D66} (2002) 103511},
  [\href{https://arxiv.org/abs/astro-ph/0205436}{{\ttfamily
  astro-ph/0205436}}].

\bibitem{Audren:2012wb}
B.~Audren, J.~Lesgourgues, K.~Benabed and S.~Prunet, \emph{{Conservative
  Constraints on Early Cosmology: an illustration of the Monte Python
  cosmological parameter inference code}},
  \href{https://doi.org/10.1088/1475-7516/2013/02/001}{\emph{JCAP} {\bfseries
  1302} (2013) 001}, [\href{https://arxiv.org/abs/1210.7183}{{\ttfamily
  1210.7183}}].

\bibitem{Kolmogorov:1960ghw}
A.~N. Kolmogorov, \emph{Foundations of the Theory of Probability}.
\newblock Chelsea Pub Co, 2~ed., June, 1960.

\bibitem{Hannestad:2016fog}
S.~Hannestad and T.~Schwetz, \emph{{Cosmology and the neutrino mass ordering}},
  \href{https://doi.org/10.1088/1475-7516/2016/11/035}{\emph{JCAP} {\bfseries
  1611} (2016) 035}, [\href{https://arxiv.org/abs/1606.04691}{{\ttfamily
  1606.04691}}].

\bibitem{Gerbino:2016ehw}
M.~Gerbino, M.~Lattanzi, O.~Mena and K.~Freese, \emph{{A novel approach to
  quantifying the sensitivity of current and future cosmological datasets to
  the neutrino mass ordering through Bayesian hierarchical modeling}},
  \href{https://doi.org/10.1016/j.physletb.2017.10.052}{\emph{Phys. Lett.}
  {\bfseries B775} (2017) 239--250},
  [\href{https://arxiv.org/abs/1611.07847}{{\ttfamily 1611.07847}}].

\bibitem{Vagnozzi:2017ovm}
S.~Vagnozzi, E.~Giusarma, O.~Mena, K.~Freese, M.~Gerbino, S.~Ho et~al.,
  \emph{{Unveiling $\nu$ secrets with cosmological data: neutrino masses and
  mass hierarchy}},
  \href{https://doi.org/10.1103/PhysRevD.96.123503}{\emph{Phys. Rev.}
  {\bfseries D96} (2017) 123503},
  [\href{https://arxiv.org/abs/1701.08172}{{\ttfamily 1701.08172}}].

\bibitem{Hannestad:2017ypp}
S.~Hannestad and T.~Tram, \emph{{Optimal prior for Bayesian inference in a
  constrained parameter space}},
  \href{https://arxiv.org/abs/1710.08899}{{\ttfamily 1710.08899}}.

\bibitem{Long:2017dru}
A.~J. Long, M.~Raveri, W.~Hu and S.~Dodelson, \emph{{Neutrino Mass Priors for
  Cosmology from Random Matrices}},
  \href{https://doi.org/10.1103/PhysRevD.97.043510}{\emph{Phys. Rev.}
  {\bfseries D97} (2018) 043510},
  [\href{https://arxiv.org/abs/1711.08434}{{\ttfamily 1711.08434}}].

\bibitem{Gariazzo:2018pei}
S.~Gariazzo, M.~Archidiacono, P.~F. de~Salas, O.~Mena, C.~A. Ternes and
  M.~Tórtola, \emph{{Neutrino masses and their ordering: Global Data, Priors
  and Models}},
  \href{https://doi.org/10.1088/1475-7516/2018/03/011}{\emph{JCAP} {\bfseries
  1803} (2018) 011}, [\href{https://arxiv.org/abs/1801.04946}{{\ttfamily
  1801.04946}}].

\bibitem{Heavens:2018adv}
A.~F. Heavens and E.~Sellentin, \emph{{Objective Bayesian analysis of neutrino
  masses and hierarchy}},
  \href{https://doi.org/10.1088/1475-7516/2018/04/047}{\emph{JCAP} {\bfseries
  1804} (2018) 047}, [\href{https://arxiv.org/abs/1802.09450}{{\ttfamily
  1802.09450}}].

\bibitem{Handley:2018gel}
W.~Handley and M.~Millea, \emph{{Maximum entropy priors with derived parameters
  in a specified distribution}},
  \href{https://arxiv.org/abs/1804.08143}{{\ttfamily 1804.08143}}.

\bibitem{Gariazzo:2018tft}
S.~Gariazzo, \emph{{Neutrino mass eigenstates and their ordering: a Bayesian
  approach}},  in \emph{{17th Incontri di Fisica delle Alte Energie (IFAE 2018)
  Milano, Italia, April 4-6, 2018}}, 2018,
  \href{https://arxiv.org/abs/1806.11344}{{\ttfamily 1806.11344}}.

\bibitem{Simpson:2017qvj}
F.~Simpson, R.~Jiménez, C.~Peña-Garay and L.~Verde, \emph{{Strong Bayesian
  Evidence for the Normal Neutrino Hierarchy}},
  \href{https://doi.org/10.1088/1475-7516/2017/06/029}{\emph{JCAP} {\bfseries
  1706} (2017) 029}, [\href{https://arxiv.org/abs/1703.03425}{{\ttfamily
  1703.03425}}].

\bibitem{Schwetz:2017fey}
T.~Schwetz, K.~Freese, M.~Gerbino, E.~Giusarma, S.~Hannestad, M.~Lattanzi
  et~al., \emph{{Comment on "Strong Evidence for the Normal Neutrino
  Hierarchy"}},  \href{https://arxiv.org/abs/1703.04585}{{\ttfamily
  1703.04585}}.

\bibitem{Skilling:2006gxv}
J.~Skilling, \emph{{Nested sampling for general Bayesian computation}},
  \href{https://doi.org/10.1214/06-BA127}{\emph{Bayesian Analysis} {\bfseries
  1} (2006) 833--859}.

\bibitem{Mukherjee:2005wg}
P.~Mukherjee, D.~Parkinson and A.~R. Liddle, \emph{{A nested sampling algorithm
  for cosmological model selection}},
  \href{https://doi.org/10.1086/501068}{\emph{Astrophys. J.} {\bfseries 638}
  (2006) L51--L54}, [\href{https://arxiv.org/abs/astro-ph/0508461}{{\ttfamily
  astro-ph/0508461}}].

\bibitem{Shaw:2007jj}
R.~Shaw, M.~Bridges and M.~P. Hobson, \emph{{Clustered nested sampling:
  Efficient Bayesian inference for cosmology}},
  \href{https://doi.org/10.1111/j.1365-2966.2007.11871.x}{\emph{Mon. Not. Roy.
  Astron. Soc.} {\bfseries 378} (2007) 1365--1370},
  [\href{https://arxiv.org/abs/astro-ph/0701867}{{\ttfamily
  astro-ph/0701867}}].

\bibitem{Feroz:2007kg}
F.~Feroz and M.~P. Hobson, \emph{{Multimodal nested sampling: an efficient and
  robust alternative to MCMC methods for astronomical data analysis}},
  \href{https://doi.org/10.1111/j.1365-2966.2007.12353.x}{\emph{Mon. Not. Roy.
  Astron. Soc.} {\bfseries 384} (2008) 449},
  [\href{https://arxiv.org/abs/0704.3704}{{\ttfamily 0704.3704}}].

\bibitem{Bassett:2004wz}
B.~A. Bassett, P.~S. Corasaniti and M.~Kunz, \emph{{The Essence of quintessence
  and the cost of compression}},
  \href{https://doi.org/10.1086/427023}{\emph{Astrophys. J.} {\bfseries 617}
  (2004) L1--L4}, [\href{https://arxiv.org/abs/astro-ph/0407364}{{\ttfamily
  astro-ph/0407364}}].

\bibitem{Bridges:2006mt}
M.~Bridges, J.~D. McEwen, A.~N. Lasenby and M.~P. Hobson, \emph{{Markov chain
  Monte Carlo analysis of Bianchi VII(h) models}},
  \href{https://doi.org/10.1111/j.1365-2966.2007.11616.x}{\emph{Mon. Not. Roy.
  Astron. Soc.} {\bfseries 377} (2007) 1473--1480},
  [\href{https://arxiv.org/abs/astro-ph/0605325}{{\ttfamily
  astro-ph/0605325}}].

\bibitem{Feroz:2008xx}
F.~Feroz, M.~P. Hobson and M.~Bridges, \emph{{MultiNest: an efficient and
  robust Bayesian inference tool for cosmology and particle physics}},
  \href{https://doi.org/10.1111/j.1365-2966.2009.14548.x}{\emph{Mon. Not. Roy.
  Astron. Soc.} {\bfseries 398} (2009) 1601--1614},
  [\href{https://arxiv.org/abs/0809.3437}{{\ttfamily 0809.3437}}].

\bibitem{Jeffreys:1939xee}
H.~Jeffreys, \emph{{The Theory of Probability}}.
\newblock Oxford Classic Texts in the Physical Sciences. 1939.

\bibitem{Kass:1995loi}
R.~E. Kass and A.~E. Raftery, \emph{{Bayes Factors}},
  \href{https://doi.org/10.1080/01621459.1995.10476572}{\emph{J. Am. Statist.
  Assoc.} {\bfseries 90} (1995) 773--795}.

\bibitem{Brooks:2011ghw}
S.~Brooks, A.~Gelman, G.~Jones and X.-L. Meng, \emph{Handbook of Markov Chain
  Monte Carlo}.
\newblock CRC press, 2011.

\bibitem{Metropolis:1953ghw}
N.~{Metropolis}, A.~W. {Rosenbluth}, M.~N. {Rosenbluth}, A.~H. {Teller} and
  E.~{Teller}, \emph{{Equation of State Calculations by Fast Computing
  Machines}}, \href{https://doi.org/10.1063/1.1699114}{\emph{J.\, Chem.\,
  Phys.} {\bfseries 21} (June, 1953) 1087--1092}.

\bibitem{Hastings:1970ghw}
W.~K. {Hastings}, \emph{{Monte Carlo Sampling Methods using Markov Chains and
  their Applications}},
  \href{https://doi.org/10.1093/biomet/57.1.97}{\emph{Biometrika, Vol.~57,
  No.~1, p.~97-109, 1970} {\bfseries 57} (Apr., 1970) 97--109}.

\bibitem{Gelman:1996ghw}
A.~Gelman, G.~O. Roberts and W.~R. Gilks, \emph{Efficient {M}etropolis jumping
  rules},  in \emph{Bayesian statistics, 5 (Alicante, 1994)}, Oxford Sci.
  Publ., pp.~599--607.
\newblock Oxford Univ. Press, New York, 1996.

\bibitem{Alam:2015mbd}
{\scshape SDSS-III} collaboration, S.~Alam et~al., \emph{{The Eleventh and
  Twelfth Data Releases of the Sloan Digital Sky Survey: Final Data from
  SDSS-III}},
  \href{https://doi.org/10.1088/0067-0049/219/1/12}{\emph{Astrophys. J. Suppl.}
  {\bfseries 219} (2015) 12},
  [\href{https://arxiv.org/abs/1501.00963}{{\ttfamily 1501.00963}}].

\bibitem{Alam:2016hwk}
{\scshape BOSS} collaboration, S.~Alam et~al., \emph{{The clustering of
  galaxies in the completed SDSS-III Baryon Oscillation Spectroscopic Survey:
  cosmological analysis of the DR12 galaxy sample}},
  \href{https://doi.org/10.1093/mnras/stx721}{\emph{Mon. Not. Roy. Astron.
  Soc.} {\bfseries 470} (2017) 2617--2652},
  [\href{https://arxiv.org/abs/1607.03155}{{\ttfamily 1607.03155}}].

\bibitem{Zhen:2015yba}
Z.~Pan and L.~Knox, \emph{{Constraints on neutrino mass from Cosmic Microwave
  Background and Large Scale Structure}},
  \href{https://doi.org/10.1093/mnras/stv2164}{\emph{Mon. Not. Roy. Astron.
  Soc.} {\bfseries 454} (2015) 3200--3206},
  [\href{https://arxiv.org/abs/1506.07493}{{\ttfamily 1506.07493}}].

\bibitem{Gerbino:2015ixa}
M.~Gerbino, M.~Lattanzi and A.~Melchiorri, \emph{{$\nu$ generation: Present and
  future constraints on neutrino masses from global analysis of cosmology and
  laboratory experiments}},
  \href{https://doi.org/10.1103/PhysRevD.93.033001}{\emph{Phys. Rev.}
  {\bfseries D93} (2016) 033001},
  [\href{https://arxiv.org/abs/1507.08614}{{\ttfamily 1507.08614}}].

\bibitem{DiValentino:2015wba}
E.~Di~Valentino, E.~Giusarma, M.~Lattanzi, O.~Mena, A.~Melchiorri and J.~Silk,
  \emph{{Cosmological Axion and neutrino mass constraints from Planck 2015
  temperature and polarization data}},
  \href{https://doi.org/10.1016/j.physletb.2015.11.025}{\emph{Phys. Lett.}
  {\bfseries B752} (2016) 182--185},
  [\href{https://arxiv.org/abs/1507.08665}{{\ttfamily 1507.08665}}].

\bibitem{DiValentino:2015sam}
E.~Di~Valentino, E.~Giusarma, O.~Mena, A.~Melchiorri and J.~Silk,
  \emph{{Cosmological limits on neutrino unknowns versus low redshift priors}},
  \href{https://doi.org/10.1103/PhysRevD.93.083527}{\emph{Phys. Rev.}
  {\bfseries D93} (2016) 083527},
  [\href{https://arxiv.org/abs/1511.00975}{{\ttfamily 1511.00975}}].

\bibitem{Cuesta:2015iho}
A.~J. Cuesta, V.~Niro and L.~Verde, \emph{{Neutrino mass limits: robust
  information from the power spectrum of galaxy surveys}},
  \href{https://doi.org/10.1016/j.dark.2016.04.005}{\emph{Phys. Dark Univ.}
  {\bfseries 13} (2016) 77--86},
  [\href{https://arxiv.org/abs/1511.05983}{{\ttfamily 1511.05983}}].

\bibitem{Huang:2015wrx}
Q.-G. Huang, K.~Wang and S.~Wang, \emph{{Constraints on the neutrino mass and
  mass hierarchy from cosmological observations}},
  \href{https://doi.org/10.1140/epjc/s10052-016-4334-z}{\emph{Eur. Phys. J.}
  {\bfseries C76} (2016) 489},
  [\href{https://arxiv.org/abs/1512.05899}{{\ttfamily 1512.05899}}].

\bibitem{Moresco:2016nqq}
M.~Moresco, R.~Jiménez, L.~Verde, A.~Cimatti, L.~Pozzetti, C.~Maraston et~al.,
  \emph{{Constraining the time evolution of dark energy, curvature and neutrino
  properties with cosmic chronometers}},
  \href{https://doi.org/10.1088/1475-7516/2016/12/039}{\emph{JCAP} {\bfseries
  1612} (2016) 039}, [\href{https://arxiv.org/abs/1604.00183}{{\ttfamily
  1604.00183}}].

\bibitem{Giusarma:2016phn}
E.~Giusarma, M.~Gerbino, O.~Mena, S.~Vagnozzi, S.~Ho and K.~Freese,
  \emph{{Improvement of cosmological neutrino mass bounds}},
  \href{https://doi.org/10.1103/PhysRevD.94.083522}{\emph{Phys. Rev.}
  {\bfseries D94} (2016) 083522},
  [\href{https://arxiv.org/abs/1605.04320}{{\ttfamily 1605.04320}}].

\bibitem{Oh:2016wls}
M.~Oh and Y.-S. Song, \emph{{Measuring neutrino mass imprinted on the
  anisotropic galaxy clustering}},
  \href{https://doi.org/10.1088/1475-7516/2017/04/020}{\emph{JCAP} {\bfseries
  1704} (2017) 020}, [\href{https://arxiv.org/abs/1607.01074}{{\ttfamily
  1607.01074}}].

\bibitem{Yeche:2017upn}
C.~Yèche, N.~Palanque-Delabrouille, J.~Baur and H.~du~Mas~des Bourboux,
  \emph{{Constraints on neutrino masses from Lyman-alpha forest power spectrum
  with BOSS and XQ-100}},
  \href{https://doi.org/10.1088/1475-7516/2017/06/047}{\emph{JCAP} {\bfseries
  1706} (2017) 047}, [\href{https://arxiv.org/abs/1702.03314}{{\ttfamily
  1702.03314}}].

\bibitem{Capozzi:2017ipn}
F.~Capozzi, E.~Di~Valentino, E.~Lisi, A.~Marrone, A.~Melchiorri and A.~Palazzo,
  \emph{{Global constraints on absolute neutrino masses and their ordering}},
  \href{https://doi.org/10.1103/PhysRevD.95.096014}{\emph{Phys. Rev.}
  {\bfseries D95} (2017) 096014},
  [\href{https://arxiv.org/abs/1703.04471}{{\ttfamily 1703.04471}}].

\bibitem{Couchot:2017pvz}
F.~Couchot, S.~Henrot-Versillé, O.~Perdereau, S.~Plaszczynski,
  B.~Rouillé~d'Orfeuil, M.~Spinelli et~al., \emph{{Cosmological constraints on
  the neutrino mass including systematic uncertainties}},
  \href{https://doi.org/10.1051/0004-6361/201730927}{\emph{Astron. Astrophys.}
  {\bfseries 606} (2017) A104},
  [\href{https://arxiv.org/abs/1703.10829}{{\ttfamily 1703.10829}}].

\bibitem{Caldwell:2017mqu}
A.~Caldwell, A.~Merle, O.~Schulz and M.~Totzauer, \emph{{Global Bayesian
  analysis of neutrino mass data}},
  \href{https://doi.org/10.1103/PhysRevD.96.073001}{\emph{Phys. Rev.}
  {\bfseries D96} (2017) 073001},
  [\href{https://arxiv.org/abs/1705.01945}{{\ttfamily 1705.01945}}].

\bibitem{Wang:2017htc}
S.~Wang, Y.-F. Wang and D.-M. Xia, \emph{{Constraints on the sum of neutrino
  masses using cosmological data including the latest extended Baryon
  Oscillation Spectroscopic Survey DR14 quasar sample}},
  \href{https://doi.org/10.1088/1674-1137/42/6/065103}{\emph{Chin. Phys.}
  {\bfseries C42} (2018) 065103},
  [\href{https://arxiv.org/abs/1707.00588}{{\ttfamily 1707.00588}}].

\bibitem{Chen:2017ayg}
L.~Chen, Q.-G. Huang and K.~Wang, \emph{{New cosmological constraints with
  extended-Baryon Oscillation Spectroscopic Survey DR14 quasar sample}},
  \href{https://doi.org/10.1140/epjc/s10052-017-5344-1}{\emph{Eur. Phys. J.}
  {\bfseries C77} (2017) 762},
  [\href{https://arxiv.org/abs/1707.02742}{{\ttfamily 1707.02742}}].

\bibitem{Upadhye:2017hdl}
A.~Upadhye, \emph{{Neutrino mass and dark energy constraints from
  redshift-space distortions}},
  \href{https://arxiv.org/abs/1707.09354}{{\ttfamily 1707.09354}}.

\bibitem{Salvati:2017rsn}
L.~Salvati, M.~Douspis and N.~Aghanim, \emph{{Constraints from thermal
  Sunyaev-Zel’dovich cluster counts and power spectrum combined with CMB}},
  \href{https://doi.org/10.1051/0004-6361/201731990}{\emph{Astron. Astrophys.}
  {\bfseries 614} (2018) A13},
  [\href{https://arxiv.org/abs/1708.00697}{{\ttfamily 1708.00697}}].

\bibitem{Nunes:2017xon}
R.~C. Nunes and A.~Bonilla, \emph{{Probing the properties of relic neutrinos
  using the cosmic microwave background, the Hubble Space Telescope and galaxy
  clusters}}, \href{https://doi.org/10.1093/mnras/stx2661}{\emph{Mon. Not. Roy.
  Astron. Soc.} {\bfseries 473} (2018) 4404--4409},
  [\href{https://arxiv.org/abs/1710.10264}{{\ttfamily 1710.10264}}].

\bibitem{Emami:2017wqa}
R.~Emami, T.~Broadhurst, P.~Jimeno, G.~Smoot, R.~Angulo, J.~Lim et~al.,
  \emph{{Evidence of Neutrino Enhanced Clustering in a Complete Sample of Sloan
  Survey Clusters, Implying $\sum m_{\nu}=0.11\pm0.03eV$}},
  \href{https://arxiv.org/abs/1711.05210}{{\ttfamily 1711.05210}}.

\bibitem{Zennaro:2017qnp}
M.~Zennaro, J.~Bel, J.~Dossett, C.~Carbone and L.~Guzzo, \emph{{Cosmological
  constraints from galaxy clustering in the presence of massive neutrinos}},
  \href{https://doi.org/10.1093/mnras/sty670}{\emph{Mon. Not. Roy. Astron.
  Soc.} {\bfseries 477} (2018) 491--506},
  [\href{https://arxiv.org/abs/1712.02886}{{\ttfamily 1712.02886}}].

\bibitem{Choudhury:2018byy}
S.~Roy~Choudhury and S.~Choubey, \emph{{Updated Bounds on Sum of Neutrino
  Masses in Various Cosmological Scenarios}},
  \href{https://doi.org/10.1088/1475-7516/2018/09/017}{\emph{JCAP} {\bfseries
  1809} (2018) 017}, [\href{https://arxiv.org/abs/1806.10832}{{\ttfamily
  1806.10832}}].

\bibitem{Choudhury:2018adz}
S.~Roy~Choudhury and A.~Naskar, \emph{{Strong Bounds on Sum of Neutrino Masses
  in a 12 Parameter Extended Scenario with Non-Phantom Dynamical Dark Energy
  ($w(z)\geq -1$) with CPL parameterization}},
  \href{https://doi.org/10.1140/epjc/s10052-019-6762-z}{\emph{Eur. Phys. J.}
  {\bfseries C79} (2019) 262},
  [\href{https://arxiv.org/abs/1807.02860}{{\ttfamily 1807.02860}}].

\bibitem{Liu:2018dsw}
J.~Liu and M.~S. Madhavacheril, \emph{{Constraining neutrino mass with the
  tomographic weak lensing one-point probability distribution function and
  power spectrum}},
  \href{https://doi.org/10.1103/PhysRevD.99.083508}{\emph{Phys. Rev.}
  {\bfseries D99} (2019) 083508},
  [\href{https://arxiv.org/abs/1809.10747}{{\ttfamily 1809.10747}}].

\bibitem{Li:2018owg}
Z.~Li, J.~Liu, J.~M.~Z. Matilla and W.~R. Coulton, \emph{{Constraining neutrino
  mass with tomographic weak lensing peak counts}},
  \href{https://doi.org/10.1103/PhysRevD.99.063527}{\emph{Phys. Rev.}
  {\bfseries D99} (2019) 063527},
  [\href{https://arxiv.org/abs/1810.01781}{{\ttfamily 1810.01781}}].

\bibitem{Coulton:2018ebd}
W.~R. Coulton, J.~Liu, M.~S. Madhavacheril, V.~Böhm and D.~N. Spergel,
  \emph{{Constraining Neutrino Mass with the Tomographic Weak Lensing
  Bispectrum}},  \href{https://arxiv.org/abs/1810.02374}{{\ttfamily
  1810.02374}}.

\bibitem{Loureiro:2018pdz}
A.~Loureiro et~al., \emph{{On The Upper Bound of Neutrino Masses from Combined
  Cosmological Observations and Particle Physics Experiments}},
  \href{https://arxiv.org/abs/1811.02578}{{\ttfamily 1811.02578}}.

\bibitem{Gariazzo:2018meg}
S.~Gariazzo and O.~Mena, \emph{{Cosmology-marginalized approaches in Bayesian
  model comparison: The neutrino mass as a case study}},
  \href{https://doi.org/10.1103/PhysRevD.99.021301}{\emph{Phys. Rev.}
  {\bfseries D99} (2019) 021301},
  [\href{https://arxiv.org/abs/1812.05449}{{\ttfamily 1812.05449}}].

\bibitem{Marques:2018ctl}
G.~A. Marques, J.~Liu, J.~M.~Z. Matilla, Z.~Haiman, A.~Bernui and C.~P. Novaes,
  \emph{{Constraining neutrino mass with weak lensing Minkowski Functionals}},
  \href{https://arxiv.org/abs/1812.08206}{{\ttfamily 1812.08206}}.

\bibitem{Gil-Marin:2015sqa}
H.~Gil-Marín et~al., \emph{{The clustering of galaxies in the SDSS-III Baryon
  Oscillation Spectroscopic Survey: RSD measurement from the LOS-dependent
  power spectrum of DR12 BOSS galaxies}},
  \href{https://doi.org/10.1093/mnras/stw1096}{\emph{Mon. Not. Roy. Astron.
  Soc.} {\bfseries 460} (2016) 4188--4209},
  [\href{https://arxiv.org/abs/1509.06386}{{\ttfamily 1509.06386}}].

\bibitem{Smith:2002dz}
{\scshape VIRGO Consortium} collaboration, R.~E. Smith, J.~A. Peacock,
  A.~Jenkins, S.~D.~M. White, C.~S. Frenk, F.~R. Pearce et~al., \emph{{Stable
  clustering, the halo model and nonlinear cosmological power spectra}},
  \href{https://doi.org/10.1046/j.1365-8711.2003.06503.x}{\emph{Mon. Not. Roy.
  Astron. Soc.} {\bfseries 341} (2003) 1311},
  [\href{https://arxiv.org/abs/astro-ph/0207664}{{\ttfamily
  astro-ph/0207664}}].

\bibitem{Takahashi:2012em}
R.~Takahashi, M.~Sato, T.~Nishimichi, A.~Taruya and M.~Oguri, \emph{{Revising
  the Halofit Model for the Nonlinear Matter Power Spectrum}},
  \href{https://doi.org/10.1088/0004-637X/761/2/152}{\emph{Astrophys. J.}
  {\bfseries 761} (2012) 152},
  [\href{https://arxiv.org/abs/1208.2701}{{\ttfamily 1208.2701}}].

\bibitem{Castorina:2013wga}
E.~Castorina, E.~Sefusatti, R.~K. Sheth, F.~Villaescusa-Navarro and M.~Viel,
  \emph{{Cosmology with massive neutrinos II: on the universality of the halo
  mass function and bias}},
  \href{https://doi.org/10.1088/1475-7516/2014/02/049}{\emph{JCAP} {\bfseries
  1402} (2014) 049}, [\href{https://arxiv.org/abs/1311.1212}{{\ttfamily
  1311.1212}}].

\bibitem{Anderson:2012sa}
L.~Anderson et~al., \emph{{The clustering of galaxies in the SDSS-III Baryon
  Oscillation Spectroscopic Survey: Baryon Acoustic Oscillations in the Data
  Release 9 Spectroscopic Galaxy Sample}},
  \href{https://doi.org/10.1111/j.1365-2966.2012.22066.x}{\emph{Mon. Not. Roy.
  Astron. Soc.} {\bfseries 427} (2013) 3435--3467},
  [\href{https://arxiv.org/abs/1203.6594}{{\ttfamily 1203.6594}}].

\bibitem{Blake:2011en}
C.~Blake et~al., \emph{{The WiggleZ Dark Energy Survey: mapping the
  distance-redshift relation with baryon acoustic oscillations}},
  \href{https://doi.org/10.1111/j.1365-2966.2011.19592.x}{\emph{Mon. Not. Roy.
  Astron. Soc.} {\bfseries 418} (2011) 1707--1724},
  [\href{https://arxiv.org/abs/1108.2635}{{\ttfamily 1108.2635}}].

\bibitem{Anderson:2013zyy}
{\scshape BOSS} collaboration, L.~Anderson et~al., \emph{{The clustering of
  galaxies in the SDSS-III Baryon Oscillation Spectroscopic Survey: baryon
  acoustic oscillations in the Data Releases 10 and 11 Galaxy samples}},
  \href{https://doi.org/10.1093/mnras/stu523}{\emph{Mon. Not. Roy. Astron.
  Soc.} {\bfseries 441} (2014) 24--62},
  [\href{https://arxiv.org/abs/1312.4877}{{\ttfamily 1312.4877}}].

\bibitem{Aghanim:2016yuo}
{\scshape Planck} collaboration, N.~Aghanim et~al., \emph{{Planck intermediate
  results. XLVI. Reduction of large-scale systematic effects in HFI
  polarization maps and estimation of the reionization optical depth}},
  \href{https://doi.org/10.1051/0004-6361/201628890}{\emph{Astron. Astrophys.}
  {\bfseries 596} (2016) A107},
  [\href{https://arxiv.org/abs/1605.02985}{{\ttfamily 1605.02985}}].

\bibitem{Adam:2016hgk}
{\scshape Planck} collaboration, R.~Adam et~al., \emph{{Planck intermediate
  results. XLVII. Planck constraints on reionization history}},
  \href{https://doi.org/10.1051/0004-6361/201628897}{\emph{Astron. Astrophys.}
  {\bfseries 596} (2016) A108},
  [\href{https://arxiv.org/abs/1605.03507}{{\ttfamily 1605.03507}}].

\bibitem{Tucci:2004zy}
M.~Tucci, E.~Martínez-González, P.~Vielva and J.~Delabrouille, \emph{{Limits
  on the detectability of the CMB B-mode polarization imposed by foregrounds}},
  \href{https://doi.org/10.1111/j.1365-2966.2005.09123.x}{\emph{Mon. Not. Roy.
  Astron. Soc.} {\bfseries 360} (2005) 935--949},
  [\href{https://arxiv.org/abs/astro-ph/0411567}{{\ttfamily
  astro-ph/0411567}}].

\bibitem{Seiffert:2006vh}
M.~Seiffert, C.~Borys, D.~Scott and M.~Halpern, \emph{{An upper limit to
  polarized submillimetre emission in Arp 220}},
  \href{https://doi.org/10.1111/j.1365-2966.2006.11186.x}{\emph{Mon. Not. Roy.
  Astron. Soc.} {\bfseries 374} (2007) 409--414},
  [\href{https://arxiv.org/abs/astro-ph/0610485}{{\ttfamily
  astro-ph/0610485}}].

\bibitem{Ade:2015fva}
{\scshape Planck} collaboration, P.~A.~R. Ade et~al., \emph{{Planck 2015
  results. XXIV. Cosmology from Sunyaev-Zeldovich cluster counts}},
  \href{https://doi.org/10.1051/0004-6361/201525833}{\emph{Astron. Astrophys.}
  {\bfseries 594} (2016) A24},
  [\href{https://arxiv.org/abs/1502.01597}{{\ttfamily 1502.01597}}].

\bibitem{Ade:2015gva}
{\scshape Planck} collaboration, P.~A.~R. Ade et~al., \emph{{Planck 2015
  results. XXVII. The Second Planck Catalogue of Sunyaev-Zeldovich Sources}},
  \href{https://doi.org/10.1051/0004-6361/201525823}{\emph{Astron. Astrophys.}
  {\bfseries 594} (2016) A27},
  [\href{https://arxiv.org/abs/1502.01598}{{\ttfamily 1502.01598}}].

\bibitem{Hamann:2010pw}
J.~Hamann, S.~Hannestad, J.~Lesgourgues, C.~Rampf and Y.~Y.~Y. Wong,
  \emph{{Cosmological parameters from large scale structure - geometric versus
  shape information}},
  \href{https://doi.org/10.1088/1475-7516/2010/07/022}{\emph{JCAP} {\bfseries
  1007} (2010) 022}, [\href{https://arxiv.org/abs/1003.3999}{{\ttfamily
  1003.3999}}].

\bibitem{Giusarma:2012ph}
E.~Giusarma, R.~De~Putter and O.~Mena, \emph{{Testing standard and nonstandard
  neutrino physics with cosmological data}},
  \href{https://doi.org/10.1103/PhysRevD.87.043515}{\emph{Phys. Rev.}
  {\bfseries D87} (2013) 043515},
  [\href{https://arxiv.org/abs/1211.2154}{{\ttfamily 1211.2154}}].

\bibitem{Pen:2004rm}
U.-L. Pen, \emph{{Beating lensing cosmic variance with galaxy tomography}},
  \href{https://doi.org/10.1111/j.1365-2966.2004.07746.x}{\emph{Mon. Not. Roy.
  Astron. Soc.} {\bfseries 350} (2004) 1445},
  [\href{https://arxiv.org/abs/astro-ph/0402008}{{\ttfamily
  astro-ph/0402008}}].

\bibitem{More:2014uva}
S.~More, H.~Miyatake, R.~Mandelbaum, M.~Takada, D.~Spergel, J.~Brownstein
  et~al., \emph{{The Weak Lensing Signal and the Clustering of BOSS Galaxies
  II: Astrophysical and Cosmological Constraints}},
  \href{https://doi.org/10.1088/0004-637X/806/1/2}{\emph{Astrophys. J.}
  {\bfseries 806} (2015) 2}, [\href{https://arxiv.org/abs/1407.1856}{{\ttfamily
  1407.1856}}].

\bibitem{Amendola:2015pha}
L.~Amendola, E.~Menegoni, C.~Di~Porto, M.~Corsi and E.~Branchini,
  \emph{{Constraints on a scale-dependent bias from galaxy clustering}},
  \href{https://doi.org/10.1103/PhysRevD.95.023505}{\emph{Phys. Rev.}
  {\bfseries D95} (2017) 023505},
  [\href{https://arxiv.org/abs/1502.03994}{{\ttfamily 1502.03994}}].

\bibitem{Giannantonio:2015ahz}
{\scshape DES} collaboration, T.~Giannantonio et~al., \emph{{CMB lensing
  tomography with the DES Science Verification galaxies}},
  \href{https://doi.org/10.1093/mnras/stv2678}{\emph{Mon. Not. Roy. Astron.
  Soc.} {\bfseries 456} (2016) 3213--3244},
  [\href{https://arxiv.org/abs/1507.05551}{{\ttfamily 1507.05551}}].

\bibitem{Pujol:2016lfe}
A.~Pujol et~al., \emph{{A new method to measure galaxy bias by combining the
  density and weak lensing fields}},
  \href{https://doi.org/10.1093/mnras/stw1612}{\emph{Mon. Not. Roy. Astron.
  Soc.} {\bfseries 462} (2016) 35--47},
  [\href{https://arxiv.org/abs/1601.00160}{{\ttfamily 1601.00160}}].

\bibitem{Singh:2016xey}
S.~Singh, R.~Mandelbaum and J.~R. Brownstein, \emph{{Cross-correlating Planck
  CMB lensing with SDSS: Lensing-lensing and galaxy-lensing
  cross-correlations}}, \href{https://doi.org/10.1093/mnras/stw2482}{\emph{Mon.
  Not. Roy. Astron. Soc.} {\bfseries 464} (2017) 2120--2138},
  [\href{https://arxiv.org/abs/1606.08841}{{\ttfamily 1606.08841}}].

\bibitem{Singh:2016edo}
S.~Singh, R.~Mandelbaum, U.~Seljak, A.~Slosar and J.~Vázquez~González,
  \emph{{Galaxy–galaxy lensing estimators and their covariance properties}},
  \href{https://doi.org/10.1093/mnras/stx1828}{\emph{Mon. Not. Roy. Astron.
  Soc.} {\bfseries 471} (2017) 3827--3844},
  [\href{https://arxiv.org/abs/1611.00752}{{\ttfamily 1611.00752}}].

\bibitem{Joudaki:2017zdt}
S.~Joudaki et~al., \emph{{KiDS-450 + 2dFLenS: Cosmological parameter
  constraints from weak gravitational lensing tomography and overlapping
  redshift-space galaxy clustering}},
  \href{https://doi.org/10.1093/mnras/stx2820}{\emph{Mon. Not. Roy. Astron.
  Soc.} {\bfseries 474} (2018) 4894--4924},
  [\href{https://arxiv.org/abs/1707.06627}{{\ttfamily 1707.06627}}].

\bibitem{Simon:2017osp}
P.~Simon and S.~Hilbert, \emph{{Scale dependence of galaxy biasing investigated
  by weak gravitational lensing: An assessment using semi-analytic galaxies and
  simulated lensing data}},
  \href{https://doi.org/10.1051/0004-6361/201732248}{\emph{Astron. Astrophys.}
  {\bfseries 613} (2018) A15},
  [\href{https://arxiv.org/abs/1711.02677}{{\ttfamily 1711.02677}}].

\bibitem{Singh:2018kmr}
S.~Singh, R.~Mandelbaum, U.~Seljak, S.~Rodríguez-Torres and A.~Slosar,
  \emph{{Cosmological constraints from galaxy-lensing cross correlations using
  BOSS galaxies with SDSS and CMB lensing}},
  \href{https://arxiv.org/abs/1811.06499}{{\ttfamily 1811.06499}}.

\bibitem{Eisenstein:2006nk}
D.~J. Eisenstein, H.-J. Seo, E.~Sirko and D.~Spergel, \emph{{Improving
  Cosmological Distance Measurements by Reconstruction of the Baryon Acoustic
  Peak}}, \href{https://doi.org/10.1086/518712}{\emph{Astrophys. J.} {\bfseries
  664} (2007) 675--679},
  [\href{https://arxiv.org/abs/astro-ph/0604362}{{\ttfamily
  astro-ph/0604362}}].

\bibitem{Padmanabhan:2012hf}
N.~Padmanabhan, X.~Xu, D.~J. Eisenstein, R.~Scalzo, A.~J. Cuesta, K.~T. Mehta
  et~al., \emph{{A 2 per cent distance to $z$=0.35 by reconstructing baryon
  acoustic oscillations - I. Methods and application to the Sloan Digital Sky
  Survey}}, \href{https://doi.org/10.1111/j.1365-2966.2012.21888.x}{\emph{Mon.
  Not. Roy. Astron. Soc.} {\bfseries 427} (2012) 2132--2145},
  [\href{https://arxiv.org/abs/1202.0090}{{\ttfamily 1202.0090}}].

\bibitem{White:2015eaa}
M.~White, \emph{{Reconstruction within the Zeldovich approximation}},
  \href{https://doi.org/10.1093/mnras/stv842}{\emph{Mon. Not. Roy. Astron.
  Soc.} {\bfseries 450} (2015) 3822--3828},
  [\href{https://arxiv.org/abs/1504.03677}{{\ttfamily 1504.03677}}].

\bibitem{Giusarma:2018jei}
E.~Giusarma, S.~Vagnozzi, S.~Ho, S.~Ferraro, K.~Freese, R.~Kamen-Rubio et~al.,
  \emph{{Scale-dependent galaxy bias, CMB lensing-galaxy cross-correlation, and
  neutrino masses}},
  \href{https://doi.org/10.1103/PhysRevD.98.123526}{\emph{Phys. Rev.}
  {\bfseries D98} (2018) 123526},
  [\href{https://arxiv.org/abs/1802.08694}{{\ttfamily 1802.08694}}].

\bibitem{Hu:2001kj}
W.~Hu and T.~Okamoto, \emph{{Mass reconstruction with cmb polarization}},
  \href{https://doi.org/10.1086/341110}{\emph{Astrophys. J.} {\bfseries 574}
  (2002) 566--574}, [\href{https://arxiv.org/abs/astro-ph/0111606}{{\ttfamily
  astro-ph/0111606}}].

\bibitem{Peiris:2000kb}
H.~V. Peiris and D.~N. Spergel, \emph{{Cross-correlating the Sloan Digital Sky
  Survey with the microwave sky}},
  \href{https://doi.org/10.1086/309373}{\emph{Astrophys. J.} {\bfseries 540}
  (2000) 605}, [\href{https://arxiv.org/abs/astro-ph/0001393}{{\ttfamily
  astro-ph/0001393}}].

\bibitem{Bleem:2012gm}
L.~E. Bleem et~al., \emph{{A Measurement of the Correlation of Galaxy Surveys
  with CMB Lensing Convergence Maps from the South Pole Telescope}},
  \href{https://doi.org/10.1088/2041-8205/753/1/L9}{\emph{Astrophys. J.}
  {\bfseries 753} (2012) L9},
  [\href{https://arxiv.org/abs/1203.4808}{{\ttfamily 1203.4808}}].

\bibitem{Pearson:2013iha}
R.~Pearson and O.~Zahn, \emph{{Cosmology from cross correlation of CMB lensing
  and galaxy surveys}},
  \href{https://doi.org/10.1103/PhysRevD.89.043516}{\emph{Phys. Rev.}
  {\bfseries D89} (2014) 043516},
  [\href{https://arxiv.org/abs/1311.0905}{{\ttfamily 1311.0905}}].

\bibitem{Bianchini:2014dla}
{\scshape Herschel ATLAS} collaboration, F.~Bianchini et~al.,
  \emph{{Cross-correlation between the CMB lensing potential measured by Planck
  and high-z sub-mm galaxies detected by the Herschel-ATLAS survey}},
  \href{https://doi.org/10.1088/0004-637X/802/1/64}{\emph{Astrophys. J.}
  {\bfseries 802} (2015) 64},
  [\href{https://arxiv.org/abs/1410.4502}{{\ttfamily 1410.4502}}].

\bibitem{Bianchini:2015yly}
F.~Bianchini et~al., \emph{{Toward a tomographic analysis of the
  cross-correlation between Planck CMB lensing and H-ATLAS galaxies}},
  \href{https://doi.org/10.3847/0004-637X/825/1/24}{\emph{Astrophys. J.}
  {\bfseries 825} (2016) 24},
  [\href{https://arxiv.org/abs/1511.05116}{{\ttfamily 1511.05116}}].

\bibitem{Schmittfull:2017ffw}
M.~Schmittfull and U.~Seljak, \emph{{Parameter constraints from
  cross-correlation of CMB lensing with galaxy clustering}},
  \href{https://doi.org/10.1103/PhysRevD.97.123540}{\emph{Phys. Rev.}
  {\bfseries D97} (2018) 123540},
  [\href{https://arxiv.org/abs/1710.09465}{{\ttfamily 1710.09465}}].

\bibitem{Bianchini:2018mwv}
F.~Bianchini and C.~L. Reichardt, \emph{{Constraining gravity at large scales
  with the 2MASS Photometric Redshift catalogue and Planck lensing}},
  \href{https://doi.org/10.3847/1538-4357/aacafd,
  10.3847/1538-4357/aacafd/meta}{\emph{Astrophys. J.} {\bfseries 862} (2018)
  81}, [\href{https://arxiv.org/abs/1801.03736}{{\ttfamily 1801.03736}}].

\bibitem{Okumura:2012xh}
T.~Okumura, U.~Seljak and V.~Desjacques, \emph{{Distribution function approach
  to redshift space distortions, Part III: halos and galaxies}},
  \href{https://doi.org/10.1088/1475-7516/2012/11/014}{\emph{JCAP} {\bfseries
  1211} (2012) 014}, [\href{https://arxiv.org/abs/1206.4070}{{\ttfamily
  1206.4070}}].

\bibitem{CasasMiranda:2001ym}
R.~Casas-Miranda, H.~J. Mo, R.~K. Sheth and G.~Börner, \emph{{On the
  distribution of haloes, galaxies and mass}},
  \href{https://doi.org/10.1046/j.1365-8711.2002.05378.x}{\emph{Mon. Not. Roy.
  Astron. Soc.} {\bfseries 333} (2002) 730--738},
  [\href{https://arxiv.org/abs/astro-ph/0105008}{{\ttfamily
  astro-ph/0105008}}].

\bibitem{Baldauf:2013hka}
T.~Baldauf, U.~Seljak, R.~E. Smith, N.~Hamaus and V.~Desjacques, \emph{{Halo
  stochasticity from exclusion and nonlinear clustering}},
  \href{https://doi.org/10.1103/PhysRevD.88.083507}{\emph{Phys. Rev.}
  {\bfseries D88} (2013) 083507},
  [\href{https://arxiv.org/abs/1305.2917}{{\ttfamily 1305.2917}}].

\bibitem{Pullen:2015vtb}
A.~R. Pullen, S.~Alam, S.~He and S.~Ho, \emph{{Constraining Gravity at the
  Largest Scales through CMB Lensing and Galaxy Velocities}},
  \href{https://doi.org/10.1093/mnras/stw1249}{\emph{Mon. Not. Roy. Astron.
  Soc.} {\bfseries 460} (2016) 4098--4108},
  [\href{https://arxiv.org/abs/1511.04457}{{\ttfamily 1511.04457}}].

\bibitem{Carlson:2012bu}
J.~Carlson, B.~Reid and M.~White, \emph{{Convolution Lagrangian perturbation
  theory for biased tracers}},
  \href{https://doi.org/10.1093/mnras/sts457}{\emph{Mon. Not. Roy. Astron.
  Soc.} {\bfseries 429} (2013) 1674},
  [\href{https://arxiv.org/abs/1209.0780}{{\ttfamily 1209.0780}}].

\bibitem{Modi:2016dah}
C.~Modi, E.~Castorina and U.~Seljak, \emph{{Halo bias in Lagrangian Space:
  Estimators and theoretical predictions}},
  \href{https://doi.org/10.1093/mnras/stx2148}{\emph{Mon. Not. Roy. Astron.
  Soc.} {\bfseries 472} (2017) 3959--3970},
  [\href{https://arxiv.org/abs/1612.01621}{{\ttfamily 1612.01621}}].

\bibitem{Modi:2017wds}
C.~Modi, M.~White and Z.~Vlah, \emph{{Modeling CMB lensing cross correlations
  with CLEFT}},
  \href{https://doi.org/10.1088/1475-7516/2017/08/009}{\emph{JCAP} {\bfseries
  1708} (2017) 009}, [\href{https://arxiv.org/abs/1706.03173}{{\ttfamily
  1706.03173}}].

\bibitem{Villaescusa-Navarro:2013pva}
F.~Villaescusa-Navarro, F.~Marulli, M.~Viel, E.~Branchini, E.~Castorina,
  E.~Sefusatti et~al., \emph{{Cosmology with massive neutrinos I: towards a
  realistic modeling of the relation between matter, haloes and galaxies}},
  \href{https://doi.org/10.1088/1475-7516/2014/03/011}{\emph{JCAP} {\bfseries
  1403} (2014) 011}, [\href{https://arxiv.org/abs/1311.0866}{{\ttfamily
  1311.0866}}].

\bibitem{Costanzi:2013bha}
M.~Costanzi, F.~Villaescusa-Navarro, M.~Viel, J.-Q. Xia, S.~Borgani,
  E.~Castorina et~al., \emph{{Cosmology with massive neutrinos III: the halo
  mass function andan application to galaxy clusters}},
  \href{https://doi.org/10.1088/1475-7516/2013/12/012}{\emph{JCAP} {\bfseries
  1312} (2013) 012}, [\href{https://arxiv.org/abs/1311.1514}{{\ttfamily
  1311.1514}}].

\bibitem{Villaescusa-Navarro:2017mfx}
F.~Villaescusa-Navarro, A.~Banerjee, N.~Dalal, E.~Castorina, R.~Scoccimarro,
  R.~Angulo et~al., \emph{{The imprint of neutrinos on clustering in
  redshift-space}},
  \href{https://doi.org/10.3847/1538-4357/aac6bf}{\emph{Astrophys. J.}
  {\bfseries 861} (2018) 53},
  [\href{https://arxiv.org/abs/1708.01154}{{\ttfamily 1708.01154}}].

\bibitem{Lesgourgues:2011re}
J.~Lesgourgues, \emph{{The Cosmic Linear Anisotropy Solving System (CLASS) I:
  Overview}},  \href{https://arxiv.org/abs/1104.2932}{{\ttfamily 1104.2932}}.

\bibitem{Lesgourgues:2011rg}
J.~Lesgourgues, \emph{{The Cosmic Linear Anisotropy Solving System (CLASS) III:
  Comparision with CAMB for LambdaCDM}},
  \href{https://arxiv.org/abs/1104.2934}{{\ttfamily 1104.2934}}.

\bibitem{Lesgourgues:2011rh}
J.~Lesgourgues and T.~Tram, \emph{{The Cosmic Linear Anisotropy Solving System
  (CLASS) IV: efficient implementation of non-cold relics}},
  \href{https://doi.org/10.1088/1475-7516/2011/09/032}{\emph{JCAP} {\bfseries
  1109} (2011) 032}, [\href{https://arxiv.org/abs/1104.2935}{{\ttfamily
  1104.2935}}].

\bibitem{Amendola:2012ys}
{\scshape Euclid Theory Working Group} collaboration, L.~Amendola et~al.,
  \emph{{Cosmology and fundamental physics with the Euclid satellite}},
  \href{https://doi.org/10.12942/lrr-2013-6}{\emph{Living Rev. Rel.} {\bfseries
  16} (2013) 6}, [\href{https://arxiv.org/abs/1206.1225}{{\ttfamily
  1206.1225}}].

\bibitem{Amendola:2016saw}
L.~Amendola et~al., \emph{{Cosmology and fundamental physics with the Euclid
  satellite}}, \href{https://doi.org/10.1007/s41114-017-0010-3}{\emph{Living
  Rev. Rel.} {\bfseries 21} (2018) 2},
  [\href{https://arxiv.org/abs/1606.00180}{{\ttfamily 1606.00180}}].

\bibitem{Perotto:2006rj}
L.~Perotto, J.~Lesgourgues, S.~Hannestad, H.~Tu and Y.~Y.~Y. Wong,
  \emph{{Probing cosmological parameters with the CMB: Forecasts from full
  Monte Carlo simulations}},
  \href{https://doi.org/10.1088/1475-7516/2006/10/013}{\emph{JCAP} {\bfseries
  0610} (2006) 013}, [\href{https://arxiv.org/abs/astro-ph/0606227}{{\ttfamily
  astro-ph/0606227}}].

\bibitem{Vagnozzi:2018pwo}
S.~Vagnozzi, T.~Brinckmann, M.~Archidiacono, K.~Freese, M.~Gerbino,
  J.~Lesgourgues et~al., \emph{{Bias due to neutrinos must not uncorrect'd
  go}}, \href{https://doi.org/10.1088/1475-7516/2018/09/001}{\emph{JCAP}
  {\bfseries 1809} (2018) 001},
  [\href{https://arxiv.org/abs/1807.04672}{{\ttfamily 1807.04672}}].

\bibitem{Raccanelli:2017kht}
A.~Raccanelli, L.~Verde and F.~Villaescusa-Navarro, \emph{{Biases from neutrino
  bias: to worry or not to worry?}},
  \href{https://doi.org/10.1093/mnras/sty2162}{\emph{Mon. Not. Roy. Astron.
  Soc.} {\bfseries 483} (2018) 734--743},
  [\href{https://arxiv.org/abs/1704.07837}{{\ttfamily 1704.07837}}].

\bibitem{Valcin:2019fxe}
D.~Valcin, F.~Villaescusa-Navarro, L.~Verde and A.~Raccanelli, \emph{{BE-HaPPY:
  Bias Emulator for Halo Power Spectrum including massive neutrinos}},
  \href{https://arxiv.org/abs/1901.06045}{{\ttfamily 1901.06045}}.

\bibitem{Parfrey:2010uy}
K.~Parfrey, L.~Hui and R.~K. Sheth, \emph{{Scale-dependent halo bias from
  scale-dependent growth}},
  \href{https://doi.org/10.1103/PhysRevD.83.063511}{\emph{Phys. Rev.}
  {\bfseries D83} (2011) 063511},
  [\href{https://arxiv.org/abs/1012.1335}{{\ttfamily 1012.1335}}].

\bibitem{LoVerde:2014pxa}
M.~LoVerde, \emph{{Halo bias in mixed dark matter cosmologies}},
  \href{https://doi.org/10.1103/PhysRevD.90.083530}{\emph{Phys. Rev.}
  {\bfseries D90} (2014) 083530},
  [\href{https://arxiv.org/abs/1405.4855}{{\ttfamily 1405.4855}}].

\bibitem{LoVerde:2016ahu}
M.~LoVerde, \emph{{Neutrino mass without cosmic variance}},
  \href{https://doi.org/10.1103/PhysRevD.93.103526}{\emph{Phys. Rev.}
  {\bfseries D93} (2016) 103526},
  [\href{https://arxiv.org/abs/1602.08108}{{\ttfamily 1602.08108}}].

\bibitem{Munoz:2018ajr}
J.~B. Muñoz and C.~Dvorkin, \emph{{Efficient Computation of Galaxy Bias with
  Neutrinos and Other Relics}},
  \href{https://doi.org/10.1103/PhysRevD.98.043503}{\emph{Phys. Rev.}
  {\bfseries D98} (2018) 043503},
  [\href{https://arxiv.org/abs/1805.11623}{{\ttfamily 1805.11623}}].

\bibitem{Chiang:2018laa}
C.-T. Chiang, M.~LoVerde and F.~Villaescusa-Navarro, \emph{{First detection of
  scale-dependent linear halo bias in $N$-body simulations with massive
  neutrinos}},
  \href{https://doi.org/10.1103/PhysRevLett.122.041302}{\emph{Phys. Rev. Lett.}
  {\bfseries 122} (2019) 041302},
  [\href{https://arxiv.org/abs/1811.12412}{{\ttfamily 1811.12412}}].

\bibitem{Fidler:2018dcy}
C.~Fidler, N.~Sujata and M.~Archidiacono, \emph{{Relativistic bias in neutrino
  cosmologies}},  \href{https://arxiv.org/abs/1812.09266}{{\ttfamily
  1812.09266}}.

\bibitem{Hannestad:2005gj}
S.~Hannestad, \emph{{Neutrino masses and the dark energy equation of state -
  Relaxing the cosmological neutrino mass bound}},
  \href{https://doi.org/10.1103/PhysRevLett.95.221301}{\emph{Phys. Rev. Lett.}
  {\bfseries 95} (2005) 221301},
  [\href{https://arxiv.org/abs/astro-ph/0505551}{{\ttfamily
  astro-ph/0505551}}].

\bibitem{Joudaki:2012fx}
S.~Joudaki, \emph{{Constraints on Neutrino Mass and Light Degrees of Freedom in
  Extended Cosmological Parameter Spaces}},
  \href{https://doi.org/10.1103/PhysRevD.87.083523}{\emph{Phys. Rev.}
  {\bfseries D87} (2013) 083523},
  [\href{https://arxiv.org/abs/1202.0005}{{\ttfamily 1202.0005}}].

\bibitem{Archidiacono:2013lva}
M.~Archidiacono, E.~Giusarma, A.~Melchiorri and O.~Mena, \emph{{Neutrino and
  dark radiation properties in light of recent CMB observations}},
  \href{https://doi.org/10.1103/PhysRevD.87.103519}{\emph{Phys. Rev.}
  {\bfseries D87} (2013) 103519},
  [\href{https://arxiv.org/abs/1303.0143}{{\ttfamily 1303.0143}}].

\bibitem{Zhang:2015uhk}
X.~Zhang, \emph{{Impacts of dark energy on weighing neutrinos after Planck
  2015}}, \href{https://doi.org/10.1103/PhysRevD.93.083011}{\emph{Phys. Rev.}
  {\bfseries D93} (2016) 083011},
  [\href{https://arxiv.org/abs/1511.02651}{{\ttfamily 1511.02651}}].

\bibitem{Wang:2016tsz}
S.~Wang, Y.-F. Wang, D.-M. Xia and X.~Zhang, \emph{{Impacts of dark energy on
  weighing neutrinos: mass hierarchies considered}},
  \href{https://doi.org/10.1103/PhysRevD.94.083519}{\emph{Phys. Rev.}
  {\bfseries D94} (2016) 083519},
  [\href{https://arxiv.org/abs/1608.00672}{{\ttfamily 1608.00672}}].

\bibitem{Zhao:2016ecj}
M.-M. Zhao, Y.-H. Li, J.-F. Zhang and X.~Zhang, \emph{{Constraining neutrino
  mass and extra relativistic degrees of freedom in dynamical dark energy
  models using Planck 2015 data in combination with low-redshift cosmological
  probes: basic extensions to $\Lambda$CDM cosmology}},
  \href{https://doi.org/10.1093/mnras/stx978}{\emph{Mon. Not. Roy. Astron.
  Soc.} {\bfseries 469} (2017) 1713--1724},
  [\href{https://arxiv.org/abs/1608.01219}{{\ttfamily 1608.01219}}].

\bibitem{Kumar:2016zpg}
S.~Kumar and R.~C. Nunes, \emph{{Probing the interaction between dark matter
  and dark energy in the presence of massive neutrinos}},
  \href{https://doi.org/10.1103/PhysRevD.94.123511}{\emph{Phys. Rev.}
  {\bfseries D94} (2016) 123511},
  [\href{https://arxiv.org/abs/1608.02454}{{\ttfamily 1608.02454}}].

\bibitem{Xu:2016ddc}
L.~Xu and Q.-G. Huang, \emph{{Detecting the Neutrinos Mass Hierarchy from
  Cosmological Data}},
  \href{https://doi.org/10.1007/s11433-017-9125-0}{\emph{Sci. China Phys. Mech.
  Astron.} {\bfseries 61} (2018) 039521},
  [\href{https://arxiv.org/abs/1611.05178}{{\ttfamily 1611.05178}}].

\bibitem{Guo:2017hea}
R.-Y. Guo, Y.-H. Li, J.-F. Zhang and X.~Zhang, \emph{{Weighing neutrinos in the
  scenario of vacuum energy interacting with cold dark matter: application of
  the parameterized post-Friedmann approach}},
  \href{https://doi.org/10.1088/1475-7516/2017/05/040}{\emph{JCAP} {\bfseries
  1705} (2017) 040}, [\href{https://arxiv.org/abs/1702.04189}{{\ttfamily
  1702.04189}}].

\bibitem{Zhang:2017rbg}
X.~Zhang, \emph{{Weighing neutrinos in dynamical dark energy models}},
  \href{https://doi.org/10.1007/s11433-017-9025-7}{\emph{Sci. China Phys. Mech.
  Astron.} {\bfseries 60} (2017) 060431},
  [\href{https://arxiv.org/abs/1703.00651}{{\ttfamily 1703.00651}}].

\bibitem{Li:2017iur}
E.-K. Li, H.~Zhang, M.~Du, Z.-H. Zhou and L.~Xu, \emph{{Probing the Neutrino
  Mass Hierarchy beyond $\Lambda$CDM Model}},
  \href{https://doi.org/10.1088/1475-7516/2018/08/042}{\emph{JCAP} {\bfseries
  1808} (2018) 042}, [\href{https://arxiv.org/abs/1703.01554}{{\ttfamily
  1703.01554}}].

\bibitem{Yang:2017amu}
W.~Yang, R.~C. Nunes, S.~Pan and D.~F. Mota, \emph{{Effects of neutrino mass
  hierarchies on dynamical dark energy models}},
  \href{https://doi.org/10.1103/PhysRevD.95.103522}{\emph{Phys. Rev.}
  {\bfseries D95} (2017) 103522},
  [\href{https://arxiv.org/abs/1703.02556}{{\ttfamily 1703.02556}}].

\bibitem{Lorenz:2017fgo}
C.~S. Lorenz, E.~Calabrese and D.~Alonso, \emph{{Distinguishing between
  Neutrinos and time-varying Dark Energy through Cosmic Time}},
  \href{https://doi.org/10.1103/PhysRevD.96.043510}{\emph{Phys. Rev.}
  {\bfseries D96} (2017) 043510},
  [\href{https://arxiv.org/abs/1706.00730}{{\ttfamily 1706.00730}}].

\bibitem{Sutherland:2018ghu}
W.~Sutherland, \emph{{The CMB neutrino mass/vacuum energy degeneracy: a simple
  derivation of the degeneracy slopes}},
  \href{https://doi.org/10.1093/mnras/sty687}{\emph{Mon. Not. Roy. Astron.
  Soc.} {\bfseries 477} (2018) 1913--1920},
  [\href{https://arxiv.org/abs/1803.02298}{{\ttfamily 1803.02298}}].

\bibitem{Guo:2018gyo}
R.-Y. Guo, J.-F. Zhang and X.~Zhang, \emph{{Exploring neutrino mass and mass
  hierarchy in the scenario of vacuum energy interacting with cold dark
  matter}}, \href{https://doi.org/10.1088/1674-1137/42/9/095103}{\emph{Chin.
  Phys.} {\bfseries C42} (2018) 095103},
  [\href{https://arxiv.org/abs/1803.06910}{{\ttfamily 1803.06910}}].

\bibitem{Zhao:2018fjj}
M.-M. Zhao, R.-Y. Guo, J.-F. Zhang and X.~Zhang, \emph{{Dark energy versus
  modified gravity: Impacts on measuring neutrino mass}},
  \href{https://arxiv.org/abs/1810.11658}{{\ttfamily 1810.11658}}.

\bibitem{Hagstotz:2019gsv}
S.~Hagstotz, M.~Gronke, D.~Mota and M.~Baldi, \emph{{Breaking cosmic
  degeneracies: Disentangling neutrinos and modified gravity with kinematic
  information}},  \href{https://arxiv.org/abs/1902.01868}{{\ttfamily
  1902.01868}}.

\bibitem{Feng:2019mym}
L.~Feng, H.-L. Li, J.-F. Zhang and X.~Zhang, \emph{{Exploring neutrino mass and
  mass hierarchy in interacting dark energy models}},
  \href{https://arxiv.org/abs/1903.08848}{{\ttfamily 1903.08848}}.

\bibitem{Chevallier:2000qy}
M.~Chevallier and D.~Polarski, \emph{{Accelerating universes with scaling dark
  matter}}, \href{https://doi.org/10.1142/S0218271801000822}{\emph{Int. J. Mod.
  Phys.} {\bfseries D10} (2001) 213--224},
  [\href{https://arxiv.org/abs/gr-qc/0009008}{{\ttfamily gr-qc/0009008}}].

\bibitem{Linder:2002et}
E.~V. Linder, \emph{{Exploring the expansion history of the universe}},
  \href{https://doi.org/10.1103/PhysRevLett.90.091301}{\emph{Phys. Rev. Lett.}
  {\bfseries 90} (2003) 091301},
  [\href{https://arxiv.org/abs/astro-ph/0208512}{{\ttfamily
  astro-ph/0208512}}].

\bibitem{Peebles:1987ek}
P.~J.~E. Peebles and B.~Ratra, \emph{{Cosmology with a Time Variable
  Cosmological Constant}},
  \href{https://doi.org/10.1086/185100}{\emph{Astrophys. J.} {\bfseries 325}
  (1988) L17}.

\bibitem{Caldwell:1997ii}
R.~R. Caldwell, R.~Dave and P.~J. Steinhardt, \emph{{Cosmological imprint of an
  energy component with general equation of state}},
  \href{https://doi.org/10.1103/PhysRevLett.80.1582}{\emph{Phys. Rev. Lett.}
  {\bfseries 80} (1998) 1582--1585},
  [\href{https://arxiv.org/abs/astro-ph/9708069}{{\ttfamily
  astro-ph/9708069}}].

\bibitem{Carroll:1998zi}
S.~M. Carroll, \emph{{Quintessence and the rest of the world}},
  \href{https://doi.org/10.1103/PhysRevLett.81.3067}{\emph{Phys. Rev. Lett.}
  {\bfseries 81} (1998) 3067--3070},
  [\href{https://arxiv.org/abs/astro-ph/9806099}{{\ttfamily
  astro-ph/9806099}}].

\bibitem{Zlatev:1998tr}
I.~Zlatev, L.-M. Wang and P.~J. Steinhardt, \emph{{Quintessence, cosmic
  coincidence, and the cosmological constant}},
  \href{https://doi.org/10.1103/PhysRevLett.82.896}{\emph{Phys. Rev. Lett.}
  {\bfseries 82} (1999) 896--899},
  [\href{https://arxiv.org/abs/astro-ph/9807002}{{\ttfamily
  astro-ph/9807002}}].

\bibitem{Amendola:1999er}
L.~Amendola, \emph{{Coupled quintessence}},
  \href{https://doi.org/10.1103/PhysRevD.62.043511}{\emph{Phys. Rev.}
  {\bfseries D62} (2000) 043511},
  [\href{https://arxiv.org/abs/astro-ph/9908023}{{\ttfamily
  astro-ph/9908023}}].

\bibitem{Linder:2002wx}
E.~V. Linder, \emph{{Probing dark energy with SNAP}},  in \emph{{Proceedings,
  4th International Workshop on The identification of dark matter (IDM 2002):
  York, UK, September 2-6, 2002}}, pp.~52--57, 2002,
  \href{https://arxiv.org/abs/astro-ph/0210217}{{\ttfamily astro-ph/0210217}}.

\bibitem{Linder:2006sv}
E.~V. Linder, \emph{{The paths of quintessence}},
  \href{https://doi.org/10.1103/PhysRevD.73.063010}{\emph{Phys. Rev.}
  {\bfseries D73} (2006) 063010},
  [\href{https://arxiv.org/abs/astro-ph/0601052}{{\ttfamily
  astro-ph/0601052}}].

\bibitem{Linder:2007wa}
E.~V. Linder, \emph{{The Dynamics of Quintessence, The Quintessence of
  Dynamics}}, \href{https://doi.org/10.1007/s10714-007-0550-z}{\emph{Gen. Rel.
  Grav.} {\bfseries 40} (2008) 329--356},
  [\href{https://arxiv.org/abs/0704.2064}{{\ttfamily 0704.2064}}].

\bibitem{Linder:2008pp}
E.~V. Linder, \emph{{Mapping the Cosmological Expansion}},
  \href{https://doi.org/10.1088/0034-4885/71/5/056901}{\emph{Rept. Prog. Phys.}
  {\bfseries 71} (2008) 056901},
  [\href{https://arxiv.org/abs/0801.2968}{{\ttfamily 0801.2968}}].

\bibitem{Corasaniti:2002vg}
P.~S. Corasaniti and E.~J. Copeland, \emph{{A Model independent approach to the
  dark energy equation of state}},
  \href{https://doi.org/10.1103/PhysRevD.67.063521}{\emph{Phys. Rev.}
  {\bfseries D67} (2003) 063521},
  [\href{https://arxiv.org/abs/astro-ph/0205544}{{\ttfamily
  astro-ph/0205544}}].

\bibitem{Linder:2005ne}
E.~V. Linder and D.~Huterer, \emph{{How many dark energy parameters?}},
  \href{https://doi.org/10.1103/PhysRevD.72.043509}{\emph{Phys. Rev.}
  {\bfseries D72} (2005) 043509},
  [\href{https://arxiv.org/abs/astro-ph/0505330}{{\ttfamily
  astro-ph/0505330}}].

\bibitem{Carroll:2003st}
S.~M. Carroll, M.~Hoffman and M.~Trodden, \emph{{Can the dark energy equation -
  of - state parameter w be less than -1?}},
  \href{https://doi.org/10.1103/PhysRevD.68.023509}{\emph{Phys. Rev.}
  {\bfseries D68} (2003) 023509},
  [\href{https://arxiv.org/abs/astro-ph/0301273}{{\ttfamily
  astro-ph/0301273}}].

\bibitem{Vikman:2004dc}
A.~Vikman, \emph{{Can dark energy evolve to the phantom?}},
  \href{https://doi.org/10.1103/PhysRevD.71.023515}{\emph{Phys. Rev.}
  {\bfseries D71} (2005) 023515},
  [\href{https://arxiv.org/abs/astro-ph/0407107}{{\ttfamily
  astro-ph/0407107}}].

\bibitem{Hu:2004kh}
W.~Hu, \emph{{Crossing the phantom divide: Dark energy internal degrees of
  freedom}}, \href{https://doi.org/10.1103/PhysRevD.71.047301}{\emph{Phys.
  Rev.} {\bfseries D71} (2005) 047301},
  [\href{https://arxiv.org/abs/astro-ph/0410680}{{\ttfamily
  astro-ph/0410680}}].

\bibitem{Caldwell:2005ai}
R.~R. Caldwell and M.~Doran, \emph{{Dark-energy evolution across the
  cosmological-constant boundary}},
  \href{https://doi.org/10.1103/PhysRevD.72.043527}{\emph{Phys. Rev.}
  {\bfseries D72} (2005) 043527},
  [\href{https://arxiv.org/abs/astro-ph/0501104}{{\ttfamily
  astro-ph/0501104}}].

\bibitem{Creminelli:2008wc}
P.~Creminelli, G.~D'Amico, J.~Noreña and F.~Vernizzi, \emph{{The Effective
  Theory of Quintessence: the w<-1 Side Unveiled}},
  \href{https://doi.org/10.1088/1475-7516/2009/02/018}{\emph{JCAP} {\bfseries
  0902} (2009) 018}, [\href{https://arxiv.org/abs/0811.0827}{{\ttfamily
  0811.0827}}].

\bibitem{Guo:2004fq}
Z.-K. Guo, Y.-S. Piao, X.-M. Zhang and Y.-Z. Zhang, \emph{{Cosmological
  evolution of a quintom model of dark energy}},
  \href{https://doi.org/10.1016/j.physletb.2005.01.017}{\emph{Phys. Lett.}
  {\bfseries B608} (2005) 177--182},
  [\href{https://arxiv.org/abs/astro-ph/0410654}{{\ttfamily
  astro-ph/0410654}}].

\bibitem{Saridakis:2010mf}
E.~N. Saridakis and S.~V. Sushkov, \emph{{Quintessence and phantom cosmology
  with non-minimal derivative coupling}},
  \href{https://doi.org/10.1103/PhysRevD.81.083510}{\emph{Phys. Rev.}
  {\bfseries D81} (2010) 083510},
  [\href{https://arxiv.org/abs/1002.3478}{{\ttfamily 1002.3478}}].

\bibitem{Carroll:2004hc}
S.~M. Carroll, A.~De~Felice and M.~Trodden, \emph{{Can we be tricked into
  thinking that w is less than -1?}},
  \href{https://doi.org/10.1103/PhysRevD.71.023525}{\emph{Phys. Rev.}
  {\bfseries D71} (2005) 023525},
  [\href{https://arxiv.org/abs/astro-ph/0408081}{{\ttfamily
  astro-ph/0408081}}].

\bibitem{Deffayet:2010qz}
C.~Deffayet, O.~Pujolàs, I.~Sawicki and A.~Vikman, \emph{{Imperfect Dark
  Energy from Kinetic Gravity Braiding}},
  \href{https://doi.org/10.1088/1475-7516/2010/10/026}{\emph{JCAP} {\bfseries
  1010} (2010) 026}, [\href{https://arxiv.org/abs/1008.0048}{{\ttfamily
  1008.0048}}].

\bibitem{Easson:2016klq}
D.~A. Easson and A.~Vikman, \emph{{The Phantom of the New Oscillatory
  Cosmological Phase}},  \href{https://arxiv.org/abs/1607.00996}{{\ttfamily
  1607.00996}}.

\bibitem{Gannouji:2006jm}
R.~Gannouji, D.~Polarski, A.~Ranquet and A.~A. Starobinsky,
  \emph{{Scalar-Tensor Models of Normal and Phantom Dark Energy}},
  \href{https://doi.org/10.1088/1475-7516/2006/09/016}{\emph{JCAP} {\bfseries
  0609} (2006) 016}, [\href{https://arxiv.org/abs/astro-ph/0606287}{{\ttfamily
  astro-ph/0606287}}].

\bibitem{Nesseris:2006er}
S.~Nesseris and L.~Perivolaropoulos, \emph{{Crossing the Phantom Divide:
  Theoretical Implications and Observational Status}},
  \href{https://doi.org/10.1088/1475-7516/2007/01/018}{\emph{JCAP} {\bfseries
  0701} (2007) 018}, [\href{https://arxiv.org/abs/astro-ph/0610092}{{\ttfamily
  astro-ph/0610092}}].

\bibitem{Caldwell:2003vq}
R.~R. Caldwell, M.~Kamionkowski and N.~N. Weinberg, \emph{{Phantom energy and
  cosmic doomsday}},
  \href{https://doi.org/10.1103/PhysRevLett.91.071301}{\emph{Phys. Rev. Lett.}
  {\bfseries 91} (2003) 071301},
  [\href{https://arxiv.org/abs/astro-ph/0302506}{{\ttfamily
  astro-ph/0302506}}].

\bibitem{BouhmadiLopez:2004me}
M.~Bouhmadi-López and J.~A. Jiménez~Madrid, \emph{{Escaping the big rip?}},
  \href{https://doi.org/10.1088/1475-7516/2005/05/005}{\emph{JCAP} {\bfseries
  0505} (2005) 005}, [\href{https://arxiv.org/abs/astro-ph/0404540}{{\ttfamily
  astro-ph/0404540}}].

\bibitem{Wei:2005fq}
H.~Wei and R.-G. Cai, \emph{{Cosmological evolution of hessence dark energy and
  avoidance of big rip}},
  \href{https://doi.org/10.1103/PhysRevD.72.123507}{\emph{Phys. Rev.}
  {\bfseries D72} (2005) 123507},
  [\href{https://arxiv.org/abs/astro-ph/0509328}{{\ttfamily
  astro-ph/0509328}}].

\bibitem{Zhang:2009xj}
X.~Zhang, \emph{{Heal the world: Avoiding the cosmic doomsday in the
  holographic dark energy model}},
  \href{https://doi.org/10.1016/j.physletb.2009.12.021}{\emph{Phys. Lett.}
  {\bfseries B683} (2010) 81--87},
  [\href{https://arxiv.org/abs/0909.4940}{{\ttfamily 0909.4940}}].

\bibitem{Frampton:2011sp}
P.~H. Frampton, K.~J. Ludwick and R.~J. Scherrer, \emph{{The Little Rip}},
  \href{https://doi.org/10.1103/PhysRevD.84.063003}{\emph{Phys. Rev.}
  {\bfseries D84} (2011) 063003},
  [\href{https://arxiv.org/abs/1106.4996}{{\ttfamily 1106.4996}}].

\bibitem{vonStrauss:2011mq}
M.~von Strauss, A.~Schmidt-May, J.~Enander, E.~Mörtsell and S.~F. Hassan,
  \emph{{Cosmological Solutions in Bimetric Gravity and their Observational
  Tests}}, \href{https://doi.org/10.1088/1475-7516/2012/03/042}{\emph{JCAP}
  {\bfseries 1203} (2012) 042},
  [\href{https://arxiv.org/abs/1111.1655}{{\ttfamily 1111.1655}}].

\bibitem{Astashenok:2012tv}
A.~V. Astashenok, S.~Nojiri, S.~D. Odintsov and A.~V. Yurov, \emph{{Phantom
  Cosmology without Big Rip Singularity}},
  \href{https://doi.org/10.1016/j.physletb.2012.02.039}{\emph{Phys. Lett.}
  {\bfseries B709} (2012) 396--403},
  [\href{https://arxiv.org/abs/1201.4056}{{\ttfamily 1201.4056}}].

\bibitem{Saridakis:2012jy}
E.~N. Saridakis, \emph{{Phantom crossing and quintessence limit in extended
  nonlinear massive gravity}},
  \href{https://doi.org/10.1088/0264-9381/30/7/075003}{\emph{Class. Quant.
  Grav.} {\bfseries 30} (2013) 075003},
  [\href{https://arxiv.org/abs/1207.1800}{{\ttfamily 1207.1800}}].

\bibitem{Myrzakulov:2013mja}
R.~Myrzakulov, L.~Sebastiani and S.~Zerbini, \emph{{Inhomogeneous viscous
  fluids in FRW universe}},
  \href{https://doi.org/10.3390/galaxies1020083}{\emph{Galaxies} {\bfseries 1}
  (2013) 83--95}, [\href{https://arxiv.org/abs/1307.4854}{{\ttfamily
  1307.4854}}].

\bibitem{Akrami:2015qga}
Y.~Akrami, S.~F. Hassan, F.~Könnig, A.~Schmidt-May and A.~R. Solomon,
  \emph{{Bimetric gravity is cosmologically viable}},
  \href{https://doi.org/10.1016/j.physletb.2015.06.062}{\emph{Phys. Lett.}
  {\bfseries B748} (2015) 37--44},
  [\href{https://arxiv.org/abs/1503.07521}{{\ttfamily 1503.07521}}].

\bibitem{Odintsov:2015zza}
S.~D. Odintsov and V.~K. Oikonomou, \emph{{Bouncing cosmology with future
  singularity from modified gravity}},
  \href{https://doi.org/10.1103/PhysRevD.92.024016}{\emph{Phys. Rev.}
  {\bfseries D92} (2015) 024016},
  [\href{https://arxiv.org/abs/1504.06866}{{\ttfamily 1504.06866}}].

\bibitem{Oikonomou:2015qha}
V.~K. Oikonomou, \emph{{Singular Bouncing Cosmology from Gauss-Bonnet Modified
  Gravity}}, \href{https://doi.org/10.1103/PhysRevD.92.124027}{\emph{Phys.
  Rev.} {\bfseries D92} (2015) 124027},
  [\href{https://arxiv.org/abs/1509.05827}{{\ttfamily 1509.05827}}].

\bibitem{Schmidt-May:2015vnx}
A.~Schmidt-May and M.~von Strauss, \emph{{Recent developments in bimetric
  theory}}, \href{https://doi.org/10.1088/1751-8113/49/18/183001}{\emph{J.
  Phys.} {\bfseries A49} (2016) 183001},
  [\href{https://arxiv.org/abs/1512.00021}{{\ttfamily 1512.00021}}].

\bibitem{Odintsov:2015ynk}
S.~D. Odintsov and V.~K. Oikonomou, \emph{{Big-Bounce with Finite-time
  Singularity: The $F(R)$ Gravity Description}},
  \href{https://doi.org/10.1142/S0218271817500857}{\emph{Int. J. Mod. Phys.}
  {\bfseries D26} (2017) 1750085},
  [\href{https://arxiv.org/abs/1512.04787}{{\ttfamily 1512.04787}}].

\bibitem{Mortsell:2017fog}
E.~Mörtsell, \emph{{Cosmological histories in bimetric gravity: A graphical
  approach}}, \href{https://doi.org/10.1088/1475-7516/2017/02/051}{\emph{JCAP}
  {\bfseries 1702} (2017) 051},
  [\href{https://arxiv.org/abs/1701.00710}{{\ttfamily 1701.00710}}].

\bibitem{Oikonomou:2018qsc}
V.~K. Oikonomou, \emph{{Is a Topology Change After a Big Rip Possible?}},
  \href{https://doi.org/10.1142/S0219887819500488}{\emph{Int. J. Geom. Meth.
  Mod. Phys.} {\bfseries 16} (2019) 1950048},
  [\href{https://arxiv.org/abs/1805.11945}{{\ttfamily 1805.11945}}].

\bibitem{Ross:2014qpa}
A.~J. Ross, L.~Samushia, C.~Howlett, W.~J. Percival, A.~Burden and M.~Manera,
  \emph{{The clustering of the SDSS DR7 main Galaxy sample – I. A 4 per cent
  distance measure at $z = 0.15$}},
  \href{https://doi.org/10.1093/mnras/stv154}{\emph{Mon. Not. Roy. Astron.
  Soc.} {\bfseries 449} (2015) 835--847},
  [\href{https://arxiv.org/abs/1409.3242}{{\ttfamily 1409.3242}}].

\bibitem{Vagnozzi:2018jhn}
S.~Vagnozzi, S.~Dhawan, M.~Gerbino, K.~Freese, A.~Goobar and O.~Mena,
  \emph{{Constraints on the sum of the neutrino masses in dynamical dark energy
  models with $w(z) \geq -1$ are tighter than those obtained in $\Lambda$CDM}},
  \href{https://doi.org/10.1103/PhysRevD.98.083501}{\emph{Phys. Rev.}
  {\bfseries D98} (2018) 083501},
  [\href{https://arxiv.org/abs/1801.08553}{{\ttfamily 1801.08553}}].

\bibitem{Itow:2001ee}
{\scshape T2K} collaboration, Y.~Itow et~al., \emph{{The JHF-Kamioka neutrino
  project}},  in \emph{{Neutrino oscillations and their origin. Proceedings,
  3rd International Workshop, NOON 2001, Kashiwa, Tokyo, Japan, December 508,
  2001}}, pp.~239--248, 2001,
  \href{https://arxiv.org/abs/hep-ex/0106019}{{\ttfamily hep-ex/0106019}}.

\bibitem{Abe:2011ks}
{\scshape T2K} collaboration, K.~Abe et~al., \emph{{The T2K Experiment}},
  \href{https://doi.org/10.1016/j.nima.2011.06.067}{\emph{Nucl. Instrum. Meth.}
  {\bfseries A659} (2011) 106--135},
  [\href{https://arxiv.org/abs/1106.1238}{{\ttfamily 1106.1238}}].

\bibitem{Ayres:2004js}
{\scshape NOvA} collaboration, D.~S. Ayres et~al., \emph{{NOvA: Proposal to
  Build a 30 Kiloton Off-Axis Detector to Study $\nu_{\mu} \to \nu_e$
  Oscillations in the NuMI Beamline}},
  \href{https://arxiv.org/abs/hep-ex/0503053}{{\ttfamily hep-ex/0503053}}.

\bibitem{Ayres:2007tu}
{\scshape NOvA} collaboration, D.~S. Ayres et~al., \emph{{The NOvA Technical
  Design Report}}, .

\bibitem{Patterson:2012zs}
{\scshape NOvA} collaboration, R.~B. Patterson, \emph{{The NOvA Experiment:
  Status and Outlook}},  \href{https://arxiv.org/abs/1209.0716}{{\ttfamily
  1209.0716}}. [Nucl. Phys. Proc. Suppl.235-236,151(2013)].

\bibitem{Beacom:2004yd}
J.~F. Beacom, N.~F. Bell and S.~Dodelson, \emph{{Neutrinoless universe}},
  \href{https://doi.org/10.1103/PhysRevLett.93.121302}{\emph{Phys. Rev. Lett.}
  {\bfseries 93} (2004) 121302},
  [\href{https://arxiv.org/abs/astro-ph/0404585}{{\ttfamily
  astro-ph/0404585}}].

\bibitem{Fardon:2003eh}
R.~Fardon, A.~E. Nelson and N.~Weiner, \emph{{Dark energy from mass varying
  neutrinos}}, \href{https://doi.org/10.1088/1475-7516/2004/10/005}{\emph{JCAP}
  {\bfseries 0410} (2004) 005},
  [\href{https://arxiv.org/abs/astro-ph/0309800}{{\ttfamily
  astro-ph/0309800}}].

\bibitem{Cirelli:2005sg}
M.~Cirelli, M.~C. González-García and C.~Peña-Garay, \emph{{Mass varying
  neutrinos in the sun}},
  \href{https://doi.org/10.1016/j.nuclphysb.2005.04.034}{\emph{Nucl. Phys.}
  {\bfseries B719} (2005) 219--233},
  [\href{https://arxiv.org/abs/hep-ph/0503028}{{\ttfamily hep-ph/0503028}}].

\bibitem{Horvat:2005ua}
R.~Horvat, \emph{{Mass-varying neutrinos from a variable cosmological
  constant}}, \href{https://doi.org/10.1088/1475-7516/2006/01/015}{\emph{JCAP}
  {\bfseries 0601} (2006) 015},
  [\href{https://arxiv.org/abs/astro-ph/0505507}{{\ttfamily
  astro-ph/0505507}}].

\bibitem{Barger:2005mh}
V.~Barger, D.~Marfatia and K.~Whisnant, \emph{{Confronting mass-varying
  neutrinos with MiniBooNE}},
  \href{https://doi.org/10.1103/PhysRevD.73.013005}{\emph{Phys. Rev.}
  {\bfseries D73} (2006) 013005},
  [\href{https://arxiv.org/abs/hep-ph/0509163}{{\ttfamily hep-ph/0509163}}].

\bibitem{Brookfield:2005bz}
A.~W. Brookfield, C.~van~de Bruck, D.~F. Mota and D.~Tocchini-Valentini,
  \emph{{Cosmology of mass-varying neutrinos driven by quintessence: theory and
  observations}}, \href{https://doi.org/10.1103/PhysRevD.73.083515,
  10.1103/PhysRevD.76.049901}{\emph{Phys. Rev.} {\bfseries D73} (2006) 083515},
  [\href{https://arxiv.org/abs/astro-ph/0512367}{{\ttfamily
  astro-ph/0512367}}]. [Erratum: Phys. Rev.D76,049901(2007)].

\bibitem{Franca:2009xp}
U.~Franca, M.~Lattanzi, J.~Lesgourgues and S.~Pastor, \emph{{Model independent
  constraints on mass-varying neutrino scenarios}},
  \href{https://doi.org/10.1103/PhysRevD.80.083506}{\emph{Phys. Rev.}
  {\bfseries D80} (2009) 083506},
  [\href{https://arxiv.org/abs/0908.0534}{{\ttfamily 0908.0534}}].

\bibitem{Wetterich:2013jsa}
C.~Wetterich, \emph{{Variable gravity Universe}},
  \href{https://doi.org/10.1103/PhysRevD.89.024005}{\emph{Phys. Rev.}
  {\bfseries D89} (2014) 024005},
  [\href{https://arxiv.org/abs/1308.1019}{{\ttfamily 1308.1019}}].

\bibitem{Geng:2015haa}
C.-Q. Geng, C.-C. Lee, R.~Myrzakulov, M.~Sami and E.~N. Saridakis,
  \emph{{Observational constraints on varying neutrino-mass cosmology}},
  \href{https://doi.org/10.1088/1475-7516/2016/01/049}{\emph{JCAP} {\bfseries
  1601} (2016) 049}, [\href{https://arxiv.org/abs/1504.08141}{{\ttfamily
  1504.08141}}].

\bibitem{Lorenz:2018fzb}
C.~S. Lorenz, L.~Funcke, E.~Calabrese and S.~Hannestad, \emph{{Time-varying
  neutrino mass from a supercooled phase transition: current cosmological
  constraints and impact on the $\Omega_m$-$\sigma_8$ plane}},
  \href{https://doi.org/10.1103/PhysRevD.99.023501}{\emph{Phys. Rev.}
  {\bfseries D99} (2019) 023501},
  [\href{https://arxiv.org/abs/1811.01991}{{\ttfamily 1811.01991}}].

\bibitem{Escudero:2015yka}
M.~Escudero, O.~Mena, A.~C. Vincent, R.~J. Wilkinson and C.~Bœhm,
  \emph{{Exploring dark matter microphysics with galaxy surveys}},
  \href{https://doi.org/10.1088/1475-7516/2015/9/034,
  10.1088/1475-7516/2015/09/034}{\emph{JCAP} {\bfseries 1509} (2015) 034},
  [\href{https://arxiv.org/abs/1505.06735}{{\ttfamily 1505.06735}}].

\bibitem{Stadler:2018dsa}
J.~Stadler and C.~Bœhm, \emph{{Is it Mixed dark matter or neutrino masses?}},
  \href{https://arxiv.org/abs/1807.10034}{{\ttfamily 1807.10034}}.

\bibitem{Kreisch:2019yzn}
C.~D. Kreisch, F.-Y. Cyr-Racine and O.~Doré, \emph{{The Neutrino Puzzle:
  Anomalies, Interactions, and Cosmological Tensions}},
  \href{https://arxiv.org/abs/1902.00534}{{\ttfamily 1902.00534}}.

\bibitem{Stadler:2019dii}
J.~Stadler, C.~Bœhm and O.~Mena, \emph{{First numerical study of Neutrino-Dark
  Matter Mixed Damping}},  \href{https://arxiv.org/abs/1903.00540}{{\ttfamily
  1903.00540}}.

\bibitem{Park:2019ibn}
M.~Park, C.~D. Kreisch, J.~Dunkley, B.~Hadzhiyska and F.-Y. Cyr-Racine,
  \emph{{$\Lambda$CDM or self-interacting neutrinos? - how CMB data can tell
  the two models apart}},  \href{https://arxiv.org/abs/1904.02625}{{\ttfamily
  1904.02625}}.

\bibitem{Takada:2005si}
M.~Takada, E.~Komatsu and T.~Futamase, \emph{{Cosmology with high-redshift
  galaxy survey: neutrino mass and inflation}},
  \href{https://doi.org/10.1103/PhysRevD.73.083520}{\emph{Phys. Rev.}
  {\bfseries D73} (2006) 083520},
  [\href{https://arxiv.org/abs/astro-ph/0512374}{{\ttfamily
  astro-ph/0512374}}].

\bibitem{Carbone:2010ik}
C.~Carbone, L.~Verde, Y.~Wang and A.~Cimatti, \emph{{Neutrino constraints from
  future nearly all-sky spectroscopic galaxy surveys}},
  \href{https://doi.org/10.1088/1475-7516/2011/03/030}{\emph{JCAP} {\bfseries
  1103} (2011) 030}, [\href{https://arxiv.org/abs/1012.2868}{{\ttfamily
  1012.2868}}].

\bibitem{Oyama:2015gma}
Y.~Oyama, K.~Kohri and M.~Hazumi, \emph{{Constraints on the neutrino parameters
  by future cosmological 21 cm line and precise CMB polarization
  observations}},
  \href{https://doi.org/10.1088/1475-7516/2016/02/008}{\emph{JCAP} {\bfseries
  1602} (2016) 008}, [\href{https://arxiv.org/abs/1510.03806}{{\ttfamily
  1510.03806}}].

\bibitem{Canac:2016smv}
N.~Canac, G.~Aslanyan, K.~N. Abazajian, R.~Easther and L.~C. Price,
  \emph{{Testing for New Physics: Neutrinos and the Primordial Power
  Spectrum}}, \href{https://doi.org/10.1088/1475-7516/2016/09/022}{\emph{JCAP}
  {\bfseries 1609} (2016) 022},
  [\href{https://arxiv.org/abs/1606.03057}{{\ttfamily 1606.03057}}].

\bibitem{Dodelson:1997hr}
S.~Dodelson, W.~H. Kinney and E.~W. Kolb, \emph{{Cosmic microwave background
  measurements can discriminate among inflation models}},
  \href{https://doi.org/10.1103/PhysRevD.56.3207}{\emph{Phys. Rev.} {\bfseries
  D56} (1997) 3207--3215},
  [\href{https://arxiv.org/abs/astro-ph/9702166}{{\ttfamily
  astro-ph/9702166}}].

\bibitem{Planck:2013jfk}
{\scshape Planck} collaboration, P.~A.~R. Ade et~al., \emph{{Planck 2013
  results. XXII. Constraints on inflation}},
  \href{https://doi.org/10.1051/0004-6361/201321569}{\emph{Astron. Astrophys.}
  {\bfseries 571} (2014) A22},
  [\href{https://arxiv.org/abs/1303.5082}{{\ttfamily 1303.5082}}].

\bibitem{Martin:2013nzq}
J.~Martin, C.~Ringeval, R.~Trotta and V.~Vennin, \emph{{The Best Inflationary
  Models After Planck}},
  \href{https://doi.org/10.1088/1475-7516/2014/03/039}{\emph{JCAP} {\bfseries
  1403} (2014) 039}, [\href{https://arxiv.org/abs/1312.3529}{{\ttfamily
  1312.3529}}].

\bibitem{Ade:2015lrj}
{\scshape Planck} collaboration, P.~A.~R. Ade et~al., \emph{{Planck 2015
  results. XX. Constraints on inflation}},
  \href{https://doi.org/10.1051/0004-6361/201525898}{\emph{Astron. Astrophys.}
  {\bfseries 594} (2016) A20},
  [\href{https://arxiv.org/abs/1502.02114}{{\ttfamily 1502.02114}}].

\bibitem{Escudero:2015wba}
M.~Escudero, H.~Ramírez, L.~Boubekeur, E.~Giusarma and O.~Mena, \emph{{The
  present and future of the most favoured inflationary models after $Planck$
  2015}}, \href{https://doi.org/10.1088/1475-7516/2016/02/020}{\emph{JCAP}
  {\bfseries 1602} (2016) 020},
  [\href{https://arxiv.org/abs/1509.05419}{{\ttfamily 1509.05419}}].

\bibitem{Gerbino:2016sgw}
M.~Gerbino, K.~Freese, S.~Vagnozzi, M.~Lattanzi, O.~Mena, E.~Giusarma et~al.,
  \emph{{Impact of neutrino properties on the estimation of inflationary
  parameters from current and future observations}},
  \href{https://doi.org/10.1103/PhysRevD.95.043512}{\emph{Phys. Rev.}
  {\bfseries D95} (2017) 043512},
  [\href{https://arxiv.org/abs/1610.08830}{{\ttfamily 1610.08830}}].

\bibitem{Array:2015xqh}
{\scshape BICEP2, Keck Array} collaboration, P.~A.~R. Ade et~al.,
  \emph{{Improved Constraints on Cosmology and Foregrounds from BICEP2 and Keck
  Array Cosmic Microwave Background Data with Inclusion of 95 GHz Band}},
  \href{https://doi.org/10.1103/PhysRevLett.116.031302}{\emph{Phys. Rev. Lett.}
  {\bfseries 116} (2016) 031302},
  [\href{https://arxiv.org/abs/1510.09217}{{\ttfamily 1510.09217}}].

\bibitem{Hou:2012xq}
Z.~Hou et~al., \emph{{Constraints on Cosmology from the Cosmic Microwave
  Background Power Spectrum of the 2500 deg$^2$ SPT-SZ Survey}},
  \href{https://doi.org/10.1088/0004-637X/782/2/74}{\emph{Astrophys. J.}
  {\bfseries 782} (2014) 74},
  [\href{https://arxiv.org/abs/1212.6267}{{\ttfamily 1212.6267}}].

\bibitem{ArkaniHamed:2003wu}
N.~Arkani-Hamed, H.-C. Cheng, P.~Creminelli and L.~Randall, \emph{{Extra
  natural inflation}},
  \href{https://doi.org/10.1103/PhysRevLett.90.221302}{\emph{Phys. Rev. Lett.}
  {\bfseries 90} (2003) 221302},
  [\href{https://arxiv.org/abs/hep-th/0301218}{{\ttfamily hep-th/0301218}}].

\bibitem{Freese:2004un}
K.~Freese and W.~H. Kinney, \emph{{On: Natural inflation}},
  \href{https://doi.org/10.1103/PhysRevD.70.083512}{\emph{Phys. Rev.}
  {\bfseries D70} (2004) 083512},
  [\href{https://arxiv.org/abs/hep-ph/0404012}{{\ttfamily hep-ph/0404012}}].

\bibitem{Bouchet:2011ck}
{\scshape COrE} collaboration, F.~R. Bouchet et~al., \emph{{COrE (Cosmic
  Origins Explorer) A White Paper}},
  \href{https://arxiv.org/abs/1102.2181}{{\ttfamily 1102.2181}}.

\bibitem{Abazajian:2016yjj}
{\scshape CMB-S4} collaboration, K.~N. Abazajian et~al., \emph{{CMB-S4 Science
  Book, First Edition}},  \href{https://arxiv.org/abs/1610.02743}{{\ttfamily
  1610.02743}}.

\bibitem{Tram:2016rcw}
T.~Tram, R.~Vallance and V.~Vennin, \emph{{Inflation Model Selection meets Dark
  Radiation}}, \href{https://doi.org/10.1088/1475-7516/2017/01/046}{\emph{JCAP}
  {\bfseries 1701} (2017) 046},
  [\href{https://arxiv.org/abs/1606.09199}{{\ttfamily 1606.09199}}].

\bibitem{Barenboim:2019tux}
G.~A. Barenboim, P.~B. Denton and I.~M. Oldengott, \emph{{Constraints on
  inflation with an extended neutrino sector}},
  \href{https://doi.org/10.1103/PhysRevD.99.083515}{\emph{Phys. Rev.}
  {\bfseries D99} (2019) 083515},
  [\href{https://arxiv.org/abs/1903.02036}{{\ttfamily 1903.02036}}].

\bibitem{Melchiorri:2008gq}
A.~Melchiorri, O.~Mena, S.~Palomares-Ruiz, S.~Pascoli, A.~Slosar and M.~Sorel,
  \emph{{Sterile Neutrinos in Light of Recent Cosmological and Oscillation
  Data: A Multi-Flavor Scheme Approach}},
  \href{https://doi.org/10.1088/1475-7516/2009/01/036}{\emph{JCAP} {\bfseries
  0901} (2009) 036}, [\href{https://arxiv.org/abs/0810.5133}{{\ttfamily
  0810.5133}}].

\bibitem{Archidiacono:2012ri}
M.~Archidiacono, N.~Fornengo, C.~Giunti and A.~Melchiorri, \emph{{Testing 3+1
  and 3+2 neutrino mass models with cosmology and short baseline experiments}},
  \href{https://doi.org/10.1103/PhysRevD.86.065028}{\emph{Phys. Rev.}
  {\bfseries D86} (2012) 065028},
  [\href{https://arxiv.org/abs/1207.6515}{{\ttfamily 1207.6515}}].

\bibitem{Archidiacono:2013xxa}
M.~Archidiacono, N.~Fornengo, C.~Giunti, S.~Hannestad and A.~Melchiorri,
  \emph{{Sterile neutrinos: Cosmology versus short-baseline experiments}},
  \href{https://doi.org/10.1103/PhysRevD.87.125034}{\emph{Phys. Rev.}
  {\bfseries D87} (2013) 125034},
  [\href{https://arxiv.org/abs/1302.6720}{{\ttfamily 1302.6720}}].

\bibitem{Mirizzi:2013gnd}
A.~Mirizzi, G.~Mangano, N.~Saviano, E.~Borriello, C.~Giunti, G.~Miele et~al.,
  \emph{{The strongest bounds on active-sterile neutrino mixing after Planck
  data}}, \href{https://doi.org/10.1016/j.physletb.2013.08.015}{\emph{Phys.
  Lett.} {\bfseries B726} (2013) 8--14},
  [\href{https://arxiv.org/abs/1303.5368}{{\ttfamily 1303.5368}}].

\bibitem{Gariazzo:2013gua}
S.~Gariazzo, C.~Giunti and M.~Laveder, \emph{{Light Sterile Neutrinos in
  Cosmology and Short-Baseline Oscillation Experiments}},
  \href{https://doi.org/10.1007/JHEP11(2013)211}{\emph{JHEP} {\bfseries 11}
  (2013) 211}, [\href{https://arxiv.org/abs/1309.3192}{{\ttfamily 1309.3192}}].

\bibitem{Archidiacono:2014apa}
M.~Archidiacono, N.~Fornengo, S.~Gariazzo, C.~Giunti, S.~Hannestad and
  M.~Laveder, \emph{{Light sterile neutrinos after BICEP-2}},
  \href{https://doi.org/10.1088/1475-7516/2014/06/031}{\emph{JCAP} {\bfseries
  1406} (2014) 031}, [\href{https://arxiv.org/abs/1404.1794}{{\ttfamily
  1404.1794}}].

\bibitem{Gariazzo:2015rra}
S.~Gariazzo, C.~Giunti, M.~Laveder, Y.~F. Li and E.~M. Zavanin, \emph{{Light
  sterile neutrinos}},
  \href{https://doi.org/10.1088/0954-3899/43/3/033001}{\emph{J. Phys.}
  {\bfseries G43} (2016) 033001},
  [\href{https://arxiv.org/abs/1507.08204}{{\ttfamily 1507.08204}}].

\bibitem{Gariazzo:2016ehl}
S.~Gariazzo, \emph{{Light Sterile Neutrinos In Cosmology}},  in
  \emph{{Proceedings, 17th Lomonosov Conference on Elementary Particle Physics:
  Moscow, Russia, August 20-26, 2015}}, pp.~469--475, 2017,
  \href{https://arxiv.org/abs/1601.01475}{{\ttfamily 1601.01475}}.

\bibitem{Archidiacono:2016kkh}
M.~Archidiacono, S.~Gariazzo, C.~Giunti, S.~Hannestad, R.~Hansen, M.~Laveder
  et~al., \emph{{Pseudoscalar—sterile neutrino interactions: reconciling the
  cosmos with neutrino oscillations}},
  \href{https://doi.org/10.1088/1475-7516/2016/08/067}{\emph{JCAP} {\bfseries
  1608} (2016) 067}, [\href{https://arxiv.org/abs/1606.07673}{{\ttfamily
  1606.07673}}].

\bibitem{Bridle:2016isd}
S.~Bridle, J.~Elvin-Poole, J.~Evans, S.~Fernandez, P.~Guzowski and
  S.~Söldner-Rembold, \emph{{A Combined View of Sterile-Neutrino Constraints
  from CMB and Neutrino Oscillation Measurements}},
  \href{https://doi.org/10.1016/j.physletb.2016.11.050}{\emph{Phys. Lett.}
  {\bfseries B764} (2017) 322--327},
  [\href{https://arxiv.org/abs/1607.00032}{{\ttfamily 1607.00032}}].

\bibitem{Abazajian:2017tcc}
K.~N. Abazajian, \emph{{Sterile neutrinos in cosmology}},
  \href{https://doi.org/10.1016/j.physrep.2017.10.003}{\emph{Phys. Rept.}
  {\bfseries 711-712} (2017) 1--28},
  [\href{https://arxiv.org/abs/1705.01837}{{\ttfamily 1705.01837}}].

\bibitem{Dentler:2018sju}
M.~Dentler, A.~Hernández-Cabezudo, J.~Kopp, P.~A.~N. Machado, M.~Maltoni,
  I.~Martínez-Soler et~al., \emph{{Updated Global Analysis of Neutrino
  Oscillations in the Presence of eV-Scale Sterile Neutrinos}},
  \href{https://doi.org/10.1007/JHEP08(2018)010}{\emph{JHEP} {\bfseries 08}
  (2018) 010}, [\href{https://arxiv.org/abs/1803.10661}{{\ttfamily
  1803.10661}}].

\bibitem{Giunti:2019aiy}
C.~Giunti and T.~Lasserre, \emph{{eV-scale Sterile Neutrinos}},
  \href{https://arxiv.org/abs/1901.08330}{{\ttfamily 1901.08330}}.

\bibitem{Viel:2005qj}
M.~Viel, J.~Lesgourgues, M.~G. Haehnelt, S.~Matarrese and A.~Riotto,
  \emph{{Constraining warm dark matter candidates including sterile neutrinos
  and light gravitinos with WMAP and the Lyman-alpha forest}},
  \href{https://doi.org/10.1103/PhysRevD.71.063534}{\emph{Phys. Rev.}
  {\bfseries D71} (2005) 063534},
  [\href{https://arxiv.org/abs/astro-ph/0501562}{{\ttfamily
  astro-ph/0501562}}].

\bibitem{Viel:2006kd}
M.~Viel, J.~Lesgourgues, M.~G. Haehnelt, S.~Matarrese and A.~Riotto, \emph{{Can
  sterile neutrinos be ruled out as warm dark matter candidates?}},
  \href{https://doi.org/10.1103/PhysRevLett.97.071301}{\emph{Phys. Rev. Lett.}
  {\bfseries 97} (2006) 071301},
  [\href{https://arxiv.org/abs/astro-ph/0605706}{{\ttfamily
  astro-ph/0605706}}].

\bibitem{Boyarsky:2008xj}
A.~Boyarsky, J.~Lesgourgues, O.~Ruchayskiy and M.~Viel, \emph{{Lyman-alpha
  constraints on warm and on warm-plus-cold dark matter models}},
  \href{https://doi.org/10.1088/1475-7516/2009/05/012}{\emph{JCAP} {\bfseries
  0905} (2009) 012}, [\href{https://arxiv.org/abs/0812.0010}{{\ttfamily
  0812.0010}}].

\bibitem{Said:2013jxa}
N.~Said, C.~Baccigalupi, M.~Martinelli, A.~Melchiorri and A.~Silvestri,
  \emph{{New Constraints On The Dark Energy Equation of State}},
  \href{https://doi.org/10.1103/PhysRevD.88.043515}{\emph{Phys. Rev.}
  {\bfseries D88} (2013) 043515},
  [\href{https://arxiv.org/abs/1303.4353}{{\ttfamily 1303.4353}}].

\bibitem{DiazRivero:2019ukx}
A.~Díaz~Rivero, V.~Miranda and C.~Dvorkin, \emph{{Observable Predictions for
  Massive-Neutrino Cosmologies with Model-Independent Dark Energy}},
  \href{https://arxiv.org/abs/1903.03125}{{\ttfamily 1903.03125}}.

\bibitem{Gerardi:2019obr}
F.~Gerardi, M.~Martinelli and A.~Silvestri, \emph{{Reconstruction of the Dark
  Energy equation of state from latest data: the impact of theoretical
  priors}},  \href{https://arxiv.org/abs/1902.09423}{{\ttfamily 1902.09423}}.

\bibitem{Peirone:2017lgi}
S.~Peirone, M.~Martinelli, M.~Raveri and A.~Silvestri, \emph{{Impact of
  theoretical priors in cosmological analyses: the case of single field
  quintessence}}, \href{https://doi.org/10.1103/PhysRevD.96.063524}{\emph{Phys.
  Rev.} {\bfseries D96} (2017) 063524},
  [\href{https://arxiv.org/abs/1702.06526}{{\ttfamily 1702.06526}}].

\bibitem{Audren:2014lsa}
B.~Audren et~al., \emph{{Robustness of cosmic neutrino background detection in
  the cosmic microwave background}},
  \href{https://doi.org/10.1088/1475-7516/2015/03/036}{\emph{JCAP} {\bfseries
  1503} (2015) 036}, [\href{https://arxiv.org/abs/1412.5948}{{\ttfamily
  1412.5948}}].

\bibitem{Baracchini:2018wwj}
{\scshape PTOLEMY} collaboration, E.~Baracchini et~al., \emph{{PTOLEMY: A
  Proposal for Thermal Relic Detection of Massive Neutrinos and Directional
  Detection of MeV Dark Matter}},
  \href{https://arxiv.org/abs/1808.01892}{{\ttfamily 1808.01892}}.

\bibitem{Betti:2019ouf}
{\scshape PTOLEMY} collaboration, M.~G. Betti et~al., \emph{{Neutrino Physics
  with the PTOLEMY project}},
  \href{https://arxiv.org/abs/1902.05508}{{\ttfamily 1902.05508}}.

\bibitem{Millar:2018hkv}
A.~Millar, G.~Raffelt, L.~Stodolsky and E.~Vitagliano, \emph{{Neutrino mass
  from bremsstrahlung endpoint in coherent scattering on nuclei}},
  \href{https://doi.org/10.1103/PhysRevD.98.123006}{\emph{Phys. Rev.}
  {\bfseries D98} (2018) 123006},
  [\href{https://arxiv.org/abs/1810.06584}{{\ttfamily 1810.06584}}].

\bibitem{Michney:2006mk}
R.~J. Michney and R.~R. Caldwell, \emph{{Anisotropy of the Cosmic Neutrino
  Background}},
  \href{https://doi.org/10.1088/1475-7516/2007/01/014}{\emph{JCAP} {\bfseries
  0701} (2007) 014}, [\href{https://arxiv.org/abs/astro-ph/0608303}{{\ttfamily
  astro-ph/0608303}}].

\bibitem{Hannestad:2009xu}
S.~Hannestad and J.~Brandbyge, \emph{{The Cosmic Neutrino Background Anisotropy
  - Linear Theory}},
  \href{https://doi.org/10.1088/1475-7516/2010/03/020}{\emph{JCAP} {\bfseries
  1003} (2010) 020}, [\href{https://arxiv.org/abs/0910.4578}{{\ttfamily
  0910.4578}}].

\end{thebibliography}\endgroup

\end{document}